\documentclass[hyperref,12pt]{cernyrep}
\usepackage{hyperref}
\usepackage{ifpdf}
\usepackage{url}
\usepackage{subfigure}
\usepackage{rotating}
\usepackage{units}
\usepackage[english]{babel}
\usepackage{feynmp,xcolor}
\usepackage{cite}
\usepackage{color}
\usepackage[utf8]{inputenc}

\DeclareMathOperator{\tr}{Tr}

\hypersetup{
  colorlinks,
  citecolor=blue,
  linkcolor=red,
  urlcolor=blue}

\usepackage{tikz}
\usetikzlibrary{trees}
\usetikzlibrary{decorations.pathmorphing}
\usetikzlibrary{decorations.markings}

  \definecolor{jblue}  {RGB}{20,50,100}
  \definecolor{npurple}  {RGB} {153, 51, 204}
  \definecolor{wred}   {RGB}{217,0,56}
  \definecolor{white}   {RGB}{255,255,255}
  
  \definecolor{korange}   {RGB}{235, 80,  43}
  \definecolor{korange2}   {RGB}{245, 100,  63}
  \definecolor{kyelloworange}   {RGB}{255, 210,  110}
  \definecolor{kyelloworange2}   {RGB}{240, 170,  90}
  \definecolor{kred}   {RGB}{204,  102, 153}
  \definecolor{kpurple}   {RGB}{153,  61, 190}
  \definecolor{kpurplelight}   {RGB}{213,  161, 230}

	\tikzset{
	  photon/.style={decorate, decoration={snake}, draw=npurple,very thick},
	  boson/.style={decorate, decoration={snake}, draw=npurple,very thick},
	  electron/.style={draw=jblue,very thick, postaction={decorate},
	           decoration={markings,mark=at position .55 with {\arrow[draw=jblue]{>}}}
	  },
	  electron2/.style={draw=jblue,very thick, postaction={decorate},
	           decoration={markings,mark=at position .55 with {\arrow[draw=jblue]{<}}}
	  },
	  fermion/.style={draw=jblue,very thick, postaction={decorate},
	            decoration={markings,mark=at position .55 with {\arrow[draw=jblue]{}}}
	  },
	  gluon/.style={decorate, draw=korange,very thick, 
	    decoration={coil,amplitude=4pt, segment length=6pt}},
	  higgs/.style={draw=wred,very thick, postaction={decorate},
	           decoration={markings,mark=at position .55 with {\arrow[draw=wred]{>}}}
	  },
	  graviton/.style={draw=wred,very thick, postaction={decorate},
	           decoration={snake}
	  },
	  nothing/.style={draw=white,very thick}
	}

\addto\captionsenglish{}

\makeatletter
\def\myaddcontentsline@toc#1#2#3{%
   \addtocontents{#1}{\protect\thispagestyle{empty}}%
   \addtocontents{#1}{\protect \vspace{-.5cm}}
   \addtocontents{#1}{\protect\contentsline{#2}{#3}{}{}}
   \addtocontents{#1}{\protect \vspace{-.3cm}}}
\def\myaddcontentsline#1#2#3{%
  \expandafter\@ifundefined{addcontentsline@#1}%
  {\addtocontents{#1}{\protect \vspace{-.5cm}}
  \addtocontents{#1}{\protect\contentsline{#2}{#3}{}{}}
  \addtocontents{#1}{\protect \vspace{-.3cm}}}
  {\csname addcontentsline@#1\endcsname{#1}{#2}{#3}}{}}
\makeatother

\makeatletter
\def\mysuperaddcontentsline@toc#1#2#3{%
   \addtocontents{#1}{\protect\thispagestyle{empty}}%
   \addtocontents{#1}{\protect \vspace{.1cm}}
   \addtocontents{#1}{\protect\contentsline{#2}{#3}{\thepage}{}}
   \addtocontents{#1}{\protect \vspace{-.1cm}}}
\def\mysuperaddcontentsline#1#2#3{%
  \expandafter\@ifundefined{addcontentsline@#1}%
  {\addtocontents{#1}{\protect \vspace{.1cm}}
  \addtocontents{#1}{\protect\contentsline{#2}{#3}{\thepage}{}}
  \addtocontents{#1}{\protect \vspace{-.1cm}}}
  {\csname addcontentsline@#1\endcsname{#1}{#2}{#3}}{}}
\makeatother

\newcommand{\superpart}[1]{\clearpage \thispagestyle{empty}  \vphantom{blabla}\vspace{5cm} \centerline{\huge #1} 
\mysuperaddcontentsline{toc}{part}{\protect\numberline{} \hspace{-2cm}#1}%
\clearpage}
   
\newcommand{\AddToContent}[1]{\myaddcontentsline{toc}{chapter}{\protect  \textit{\hspace{1.5cm} \small#1}}}

\def\etmiss{E_T^{miss}}

\catcode`\@=11
\font\manfnt=manfnt
\def\Watchout{\@ifnextchar [{\W@tchout}{\W@tchout[1]}}
\def\W@tchout[#1]{{\manfnt\@tempcnta#1\relax%
  \@whilenum\@tempcnta>\z@\do{%
    \char"7F\hskip 0.3em\advance\@tempcnta\m@ne}}}
\let\foo\W@tchout
\def\dubious{\@ifnextchar[{\@dubious}{\@dubious[1]}}

\def\@dubious[#1]{%
  \setbox\@tempboxa\hbox{\@W@tchout#1}
  \@tempdima\wd\@tempboxa
  \list{}{\leftmargin\@tempdima}\item[\hbox to 0pt{\hss\@W@tchout#1}]}
\def\@W@tchout#1{\W@tchout[#1]}
\catcode`\@=12

\newcommand{\leer}[1]{}

\newcommand{\mt}[1]{\ensuremath{m_{t'}}}
\newcommand{\mb}[1]{\ensuremath{m_{b'}}}
\newcommand{\me}[1]{\ensuremath{m_{\tau'}}}
\newcommand{\mv}[1]{\ensuremath{m_{\nu'}}}

\begin{document}

\setcounter{tocdepth}{0}
\thispagestyle{empty}

\vspace{1cm}

\begin{center}
{\Large {\bf LES HOUCHES 2015: PHYSICS AT TEV COLLIDERS \\[4mm]}}
{\Large {\bf NEW PHYSICS WORKING GROUP REPORT}}
\end{center}

\vspace{0.1cm}
\begin{center}
\textbf{G.~Brooijmans}$^{1}$, 
\textbf{C.~Delaunay}$^{2}$, %
\textbf{A.~Delgado}$^{3}$,
\textbf{C.~Englert}$^{4}$,
\textbf{A.~Falkowski}$^{5}$, %
\textbf{B.~Fuks}$^{6,7}$,
\textbf{S.~Nikitenko}$^{8}$
\textbf{and}
\textbf{S.~Sekmen}$^{9}$,
\textbf{(convenors)}\\
D.~Barducci$^{2}$,
J.~Bernon$^{10}$,
A.~Bharucha$^{11}$,
J.~Brehmer$^{12}$,
I.~Brivio$^{13}$,
A.~Buckley$^{4}$,
D.~Burns$^{14}$,
G.~Cacciapaglia$^{15}$,
H.~Cai$^{15}$,
A.~Carmona$^{16}$,
A.~Carvalho$^{17}$,
G.~Chalons$^2$,
Y.~Chen$^{18}$,
R.~S.~Chivukula$^{19}$,
E.~Conte$^{20}$,
A.~Deandrea$^{15}$,
N.~De Filippis$^{21}$,
N.~Desai$^{12}$,
T.~Flacke$^{22}$,
M.~Frigerio$^{23}$,
M.~Garcia-Pepin$^{24}$,
S.~Gleyzer$^{25}$,
A.~Goudelis$^{26}$,
F.~Goertz$^{16}$,
P.~Gras$^{27}$,
S.~Henrot-Versill\'e$^{28}$,
J.L.~Hewett$^{29}$,
P.~Ittisamai$^{30}$,
A.~Katz$^{16,31}$,
J.~Kopp$^{32}$,
S.~Kraml$^{10}$,
M.E.~Krauss$^{33,34}$,
S.~Kulkarni$^{26}$,
U.~Laa$^{2,10}$,
S.~Lacroix$^{35}$,
K.~Lane$^{36}$,
D.~Majumder$^{37}$,
A.~Martin$^{3}$,
K.~Mawatari$^{10}$,
K.~Mohan$^{20}$,
D.M.~Morse$^{38}$,
K.~Mimasu$^{39}$,
M.~M\"uhlleitner$^{40}$,
M.~Nardecchia$^{41}$,
J.~M.~No$^{39}$,
R.~D.~Orlando$^{42}$,
P.~Pani$^{43}$,
M.~Papucci$^{44,45}$,
G.~Polesello$^{46}$,
C.~Pollard$^{4}$,
W.~Porod$^{33}$,
H.~B.~Prosper$^{10}$,
M.~Quir\'{o}s$^{24,47}$,
T.~Rizzo$^{29}$,
K.~Sakurai$^{48}$,
J.~Santiago$^{49}$,
V.~Sanz$^{39}$,
T.~Schmidt$^{50}$,
D.~Schmeier$^{34}$,
D.~Sengupta$^{10}$,
H.-S.~Shao$^{16}$,
E.~H.~Simmons$^{20}$,
J.~Sonneveld$^{51}$,
T.~Spieker$^{28}$,
M.~Spira$^{52}$,
J.~Tattersall$^{53}$,
G.~Unel$^{54}$,
R.~Vega-Morales$^{49}$,
W.~Waltenberger$^{26}$,
A.~Weiler$^{55}$,
T.~You$^{41,56}$,
O.~A.~Zapata$^{57}$,
D.~Zerwas$^{28}$
\end{center}

 \vspace{1cm}
\begin{center}
{\large {\bf Abstract}}\\[.2cm]
\end{center}
We present the activities of the `New Physics' working group for the `Physics at
TeV Colliders' workshop (Les Houches, France, 1--19 June, 2015). Our report
includes new physics studies connected with the Higgs boson and its properties,
direct search strategies, reinterpretation of the LHC results in the building
of viable models and new computational tool developments. Important signatures
for searches for natural new physics at the LHC and new assessments of the
interplay between direct dark matter searches and the LHC are also considered.
\vspace{1cm}
\begin{center}
{\bf Acknowledgements}\\[.2cm]
\end{center}
We would like to heartily thank all funding bodies, the organisers
(G.~B\'elanger, N.~Berger, F.~Boudjema, M.~Delmastro, S.~Gascon, P.~Gras,
D.~Guadagnoli, J.P.~Guillet, B.~Herrmann, S.~Kraml, G.~Moreau, E.~Pilon,
P.~Slavich and D.~Zerwas), the staff and all participants of the Les
Houches workshop for providing a stimulating and lively atmosphere in which to
work.

\newpage

\vspace{1cm}

\thispagestyle{empty}
\setcounter{page}{2}

\begin{center}
{\footnotesize
$^1$ Physics Department, Columbia University, New York, NY 10027, USA\\ 
$^{2}$ LAPTh, Universit\'e Savoie Mont Blanc, CNRS, B.P. 110, F-74941 Annecy-le-Vieux, France\\
$^{3}$ Physics Department, University of Notre Dame, Notre Dame, IN 46556, USA\\
$^{4}$ SUPA, School of Physics and Astronomy, University of Glasgow, Glasgow G12 8QQ, UK\\
$^{5}$ Laboratoire de Physique Th\'eorique, CNRS -- UMR 8627,  Universit\'e de Paris-Sud 11, F-91405 Orsay Cedex, France\\
$^{6}$ Sorbonne Universit\'es, UPMC Univ.~Paris 06, UMR 7589, LPTHE, F-75005 Paris, France\\
$^{7}$ CNRS, UMR 7589, LPTHE, F-75005 Paris, France\\
$^{8}$ Imperial College, London, UK\\
$^{9}$ Center for High Energy Physics, Kyungpook National University, Daegu, South Korea\\
$^{10}$ Laboratoire de Physique Subatomique et de Cosmologie, Universit\'e Grenoble-Alpes, CNRS/IN2P3,\\ 53 Avenue des Martyrs, 38026 Grenoble, France\\
$^{11}$ CNRS, Aix Marseille U., U. de Toulon, CPT, UMR 7332, F-13288, Marseille, France \\
$^{12}$ Institut f\"ur Theoretische Physik, Universit\"at Heidelberg, Germany \\
$^{13}$ Departamento de F\'isica Te\'orica and IFT, Universidad Aut\'onoma de Madrid, Madrid, Spain\\
$^{14}$ University of California, Davis, USA\\
$^{15}$ Univ.~Lyon, Universit\'e Lyon 1, CNRS/IN2P3, IPNL, F-69622, Villeurbanne, France\\
$^{16}$ Physics Department, CERN, CH-1211 Geneva 23, Switzerland\\
$^{17}$ Dipartimento di Fisica e Astronomia and INFN, Sezione di Padova, Via Marzolo 8, I-35131 Padova, Italy\\
$^{18}$ Lauritsen Laboratory for High Energy Physics, California Institute of Technology, Pasadena,~USA\\
$^{19}$ Groupe de Recherche de Physique des Hautes \'Energies (GRPHE), Universit\'e de Haute-Alsace, IUT Colmar, 34 rue du Grillenbreit BP 50568, 68008 Colmar Cedex, France\\
$^{20}$ Department of Physics and Astronomy,  Michigan State University, East Lansing, MI 48824, USA\\
$^{21}$ Dipartimento Interateneo di Fisica, Politecnico and INFN Bari, 70125 Bari, Italy \\
$^{22}$ Department of Physics, Korea University, Seoul 136-713, Korea \\
$^{23}$ Laboratoire Charles Coulomb (L2C), UMR 5221 CNRS-Universit\'e de Montpellier, F-34095 Montpellier, France \\
$^{24}$ Institut de F\'{i}sica d'Altes Energies (IFAE), The Barcelona Institute of  Science and Technology (BIST)\\ Universitat Aut$\grave{o}$noma de Barcelona, Barcelona, Spain\\
$^{25}$ Department of Physics, University of Florida, USA\\
$^{26}$ Institute of High Energy Physics, Austrian Academy of Sciences, 1050 Vienna, Austria\\
$^{27}$ DSM/IRFU, CEA/Saclay, Gif-sur-Yvette, France \\
$^{28}$ LAL, Univ. Paris-Sud, CNRS/IN2P3, Universit\'e Paris-Saclay, Orsay, France\\
$^{29}$ SLAC National Accelerator Laboratory, Menlo Park, CA 94025, USA \\
$^{30}$ Department of Physics, Faculty of Science, Chulalongkorn University, Bangkok, Thailand \\
$^{31}$ Universit\'e de Gen\`eve, Department of Theoretical Physics and Center for Astroparticle Physics (CAP), CH-1211 Geneva 4, Switzerland \\
$^{32}$ PRISMA Cluster of Excellence, 55099 Mainz, Germany, and \\
Mainz Institute for Theoretical Physics, Johannes Gutenberg-Universit\"{a}t Mainz, 55099 Mainz, Germany \\
$^{33}$ Institut f\"ur Theoretische Physik und Astrophysik, Universit\"at  W\"urzburg, 97074  W\"urzburg, Germany\\
$^{34}$ Bethe Center for Theoretical Physics \& Physikalisches Institut der Universit\"at Bonn, 
Nussallee 12, 53115 Bonn, Germany\\
$^{35}$ Laboratoire de Physique, ENS de Lyon et CNRS UMR 5672, Universit\'e de Lyon, 69364 LYON Cedex 07, France \\
$^{36}$ Department of Physics, Boston University, Boston, MA 02215, USA \\
$^{37}$ The University of Kansas, Lawrence, USA \\
$^{38}$ Department of Physics, Northeastern University, Boston MA, USA \\
$^{39}$ Department of Physics and Astronomy, University of Sussex, Brighton BN1 9QH, UK \\
$^{40}$ Institute for Theoretical Physics, Karlsruhe Institute of Technology, D--76128 Karlsruhe, Germany \\
$^{41}$ DAMTP, University of Cambridge, Wilberforce Road, Cambridge, CB3 0WA, UK\\
$^{42}$ Department of Physics, Florida State University, Tallahassee, Florida 32306, USA\\
$^{43}$ Stockholm University, Department of Physics, AlbaNova University Center, 106 91 Stockholm, Sweden \\
$^{44}$ Berkeley Center for Theoretical Physics, University of California, Berkeley, CA 94720, USA\\
$^{45}$ Theoretical Physics Group, Lawrence Berkeley National Laboratory, Berkeley, CA 94720, USA\\
$^{46}$ INFN, Sezione di Pavia, 27100 Pavia, Italy \\
$^{47}$ Instituci\'{o} Catalana de Recerca i Estudis Avan\c{c}ats (ICREA), Barcelona, Spain\\
$^{48}$ Institute for Particle Physics Phenomenology, Department of Physics, University of Durham, Science Laboratories, South Road, Durham, DH1 3LE, UK \\
$^{49}$ Departamento de F\'{i}sica Te\'{o}rica y del Cosmos and CAFPE, Universidad de Granada, Campus de Fuentenueva, E-18071 Granada, Spain\\
$^{50}$ Albert-Ludwigs-Universit\"at Freiburg, Physikalisches Institut, D--79104 Freiburg, Germany \\
$^{51}$ Institut f\"{u}r Experimentalphysik, Universit\"{a}t Hamburg, 22761 Hamburg, Germany \\
$^{52}$ Paul Scherrer Institut, CH--5232 Villigen PSI, Switzerland \\
$^{53}$ Institute for Theoretical Particle Physics and Cosmology, Sommerfeldstr.\ 16, 52074 Aachen, Germany \\
$^{54}$ Department of Physics and Astronomy, University of California Irvine, USA \\
$^{55}$ Physik Department T75, James-Franck-Strasse 1, Technical University of Munich, 85748 Garching, Germany \\
$^{56}$ Cavendish Laboratory, University of Cambridge, J.J. Thomson Avenue, Cambridge, CB3 0HE, UK\\
$^{57}$ University of Antioquia and Metropolitan Institute of Technology, Colombia
}

\end{center}

\newpage

\tableofcontentscern

\newpage

\noindent {\Large {\bf Introduction}}
\vspace{.5cm}

{\it G.~Brooijmans, C.~Delaunay, A.~Delgado, C.~Englert, A.~Falkowski, B.~Fuks,
   S.~Nikitenko and S.~Sekmen}\\[.4cm]

This document is the report of the New Physics session of the 2015 Les Houches
Workshop `Physics at TeV Colliders'. The workshop brought together theorists
and experimenters who discussed a significant number of novel ideas related to Higgs and
beyond the Standard Model physics. New computational methods
and techniques were considered, with the aim of improving the technology
available for theoretical, phenomenological and experimental new physics studies.
More precisely, one set of studies undertaken during
the workshop
concerns investigations associated with specific new physics models either
constructed from a top-down approach or built following a bottom-up path. A
second set of studies is connected to the Higgs boson discovered a few years
ago. Its properties are now measured with increasing accuracy at the
LHC, constraining the contruction of any realistic
new physics theory correspondingly. Finally,
detector fast simulator and recasting techniques are the subject of a third
series of contributions, including details on the way experimental
information could be presented, together with the introduction of new developments 
of a package dedicated to a generic way to constrain effective
field theories.

In the first section of these proceedings, in a first instance the
phenomenological properties
of models constructed on the basis of  extensions of the Standard Model
symmetries are considered. The LHC sensitivity to a four-jet signature of a sgluon particle
that arises, for instance, in non-minimal supersymmetric models, is
estimated, and the LHC Run--I constraints on Pati-Salam-inspired supersymmetric
theories where the lightest new physics particle is a sneutrino are derived. In
addition, simplified models are examined as they represent efficient handles on new physics.
Dark matter-related constructions are studied via their monojet, monohiggs
and $t\bar{t}$ plus missing energy signatures. A new approach for tagging dijet
resonances is also proposed, and an overview is given of
of the diboson excesses observed by the LHC
experiments and their potential implications. Finally, a broader classification of vector-like
quark signatures is built, since such particles appear
in many extensions of the Standard Model.

The second section of this document is dedicated to new physics studies linked
to the Higgs boson. Two contributions focus on the effective field theory approach to
study deviations from the Standard Model. They first address precision
predictions for Higgs-boson production via gluon fusion and second underline
the uncertainties related
to any possible measurement of the effective Wilson coefficients. Another study
focuses on new physics contributions to Higgs-pair production and show the
effects of new particles on the corresponding cross section and differential
distributions. This
includes a first calculation of the next-to-leading order corrections in QCD to
several channels. Complete models, such as the Two-Higgs-Doublet-Models, have
are moreover
considered from the viewpoint of the most recent LHC constraints.  Diphoton
and diboson probes of non-minimal Higgs weak isospin representations are
discussed in another contribution, and multi-dimensional matrix element
techniques have been developed in order to get new handles on Higgs
anomalous interactions.

Finally, the third section presents progress specific to software tools and
methods that are crucial for any new physics investigation. A first contribution
addresses the development of a super-fast simulation of the response of an
LHC-like detector such as
ATLAS or CMS; the second includes a first proposal on the way experimental
analyses could be released to be used in a more
efficient way by the community, and a last contribution focuses on a
platform dedicated to Higgs effective field theories and how to relate
existing constraints to effective operator basis choices.

The meeting in Les Houches has fostered a large number of discussions between
theorists and experimenters. In-depth studies could however only be completed
for several of the generated ideas on the required timescale. It is clear that
even those that could not converge to a written contribution have paid off
through the breadth of searches conducted by experimenters and the understanding of
the conditions imposed on an experiment by the theory community. We expect that
many more future results will  benefit from the discussions held at the
workshop.

\addtocontents{toc}{\protect\contentsline{part}{\protect\numberline{} \hspace{-2cm}Introduction}{6}{}}
\AddToContent{G.~Brooijmans, C.~Delaunay, A.~Delgado, C.~Englert, A.~Falkowski, B.~Fuks, S.~Nikitenko and S.~Sekmen}

\setcounter{figure}{0}
\setcounter{table}{0}
\setcounter{section}{0}
\setcounter{equation}{0}
\setcounter{footnote}{0}
\clearpage


\superpart{ New physics }

\graphicspath{{Diquarks-Final/}}

\chapter{Dijet resonance discrimination at LHC}

{\it R.S.~Chivukula, P.~Ittisamai, K.~Mohan and E.H.~Simmons}


\begin{abstract}
The LHC is actively searching for narrow dijet resonances corresponding to physics beyond the Standard Model, including colored vector bosons, scalars, and fermions. A dimensionless ``color discriminant variable'', $D_{col}$ can distinguish among various dijet resonances. Here, we summarize two  extensions of the color discriminant variable technique first presented in \cite{Chivukula:2015zma}.
\end{abstract}

\section{INTRODUCTION}

The LHC is actively searching for narrow dijet resonances corresponding to physics beyond the Standard Model, including colored vector bosons, scalars, and fermions.  Most recently, ATLAS has set a lower bound on the mass of a leptophobic $Z^{\prime}$ boson of 1.5 TeV \cite{ATLAS:2015nsi}, and CMS has set lower bounds on the masses of color-triplet scalar diquarks and colorons of 6 TeV and 5.1 TeV, respectively \cite{Khachatryan:2015dcf}.  

When the LHC discovers a new dijet resonance, it will be crucial to determine the spin, color, and other properties of the resonance in order to understand what kind of BSM context it represents. We have previously shown that a dimensionless ``color discriminant variable'', $D_{col}$ can distinguish among various dijet resonances. The variable is constructed from the dijet cross-section for the resonance ($\sigma_{jj}$), its mass ($M$), and its total decay width ($\Gamma$), observables that will be available from the dijet channel measurements of the resonance:
	\begin{equation}
		D_{col} \equiv \frac{M^3}{\Gamma} \sigma_{jj}~.
	\label{eq:dcol}
	\end{equation}
For a narrow-width resonance, the color discriminant variable is independent of the resonance's overall coupling strength. We have applied the color discriminant variable technique both to flavor universal vector resonances with  identical couplings to all quarks~\cite{Atre:2013mja} and also to more generic flavor non-universal vector resonances~\cite{Chivukula:2014npa} whose couplings to quarks vary by electric charge, chirality, or generation.  In the latter case, combining the color discriminant variable with  information from resonance decays to heavy top ($t\bar{t}$) or bottom ($b\bar{b}$) flavors still enables one to determine what type of resonance has been discovered.  We have also shown~\cite{Chivukula:2014pma} that the method can be used to separate fermionic or scalar dijet resonances from vector states.

Here, we summarize two further extensions of the color discriminant variable technique first presented in \cite{Chivukula:2015zma}. First, the theoretical calculation of the variable has been generalized to show its broader applicability and its relationship to the properties of the partons involved in production and decay of a narrow resonance.  Second, it has been demonstrated that $D_{col}$ can distinguish a color-triplet or color-sextet scalar diquark (a weak-singlet state coupling to two quarks) from weak-singlet vector dijet resonances that couple to a quark/anti-quark pair, such as a coloron (color-octet) or $Z'$ (color-singlet).

\section{THE RESONANCES}

\subsection{Scalar Diquarks}
\label{sec:diquark-intro}

The diquarks we consider are weak-singlet scalar resonances coupling to two {\it quarks}; references to the literature on these states are given in \cite{Chivukula:2015zma}.  Note that the canonical benchmark diquarks considered by CMS \cite{Harris:2011bh,Khachatryan:2015dcf} are color triplets.  In the absence of flavor symmetries \cite{Chivukula:1987py,Arnold:2009ay}, there would be strong constraints on the couplings of these particles~\cite{Ma:1998pi}. We avoid this by assigning appropriate flavor quantum-numbers \cite{Chivukula:1987py} to the specific color-triplet or color-sextet states as discussed in \cite{Arnold:2009ay}; essentially, one has a full mass-degenerate flavor multiplet of any of the particles present. 

To start, let us use the classification system from  \cite{Arnold:2009ay} to describe the possible weak-singlet, color-triplet diquarks associated with a weak doublet quark field (${\bf\rm Q_L}$) and weak singlet quark fields (${\bf\rm u_R}$ and ${\bf\rm d_R}$) of any generation.  Since the color-triplet state of two quarks will be anti-symmetric in color and in Lorentz spinor indices, it must also be anti-symmetric under the combination of flavor and SU(2) indices.  The two states most readily produced at LHC \cite{Chivukula:2015zma} are [a] 
 a weak-singlet, charge $1/3$ diquark: the $u_L d_L \equiv \omega_3$ state, and [b] a charge $1/3$ diquark: the $u_R d_R \equiv \tilde{\omega}_3$ state.
Following the notation in~\cite{Han:2009ya}, we write the interactions of these diquark states with quarks as
 \begin{equation}
 {\cal L} = 2\sqrt{2}\left(\bar{K}_3\right)^{ab}_c\left[
 \lambda_\omega \omega_3^c \bar{u}_{La} d^C_{Rb} + \lambda_{\tilde{\omega}} \tilde{\omega}_3^c \bar{u}_{Ra} d^C_{Lb}\right]+h.c.~,
 \end{equation}
where $a,b$ and $c$ are color (triplet) indices, $\bar{K}_3$ is the color Clebsch-Gordan coefficient connecting~\cite{Han:2009ya} two triplets to an anti-triplet (related to $\epsilon_{abc}$), and $\lambda_{\omega,\tilde{\omega}}$ are unknown coupling constants. As the forms of the decay width and production cross-section for these two states turn out to be identical, we will show results only for the $\omega_3$ state. 

We can similarly enumerate the possible color sextet scalar diquark states.  Because the color-sextet state is symmetric in color and anti-symmetric in Lorentz spinor indices, it must be {\it symmetric} under the combination of flavor and SU(2) indices.  Accordingly, as discussed in~\cite{Chivukula:2015zma}, the most relevant states for our analysis of LHC phenomenology are [a] a weak-singlet, charge $1/3$ diquark: $\delta_6 \equiv (u_L s_L - c_L d_L)$, [b] a charge 1/3 diquark: $\Delta_6 \equiv u_R d_R$, [c] 
a charge 4/3 diquark: $\Phi_6 \equiv u_R u_R$, and [d] a charge -2/3 diquark: $\phi_6 \equiv d_R d_R$.
We may denote \cite{Han:2009ya} the corresponding interactions between the diquarks and light-generation fermions as
\begin{equation}
{\cal L} = 2 \sqrt{2} (\bar{K}_6)^{ab}_\gamma \left[
\lambda_\delta \delta^\gamma_6 (\bar{u}_{La} s^C_{Rb}-\bar{c}_{La}d^C_{Rb})
+\lambda_\Phi \Phi^\gamma_6 \bar{u}_{Ra}u_{Lb} + \lambda_\phi \phi^\gamma_6 \bar{d}_{Ra}d_{Lb} + \lambda_\Delta \Delta^\gamma_6 \bar{u}_{Ra} d_{Lb}
\right] + h.c.~,
\end{equation}
where $a,b$ are triplet color indices and $\gamma$ is a sextet color index,
$\bar{K}_6$ is the Clebsch-Gordan coefficient connecting two $SU(3)$ triplets to a sextet, and $\lambda_{\delta, \Phi,\phi, \Delta}$ are unknown couplings.

\subsection{Vector Bosons}
\label{sec:vector-intro}

A color-octet vector boson (coloron) arises from extending the gauge group of the strong sector;  Likewise, an electrically neutral color-singlet vector boson ($Z^\prime$) often originates from extending the electroweak $U(1)$ or $SU(2)$ gauge group. Ref. \cite{Chivukula:2015zma} provides references to the many coloron and $Z^\prime$ models and analyses in the literature. While a typical $Z^\prime$ couples to leptons as well as quarks, it is possible (see, e.g.,~\cite{Harris:1999ya}) to have a ``leptophobic'' $Z^\prime$ that does not decay to charged leptons  and would appear experimentally as a dijet final state.

A coloron ($C$) or a $Z^\prime$ manifesting as a dijet resonance is produced at hadron colliders via quark-antiquark annihilation. The interaction of a $C$ with the SM quarks $q_i$ is described by
	\begin{equation}
		\mathcal{L}_C  = i g_{QCD} C_\mu^a \sum_{i=u,d,c,s,t,b}
		\bar{q}_i\gamma^\mu t^a \left( g_{C_L}^i P_L + g_{C_R}^i P_R \right) q_i , \\
	\label{eq:colcoupl}
	\end{equation}
where $t^a$ is an $\rm{SU}(3) $ generator, while $g_{C_L}^i$ and $g_{C_R}^i$ denote left and right chiral coupling strengths (relative to the strong coupling $g_{QCD}$) of the color-octet to the SM quarks. The projection operators have the form $P_{L,R} = (1 \mp \gamma_5)/2$. Similarly, the interactions of a leptophobic $Z^\prime$ with the SM quarks are given by
	\begin{equation}
		\mathcal{L}_{Z^\prime}  = i g_w  Z^\prime_\mu \sum_{i=u,d,c,s,t,b}
		\bar{q}_i \gamma^\mu\left( g_{Z^\prime_L}^i P_L + g_{Z^\prime_R}^i P_R \right) q_i,
	\label{eq:zpcoupl}
	\end{equation}
where $g_{Z^\prime_L}^i$ and $g_{Z^\prime_R}^i$ denote left and right chiral coupling strengths of the leptophobic $Z^\prime$ to the SM quarks, relative to the weak coupling $g_w = e/\sin\theta_W$.
This analysis only includes vector resonances with flavor-universal couplings to quarks.

\section{GENERALIZING THE COLOR DISCRIMINANT VARIABLE}
\label{sec:coldis}
Here, we summarize a more general formulation of $D_{col}$, drawing on Ref. \cite{Harris:2011bh} and use it to evaluate $D_{col}$ for various diquark states.

A scalar or vector resonance coupled to quarks in the standard model can be abundantly produced at a hadron collider of sufficient energy. Then it decays to a final state of simple topology: a pair of jets (including $b$-jets) or top quarks, both of which are highly energetic and clustered in the central region of the detector. In a large data sample, a resonance with a relatively small width will appear as a distinct bump over a large, but exponentially falling, multijet background. These features make the hadronic decay channels favorable for discovery.

Searches for new particles currently being conducted at the LHC are focused on resonances having a narrow width. So one can expect that if a new dijet resonance is discovered, the  dijet cross section, mass,  and width of the resonance will be measured. These three observables are exactly what is needed to construct the color discriminant variable \cite{Atre:2013mja}, as defined in (\ref{eq:dcol}) that can distinguish between resonances of differing color charges.

There is a particular formulation of the tree-level $s$-channel resonance cross section that makes the properties of the color-discriminant variable more transparent and makes $D_{col}$ easier to calculate for diverse types of resonances. Following Eq.~(44) of \cite{Harris:2011bh}, the spin- and color-averaged partonic tree-level $s$-channel cross section for the process $i + k \to R \to x + y$ is written
\begin{equation}
\hat{\sigma}_{ik\to R\to xy}(\hat{s}) = 16 \pi \cdot {\cal N} \cdot (1 + \delta_{ik}) \cdot
\frac{\Gamma(R\to ik) \cdot \Gamma(R\to xy)}
{(\hat{s}-m^2_R)^2 + m^2_R \Gamma^2_R} ~,
\end{equation}
where $(1 + \delta_{ik})$ accounts for the possibility of identical incoming partons.  The factor ${\cal N}$ is a ratio of spin and color counting factors
\begin{equation}
{\cal N} = \frac{N_{S_R}}{N_{S_i} N_{S_k}} \cdot
\frac{C_R}{C_i C_k}~,
\end{equation}
where $N_S$ and $C$ count the number of spin- and color-states for initial state partons $i$ and $k$. In the narrow-width approximation, we also have
\begin{equation}
\frac{1}
{(\hat{s}-m^2_R)^2 + m^2_R \Gamma^2_R}
\approx \frac{\pi}{m_R \Gamma_R} \delta(\hat{s} - m^2_R)~.
\end{equation}

Integrating over parton densities, and summing over incoming partons, as well as the outgoing partons that produce hadronic jets ($jj$), we then find the tree-level hadronic cross section to be
\begin{equation}
\sigma_R =
16\pi^2 \cdot {\cal N} \cdot \frac{ \Gamma_R}{m_R} \cdot
\left(\sum_{xy = jj} BR(R\to xy)\right)
\left( \sum_{ik} (1 + \delta_{ik}) BR(R\to ik) \left[\frac{1}{s} \frac{d L^{ik}}{d\tau}\right]_{\tau = \frac{m^2_R}{s}}\right)~,
\end{equation}
where the parton luminosity function $\tau d{\cal L}/d\tau$ for production of the vector resonance with mass $M_R$ via collisions of partons $i$ and $k$ at the center-of-mass energy squared $s$, is defined by
	\begin{equation}
	\tau \left[ \frac{d{\cal L}^{ik}}{d\tau}\right] \equiv 
	\frac{\tau}{1 + \delta_{ik}} \int_{\tau}^{1} \frac{dx}{x}
			\left[ f_i\left(x, \mu_F^2\right) f_k\left( \frac{\tau}{x}, \mu_F^2 \right) +
			f_{k}\left(x, \mu_F^2\right) f_i\left( \frac{\tau}{x}, \mu_F^2 \right) \right]  \,,
	\label{eq:lumi-fun}
	\end{equation}
where $f_{i}\left(x,\mu_F^2\right)$ is the parton distribution function at the factorization scale $\mu_F^2$ and we set the factorization scale equal to the resonance mass.

Hence, for the color discriminant variable we find the general expression
\begin{equation}
D_{col}=  16\pi^2 \cdot {\cal N}  \cdot
\left(\sum_{xy=jj} BR(R\to xy)\right)
\left( \sum_{ik} (1 + \delta_{ik}) BR(R\to ik) \left[\tau \frac{d L^{ik}}{d\tau}\right]_{\tau = \frac{m^2_R}{s}}\right)~,
\label{eq:dcol-expression}
\end{equation}
which illustrates the dependence of the color discriminant variable on the properties of the incoming and outgoing partons, and can easily be applied to any narrow resonance.

For example, for the classic flavor-universal coloron resonance, we note: the coloron has $C_R = 8$ and $N_{S_R} = 3$; the incoming $xy$ and outgoing $ik$ states are a light quark $q = u,d,c,s$ and its anti-quark $\bar{q}$; each incoming quark has $N_{S_i} = N_{S_k} = 2$ and $C_i = C_k = 3$; the sum over outgoing branching ratios ($Br(C \to xy)$) is 4/6; and each incoming branching ratio is ($Br(C\to ik) = 1/6$).  Then Eq.~(\ref{eq:dcol-expression}) becomes
\begin{equation}
D_{col}^C=  \frac{32\pi^2}{27}
\left( \sum_{q=u,c,d,s} \left[\tau \frac{d L^{q\bar{q}}}{d\tau}\right]_{\tau = \frac{m^2_C}{s}}\right)~,
\label{eq:dcol-expression-C-alt}
\end{equation}
which is identical to the result in \cite{Atre:2013mja}.  The expressions for $D_{col}$ for each of the other resonances discussed here are given in detail in \cite{Chivukula:2015zma}.

\section{DISTINGUISHING DIQUARKS FROM VECTOR RESONANCES}
\label{sec:result-dcols}
Let us see how well the color discriminant variable $D_{col}$  can distinguish whether a newly discovered dijet resonance is a scalar diquark, a coloron or a leptophobic $Z^{\prime}$.  We will focus on resonances with masses of $3-7\,TeV$ at the $\sqrt{s} = 14\,TeV$ LHC with integrated luminosities up to $1000\, fb^{-1}.$ The values of $D_{col}$ and other observables have been evaluated using the uncertainties discussed in \cite{Atre:2013mja,Chivukula:2014npa,Chivukula:2014pma} and the applicable region of parameter space is as in \cite{Chivukula:2015zma}.

\begin{figure}[h]
{
\includegraphics[width=0.49\textwidth, clip=true]{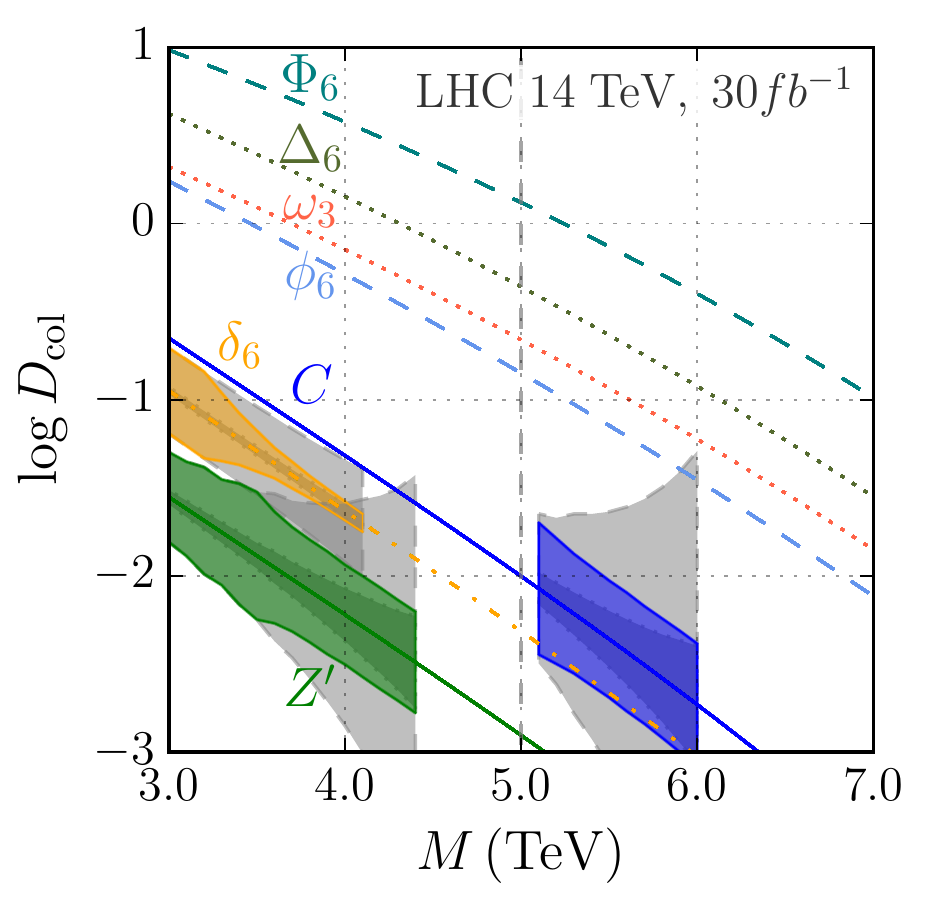}
\includegraphics[width=0.49\textwidth, clip=true]{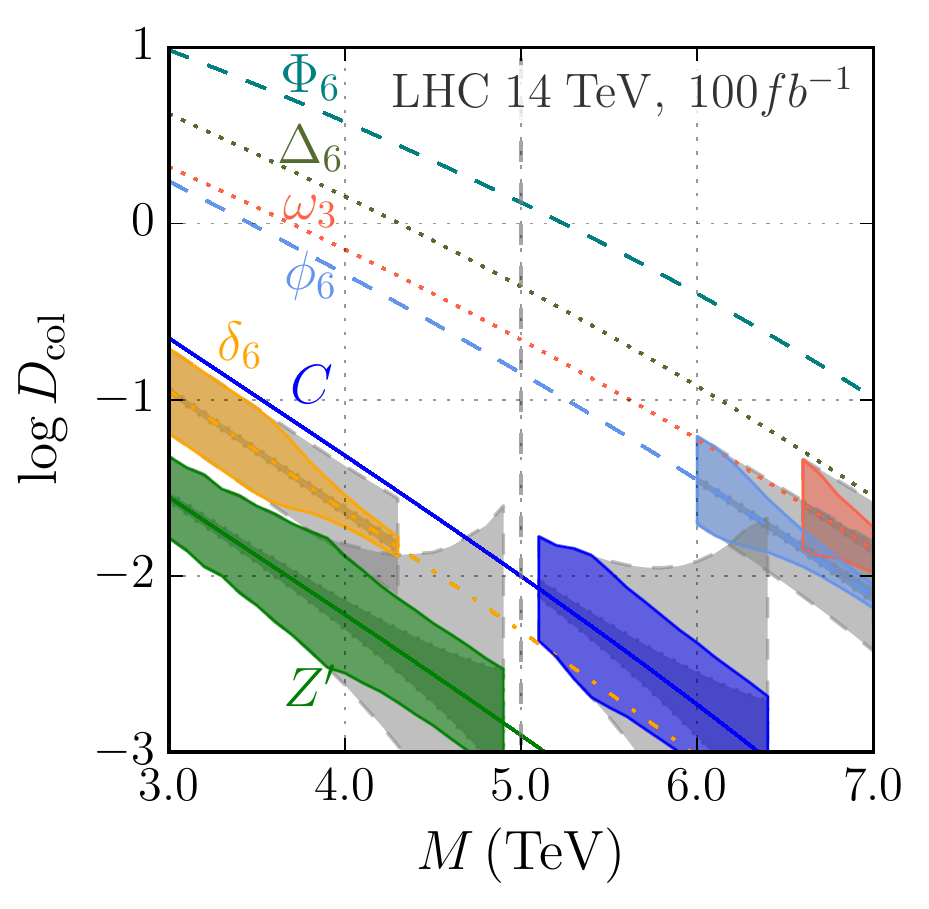}
}
\caption{\small
Color discriminant variables calculated for LHC-14 assuming the integrated luminosities $30\,~fb^{-1}$ (Left) and $100\,~fb^{-1}$ (Right).  The central value of $D_{col}$ for each particle is shown as, from top to bottom: $\Phi_6$ (dashed green), $\Delta_6$ (dotted black), $\omega_3$ (dotted red), $\phi_6$ (dashed blue), $C$ (solid blue), $\delta_6$ (dotted-dash yellow), $Z^{\prime}$ (solid green). The uncertainty in $D_{col}$ due to the uncertainties in the resonance's cross section, mass and width is indicated by gray bands. The outer (lighter gray) band shows the uncertainty in $D_{col}$ when the width is equal to the experimental mass resolution; the inner (darker gray) band shows the case where the width $\Gamma= 0.15M$. Resonances with widths between those extremes will fall between the outer and inner gray bands. (Note that no region of sensitivity is plotted for $\Phi_6$ or $\Delta_6$, see footnote 1.)}
\label{fig:pcombined-dcols-sys-stat-errors_30_100}
\end{figure}

\begin{figure}[h]
{
\includegraphics[width=0.49\textwidth, clip=true]{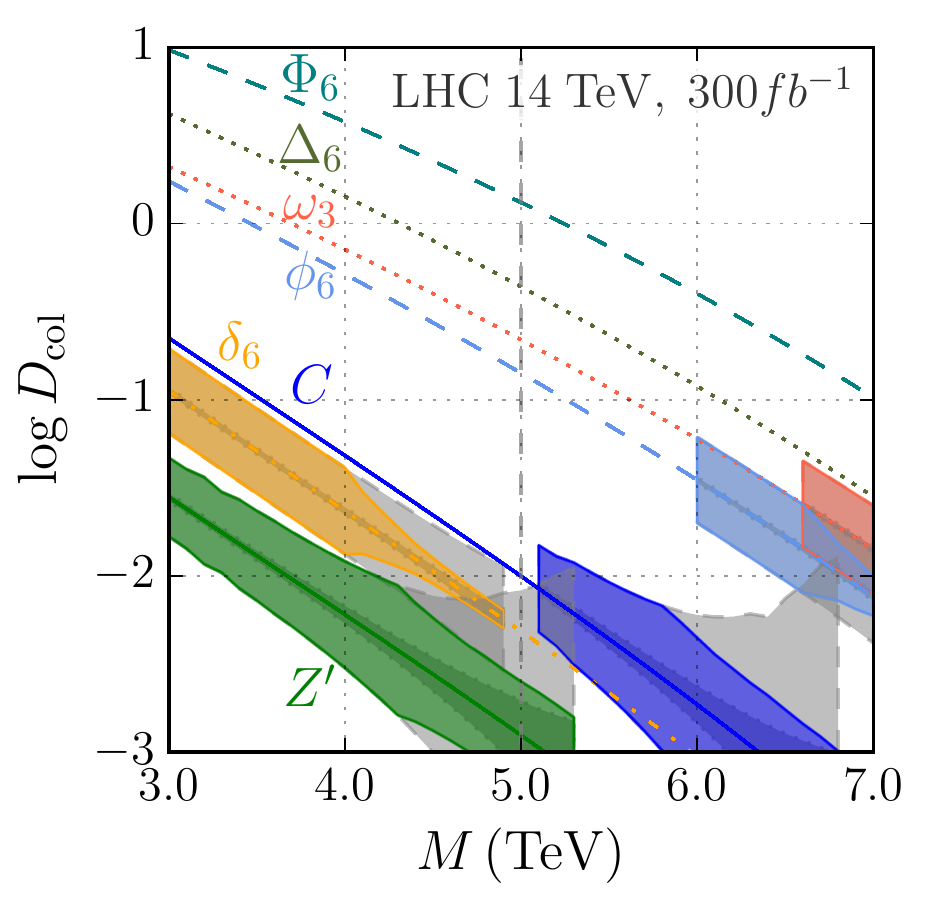}
\includegraphics[width=0.49\textwidth, clip=true]{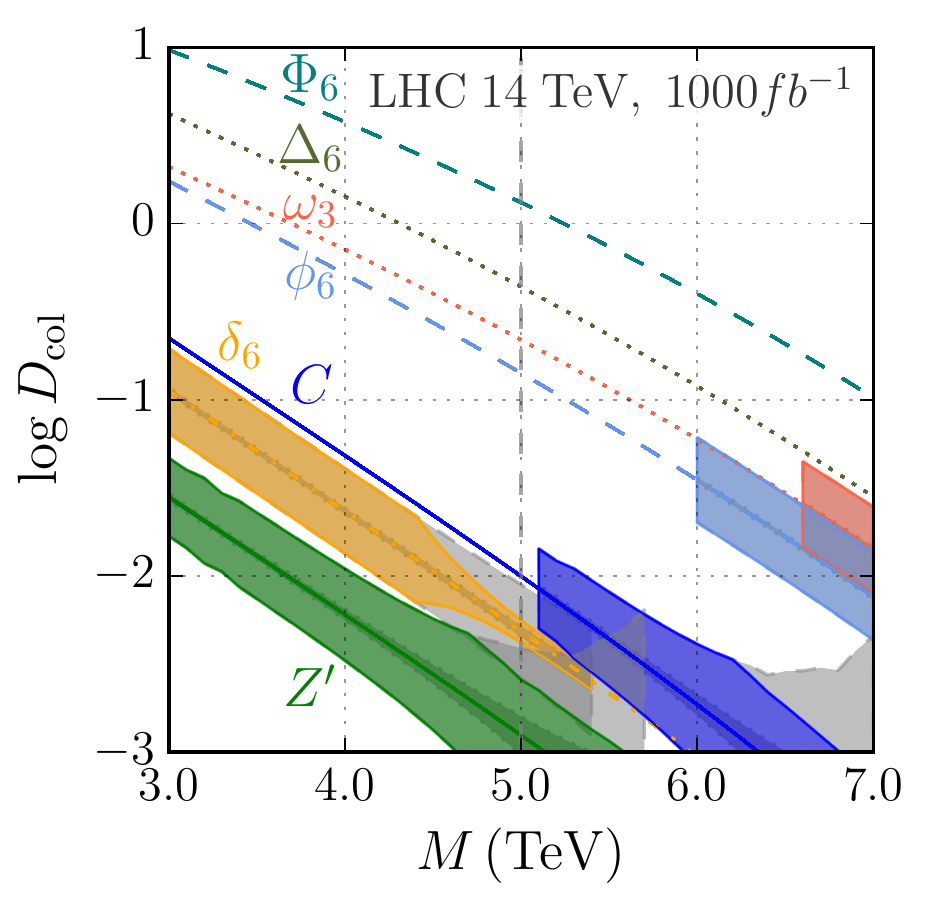}
}
\caption{\small 
Same as Fig.~\ref{fig:pcombined-dcols-sys-stat-errors_30_100}, but for integrated luminosities $300\,~fb^{-1}$ (Left) and $1000\,~fb^{-1}$ (Right).
}
\label{fig:pcombined-dcols-sys-stat-errors_300_1000}
\end{figure}

Figures~\ref{fig:pcombined-dcols-sys-stat-errors_30_100} and~\ref{fig:pcombined-dcols-sys-stat-errors_300_1000} compare the value of $D_{col}$ as a function of resonance mass for  colorons, $Z^{\prime}$ bosons, the color-triplet diquark $\omega_3$ and the color-sextet diquarks $\Phi_6$, $\phi_6$, $\Delta_6$ and $\delta_6$.  Fig.~\ref{fig:pcombined-dcols-sys-stat-errors_30_100} focuses on integrated luminosities of 30 and 100 $fb^{-1}$, while Fig.~\ref{fig:pcombined-dcols-sys-stat-errors_300_1000} displays results for 300 and 1000 $fb^{-1}$.  In each plot, a given colored band shows the mass range in which the corresponding resonance is viable and accessible. The appropriate exclusion limit from \cite{Khachatryan:2015dcf} delimits the left-hand edge of each band.\footnote{The $\Phi_6$ and $\Delta_6$ diquarks are so strongly constrained by recent data \cite{Khachatryan:2015dcf}
that they cannot be lighter than the 7 TeV shown in these figures, and the phenomenologically allowed couplings for the $\omega_3$ and $\phi_6$ (if lighter than 7 TeV) are too small to allow for a measurement of $D_{col}$ with only 30 $fb^{-1}$ at 14 TeV. All estimates are based on the current or projected sensitivities of the LHC detectors as described in \cite{Chivukula:2015zma}. If these estimates are exceeded, of course, the sensitive region is increased.}
The appropriate integrated luminosity curve from \cite{Chivukula:2015zma} delimits the right-hand edge, beyond which there will not be enough data to allow discovery at a given mass.  The width of each band relates to measurement uncertainties as noted in the figure caption.

For a given dijet resonance mass, some resonance types may already be excluded (e.g., one found below about 5.1 TeV cannot be a coloron), while others may lie beyond the LHC's discovery reach at a given integrated luminosity, because too few events would be produced (e.g., a leptophobic $Z^{\prime}$ will not be seen above about 4.8 TeV with only 100 $fb^{-1}$).  But for any resonance mass between about 3.5 and 7 TeV, there are generally several dijet resonances that remain viable candidates For example, a resonance discovered at 4.0 TeV could be a $Z^{\prime}$ or $\delta_6$, while one found at 6.1 TeV could be a $\phi_6$ or coloron.  

In many situations where a dijet resonance of a given mass discovered at LHC could correspond to more than one class of particle, measuring $D_{col}$ will distinguish among them.  A leptophobic $Z^{\prime}$ would not be confused with any of the weak-singlet scalar diquarks (except, possibly, the $\delta_6$ near the top of the mass range for a given integrated luminosity).  Nor would any of the color-sextet diquarks be mistaken for one another.  The color-triplet diquark and the coloron are, likewise, distinct by this measure.

In some cases, the value of $D_{col}$ will suffice to show that a resonance is a diquark, but not to determine which kind of diquark state has been found.  For instance, at masses of order 6.5 TeV, the $\omega_3$ overlaps the $\phi_6$. Measuring the color flow~\cite{Maltoni:2002mq,Kilian:2012pz,Curtin:2012rm,Gallicchio:2010sw} in the events may be of value here.  

In other cases, the measurement of $D_{col}$ may leave us unsure as to whether a vector boson (coloron) or a color-sextet $\delta_6$ diquark has been discovered. In this case, measuring the angular distributions of the final state jets may assist in further distinguishing the possibilities \cite{Harris:2011bh}.

\section*{CONCLUSIONS}
The LHC has the potential to discover new dijet resonances characteristic of many theories beyond the Standard Model. Because the color discriminant variable~\cite{Atre:2013mja}, $D_{col}$, is constructed from measurements available directly after the discovery of the resonance via the dijet channel, namely, its mass, its total decay width, and its dijet cross section, this variable can be valuable in identifying the nature of a newly discovered state ~\cite{Atre:2013mja,Chivukula:2014npa,Chivukula:2014pma}.
Here, we have summarized recent work \cite{Chivukula:2015zma} that extends the color discriminant variable technique in two directions. First, it places the theoretical discussion of the variable in more general language that shows its broader applicability and its relationship to the properties of the partons involved in production and decay of the resonance.  Second, it shows that $D_{col}$ may be used both to identify scalar diquark resonances as color triplet or color sextet states and to distinguish them from color-neutral or color-octet vector bosons.  We look forward to seeing this discriminant put to use in the wake of an LHC discovery.

\section*{ACKNOWLEDGEMENTS}
This material is based upon work supported by the National Science Foundation under Grant No. PHY-0854889. We wish to acknowledge the support of the Michigan State University High Performance Computing Center and the Institute for Cyber Enabled Research. 
EHS and RSC thank the Aspen Center for Physics and the NSF Grant \#1066293 for hospitality during the writing of this paper.



\AddToContent{R.S.~Chivukula, P.~Ittisamai, K.~Mohan and E.~H.~Simmons}
\renewcommand{\thesection}{\arabic{section}}

\newcommand{\be}{\begin{equation}}
\newcommand{\ee}{\end{equation}}
\newcommand{\amc}{{\sc MadGraph5\textunderscore}a{\sc MC@NLO}}
\def\bsp#1\esp{\begin{split}#1\end{split}} 
\graphicspath{{mdmonojets-proceedings1/}}


\chapter{Momentum-dependent dark matter interactions and monojets at the LHC}

{\it D.~Barducci, A.~Bharucha, N.~Desai, M.~Frigerio, B.~Fuks, A.~Goudelis, S.~Kulkarni, S.~Lacroix, G.~Polesello
and D.~Sengupta}



\begin{abstract}
We consider scenarios with momentum-dependent interactions between the dark matter and 
the Standard Model, induced by non-renormalizable derivative operators. We
derive the constraints  from the 8~TeV LHC monojet
searches and from dark matter experiments, and we
estimate the sensitivity of the  future 13~TeV LHC monojet searches.
We compare these constraints to those arising in the case where the interactions are, more conventionally, momentum-independent.  For a given dark
matter mass, we find that the LHC has greater sensitivity to smaller monojet cross-sections in the momentum-dependent case, by virtue of the harder monojet transverse-momentum distribution. 
Finally, making use of our estimate for the 13 TeV sensitivity, we briefly comment on the prospectives of distinguish the two cases in the future.
\end{abstract}

\section{INTRODUCTION}

Monojet events at the LHC \cite{Aad:2014nra,Aad:2015zva,Khachatryan:2014rra,Khachatryan:2015wza} allow one to search for new
particles that escape invisibly from the detector after having been produced in
association with one hard jet and possibly additional softer jets. Such new
states may be for instance long-lived or stable on detector, or even
cosmological, scales. In the latter case, they can further constitute (all or part of)
the dark matter (DM) energy density in the Universe. From the collider point of view, one interesting question is to what extent any future monojet-related measurements would shed light on
the underlying new physics. Of all the jet properties that could be
reconstructed, the jet transverse momentum
$p_T$ would play a key role, and the understanding of the dependence of the differential cross-section
${\rm d}\sigma/{\rm d}p_T$ on the new physics  masses and couplings is therefore
crucial. Along these lines, we attempt to compare 
derivative and non-derivative couplings of the Standard Model (SM)
to the new physics sector, 
which correspond to momentum-dependent (MD) and momentum-independent (MI) interactions, respectively.
As a consequence, the shape of the differential cross-section is different in the two cases. 

A theoretical motivation for building models involving derivative couplings is
provided by pseudo-Nambu-Goldstone bosons (pNGBs), or equivalently by light scalar fields
associated with the spontaneous breaking of a global symmetry at some scale $f$.
A relevant and concrete example is provided in composite Higgs scenarios where
the set of pNGBs includes the Higgs boson and possibly extra dark scalars. In this case
only derivative (momentum-dependent) interactions of the pNGBs suppressed by powers of $f$ are allowed by the shift symmetry related to the pNGBs.
An explicit weak breaking of the shift symmetry, parameterized by a small
coupling strength $\epsilon$, is however necessary in order to induce pNGB
masses, which additionally generates non-derivative momentum-independent couplings proportional to
$\epsilon/f$. The parameterization of our effective
Lagrangian is inspired by these scenarios, but we refrain from imposing any specific and model-dependent assumptions in connection to the new physics masses and couplings.

Most ultraviolet-complete models of dark matter predict the existence of additional
particles, many of them carrying Standard Model quantum numbers. Depending on
the specific details of each construction, dedicated searches at the
LHC could detect these additional states. In Sec.~\ref{sec:models}, we instead study a
minimal setup where the only new states accessible at the LHC are the
dark matter particle
itself and if necessary the particle mediating its interactions with the Standard Model sector.
Concretely, we focus on an invisible sector comprised of a
SM-singlet real scalar $\eta$. We impose a ${\cal Z}_2$ parity symmetry under
which the Standard Model fields are even and $\eta$ is odd. Consequently, the $\eta$ particle cannot
decay into Standard Model particles and is thus a potential dark matter candidate. We
will discuss two possibilities that allow us to couple the $\eta$ field to the Standard Model. In the
minimal scenario, the mediator is the Standard Model Higgs field $H$ that has a quartic coupling to $\eta$ at the renormalizable level,
as well as a non-renormalizable derivative coupling to $\eta$. 
However, the LHC measurements of the Higgs
boson properties  turn out to be overly constraining.
Alternatively, one needs to introduce an additional mediator $s$,
and we will more precisely consider the case where $s$ is a real gauge-singlet
scalar even under the ${\cal Z}_2$ symmetry.

The most standard LHC search channel related to those models is the
monojet one (and to a smaller extent the monophoton channel that will be ignored
here). The corresponding analysis requires a hard jet (presumably issued
from initial state radiation) recoiling against a pair of invisible particles. 
In what follows we examine in detail the constraints from the
currently published monojet search results in proton-proton collisions
at a center-of-mass energy of 8~TeV and make predictions for the
corresponding sensitivities that are expected for the 13~TeV LHC run. We
moreover compare the behaviour of MD and MI scenarios for
the new physics couplings.
Since mediator production via gluon fusion will be considered, we additionally
comment on constraints that could arise from dijet searches at past and present
hadron colliders. Finally, we entertain the possibility that the $\eta$ particle
could constitute the dark matter in the Universe, and study the related experimental constraints.

\section{MODELS AND EXPERIMENTAL CONSTRAINTS}
\label{sec:models}

\subsection{The minimal scenario: the Higgs portal}

The simplest model that predicts the production of a substantial number of monojet events is
obtained by adding to the Standard Model a gauge-singlet real scalar field $\eta$ that is odd under a ${\cal Z}_2$ symmetry, the SM fields being taken to be even.
The interactions of the $\eta$ particle with the Standard Model then arise through the
multiscalar couplings of the Higgs doublet $H$ to the $\eta$ field. This setup can be described by a 
Lagrangian of the form
\be
{\cal L}_\eta = {\cal L}_{SM} +
\frac 12 \partial_\mu\eta \partial^\mu\eta -\frac 12 \mu_\eta^2 \eta^2 -\frac 14 \lambda_\eta \eta^4 - \frac 12 \lambda \eta^2 H^\dagger H 
+\frac {1}{2f^2} (\partial_\mu \eta^2) \partial^\mu(H^\dagger H)~,
\label{eq:mdmonojets_lag}\ee
which contains a renormalizable part compatible with the ${\cal Z}_2$
symmetry $\eta\rightarrow -\eta$ and an independent dimension-six operator
that involves derivatives. Several non-derivative dimension-six operators are
additionally allowed by the symmetries of the model, but their effect, not enhanced at
large momentum transfer, is expected to be negligible in the context of monojet
searches. These operators have therefore been omitted from Eq.~\eqref{eq:mdmonojets_lag}.
The scalar field $\eta$ may arise as a pNGB in the context of composite Higgs
models and $f$ then would play the role of the pNGB decay constant. This minimal
model and the associated dark matter phenomenology, in particular the role of
the derivative operator, has been studied in Ref.~\cite{Frigerio:2012uc}.
Additional relevant analyses can also be found in Refs.~\cite{Marzocca:2014msa,%
Fonseca:2015gva,Brivio:2015kia}.

After the breaking of the electroweak symmetry, the interactions of the $\eta$
particle with the physical Higgs boson $h$ take the form
\be
{\cal L}_\eta \supset 
- \frac 14 (v+h)^2 \left( \lambda \eta^2 + \frac {1}{f^2} \partial_\mu\partial^\mu\eta^2 \right)~,
\label{eq:mdmonojets_int1}\ee
and the $\eta$-mass $m_\eta$ satisfies
\be
 m_\eta^2 = \mu_\eta^2 + \lambda v^2/2~.
\ee
While the trilinear scalar interaction of Eq.~\eqref{eq:mdmonojets_int1} induces the
production of monojet events via, for instance, gluon fusion
$gg\to g h^{(*)}\to g\eta\eta$, the quartic interactions will allow for the
production of mono-Higgs events $gg\to h^* \to  h \eta\eta$ that will not be
considered in this work. In the case where $2m_\eta < m_h$, the Higgs boson
is essentially produced on-shell so that the strength of the derivative
interaction vertex is proportional to $p_h^2/f^2 =m_h^2/f^2$. Its
momentum-dependence thus reduces to a constant so that the MD and MI cases become indistinguishable. 
In this regime, monojet searches yield weaker bounds with respect to the strongest collider constraints provided by the
Higgs invisible width results~\cite{Aad:2015pla,Khachatryan:2015vta,%
Khachatryan:2014jba},
\be
\Gamma(h\rightarrow \eta\eta)=\frac{v^2}{32\pi m_h}\left(\frac{m_h^2}{f^2} - \lambda\right)^2 \sqrt{1-\dfrac{4m_\eta^2}{m_h^2}} ~\theta(m_h^2-4m_\eta^2) 
\lesssim 0.15 \Gamma_h^{SM} \simeq  0.7{\rm~MeV} ~,
\label{eq:mdmonojets_invC}\ee
at the 95\%~ confidence level (CL).

Instead, we are interested in the complementary region where $2m_\eta > m_h$. 
Here, the monojet signal will arise from off-shell Higgs
production and the derivative interactions of the $\eta$ particle alter the momentum dependence of the
differential cross-section. The monojet $p_T$ distribution would then possibly allow
one to distinguish between the derivative and non-derivative
couplings in Eq.~\eqref{eq:mdmonojets_int1}. The price to pay is however a suppression of the
monojet signal, since the relevant partonic cross-section $\hat \sigma$
depends on the Higgs virtuality $p_h^2$ via
\be
\hat\sigma(gg\rightarrow gh^*\rightarrow g\eta\eta) \propto
\frac{\theta(p_h^2-4m_\eta^2)}{(p_h^2-m_h^2)^2+\Gamma_h^2 m_h^2} \left(\dfrac{p_h^2}{f^2} - \lambda\right)^2\sqrt{1-\dfrac{4m_\eta^2}{p_h^2}} ~,
\ee
where $\Gamma_h$ is the Higgs total width. The denominator is thus larger when
the Higgs is off-shell, or equivalently when \mbox{$p_h^2>4m_\eta^2>m_h^2$}.

A preliminary analysis of the monojet signature in this model was presented
in Ref.~\cite{lacroix}, and the collider signatures of the off-shell Higgs
portal were discussed in Ref.~\cite{Craig:2014lda}. However our numerical analysis
shows that in the off-shell region the signal is too weak to be
observed at the LHC. The LHC experiments have not only determined the
Higgs mass precisely, but also placed significant constraints on the production
cross-section and decay width of the Higgs. This means that the only free parameters of the model must
fulfill $m_\eta \gtrsim m_h/2$, $\lambda\lesssim 1$ and $f\gtrsim 500$ GeV. The
total monojet cross-section with $p_T^{\rm jet} > 20$~GeV is in this case always smaller than 1~fb for MD and 0.5~fb for MI couplings respectively.

\subsection{A pragmatic scenario with a scalar singlet mediator}\label{sec:mdmonojets_ourmodel}

We extend the model introduced in the previous section by considering a scenario where, 
in addition to the dark, stable ({\textit i.e.}~ ${\cal{Z}}_2$ odd) $\eta$ particle, another mediator 
links the SM to the dark sector: a ${\cal{Z}}_2$-even scalar
singlet $s$. We assume as usual that the scalar potential does not break
the ${\cal{Z}}_2$ symmetry spontaneously, that is, $\eta$ does not acquire a
non-vanishing vacuum expectation value (vev). With no loss of generality, we
also impose that the vev of the $s$ field vanishes, as the latter could always
be absorbed in a redefinition of the couplings. The relevant Lagrangian reads
\be\bsp
{\cal L}_{\eta,s} = &\ {\cal L}_{\rm SM} +
\frac{1}{2}\partial_\mu\eta \partial^\mu\eta - \frac{1}{2} m_\eta^2 \eta\eta
+ \frac{1}{2} \partial_\mu s \partial^\mu s - \frac{1}{2} m_s^2 ss \\
& \ + \frac{c_{s\eta}f}{2}  s \eta\eta + \frac{c_{\partial s\eta}}{f} (\partial_\mu s) (\partial^\mu \eta)\eta + \frac{\alpha_s}{16\pi} \frac{c_{sg}}{f} sG^a_{\mu\nu}G^{a\mu\nu}~,
\label{eq:mdmonojets_thelagrangian}
\esp\ee
where we have included an effective coupling of $s$ to gluons, that allows 
it to be produced at the LHC via gluon fusion and leave a monojet signal in a
detector
via $gg\rightarrow gs^* \to g \eta\eta$. A similar model, but with a fermionic
dark matter candidate and focusing on the mono-Higgs signature, has been
considered in Ref.~\cite{Carpenter:2013xra}. The $c_{sg}$ coupling could be
induced
by the presence of extra particles in the new physics sector. For instance, in
an ultraviolet-complete model featuring a vector-like color-triplet fermion
$\psi$ of mass $M_\psi \gg m_s$ and a Yukawa coupling $y_\psi \bar{\psi}\psi s$,
$c_{sg} = (4/3)(y_\psi f/M_\psi)$ would be generated by a triangle loop diagram. The Lagrangian in
Eq.~\eqref{eq:mdmonojets_thelagrangian}, only contains those interactions that are
relevant for our analysis. The non-derivative coupling $c_{s\eta}$ determines
the strength of the MI interaction between $s$ and
$\eta$, while the derivative coupling $c_{\partial s\eta}$ describes the leading
MD interactions. The associated operator is moreover the
unique independent dimension-five operator containing derivatives that couples $s$ to
$\eta$.

This simple setup is described by six parameters,
\begin{equation}
  m_{s},\quad
  m_{\eta},\quad
  f, \quad
  c_{s\eta},\quad
  c_{\partial s\eta}\quad\text{and}\quad
  c_{sg} .
\end{equation}
Strictly speaking, there are only 5 independent parameters as one can fix,
\textit{e.g.}, $c_{\partial s\eta}=1$ and determine the strength of the MD
interaction by varying $f$ only. In models where $s$, $\eta$ and the Higgs are
pNGBs associated with a spontaneous symmetry breaking at a scale $f$, one indeed
expects $c_{\partial s\eta}$ to be of order one. The value of the $f$ parameter is
however constrained by other sectors of the theory, and more precisely by
precision Higgs and electroweak measurements that roughly impose
$f\gtrsim 500$~GeV 
(see, \textit{e.g.}, Ref.~\cite{Panico:2015jxa} and references therein).

The model described by the Lagrangian of Eq.~\eqref{eq:mdmonojets_thelagrangian} is subject
to constraints arising from several sources. In particular, collider searches
for dijet resonances could play a role when the mediator is singly produced by
gluon fusion and then decays back into a pair of jets ($g g\to s^{(*)} \to gg$).
Moreover, if $\eta$ constitutes a viable dark matter candidate, it must yield a
relic density in agreement with the dark matter abundance measurements and its properties
must agree with bounds stemming from direct dark matter detection. Before
investigating those constraints in details, we perform a quick study of the
$s$ mediator properties.

Ignoring additional potential couplings of the mediator $s$ to other SM or new physics particles, the Lagrangian of Eq.~\eqref{eq:mdmonojets_thelagrangian} predicts that
the partial decay widths associated with all the decay modes of the $s$ particle
are
\begin{align}
\Gamma(s \rightarrow gg) & = \frac{\alpha_s^2 c_{sg}^2 m_s^3}{128 \pi^3 f^2}~, \\
\Gamma(s \rightarrow \eta \eta) 
& = \frac{f^2}{32\pi m_s} \left( c_{\partial s\eta} \frac{m_s^2}{f^2} + c_{s\eta} \right)^2  \sqrt{1 - \frac{4 m_\eta^2}{m_s^2}}\ \theta(m_s^2-4m_\eta^2)~,
\end{align}
these results having been verified with the decay module of
{\sc FeynRules}~\cite{Alwall:2014bza,Alloul:2013bka}.
For the choices of couplings which we will adopt in our analysis, we find that the total width $\Gamma_s$ is always
relatively small, which implies that we can safely work within the narrow width
approximation. Throughout the subsequent analysis we will consider four
representative values of $m_s$ that we fix to 50, 250, 500 and 750~GeV. These
choices allow us to cover a wide range of mediator masses, whereas the last
value is motivated by the tantalizing hints of an excess in the
diphoton invariant mass distribution  observed in LHC data at a
center-of-mass energy of 13~TeV~\cite{ATLAS-CONF-2015-081,CMS:2015dxe}.

Coming to the bounds on dijets, 
the $s$-resonance contribution to the dijet signal reads, in the narrow width approximation,
\be
\sigma(pp \rightarrow s \rightarrow gg) = 
  \int_0^1 {\rm d}x_1 \int_0^1{\rm d} x_2 \ f_g(x_1,m_s) f_g(x_2,m_s) 
     \frac{ \alpha_s^2 c_{sg}^2 m_s^2}{1024 \pi f^2} \ \delta(\hat s-m_s^2)
    {\rm BR}\Big(s\to gg\Big)\ ,
\ee
where $\hat{s}$ is the partonic center-of-mass energy and $f_g(x,\mu)$ denotes
the universal gluon density which depends on the longitudinal momentum fraction
$x$ of the gluon in the proton and is evaluated at a factorization scale $\mu$.
For our choices of values of $m_s$, the most stringent dijet constraints
originate from the Sp$\bar{\rm p}$S~\cite{Alitti:1993pn} and
Tevatron~\cite{Aaltonen:2008dn} collider data that provides upper limits on
the new physics cross section $\sigma$ for mediator masses of \mbox{140 -- 300~GeV}
and \mbox{200 -- 1400~GeV}, respectively. LHC Run I results further extend the
covered mediator masses up to 4.5~TeV~\cite{Khachatryan:2015sja,Aad:2014aqa}.
Our analysis has shown that after fixing $f=1000$~GeV, a coefficient
as large as $c_{sg}\simeq 100$ (that corresponds to an effective $sGG$ coupling
of about $10^{-3}$) is allowed, regardless of the other model
parameters. This value will be used as an upper limit in the rest of this study.


For dark matter direct detection, the MD interaction can be neglected, as the dark matter -- nucleus momentum transfer is tiny. 
The MI couplings in Eq.~\eqref{eq:mdmonojets_thelagrangian} give
rise to an effective interaction between $\eta$ particles and gluons which,
after integrating out the mediator $s$, is given by
\begin{equation}
\label{eq:mdmonojets_DDlaggluons}
{\cal{L}}_{\eta g} = f_G ~ \eta^2 ~ G_{\mu\nu}G^{\mu\nu}
\qquad \text{with}\qquad
f_G = \frac{\alpha_s c_{sg} c_{s\eta}}{32 \pi} \frac{1}{m_s^2} \ .
\end{equation}
The spin-independent dark matter scattering cross section $\sigma_{\rm SI}$
can then be computed as~\cite{Hisano:2010ct,Chu:2012qy}
\begin{equation}
\sigma_{\rm SI} = \frac{1}{\pi}\
    \bigg(\frac{m_\eta m_p}{m_\eta + m_p}\bigg)^2\
    \left| \frac{8\pi}{9 \alpha_s} \frac{m_p}{m_\eta} f_G f_{TG} \right|^2\ ,
\end{equation}
where the factor in brackets is the DM-nucleon reduced mass,
and
the squared matrix element depends on the gluon form factor $f_{TG}$ that
can be expressed as a function of the quark form factors
$f_{Tq}$~\cite{Shifman:1978zn},
\be
  f_{TG} = 1 - \sum_{q=u,d,s} f_{Tq}\ ,
\ee
for which we adopt the values $f_{Tu} = 0.0153$,
$f_{Td} = 0.0191$ and $f_{Ts} = 0.0447$~\cite{Belanger:2014vza}.  The value of $f_{TG}$
can however be modified
if one introduces additional $s$-couplings to the quarks.
In our model, such interactions can arise at
the non-renormalizable level only, and will be ignored in the following. In our analysis
presented below, we confront the above predictions to the latest limits
extracted from LUX data~\cite{Akerib:2015rjg}.


For the computation of the $\eta$ relic abundance, we have implemented our model
in the {\sc Mi\-crO\-me\-gas} package~\cite{Belanger:2008sj} via
{\sc FeynRules}. For the sake of completeness, we
nonetheless present approximate expressions for the total thermally-averaged
self-annihilation cross section of $\eta$ pairs. Keeping only the leading
($S$-wave) component and ignoring special kinematic configurations like
those originating from the presence of intermediate resonances, the annihilation
of the $\eta$ dark matter particle into gluon pairs is approximated by
\begin{align}
\label{eq:mdmonojets_RDgg}
\left\langle \sigma v \right\rangle_{gg} \simeq \frac{\alpha_s^2 c_{sg}^2 \left( c_{s\eta} f^2 + 4 c_{\partial s\eta} m_s^2 \right)^2}{256\pi^3 f^4 \left( m_s^2 - 4 m_\eta^2 \right)^2}~. 
\end{align}
When $m_\eta > m_s$, there is an additional $2\leftrightarrow 2$ annihilation channel, $\eta\eta \leftrightarrow s s$ for which the leading (again $S$-wave) contribution to $\left\langle \sigma v \right\rangle$ reads
\begin{align}\label{eq:mdmonojets_RDss}
\left\langle \sigma v \right\rangle_{ss} \simeq 
\frac{\sqrt{1 - \frac{m_s^2}{m_\eta^2}} \left( c_{\partial s\eta} m_s^2 + c_{s\eta} f^2 \right)^4}{16 \pi f^4 m_\eta^2 \left( m_s^2 - 2 m_\eta^2 \right)^2}\ .
\end{align}
The leading contributions to the relic density are different in 
the case that either the MI or MD couplings dominate. In the former the coupling appears in conjuction 
with $f^2$, while in the latter the coupling appears with $m_s^2$. 
We are interested in determining the regions of parameter space where the relic density does not 
exceed the measured value from Planck $\Omega h^2\rvert_\textrm{exp} = 0.1188\pm 0.0010$~\cite{Ade:2015xua}.
 As a rule of thumb, the thermal freeze-out relic density of dark matter candidates that can 
be probed at the LHC tends to be below this measured value
 (see, \textit{e.g.}, Ref.~\cite{Feng:2005gj}), but this is not without exceptions \cite{Busoni:2014gta}. 

\section{Results}\label{sec:mdmonojets_lhc}

From the Lagrangian given by Eq.~\eqref{eq:mdmonojets_thelagrangian}, the
production of dark matter at the LHC is detectable via so called missing
transverse energy ($\slashed{E}_T$) signatures, where the invisible dark matter
particle recoils against a Standard Model particle resulting in a non zero value
of the vector sum of the transverse momenta of the reconstructed physics objects
in the event. Typically, the greatest sensitivity arises from searches where the
recoiling object is a QCD jet, giving rise to a monojet signature. The
primary Standard Model backgrounds for this process consist of irreducible
$Z(\rightarrow \nu \nu) + \mathrm{jets}$ events along with smaller contributions
from $W + \mathrm{jets}$, QCD multijet, $t\bar t$, single top and diboson processes.

The goal of our study is to estimate the capacity of the LHC searches to probe
the structure of the dark matter couplings and in particular whether the momentum-(in)dependent nature of the couplings can be identified.  To this end, we perform our analysis with only one of either the MI ($c_{ s\eta}$) or the MD ($c_{\partial s\eta}$) coupling being non-zero at a time. The scale $f$ is set to  1 TeV throughout the course of the subsequent discussion while for the coupling to the gluon field strength $c_{sg}$ we adopt three distinct values, $c_{sg} =$ 10, 50 and 100, following the discussion on dijet constraints of Section~\ref{sec:mdmonojets_ourmodel}. 
\subsection{Analysis setup}\label{sec:mdmonojets_setup}

In order to quantify the monojet limits arising from 8 TeV LHC data, we use a
validated (recast) implementation of the ATLAS monojet
search~\cite{Aad:2014nra} (search for top squarks that decay via the
$\tilde t \rightarrow c \tilde \chi^1_0$ mode in the case where the stop and the
neutralino are compressed) that has been implemented in the {\sc MadAnalysis 5}
framework~\cite{Conte:2012fm, Conte:2014zja}. Details of the validation, as well
as the analysis code~\cite{atlasmonojet}, are publicly available
on the {\sc MadAnalysis 5} Public Analysis Database~\cite{Dumont:2014tja}
website,\\
\hspace*{0.5cm}\url{https://madanalysis.irmp.ucl.ac.be/wiki/PublicAnalysisDatabase}.\\
The ATLAS monojet search includes a baseline selection in which the leading jet
transverse momentum has to satisfy $p_T > 120$~GeV and the event fulfills
$\slashed{E}_{T}>150$~GeV. It makes use of three signal regions in which the
jet-$p_{T}$ and the missing transverse energy have to be above the (280, 220)~GeV,
(340, 340)~GeV and (450, 450)~GeV thresholds, respectively (the first number
being associated with the jet and the second one with the missing energy selections).

To set up our analysis, the Lagrangian of Eq.~\eqref{eq:mdmonojets_thelagrangian}
has been implemented into the {\sc FeynRules}~\cite{Alloul:2013bka} package and
imported to \amc~\cite{Alwall:2014hca} via the
UFO~\cite{Degrande:2011ua} interface of {\sc FeynRules}. We have generated events
describing the process $pp \rightarrow \eta \eta + j$ for a collision center-of-mass
energy of 8~TeV. In our event-generation process, we include a generator-level
selection of 80~GeV on the jet-$p_T$. For a proper description of the QCD
environment, we have matched the hard-scattering sample to parton showering and
simulated the hadronization process via {\sc Pythia 6}~\cite{Sjostrand:2006za}.
A fast simulation of the detector response has next been achieved by using the
{\sc MadAnalysis 5}-tuned~\cite{Dumont:2014tja} version of
{\sc Delphes 3}~\cite{deFavereau:2013fsa}. Jets are reconstructed using the
anti-$k_T$ algorithm~\cite{Cacciari:2008gp}  with a cone size of 0.4 and a
jet $p_{T}$ threshold of 20~GeV, as implemented in
{\sc Fastjet}~\cite{Cacciari:2011ma}. All parameters for the detector simulation
are finally kept the same as in the publicly available implemented analysis.
We have then computed the 95\% confidence level (CL) upper limits on the monojet
cross section (calculated with a generator-level selection on the jet
transverse-momentum of $p_T> 80$~GeV) using the dedicated {\sc MadAnalysis 5}
method. The latter is based on the CLs technique~\cite{Read:2000ru,Read:2002hq}
and determines, given the background rate, its uncertainty, the observed number
of events and the signal selection acceptance for each signal region (SR), the
upper limit on the cross section for the most sensitive of the SRs defined in
the analysis. This has been performed for all mediator masses defined in
Section~\ref{sec:mdmonojets_ourmodel} and for various dark matter mass
combinations. These upper limits however only depend on the kinematics of the
events, and not on the overall event rate, and are independent of the actual
values of the MI and MD coefficients (under the assumption that just one of the
two is non-vanishing at a time) so that they depend only on $m_\eta$.
We have consequently fixed ($c_{s\eta}$,$c_{\partial s \eta}$) to nominal values
of (1,0) and (0,1) for the MI and MD cases respectively. This choice moreover
allows for an easy rescaling of the monojet cross section when computing
the rates for different values of these coefficients, which will in turn allow us
to set limits on the MI and MD coefficients for a given mass combination.

In order to project the monojet search sensitivity for the expected luminosity
of the 13 TeV LHC run, we have generated hard-scattering events describing the dominant
background contribution related to $Z(\rightarrow \nu \nu)$ plus jets with
\amc, and
merged matrix elements containing at least one and up to three additional jets.
To this aim, parton showering has been performed with
{\sc Pythia~8}~\cite{Sjostrand:2014zea} that has also taken care of the merging
procedure following the CKKW-L prescription~\cite{Lonnblad:2011xx}. The merged event
sample has then been normalized to the NLO \mbox{$Z+1$}~jet result, as returned
by \amc. Signal generation has been performed in the same
way as for the 8~TeV study, except for parton showering and hadronization for
which we have made use of {\sc Pythia 8}. Jets are then reconstructed using the
anti-$k_T$ jet algorithm as implemented in {\sc FastJet}, with a $p_{T}$
threshold of 20~GeV and a cone size of 0.4. The missing momentum is built
as the vector sum of the transverse momenta of all the non-interacting
particles. Detector effects have finally been simulated by applying efficiency
factors and smearing functions to the momenta of the physics objects, so that the
performance of the ATLAS detector during the first run of the LHC are
mimicked~\cite{Aad:2009wy}.
Our selection demands the events to feature at least one reconstructed central
jet with a transverse momentum $p_T>250$~GeV and a pseudorapidity $|\eta|<2$ and
not to contain any lepton. Additionally, no jet should point along the missing
momentum four-vector. The signal regions have been defined by applying
progressive thresholds for a $\slashed{E}_T$ selection, using steps of 100~GeV
between 400~GeV and 1.7~TeV. For each signal region $i$, the number of $Z$+jets
events $N_Z^i$ has been calculated and the total number of background events has
been assumed to be $1.5 N_Z^i$ in order to account for the non-simulated
background, as suggested by Ref.~\cite{Aad:2015zva} where the total number of
background events  is estimated to be within $(1.3-1.6)\times N_Z^i$.

The relevant quantity for the sensitivity evaluation is the uncertainty on the
background estimate, a quantity that depends on the techniques used by the
experiments. In the absence of any available 13~TeV experimental monojet
analysis, it can only be roughly approximated on the basis of published 8~TeV
data. Concerning the $Z(\rightarrow\nu\nu)$+jets contribution, this is performed
in Ref.~\cite{Aad:2015zva} using a straightforward extrapolation
from events where leptonic decays of vector bosons are identified.
The evolution of $\sigma(N_Z^i)$ with $N_Z^i$ can be reasonably
approximated  by the formula
\be
\sigma(N_Z^i)^2=(k_1\times\sqrt{N_Z^i})^2+(k_2\times N_Z^i)^2\ ,
\ee
where the first term represents the statistical error on
the control regions used for the background estimate,
and where the second term is the systematic uncertainty of the
extrapolation from the control regions to the signal regions.
The latter varies, as shown in Ref.~\cite{Aad:2015zva}, between 3.5 and 5\%
and increases with the $\slashed{E}_T$ selection threshold. For the
present study the values
$k_1=1.8$ and $k_2=0.05$ are adopted as a conservative choice
which reasonably approximates the uncertainties given in Ref.~\cite{Aad:2015zva},
and the error is scaled up by a factor 1.2 to take into
account contributions from the minor background components which
have not been not simulated.

The 95\% CL upper limit on the signal cross-section for each signal region has been
calculated on the basis of a Poisson modeling with Gaussian constraints, using
the CLs prescription and the asymptotic calculator
implemented in the {\sc RooFit} package~\cite{Moneta:2010pm}. The assumed
integrated luminosity is taken to be 300 fb$^{-1}$, and for each signal sample, the analysis
efficiency in each signal region and the lowest upper limit on the
production cross-section for $pp \rightarrow \eta\eta + j$, with $p_T^{\rm jet}>80$~GeV
are calculated.

\subsection{MD/MI operators and cross section upper limits}\label{sec:mdmonojets_cslimits}

\begin{figure}[t]
\begin{center}
\includegraphics[width=0.45\textwidth]{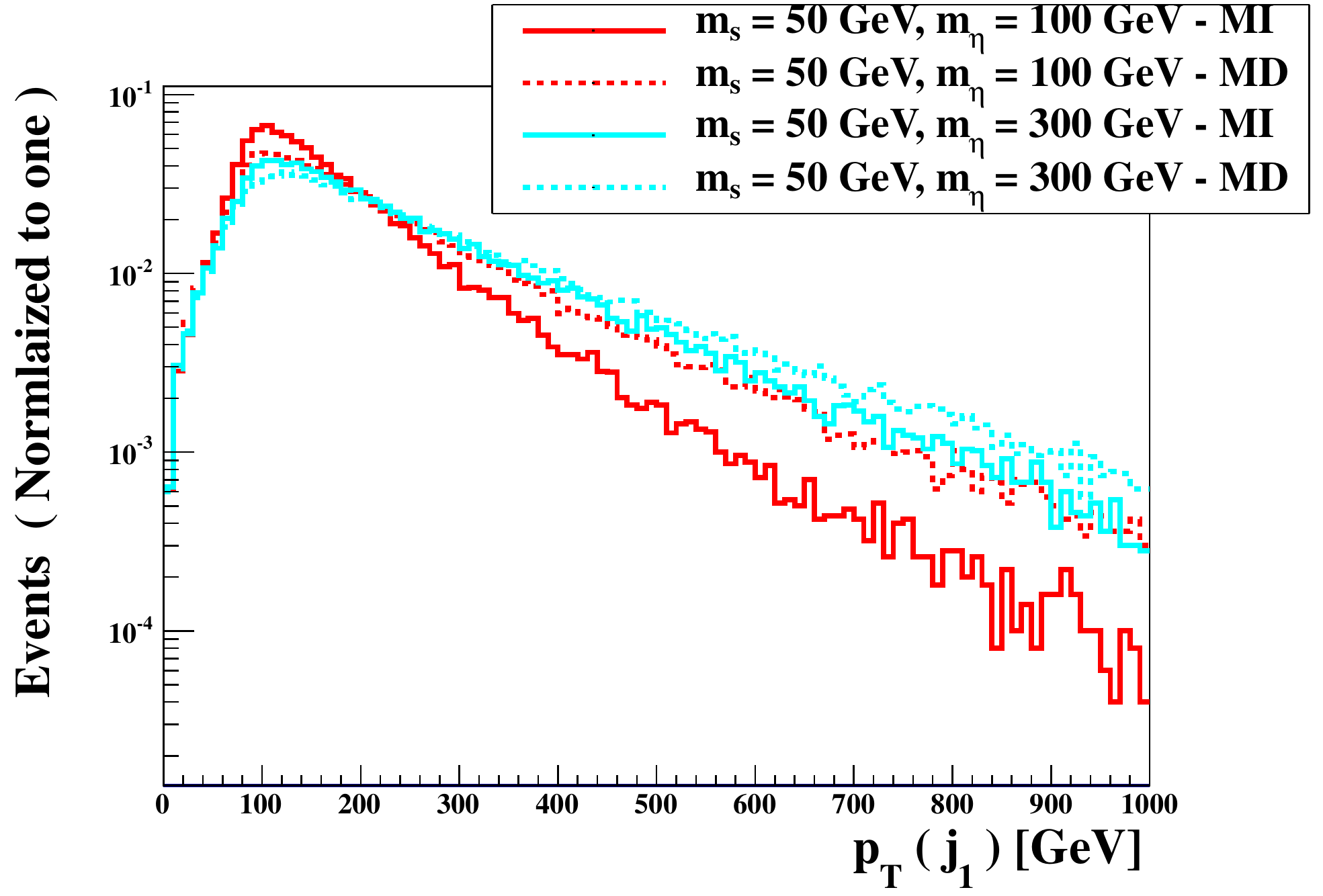}
\includegraphics[width=0.45\textwidth]{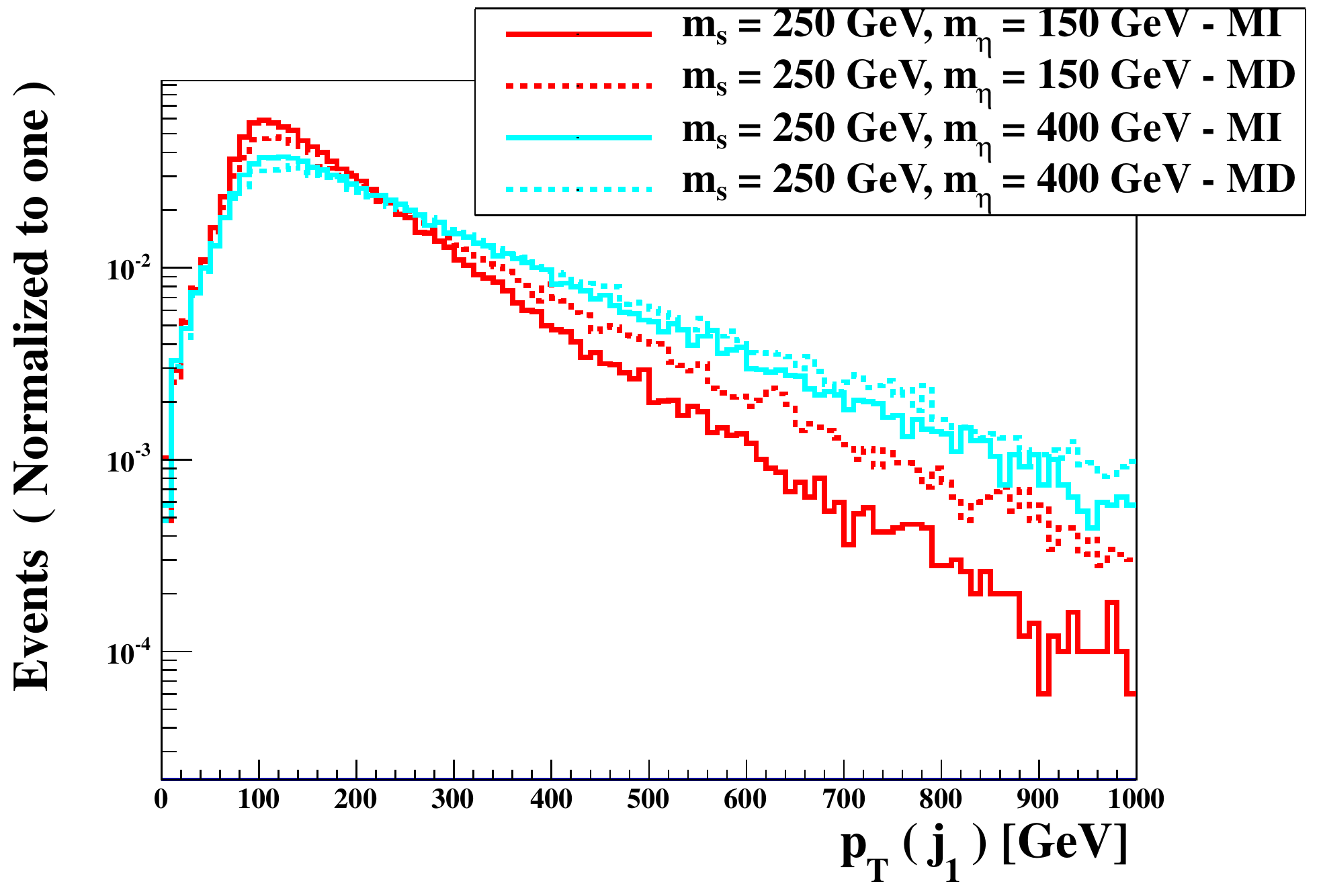}
 \caption{Normalized hadron-level $p_T$ distributions for the leading jet in the
   case of a mediator matter mass of 50~GeV  on the left panel, and of 250~GeV
    on the right panel. We adopt a dark matter mass of $m_\eta  = 100,300$~GeV for a mediator 
    mass of 50~GeV and  $m_\eta  = 100,400$~GeV for a mediator mass of 250 GeV. The
   red (cyan) solid lines indicate momentum-independent interactions while the red
   (cyan) dashed lines represent momentum-dependent interactions.}
\label{fig:mdmonojets_monopt}
\end{center}
\end{figure}

As a first illustration of the differences between the MI and MD scenarios, 
in Figure~\ref{fig:mdmonojets_monopt} we show the jet $p_T$ distributions for LHC
proton-proton collisions at 13 TeV (as returned by {\sc MadAnalysis} 5~\cite{Conte:2012fm}) 
for the representative mass combinations $(m_{s}, m_\eta) = (50, 100/300)$ GeV
(left panel) and $(m_{s}, m_\eta) = (250, 150/400)$ GeV (right panel). 
In order to perform a meaningful comparison, the two distributions have been
normalized to one, and 100000 events have been generated in both cases. We have
found that the MD operator induces a harder spectrum, which is expected to lead
to a larger fraction of selected events compared to the MI case. We moreover
observe that the difference between the MD and MI operators depends on the mass
of the dark matter particle. For a fixed mediator mass, heavier dark matter
leads to smaller differences between the jet $p_T$ distribution originating from
non-vanishing MI and MD operators. 

We thus expect that for a given cross section and for low dark matter masses, MD
operators will be more efficiently constrained by the LHC searches than their MI
counterparts. Keeping constant $c_{gs} = 100$ and $f =1$ TeV, we choose the
couplings to be $c_{\partial s\eta} = 2.5$ for the MD case and $c_{s\eta}= 0.5$
for the MI case, which both yield a cross section of 2.9 pb once a generator-level
selection on the leading jet $p_T$ of 80~GeV is enforced. After imposing that the transverse-momentum
of the leading jet satisfies $p_T > 300$~GeV, one retains 131300 and 196533 events in the MI and
MD cases, respectively, for a luminosity of 300~fb$^{-1}$. The MD case is thus expected to
yield a better sensitivity by about 50 $\%$.

\begin{figure}[h!]
\begin{center}
\includegraphics[width=0.4\textwidth]{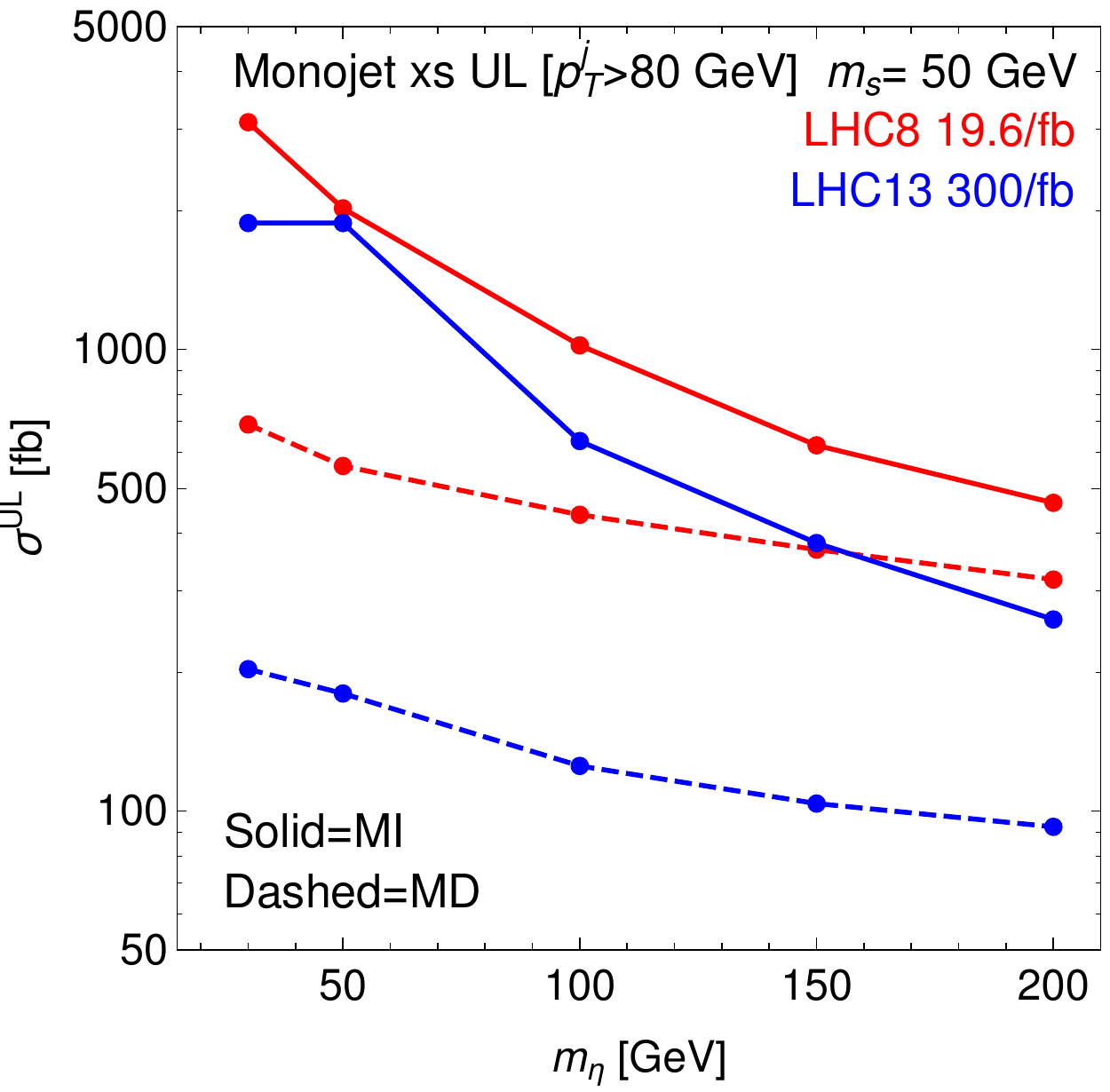}
\includegraphics[width=0.4\textwidth]{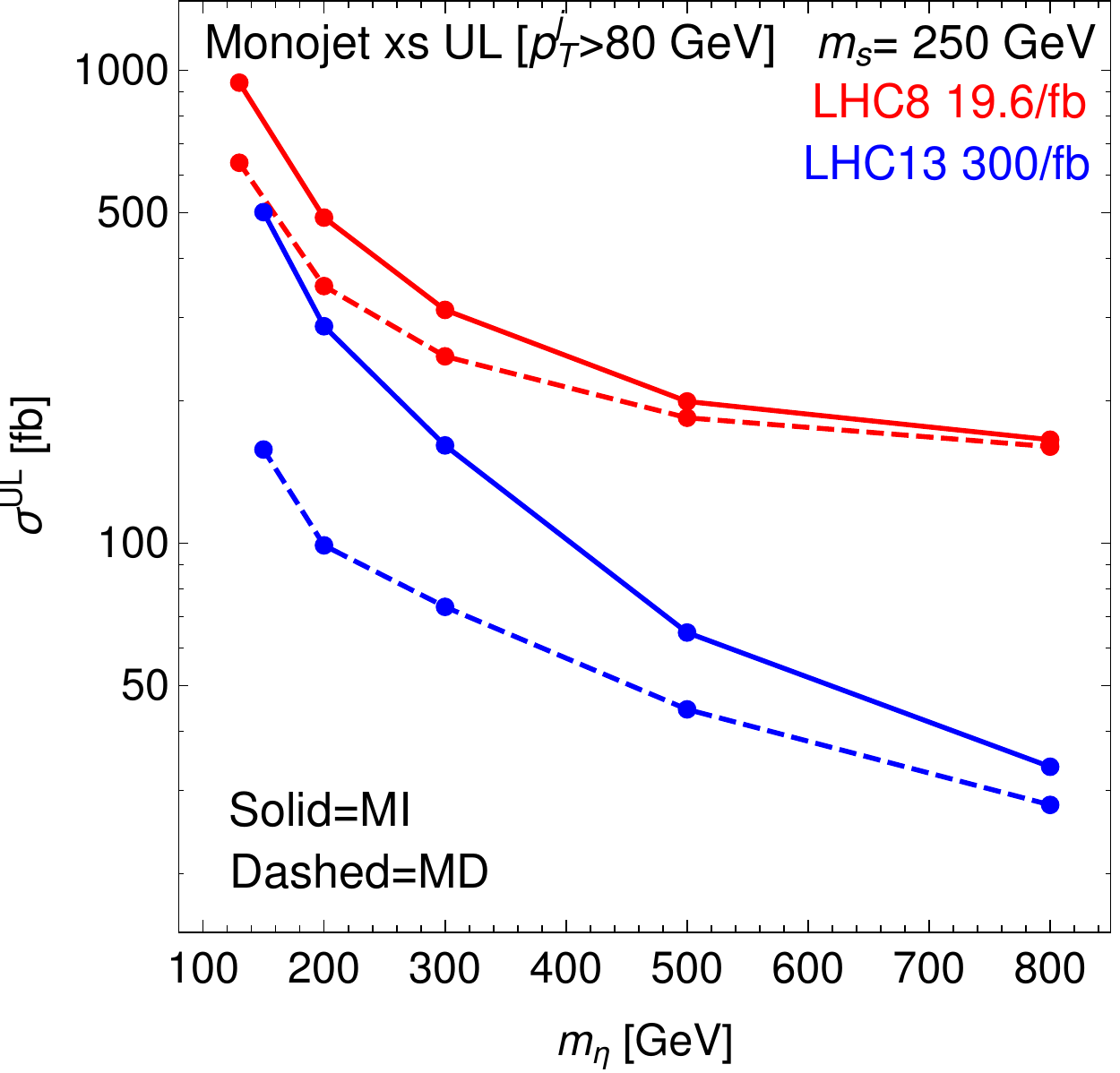}
\includegraphics[width=0.4\textwidth]{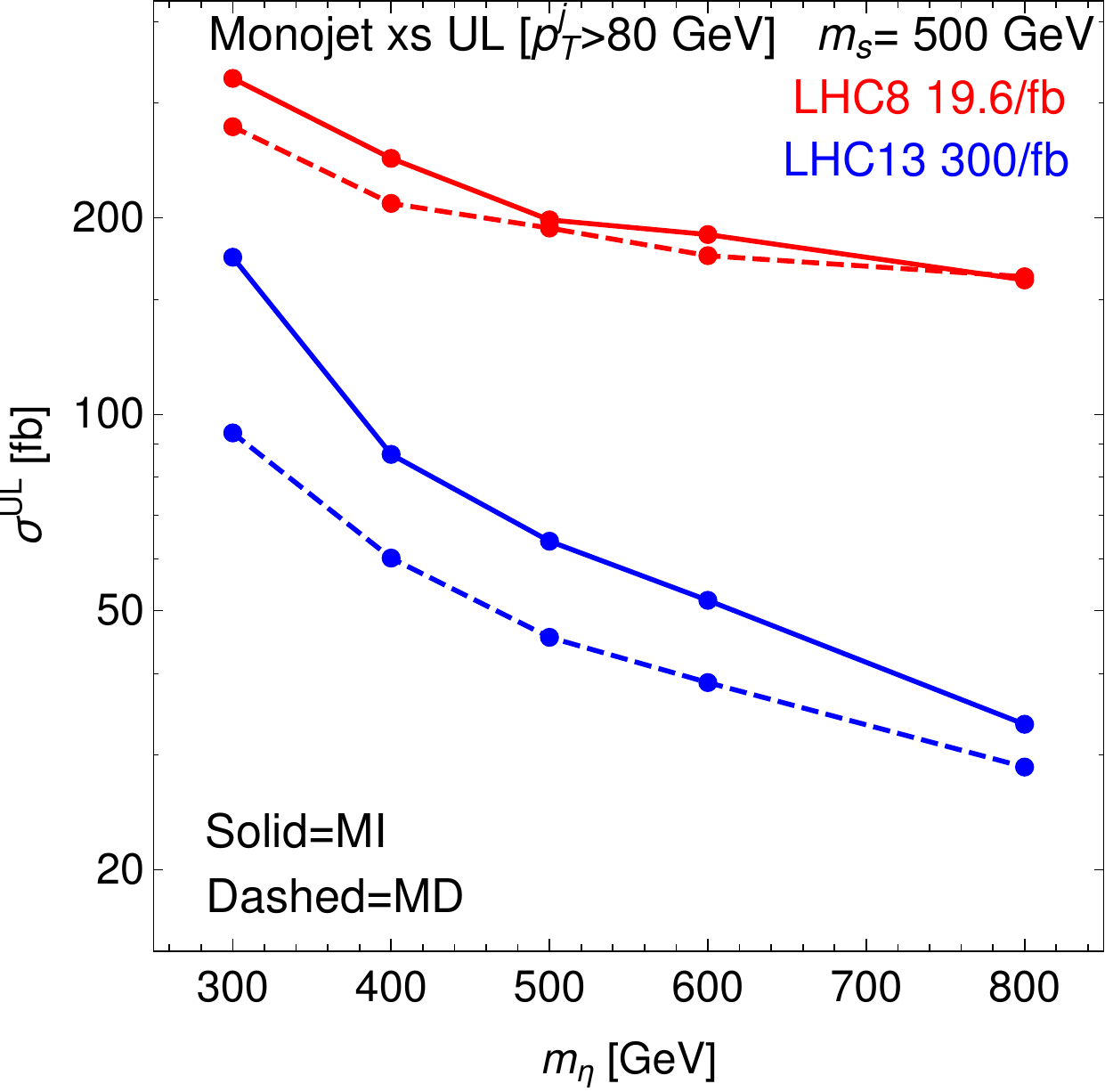}
\includegraphics[width=0.4\textwidth]{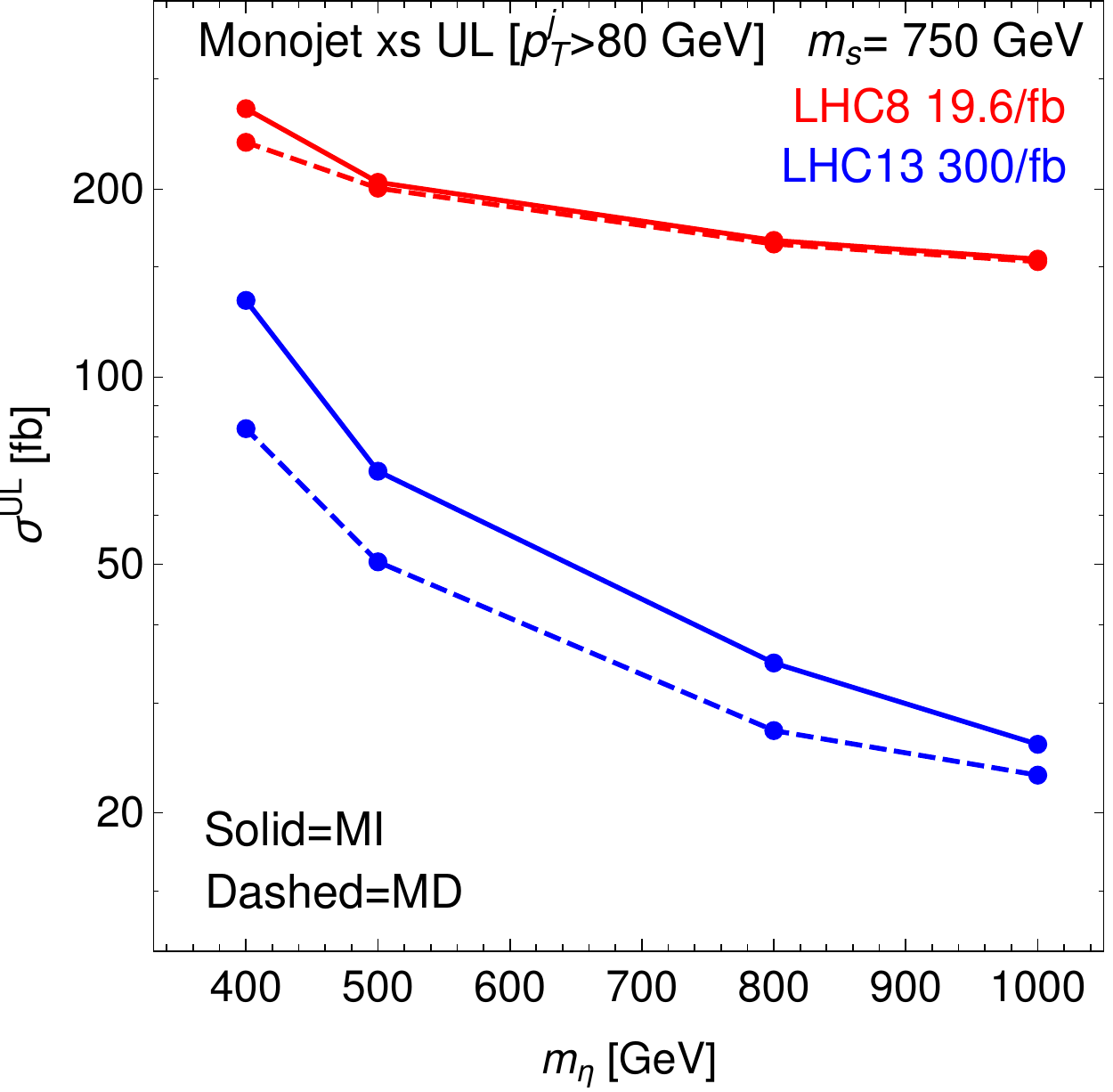}
 \caption{95\% CL upper limits (UL) on the monojet production cross section after
   including a generator-level selection of $p_T > 80$~GeV on the leading jet
   for proton-proton collisions at a center-of-mass energy of 8~TeV (recasting,
   red lines) and 13~TeV (projections, blue lines) for $m_s = 50$~GeV (top left),
   250~GeV (top right), 500~GeV (bottom left) and 750~GeV (bottom right) as a
   function of $m_\eta$. The solid lines correspond to the momentum-independent
   case, whereas the dashed lines correspond to the momentum-dependent case.
   \label{fig:mdmonojets_ulimits}}
\end{center}
\end{figure}

As explained in Section~\ref{sec:mdmonojets_setup}, the upper limits on the
cross section only depend only on $m_\eta$. In
Figure~\ref{fig:mdmonojets_ulimits}, we show the cross section upper limits for
$p p \rightarrow \eta \eta j$ with a generator selection of $p_T>80$~GeV on the
leading jet.
The 8-TeV constraints are depicted by red lines for the MI (solid) and MD
(dashed) cases. As anticipated, we see that the excluded cross sections are
consistently smaller in the MD scenario than in the MI one, \textit{i.e.}, the
former case is more efficiently constrained than the latter one. We moreover
observe that the exclusion bounds become stronger with increasing $m_\eta$. This
can be understood by the fact that as long as enough phase space is available,
larger $\eta$ masses imply a larger amount of missing energy which, in turn,
renders the monojet bounds stronger. For $m_{\eta} > 200$ GeV, the upper limits
become largely insensitive of the $\eta$ mass.

We have moreover found that the differences between the MI and MD cases become
maximal for small values of $m_\eta$. This behavior is in accordance with the
jet $p_T$-distribution illustrated in Figure~\ref{fig:mdmonojets_monopt} and can
be understood from the fact that as $m_\eta$ increases, the $\eta$ particles
become less and less boosted while at the same time, the amount of
$\slashed{E}_T$ increases for both the MI and MD cases. Eventually, for dark
matter masses of about 1~TeV, the limits obtained on the strengths of the MI and
MD interactions become identical. The LHC however looses sensitivity for such
heavy dark matter scenarios.

As already noted in
Ref.~\cite{Abercrombie:2015wmb}, in the case where $m_s<2 m_\eta$ and for a given
value of $m_\eta$, the cross section upper limits appear to be roughly
independent of the mediator mass. In order to further quantify this behavior, we
report in Table~\ref{tab:mdmonojets_accXeff} the acceptance ($\mathcal{A}$)
$\times$ efficiency ($\epsilon$) obtained in the case of the three different
regions of the analysis, for 8~TeV collisions and for a dark matter mass of
200~GeV. In our results, we adopt two mediator mass choices of 50 and 250~GeV.
This illustrates that the $\mathcal{A}\times \epsilon$ in all three signal
regions is very much independent of $m_s$ and the corresponding upper limits are
thus unaffected by the mediator mass.
\begin{table}[t]
\centering
\begin{tabular}{|c|c|c|c|c|c|c|c|c|c|} 
\hline \hline 
$m_{\eta}$ & $m_s$ & \multicolumn{2}{|c|}{$\mathcal{A}\times \epsilon$ (SR1)} & \multicolumn{2}{|c|}{$\mathcal{A}\times \epsilon$ (SR2)} & \multicolumn{2}{|c|}{$\mathcal{A}\times \epsilon$ (SR3)} & \multicolumn{2}{|c|}{$\sigma^{95\% {\rm CL}}_{UL}$[pb]}\\
& & MD & MI & MD & MI & MD & MI & MD & MI \\
\hline \hline 
200 & 50 & 0.123 & 0.101 & 0.073 & 0.056 & 0.033 & 0.023 & 0.317 & 0.465\\ 
200 & 250 & 0.124 & 0.104 & 0.069 & 0.054 & 0.031 & 0.022 & 0.349 & 0.487\\
\hline \hline
\end{tabular} 
\caption{Acceptance (${\cal{A}}$) $\times$ efficiency ($\epsilon$) of the three
  signal regions of the 8~TeV monojet analysis, for a dark matter mass of
  200~GeV and for two different mediator masses in the case of MI and MD
  operators. The jet-$p_T$ and $\slashed{E}_T$ requirements defining these
  regions are (280,220)~GeV, (340,340)~GeV and (450,450)~GeV respectively.}
\label{tab:mdmonojets_accXeff}
\end{table}

Moving on with 13~TeV projections, we also present in
Figure~\ref{fig:mdmonojets_ulimits} upper limits on the the signal production
cross-section that result from the procedure previously described. Blue solid
and blue dashed lines represent the MI and MD cases respectively. Similarly to
the 8~TeV case, the acceptance related to momentum-dependent dark matter
couplings for high values of the missing energy selection threshold is better
than in the momentum-independent case, and the distinction between the MI and MD
operators can be performed to a much larger extent than at 8~TeV. Furthermore,
the degeneracy of the limits at high values of the dark matter mass appears at
much higher values.

\begin{figure}[tb]
\begin{center}
\includegraphics[width=0.4\textwidth]{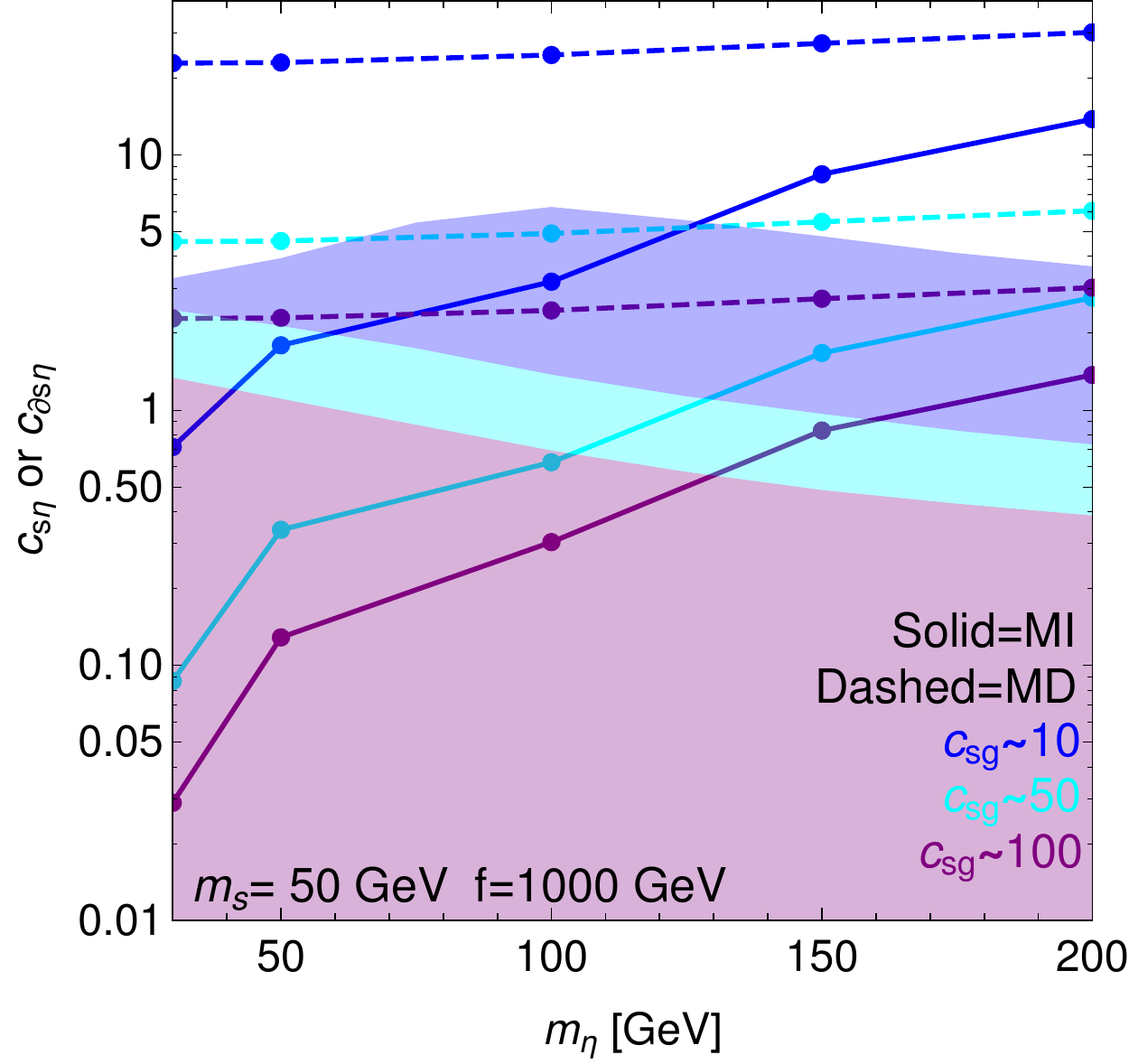}
\includegraphics[width=0.4\textwidth]{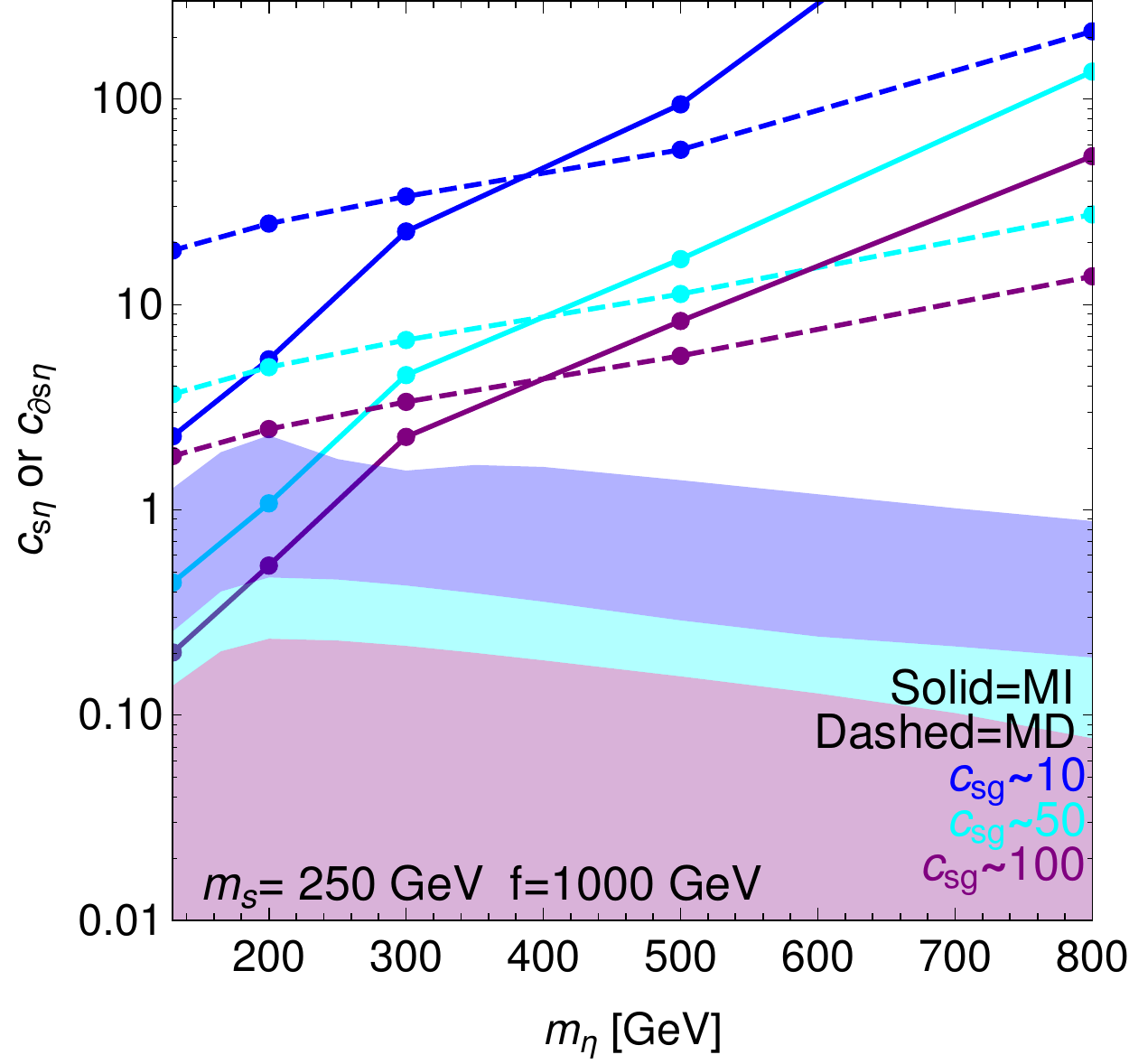}
\includegraphics[width=0.4\textwidth]{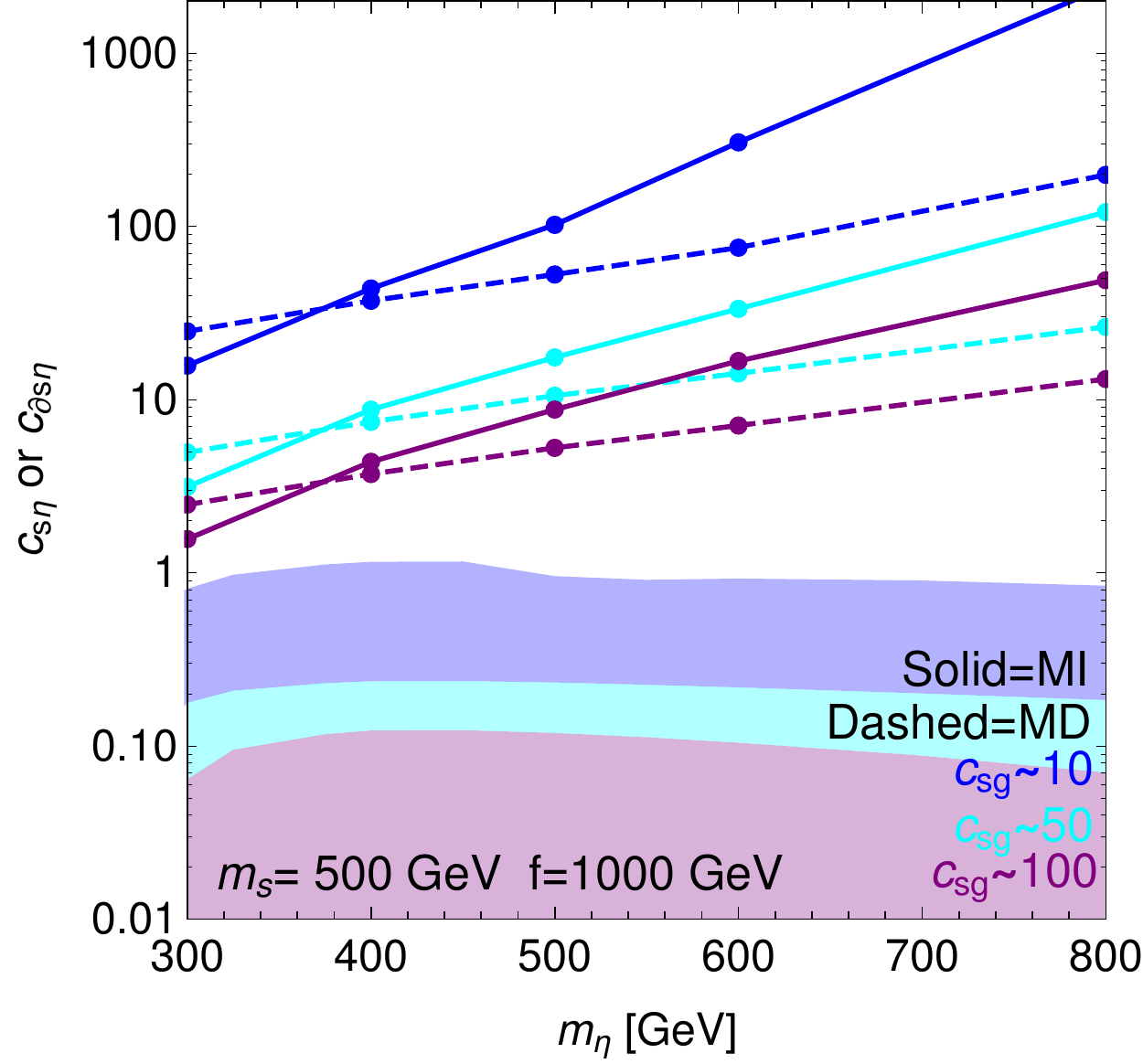}
\includegraphics[width=0.4\textwidth]{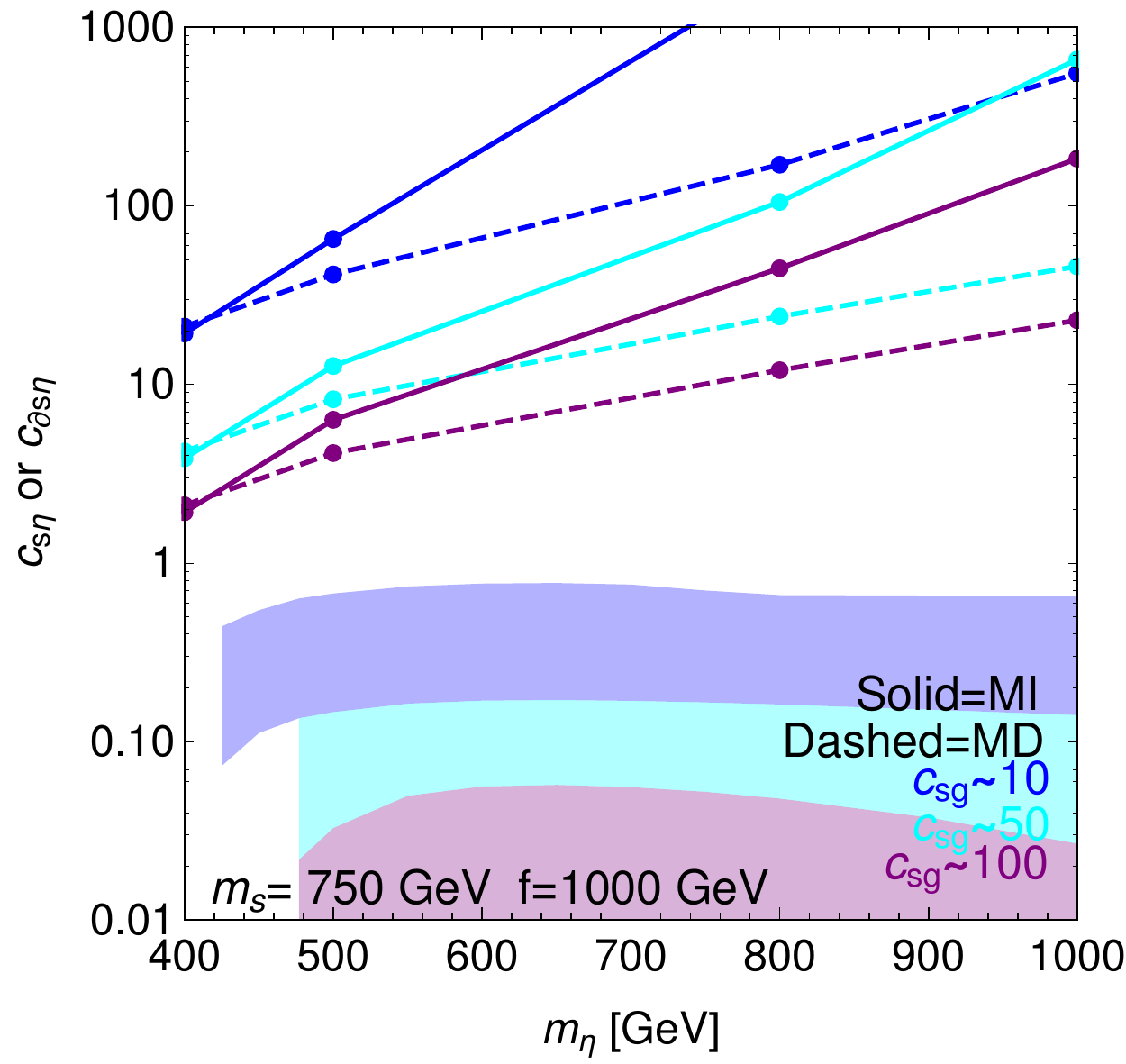}
 \caption{Constraints on the couplings $c_i$ defined in Eq.~\eqref{eq:mdmonojets_thelagrangian} from monojet searches (solid and dashed lines for the MI and MD cases respectively) for $m_s = 50$ (top left), 250 (top right), 500 (bottom left) and 750 GeV (bottom right) as a function of $m_\eta$, in each case for three distinct values of $g_s$. The shaded regions correspond to MD coupling values for which the universe is overclosed. \label{fig:mdmonojets_bounds}}
\end{center}
\end{figure}

\subsection{Combination of relic density, Direct detection and LHC constraints}  

Finally, we turn to the actual parameter space of our model and study the
interplay of the LHC monojet bounds presented in the previous section with the
dark matter constraints discussed in Sec.~\ref{sec:mdmonojets_ourmodel}. As a
preliminary remark, our numerical analysis has shown that in the MI scenario, the LUX bound already excludes the region of parameter space that is probed by the 8 TeV LHC monojet searches. Concretely, for a dark matter mass of
50~GeV, close to where the LUX sensitivity peaks, we find that assuming the minimal value $c_{sg} = 10$ the maximal allowed values of the coupling $c_{s\eta}$ are
of the order of $1.2\times 10^{-3}$, $0.03$, $0.13$ and $0.28$ for $m_s = 50$, 250, 500 and 750 GeV respectively. Going to a slightly higher mass $m_{\eta} = 200$ GeV, which is still expected to be within the LHC reach, these numbers translate to $0.008$, 0.2, 0.5 and 0.9. We will, therefore, not discuss the dark matter phenomenology of the momentum independent scenario any further.

In Figure~\ref{fig:mdmonojets_bounds}, we superimpose the 8 TeV LHC monojet
constraints on the MI and MD couplings $c_{s\eta}$ and $c_{\partial s \eta}$ for
a fixed value of the scale $f=1000$~GeV, and the predicted relic abundance for
the MD scenario according to standard thermal freeze-out. While deriving the
constraints, we have factored out the dependence on $f$. The limits on the
coupling $c_{s\eta}$ are however stronger for larger values of $f$, while those
on $c_{\partial s \eta}$ are correspondingly weaker. The cross-over between the
MI and MD coupling limits is hence an artefact of the choice of $f$.

In the shaded regions $\eta\eta$ annihilation is not efficient enough and the
Universe is overclosed, whereas along the borders of these regions the relic density limit is exactly reproduced. The shape of these
borders is well described by Eqs.~\eqref{eq:mdmonojets_RDgg} and \eqref{eq:mdmonojets_RDss}. We observe that as long as no resonance configuration or threshold is attained, the $c_{\partial s\eta}$ values required in order to satisfy the dark matter abundance bounds vary relatively mildly with the dark matter and mediator mass. The small features apparent especially in the $m_{s} = 250$ and $500$ GeV scenarios are due to the opening of the additional annihilation channel into $s$ pairs, although we find that for our choices of parameters the relic density is mostly driven by annihilation into gluons (the maximal contributions from the $\eta\eta \rightarrow ss$ channel being of the order of $15\%$). Then, since for $c_{s\eta} = 0$ the cross section $\left\langle \sigma v \right\rangle_{gg}$ scales as $(c_{\partial s\eta} \times c_{sg})^2$, smaller values of $c_{sg}$ imply almost proportionally larger values of $c_{\partial s\eta}$ so that
the Planck bound is saturated.

We observe that existing monojet searches are not yet efficient enough to probe
the regions where both the upper and the lower relic density limits are satisfied. On
the other hand, a significant fraction of the parameter space where only a part
of the dark matter in the Universe can be accounted for is excluded. While the
LHC searches probe large coupling values, the requirement of not overclosing the
universe excludes the opposite regime. In this sense, there is an interesting
complementarity between LHC and Planck observations.

\section{Conclusions}

In this work we studied a scenario in which the interactions of dark matter with
the SM are mediated by non-renormalizable derivative operators. We considered a
minimal model, in which a pair of dark matter particles are produced via
gluon-fusion, via a SM-singlet scalar mediator. We computed the 8 TeV LHC upper
limits on the monojet production cross section in presence of a MD interaction,
and compared them to the conventional scenario of MI interactions.
We highlighted the different behaviour of the signal in the two cases, and we
estimated the projected monojet limit at the 13 TeV LHC run. Moreover, dijet
bounds from past and present hadron colliders have been carefully taken into
account. We furthermore investigated the interplay of the LHC exclusion bounds
with the requirement that $\eta$ constitutes (part of) the dark matter of the
Universe, and computed the bounds that were stemming from dark matter direct
detection.

We observed that, for a given mass of the dark matter particle, MD scenarios can
be probed more efficiently at the LHC, as the latter is sensitive to smaller
cross sections with respect to the MI case due to the different jet $p_T$
distribution. We showed that, in MD scenarios, the LHC did not probe yet the
regions of parameter space where the dark matter relic density is exactly reproduced, 
whereas in MI scenarios the regions with a sizable monojet signal
are in severe conflict with dark matter direct detection constraints.

The minimal scenarios that we have investigated could be extended to cases where
the dark matter particles have additional couplings to the Standard Model. For
example, along with the coupling to gluons, the mediator may couple to the
electroweak field strength tensors and thus decay into $W$, $Z$ or $\gamma$
pairs. As long as the mediator width remains small, our upper limits on the
monojet production cross section are robust with respect to additional
couplings. On the other hand, monojet constraints on the size of the (effective)
$\eta\eta gg$ coupling become weaker as soon as the mediator is allowed to decay
via additional channels. A similar remark applies to the interplay between the
monojet limit and the dark matter relic abundance constraint, since smaller
couplings to gluons are required in order to saturate the observed relic density
as soon as additional $\eta\eta$ annihilation channels are turned on. Additional
couplings also imply the existence of additional dark matter search channels,
such as mono-$Z$ and mono-$W$, as well as additional possibilities to probe the
mediator of the DM-SM interactions, \emph{e.g.}, through dilepton, diphoton and
four-lepton searches.

Perhaps the most interesting question is whether the LHC can actually
\textit{distinguish} the MI or MD nature of the dark matter couplings. In this
manuscript we supplied some preliminary analyses to address this issue, that we
intend to investigate in detail in a forthcoming work.
\section*{ACKNOWLEDGEMENTS}
AG and SK are supported by the New Frontiers program of the Austrian Academy of
Sciences. The work of DS is supported by the French ANR, project DMAstroLHC,
ANR-12-BS05-0006, and by the Investissements d`avenir, Labex ENIGMASS.
MF is partially supported by the OCEVU Labex (ANR-11-LABX-0060) and the A*MIDEX
project (ANR-11-IDEX-0001-02), and by the European Union FP7 ITN INVISIBLES
(Marie Curie Actions, PITN-GA-2011-289442). BF has been supported by the
Th\'eorie LHC France initiative of the CNRS. We would like to thank the
Organizers of the 2015 ``Les Houches - Physics at TeV colliders" workshop, where
this work was initiated.



\AddToContent{D.~Barducci, A.~Bharucha, N.~Desai, M.~Frigerio, B.~Fuks, A.~Goudelis, S.~Kulkarni, S.~Lacroix, G.~Polesello and D.~Sengupta}
\renewcommand{\thesection}{\arabic{section}}

\graphicspath{{sgluon/}}

\chapter{Search for sgluons at the LHC}
{\it S.~Henrot-Versill\'e, T.~Spieker and D.~Zerwas}


\begin{abstract}
Extensions of the Standard model predict a massive scalar color-octet electroweak-singlet particle (scalar gluon or sgluon). In proton-proton collisions sgluons are predicted to be primarily produced in pairs and give rise to 4-jet final states.\\
This work analyses the discovery potential of sgluons in such final states for the 13 TeV run. The Monte Carlo based analysis is cross checked by reproducing the previous analysis from the ATLAS collaboration at 7 TeV.  Assuming an integrated luminosity of 10 fb$^{-1}$, the expected upper limit for sgluon masses is of the order of 550 GeV, which is an improvement of the previous exclusion limits by almost a factor of 2.
\end{abstract}

With the increase in center of mass energy of the Large Hadron Collider (LHC) to 13 TeV, the detection potential for new physics has been extended. 
This is motivated by extentions to the Standard Model (SM) like the Minimal Supersymmetric Standard Model (MSSM), which attempt to solve some of the open questions the Standard Model does not address. Like most new physics scenarios, the MSSM also predicts new particles which need to be added to the Standard Model particle content.

In certain theories, a scalar color-octet electroweak-singlet particle (called sgluon) is introduced which couples to the Standard Model and SUSY particles carrying a color charge. Its decay into gluon- and quark-pairs may be observable at the LHC as sgluon-pair production in a 4 gluon-jet (quark-jet) final state.
Such final states have been investigated by~\cite{Kilic:2008ub} and were further analysed in~\cite{Schumann:2011ji}. ATLAS has set an exclusion limit of sgluons in the mass range of 100 $< m_G <$ 287 GeV~\cite{Aad:2011yh,ATLAS:2012ds}. The CMS collaboration analysed 4 jet final states, setting limits in squark and coloron production \cite{Khachatryan:2014lpa}. Furthermore the CDF collaborations extended the mass limits down to 50 GeV \cite{Aaltonen:2013hya}. So far, all measurements are consistent with the Standard Model. 
An explicit realization of the R-symmetric MSSM is analysed in \cite{Kribs:2007ac,Plehn:2008ae,Choi:2008ub}, which introduces sgluons and allows to calculate the production cross sections as functions of the sgluon mass and the center of mass energy.  In this study, signal samples between 100 GeV and 1\,000 GeV corresponding to sgluon-pair production cross sections between 10$^{4}$ pb and 10$^{-3}$ pb are generated. The signal properties and the dependence of the signal signature on the center of mass energy are investigated. The sgluons are produced centrally decaying into jets. 
For the background, only hard QCD events are considered, as all other sources of background are found to be negligible.

The background at 7 TeV is studied in the chosen signal region to calculate excluded cross sections as a function of the sgluon mass. Predictions obtained from a QCD MC sample at 7 TeV corresponding to 4.6 fb$^{-1}$ are compared to the ATLAS analysis results to give confidence in the extrapolation of this analysis to 13 TeV. The cross section limits obained for 13 TeV and 10 fb$^{-1}$ are then translated into a prediction for the new mass limit.

Signal and background events are generated using the event generator Pythia 8, tuned according to a publicly available 7 TeV ATLAS tune \cite{ATL-PHYS-PUB-2014-021}. The generator was first validated by comparing to Pythia 6, which was used in the ATLAS analysis. Both generators provide physically reasonable distributions of the main observables and are compatible with each other.
QCD background samples for 7 TeV and 13 TeV were generated corresponding to 8.44 pb$^{-1}$ and 13.7 pb$^{-1}$ respectively. Here a filter on generator level was introduced, requiring a minimal 4th-largest-jet-$p_T$ of 60 GeV for the 7 TeV sample as well as a minimum of 100 GeV for the 13 TeV sample. Furthermore the QCD background was generated with a minimal particle $p_T$ of 50 GeV and 80 GeV for 7 and 13 TeV respectivly. Both measures are used to reduce the amount of diskspace needed and were checked for biases, where none were found. The different cuts are due to the trigger configurations. At 7 TeV a $p_T > 45$ GeV was required on the 4th-largest-$p_T$-jet, while this value is increased to 100 GeV for 13 TeV.
The 7 TeV and 13 TeV QCD background samples are rescaled by factors of 545 and 729 respectively to correspond to the desired integrated luminosities. Furthermore, the QCD background samples are normalized to the publicly available ATLAS data in the signal region \cite{ATLAS:2012ds}. For the 7 TeV QCD MC sample a correction factor $C = 1/1.1$ was found, which was also applied to the 13 TeV background.
The fast detector simulator Delphes~\cite{deFavereau:2013fsa} is used for detector uncertainties and jet reconstruction. For jet reconstruction, Delphes provides an interface for the FastJet library~\cite{Cacciari:2011ma}. The anti-$k_T$ algorithm with a cone radius of R = 0.4 is used~\cite{Cacciari:2008gp} as in the ATLAS analyses.

The sgluon candidates are selected by using a jet pairing algorithm. This uses the opening angle $\Delta R = \sqrt{(\Delta \phi)^2 + (\Delta \eta)^2}$, between the gluon jets, which are taken to be the 4 largest-$p_T$-jets. For the correct gluon pair, this opening angle tends towards 1 after a boost on the sgluons is ensured through a $p_T$ cut on the 4th-largest-$p_T$-jet. Hence the sgluon canidates are found by minimizing the function $|\Delta R_{pair1} - 1| + |\Delta R_{pair2} - 1|$. In the signal region 72\% of sgluons are reconstructed correctly.

\begin{figure}[htbp]
\begin{center}
\includegraphics[scale=0.45]{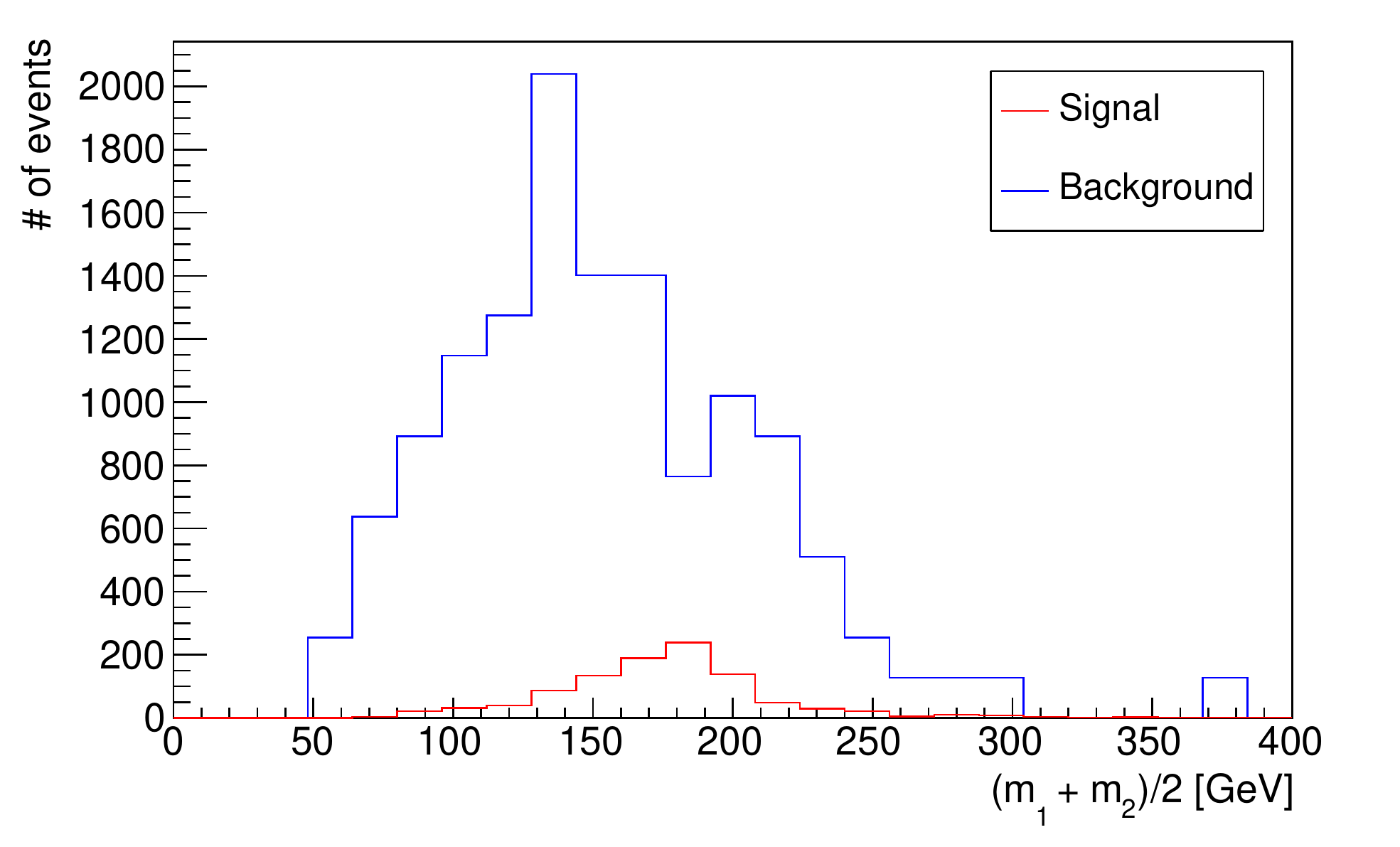}
\caption[Signal + Background at 7 TeV]{Signal and Background for a $m_G = 200$ GeV sgluon at a center of mass energy of 7 TeV. The background was normalized to match the signal integrated luminosity of $\sim 1$ fb$^{-1}$.}
\label{fig:SgluonLAL_fig1}
\end{center}
\end{figure}

The event selection exploits the resonant, central production of the massive sgluons. For the 4 jet final state, a minimum of 4 reconstructed jets is required, followed by a mass dependent cut on the jet transverse momenta. The $p_T$ of the 4th-largest-$p_T$-jet is required to be larger than 0.3 times the nominal sgluon mass plus 30 GeV, with a minimum of 80 GeV ($p_T $(4th jet)$ > $max$(0.3 \times m_G + 30 , 80)$). This ensures a boosted sgluon whose decay products are collimated. Additionally, the absolute value in pseudorapidity of all 4 jets needs to be smaller than 1.4 in order to exploit the central production ($|\eta| < 1.4$).  
The opening angle $\Delta R$ between a matched pair of gluon jets is required to be smaller than 1.6 to reject badly reconstructed events. The invariant mass of correctly reconstructed sgluons should be almost identical, so that a relative invariant sgluon mass difference $|m_1 - m_2|/(m_1+m_2) <$ 0.15 is required. Finally the centrality of the signal is exploited again by requiring the cosine of the scattering angle $\theta^*$ to be smaller than 0.5. Here $\theta^*$ is defined as the angle between the direction of the sgluon in the center of mass frame of the 4 highest $p_T$ jets and the boost direction between the center of mass frame and the rest frame of the 4 highest $p_T$ jets.
The signal efficiency for a 200 GeV sgluon at 7 TeV is found to be 0.63, which is in good agreement with the ATLAS results. The background in this region is suppressed with respect to the signal by roughly a factor of 4500. The average mass distribution $(m_1 + m_2)/2$ (which is the final observable considered) for signal and background for a sgluon with mass 200 GeV is displayed in Figure \ref{fig:SgluonLAL_fig1}.
The signal in the average mass distribution is sharply peaked, while the width is dominated completely by the detector resolution. For a higher sensitivity, only a mass window around the mean value of the reconstructed sgluon signal ($m_G \pm $RMS), which includes the entire core of the signal, is considered.

To obtain the correct background normalization and background shape in the signal region, the ABCD method is used. Here 4 orthogonal regions are defined, using the last two cuts of the signal region (see Table \ref{tab:SgluonLAL_tab1}). Region A remains the signal region, while regions B, C and D are background dominated.
\begin{table}[htbp]
\begin{center}
\begin{tabular}{c||c|c}
Region &  $|\cos\theta^*|$ & $|m_1 - m_2|/(m_1 + m_2)$\\ \hline
A & $< 0.5$ & $< 0.15$ \\
B & $> 0.5$ & $< 0.15$ \\
C & $< 0.5$ & $> 0.15$ \\
D & $> 0.5$ & $> 0.15$ \\
\end{tabular}
\caption[Definition of the background regions]{Definition of the 4 orthogonal regions needed for the background estimation. Region A is the signal region \cite{ATLAS:2012ds}.}
\label{tab:SgluonLAL_tab1}
\end{center}
\end{table}
The $|\cos\theta^*|$ cut is geometrical, wherefore the background shape in region B is the same as in region A (Kolmogorov-Smirnov test gives a value of 0.75). Furthermore the two variables are uncorrelated (shown by their correlation factor of -5.3\%). Therefore the number of background events in region A can be obtained via the ratio $N'_A = N_B \frac{N_C}{N_D}$. A closure test was performed, predicting the correct number of events within 1 standard deviation.

The measured signal region will have a statistical error $\delta N_A$, while the background prediction using the ABCD method will have an error of $\delta N'_A$ obtained from propagating the Possionian errors of regions B, C and D. The cross section limit is the minimal sgluon production cross section necessary in order to be discernable from background at a given confidence level.  The confidence level of 95\% is calculated as in ~\cite{Schumann:2011ji} by using the background in the signal region (imitating a background only measurement) and its ABCD-prediction together with their errors, as well as the signal selection efficiency. As a simplification the total error is obtained by propagating the Poissonian errors of the latter using Gaussian error propagation and adding it to the error on the background estimate. The one sided Gaussian point of 95\% area on this combined error is then defined as the excluded cross section as in ~\cite{Schumann:2011ji}.
This method was first validated at 7 TeV, where the cross section limits were found to be slightly too obtimistic, but only a factor of 1.5 smaller than the ATLAS results. 
\begin{figure}[htbp]
\begin{center}
\includegraphics[scale=0.45]{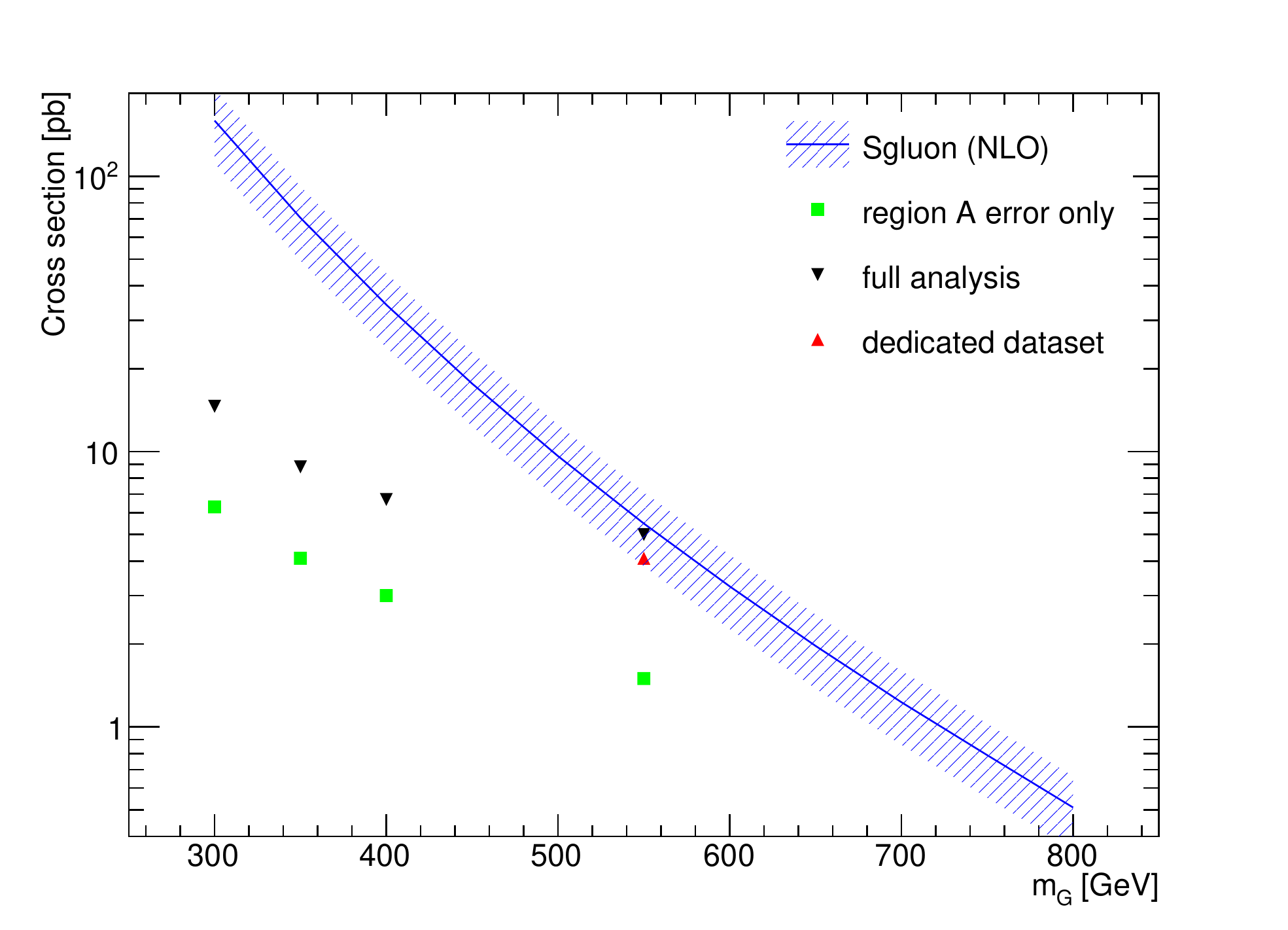}
\caption[Expected upper exclusion limit on sgluon masses]{Excluded cross sections using 1) summation in quadrature of the error on region A and the ABCD method; 2) the full analysis as in 1) with the new dedicated dataset for a 550 GeV sgluon mass. The Sgluon production cross sections are displayed at NLO \cite{GoncalvesNetto:2012yt,GoncalvesNetto:2012nt}.}
\label{fig:SgluonLAL_fig2}
\end{center}
\end{figure}
Together with the NLO production cross sections calculated in \cite{GoncalvesNetto:2012yt,GoncalvesNetto:2012nt} the new prediction on the excluded sgluon mass limit is found. The intersection of the excluded cross sections with the production cross section minus one standard deviation is at $\sim 550$ GeV (see Figure \ref{fig:SgluonLAL_fig2}). 

In summary, extrapolation of the previous ATLAS sgluon analysis to 13 TeV is possible. Assuming 10 fb$^{-1}$ of 13 TeV data, a full analysis will set new sgluon mass limits, improving the current limits roughly by a factor of two.

\AddToContent{S.~Henrot-Versill\'e, T.~Spieker and D.~Zerwas}
\renewcommand{\thesection}{\arabic{section}}

\graphicspath{{DMtt/}}

\def\etmiss{\ensuremath{E_{\mathrm{T}}^{\mathrm{miss}}}\xspace}
\def\EtMiss{\ensuremath{E_{\mathrm{T}}^{\mathrm{miss}}}\xspace}
\def\ptmiss{\ensuremath{\vec p^{\mathrm{\ miss}}_\mathrm{T}\xspace}}
\def\dphimin{\ensuremath{\Delta\phi_{\mathrm{min}}}}
\newcommand{\mttwo}{\ensuremath{m_{\mathrm{T2}}}\xspace}
\newcommand{\pbll}{\ensuremath{p_{\mathrm{b}}^{ll}}\xspace}
\newcommand{\ptll}{\ensuremath{p_{\mathrm{T}}^{ll}}\xspace}
\newcommand{\ttbarW}{\ensuremath{t\bar{t}W}\xspace}
\newcommand{\ttbarZ}{\ensuremath{t\bar{t}Z}\xspace}
\newcommand{\mTlep}{\ensuremath{m_\mathrm{T}^{lep}}}
\def\ttbar{\ensuremath{t\bar{t}}}
\newcommand{\Wjets}{\ensuremath{W}+jets}
\newcommand{\W}{\ensuremath{W}}

\def\TeV{\ifmmode {\mathrm{\ Te\kern -0.1em V}}\else
                   \textrm{Te\kern -0.1em V}\fi}%
\def\GeV{\ifmmode {\mathrm{\ Ge\kern -0.1em V}}\else
                   \textrm{Ge\kern -0.1em V}\fi}%



\chapter{LHC sensitivity to associated production of dark matter and
$t\bar{t}$ pairs}

{\it P.~Pani and G.~Polesello}


\begin{abstract}
The sensitivity of the LHC to the associated production 
of dark matter and $t\bar{t}$ pairs is studied in the framework of
simplified models with scalar or pseudoscalar mediator.
An analysis strategy is developed for final states with
two leptons, and projected constraints are assessed
on the parameter space of the model for the LHC experimental
conditions corresponding to 300 fb$^{-1}$ of integrated 
luminosity at the LHC.
\end{abstract}

\section{Introduction}

Astrophysical observations have provided compelling proof for the
existence of a non-baryonic dark component of the universe:
Dark Matter (DM). The DM abundance has been precisely measured
\cite{Komatsu:2010fb,Ade:2013zuv} and
corresponds to 27\% of the total universe content.  
The nature of DM is not known, but from
the theoretical point of view the most studied candidate is
represented by a WIMP: a neutral particle with weak-scale mass
and weak interactions, whose thermal relic density may naturally fit
the observed DM abundance.  
A wide range of experimental searches are focused on searching for
such a WIMP candidate at the LHC, using mainly three
different approaches to model DM signals: 
\begin{itemize}
  \item UV-complete models,
  \item effective field theories (EFT),
  \item Simplified models.
\end{itemize}

\noindent
In this paper, we focus on the third approach, which has the advantage
of introducing only minimal additional parameters for the model with
respect to the SM, while not suffering of the limited
applicability of the EFT. 
\begin{figure}
  \begin{center}
    \includegraphics[width=0.3\textwidth]{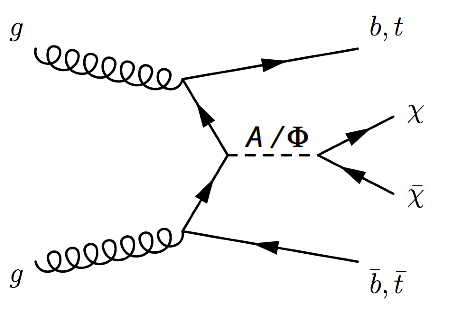}
    \caption{ Representative Feynman diagram for Dark Matter associated production with top pairs via a spin-0 mediator to the dark sector. }
    \label{DMtt_diagram}
  \end{center}
\end{figure}
The particular simplified model that is the focus of this study is 
a spin-0 scalar or pseudoscalar mediator to the dark sector, which allows
s-channel production of dark matter from Standard Model partons at the
LHC. The model was introduced and discussed already in
Ref.~\cite{Buckley:2014fba,Haisch:2015ioa,Abercrombie:2015wmb}. 
In order to fulfill precision constraints from flavor measurements,
this model assumes Yukawa-like couplings between the dark sector
mediator and the SM fermions. The final state of interest,
depicted in Figure~\ref{DMtt_diagram} is characterised by a
top pair and missing transverse momentum originating from the mediator
decay into a pair of DM particles. The final state signature 
depends on the decays assumed for the $W$ from top.
In the present study the signature with two leptons ($e,\mu$), possibly
of different flavors, 
in the final state is considered.

\section{Signal and background simulation}
\subsection{Signal}
Following the notation of \cite{Buckley:2014fba}, the model 
has 5 parameters: $m_\phi$, the mass of the mediator, 
$m_\chi$, the mass of the dark matter particle, $g_\chi$, the dark matter-
mediator coupling, $g_v$, the universal SM-mediator coupling and
$\Gamma_\phi$, the mediator width. If no additional particles 
are present, $\Gamma_\phi$ can be calculated from the other parameters.
For the present study, samples were generated for different values
of $m_\phi$, varying from 10~GeV to 500~GeV. The generation was performed
for $g_v=g_\chi=1$ and $m_\chi=1$~GeV. It was shown in previous studies \cite{Abercrombie:2015wmb}
that the acceptance of experimental cuts is insensitive to $\Gamma_\phi$,
therefore a sample with a single values of the couplings can 
be used. The signal samples were generated with {\sc Madgraph5} \cite{Alwall:2014hca}
and showered with {\sc PYTHIA8} \cite{Sjostrand:2007gs} 
according to the generator settings suggested in \cite{Abercrombie:2015wmb}. \par

\subsection{Background}
For the two-lepton analysis all the Standard Model backgrounds
involving at least two leptons coming from the decay of 
vector bosons are generated. Backgrounds either with fake electrons
from jet misidentification or with real non-isolated leptons 
from the decay of heavy flavours are not considered in the analysis,
as a reliable estimate of these would require a simulation of 
detector effects beyond the scope of this work. It has been shown
by ATLAS with analyses based on kinematic variables similar
to the ones used in this study \cite{Aad:2014qaa} 
that in the relevant signal region
the background from non-prompt leptons is negligible.
The backgrounds considered are $t\bar{t}$, $Wt$, $WW$, $WZ$, $ZZ$, 
all produced with {\sc POWHEG BOX} \cite{Alioli:2010xd} 
and showered with {\sc PYTHIA8}; $Z+\mathrm{jets}$,
produced with a multiplicity of up to four jets with {\sc Madgraph5} 
and showered with  {\sc PYTHIA8}; $t\bar{t}W$ and $t\bar{t}Z$ produced with 
a multiplicity of up to two jets with {\sc Madgraph5} and showered with {\sc PYTHIA8}.
The cross-sections are normalised to the calculation of the
highest perturbative order available in literature.\par
\subsection{Detector smearing}
From the stable particles produced from the generators
the following physics objects are built: electrons, 
muons, photons jets and \etmiss. Jets are built from the
true momenta of particles interacting in the calorimeters
except muons,with an anti-$k_\mathrm{T}$ algorithm with a parameter $R=0.4$, as implemented in  {\sc Fastjet} \cite{Cacciari:2008gp}. 
The variable \ptmiss, with magnitude \etmiss, is built as 
the vector sum of the transverse momenta of all the non-interacting particles, 
i.e. neutrinos and dark matter particles.
The effect of the detector on the kinematic quantities of interest
is simulated by applying a gaussian smearing to the momenta of 
the different reconstructed objects, and reconstruction 
efficiency factors. The parametrisation of the smearing 
and reconstruction efficiency 
as a function of momentum and pseudorapidity of the objects
is tuned to mimic the performance reported by ATLAS for Run 1 
at the LHC \cite{Aad:2009wy}. 


\section{Analysis strategy}
For the two-lepton analysis, the considered 
final state includes two leptons and significant hadronic activity
from the decay of top, plus \etmiss from the neutrinos from $W$ decay,
from dark matter production and from mismeasurement of the hadronic part 
of the event.\par
The main discriminant variable against all of the backgrounds
where the two leptons are produced in the decay of two $W$ bosons
is the \mttwo variable \cite{Lester:1999tx, Barr:2003rg}
calculated using the two leptons and \ptmiss, defined as:
\begin{equation}
m^2_\mathrm{T2} (\vec p^{\ \alpha}_\mathrm{T}, \vec p^{\ \beta}_\mathrm{T}, \vec
p^{\mathrm{\ miss}}_\mathrm{T}) = \min_{\vec q^{\ 1}_\mathrm{T} + \vec q^{\ 2}_\mathrm{T} = \vec p^{\mathrm{\ miss}}_\mathrm{T}}  \max ( 
    m^2_\mathrm{T}( \vec p^{\ \alpha}_\mathrm{T}, \vec q^{\ 1}_\mathrm{T}),  m^2_\mathrm{T}( \vec p^{\ \beta}_\mathrm{T},
    \vec q^{\ 2}_\mathrm{T} ) ) 
\end{equation}

\begin{figure}[htb]
\begin{center}
\includegraphics[width=0.49\textwidth]{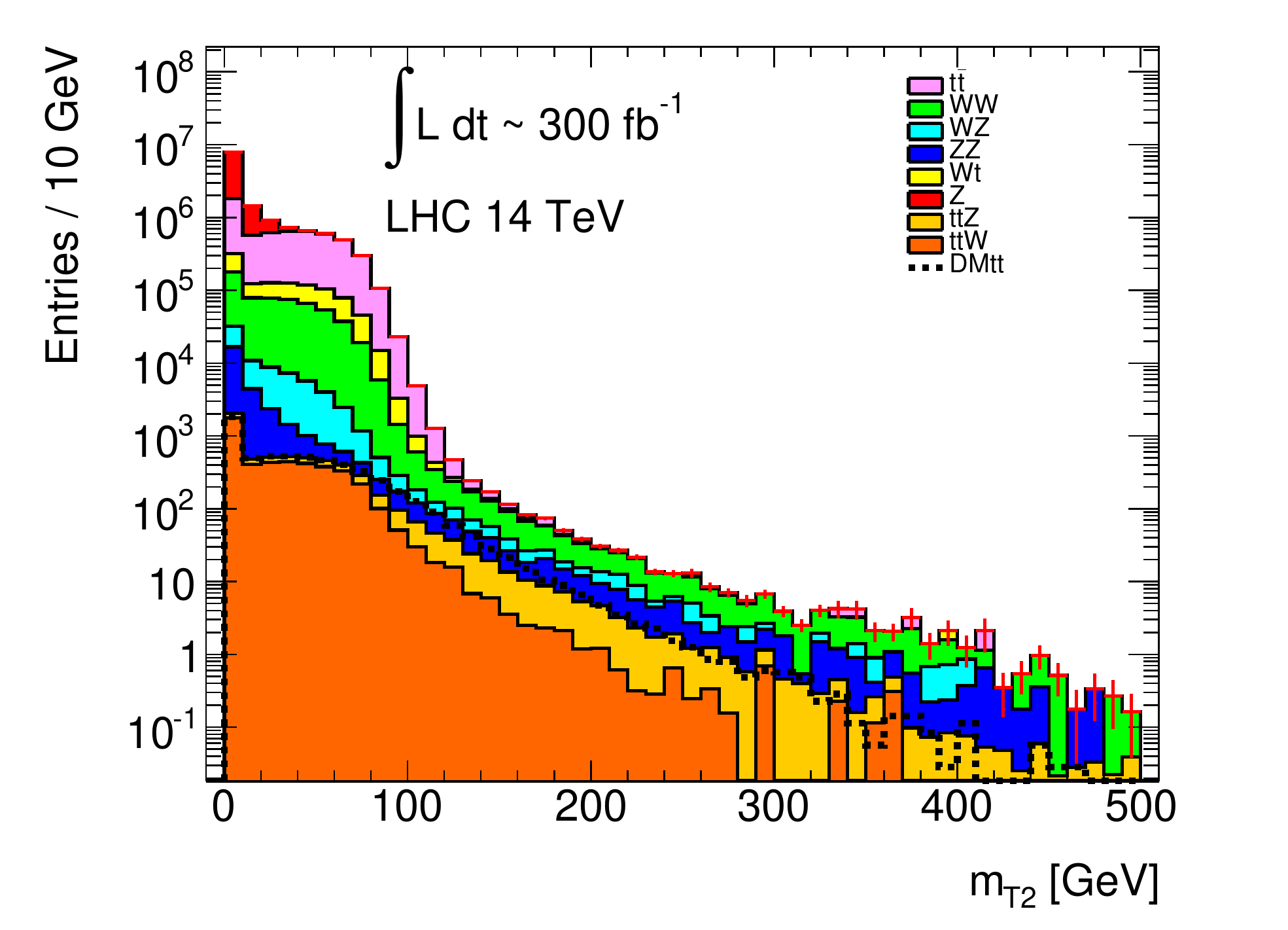}
\includegraphics[width=0.49\textwidth]{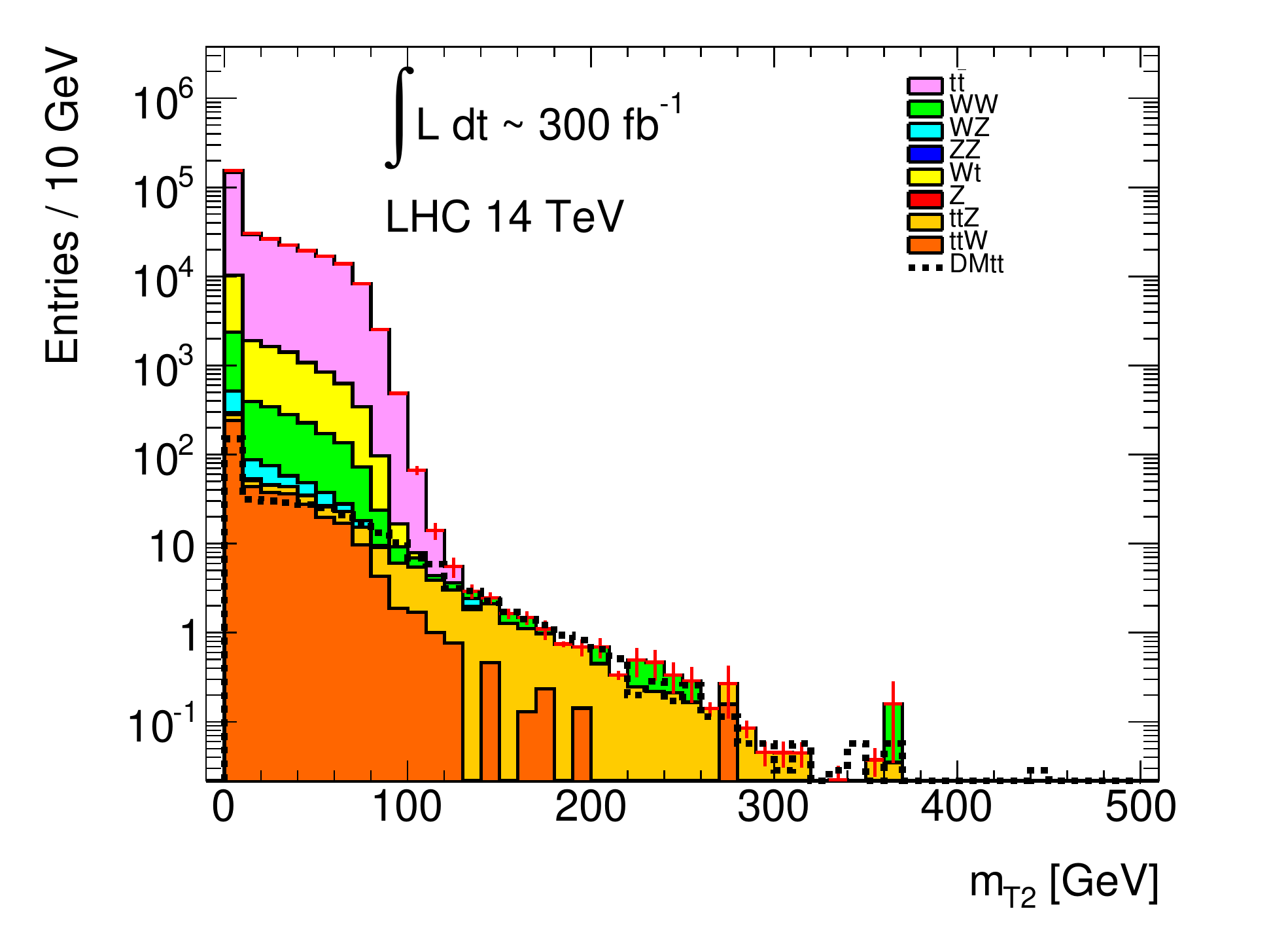}
\caption{Distribution of the $\mttwo$ variable after preselection (left),
and after all cuts (right), for the SM backgrounds 
and for a benchmark model  with $m_\phi=100$~GeV, $m_\chi=1$~GeV, $g_v=g_\chi=1$.}
\label{DMtt_mttwo}
\end{center}
\end{figure}

\begin{figure}[htb]
\begin{center}
\includegraphics[width=0.49\textwidth]{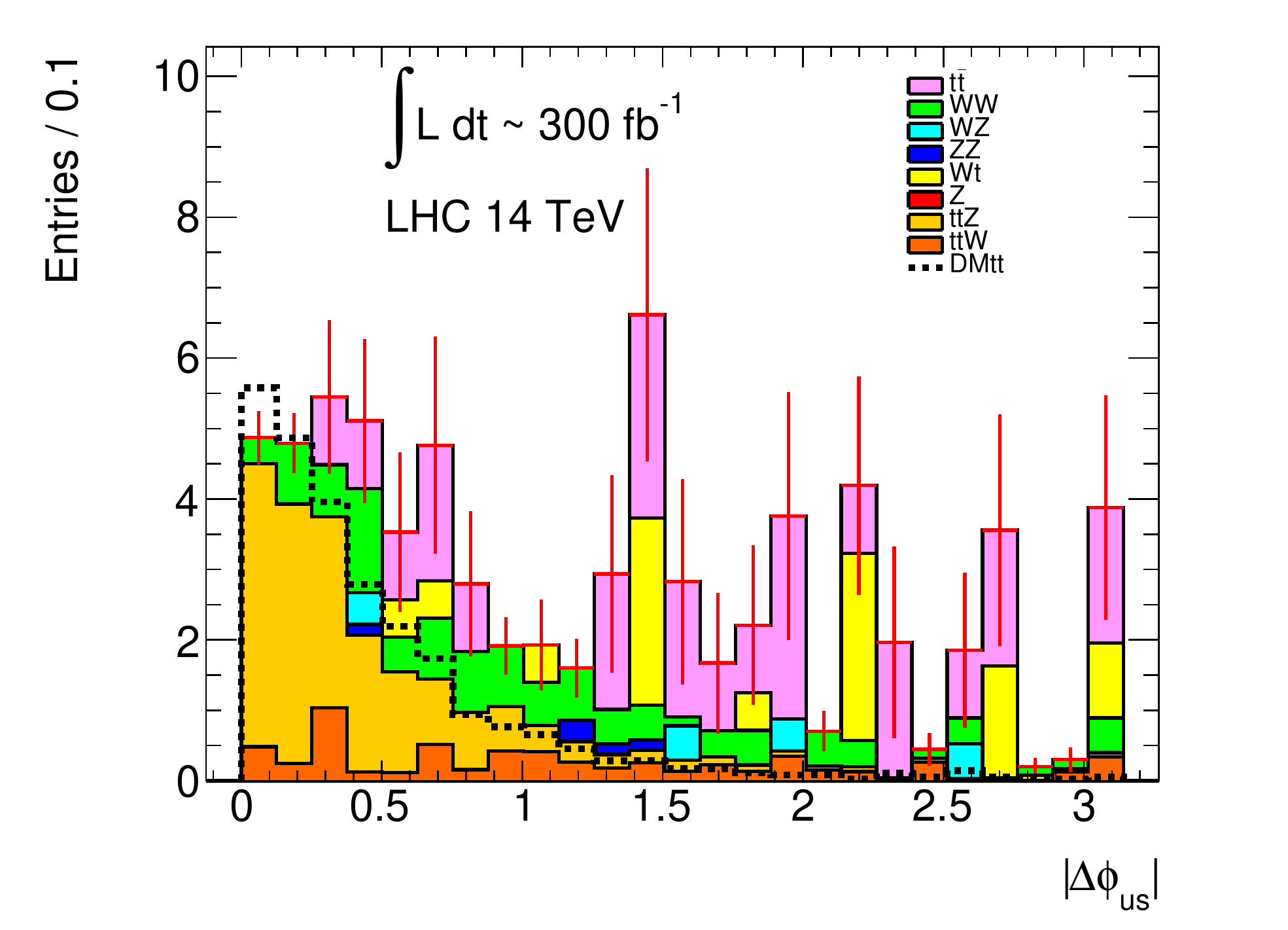}
\includegraphics[width=0.49\textwidth]{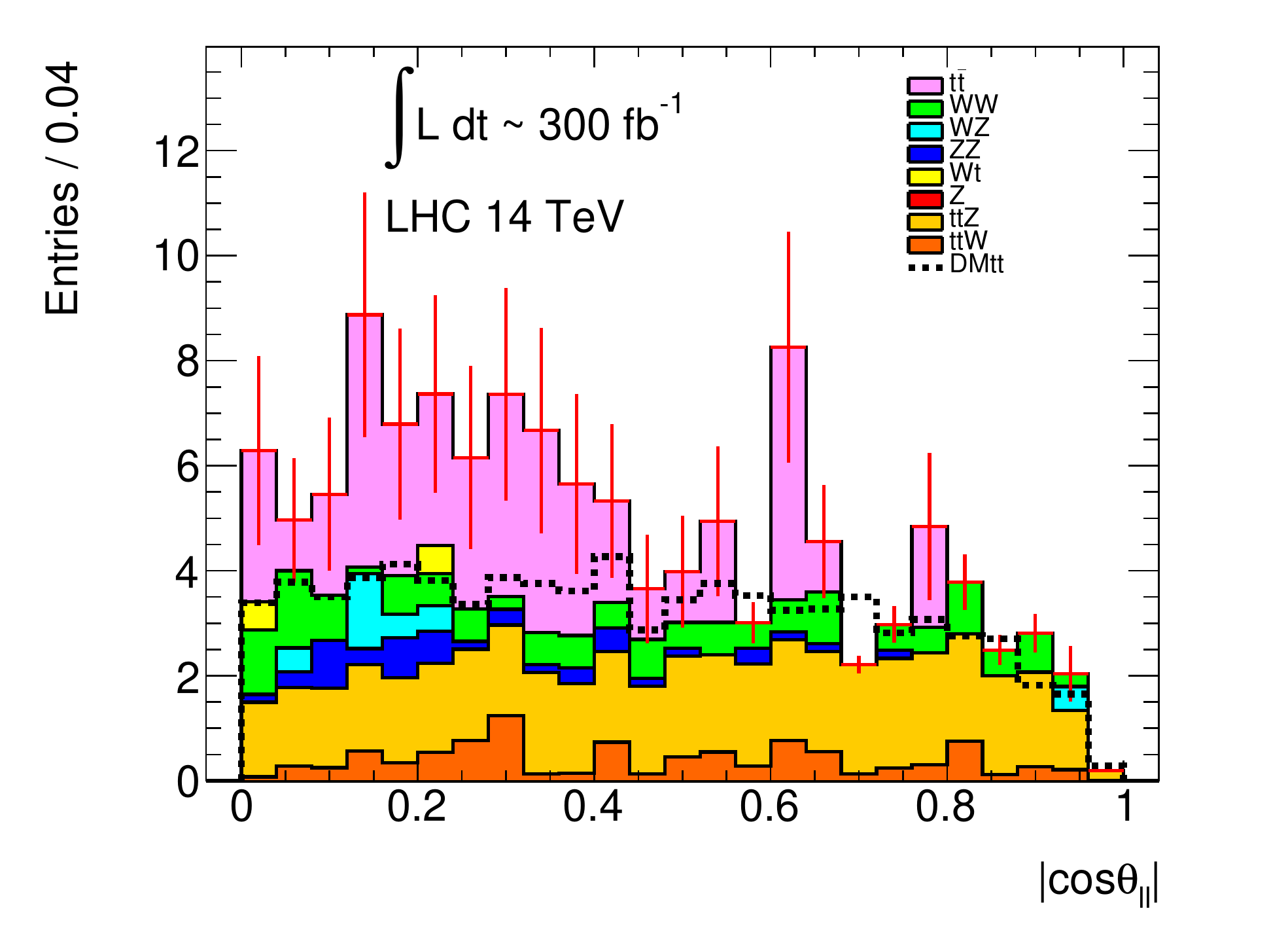}
\caption{Distribution of the $\Delta\phi_{us}$ variable (left) and of the 
$\cos\theta_{\ell\ell}$ variable (right), for the SM backgrounds and 
for a benchmark model with $m_\phi=100$~GeV, $m_\chi=1$~GeV, $g_v=g_\chi=1$. 
The full set of cuts 
described in the texts are applied, except for the cut on the
plotted variable, and with the additional requirement $\mttwo>120$~GeV.}
\label{DMtt_cutm1}
\end{center}
\end{figure}

Here, $m_\mathrm{T}$ indicates the transverse mass, $\vec p^{\ \alpha}_\mathrm{T}$ and
$\vec p^{\ \beta}_\mathrm{T}$ are the momenta of the two leptons, and $\vec
q^{\ 1}_\mathrm{T}$ and $\vec q^{\ 2}_\mathrm{T}$ are vectors which satisfy
$\vec q^{\ 1}_\mathrm{T} + \vec q^{\ 2}_\mathrm{T} = \vec p^{\mathrm{\ miss}}_\mathrm{T}$. The minimum
is taken over all the possible choices of  $\vec q^{\ 1}_\mathrm{T}$ and $\vec q^{\ 2}_\mathrm{T}$.
This variable has an upper limit at the $W$ mass for $W$-induced backgrounds,
whereas for the signal the presence of additional \etmiss from 
DM production generates tails in the \mttwo distribution.
Other backgrounds, in particular the ones including the leptonic
decay of the $Z$ bosons have rapidly falling \mttwo and \etmiss 
distributions. In addition backgrounds from $Z$+jets and diboson
production have much lower hadronic activity than the signal
which includes the $b$-jets from the top decay, and can be thus 
significantly reduced  by requiring a moderate amount of hadronic activity.
An alternative approach for these backgrounds would be the requirement
of at least one b-tagged jet. 
The basic signal selections require:
\begin{itemize}
\item
Two opposite-sign (OS) isolated leptons, 
$p_\mathrm{T}(\ell_1)>25$~GeV and $p_\mathrm{T}(\ell_2)>10$~GeV $|\eta_{\ell}|<2.5$,
\item
$m(\ell\ell)>20$ GeV, if same-flavour $m(\ell\ell)<71$~GeV or $m(\ell\ell)>111$~GeV to veto $Z$ bosons.
\item
For the same-flavour background additional requirements 
are applied in \cite{Aad:2014qaa} to suppress the residual $Z$ backgrounds.
These are $\Delta\phi_{min}>1$, where $\Delta\phi_{min}$ is the 
minimal azimuthal angle between a reconstructed
jet and $\etmiss$, and $\Delta\phi_{us}<1.5$, where $\Delta\phi_{us}$
is the azimuthal angle between $\vec p_\mathrm{T}^{\ miss} $ and $\vec p_\mathrm{T}^{\ us} \equiv-(\vec p_\mathrm{T}^{\ \ell1 }+\vec p_\mathrm{T}^{\ \ell2 }+\vec p_\mathrm{T}^{\ miss})$.
\end{itemize}
The distribution of the $\mttwo$ variable for the backgrounds and 
a benchmark signal is shown in Figure~\ref{DMtt_mttwo} left for an integrated
luminosity of 300~fb$^{-1}$. The selection of events with $\mttwo$ between 100 and 150 GeV
is required to reduce the backgrounds at a manageable level. The dominant
backgrounds at high $\mttwo$ are from diboson production. 
To reduce these backgrounds the following requirement is applied:
\begin{itemize}
\item
at least one jet with $p_\mathrm{T}>60$~GeV; at least two jets with $p_\mathrm{T}>40$~GeV.
\end{itemize}
For further optimisation two variables have been found useful:
$\Delta\phi_{us}$ defined above, and $\cos\theta_{\ell\ell}$, 
introduced in 
\cite{Barr:2005dz} as $\cos\theta_{\ell\ell}=\tanh(\delta\eta_{\ell\ell}/2)$.
The usage of both variables allows for a further reduction of the remaining $t\bar{t}$
background, and the requirements:
\begin{itemize}
\item
$|\Delta\phi_{us}|<0.5$,
\item
$|\cos\theta_{\ell\ell}|>0.7$,
\end{itemize}
are applied.
The choice of the cut value is justified by Figure~\ref{DMtt_cutm1},
where the distributions of these two variables are shown
after the cuts described above, except the cut on the plotted variable,
and with a requirement of $\mttwo>120$~GeV.\par
The $\mttwo$ distribution after all analysis cuts  is shown in the right
side of Figure~\ref{DMtt_mttwo}.\par
After all  the requirements are applied the dominant background
is the irreducible $t\bar{t}Z$ background with $Z\rightarrow\nu\nu$.
The final optimisation is performed by scanning values for the lower
cut in $\mttwo$  in the range of 40-200~GeV.
For each signal point, values of the dimensionless
couplings couplings $g_v$ and $g_\chi$, for $g_v=g_\chi$ between 0.1 and 3.5
are considered, and for each point and each value of the 
couplings, the value of the threshold in $\mttwo$ is 
found which optimises signal significance. The significance is
defined  through the $Z_n$ variable \cite{Linnemann:2003vw},
based on the convolution of a poisson function with a gaussian function
modelling the systematic uncertainty on background evaluation.
The value of such uncertainty is assumed to be 20\%  of the
estimated background contribution.\par
\section{Results}
\begin{figure}
\begin{center}
\includegraphics[width=0.49\textwidth]{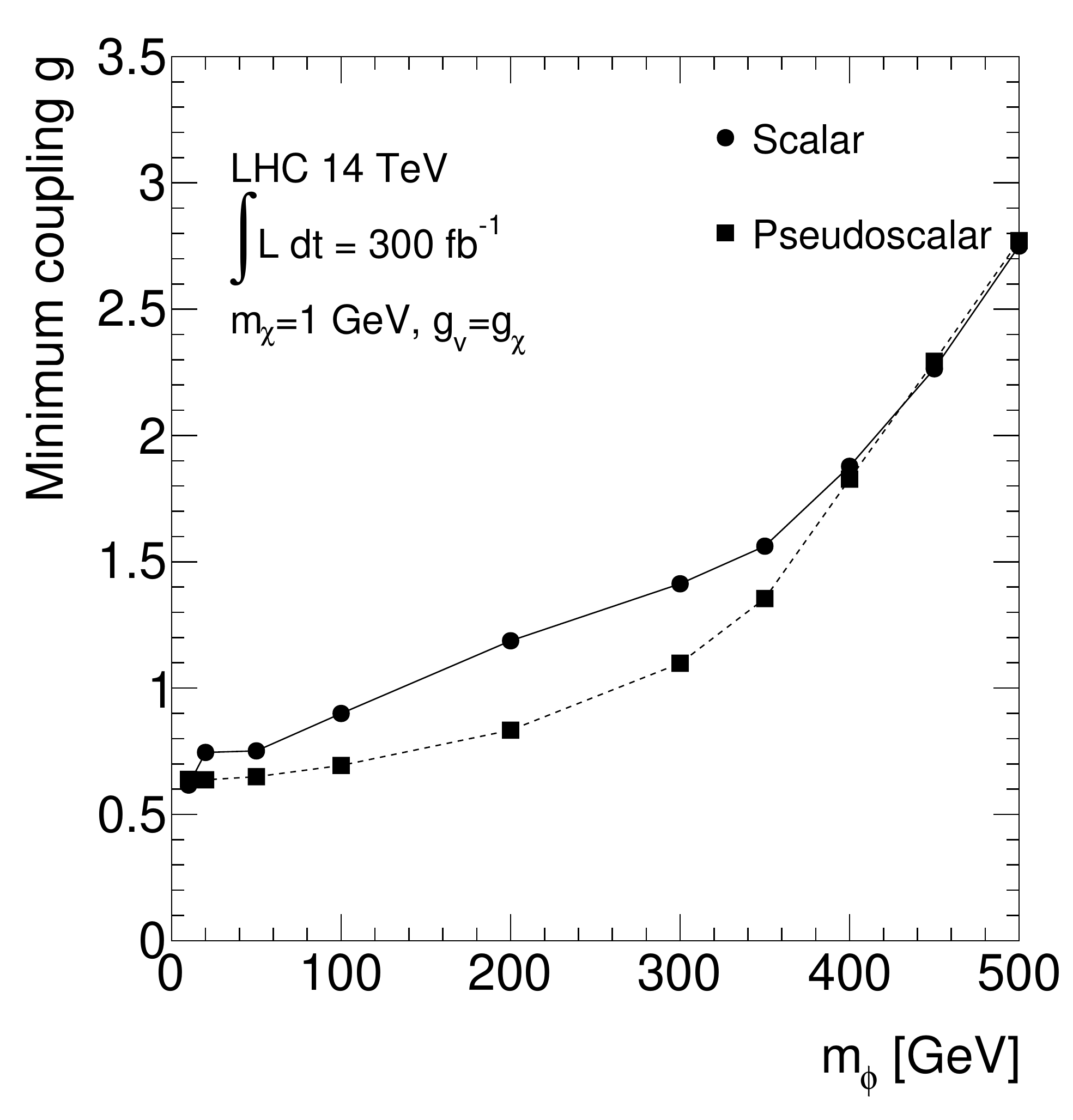}
\caption{Minimal value of the couplings  $g_v$ $g_\chi$, for $g_v=g_\chi$
and $m_\chi=1$~GeV yielding a 2-sigma significance as a function of 
the mass of the mediator $m_\phi$, for an integrated luminosity
of 300~fb$^{-1}$ proton-proton collisions at 14~TeV. 
The round bullets connected by a full line are for a scalar mediator, 
squre bullets connected by a dashed line for a ppsudoscalar mediator.}
\label{DMtt_2lep}
\end{center}
\end{figure}

The sensitivity of the analysis, for each model point given in terms
of the mediator mass $m_\phi$ and of the DM mass $m_\chi$, is given 
by the minimum value of the couplings $g_v=g_\chi$ which gives 
a significance of 2 
above the SM backgrounds for a given model point. Such a value is shown
in Figure~\ref{DMtt_2lep} as a function of $m_\phi$, for
a $m_\chi=1$~GeV and for an integrated luminosity of 
300~fb$^{-1}$ for a 14 TeV LHC. The analysis is sensitive to 
a $g_v=g_\chi<1$ for $m_\phi$ below $\sim$150 GeV
for a scalar mediator and below $\sim$250~GeV for a pseudoscalar mediator.
The difference of the two curves is due to the fact 
that for equal values of the parameters, the pseudoscalar mediator 
has a smaller cross-section but a harder $\etmiss$ spectrum than the 
scalar mediator, and thus the sensitivity evolves differently
with $m_\phi$.\par

\section*{Conclusions}
The sensitivity of a 14~TeV LHC to the associated production 
of $t\bar{t}$ and dark matter  has been studied in the framework of 
a simplified model with scalar and pseudoscalar mediators. 
An analysis for an integrated luminosity
of 300~fb$^{-1}$ has been developed based on the signature 
with two leptons in the final state, showing a sensitivity 
to values of the couplings below 1 for masses below $\sim$150~GeV for
a scalar mediator and $\sim$250~GeV for a pseudoscalar mediator.\par
 
\section*{ACKNOWLEDGEMENTS}
We would like to thank the
Organizers of the 2015 ``Les Houches - Physics at TeV colliders" workshop, where
this work was initiated for the excellent organisation, and the very friendly
and productive atmosphere.

%


\AddToContent{P.~Pani and G.~Polesello}
\renewcommand{\thesection}{\arabic{section}}

\graphicspath{{monohiggs/}}

\newcommand{\F}{\mathcal{F}}
\newcommand{\s}{\sigma}
\newcommand{\A}{\mathcal{A}}
\newcommand{\U}{\mathbf{U}}
\newcommand{\T}{\mathbf{T}}
\newcommand{\V}{\mathbf{V}}
\newcommand{\Y}{\mathbf{Y}}
\newcommand{\LL}{\mathcal{L}}
\newcommand{\de}{\partial}



\chapter{Search for dark matter via mono-Higgs production at the LHC}

{\it I.~Brivio, D.~Burns, N.~De Filippis, N.~Desai, J.M.~No, H.~Prosper, S.~Sekmen, D.~Schmeier and J.~Sonneveld}



\begin{abstract}
A  study of searches for dark matter in association with a Higgs boson at the Large Hadron Collider (LHC) is presented.  
In particular, the study focused 
on the decay of the Higgs boson into two Z bosons, which subsequently decay into charged leptons (electron or muons). 
Several Effective Field Theories (EFTs) and Simplified Models  predicting Dark Matter-Higgs interactions are investigated.  
The experimental signature of large missing transverse momentum and the visible products of the Higgs boson decay, the so-called ``mono-Higgs'' signature, is 
studied  using a fast simulation of the ATLAS and CMS detectors. 
Monte Carlo samples for the signal and the background are used to model the most relevant observables and develop 
cut-based analyses, optimized using a simple measure of significance.  
Results are given in terms of the predicted number of events for signal and background assuming a data
set of
3000\,fb$^{-1}$,
the maximum anticipated integrated luminosity of the LHC. It is found that for several
well-motivated models, a discovery of dark matter particles is possible. For the other models, 
 a Bayesian approach is used to determine the reach of these searches, in terms of 95\% C.L. upper limits on the production 
cross section.
\end{abstract}

\section{INTRODUCTION}
The Standard Model (SM) of particle physics~\cite{Glashow:1961tr, Weinberg:1967tq, salam} 
currently gives the best description of fundamental interaction phenomena in high energy physics. However, despite the excellent agreement between theoretical
predictions and experimental results, it is known that the SM needs to be extended in order 
to explain 
several physical phenomena such as the existence of dark matter, one of the most enduring mysteries in science. Indeed, new data from the Planck satellite launched by the European Space Agency~\cite{Ade:2013zuv} and several other observations and measurements suggest that only about 5\% of our Universe is made of visible matter, while the largest component consists of dark matter (DM) and dark energy, whose nature and composition is unknown. 
If dark matter is particulate in nature, it is expected to be long-lived, non-relativistic, with no electric or color charge, very weakly interacting with SM particles and subject to the gravitational interaction. Any progress towards the discovery of dark matter would be a dramatic breakthrough in our field.

The discovery of a new boson at the LHC with a mass of 125\,GeV by the ATLAS~\cite{Aad:2012tfa} and CMS~\cite{Chatrchyan:2012xdj} collaborations in 2012, with properties consistent with those of  the SM Higgs boson, not only sheds light on the Brout-Englert-Higgs (BEH) 
mechanism~\cite{Englert:1964et,Higgs:1964ia,Higgs:1964pj,Guralnik:1964eu,Higgs:1966ev,PhysRev.155.1554} within the SM but also provides an additional probe of physics beyond the Standard Model in the DM sector.  DM particles could be produced in pairs at the LHC, but would escape detection being stable and weakly interacting. Tagging such events  is extremely challenging. For that reason, collider searches for weakly interacting massive particles (WIMPs) rely on their recoil off a visible SM particle produced by initial state radiation, either a jet, a photon, a Z boson, or a Higgs  boson.
In the SM, however, the probability to radiate a Higgs boson is small because of the small parton-Higgs boson couplings. On the other hand, 
a sizable effective coupling of the Higgs boson to DM particles  is predicted in well-motivated theoretical models, as reported in the next sections.

Experimentally, such ``mono-Higgs''  events are characterized by the presence of a Higgs boson and non negligible missing transverse momentum due to the 
undetected dark matter particles.

\section{MODELS and SIGNATURES}

We assume for this study that the DM particle is either a scalar or a Dirac fermion.  The mono-Higgs signatures studied here can be classified into (1) EFT 
operators up to dim-8 (2) Simplified models with a (pseudo-) scalar mediator and (3) Simplified models with a vector mediator.  

\subsection{Effective Field Theory}
\label{SectionEFT}
We consider operators up to dimension-8 for the EFT study~\cite{Carpenter:2013xra}. We start with the simplest case of scalar DM which couples to the Higgs sector via a portal term
\begin{equation}\label{Higgs_portal}
\mathcal{L}_4 = \lambda |H|^2 \chi^2
\end{equation}

For fermionic DM, the lowest addition we can make to the SM is at dim-5.  The terms relevant for mono-Higgs searches are:
\begin{equation}
\mathcal{L}_5 \supset \frac{c_1}{\Lambda} |H|^2 \bar \chi \chi + \frac{c_2}{\Lambda} |H|^2 \bar \chi i \gamma_5 \chi
\end{equation}

\noindent At dim-6, we have:
\begin{equation}
\mathcal{L}_6 \supset \frac{c_3}{\Lambda^2} \bar \chi \gamma^\mu \chi H^\dag iD_\mu H + \frac{c_4}{\Lambda^2} \bar \chi i \gamma^\mu \gamma_5  \chi H^\dag iD_\mu H 
\end{equation}

\noindent Finally, we investigate the minimal operator that results in a mono-Higgs signature where the Higgs and the DM are produced through coupling of the Higgs and 
the DM with an EW field tensor.  The simplest of these operators appears at dim-8 and is:
\begin{equation}
\mathcal{L}_8 \supset \frac{c_5}{\Lambda^4} \bar \chi \gamma^\mu B_{\mu \nu} \chi H^\dag iD^\nu H\,.
\end{equation}

\subsection{Scalar mediator model}
\label{sec:scalar}
The simplest model with a scalar mediator simply contains a Yukawa-like coupling of the new scalar to the DM
\begin{equation}
\mathcal{L} \supset - y_\chi \bar \chi \chi S
\end{equation}

\noindent The coupling of S with the SM is through the full potential of the scalar sector which also contains the SM Higgs.
\begin{equation}
V \supset a |H|^2 S + b |H|^2 S^2 + \lambda_h |H|^4 + ...
\end{equation}
This results in mixing between the SM Higgs and the scalar S. The scalar then couples to the SM particles only through mixing with the Higgs
(see e.g.~\cite{Carpenter:2013xra}).

\subsection{Pseudoscalar Portal to Dark Matter - Two Higgs Doublet Model (2HDM)}
\label{Section2HDM}

Dark matter-SM particle interactions mediated by a pseudoscalar particle---a possible portal to a dark matter sector---are highly motivated, evading constraints from 
direct detection experiments
as well as providing a potential explanation of the observed Galactic Center gamma ray excess. It has been recognized that such a pseudoscalar dark matter
portal could yield mono-Higgs signatures at the LHC \cite{Berlin:2014cfa,Berlin:2015wwa,No:2015xqa} as a primary discovery avenue.
Here we discuss the main features of these models in connection to LHC mono-Higgs signatures. 

The minimal renormalizable model featuring a pseudoscalar state is the 2HDM~\cite{Nomura:2008ru},
extending the SM Higgs sector to include two scalar $SU(2)_L$ doublets $H_i$ ($i=1,2$).
The scalar potential of the 2HDM reads 
\begin{eqnarray}	
\label{2HDM_potential}
V_{\mathrm{2HDM}} &= &\mu^2_1 \left|H_1\right|^2 + \mu^2_2\left|H_2\right|^2 - \mu^2\left[H_1^{\dagger}H_2+\mathrm{h.c.}\right] 
+\frac{\lambda_1}{2}\left|H_1\right|^4 +\frac{\lambda_2}{2}\left|H_2\right|^4 \\ 
&+& \lambda_3 \left|H_1\right|^2\left|H_2\right|^2 + \lambda_4 \left|H_1^{\dagger}H_2\right|^2+ \frac{\lambda_5}{2}\left[\left(H_1^{\dagger}H_2\right)^2+\mathrm{h.c.}\right]\nonumber 
\end{eqnarray}
where CP conservation and a $Z_2$ 
symmetry, softly broken by $\mu^2$, are assumed. The spectrum of the 2HDM contains a charged scalar $H^{\pm}$, two neutral CP-even scalars $h$, $H_0$, and a pseudoscalar $A_0$ 
(see e.g.~\cite{Branco:2011iw} for a review of 2HDM). We identify $h$ with the 125 GeV Higgs 
boson state.

There are various possible ways in which a 2HDM can yield a pseudoscalar portal to a dark sector,
which we discuss in the following. A minimal embedding of dark matter into the pseudoscalar portal scenario~\cite{Ipek:2014gua}
corresponds to considering dark matter as a singlet Dirac fermion $\chi$ with mass $m_{\chi}$, coupled to a real singlet pseudoscalar state $a_0$
\begin{equation}
\label{Ldark}
 V_{\mathrm{dark}} = \frac{m^2_{a_0}}{2}\,a_0^2 + m_{\chi}\, \bar{\chi}\chi + y_{\chi}\,a_0 \,\bar{\chi} i\gamma^{5} \chi\, .
\end{equation}
The portal between the visible and dark sectors occurs via~\cite{Nomura:2008ru,Ipek:2014gua}
\begin{equation}
\label{Vportal}
V_{\mathrm{portal}} = i\,\kappa\,a_0 \,H_1^{\dagger}H_2 + \mathrm{h.c.}
\end{equation}
which induces a mixing between the two pseudoscalar states $A_0$ and $a_0$ (we stress that $A_0$ cannot couple to the dark matter state $\chi$ in the absence of this mixing),
yielding two mass eigenstates $A$, $a$ ($m_A > m_a$). In this context, gauge interactions of the two doublets $H_i$ yield the interactions $aZh$ and $AZh$, while 
$V = V_{\mathrm{2HDM}} + V_{\mathrm{dark}} + V_{\mathrm{portal}}$ yield the interactions $aAh$, $a\bar{\chi}\chi$ and $A\bar{\chi}\chi$ (see~\cite{No:2015xqa,Ipek:2014gua} for details).

\begin{figure}[t!]\centering
\input{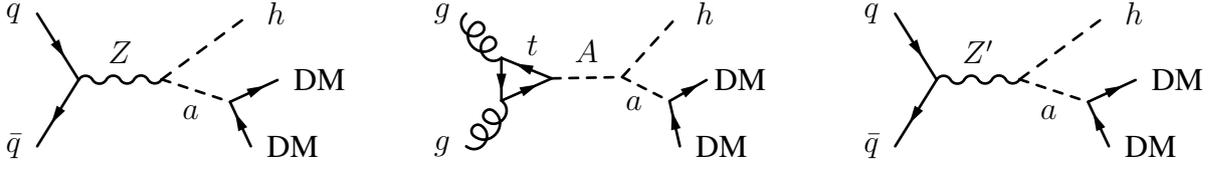}
\caption{Feynman diagrams contributing to mono-$h$ production in pseudoscalar (here denoted $a$) portal dark matter scenarios: 
Left: via an off-shell s-channel $Z$-boson. Middle: via a resonantly produced heavy pseudoscalar $A$. Right: Via a resonantly produced $Z'$ boson.
}\label{fig:PseudoscalarDM}
\end{figure}

The interactions above lead to two kinds of mono-$h$ signatures: {\it (i)} $p p \to Z^* \to a\,h \,\,(a \to \bar{\chi}\, \chi)$, 
$p p \to Z^* \to A\,h \,\,(A \to \bar{\chi}\, \chi)$, with the production of $a\,h$ mediated by an off-shell $Z$ boson (Figure \ref{fig:PseudoscalarDM}-Left). 
{\it (ii)} $p p \to A \to a\,h\,\,(a \to \bar{\chi}\, \chi)$, with the production of $a\,h$ mediated by an on-shell $A$ state (produced in gluon fusion)
(Figure \ref{fig:PseudoscalarDM}-Middle) for $m_A > m_h + m_a$~\cite{No:2015xqa}.

Alternatively, a pseudoscalar portal to dark matter within the 2HDM may be achieved if dark matter (or some field(s) in the dark sector) has $SU(2)_L$ 
quantum numbers~\cite{Berlin:2015wwa}, allowing it to directly couple to $A_0$. While dark matter direct detection constraints pose a challenge 
to this type of scenario, due to a potential direct coupling of dark matter to the $Z$ boson, it it possible to evade this constraint in specific setups
(see~\cite{Berlin:2015wwa} for details). Through an interaction $A_0Zh$, these models yield a mono-Higgs signature of the kind {\it (i)} discussed above:
$p p \to Z^* \to A_0\,h \,\,(A_0 \to \bar{\chi}\, \chi)$, $\chi$ being the dark matter particle (Figure \ref{fig:PseudoscalarDM}-Left).

Finally, in the presence of an extra spontaneously broken $U(1)$ gauge symmetry, it is possible for a coupling between the $Z'$ gauge boson, the Higgs $h$ and the pseudoscalar 
$a_0$ to arise~\cite{Berlin:2014cfa}. For $m_{Z'} > m_h + m_{A_0}$ this gives rise to a mono-Higgs signature (assuming that $a_0$ mediates interactions between the SM and 
the dark sector) with $a_0$ and $h$ coming from the decay of a resonantly produced $Z'$ (Figure \ref{fig:PseudoscalarDM}-Right). We however consider 
this scenario as part of the general $Z'$ scenarios of Section~\ref{ZprimeModels}, and discuss it in detail in Section~\ref{Zprime2HDM}.

\vspace{2mm}

We stress that the kinematical features of the mono-Higgs signature are qualitatively different between off-shell s-channel production via a $Z$ boson 
and resonant production. In the latter, denoting by $X$ the s-channel resonance (e.g. an $A$ or $Z'$ state as discussed above) 
and the pseudoscalar portal mediator by $Y$ (we assume $m_X > m_h + m_Y$), the decay of $X$ fixes the    
4-momentum of $h$ and $Y$, such that if $X$ is produced at rest, the $E_{T}\hspace{-4mm}/$\hspace{2mm} distribution 
is a steeply rising function with a sharp cut-off at $E^{\mathrm{max}}_{T}\hspace{-7.5mm}/$\hspace{5.5mm}, 
\begin{equation}
\label{ETmax}
E^{\mathrm{max}}_{T}\hspace{-7.5mm}/ \hspace{5.5mm}= \frac{1}{2\, m_{X}} \sqrt{(m^2_X - m^2_h - m^2_Y)^2 - 4 \,m^2_h\,m^2_Y}\, .
\end{equation}
a very distinct feature of these scenarios~\cite{No:2015xqa}.

\vspace{2mm}

In the following, we consider two benchmark scenarios for mono-Higgs signatures at the LHC based on the 2HDM pseudoscalar portal to dark matter models discussed above, and 
defined via eqs.~(\ref{2HDM_potential})-(\ref{Vportal}). We consider a Type II 2HDM (see~\cite{Branco:2011iw} for details) 
with $\mathrm{tan}\beta = 3$, $\mathrm{cos} (\beta-\alpha) = 0.05$ (close to the 2HDM alignment limit) and $m_{H^{\pm}} = m_{H_0} = m_{A}$. 
The mixing between the visible and dark sectors (between $A_0$ and $a_0$, see the discussion above) is set to $\mathrm{sin} \theta = 0.3$, 
and we fix $y_{\chi} = 0.2$, $m_a = 80$\,GeV, $m_{\chi} = 30$\,GeV (see~\cite{No:2015xqa,Ipek:2014gua} for details). 
The two benchmark scenarios correspond to $m_A = 500$ GeV and $m_A = 700$\,GeV, respectively, and we analyze in this work the mono-Higgs prospects of the second 
benchmark (see Table~\ref{nevents2HDM}).



\subsection{Z$^\prime$ Models}
\label{ZprimeModels}

Another simple model that predicts the mono-Higgs signature is the vector mediator model where a neutral vector $Z^{\prime}$ serves as a mediator between 
the dark sector and the SM.  In the simplest case, the $Z^{\prime}$ corresponds to a $\mathrm{U(1)}$ under which the DM is charged.  This ensures its stability. 
The coupling of the new $Z^{\prime}$ to the SM is model dependent and several options have been considered. e.g. that the U(1) results from gauging baryon number 
or that the $Z^{\prime}$ is hidden and mixes with the SM $Z$~\cite{Carpenter:2013xra}. A non-minimal option is to also add a pseudoscalar $a_0$ which then couples 
to the DM candidate~\cite{Berlin:2014cfa}.  

Assuming the DM is a Dirac fermion, the part of the Lagrangian with couplings of the $Z^{\prime}$ can be written in general as
\begin{equation}
\label{LZp}
\mathcal{L} \supset g_q \bar q \gamma^\mu q Z^{\prime}_\mu + g_\chi \bar \chi \gamma^\mu \chi Z^{\prime}_\mu
\end{equation}

\begin{figure}\centering
\input{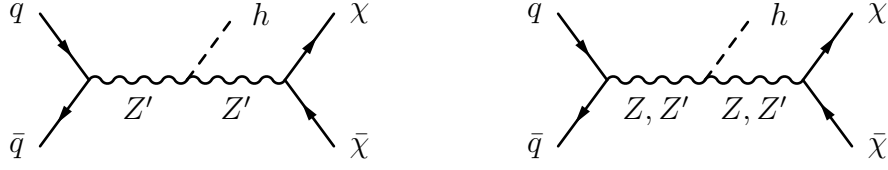}
\caption{\label{fig:zprime} Feynman diagrams for mono-Higgs final states obtained via associated production in the baryonic-$Z^{\prime}$ model (left) and 
the hidden--$Z^{\prime}$ model (right).}
\end{figure}

A mono-higgs signature is obtained mainly via associated production with the $Z^{\prime}$ (see fig.\ref{fig:zprime}).  Therefore, besides the fermionic 
couplings above, we need the Z' to couple to the Higgs.  Thus, the main parameters of the model are the couplings $g_q, g_\chi$ and the mass $m_{Z^{\prime}}$ 
and the coupling $g_{hZ^{\prime}Z^{\prime}}$.

\subsubsection{Baryonic $Z^{\prime}$}
\label{ZpBModel}

The Baryonic $Z^{\prime}$, first suggested in~\cite{Carpenter:2013xra}, assumes that the $Z^{\prime}$ is the gauge boson resulting from gauging the $\mathrm{U(1)}_B$ via 
a ``baryonic higgs'' $h_B$ with a vev $v_B$.  The baryonic Higgs mixes with the SM higgs with a mixing angle $\theta$ which results in a term of the form 
\begin{equation}
\label{LZpB}
\mathcal{L} \supset -g_{hZ^{\prime}Z^{\prime}} hZ^{\prime}Z^{\prime},~~g_{hZ^{\prime}Z^{\prime}} \equiv  \frac{m^2_{Z^{\prime}} \sin \theta} {v_B}
\end{equation}

The couplings to the fermions are given by $g_q = g_B/3$, where $g_B$ is the gauge coupling and $g_\chi = B_\chi g_B$ where $B_\chi$ is the baryon number of the DM.  

In the following we consider benchmark scenarios for signatures based on the Baryonic $Z^{\prime}$ portal to dark matter, defined via eqs.~(\ref{LZp})-(\ref{LZpB}), 
with $g_B = B_\chi = 1$ and $g_{hZ^{\prime}Z^{\prime}}/m_{Z^{\prime}}=0.3$. The dark matter and $Z^{\prime}$ masses are scanned over the ranges 1-1000 GeV and 10-10000 GeV, 
respectively. Benchmarks corresponding to a subset of these mass scans are summarized in Table~\ref{tab:MMVector}.

\subsubsection{Hidden $Z^{\prime}$}
\label{ZpHModel}

The Hidden $Z^{\prime}$ model simply assumes an extra $\mathrm{U(1)}$ gauge boson ($Z_H$) that does not directly couple to the SM.  The $Z_H$ mixes with the $Z_{SM}$ to 
form mass eigenstates $Z$ and the heaver $Z^{\prime}$
\begin{equation}
Z_{SM}= \cos \theta Z + \sin \theta Z^{\prime}, ~ Z_H= \sin \theta Z + \cos \theta Z^{\prime}
\end{equation}

\noindent The mono-Higgs production is then again through ``associated'' production with $Z/Z^{\prime}$.  The coupling of the Higgs to the $Z^{\prime}$ is then given by
\begin{equation}
\mathcal{L} \supset \frac{m_Z^2 \sin^2 \theta}{v} Z^{\prime}_\mu Z{\prime}^\mu
\end{equation}

\noindent This results also in an SU(2) violating coupling of the form
\begin{equation}
\mathcal{L} \supset \frac{m_Z^2 \sin \theta}{v} Z^{\prime}_\mu Z^\mu
\end{equation}

\noindent The fermionic couplings can be obtained from the original couplings of the $Z_{SM}$ and the $Z_H$ 
\begin{equation}
\mathcal{L}  \supset \frac{g_2}{2 \cos \theta_w} J_\mu^{NC} Z_{SM}^\mu + g_\chi \bar \chi \gamma_\mu \chi Z_H^\mu
\end{equation}
where $J_\mu^{NC}$ is the SM neutral current.

In the following we consider benchmark scenarios for signatures based on the Hidden $Z^{\prime}$ portal to dark matter, defined above, with $g_\chi = 1$ and $\sin \theta = 0.1$. 
The dark matter and $Z^{\prime}$ masses are scanned over the ranges 1-1000 GeV and 10-10000 GeV, respectively. 

\subsubsection{Non-minimal $Z^{\prime}$ model with pseudoscalar $a_0$}
\label{Zprime2HDM}

This model combines the hidden-$Z^{\prime}$ and 2HDM models described above.  The mono-Higgs signature then arises from the right-most diagram of 
fig.~\ref{fig:PseudoscalarDM} rather than the left-most.  We therefore have an added parameter, the coupling of the $a_0$ to the DM ($y_\chi$).
\begin{equation}
\mathcal{L} \supset i y_\chi \bar \chi \gamma^5 \chi a_0
\end{equation}

The two Higgs fields $\Phi_u$ and $\Phi_d$ are also charged under the new U(1) with charges $z_u$ and $z_d$ respectively.  We also rewrite the $Z$-$Z^{\prime}$ 
mixing in terms of the mixing parameter $\epsilon$~\cite{Berlin:2014cfa}.

\begin{eqnarray}
Z^\mu  & \approx & W^{3\mu} \cos \theta_w - B_Y^\mu \sin \theta_w + \epsilon Z_H \nonumber \\
Z^{' \mu} & \approx & Z_H - \epsilon (W^{3\mu} \cos \theta_w - B_Y^\mu \sin \theta_w)
\end{eqnarray}

\subsection{Non-linear Higgs Portal to Scalar Dark Matter}
\label{scalar}
Another pertinent framework for the generation of mono-Higgs signatures is that of a non-linear Higgs portal~\cite{Brivio:2015kia}. This scenario 
generalizes the Higgs portal of Eq.~\eqref{Higgs_portal} to the context of non-linearly realized electroweak symmetry breaking, in which the 
lightness of the Higgs particle results from its being a pseudo-Goldstone boson of some global symmetry, spontaneously broken by strong dynamics at a high energy 
scale $\Lambda$. 
Here we review the formulation of this model and comment on its main mono-Higgs production modes. The experimental simulation analysis for this particular EFT, however, has been deferred to a future work, as it deserves a more detailed study.

Within the non-linear EWSB framework, the longitudinal components of the $W^\pm$ and $Z$  gauge bosons, denoted here by  $\pi(x)$, can be described at the EW scale 
by a dimensionless unitary matrix $\U(x)\equiv e^{i\sigma_a \pi^a(x)/v}$ that transforms as a bi-doublet under a (global) $SU(2)_L\times SU(2)_R$ symmetry:
$\U(x)\mapsto L\U(x) R^\dagger$.
At the same time, the physical Higgs particle does not behave as an exact  $SU(2)_L$ doublet and can be parametrised as a generic SM scalar singlet. Its couplings 
are therefore arbitrary, and thery are customarily encoded into generic functions
$ \F_i(h) = 1+2a_i\frac{h}{v}+b_i\frac{h^2}{v^2}+\dots$
that replace the typical SM dependence on $(v+h)^n$.
While in linear BSM scenarios, $h$ and $\U(x)$ are components of the same object, i.e. the $SU(2)_L$ Higgs doublet
$\Phi \sim(v+h) \, \U \binom{0}{1}$,
$h$ and $\U(x)$ are independent in the non-linear formalism, and this induces a different pattern of dominant couplings.
At the leading order, the effective Lagrangian for the $\chi $ field contains the terms 
\begin{equation}
\LL_\chi \supset\frac{1}{2}\de_\mu \chi  \de^\mu \chi - \dfrac{m_\chi ^2}{2}  \chi ^2 
-\lambda_{\chi }\chi ^2\left(2vh+b h^2\right)+\sum_{i=1}^5 c_i\A_i(h) 
\label{L0S}
\end{equation}
The parameter $b$ ensures an arbitrary relative weight between the $\chi \chi h$ and $\chi \chi hh$ couplings.
The coefficients $c_i$ are real and of order one, and the operators $\A_i$ form a complete basis:
\begin{equation}
\begin{aligned}
\A_1&=\tr(\V_\mu\V^\mu)\chi ^2\F_1(h)&
\A_2&=\chi ^2 \square \F_{2}(h)\\
\A_3&=\tr(\T\V_\mu)\tr(\T\V^\mu ) \chi ^2 \F_{3}(h)\quad&
\A_4&=i\tr(\T\V_\mu) (\de^\mu \chi ^2) \F_{4}(h)\\
\A_5&=i\tr(\T\V_\mu)\chi ^2\de^\mu \F_{5}(h)
\end{aligned}
\label{scalar.op}
\end{equation} 
where the scalar and vector chiral fields are defined respectively as
${\T\equiv \U \sigma_3 \U^\dagger}$,  ${\V_\mu\equiv \left(D_\mu \U\right)\U^\dagger}$
and transform in the adjoint of $SU(2)_L$. The covariant derivative of the Goldstone bosons matrix is
$D_\mu \U(x) \equiv \de_\mu \U(x) +igW_{\mu}^a(x)\sigma_a\U(x)/2 - ig' 
B_\mu(x) \U(x)\sigma_3/2$.

The five effective operators in Eq.~\eqref{scalar.op} describe interactions between two $\chi $ particles and either two $W$ bosons, one or two $Z$ or $h$ bosons, or a 
$Z$ and a $h$ boson, providing a larger parameter space and a much richer phenomenology with respect to the linear Higgs portal in Eq.~\eqref{Higgs_portal} 
(see Ref.\cite{Brivio:2015kia} for details). $\A_1$ and $\A_2$ are custodial invariant couplings, while $\A_3$, $\A_4$ and $\A_5$  violate the custodial 
symmetry, due to the presence of the spurion $\T$. 
Finally, the operators $\A_1$, $\A_2$ and $\A_3$ are CP-even, while $\A_4$ and $\A_5$ are CP-odd. 


\begin{figure}[t!]\centering
\input{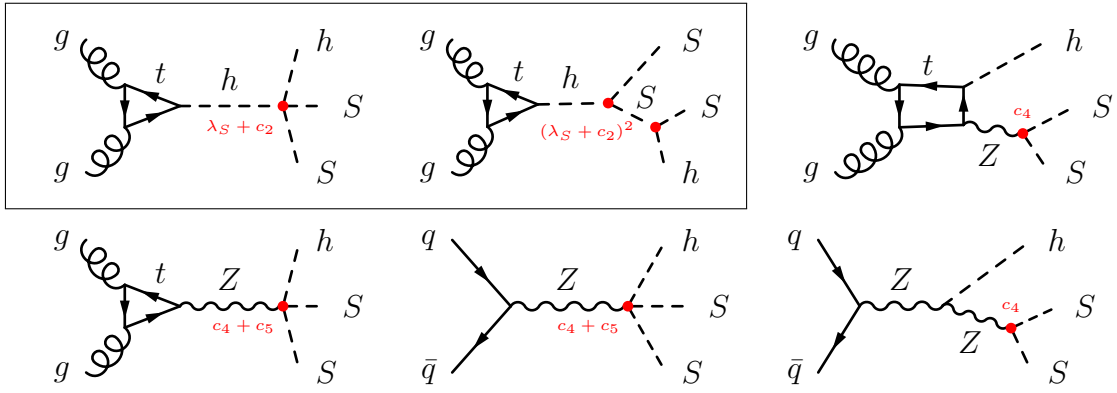}
\caption{Sample of the main Feynman diagrams contributing to mono-$h$ production for the non-linear Higgs portal model. In the non-linear scenario 
both $gg$- and $\bar{q}q$-initiated processes are included, while in the standard Higgs case (Eq.~\eqref{Higgs_portal}) only those inside the frame are present: the process is entirely $gg$-initiated, 
with contributions proportional to $\lambda_\chi $ and to $\lambda_\chi ^2$. The proportionality of each diagram to the non-linear parameters is indicated in the figure (overall 
factors and numerical coefficients are not specified).
}\label{diagram_monoHprod}
\end{figure}
A sample of the Feynman diagrams contributing to mono-$h$ in this model is shown in Figure \ref{diagram_monoHprod}. 
The most relevant features of the non-linear Higgs portal scenario are 
\begin{itemize}
 \item the presence of direct couplings of the DM to $Z$-bosons (from $\A_4$ and $\A_5$), which yield a $\bar{q}q$-initiated mono-$h$ contribution 
 \item the explicit dependence on the momentum in the $\chi $-$h$, $\chi $-$Z$ and $\chi $-$h$-$Z$ interactions (from $\A_2$, $\A_4$ and $\A_5$).
\end{itemize}
 In presence of at least one among the operators $\A_2$, $\A_4$, $\A_5$ the total cross section $\s(pp\to h \chi \chi )$ at $\sqrt{s}=13$ TeV would lay 
 between $10^{-2}\times c_i^2$ fb and $10\times c_i^2$ fb, depending on the DM mass. This corresponds to an enhancement of four orders 
 of magnitude {\it w.r.t.}~the standard Higgs portal scenario  for $m_\chi \gg v$. At the same time, the differential distribution of the Higgs boson transverse 
 momentum $P^h_T$ would be shifted towards larger values. This holds particularly for the case of $\A_5$ and represents a landmark signature of non-linear 
 Higgs portals, that may also allow for a much better signal extraction from the SM background. 
 In addition, a more modest but still appreciable enhancement of the total mono-$h$ rate may also come from the $b$ parameter (defined in Eq.~\eqref{L0S}) taking a value 
larger than unity~\cite{Brivio:2015kia}.

\section{SIMULATION AND CUT-BASED ANALYSIS}

\subsection{Data Samples and Tools}

In order to assess the feasibility of testing the prediction of each model introduced in the previous section, simulated events for signal and background were generated with Monte Carlo programs.
A summary of the cross sections values for all the signal models and the values of the parameters used for the simulation are reported in the
 Tables~\ref{tab:efts}-- \ref{tab:scalarxsecs}. 

\begin{table}[hp]
\renewcommand{\arraystretch}{1.2}
\begin{center}
\caption{Effective Field Theory Models.}\label{tab:efts}
\begin{tabular}{| l | c | l | c | c |}
\hline
Name & Operator & Param. & Dim. & $S_{\chi}$ \\
\hline
EFT$\_$HHxx$\_$scalar & $\lambda |H|^{2} \chi^{2}$ & $m_{\chi}, \lambda = 0.1$ & 4 & 0 \\
EFT$\_$HHxx$\_$combined & $\frac{1}{\Lambda} |H|^{2} \bar{\chi} \chi$ & $m_{\chi}, \Lambda = 1000$ GeV & 5 & 1/2 \\
EFT$\_$HHxxg5x & $\frac{1}{\Lambda} |H|^{2} \bar{\chi} i \gamma_{5} \chi$ & $m_{\chi}, \Lambda = 100$ GeV & 5 & 1/2\\
EFT$\_$xdxHDHc & $\frac{1}{\Lambda^{2}} \chi^{\dag} i \partial^{\mu} \chi H^{\dag} i D_{\mu} H $ & $m_{\chi}, \Lambda = 100,1000$ GeV & 6 & 0\\
EFT$\_$xgxFHDH & $\frac{1}{\Lambda^{4}} \bar{\chi} \gamma^{\mu} \chi B_{\mu\nu} H^{\dag} D^{\nu} H$ & $m_{\chi}, \Lambda = 200$ GeV & 8 & 1/2\\
\hline
\end{tabular}

\smallskip

%
\caption{Mass points for models with a vector mediator.}\label{tab:MMVector}
\begin{tabular}{| l | c | c | c | c | c | c | c | c | c | c|}
\hline
$M_\chi$ [GeV] & \multicolumn{10}{c|}{$M_{Z'}$ [GeV]} \\
\hline
1 & 10 & 20 & 50 & 100 & 200 & 300 & 500 & 1000 & 2000 & 10000 \\
10 & 10 & 15 & 50 & 100 & & & & & & 10000 \\
50 & 10 & & 50 & 95  & 200 & 300 & & & & 10000 \\
150 & 10 & & & & 200 & 295 & 500 & 1000 & & 10000 \\
500 & 10 & & & & & & 500 & 995 & 2000 & 10000 \\
1000 & 10 & & & & & & & 1000 & 1995 & 10000 \\
\hline
\end{tabular}

\smallskip

%
\caption{Mass points for Zp2HDM.} \label{tab:MM2HDM}
\begin{tabular}{ | l | c | c | c | c | c | c | c | c |}
\hline
$M_{A^0}$ [GeV] & \multicolumn{8}{c|}{$M_{Z'}$ [GeV]} \\
\hline
300 & 600 & 800 & 1000 & 1200 & 1400 & 1700 & 2000 & 2500 \\
400 & 600 & 800 & 1000 & 1200 & 1400 & 1700 & 2000 & 2500 \\
500 & & 800 & 1000 & 1200 & 1400 & 1700 & 2000 & 2500 \\
600 & & 800 & 1000 & 1200 & 1400 & 1700 & 2000 & 2500 \\
700 & & & 1000 & 1200 & 1400 & 1700 & 2000 & 2500 \\
800 & & & 1000 & 1200 & 1400 & 1700 & 2000 & 2500 \\
\hline
\end{tabular}

\smallskip

%

\caption{Mass points for models with a scalar mediator.}\label{tab:MMScalar}
\begin{tabular}{| l | c | c | c | c | c | c | c | c | c|}
\hline
$M_\chi$ [GeV] & \multicolumn{9}{c|}{$M_{S}$ [GeV]} \\
\hline
1 & 10 & 20 & 50 & 100 & 200 & 300 & 500 & 1000 & 10000 \\
10 & 10 & 15 & 50 & 100 & & & & & 10000 \\
50 & 10 & & 50 & 95  & 200 & 300 & & & 10000 \\
150 & 10 & & & & 200 & 295 & 500 & 1000 & 10000 \\
500 & 10 & & & & & & 500 & 995 & 10000 \\
1000 & 10 & & & & & & & 1000 & 10000 \\
\hline
\end{tabular}

\smallskip

%
\caption{Zp2HDM model (sect. \ref{Zprime2HDM}) production cross sections [fb] corresponding to mass points in Table~\ref{tab:MM2HDM} \label{tab:zp2hdmxsecs}}
\begin{tabular}{| l | c | c | c | c | c | c | c | c |}
\hline
$M_{A^0}$ [GeV] & \multicolumn{8}{c|}{$M_{Z'}$ [GeV]} \\
\hline
300 & 42.386 & 45.097 & 35.444 & 26.07 & 18.942 & 11.778 & 7.4456 & 3.6446 \\
400 & 5.8513 & 14.847 & 14.534 & 11.792 & 9.029 & 5.851 & 3.7819 & 1.8758 \\
500 & & 5.9605 & 8.4961 & 7.9575 & 6.5515 & 4.5063 & 3.0028 & 1.5235 \\
600 & & 1.5853 & 4.6972 & 5.4808 & 4.9946 & 3.7044 & 2.5694 & 1.3447 \\
700 & & & 2.1092 & 3.4848 & 3.6766 & 3.0253 & 2.2023 & 1.1984 \\
800 & & & 0.65378 & 1.9638 & 2.5511 & 2.4077 & 1.8689 & 1.0692 \\
\hline
\end{tabular}
\end{center}
\end{table}

\begin{sidewaystable}[hp]
\begin{center}
\caption{EFT model production cross sections [pb]\label{tab:eftxsecs}}
\begin{tabular}{| l | c | c | c | c | c | }
\hline
$m_\chi$ [GeV] & 1 & 10 & 50 & 65 & 100 \\
\hline
EFT$\_$HHxx$\_$scalar    & & & &  &  \\
                           & 0.10071 $\times 10^{1}$ & 0.99793 $\times 10^{0}$ & 0.60671 $\times 10^{0}$ & 0.48291 $\times 10^{-4}$ & 0.22725 $\times 10^{-5}$  \\
EFT$\_$HHxx$\_$combined &   & & & &   \\
                           & 0.15731 $\times 10^{1}$ & 0.15194 $\times 10^{1}$ & 0.34134 $\times 10^{0}$ & 0.41039 $\times 10^{-4}$ & 0.10581 $\times 10^{-4}$  \\
EFT$\_$HHxg5x              & & & &  & \\
                           & 0.15735 $\times 10^{3}$ & 0.15594 $\times 10^{3}$ & 0.94804 $\times 10^{2}$ & 0.12990 $\times 10^{-1}$ & 0.23075 $\times 10^{-2}$ \\
EFT$\_$xdxHDHc             & & &  &  &\\
\quad $\Lambda = 100$ GeV  & 0.29530 $\times 10^{0}$ & 0.29067 $\times 10^{0}$ & 0.10540 $\times 10^{0}$ & 0.89849 $\times 10^{-1}$ & 0.64959 $\times 10^{-1}$ \\
\quad $\Lambda = 1000$ GeV & 0.16306 $\times 10^{-4}$ & 0.15508 $\times 10^{-4}$ & 0.12088 $\times 10^{-5}$ & 0.88288 $\times 10^{-6}$ & 0.53312 $\times 10^{-6}$ \\
EFT$\_$xgxFHDH             & & & &  &\\
                           & 0.57027 $\times 10^{0}$ & 0.57001 $\times 10^{0}$ & 0.56025 $\times 10^{0}$ & 0.55337 $\times 10^{0}$ & 0.53270 $\times 10^{0}$ \\
\hline
$m_\chi$ [GeV] &  200 & 400 & 800 & 1000 & 1300 \\
\hline
EFT$\_$HHxx$\_$scalar      & & &  & & \\
                           & 0.11059 $\times 10^{-6}$ & 0.36569 $\times 10^{-8}$ & 0.40762 $\times 10^{-10}$ & 0.64956 $\times 10^{-11}$ & 0.51740 $\times 10^{-12}$ \\
EFT$\_$HHxx$\_$combined    & & &  & & \\
                           & 0.16553 $\times 10^{-5}$ & 0.14628 $\times 10^{-6}$ & 0.40608 $\times 10^{-8}$ & 0.85950 $\times 10^{-9}$ & 0.96480 $\times 10^{-10}$ \\
EFT$\_$HHxg5x              & & & &  & \\
                           & 0.41820 $\times 10^{-3}$ & 0.45743 $\times 10^{-4}$ & 0.16734 $\times 10^{-5}$ & 0.39327 $\times 10^{-6}$ & 0.49769 $\times 10^{-7}$ \\
EFT$\_$xdxHDHc             & & & & & \\
\quad $ \Lambda = 100$ GeV  & 0.30639 $\times 10^{-1}$ & 0.88644 $\times 10^{-2}$ & 0.97986 $\times 10^{-3}$ & 0.33847 $\times 10^{-3}$ & 0.68674 $\times 10^{-4}$ \\
\quad $\Lambda = 1000$ GeV & 0.18046 $\times 10^{-6}$ & 0.34918 $\times 10^{-7}$ & 0.27514 $\times 10^{-8}$ & 0.90662 $\times 10^{-9}$ & 0.19313 $\times 10^{-9}$ \\
EFT$\_$xgxFHDH             & & & & & \\
                           & 0.45792 $\times 10^{0}$ & 0.29777 $\times 10^{0}$ & 0.10288 $\times 10^{0}$ & 0.57444 $\times 10^{-1}$ & 0.23260 $\times 10^{-1}$ \\
\hline

\end{tabular}

\end{center}
\end{sidewaystable}

\begin{sidewaystable}[h]
\begin{center}
\caption{ZpHS model (sect. \ref{ZpHModel}) production cross sections [pb] corresponding to mass points in Table~\ref{tab:MMVector} \label{tab:zphsxsecs}}
\begin{tabular}{| l | p{1.5cm} | p{1.5cm} | p{1.5cm} | p{1.5cm} | p{1.5cm} | p{1.5cm} | p{1.5cm} | p{1.5cm} | p{1.5cm} | p{1.5cm} | p{1.5cm} | p{1.5cm} |}
\hline
$M_\chi$ [GeV] & \multicolumn{10}{c|}{$M_{Z'}$ [GeV]} \\
\hline
1 & 0.61935 $\times 10^{-2}$ & 0.63192 $\times 10^{-2}$ & 0.82991 $\times 10^{-2}$ & 0.11942 $\times 10^{-1}$ & 0.19171 $\times 10^{-1}$ & 0.21560 $\times 10^{-1}$ & 0.16010 $\times 10^{-1}$ & 0.64416 $\times 10^{-2}$ & 0.56526 $\times 10^{-2}$ & 0.58902 $\times 10^{-2}$ \\
 & & & & & & & & & & \\
10 & 0.58781 $\times 10^{-2}$ & 0.58938 $\times 10^{-2}$ & 0.82944 $\times 10^{-2}$ & 0.11937 $\times 10^{-1}$ & & & & & & 0.58805 $\times 10^{-2}$\\
 & & & & & & & & & & \\
50 & 0.10294 $\times 10^{-3}$ & & 0.77820 $\times 10^{-4}$ & 0.15258 $\times 10^{-4}$  & 0.12066 $\times 10^{-1}$ & 0.12105 $\times 10^{-1}$ & & & & 0.10387 $\times 10^{-3}$ \\
 & & & & & & & & & & \\
150 & 0.28382 $\times 10^{-6}$ & & & & 0.83917 $\times 10^{-5}$ & 0.72889 $\times 10^{-3}$ & 0.65401 $\times 10^{-2}$ & 0.68337 $\times 10^{-3}$ & & 0.29033 $\times 10^{-6}$ \\
 & & & & & & & & & & \\
500 & 0.34689 $\times 10^{-9}$ & & & & & & 0.43355 $\times 10^{-6}$ & 0.87799 $\times 10^{-4}$ & 0.28292 $\times 10^{-5}$ & 0.36327 $\times 10^{-9}$ \\
 & & & & & & & & & & \\
1000 & 0.20703 $\times 10^{-11}$ & & & & & & & 0.44782 $\times 10^{-7}$ & 0.19974 $\times 10^{-6}$ & 0.27001 $\times 10^{-11}$ \\
\hline
\end{tabular}

\end{center}
\end{sidewaystable}

\begin{sidewaystable}[h]
\begin{center}
\caption{Scalar model production cross sections [pb] corresponding to mass points in Table~\ref{tab:MMScalar} \label{tab:scalarxsecs}}
\begin{tabular}{| l | *9{p{2.0cm}|}}
\hline
 $M_\chi$ [GeV] & \multicolumn{9}{c|}{$M_{S}$ [GeV]} \\
\hline
1 & 0.21915 $\times 10^{1}$ & 0.20798 $\times 10^{1}$ & 0.19192 $\times 10^{1}$ & 0.18118 $\times 10^{1}$ & 0.16735 $\times 10^{1}$ & 0.52244 $\times 10^{1}$ & 0.41877 $\times 10^{1}$ & 0.28732 $\times 10^{1}$ & 0.18028 $\times 10^{1}$\\
 & & & & & & & & & \\
10 & 0.17416 $\times 10^{1}$ & 0.17420 $\times 10^{1}$ & 0.18581 $\times 10^{1}$ & 0.17510 $\times 10^{1}$ & & & & & 0.17398 $\times 10^{1}$\\
 & & & & & & & & & \\
 50 & 0.39053  & & 0.38877  & 0.38409  & 0.37097  & 0.12861 $\times 10^{1}$ & & & 0.39096 \\
 & & & & & & & & & \\
 150 & 0.24136 $\times 10^{-5}$ & & & & 0.38372 $\times 10^{-5}$ & 0.21922 $\times 10^{-4}$ & 0.42337 $\times 10^{-3}$ & 0.57124 $\times 10^{-4}$ & 0.11105 $\times 10^{-4}$\\
 & & & & & & & & & \\
 500 & 0.34099 $\times 10^{-8}$ & & & & & & 0.49399 $\times 10^{-8}$ & 0.25206 $\times 10^{-6}$ & 0.36823 $\times 10^{-6}$\\
 & & & & & & & & & \\
 1000 & 0.17012 $\times 10^{-10}$ & & & & & & & 0.55260 $\times 10^{-10}$ & 0.11067 $\times 10^{-7}$\\
\hline
\end{tabular}
\end{center}
\end{sidewaystable}
 
The signal models were interfaced to the MadGraph5\_aMC@NLO version 2.3.2.2 ~\cite{Alwall:2014hca} program that was used to generate multi-parton amplitudes and events 
and the calculation of the cross sections.  The decay of he SM-like Higgs boson was handled by the general multi-purpose Monte Carlo event generator 
PYTHIA 8.212~\cite{Sjostrand:2006za}; that program also served either to generate a given hard process at leading order (LO),  or, 
in cases where the hard processes are generated at higher orders, only for
parton  showering, hadronization, and for adding the underlying event.

We study the decay of the SM-like Higgs boson to Z bosons both decaying to charged leptons (electron and muons), which is a clean final state with four
charged leptons in addition to the missing transverse momentum coming from the undetected dark matter particles.  The most important source of background comes from 
the associated production of the SM Higgs boson with the Z boson in the following decay chains:

\begin{itemize}
\item  the Higgs boson decays to two Z bosons, which subsequently
  decay to leptons, while the associated Z boson decays to a neutrino
  pair ($Z\rightarrow \nu \nu$, $H \rightarrow ZZ \rightarrow 4l$); 
\item the Higgs boson decays to two Z bosons, one to charged leptons
  and the other to a neutrino pair, while the accompanying Z decays to
  leptons ($Z\rightarrow ll $, $H \rightarrow ZZ \rightarrow 2l 2\nu$).
\end{itemize}

The SM Higgs boson production in association with top quarks can also contribute as a background. 
The production of the SM Higgs boson in association with top quarks ($ttH$), via the gluon fusion ($gg \rightarrow H$) and vector boson fusion ($q q \rightarrow q q H$) 
mechanisms and the associated production with gauge bosons ($VH$, $V = Z,  W$) were simulated by using the 
POWHEG-BOX 2.0~\cite{Nason:2004rx,Frixione:2007vw,Alioli:2010xd,Alioli:2008tz,Nason:2009ai,Luisoni2013} program. 
The $ZH$ samples was also simulated by using MadGraph5\_aMC@NLO version 2.3.2.2 to increase the statistics of events for the final state studied. 

Background events coming from $t\bar{t}$ and $q q \rightarrow ZZ$ production were also simulated using POWHEG-BOX 2.0 and POWHEG-BOX 1.0, respectively. 
The missing transverse momentum was studied using a large sample of W+jets events simulated with MadGraph5\_aMC@NLO version 2.2.2.
The cross sections of all the background processes were computed at NLO
accuracy. For Higgs boson production via  gluon fusion, vector boson fusion, and 
associated production, the most recent calculations of the cross sections were used~\cite{Heinemeyer:2013tqa}.

The simulation of the response of the ATLAS and CMS detectors, and the event reconstruction, 
was done with the DELPHES 3.3.2 program~\cite{deFavereau:2013fsa}. 
This program simulates tracking in a magnetic field, electromagnetic and hadronic calorimeters, and a muon identification system. 
Reconstructed objects are simulated from the parametrized detector response and includes tracks, calorimeter deposits, and high level
objects such as isolated electrons, jets, taus, and missing transverse momentum. 
We were particularly careful to implement a realistic parametrization of the 
reconstruction efficiency, 
identification, and isolation 
of electron and muons, as measured by ATLAS and CMS with real data from Run1~\cite{Aad:2014zya,Aad:2014fxa,Chatrchyan:2012xi,Khachatryan:2015hwa}.

\subsection{Event Selection}
\label{sec:selection}
Signal events include four isolated leptons with large transverse momentum ($\pT$), and large missing transverse momentum from the dark matter particles that remain undetected.  The selection was designed to discriminate the signal from the background using a cut-based approach.

Events are first required to have at least four reconstructed leptons ($e^\pm, \mu^\pm$), which are spatially separated by at least $\Delta R \geq 0.02$\footnote{$\Delta R =\sqrt{\Delta\eta^2 + \Delta\phi^2}$,
where $\Delta\eta$ and $\Delta\phi$ are the differences in pseudorapidity and azimuthal angle,
respectively.}.  
Electrons are required to have a minimum $\pT$ of \unit[7]{\GeV} and need to be in the pseudorapidity range $|\eta|<2.5$ (the geometrical acceptance of the ATLAS and CMS experiments). Selected muons need to be reconstructed with $\pT \geq \unit[5]{\GeV}$ and be in the geometrical acceptance.  

All four leptons have to be isolated. The isolation variable is defined as the sum of the transverse momentum of the tracks inside a cone of opening 
$\Delta R \geq 0.3$ around the lepton. This variable is known to be robust against the increase in the number of pileup interactions.  
The cut on the isolation variable was optimized by using the lowest $\pT$ lepton for each signal sample and the $t \bar{t}$ sample; a receiver operating characteristic (ROC) curve was derived for each combination of signal and background samples, and a working point was chosen corresponding to the point of the largest change of the curvature of that curve, and that corresponds to a signal efficiency in between 85\% and 95\%. 

Leptons of opposite sign and same flavor are paired, and the dilepton is required to have an invariant mass larger than \unit[4]{\GeV} in order to suppress the light-jet QCD background. 
A b-tag veto was used to reduce the contribution of $t \bar{t}$ events. 
If more than two dileptons can be formed, ambiguities are resolved as follows: the dilepton with total invariant mass closest to the Z boson mass is chosen as the first Z boson. Among all valid, same flavor opposite sign, dileptons that can be formed from the remaining leptons, we choose as the second Z boson the dilepton with the highest $\pT$ whose total three-momentum vector is at least 
$\Delta R \geq 0.05$ away from the first dilepton. 

The first selected dilepton is required to have invariant mass in the range $[\unit[40]{\GeV}, \unit[120]{\GeV}]$ and the second to
have invariant mass in the range $[\unit[12]{\GeV}, \unit[120]{\GeV}]$. 
The leading and sub-leading leptons of the four selected leptons are required to have $\pT \geq \unit[20]{\GeV}$ and $\pT \geq \unit[10]{\GeV}$, respectively. 

Events are rejected if the four-lepton invariant mass is below \unit[100]{\GeV}. 

\begin{figure}
\begin{center}
\includegraphics[width=\textwidth]{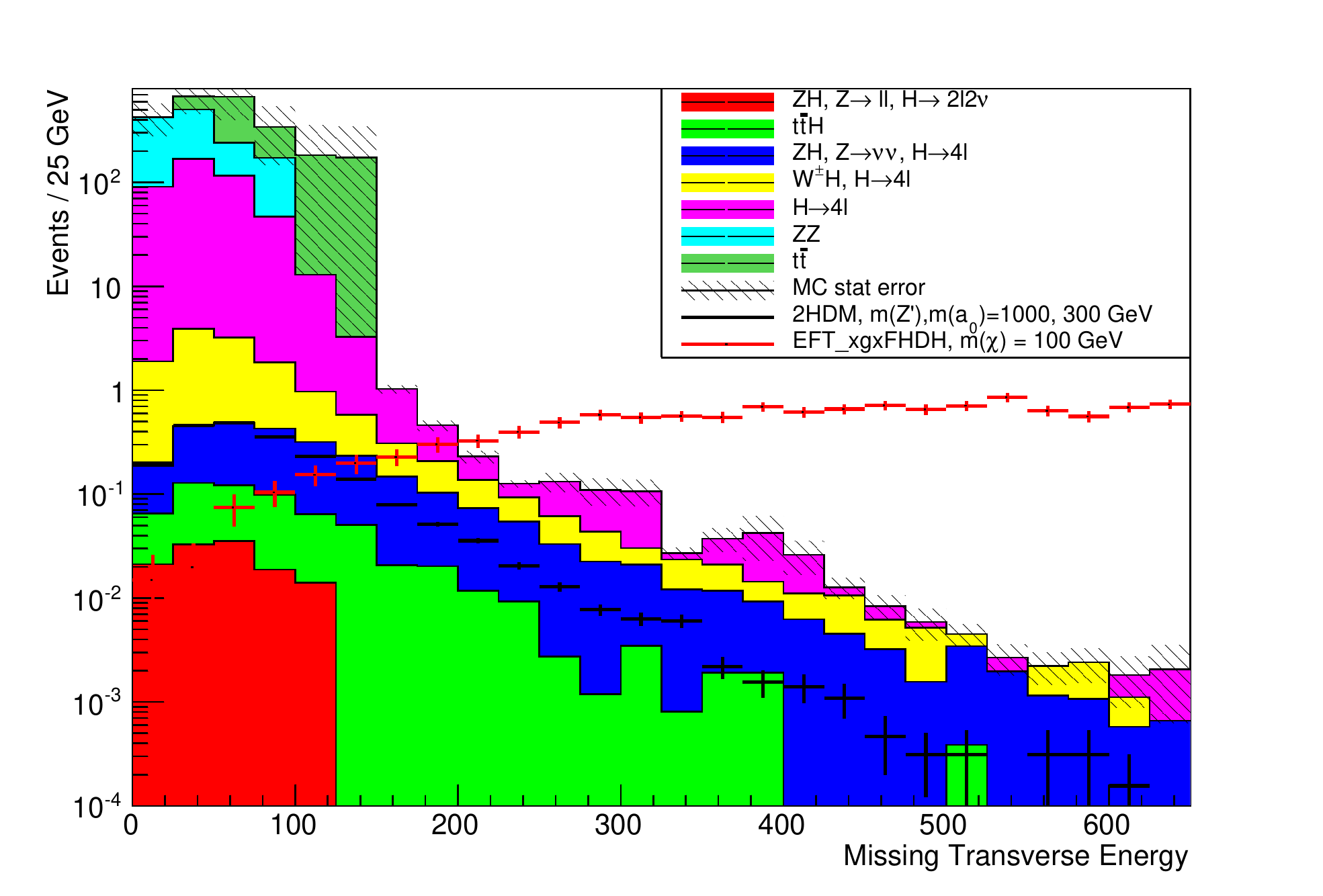}
\caption{Distribution of the missing transverse energy for the background processes and two signal samples: one simulated according to the non minimal Z$^\prime$ model with pseudoscalar $a_0$ with $m_{\textrm{Z}^\prime}=1$\,TeV and $m_{a_0}=300$\,GeV, and the other simulated according to the EFT model referred as EFT\_xdxHDHc in Table~\ref{tab:efts}. The distributions are normalized to the expected number of events according 
for an integrated luminosity of $3000 \, fb^{-1}$.}
\label{kinem}
\end{center}
\end{figure}

The most powerful observable for suppressing the background is the missing transverse 
momentum of the events, that is, the magnitude $\mathrm{MET} =  |\vec{E}^{miss}_T|$ of
the missing transverse momentum 
vector $\vec{E}^{miss}_T$. The latter is defined as the negative sum of the transverse momenta of all reconstructed objects. Large missing transverse momentum, in addition to four charged leptons, 
also characterizes  
background events from the associated production of the Higgs boson with a Z boson in 
final states in which one of the Z bosons decays to neutrinos. A cut on the four-lepton mass 
of $m_{4l}<140 \, \mathrm{\GeV}$ was used to reduce the contribution 
from  $Z\rightarrow ll $, $H \rightarrow ZZ \rightarrow 2l 2\nu$ events. This cut
is effective because for these events there is no peak around 125\,GeV. 

The distribution of the missing transverse momentum of the events is shown in Figure~\ref{kinem} after applying all the previous cuts, for the background samples and a couple of signal samples.  All distributions are normalized to the predicted number of events for an integrated luminosity of $3000 \, fb^{-1}$.

For simplicity, the missing transverse momentum cut was chosen by maximizing $S / \sqrt{B}$, where $S$ and $B$ are
the predicted signal and background counts, respectively\footnote{We are aware that this is a 
simplistic measure and overly optimistic when the background estimate is small, but a more
refined analysis was beyond the scope of this preliminary study.}.  Since the 
missing transverse momentum distribution depends on the kinematics of the production mechanism and on the particular model used to simulate the signal event, four signal regions for
the missing transverse momentum were defined: 
 $\mathrm{MET} > 150$\,GeV, $\mathrm{MET} > 300$\,GeV, $\mathrm{MET} > 450$\,GeV, $\mathrm{MET} > 600$\,GeV. We note that these signal regions are not disjoint; therefore, the results
 obtained using them cannot be readily combined unless the regions are rendered disjoint in some
 way. 

Another variable useful to reduce the background further and enhance the signal is the transverse mass ($M_T$) of the four-lepton system defined as:

\begin{eqnarray}        
\label{mT4l + 2 p} 
M_T=\sqrt{m^2_{4l} + 2 \times p_{T,4l} \times MET - \vec{p}_{T,4l} \cdot \vec{E}^{miss}_T}
\end{eqnarray}

where $m_{4l}$ and $p_{T,4l}$ are the invariant mass and the transverse momentum of the four lepton candidate, respectively. $ \vec{p}_{T,4l} $ is the vector of 
four lepton momentum projected on the transverse plane. 

\begin{figure}
\begin{center}
\includegraphics[width=\textwidth, height=0.42\textheight]{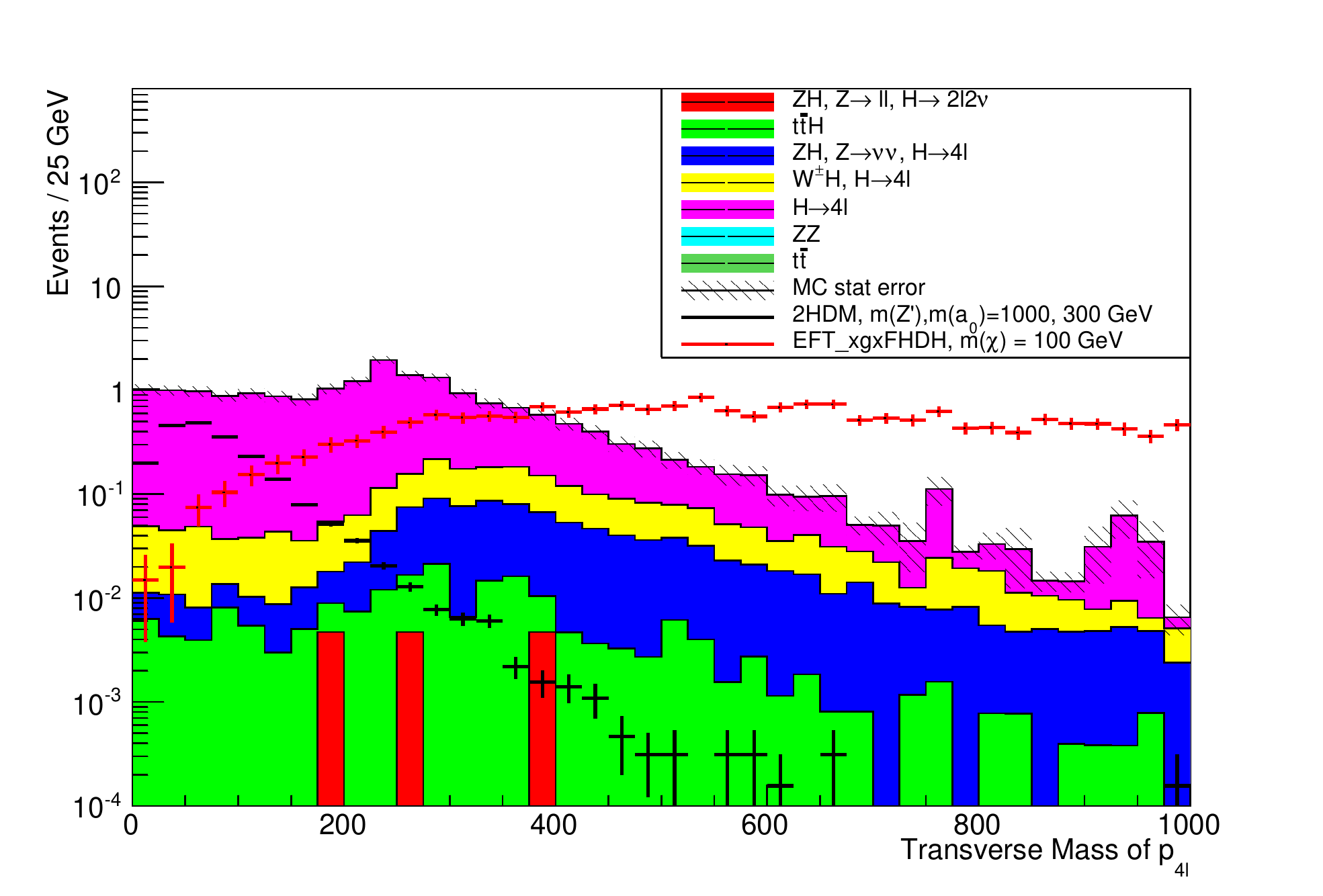} \\
\includegraphics[width=\textwidth, height=0.42\textheight]{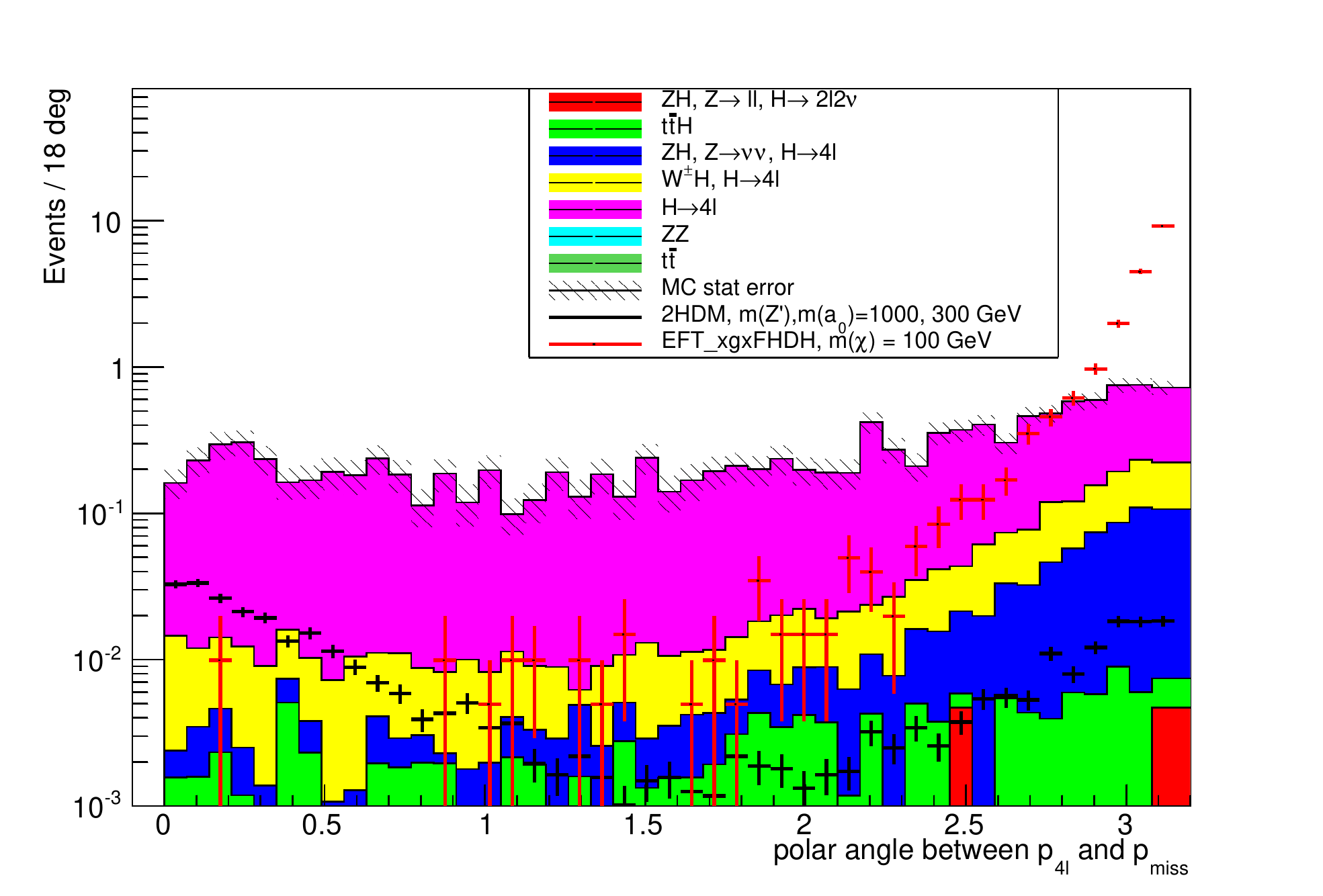}
\caption{(Top) Distribution of the transverse mass of the four-lepton system and the missing transverse momentum for the background processes and a couple of signal samples. (Bottom) Distribution of the angle between the missing transverse momentum and the four lepton momentum for the background processes and two signal samples: one simulated according to the non minimal Z$^\prime$ model with pseudoscalar $a_0$ with $m_{\textrm{Z}^\prime}=1$\,TeV and $m_{a_0}=300$\,GeV, and the other simulated according to the EFT model referred as EFT\_xgxFHDH in Table~\ref{tab:efts}. All distributions are normalized to the expected number of events according for an integrated luminosity of 3000\,fb$^{-1}$.}
\label{kinem_b}
\end{center}
\end{figure}

For some models vetoing 
the transverse mass range between 200 and 400\,GeV improves the signal to background ratio. Therefore, 
two different signal regions were defined by either vetoing or not vetoing this region.  
The distribution of $m_T$ for a 
specific signal model and the background events is shown in Figure~\ref{kinem_b}. 

Further studies have shown that another useful variable is the angle between $\vec{E}^{miss}_T$ and the four-lepton momentum in the transverse plane ($\Delta \phi _ {4l-MET}$). For some models these vectors tend to be back-to-back depending on the kinematics of the production. For these cases, we therefore
imposed the additional requirement, 
$\Delta \phi _ {4l-MET}<\pi/2$.  The distribution of 
$\Delta \phi _ {4l-MET}$ for a specific signal model and the background events is 
shown in Figure~\ref{kinem_b}. 

\section{STATISTICAL ANALYSIS}
\label{sec:stats}
The results of this study are estimates, and uncertainties, of the signal  efficiencies (including 
acceptance), $\hat{\epsilon} \pm\delta \epsilon$,  for different mono-Higgs signal models,
together with
the background estimates and associated uncertainties, $B \pm \delta B$. In an analysis using real data,
we would have the observed number of events, $N$,  the likelihood for which
we take to be
\begin{equation}
        p(N | \sigma, a, b) = \textrm{Poisson}(N, a \, \sigma + b),
\end{equation}
where $a \equiv \epsilon \times {\cal L}$ is the effective integrated luminosity, that is, 
the signal efficiency times the integrated luminosity, $\sigma$ is the 
cross section, and $b$ is the expected (that is, mean) background of
which $B \pm \delta B$ is the estimate. 
We adopt a Bayesian approach and incorporate 
the estimate of $a$, which we denote by
$A \pm \delta A$, 
as well as the estimate of $b$ into the statistical model using priors, each 
modeled as a gamma density
\begin{equation}
\textrm{gamma}(x, \gamma, \beta) = \beta^{-1} (x / \beta)^{\gamma-1} \exp(-x/\beta) / \Gamma(\gamma),
\end{equation}
with the mode and variance set to $A$ and $\delta A^2$, respectively, for
the effective luminosity, or to $B$ and $\delta B^2$, for the background.
Writing $k = (c / \delta c)^2$, where $c = A$ or $B$ and $\delta c = \delta A$ or $\delta B$, the gamma
density parameters are
\begin{eqnarray}
        \gamma & = & [k + 2  + \sqrt{(k+2)^2 - 4}]/2, \\
        \beta & = & [\sqrt{c^2 + 4 \delta c^2} - c] / 2.
\end{eqnarray}
These priors are appropriate for signal and background estimates that are the results of scaled counts. 

In order to make an inference about the cross section, $\sigma$, we first calculate the marginal likelihood
\begin{equation}
        p(N| \sigma) = \int da \int db \, p(N | \sigma, a, b) \, \pi(a) \, \pi(b),
\end{equation} 
where $\pi(a) = \textrm{gamma}(a, *, *)$ and $\pi(b) = \textrm{gamma}(b, *, *)$ are the priors
for the (expected) effective luminosity and the expected background, respectively, 
and then calculate
the posterior density 
\begin{equation}
        p(\sigma | N) = p(N | \sigma) \, \pi(\sigma) / p(N). 
\end{equation}
In this study, we take the cross section prior $\pi(\sigma)$ to be flat\footnote{More sophisticated
choices are possible. But experience shows that this is a reasonable choice for this problem.}.

Given an upper limit on the cross section, or equivalently on the expected signal count,  we can infer a lower
limit on the mass of the hypothesized dark matter particle.
A 95\% upper limit on the cross section, $\sigma_{\textrm{up}}$, is found by solving
\begin{equation}
        \int_0^{\sigma_\textrm{up}} p(\sigma | N) \, d\sigma = 0.95. 
\end{equation} 
Since this is a study
of simulated data, we do not have observed counts. We therefore 
define the expected limit on the cross section as the limit  obtained by
setting $N  = B$. 

By the end of the high-luminosity era of the LHC, the LHC detectors will be
very well understood and the state-of-the-art in theoretical calculations will have 
advanced considerably. Systematic uncertainties may well be at the few percent level or smaller.
In the limit calculations, we assume a  5\% uncertainty in the
effective luminosities and backgrounds.

\section{RESULTS}
The results of this study are the predicted signal and background counts, 
after the event selection described
in Section~\ref{sec:selection}, for each of the 175 models investigated. We find 61 models 
that predict
a signal yield between 1 and  1660 events assuming an integrated luminosity of 3000\,fb$^{-1}$,
and of these 38 predict signal yields greater than 10 events. These include the 
Effective Field Theory models listed in Table~\ref{tab:efts}, the Scalar models, and the Z$^\prime$
baryonic models. For these models and our event selection, the backgrounds
are found to be negligible. In
Tables~\ref{nevents2HDM} to \ref{neventsEFT}, we report the event yields for the
most promising models. For the models that predict at least one event but fewer than ten, we
report the  95\% C.L. upper limits on the cross sections in Table~\ref{tab:limits} along with
the predicted cross sections.

\begin{sidewaystable}[h]
\centering
\caption{Number of events for 2HDM signal  model (sect. \ref{Section2HDM}) and total background  (right) after the full selection, assuming an integrated luminosity of 3000\,fb$^{-1}$.}
\label{nevents2HDM}

\begin{tabular}{||l|c|c|c|c||}
	\hline 
	& & & & \\
	Model & $\sigma [pb] $ & opt. cuts & n. sig.  & n. bkg \\
	& & & & \\
	\hline 
	\hline 
	2HDM (sect. \ref{Section2HDM}) & & & &\\
	\hline 
	($m_{\chi},m_{a},m_{A}$)= (30,80,700) GeV & $4.33\times 10^{-6}$ & $MET>150,M_T$                   & $1.37  \pm 0.01$ & $1.92 \pm 0.11 $\\         
 	\hline 
	
\end{tabular}

\smallskip

\centering
\caption{Number of events for ZpBaryonic signal model (sec. \ref{ZpBModel}) and total background  (right) after the full selection, assuming an integrated luminosity of 3000\,fb$^{-1}$.}
\label{neventsZpBarionic}

\begin{tabular}{||l|c|c|c|c||}
\hline 
	& & & & \\
	Model & $\sigma [pb] $ & opt. cuts & n. sig. & n. bkg\\
	& & & & \\
	\hline 
	\hline 
	$Z^\prime$ Baryonic (sect. \ref{ZpBModel}) & & & &\\
	\hline 
	($M_{Z^\prime},M_{\chi}$)=(1000,1) & $2.32\times 10^{-5}$ & $MET>300$ & $1.68 \pm 0.07$ & $0.31  \pm 0.04 $\\
	($M_{Z^\prime},M_{\chi}$)=(100,10) & $4.00\times 10^{-4}$ & $MET>300$  & $3.27 \pm 0.39$ & $0.31  \pm 0.04 $\\
	($M_{Z^\prime},M_{\chi}$)=(100,1) & $4.01\times 10^{-4}$ & $MET>150,\Delta \phi,M_{T}$ & $4.51 \pm 0.47 $ & $0.78  \pm 0.07 $\\
	($M_{Z^\prime},M_{\chi}$)=(10,1) & $3.30\times 10^{-4}$ & $MET>150,M_{T}$ & $13.5 \pm 0.74 $ & $1.92 \pm 0.11 $\\
	($M_{Z^\prime},M_{\chi}$)=(20,1) & $3.45\times 10^{-4}$ & $MET>150,M_{T}$ & $13.0 \pm 0.74 $ & $1.92 \pm 0.11 $\\	
	($M_{Z^\prime},M_{\chi}$)=(300,1) & $2.88\times 10^{-4}$ & $MET>150,M_{T}$ & $13.1 \pm 0.69 $ & $1.92 \pm 0.11 $\\
	($M_{Z^\prime},M_{\chi}$)=(500,1) & $1.34\times 10^{-4}$ & $MET>150$ & $13.4 \pm 0.47$ & $2.43 \pm 0.12 $\\
	($M_{Z^\prime},M_{\chi}$)=(50,1) & $4.12\times 10^{-4}$ & $MET>300,M_{T}$ & $2.75 \pm 0.38 $ & $0.31  \pm 0.04$\\
\hline

\end{tabular}
\end{sidewaystable}


\begin{sidewaystable}[h]
\centering
\caption{Number of events for Scalar signal model (sect.~\ref{sec:scalar}) and total background after the full selection, assuming an integrated luminosity of 3000\,fb$^{-1}$.}
\label{neventsScalar}

\begin{tabular}{||l|c|c|c|c||}
\hline 
	& & & & \\
	Model & $\sigma [pb] $ & opt. cuts & n. sig. & n. bkg  \\
	& & & & \\
	\hline 
	\hline 
	Scalar model (sect. \ref{sec:scalar}) & & & & \\
	\hline
	($M_{S},M_{\chi}$)=(10000,10) & $2.16\times 10^{-4}$ & $MET>450$ 	        & $17.3 \pm 0.69$ & $0.05  \pm 0.02 $ \\
	($M_{S},M_{\chi}$)=(10000,1) & $2.24\times 10^{-4}$ & $MET>450$ 	       	& $16.0 \pm 0.67 $ & $0.05  \pm 0.02 $ \\
	($M_{S},M_{\chi}$)=(10000,50) & $4.85\times 10^{-5}$ & $MET>150,M_{T}$ & $12.5 \pm 0.27 $ & $1.92 \pm 0.11 $ \\
	($M_{S},M_{\chi}$)=(1000,1) & $3.56\times 10^{-4}$ & $MET>300$ & $55.0 \pm 1.57$ & $0.31  \pm 0.04 $ \\
	($M_{S},M_{\chi}$)=(100,10) & $2.17\times 10^{-4}$ & $MET>450$ & $19.0 \pm 0.72$ & $0.05  \pm 0.02 $ \\
	($M_{S},M_{\chi}$)=(100,1) & $2.25\times 10^{-4}$ & $MET>450$ & $20.1 \pm 0.76$ & $0.05  \pm 0.02 $ \\
	($M_{S},M_{\chi}$)=(10,10) & $2.16\times 10^{-4}$ & $MET>450$ & $19.4 \pm 0.72$ & $0.05  \pm 0.02 $ \\
	($M_{S},M_{\chi}$)=(10,1) & $2.72\times 10^{-4}$ & $MET>450$  & $19.4 \pm 0.81$ & $0.05 \pm 0.02$ \\
	($M_{S},M_{\chi}$)=(10,50) & $4.84\times 10^{-5}$ & $MET>150,M_{T}$ & $13.0 \pm 0.28 $ & $1.92 \pm 0.11 $ \\
	($M_{S},M_{\chi}$)=(15,10) & $2.16\times 10^{-4}$ & $MET>450$ & $18.4 \pm 0.70 $ & $0.05  \pm 0.02 $ \\
	($M_{S},M_{\chi}$)=(200,1) & $2.08\times 10^{-4}$ & $MET>450$ & $19.1 \pm 0.70 $ & $0.05  \pm 0.02 $ \\
	($M_{S},M_{\chi}$)=(200,50) & $4.60\times 10^{-5}$ & $MET>150,M_{T}$ & $12.4 \pm 0.27 $ & $1.92 \pm 0.11$ \\
	($M_{S},M_{\chi}$)=(20,1) & $2.58\times 10^{-4}$ & $MET>450$ & $19.4 \pm 0.7.9 $ & $0.05  \pm 0.02 $ \\
	($M_{S},M_{\chi}$)=(300,1) & $6.48\times 10^{-4}$ & $MET>450$ & $19.3 \pm 1.25$ & $0.05  \pm 0.02 $ \\
	($M_{S},M_{\chi}$)=(300,50) & $1.59\times 10^{-4}$ & $MET>150,M_{T}$ & $13.6 \pm 0.52 $ & $1.92 \pm 0.11 $ \\
	($M_{S},M_{\chi}$)=(500,1) & $5.19\times 10^{-4}$ & $MET>450$  & $17.6 \pm 1.07$ & $0.05  \pm 0.02 $ \\
	($M_{S},M_{\chi}$)=(50,10) & $2.30\times 10^{-4}$ & $MET>450$ & $17.2 \pm 0.70$ & $0.05  \pm 0.02 $ \\
	($M_{S},M_{\chi}$)=(50,1) & $2.38\times 10^{-4}$ & $MET>450$ & $19.8 \pm 0.77$ & $0.05  \pm 0.02 $ \\
	($M_{S},M_{\chi}$)=(50,50) & $4.82\times 10^{-5}$ & $MET>150,M_{T}$ & $12.7 \pm 0.27 $ & $1.92 \pm 0.11$ \\
	($M_{S},M_{\chi}$)=(95,50) & $4.76\times 10^{-5}$ & $MET>150,M_{T}$ & $13.1 \pm 0.28 $ & $1.92 \pm 1.10$ \\
	\hline
\end{tabular}
\end{sidewaystable}

\begin{sidewaystable}[h]
\centering
\caption{Number of events events for EFT signal model (sect. \ref{SectionEFT}) and total background  after the full selection, assuming an integrated luminosity of 3000\,fb$^{-1}$.}
\label{neventsEFT}

\begin{tabular}{||l|c|c|c|c|c||}
\hline 
	& & & &  \\
	Model & $\sigma [pb] $ & opt. cuts & n. sig. & n. bkg \\
	& & & & \\
	\hline 
	\hline 
	EFT\_HHxg5x, $M_{\chi}=10$ & $1.93\times 10^{-2}$ & $MET>450,M_T$ & $1660 \pm 63.1$ & $0.05  \pm 0.02 $\\
	EFT\_HHxg5x, $M_{\chi}=1$ & $1.95\times 10^{-2}$ & $MET>450$           & $1620 \pm 63.0 $ & $0.05  \pm 0.02 $\\
	EFT\_HHxg5x, $M_{\chi}=50$ & $1.18\times 10^{-2}$ & $12<=mZ2<=120$ & $4870 \pm 84.3$ & $28700 \pm 6030 $ \\
	EFT\_HHxx\_combined, $M_{\chi}=10$ & $1.88\times 10^{-4}$ & $MET>450$           & $15.2 \pm 0.60 $ & $0.05  \pm 0.02 $ \\
	EFT\_HHxx\_combined, $M_{\chi}=1$ & $1.95\times 10^{-4}$ & $MET>450$ & $15.3 \pm 0.61 $ & $0.05  \pm 0.02 $\\
	EFT\_HHxx\_combined, $M_{\chi}=50$ & $4.23\times 10^{-5}$ & $MET>150,M_T$ & $10.9 \pm 0.24 $ & $1.92 \pm 0.11 $\\
	EFT\_HHxx\_scalar, $M_{\chi}=10$ & $1.24\times 10^{-4}$ & $MET>150,\Delta \phi,M_T$  & $8.62 \pm 0.36 $ & $0.78  \pm 0.07 $\\
	EFT\_HHxx\_scalar, $M_{\chi}=50$ & $7.52\times 10^{-5}$ & $MET>150,\Delta \phi,M_T$ & $5.92 \pm 0.23 $ & $0.78  \pm 0.07 $\\
	EFT\_xdxHDHc, $M_{\chi}=10$, $\Lambda=10^{-4}$ & $3.60\times 10^{-5}$ & $MET>450$ & $1.64 \pm 0.09 $ & $0.05  \pm 0.02 $\\
	EFT\_xdxHDHc, $M_{\chi}=1$, $\Lambda=10^{-4}$ & $3.66\times 10^{-5}$ & $MET>450$ & $1.30 \pm 0.08 $ & $0.05  \pm 0.02 $\\
	EFT\_xgxFHDH, $M_{\chi}=1000$ & $7.12\times 10^{-6}$ & $MET>450,M_T$ & $2.24 \pm 0.05 $ & $0.05 \pm 0.02 $\\
	EFT\_xgxFHDH, $M_{\chi}=100$ & $6.61\times 10^{-5}$ & $MET>450,M_T$ & $19.2 \pm 0.40 $ & $0.05 \pm 0.02 $\\
	EFT\_xgxFHDH, $M_{\chi}=10$ & $7.07\times 10^{-5}$ & $MET>450,M_T$ & $19.7 \pm 0.42$ & $0.05 \pm 0.02$\\	
	EFT\_xgxFHDH, $M_{\chi}=400$ & $3.69\times 10^{-5}$ & $MET>150$ & $14.2 \pm 0.26 $ & $2.43 \pm 0.12 $\\
	EFT\_xgxFHDH, $M_{\chi}=50$ & $6.95\times 10^{-5}$ & $MET>450,M_T$ & $19.9 \pm 0.42 $ & $0.05 \pm 0.02 $\\
	EFT\_xgxFHDH, $M_{\chi}=65$ & $6.86\times 10^{-5}$ & $MET>450,M_T$ & $19.9 \pm 0.42 $ & $0.05  \pm 0.02 $\\
	\hline	
\end{tabular}

\end{sidewaystable}

\begin{sidewaystable}[h]
\centering
\caption{95\% upper limits on the cross section, for models that yield between 1 and 10 events, assuming an integrated luminosity of 3000\,fb$^{-1}$.}
\label{tab:limits}

\begin{tabular}{||l|c|c|c||}
\hline 
	& &  \\
	Model & $\sigma [pb] $ & $\sigma^\textrm{95\%}$ [pb]\\
	& &  \\
	\hline 
	\hline 
2HiggsDM\_m\_DM\_m\_a\_m\_A\_1\_500               	& $7.49\times10^{-2}$ & $1.94\times10^{-1}$\\
EFT\_HHxx\_combined ($M_\chi = 50$)               	& $3.41\times10^{-1}$ & $3.22\times10^{-1}$\\
EFT\_HHxx\_scalar ($M_\chi = 10$)                 	& $9.98\times10^{-1}$ & $3.28\times10^{-1}$\\
EFT\_HHxx\_scalar ($M_\chi = 50$)                 	& $6.07\times10^{-1}$ & $3.12\times10^{-1}$\\
EFT\_xdxHDHc ($M_\chi = 100$, $\Lambda = 10^{-4}$)	& $6.50\times10^{-2}$ & $1.79\times10^{-1}$\\
EFT\_xdxHDHc ($M_\chi = 10$, $\Lambda = 10^{-4}$)	& $2.91\times10^{-1}$ & $5.50\times10^{-1}$\\
EFT\_xdxHDHc ($M_\chi = 1$, $\Lambda = 10^{-4}$)	& $2.95\times10^{-1}$ & $7.03\times10^{-1}$\\
EFT\_xdxHDHc ($M_\chi = 1$, $\Lambda = 10^{-6}$)	& $1.63\times10^{-1}$ & $7.43\times10^{-1}$\\
EFT\_xdxHDHc ($M_\chi = 50$, $\Lambda = 10^{-4}$)	& $1.05\times10^{-1}$ & $2.43\times10^{-1}$\\
EFT\_xdxHDHc ($M_\chi = 65$, $\Lambda = 10^{-4}$)	& $8.98\times10^{-2}$ & $2.35\times10^{-1}$\\
EFT\_xgxFHDH ($M_\chi = 1000$)                    	& $5.74\times10^{-2}$ & $9.39\times10^{-2}$\\
EFT\_xgxFHDH ($M_\chi = 800$)                     	& $1.03\times10^{-1}$ & $9.41\times10^{-2}$\\
Scalar ($M_{Z^\prime} = 10000$, $M_\chi = 50$)    	& $3.91\times10^{-1}$ & $3.40\times10^{-1}$\\
Scalar ($M_{Z^\prime} = 10$, $M_\chi = 50$)       	& $3.91\times10^{-1}$ & $3.18\times10^{-1}$\\
Scalar ($M_{Z^\prime} = 200$, $M_\chi = 50$)      	& $3.71\times10^{-1}$ & $2.92\times10^{-1}$\\
Scalar ($M_{Z^\prime} = 300$, $M_\chi = 50$)      	& $1.29$ & $9.31\times10^{-1}$\\
Scalar ($M_{Z^\prime} = 50$, $M_\chi = 50$)       	& $3.89\times10^{-1}$ & $3.04\times10^{-1}$\\
Scalar ($M_{Z^\prime} = 95$, $M_\chi = 50$)       	& $3.84\times10^{-1}$ & $2.91\times10^{-1}$\\
ZpBaryonic ($M_{Z^\prime} = 1000$, $M_\chi = 150$)	& $1.81\times10^{-1}$ & $2.00\times10^{-1}$\\
ZpBaryonic ($M_{Z^\prime} = 1000$, $M_\chi = 1$)  	& $1.87\times10^{-1}$ & $3.89\times10^{-1}$\\
ZpBaryonic ($M_{Z^\prime} = 295$, $M_\chi = 150$) 	& $1.80\times10^{-1}$ & $3.86\times10^{-1}$\\
ZpBaryonic ($M_{Z^\prime} = 500$, $M_\chi = 150$) 	& $6.73\times10^{-1}$ & $3.98\times10^{-1}$\\
ZpBaryonic ($M_{Z^\prime} = 95$, $M_\chi = 50$)   	& $3.13\times10^{-1}$ & $8.23\times10^{-1}$\\

\hline
\end{tabular}

\end{sidewaystable}

\section{CONCLUSIONS}

A strategy to search for dark matter particles produced in association with a Higgs boson at the LHC collider was presented in the context of 
several Effective Field Theories (EFTs) and simplified models that predict Dark Matter-Higgs 
boson interactions.   We studied simulated samples in which the Higgs boson decays into four charged leptons  (electron and muons), via two Z bosons, and recoils against dark matter particles. 
For several models, including some $Z^\prime$ baryonic models,  Scalar models, and EFT models, a discovery is possible by the end of the high-luminosity era of
the LHC, while for some variations of these models useful limits on the cross section
can be set.
We therefore conclude that an analysis by ATLAS and CMS, along the lines described in this paper, has significant  discovery potential.

\section*{ACKNOWLEDGEMENTS}
We would like to thank the organizers of the  the Les Houches Workshop Series  ``Physics at TeV Colliders  2015" for the wonderful atmosphere, 
the productive environment and the effective collaboration between theorists and experimentalists, so important for the progress in our field.
We found the organization of the round-tables and discussions very productive. 
New ideas emerged and we were all inspired to pursue the current and future activities.




\AddToContent{I.~Brivio, D.~Burns, N.~De Filippis, N.~Desai, J.~M.~No, H.~Prosper, S.~Sekmen, D.~Schmeier
and J.~Sonneveld}
\renewcommand{\thesection}{\arabic{section}}


\def\etmiss{\ensuremath{E_{\mathrm{T}}^{\mathrm{miss}}}\xspace}


\chapter{Vector-like Quark Decays}

{\it G.~Brooijmans and G.~Cacciapaglia}



\begin{abstract}
Vector-like quarks are present in many extensions of the Standard Model and the subject of active 
searches by the LHC experiments.  These searches focus on decays to a quark, most often of the third generation, 
and a $W$, $Z$ or Higgs boson.  However, models with vector-like quarks often include additional 
new particles, e.g.~new gauge bosons or a dark matter candidate in the case of a conserved parity.
A matrix of possible final states is built, and categorized as a function of the number of final state
leptons and bosons.  The coverage of such final states by experimental searches is briefly 
examined.
\end{abstract}

\section{Introduction}

Vector-like quarks (VLQs) are fermions for which both chiralities have the same quantum numbers.  Such quarks can 
be naturally heavier than the electroweak symmetry breaking scale without introducing large dimensionless
couplings, and, in particular, they decouple as their mass becomes large.  They are therefore only 
weakly constrained by electroweak precision tests.  They can be produced in pairs at the LHC through the strong 
interaction, or in association with a Standard Model (SM) quark, but in the latter case the production cross-section
is model-dependent.  The LHC experiments have searched for VLQs through their decays to a quark, most often of the 
third generation, and a $W$, $Z$ or Higgs boson~\cite{Aad:2014efa,Aad:2015gdg,Aad:2015kqa,Aad:2015mba,Aad:2015tba,Aad:2015voa,Aad:2016shx,Aad:2016qpo,Chatrchyan:2013uxa,Chatrchyan:2013wfa,Khachatryan:2015axa,Khachatryan:2015gza,Khachatryan:2015oba}.  These searches typically assume that the sum of the 
branching ratios of the three considered decays, e.g. $T\to tH$, $T \to tZ$ and $T \to bW$, where $T$ is the VLQ and 
$H$ the Higgs boson, equals one.  However, in most models with VLQs, additional new particles are present and may appear in the decay products of VLQs, thus suppressing
the decays considered to the benefit of other channels. For recent studies, see e.g.~\cite{Serra:2015xfa,Anandakrishnan:2015yfa}. In this note we propose a classification of the possible final states, and characterize their signatures and the coverage by existing searches.

\section{Vector-like Quarks: General Decays} 

In this note we will focus on VLQs with standard charges of up ($+2/3$, $T$) and down ($-1/3$, $B$) quarks, and exotic charges of $+5/3$ ($X$) and $-4/3$ ($Y$): these 4 types of states allow for direct decays of the VLQ to a SM quark plus a single electroweak (EW) boson ($W$, $Z$ or the Higgs $H$).
We will go beyond the standard decay channels by considering all motivated final states allowed by conservation of quantum numbers that include either SM particles or a Dark Matter (DM) candidate. It is convenient to list decay products containing 3 particles: in fact, new decays typically involve one light SM state plus one heavy state that further decays into two SM particles.  Since $W$, $Z$ and $H$ also decay to two SM states, this approach allows us to cover the standard decays as well.
The nature of the intermediate state in the decay will then affects the kinematic properties of the event: for instance, in the decay $T \to t\ l^+ l^-$, the di-lepton pair may come from a $Z$ or heavy $Z'$ boson.
In this classification we leave out decays that can be detected as a two-particle resonance, like $Q \to q \gamma/g$: one reason is that these couplings are generated by loops or higher-dimensional operators and they are associated in general with very small branching ratios (see~\cite{Cacciapaglia:2010vn} for an example of explicit calculation in a VLQ model). On the other hand, the coupling with gluons can be more efficiently tested via single production~\cite{Cai:2012ji}.

While the intermediate state is of secondary importance for many experimental signatures, we list here a selection of possibilities that arise in known new physics scenarios. This list is, by its own nature, incomplete and it is presented to give a flavor of the range of models covered by our classification.
\begin{itemize}
\item[]
\item[i)] EW bosons, $W$, $Z$ and $H$: they appear in the standard decays of VLQs in association with a SM quark. We then classify the decays further: to charged leptons, $Z/H\to l^+ l^-$ and $W^\pm \to l^\pm \nu$; to quarks, $Z/H \to q\bar{q}$, $W \to q \bar{q}'$ and $H \to b\bar{b}$; to gauge bosons, $H \to W^+ W^-/ZZ$; to invisible particles, $Z \to \nu \nu$ and $H \to$ Dark Matter.  As the Higgs has rare but clean decays, like $H\to \gamma \gamma$, we will also keep the final state $qH$ in the classification.
\item[ii)] New spin-1 resonances, $Z'$ and $W'$, appear in association with VLQs in many models, like composite scenarios (CHMs)~\cite{Contino:2011np} and Little Higgs~\cite{ArkaniHamed:2002qx,Burdman:2002ns}. Their decays are similar to the decays of the EW bosons, with the addition of top quarks in the final states, and also di-boson channels $Z' \to W^+ W^-$ and $W' \to W Z$.
\item[iii)] New color-neutral scalars are also predicted in non-minimal CHMs~\cite{Dugan:1984hq,Katz:2005au,Gripaios:2009pe} and may appear in the decays of VLQs, e.g.~see~\cite{Serra:2015xfa}. In general, one can have both neutral ($\eta_0$) and charged ($\eta^\pm$) states. They have two main decay channels: via the Wess-Zumino-Witten anomaly, giving rise to di-boson final states $\eta_0 \to W^+ W^-/ZZ/Z\gamma/\gamma\gamma$ and $\eta^\pm \to W^\pm Z/W^\pm\gamma$; or via direct couplings to quarks induced by the Yukawa interactions, thus inducing decays preferentially to third generation quarks, $\eta_0 \to t\bar{t}$ and $\eta^+ \to t \bar{b}$. This class also includes extended Higgs sectors~\cite{Kearney:2013oia,Kearney:2013cca}.
\item[iv)] Colored scalar resonances, $\phi_c$, are also a typical prediction in CHMs with top partners, as for instance in~\cite{Barnard:2013zea,Cacciapaglia:2015eqa,Ferretti:2014qta}. They can have various color assignments, however they may always appear in decays of the VLQ into a SM quark. They then further decay into a pair of quarks.
\item[v)] Leptoquarks (LQ) have been widely studied in the context of GUTs~\cite{Pati:1974yy}, however scalar LQs may also appear in CHMs as color-triplet scalar resonances that couple to quarks and leptons~\cite{Wudka:1985ef,Gripaios:2009dq}. They may appear in decays of the VLQ into a LQ plus a lepton, thus their final state will be similar to the standard leptonic channels, however with different kinematical distributions.
\item[vi)] Dark Matter can also appear in the decay products: if the VLQ is odd under the DM parity, it will decay into a quark plus the DM particle, leading to missing transverse energy (\etmiss), as in~\cite{Anandakrishnan:2015yfa,Giacchino:2015hvk}. Another possibility is that the VLQ decays into a mediator that further decays into a pair of DM. While the experimental signature is the same, the former case is characterized by the absence of mixed final states, i.e. both VLQs in pair production need to decay in this channel.
\item[vii)] Finally, we will consider the possibility of cascade decays of a VLQ into a lighter one that further decays into a quark plus an EW boson: $Q'' \to Q' V \to q V V'$, where \mbox{$V^{(')} = W, Z, H$}.
\item[]
\end{itemize}

\begin{table}
\center
{\footnotesize
\begin{tabular}{l|c|c|c|c|c|c|c|c|c|c|}
$T/B$ & $qH$ & $ql^+l^-$ & $q~$ \etmiss & $q l^+ \nu$ & $q q q$ & $qW^+W^-$ & $q Z H/Z$ & $q H H$ & $q W^+ Z$ & $q W^+ H$ \\
\hline
res. &  $\eta_0$ &$Z$, LQ &  $Z$, $H_{\rm inv}$ & $W$, LQ & $Z/W/H$              & $H$, VLQ        & $H$, VLQ & VLQ & $W'$, VLQ  & VLQ \\
       &                 &                     &   LQ, DM                          &                       &  $\eta_0/\eta^\pm/\phi_c$  & $Z'$, $\eta_0$ &                  &         &  $\eta^\pm$ &         \\
 \hline
 tops $T/B$  &  1/0  &  1/0  &  1/0  &  0/1  &  3/2  & 1/0 & 1/0 & 1/0 & 0/1 & 0/1 \\
 \hline \hline
 single & D & A & C & B & A/E & B & A & D & A & B \\
 \hline\hline
$qH$ & D & - & & & & & & & & \\ 
\hline
$ql^+l^-$ & A & A & - & & & & & & & \\ 
\hline
$q~$ \etmiss & C & A & C  &-  & & & & & & \\ 
\hline
$ql^- \nu$ & D & A & C & A & - & & & & & \\ 
\hline
$q q q$ & E & A & B/C  & A/B & A/E & - & & & & \\ 
\hline
$q W^+W^-$ & B & A & B & A & A/B & A & - & & & \\ 
\hline
$q Z H/Z$ & A & A & A & A & A & A & A & - & & \\ 
\hline
$q H H$ & D & A & C & B & B/D & B & A & D & - & \\ 
\hline
$q W^- Z$ & A & A & A & A & A & A & A & A & A & - \\ 
\hline
$qW^- H$ & D & A & C & B & A/D & A & A & D & A & B \\ 
\hline
\end{tabular} }
\caption{Classification of final states for VLQs of standard charge, $T$/$B$. The second row contains the eventual presence of a resonance (here, $Z$ and $W$ always include the case of $Z'$/$W'$, unless the $'$ is explicitly indicated). In the third row we indicate the maximum number of tops for the two types.  The fourth row describes the final state in single production, where we do not include the presence of additional particles, which is model dependent. Finally, the rest of the table describes pair production. The experimental signatures are: A: $\geq 2$ leptons ($\geq 1\,Z$, $\geq 3\,W$, $l + W$); B: 1 lepton ($2 W$); C: \etmiss; D: $VH$, $HH$; E: $W$/$H$/$t$+jets, all jets.} \label{tab:TB}
\end{table}

\begin{table}
\center
\begin{tabular}{l|c|c|c|c|}
$X$, $Y$ & $q l^+ \nu$ & $q q q$ & $q W^+ Z$ & $q W^+ H$  \\
\hline
res. &  $W$, LQ   & $W$, $\eta^\pm$ & $W'$, $\eta^\pm$, VLQ & VLQ  \\
 \hline 
top $X/Y$ & 1/0 & 2/1 & 1/0 & 1/0 \\
 \hline\hline
single  & B & B/E & A & D \\ 
 \hline\hline
$q l^- \nu$ & A & - & & \\
\hline
$q q q$ & A/B & A/E & - & \\
\hline
$q W^- Z$ & A & A & A & - \\
\hline
$q W^- H$ & A & D & A & B\\
\hline
\end{tabular}
\caption{Classification for the exotic charged states $X$/$Y$, following the same structure as Table~\ref{tab:TB}.} \label{tab:XY}
\end{table}

The the final states are classified in Table~\ref{tab:TB} for the standard charges, $T$ and $B$, and in Table~\ref{tab:XY} for exotic charges, $X$ and $Y$. 
In each table, the second row shows the possible intermediate resonances that may appear in the decay process, as detailed above. As $q$ stands for any SM quark, we also indicate the maximal number of tops that may appear in the final state (which is equivalent to the maximum number of up-type quarks): this is experimentally important as tops are a source of additional leptons that improve background rejection.
The rest of the table contains all the possible final states for single and pair production. For single production, the VLQ is typically produced in association with other particles that can be used to tag the events. However, we only use the decay products of the VLQ to characterize the event because the nature of the additional particles is inherently model dependent. 
The letters A--E categorize the finals states according to typical experimental selections: 
\begin{itemize}
\item[]
\item[A:] includes final states with at least two leptons (including final states with at least one $Z$, one lepton and at least one $W$/top, or at least three $W$/top's, which have sizeable branching ratios into two or more leptons). Final states with two or more leptons are probed in many LHC searches, as for example~\cite{Aad:2014efa,Aad:2015gdg,Chatrchyan:2013uxa}, and at high mass the final state is sufficiently well understood that explicit reconstruction of the resonance is not always necessary for discovery, in particular if events contain substantial \etmiss~\cite{Aad:2015wqa}.
\item[B:] includes final states with a single lepton (sizeable rates also come from final states with two $W$/top's).  In this case, because of the large backgrounds from $t\bar{t}$ and $W$+jets production, explicit resonance reconstruction is probably necessary to observe a signal.  While some of these final states have not been probed in direct VLQ searches like for instance, $qW^+W^-q$\etmiss, if either branching ratio is substantial, other final states in same row or column will yield multiple leptons or bosons.  It is therefore likely that multi-boson differential measurements, for example searches for anomalous gauge couplings~\cite{Aad:2014mda} or multi-Higgs production~\cite{Aad:2015xja}, offer sufficient coverage.
\item[C:] final states with sizeable \etmiss are covered by the many experimental searches for supersymmetry.
\item[D:] final states containing one Higgs in association with one or more bosons ($W/Z/H$).  If the $q$'s are light quarks, sensitivity to these channels could be enhanced by exploiting the invariant mass distributions of the candidate VLQ decay products in searches for $VH$ and $HH$ production. The $qHqH$ case is also discussed in detail in another contribution to these proceedings.
\item[E:] hadronic final states. In this case, the presence of hadronically decaying bosons or tops may facilitate the experimental searches. If the VLQ mass is large enough, searches for pairs of 3-jet resonances~\cite{Chatrchyan:2013gia} have efficient coverage.  Three-jet resonance searches should be undertaken in addition to the dijet resonance searches to have sensitivity to single production of VLQs that dominantly decay to three quarks. 
\item[]
\end{itemize}
Most of the columns and rows in Tables~\ref{tab:TB} and \ref{tab:XY} have multiple cells that are covered either by dedicated VLQ searches or other searches for new physics.  However, in the latter cases, the sensitivity could be enhanced by adding the appropriate invariant mass distributions to those studied.  This is particularly true if the SM quark in the VLQ decay is a light quark.

\section*{CONCLUSIONS}

Vector-like quarks are present in many extensions of the Standard Model and the subject of active 
searches by the LHC experiments.  These searches focus on decays to a quark, most often of the third generation, 
and a $W$, $Z$ or Higgs boson.  However, models with vector-like quarks often include additional 
new particles, e.g.~new gauge bosons or a dark matter candidate in the case of a conserved parity.
We present a matrix of possible final states, categorized as a function of the number of final state
leptons and bosons.  While most final states seem to be well covered by existing experimental searches, there
are cases, in particular in the absence of third generation quarks, where experimental coverage could
be enhanced by the simple addition of the appropriate invariant mass distributions to other existing searches 
for new physics.  For single VLQ production, in a few cases a dedicated new search is necessary.

\section*{ACKNOWLEDGEMENTS}

GC acknowledges partial support from the Labex-LIO (Lyon Institute of Origins) under grant ANR-10-LABX- 66 and FRAMA (FR3127, F\'ed\'eration de Recherche ``Andr\'e Marie Amp\`ere'') and from the Th\'eorie LHC France initiative of the CNRS.


\AddToContent{G.~Brooijmans, G.~Cacciapaglia}
\renewcommand{\thesection}{\arabic{section}}

\chapter{The Diboson Excess: Experimental Situation and Classification of Explanations}

{\it J.~Brehmer,
G.~Brooijmans,
G.~Cacciapaglia,
A.~Carmona,
A.~Carvhalo,
R.~S.~Chivukula,
A.~Delgado,
F.~Goertz,
J.L.~Hewett,
A.~Katz,
J.~Kopp,
K.~Lane,
A.~Martin,
K.~Mohan,
D.M.~Morse,
M.~Nardecchia,
J.M.~No,
C.~Pollard,
M.~Quiros,
T.G.~Rizzo,
J.~Santiago,
V.~Sanz,
E.H.~Simmons
and J.~Tattersall}

\begin{abstract}
We examine the `diboson' excess at $\sim 2$\, TeV seen by the LHC experiments in various channels. We provide a comparison of the excess significances as a function of the mass of the tentative resonance and give the signal cross sections needed to explain the excesses.
We also present a survey of available theoretical explanations of the resonance, classified in three main approaches. Beyond that, we discuss methods to verify the anomaly, determining the major properties of the various surpluses and exploring how different models can be discriminated.
Finally, we give a tabular summary of the numerous explanations, presenting their main phenomenological features.
\end{abstract}

This contribution was made public~\cite{Brehmer:2015dan} earlier than the rest of the proceedings to match 
the release 
of $\sqrt{s}$ = 13 TeV results by the ATLAS and CMS collaborations.

\AddToContent{J.~Brehmer, G~.Brooijmans, G.~Cacciapaglia, A.~Carmona, A.~Carvalho,
R.S.~Chivukula, A.~Delgado, F.~Goertz,
J.L.~Hewett, A.~Katz, J.~Kopp, K.~Lane, A.~Martin, K.~Mohan, D.M.~Morse, M.~Nardecchia, J.M.~No,
C.~Pollard, M.~Quiros, T.G.~Rizzo, J.~Santiago, V.~Sanz, E.~H.~Simmons and J.~Tattersall}
\renewcommand{\thesection}{\arabic{section}}



\graphicspath{{PatiSalamSUSY/}}

\def\bpm{\begin{pmatrix}} 
\def\epm{\end{pmatrix}} 
\def\bea{\begin{eqnarray}}
\def\eea{\end{eqnarray}}

\def\SK#1{{ \textcolor{blue}{{[SK:~#1]}}}}
\def\JT#1{{ \textcolor{red}{{[JT:~#1]}}}}
\newcommand{\cyan}{\color{cyan}}
\newcommand{\WP}[1]{{\cyan WP: #1}}
\newcommand{\WPadd}[1]{{\cyan #1}}
\newcommand{\WPout}[1]{{ \st{#1}}}
\newcommand{\WPrep}[2]{{\cyan \st{#1}} {\cyan #2}}


\chapter{Collider constraints on Pati-Salam inspired SUSY models with a sneutrino LSP}

{\it S.~Kraml, M.E.~Krauss, S.~Kulkarni, U.~Laa, W.~Porod and J.~Tattersall}

\begin{abstract}
We study Pati-Salam inspired supersymmetric models which feature a somewhat
more compressed spectrum than the usual constrained Minimal Supersymmetric
Standard Model. Neutrino data are explained via an inverse seesaw mechanism 
and the corresponding scalar partners of this sector can be the lightest supersymmetric states,   
accounting for the dark matter of the universe. 
Our particular focus lies on the question how the LHC SUSY searches constrain this setup.
\end{abstract}

\section{INTRODUCTION}

After the highly successful LHC operation at 7 and 8~TeV in 2010--2012, the exploration of the TeV energy scale now continues at 13 TeV.  One of the main experimental goals of this Run~2 of the LHC, which begun last year, is to discover signs for new physics.  So far, however, no conclusive signal has been established. Instead, the mass limits for new particles, in particular  supersymmetric ones, are being pushed higher and higher, presently reaching 1.5--1.7~TeV for gluino masses in certain scenarios~\cite{dec15seminar,ATLAS-CONF-2015-062,CMS-PAS-SUS-15-003}.
This is a somewhat paradoxical situation as theoretical arguments like the unification of the gauge couplings,
stability of the Higgs potential or the stability of the electroweak scale against
huge radiative corrections suggest that there should be new physics at the TeV scale. 
An important aspect in this context is that the observed Higgs mass $m_h \simeq 125.1$~GeV \cite{Aad:2015zhl}
is somewhat larger than predicted in  the `natural' part of the Minimal Supersymmetric Standard Model (MSSM), or constrained versions thereof like the CMSSM, implying the necessity of large trilinear parameters and/or huge stop masses.

However, this does not imply that supersymmetry (SUSY) itself is ruled out --- non-minimal SUSY models are still in good shape.
In particular non-minimal models offer the possibility of additional tree-level contributions to the
mass of the light Higgs boson, either due to new F-term contributions, like in the 
Next-to-MSSM~\cite{Ellwanger:2006rm}, or due to additional D-term 
contributions in case of extended gauge symmetries~\cite{Ma:2011ea,Hirsch:2011hg}. 
A particular interesting class are $SO(10)$-inspired models with left-right symmetry at the TeV scale as
they can also naturally explain the measured neutrino data. The $SO(10)$ gauge group can be broken
in various ways down to the Standard Model gauge group, see e.g.~\cite{Mohapatra:1999vv} and references therein.
For example, it can be broken directly to $SU(2)_L \times SU(2)_R \times U(1)_{B-L}$~\cite{Hirsch:2015fvq} or to $SU(2)_L \times U(1)_{R}\times U(1)_{B-L}$~\cite{Hirsch:2011hg,Hirsch:2012kv} at the scale of the grand unified 
theory (GUT). An equally justified alternative which is less well studied in recent literature is to have an  intermediate Pati-Salam~\cite{Pati:1974yy} stage where the gauge group is $SU(4)_c \times SU(2)_L \times SU(2)_R$. 
This scenario is of particular interest because sleptons and squarks belong to the same $SU(4)_c$ multiplet in the unbroken phase, allowing for a rather compressed spectrum
between the sfermions at accessible LHC energies if the scale of the Pati-Salam breaking is in the order of 10--100 TeV \cite{DeRomeri:2011ie}. 
If moreover the gauginos and higgsinos are heavier than the sfermions, 
the LHC bounds are considerably weakened \cite{Dreiner:2012gx}.
In such a scenario the lightest supersymmetric particle (LSP) would be the lightest of the scalar partners of the extended neutrino sector, i.e.\ a sneutrino. 
Sneutrinos as dark matter (DM) candidates have been discussed extensively in the literature (see e.g.~\cite{Arina:2007tm} for a review). 
LHC limits on scenarios with sneutrino LSPs were studied recently in the context of so-called SMS (simplified model spectra) constraints, in \cite{Arina:2015uea} for the MSSM+$\tilde\nu_R$ and in \cite{Belanger:2015cra} for the $U(1)$-extended MSSM.

In this contribution, we focus on Pati-Salam inspired SUSY with a rather compressed sfermion spectrum, heavy gauginos and higgsinos, and a sneutrino LSP.  An additional interesting feature of such a model is that it can also explain neutrino masses and mixings by an inverse seesaw mechanism~\cite{Mohapatra:1986bd}. In the present work, we are mainly interested in the  question to what extent the present SUSY search results from Run~1 of the LHC constrain this model, and what scenarios evade present (and possibly future) bounds. 
The complementarity with dark matter constraints will be discussed in a subsequent paper publication.

\section{THE MODEL}\label{sec:PatiSalam-model}

\subsection{Model description}

We assume that $SO(10)$ is broken at the GUT scale to a Pati-Salam subgroup 
$\mathcal G_{PS} = SU(4)_c \times SU(2)_L\times SU(2)_R$ which itself survives 
down to a scale $M_{PS} \sim 10^4 - 10^5~$GeV. At $M_{PS}$, $\mathcal G_{PS}$ is broken to 
$SU(3)_c\times SU(2)_L\times U(1)_R\times U(1)_{B-L}$, and we will work in this phase. 
The matter sector can be embedded in a complete {\bf{16}}-plet under $SO(10)$ for each flavour
which adds three copies of right-handed neutrino superfields $\hat \nu^c$ with respect to the MSSM.
The Higgs sector contains, in addition to the usual Higgs doublets $\hat H_u$ and $\hat H_d$, two 
superfields $\hat \chi_R$ and $\hat{\bar{\chi}}_R$ originating from $SU(2)_R$ doublets 
in the Pati-Salam phase and
are responsible for the breaking of $U(1)_R \times U(1)_{B-L}$ as soon as their scalar components 
receive vacuum expectation values (vevs). $\hat \chi_R$ ($\hat{\bar{\chi}}_R$) carry
the opposite (same) $U(1)$ charges as $\hat \nu^c$, enabling an inverse seesaw mechanism 
once a gauge singlet superfield $\hat S$ is added. 
In Table~\ref{PatiSalam:tab:matter_content}, we list the particle content of 
our model with the respective quantum
numbers under the considered gauge group. Here, we have normalized the $U(1)$ charges such that, after 
the breaking $U(1)_{R} \times U(1)_{B-L} \to U(1)_Y$, the hypercharge operator reads 
\begin{align}
Y = T_R + T_{B-L}\,, \quad {\rm where} \quad
T_{B-L} \, \Phi = \frac{B-L}{2} \, \Phi\,.
\end{align} 
Note that the $U(1)_R\times U(1)_{B-L}$ basis can be rotated into a basis featuring the hypercharge $U(1)_Y$ and an orthogonal $U(1)_\chi$ \cite{Krauss:2013jva}. The corresponding quantum numbers 
are also displayed in Table~\ref{PatiSalam:tab:matter_content}.

As $\chi_R$, $\bar \chi_R$ carry $B-L$ quantum numbers of $\pm 1$, the
usual $R$-parity gets spontaneously broken together with the breaking of $U(1)_R\times U(1)_{B-L}$. 
We therefore follow Refs.~\cite{Dev:2009aw,BhupalDev:2010he} by introducing a discrete $Z_2^M$ 
matter parity under which the matter superfields are odd and the Higgs superfields even, thereby 
forbidding the terms leading to proton decay as well as the LSP decay.\footnote{The constraints due 
to the discrete gauge symmetry anomalies are fulfilled by this type of matter parity 
\cite{Ibanez:1991hv,Dreiner:2005rd}. } 
\begin{table}[t]
\centering
\begin{tabular}{|c|c|c|c|c|c|c|} 
\hline \hline 
{\small{Superfield}} & {\small{Spin 0}} & {\small{Spin \(\frac{1}{2}\)}} & \hspace{-.1cm}{\small{\# Gen.}}\hspace{-.1cm} &
\hspace{-.1cm}{\scriptsize{$SU(3)_c \times SU(2)_L$}}\hspace{-.1cm} & \hspace{-.1cm}{\scriptsize{$U(1)_R \times U(1)_{B-L}$}}\hspace{-.1cm} &  \hspace{-.1cm}{\scriptsize{$U(1)_Y \times U(1)_{\chi}$}}\hspace{-.1cm} \\ 
\hline \hline
\(\hat{Q} \) & \((\tilde u_L, \tilde d_L)\) & \((u_L,d_L)\) & 3
 & \(({\bf 3},{\bf 2}) \) & \((0,\frac{1}{3}) \) & \((\frac{1}{6},\frac{1}{4}) \) \\ 
\({\hat{d}}^{c}\) & \(\tilde{d}_R^c\) & \(d_R^c\) & 3
 & \(({\bf \overline{3}},{\bf 1})\) & \((\frac{1}{2},-\frac{1}{3}) \) & \((\frac{1}{3},-\frac{3}{4}) \)\\ 
\({\hat{u}}^{c}\) & \(\tilde{u}_R^c\) & \(u_R^c\) & 3
 & \(({\bf \overline{3}},{\bf 1})\) & \((-\frac{1}{2},-\frac{1}{3}) \) & \((-\frac{2}{3},\frac{1}{4}) \) \\ 
\(\hat{L} \) & \((\tilde \nu_L,\tilde e_L)\) & \((\nu_L,e_L)\) & 3
 & \(({\bf 1},{\bf 2})\) &\((0,-1) \) &\((-\frac{1}{2},-\frac{3}{4}) \)\\ 
\({\hat{e}}^{c}\) & \(\tilde{e}_R^c\) & \(e_R^c\) & 3
 & \(({\bf 1},{\bf 1})\) &\((\frac12,1) \) &\((1,\frac{1}{4}) \)\\ 
\({\hat{\nu}^{c}}\) & \(\tilde{\nu}_R^c\) & \(\nu_R^c\) & 3
 & \(({\bf 1},{\bf 1})\) &\((-\frac12,1) \) &\((0,\frac{5}{4}) \)\\ 
 \({\hat{S}}\) & \(\tilde{S}\) & \(S\) & 3
 & \(({\bf 1},{\bf 1})\) & \((0,0) \) & \((0,0) \)\\ 
\(\hat{H}_d\) & \((H_d^0,H_d^-)\) & \((\tilde H_d^0,\tilde H_d^-)\) & 1
 & \(({\bf 1},{\bf 2})\) &\((-\frac{1}{2},0) \) &\((-\frac{1}{2},\frac{1}{2}) \)\\ 
\(\hat{H}_u\) & \((H^+_0,H^0_u)\) & \((\tilde H^+_0,\tilde H^0_u)\) & 1
 & \(({\bf 1},{\bf 2})\) & \((\frac{1}{2},0) \) & \((\frac{1}{2},-\frac{1}{2}) \)\\ 
\(\hat{\chi}_R\) & \(\chi_R\) & \(\tilde{\chi}_R\) & 1
 & \(({\bf 1},{\bf 1})\) &\((\frac12,-1) \) &\((0,-\frac{5}{4}) \)\\ 
\(\hat{\bar{\chi}}_R\) & \(\bar{\chi}_R\) & \(\tilde{\bar{\chi}}_R\) & 1
 & \(({\bf 1},{\bf 1})\) &\((-\frac12,1) \) &\((0,\frac{5}{4}) \)\\ 
\hline \hline
\end{tabular} 
\caption{Chiral superfields and their quantum numbers with
respect to $ SU(3)_c\times\, SU(2)_L\times\,  U(1)_R\times\, U(1)_{B-L}
$ as well as in the rotated basis $ SU(3)_c\times\, SU(2)_L\times\,  U(1)_Y\times\, U(1)_{\chi}
$.}
\label{PatiSalam:tab:matter_content}
\end{table}
The superpotential of the model then reads 
\cite{Hirsch:2012kv,Hirsch:2011hg,Krauss:2013jva}
\begin{align}
\notag  W =& Y^{ij}_u\, {\hat{u}}^{c}_i \, \hat{Q}_j \cdot \hat{H}_u
- Y_d^{ij}\, {\hat{d}}^{c}_i\,\hat{Q}_j\cdot \hat{H}_d
- Y^{ij}_e\, {\hat{e}}^{c}_i\,\hat{L}_j\cdot \hat{H}_d
+\mu\,\hat{H}_u \cdot \hat{H}_d \\
& \label{PatiSalam:eq:superpotential}
+Y^{ij}_{\nu}\,{\hat{\nu}}^{c}_i\,\hat{L}_j \cdot \hat{H}_u 
+Y_S^{ij}\, \hat \nu^c_i\, \hat S_j\, \hat \chi_R - 
\mu_R\, \hat{ \bar{ \chi}}_R \, \hat \chi_R + \frac{1}{2} \mu_S^{ij}\,  \hat S_i\, \hat S_j\,,
\end{align}
where the $\cdot$ indicates the invariant $SU(2)_L$ product. 
Below $M_{PS}$ we have a 
product of two $U(1)$ factors which in general leads to a gauge kinetic mixing. We absorb the corresponding mixing term
$\chi F^{\mu \nu}_{R} F_{B-L,\, \mu\nu}$ into a noncanonical structure of the covariant derivative, resulting in a non-diagonal $U(1)$ gauge coupling matrix,  
and also account for that effect in the RGE running using the formalism given in Ref.~\cite{Fonseca:2011vn}. Depending on the particle content and the separation of the different symmetry breaking scales, this mixing can turn out to be large and have
non-negligible effects for the collider phenomenology of a model \cite{Krauss:2012ku,Staub:2016dxq}. 
However, due to the closeness
of the scales in the model considered here, the phenomenological effect is small
and, thus, we omit it for simplicity in the formulas below.
As 
usual in inverse seesaw scenarios, 
$\mu_S$ is a small
dimensionful parameter, $\mu_S \ll M_Z$.
The Yukawa coupling between the singlet and the right-handed sneutrino superfields, $Y_S$, is considered to be of 
$\mathcal O(1)$ and will become important
later when we discuss the (s)neutrino masses.

Finally, the soft SUSY-breaking Lagrangian is given by
\begin{align}
 -&\mathcal L_{soft} =  m^2_{ij} \phi_i^* \phi_j +\bigg( \frac{1}{2} M_{ab} \lambda_a \lambda_b
 + B_\mu\, H_u \cdot H_d - B_{\mu_R}\, \bar{{\chi}}_R\, \chi_R +B_{\mu_S}\, \tilde S\, \tilde S 
 + T_S^{ij}\, \chi_R\, \tilde \nu^c_i\, \tilde S_j \nonumber \\
  &
  - T_d^{ij}\,  \tilde d^c_i\, \tilde Q_j \cdot H_d
  + T_u^{ij}\,  \tilde u^c_i\, \tilde Q_j \cdot H_u
  - T_e^{ij}\, \tilde e^c_i\, \tilde L_j \cdot H_d
 + T_\nu^{ij}\, \tilde \nu^c_i\, \tilde L_j \cdot H_u
 + {\rm h.c.} \bigg)\,.
\end{align}
Here, $\phi$ stands for all kinds of scalar particles, $i$ and $j$ are generation indices,
 and $\lambda_a$ denotes the gaugino of gauge group $a$. 
The mixed term $M_{R,B-L} \lambda_R \lambda_{B-L}$ appears because of gauge kinetic mixing.

\subsection{Relevant mass spectrum}

\subsubsection{Higgs sector}
 
We split the neutral Higgs fields into their $CP$-even and $CP$-odd components as well as a vev according to
\begin{align}
 \notag &H_u^0 = \frac{1}{\sqrt{2}}(\phi_u + i \sigma_u + v_u), &
 H_d^0 = \frac{1}{\sqrt{2}}(\phi_d + i \sigma_d + v_d), \\
 &\chi_R = \frac{1}{\sqrt{2}}(\phi_R + i \sigma_R + v_{\chi_R}),  &\bar \chi_R = \frac{1}{\sqrt{2}}(\bar \phi_R + i \bar \sigma_R + v_{\bar \chi_R})\,.
\label{PatiSalam:eq:higgs_states}
 \end{align}
Moreover, we define the ratios 
\begin{align}
\tan \beta = \frac{v_u}{v_d}\,,\qquad \tan \beta_R = \frac{v_{\chi_R}}{v_{\bar \chi_R}}\,,
\end{align}
as well as $v^2=v_u^2+v_d^2  $ and $ v_R^2=v_{\chi_R}^2 +v_{\bar \chi_R}^2 $. Due to the constraints on dilepton resonances \cite{atlasdilepton}, $v_R$ needs to be larger than $\sim 7~$TeV,
so that $v \ll v_R$ generically holds.
As discussed in some detail in Ref.~\cite{Krauss:2013jva}, the minimization conditions for the scalar potential 
only allow small deviations of $\tan \beta_R$ from one. 
In the limit of vanishing gauge kinetic mixing and $\tan \beta \gg 1$, $\tan \beta_R \to 1$,
the lightest eigenstate that corresponds to a $SU(2)_L$ doublet features the tree-level mass of 
\begin{align}
m_h^2  = \frac{1}{4} (g_L^2 + g_R^2) \, v^2\,,
\end{align}
corresponding to a tree-level enhancement with respect to the MSSM\footnote{In the alternative $U(1)$ basis $Y\chi$, this corresponds to $m_h^2  = M_Z^2 + \frac{1}{4} (g_\chi - g_{Y\chi})^2\, v^2$ \cite{Krauss:2013jva}.}  and therefore alleviating the need for large loop corrections from the stop sector.\footnote{Note that, in this limit, the lightest $\bar \phi_R/\phi_R$ eigenstate is massless at tree level and only acquires a mass of $\sim 50~$GeV radiatively \cite{Hirsch:2012kv,Hirsch:2011hg,Krauss:2013jva}. As the mixing with the doublet state is small, there is no constraint coming from LEP data.} The full mass matrix in both $U(1)$ bases including further details is given in Ref.~\cite{Krauss:2013jva}.

\subsubsection{Neutrinos and sneutrinos}

The (s)neutrino sector consists of the left- and right-handed superfields $\hat \nu_{L/R}$ as well as the singlet superfield $\hat S$. After $U(1)_R\times U(1)_{B-L}$ and electroweak symmetry breaking, the gauge eigenstates mix among each other according to the Yukawa interactions defined in Eq.~(\ref{PatiSalam:eq:superpotential}).

In the seesaw approximation, $\mu_S \ll \frac{1}{\sqrt{2}} v_u Y_\nu \ll \frac{1}{\sqrt{2}} v_{\chi_R} Y_S$, the masses of the light neutrinos read 
 \cite{Mohapatra:1986bd,GonzalezGarcia:1988rw}
\begin{align}
 M_{\rm{light}} \simeq \frac{v_u^2}{v^2_{\chi_R}} Y_\nu^T (Y_S^T)^{-1} \mu_S Y_S^{-1} Y_\nu\,.
 \label{PatiSalam:eq:light_neutrino_masses}
\end{align}
The heavy neutrinos form three quasi-Dirac pairs whose mass can be approximated by
\begin{align}
m_{\nu_h} \simeq \frac{1}{\sqrt{2}} v_{\chi_R} Y_S\,.
\end{align}
Typical masses for the heavy neutrinos hence range around $\mathcal O(100~{\rm GeV} - 1~{\rm TeV})$.
As one can always choose a basis in which $Y_S$ is diagonal, one can fit the masses of the light neutrinos by a non-diagonal structure of $Y_\nu$ and/or $\mu_S$. We choose to adapt $Y_\nu$ following the parametrisation given in Refs.~\cite{Basso:2012ew,Abada:2012mc,Abada:2014kba}.

Analogously to the Higgs sector, we decompose the sneutrinos into their $CP$-even ($S$) and $CP$-odd  ($P$) components,
\begin{align}
\tilde \nu_L = \frac{1}{\sqrt{2}} (\nu_L^S + i\,\nu_L^P)\,,\quad \tilde \nu_R = \frac{1}{\sqrt{2}} (\nu_R^S + i\,\nu_R^P) \,,\quad \tilde S =  \frac{1}{\sqrt{2}} (\nu_S^S + i\,\nu_S^P) \,.
\end{align}
In the limit of vanishing gauge kinetic mixing, the mass matrix of the scalar/pseudoscalar sneutrinos reads in the basis ($\tilde \nu_L^{S/P},\tilde \nu_R^{S/P}, \tilde S^{S/P}$):
\begin{align}
 (m^{S/P}_{\tilde \nu})^2 = \begin{pmatrix}
                     m^2_{\tilde \nu L} & \frac{v}{\sqrt{2}} \big( T_\nu^T s_{\beta} -\mu Y_\nu^T c_{\beta} \big)  
                     &  \frac{1}{2}v v_R (Y_\nu^T Y_S) s_{\beta} s_{\beta_R}\\
                     \frac{v}{\sqrt{2}}  \big( T_\nu s_{\beta} -\mu Y_\nu c_{\beta} \big) 
                     & m^2_{\tilde \nu R}  & (m^2_{32})^T \\
                     \frac{1}{2}v v_R ( Y_S^T Y_\nu) s_{\beta} s_{\beta_R} 
                     & m^2_{32}
                     & ~m^2_S + \frac{v_R^2 s_{\beta_R}^2}{2} Y_S^T Y_S + \mu_S^T \mu_S \pm B_{\mu_S}
                    \end{pmatrix}\,,
\end{align}
with
\begin{align}
\notag m^2_{32} =& \frac{v_R}{\sqrt{2}} \big(T_S^T s_{\beta_R} - \mu_R Y_S^T c_{\beta_R} \pm \mu_S Y_S^T s_{\beta_R}\big)\,,\\
\notag m^2_{\tilde \nu L} =& m_L^2 + \frac{v^2}{2} s_{\beta}^2 Y_\nu^T Y_\nu + \frac{1}{8} 
\Big( v^2 c_{2 \beta} g_L^2  - 
 v_R^2 c_{2 \beta_R} g_{BL}^2 \Big) {\bf 1}\,,\\  
 m^2_{\tilde \nu R} =& m^2_{\nu^c} + \frac{v_{R}^2}{2} s^2_{\beta_R} Y_S Y_S^T + \frac{v^2}{2} s_{\beta}^2 Y_\nu Y_\nu^T + 
 \frac18 \Big(  v^2 c_{2 \beta}  g_R^2 + v_R^2 c_{2 \beta_R}\big( g_R^2 + g_{BL}^2 \big)
 \Big){\bf 1}\,,
\end{align}
where $m^2_{L,\nu^c,S}$ are the soft SUSY-breaking sneutrino mass parameters, and we have used the abbreviations $\cos \alpha,\sin \alpha \to c_\alpha,s_\alpha$. Furthermore, we have used real parameters $T_i,Y_i,\mu_i$.
While the mixing of the left sneutrinos with $\tilde \nu_R$ and $\tilde S$ scale with the 
Dirac neutrino mass and is therefore small, there is in general a sizeable mixture in the $\tilde \nu_R - \tilde S$ submatrix because of the potentially large $m^2_{32}$ which scales with the
heavy neutrino mass $m_{\nu_h}$. Assuming a common soft mass $m^2$, $\tan \beta_R\to 1$ and writing $v_R T_S/\sqrt{2} = m_{\nu_h} A_S$, the mass of the lightest eigenstate can be approximated 
by 
\begin{align}
m^2_{\tilde \nu_1} = m^2 + m_{\nu_h} \, \left|m_{\nu_h} -(A_S-\mu_R)\right|
\end{align}
Therefore, due to the 
inverse seesaw mechanism at work, the sneutrino spectrum can be spread rather than compressed (as  is the case for the charged sfermions), and the lightest eigenstate can be the dark matter candidate if it is the LSP, i.e. if $m_{\tilde \nu_1}<\mu,\mu_R,M_{BL},M_{R},M_L$ ($M_i$ being the gaugino masses). 

The terms $\mu_S$ and $B_{\mu_S}$ induce a small splitting 
between the $CP$ eigenstates. As $\mu_S$ must be a small parameter for a 
successful seesaw mechanism, and thus much smaller than $\mu_R$, this splitting is tiny and irrelevant for collider phenomenology. It does become cosmologically relevant, however,
in case the lightest sneutrino is the LSP 
as it induces a decay of the second-lightest sneutrino eigenstate; that way, the 
dark matter candidate is a real scalar/pseudoscalar state instead of a complex scalar.

\subsubsection{Squarks}

The mass matrices of the squarks read, in the limit of vanishing gauge kinetic mixing,
\begin{align}
m^2_{X} = \begin{pmatrix}
m_Q^2 + \frac{v_X^2}{2} Y_X^\dagger Y_X + D_{X,L} & \frac{1}{\sqrt{2}} 
\left( v_X T^\dagger_X - v_{X'} \mu Y^\dagger_X \right) \\ 
\frac{1}{\sqrt{2}} \left( v_X T_X - v_{X'} \mu^* Y_X \right) & m^2_X + \frac{v_X^2}{2} Y_X Y_X^\dagger + D_{X,R}
\end{pmatrix}\,,
\end{align}
where $X=u,d$, $X'\ne X$ and 
\begin{align}
D_{u/d,L} &=  -\frac{1}{24}\left( (g_{BL}^2 + g_R^2) (v_{\chi_R}^2 - v_{\bar \chi_R}^2) \pm 3 g_L^2 (v_u^2-v_d^2) \right)\, \mathbf{1}\,,\\ \notag 
D_{u/d,R} &=  \frac{1}{24}\left( (g_{BL}^2 \mp 3 g_R^2) (v_{\chi_R}^2 - v_{\bar \chi_R}^2) \mp 3 g_R^2 (v_u^2-v_d^2) \right)\, \mathbf{1}\,.
\end{align}
The full mass matrices including kinetic mixing are given in Ref.~\cite{Hirsch:2012kv}.

\subsection{Technical setup}

We have used the implementation of the model in the  {\tt SARAH} package~\cite{Staub:2008uz,Staub:2010jh,Staub:2012pb,Staub:2013tta,Staub:2015kfa} presented in Ref.~\cite{Hirsch:2012kv} to produce model-specific code for {\tt SPheno} \cite{Porod:2003um,Porod:2011nf}. {\tt SPheno} provides a precise calculation of the complete mass spectrum 
at the one-loop level, and of the $CP$-even Higgs sector at the two-loop level~\cite{Goodsell:2015ira}. The complete RGE running is performed at two loops, including the full gauge kinetic mixing information. The {\tt SPheno} modules were created such as to reproduce Pati-Salam inspired boundary conditions at the scale $M_{PS}$.
In particular, the soft SUSY-breaking mass parameters $m^2_i$ of the sfermions are taken degenerate for each generation $i$, and $g_{BL}$ is to (approximately) unify with $g_3$. The $U(1)_R$ gauge coupling is then calculated by requiring the relation\footnote{Here, the $U(1)$ gauge couplings are not GUT-normalized.}
\begin{align}
g_R &= \frac{g_{BL}\, g_Y}{\sqrt{g_{BL}^2 - g_Y^2}}
\end{align}
at the scale of $U(1)_R\times U(1)_{B-L}$ breaking.
Further, we use a common SUSY-breaking trilinear parameter $A_0$ at $M_{PS}$ so that  $T_k = A_0\,Y_k$. 
With this ansatz, the mixing between the generations is small and, thus, 
we can use the usual notation of $\tilde t_{1,2}$, $\tilde b_{1,2}$ and so on.
 
 The tadpole equations are solved for
 the soft SUSY-breaking Higgs mass parameters. 
  The free input parameters for the subsequent random scan are summarized in Table~\ref{PatiSalam:tab:scan}. As a further selection criterium, we demand the lightest sneutrino to be the LSP and the doublet-like Higgs mass to be within $122 < m_h/{\rm GeV} < 129$ in order to account for the estimated uncertainty on the mass calculation.\footnote{For light stop  masses, this roughly selects the region $-1.5 < A_0/{\rm TeV} < 1$.}

\begin{table}[t]
\centering
\begin{tabular}{l| c || c | c }
\hline \hline
input at $M_{PS}$ &  range & input at $M_{SUSY}$ & range\\ \hline
 $m^2_{1,2}$ [GeV$^2$] &  $2\times 10^6 \, \dots \, 4\times 10^6$ & $v_R$~[TeV] & $7\, \dots \, 10$\\
 $m^2_{3}$ [GeV$^2$] &  $5\times 10^4 \, \dots \, m^2_{1,2}/{\rm GeV}^2$ & $\tan \beta$ & $2\, \dots \, 40$\\
 $M_3$ [GeV] &  $900\, \dots \, 1200 $ & $\tan \beta_R $ & $1\, \dots \, 1.1$\\
 $A_0$~[GeV] &  $-1500 \, \dots \, 1500 $ & $Y_S^{ii}$ &  $0\, \dots \, 0.7$\\
 $\mu$~[GeV] & $500 \, \dots\, 1500$ & $m_{A}$~[GeV] & $500\, \dots \, 1500$\\
 $\mu_R$~[GeV] & $300 \, \dots\, 3000$ & $m_{A_R}$~[GeV] & $m_{A}\, \dots \, 6000$
 \\ \hline \hline
\end{tabular}
\caption{Free parameters and the input values at the respective scale. The Pati-Salam scale itself has been varied between $10^4$ and $10^5$~GeV, and the SUSY scale $M_{SUSY}$ is defined as the average 
 up squark mass $\sqrt{m_{\tilde u_1} m_{\tilde u_6}}$. We assume $1~$TeV for the other gaugino masses $M_i$. $\mu_S^{ii}$ has been fixed to diag$(10^{-5}~$GeV$)$, and we have assumed $M_{RBL},~B_{\mu_S}=0$ at $M_{PS}$. $m_{A_{(R)}} = \sqrt{2 B_{\mu_{(R)}}/s_{2 \beta_{(R)}}}$ is the mass of the physical ($R$-) pseudoscalar.
 The criterion for $m_{A_R}$ is inspired by the hierarchy $v_R>v$ while $\mu_R$ can turn out to be required
 smaller than $\mu$ due to the constraints from the tadpole equations, see the discussion in Ref.~\cite{Krauss:2013jva}.}\label{PatiSalam:tab:scan}
\end{table}

For the evaluation of the collider constraints, we use the same SARAH setup as above to produce model files in 
the {\tt UFO} format \cite{Degrande:2011ua}. This model is used along with the parameter point specified as an SLHA~\cite{Allanach:2008qq} 
spectrum files by {\tt MadGraph\,5\,v2.2.3} \cite{Alwall:2011uj,Alwall:2014hca} to generate events for the
hard matrix element and subsequent particle
decays, showering and hadronisation are performed by {\tt Pythia\,6} \cite{Sjostrand:2006za}. 
The events are normalised to 
the NLO+NLL cross-section using {\tt NLLFast} \cite{Beenakker:1996ch,Beenakker:1997ut,Kulesza:2008jb,Kulesza:2009kq,Beenakker:2010nq,Beenakker:2011fu} 
with the PDF set {\tt CTEQ6L1} at LO and {\tt CTEQ6.6M} at NLO \cite{Pumplin:2002vw}. The events are
then passed to {\tt CheckMATE\,v.1.2.2} \cite{Drees:2013wra,Kim:2015wza} which makes use of the 
{\tt Delphes\,3} detector simulation \cite{deFavereau:2013fsa} and the anti-kT jet algorithm \cite{Cacciari:2008gp}
provided by {\tt FastJet} \cite{Cacciari:2005hq,Cacciari:2011ma}. All 8~TeV ATLAS analyses implemented
in {\tt CheckMATE} are considered which includes those impemented externally 
\cite{Cao:2015ara}. In the results section we will detail those that are most sensitive to
the parameter points tested.

\section{RECASTING LHC SEARCHES}\label{sec:PatiSalam-LHCresults}

For the collider constraints we are most interested in scenarios with relatively 
light third-generation squarks which do not substantially mix with the squarks of the first two generations. 
We thus select points for which the predominantly third-generation states (which we denote as stops and 
sbottoms for simplicity) are lighter than the other squarks and the admixture from other flavours is at most 
one percent. The mixing between the first two generations will not impact our results and is thus not restricted.
Currently, parameter points yielding displaced vertices or heavy stable charged particle (HSCP) signatures cannot 
be tested within the current version of {\tt CheckMATE} and are discarded.
We consider scenarios to contain a  displaced vertex when a particle has a lifetime with $c\tau$ between $0.01$ and $100$m. 
Invisible particles with lifetimes larger than $c\tau=100$m will be considered stable for the purpose of the collider study.

These parameter points (1920 spectra in the analysed sample) are passed to {\tt CheckMATE} which computes the
the $r$ value, defined as the ratio of predicted signal events (taking into account 
uncertainties) over the experimentally measured 95\% CL upper limit on the number of signal events.
Thus a model can be considered excluded by a search if $r \geq 1$.
For a consistent statistical treatment, for each parameter point, $r$ is evaluated based on the most sensitive 
search and signal region, i.e.\ the signal region with the maximum {\em expected} $r$ value.

\subsection{Scan results}

\begin{figure}[t!]
\centering
\hspace*{-2mm}\includegraphics[width=0.52\textwidth]{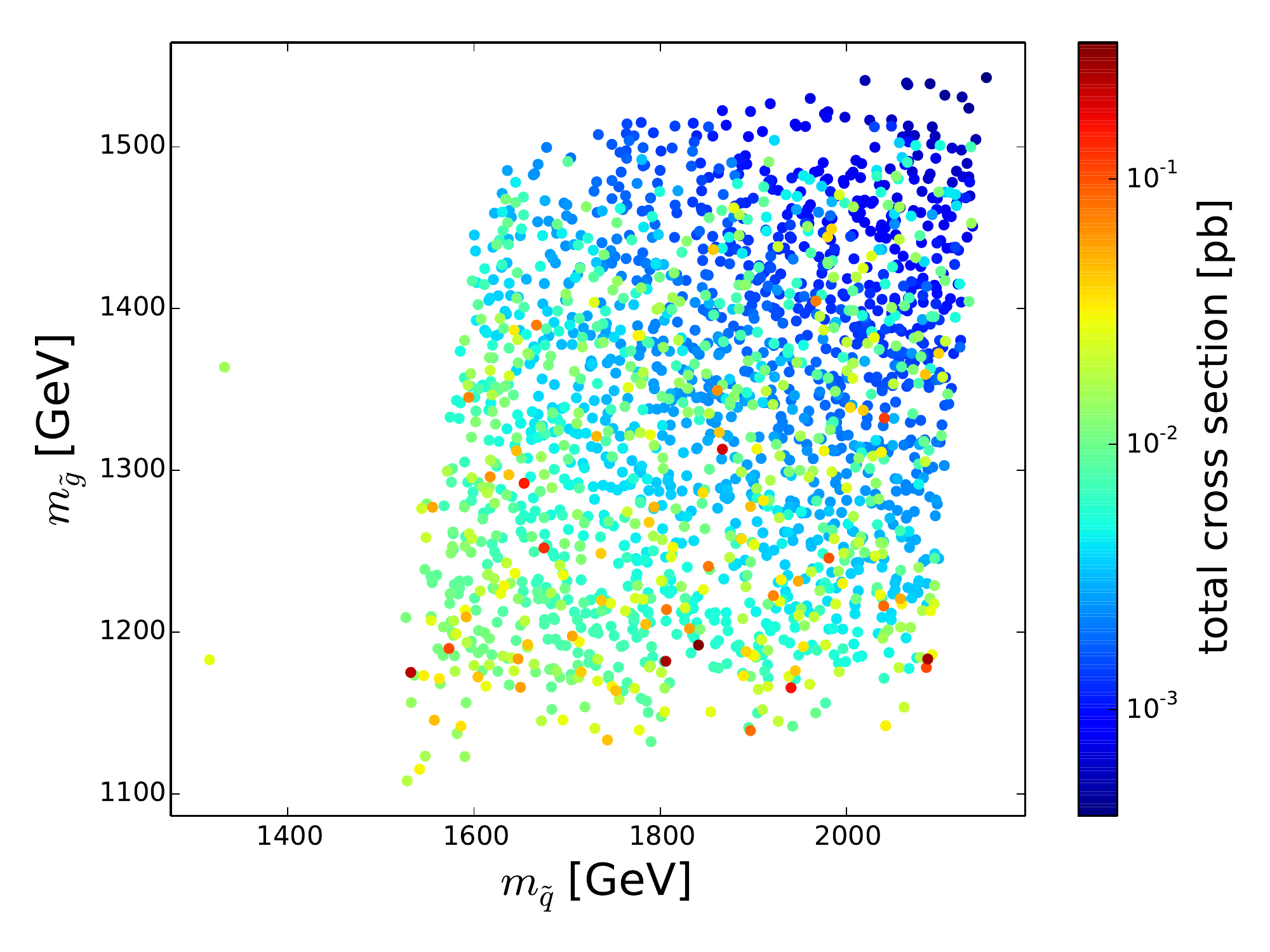}\includegraphics[width=0.52\textwidth]{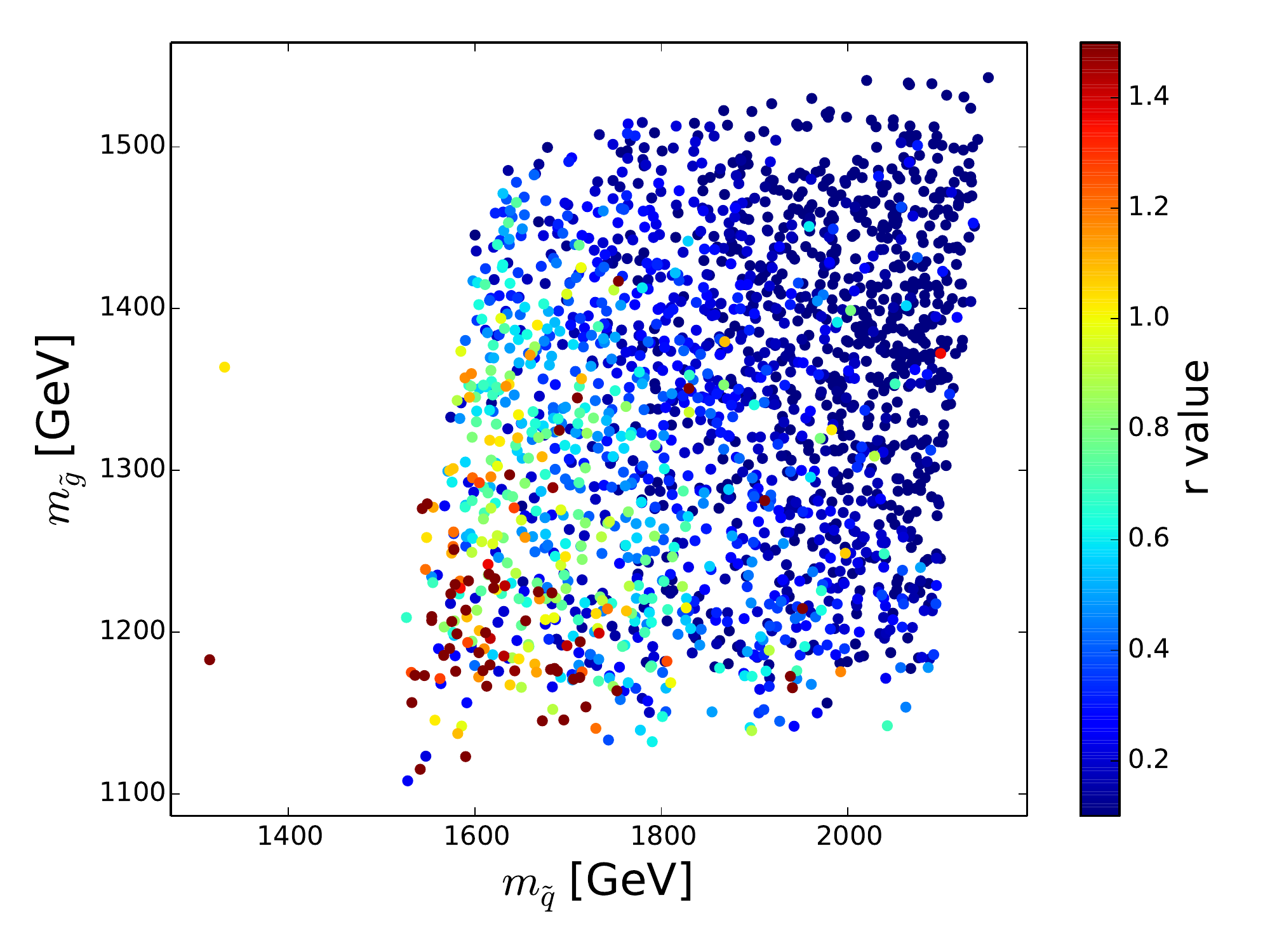}
\hspace*{-2mm}\includegraphics[width=0.52\textwidth]{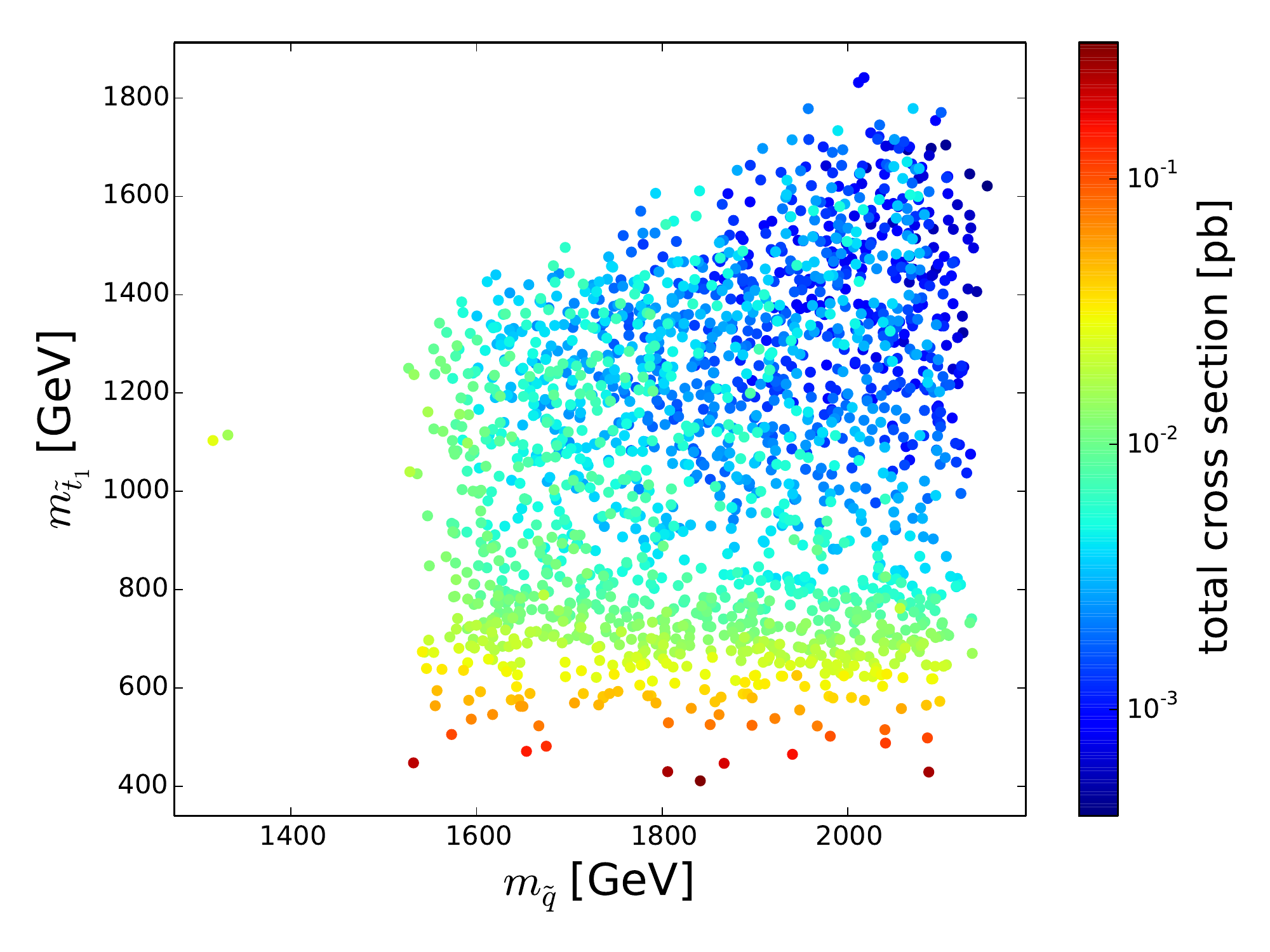}\includegraphics[width=0.52\textwidth]{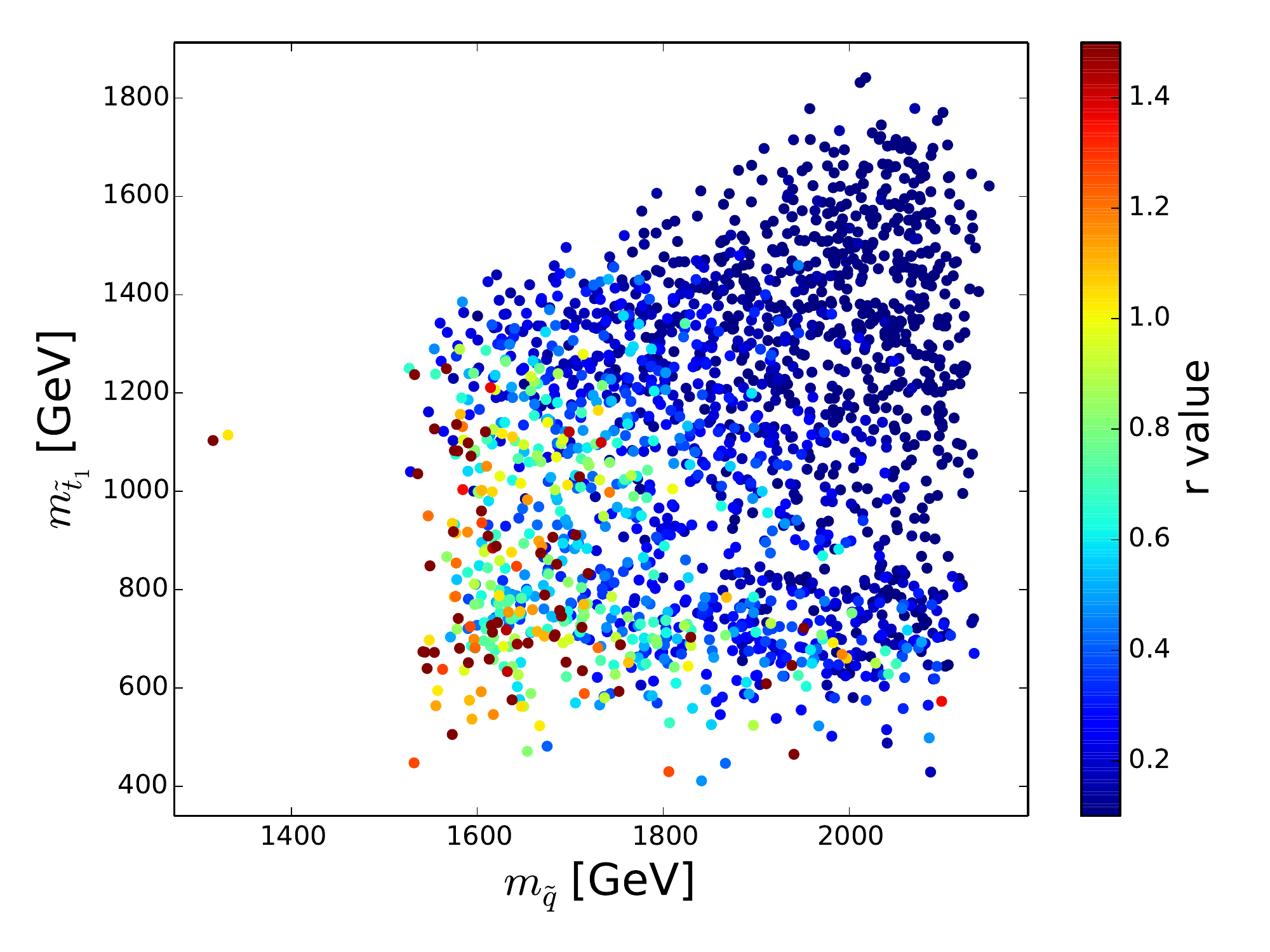}
\hspace*{-2mm}\includegraphics[width=0.52\textwidth]{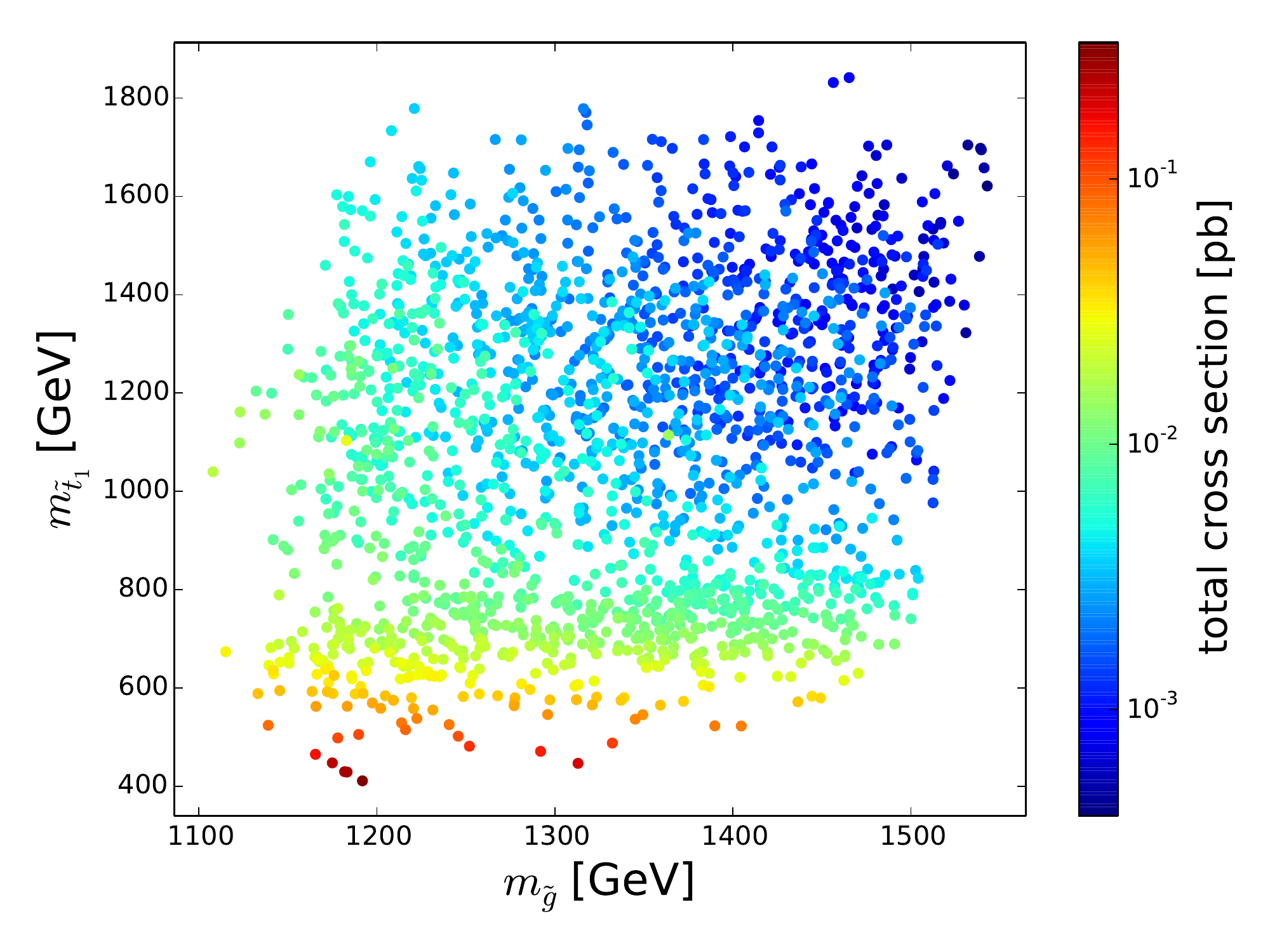}\includegraphics[width=0.52\textwidth]{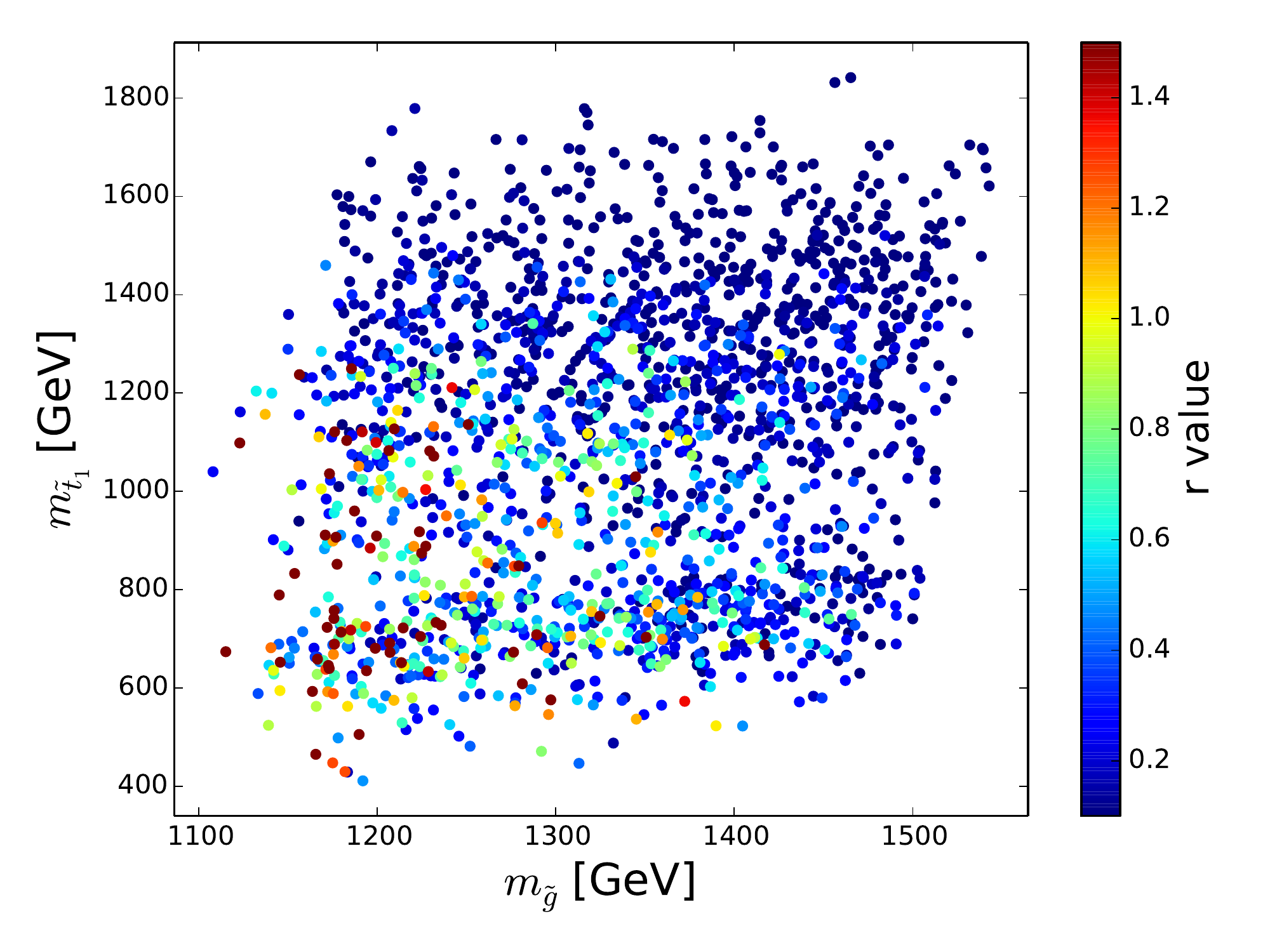}
\caption{\label{fig:PatiSalam-xsectandR} Total SUSY production cross section in pb (right) and $r$ value reported by {\tt CheckMATE} (left) in various mass planes. Top row: gluino ($\tilde g$) vs.\ average light-flavour squark ($\tilde q$) mass, middle row:  $\tilde t_1$ vs.\ $\tilde g$ mass, bottom row: $\tilde t_1$ vs.\ $\tilde g$ mass.}
\end{figure}

To begin with, we contrast in Fig.~\ref{fig:PatiSalam-xsectandR} the total SUSY production cross 
sections with the observed $r$ values in three different mass planes: 
gluino $\tilde g$ vs.\ average light-flavour squark $\tilde q$ mass,\footnote{Due to their near-degeneracy, 
first and second generation squark masses are characterized by the arithmetic mean (including left and right
states), $m_{\tilde q}=\frac{1}{8}(m_{{\tilde u}_L} + m_{{\tilde u}_R} + m_{{\tilde d}_L} + m_{{\tilde d}_R} + m_{{\tilde c}_L} + m_{{\tilde c}_R} + m_{{\tilde s}_L} + m_{{\tilde s}_R})$.} 
lightest stop $\tilde t_1$ vs.\ squark $\tilde q$ mass, and $\tilde t_1$ vs.\ $\tilde g$ mass. 
As expected we observe a strong dependence of the observed $r$ value on the 
gluino and $\tilde q$ masses which is correlated directly to the total cross section in the same mass plane.
Nonetheless, points with very heavy gluino and 1st and 2nd generation squarks may still be excluded if the third generation squarks 
are light, see the middle and bottom rows of plots in Fig.~\ref{fig:PatiSalam-xsectandR} for comparison.
On the other hand, we also find points with lightish gluinos and/or $\tilde q$ (and thus large cross section) for which
the ATLAS searches have hardly any sensitivity ($r$ values of $0.2$ and lower). The differences in the LHC sensitivity
to points with similar masses will be discussed below by means of a few representative benchmark points.

\begin{figure}[t]
\centering
\hspace*{-3mm}\includegraphics[width = 0.51\textwidth]{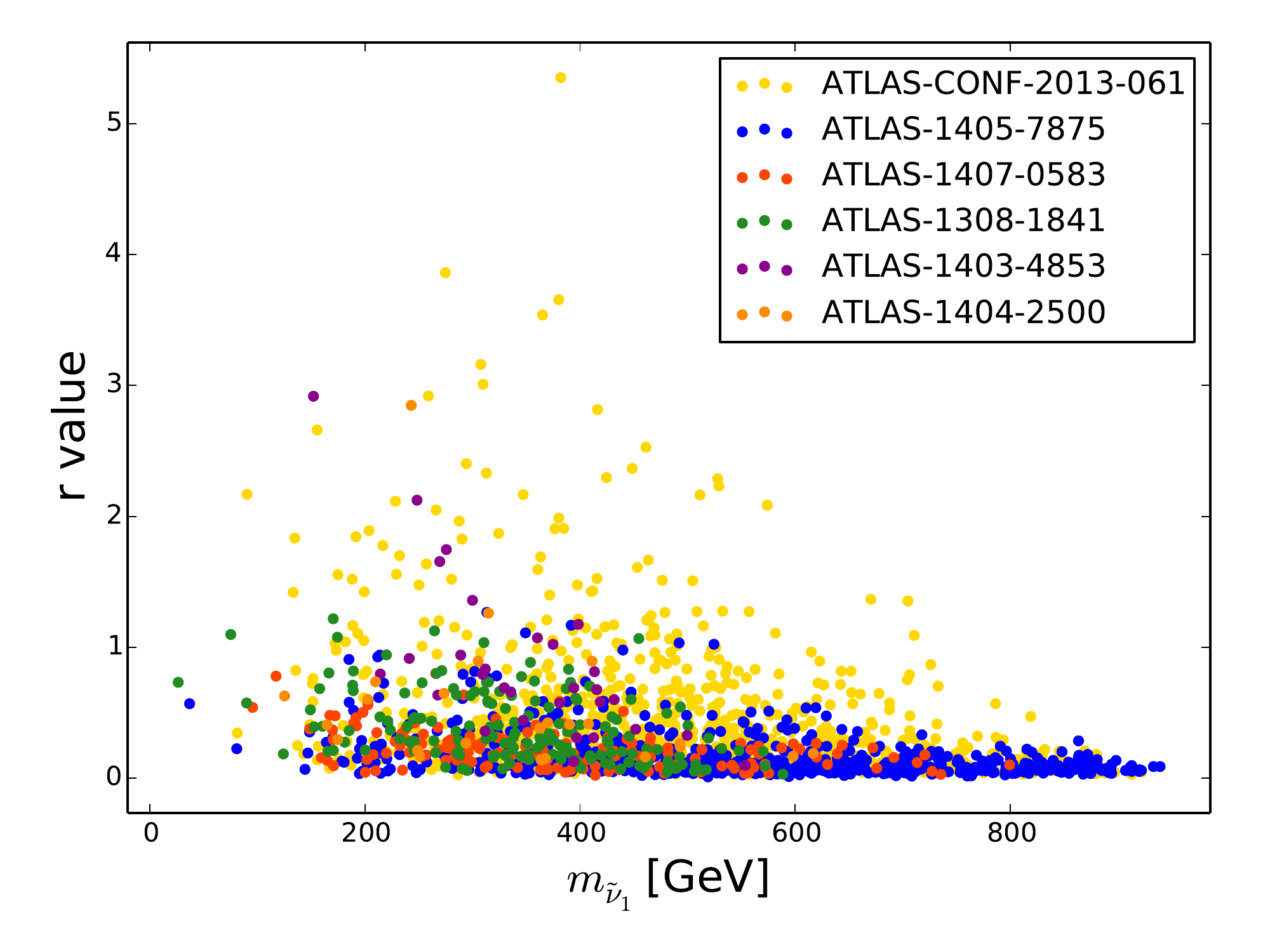}\includegraphics[width = 0.51\textwidth]{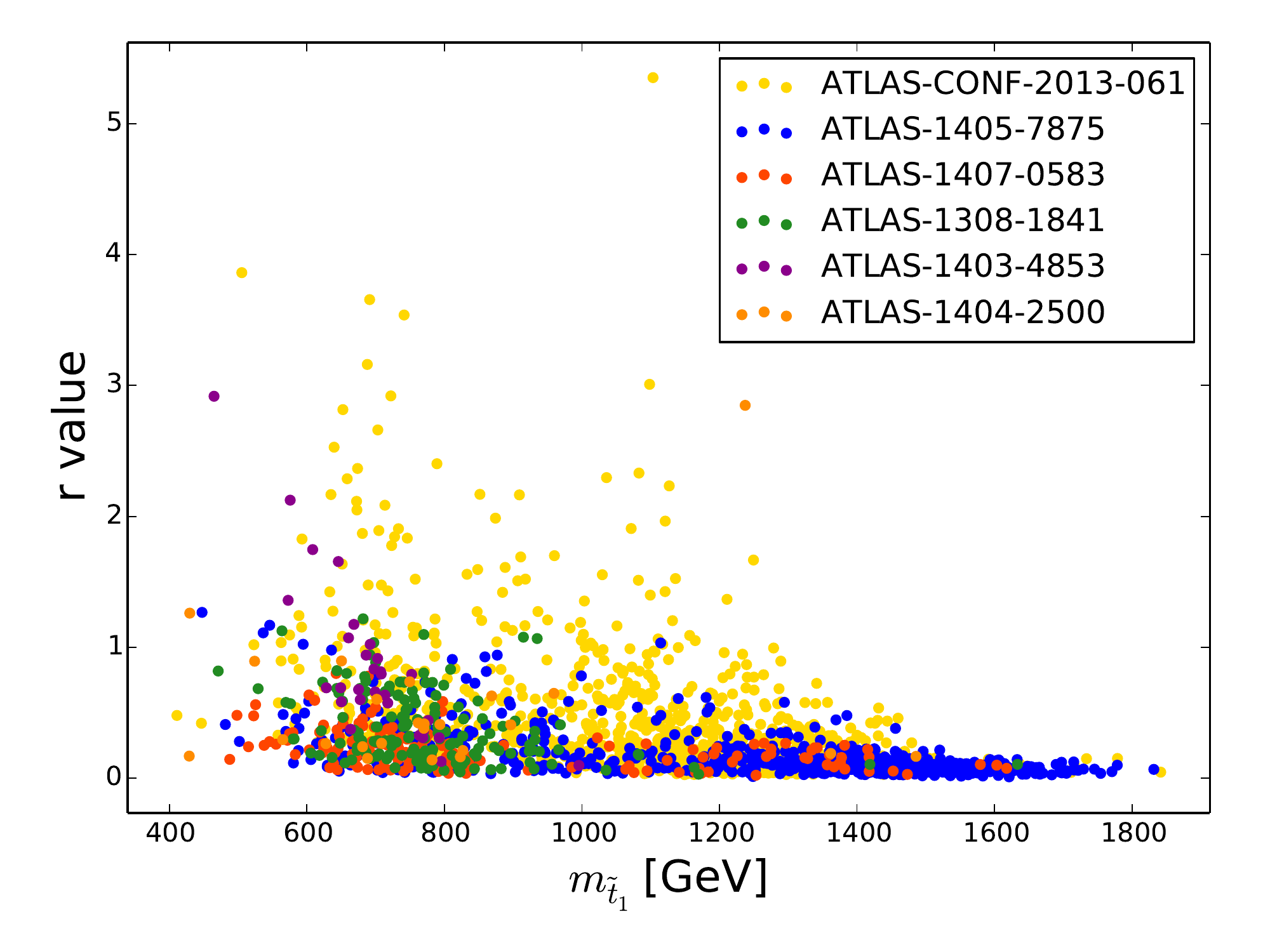}
\hspace*{-3mm}\includegraphics[width = 0.51\textwidth]{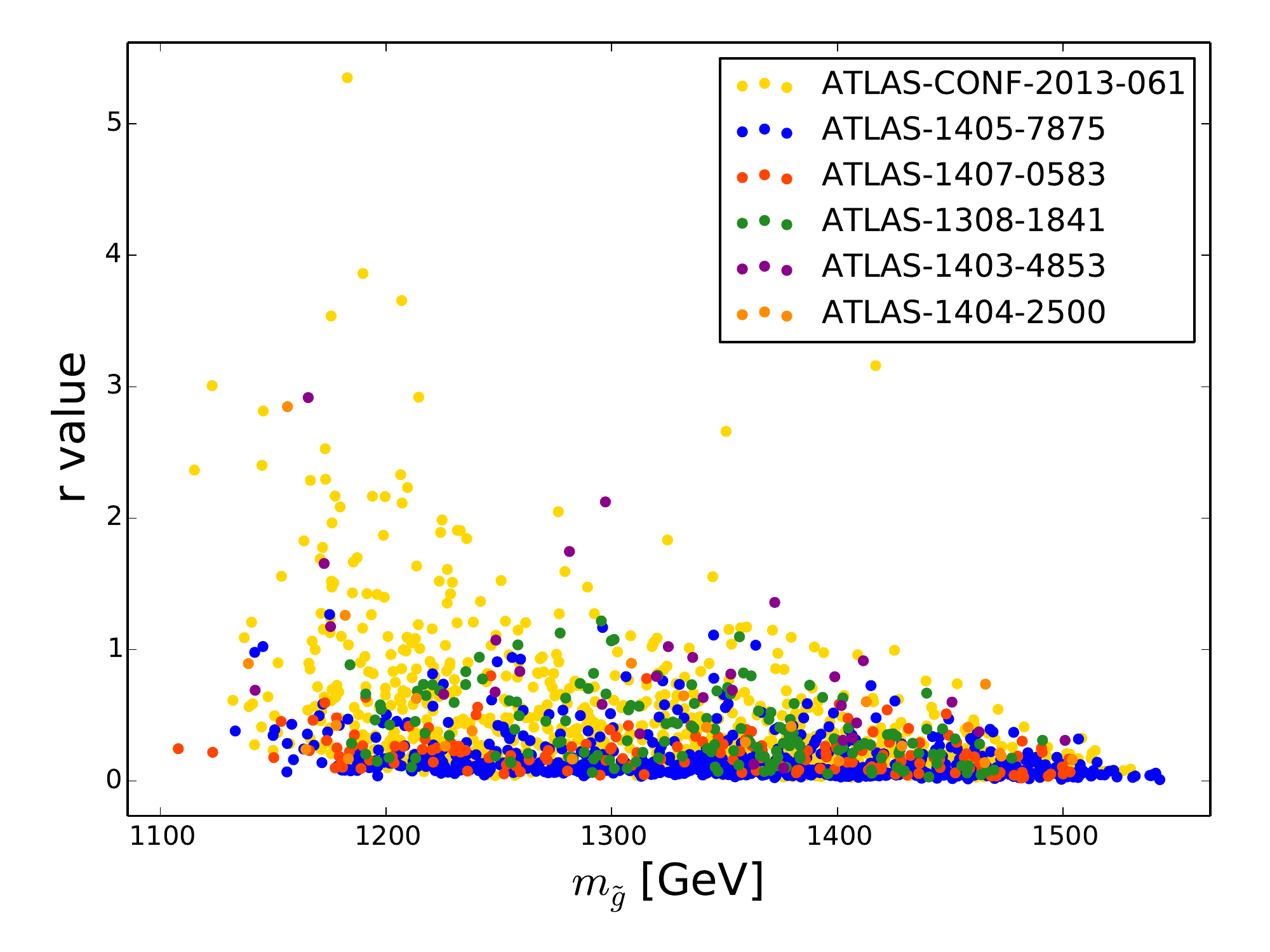}\includegraphics[width = 0.51\textwidth]{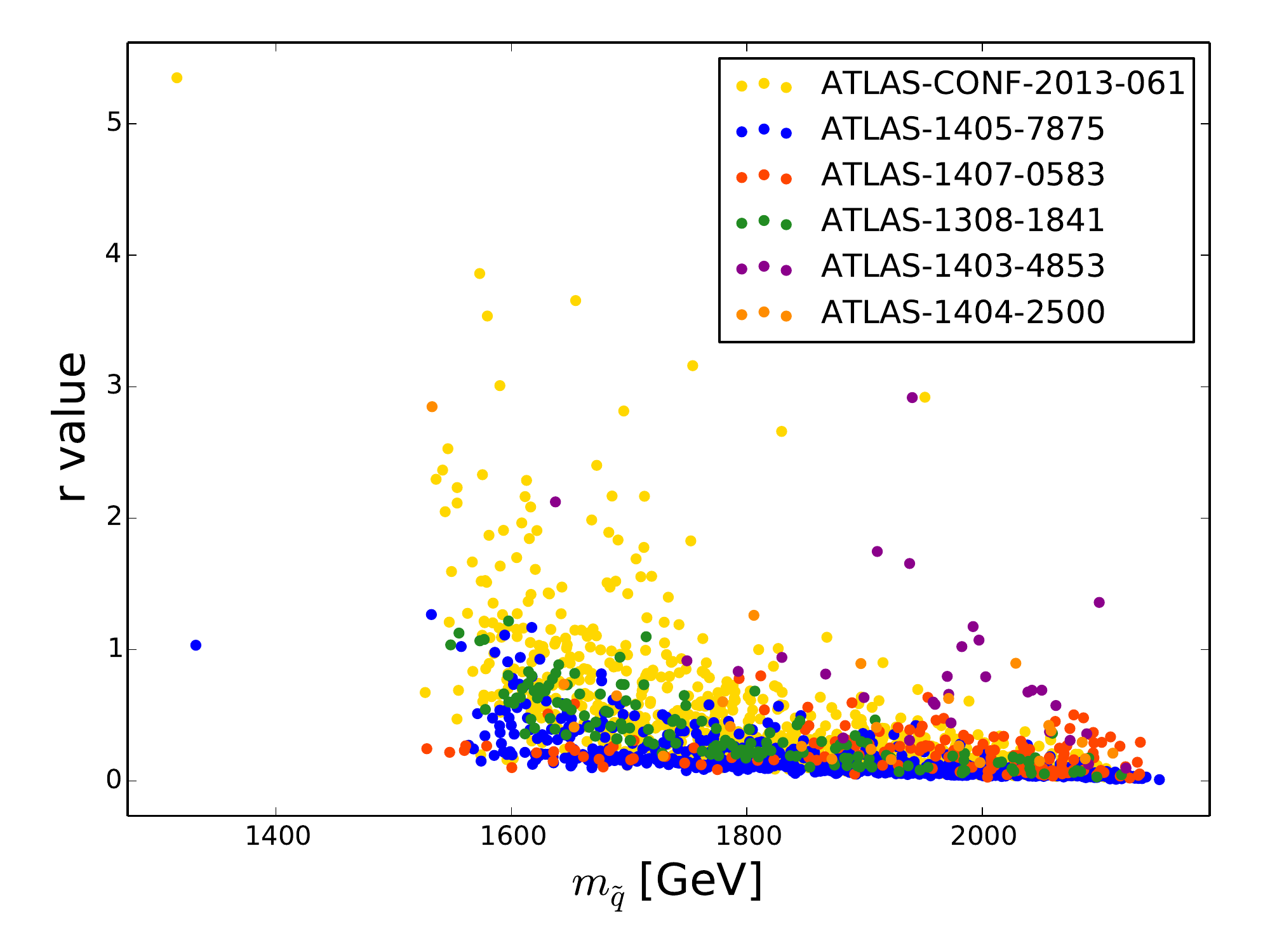}
\caption{\label{PatiSalam:fig:rVal} $r$ value versus sparticle masses with the colour code denoting the most 
sensitive analysis for each point.}
\end{figure}

Before analysing these points in detail, let us ask which analyses are actually constraining the scenarios that 
we consider. 
To address this question, Fig.~\ref{PatiSalam:fig:rVal} shows the current $r$ value from the 8~TeV searches 
versus different sparticle masses. For each point, the most sensitive analysis is denoted by the colour code indicated in the plot labels. As mentioned, the exclusion limit from the 8~TeV searches is $r\ge 1$; the plots allow to roughly depict the increase in sensitivity that would be necessary to extend the exclusion limits assuming the same types of analyses 
will be performed in Run~2 with similar signal/background selection as in Run~1.
As can be seen, the relevant searches are:

\begin{itemize}
\item ATLAS-CONF-2013-061 \cite{TheATLAScollaboration:2013tha}: 0--1 lepton, $\geq 3$ b jets, MET (gluino + 3rd gen)
\item ATLAS-1405-7875 \cite{Aad:2014wea}: 2--6 jets, lepton veto, MET (inclusive squarks and gluinos)
\item ATLAS-1407-0583 \cite{Aad:2014kra}: $1$ lepton, (b) jets, MET (1-lepton stop search)
\item ATLAS-1308-1841 \cite{Aad:2013wta}: $\geq 6$ jets, lepton veto, MET (inclusive squarks and gluinos)
\item ATLAS-1403-4853 \cite{Aad:2014qaa}: $2$ leptons, (b) jets, MET (dilepton stop search)
\item ATLAS-1404-2500 \cite{Aad:2014pda}: 2--3 leptons, (b) jets, MET (inclusive squarks and gluinos)
\end{itemize}

\noindent
We also see that the excluded points are dominated by the gluino search that targets 
third generation quarks in the decay chain. This is to be expected since our mass spectrum 
generically contains lighter stops and sbottoms and thus the gluino often decays via 
these states.

\begin{figure}[ht!]
\centering
\hspace*{-3mm}\includegraphics[width = 0.52\textwidth]{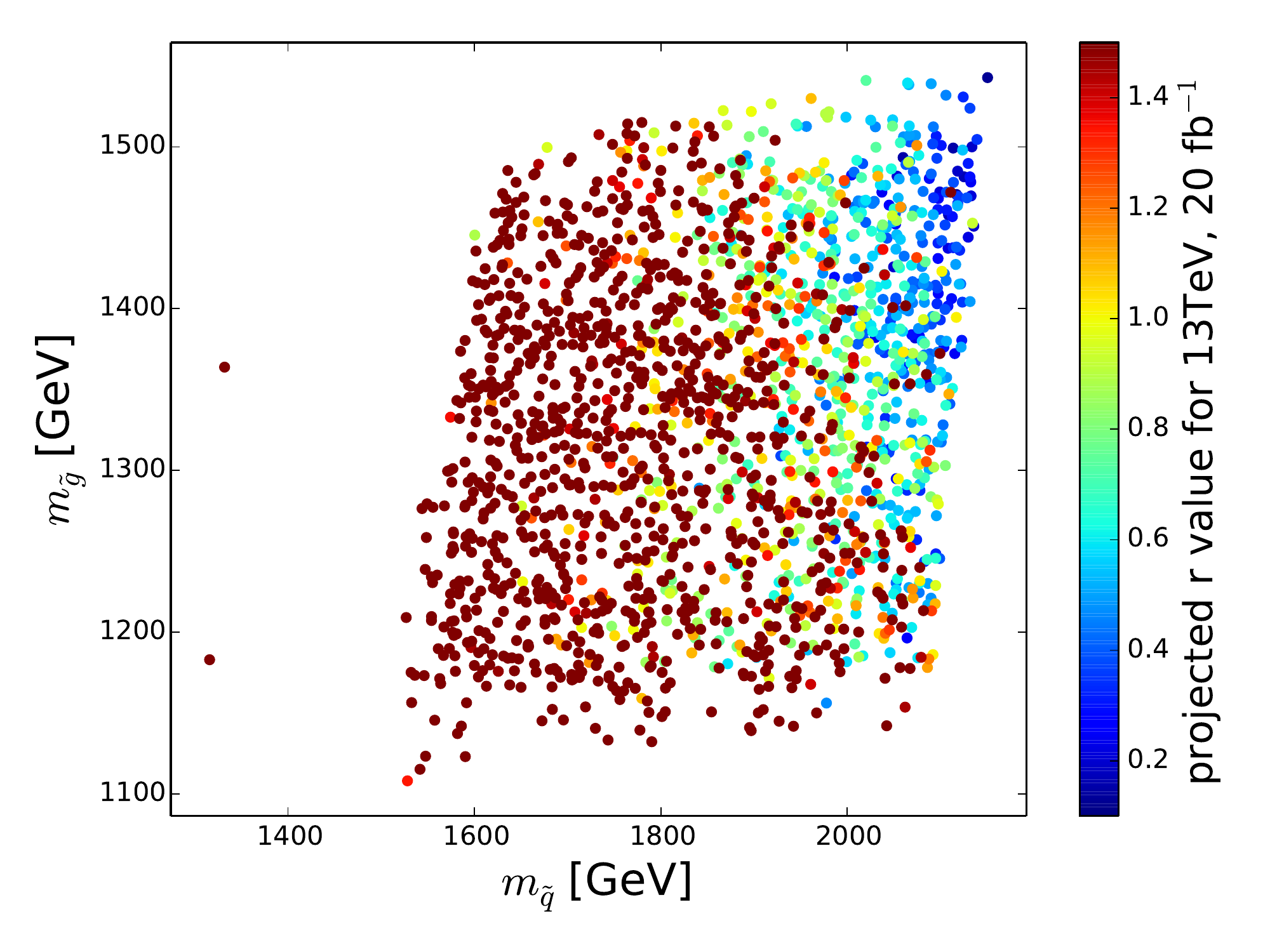}\includegraphics[width = 0.52\textwidth]{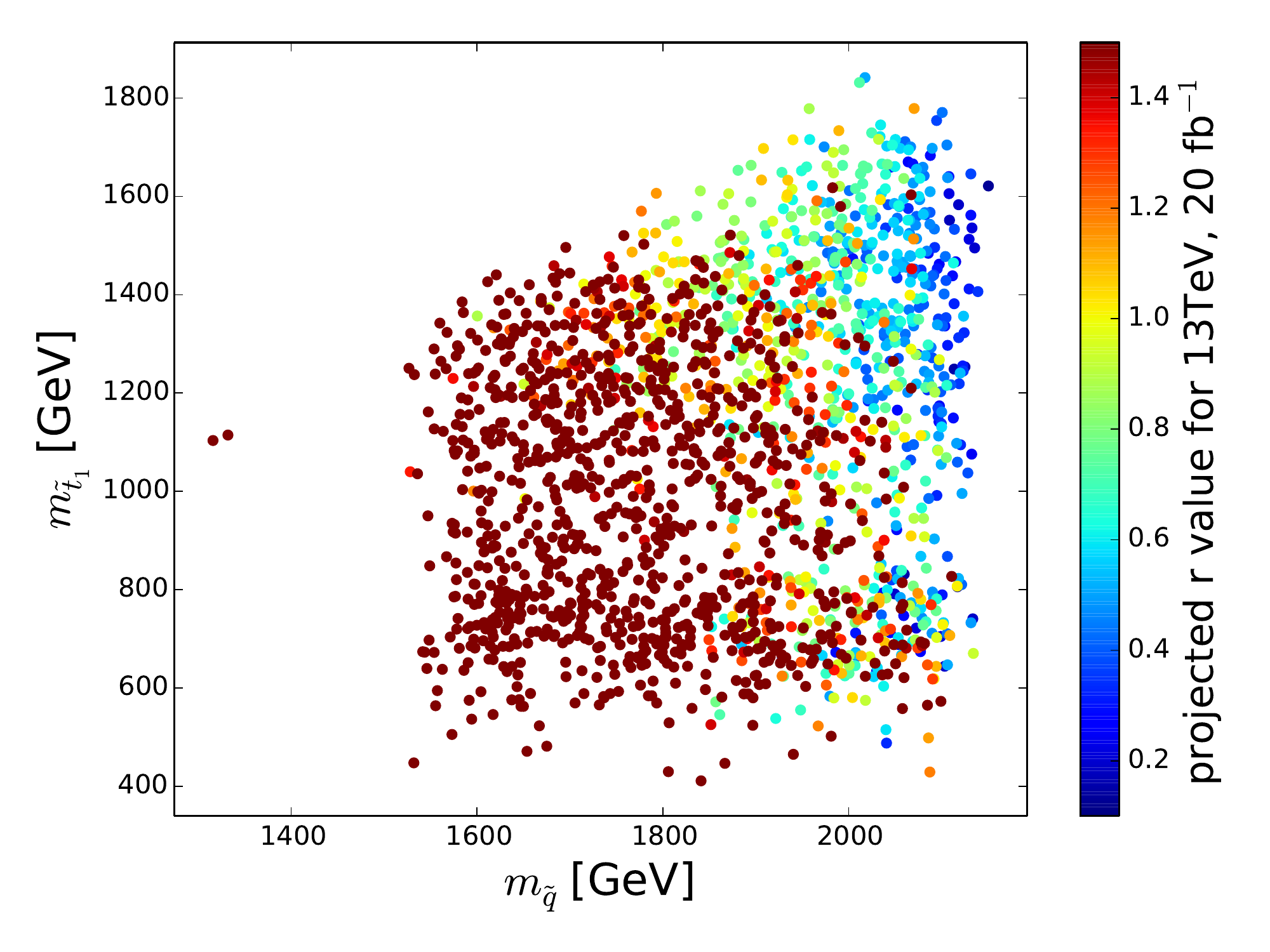}
\hspace*{-3mm}\includegraphics[width = 0.52\textwidth]{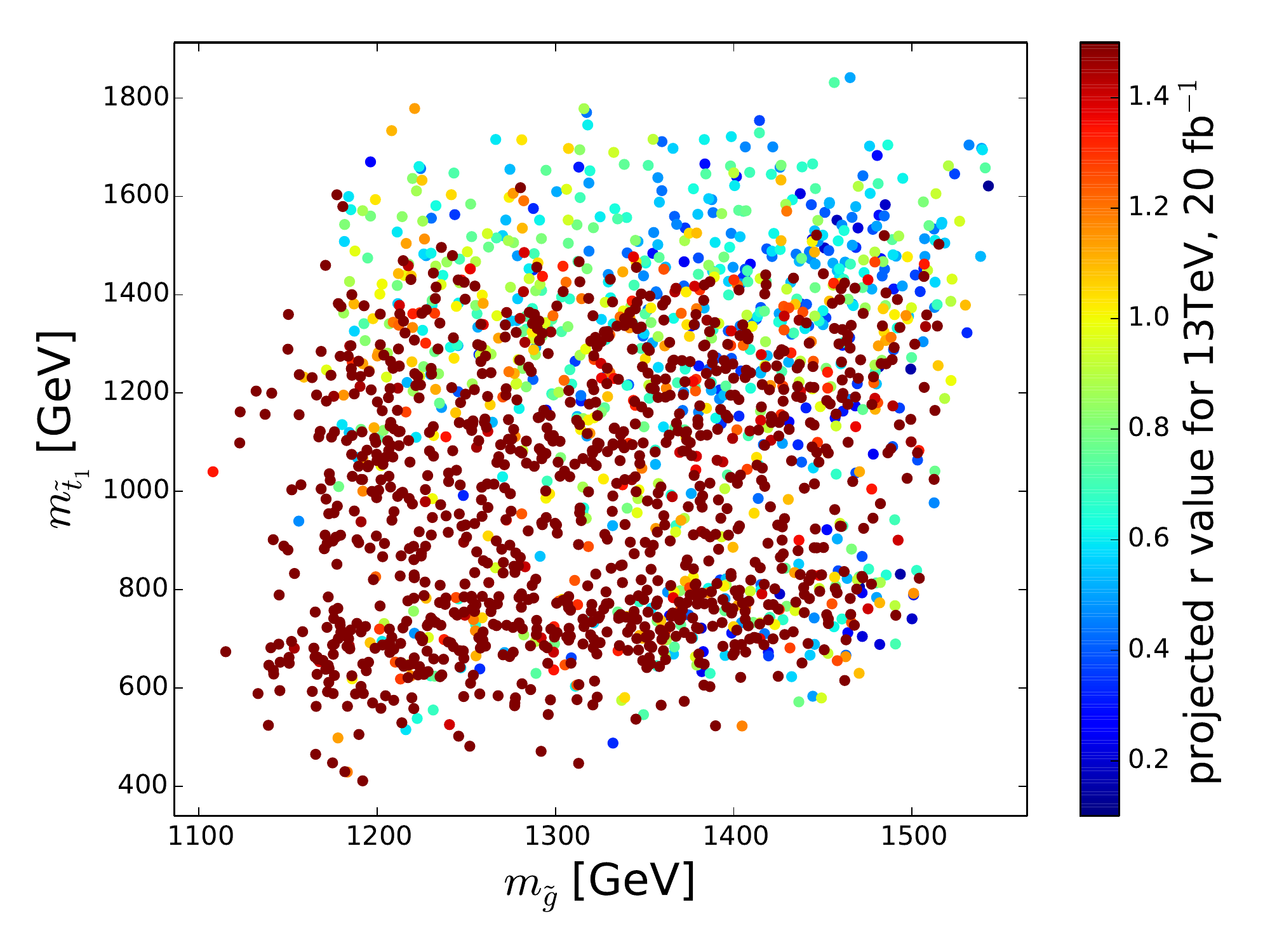}\includegraphics[width = 0.52\textwidth]{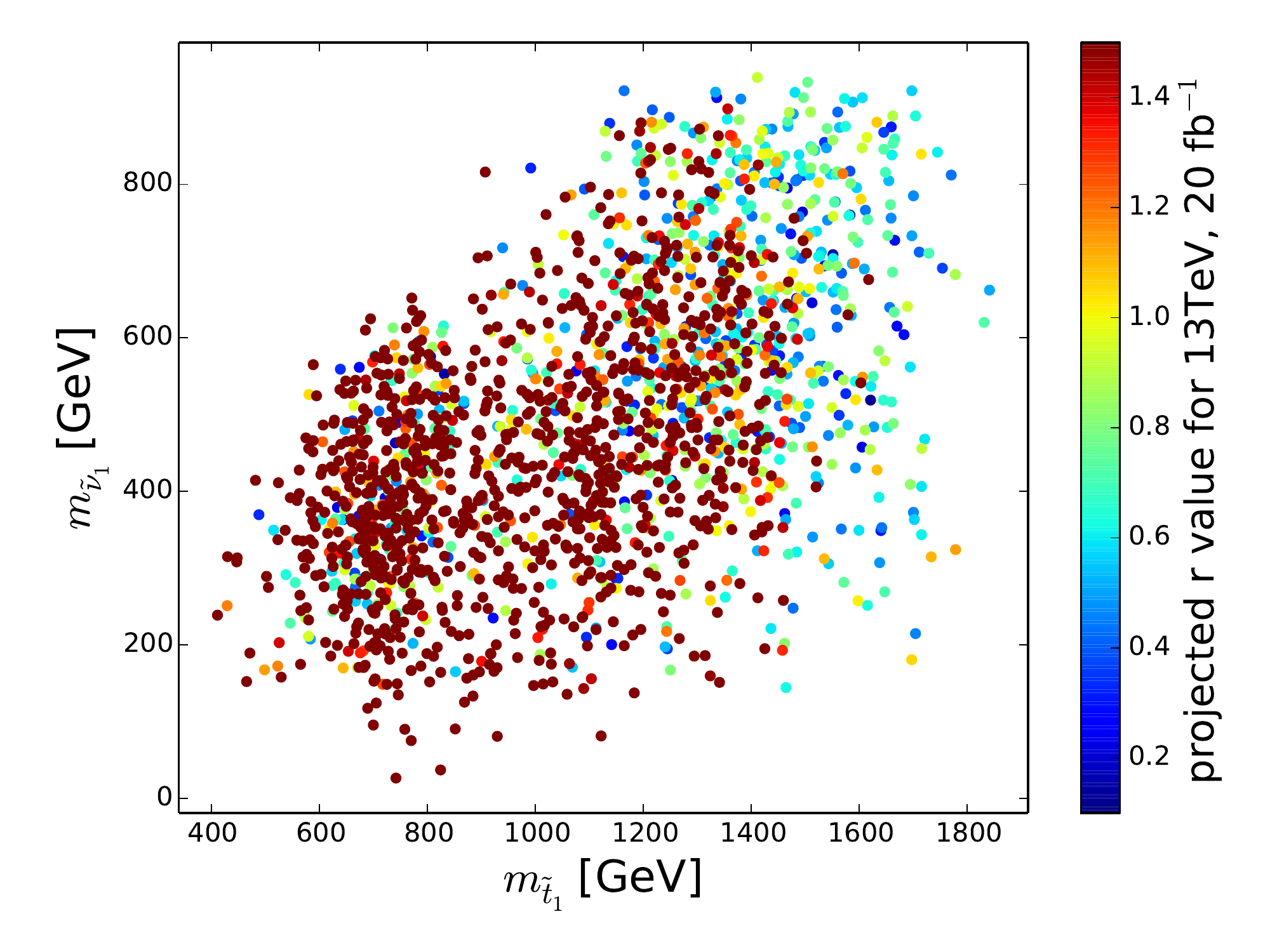}
\caption{\label{PatiSalam:fig:rProj} Projected $r$ value in various mass planes for the LHC at 13~TeV
with 20~fb$^{-1}$.}
\end{figure}

Since only a small proportion of the considered points are excluded we also estimate 
in Fig.~\ref{PatiSalam:fig:rProj} the corresponding reach of the LHC at 13~TeV with 20~fb$^{-1}$ 
(to be compared with the right panels in Fig.~\ref{fig:PatiSalam-xsectandR}).
This is done using {\tt Collider-Reach} \cite{ColliderReach,Salam:2008qg} to estimate 
the sensitivity at the higher energy and recalculating the $r$-values using the relevant
cross-sections. We see the striking feature that almost all points with 1st and 2nd
generation squarks $m_{\tilde{q}_{1,2}}<1.7$~TeV will be excluded. The effect is less
clear in terms of the gluino mass, $m_{\tilde{g}}$ which is due to the fact that the dominant 
production process is associated squark-gluino production. Since all our scenarios already
contain a relatively light gluino, $m_{\tilde{g}}<1.6$~TeV we only see an important reduction in 
production cross-section when the squarks are made heavier.
In addition we see that light stop
scenarios remain elusive as an important fraction of points with $m_{\tilde{t}_1}<800$~GeV has 
a projected $r$ value below 1, in particular when the 1st/2nd generation is heavy.

\subsection{Benchmark points}
\label{PatiSalam:sec:benchmarks}

In order to understand why some parameter points are excluded while others with similar masses are allowed, 
we now examine four selected benchmark points in more detail. The mass spectra of these points, denoted BP1 to BP4, 
and their current and projected $r$ values are given in Tab.~\ref{PatiSalam:tab:benchmarkSpectra}.

Our first benchmark scenario (BP1) is excluded by the ATLAS search for gluinos
with at least 3 $b$-jets in the final state (ATLAS-CONF-2013-061 \cite{TheATLAScollaboration:2013tha})
which specifically targets decays via third generation squarks. In our study we see that
the associated production of gluinos with 1st and 2nd generation squarks is actually dominant.
However, the heavier squarks ($m_{\tilde{q}}=1672$~GeV) then predominantly decay to the lighter 
gluino ($m_{\tilde{g}}=1308$~GeV) and a hard jet leading to the pair of gluinos that are targeted. Subsequently the gluinos
decay democratically through the different 3rd generation squarks to give the $b$-jet
rich signature that is searched for. In addition, leptons originating from the top decay when $\tilde{g} \to \tilde{t}_1 t$ 
can help enhance the signal.

The second benchmark point we choose (BP2) actually has slightly lighter gluinos 
($m_{\tilde{g}}=1250$~GeV) and squarks ($m_{\tilde{q}}=1622$~GeV) but in contrast to BP1,
we find that this point is not excluded. The reason for this is two-fold. Firstly, 
the sneutrino LSP is significantly
heavier in this scenario compared to BP1 ($m_{\tilde \nu_1}=753$~GeV vs.\ $m_{\tilde \nu_1}=193$~GeV)
which means we have a far more compressed spectrum and much softer decay products. Secondly, 
the lighter stop $\tilde t_1$ is also heavier ($m_{\tilde{t}_1}=1198$~GeV which is in 
fact close to the gluino mass) 
and this prevents the two body
decay, $\tilde{g} \to \tilde{t}_1 t$. As a result, the two body decay into the sbottom,
$\tilde{g} \to \tilde{b}_1 b$, accounts for almost all the gluino decays. However, since
there is less than 40~GeV mass difference between the gluino and sbottom, the $b$-jets
that emerge are relatively soft and hard to reconstruct.
In addition, the final state will in general have a lower lepton and jet multiplicity 
than we can expect from BP1 where $t$-quarks are dominantly produced in the decay chains. 
These lower multiplicity signals have larger SM backgrounds and thus the resulting limit on the 
model is weaker.

Moving on to the third benchmark point (BP3), we find a scenario that is also not excluded despite
the relatively light gluino ($m_{\tilde{g}}=1232$~GeV) and light sneutrino 
LSP ($m_{\tilde \nu_1}=244$~GeV). Here, the reason is that the 1st and 2nd generation squarks 
are rather heavy ($m_{\tilde{q}}=1930$~GeV) and thus the 
associated production cross-section with gluinos
is substantially reduced. We note that even with such heavy squarks, the cross-section of associated
squark-gluino production (1.44~fb) is still comparable to that of gluino pair production (2.31~fb)
due to the large valence quark contribution.
In actual fact, the reduction in cross section of the heavier
states is large enough to result in the direct stop search (ATLAS-1407-0583 \cite{Aad:2014kra})
becoming the most sensitive channel. However, examining the stop decay in this scenario
we find the following cascade is dominant,
\begin{align}
 \tilde{t}_1 &\to \tilde{\tau}_1 \nu b\,, \\
 \tilde{\tau}_1 &\to \tilde{\nu} W^* (\to q\bar{q}, \ell\bar{\nu})\,.
\end{align}
Examining the decay chain in more detail, we find two reasons why the stop searches
have limited sensitivity in this case. Firstly, the relatively small leptonic branching 
fraction of the off-shell $W^{\pm}$ reduces the effective cross-section for the searches
that rely on final state leptons. Secondly, the mass difference between the lightest stau, $\tilde{\tau}_1$, 
and the sneutrino LSP, $\tilde{\nu}_1$, is only $\sim70$~GeV which must be distributed across the three body decay.
Thus a sizeable proportion of the final state leptons are too soft to be reconstructed at the LHC.

A contrast to the above scenario is provided by our fourth benchmark (BP4), where a parameter
point with a stop which is only slightly lighter 
($m_{\tilde{t}_1}=573$~GeV vs.\ $m_{\tilde{t}_1}=624$~GeV) is now excluded. The reason has nothing to
do with the mass difference but rather the fact that the stop now decays dominantly via 
\begin{align}
 \tilde{t}_1 &\to \tilde{\chi}^{\pm} b\,, \\
 \tilde{\chi}^{\pm} &\to \tilde{\nu} \ell^{\pm}\,. 
\end{align}
Consequently, for BP4 every stop decay produces a final-state lepton and in addition a relatively
large mass splitting ($\sim210$~GeV) is present between the 
chargino, $\tilde{\chi}^{\pm}$, and sneutrino LSP,  
$\tilde{\nu}$. Since this is a two body decay, the vast majority of leptons will have sufficient 
energy to pass the cuts in the direct stop searches.
As a result, the dileptonic stop search (ATLAS-1403-4853 \cite{Aad:2014qaa}) strongly constrains this benchmark
point.

\begin{table*}[htb!]
\centering
\begin{tabular}{|l|c|c|c|c|}
\hline \hline
 & BP1 & BP2  & BP3 & BP4 \\
\hline \hline
$m_{\tilde \nu_1}$ & 193 & 752   & 244 & 300 \\
$m_{\tilde \tau_1}$ & 249 & 1044  & 315 & 553\\
$m_{\tilde \chi^0_1}$ & 636 & 923   & 491 & 504 \\
$m_{\tilde \chi^0_2}$ & 918 & 977   & 929 & 521 \\
$m_{\tilde \chi^{\pm}_1}$ & 981 & 975  & 985 & 510 \\
$m_{\tilde t_1}$ & 705 & 1198  & 624 & 573 \\
$m_{\tilde b_1}$ & 810 & 1212  & 709 & 822 \\
$m_{\tilde t_2}$ & 985 & 1264  & 885 & 925 \\
$m_{\tilde b_2}$ & 995 & 1269  & 858 & 849 \\
$m_{\tilde g}$ & 1308 & 1250  & 1232 & 1372 \\
$m_{\tilde q}$ & 1672 & 1622  & 1930 & 2099 \\
\hline
$r_{\rm obs}$ & 1.10 & 0.20   & 0.26 & 1.36 \\
$r_{\rm proj}$ (3.2\,fb$^{-1}$) & 3.82 & 0.65   & 0.31 & 1.46 \\
$r_{\rm proj}$ (20\,fb$^{-1}$) & 9.54 & 1.62  & 0.77 & 3.65 \\
\hline \hline
\end{tabular}
\caption{Relevant masses (in GeV) for the four benchmark scenarios discussed in the text. The observed $r$ value at 8 TeV and projections to 3.2\,fb$^{-1}$ and 20\,fb$^{-1}$ at 13 TeV are also given.}
\label{PatiSalam:tab:benchmarkSpectra}
\end{table*}

\section{CONCLUSIONS}\label{sec:PatiSalam-conclusions}

We considered a supersymmetric model based on the gauge group 
$SU(3)_c \times SU(2)_L \times U(1)_R \times U(1)_{B-L}$ for which we impose Pati-Salam inspired boundary 
conditions at the scale 10--100~TeV. This leads to a more compressed sfermion spectrum within one generation than 
in more common scenarios with, e.g., CMSSM-like boundary conditions. The lightest coloured sparticles in these scenarios are usually stops and sbottoms which are close in mass.
For the explanation of the neutrino masses and mixings, we evoked an inverse seesaw mechanism. 
As a consequence, the lightest sneutrinos can be the lightest supersymmetric particles, 
with varying mass differences to the lightest coloured sparticles. 
We recasted existing LHC analyses searching for supersymmetric particles at $\sqrt{s}=8~$TeV, examining the current LHC exclusions and contrasting excluded/not excluded scenarios with similar mass patterns. Moreover, we estimated the projected reach of the 13~TeV run. 
We found that apart from the overall mass scale and mass splitting between the different squark  generations and the gluino, the LHC exclusions strongly depend on the mass difference between the stop and the sneutrino which, if small, leads to soft 
decay products. The ``conventional'' SUSY mass limits from the LHC searches can thus  be avoided even for rather light gluinos and 3rd generation squarks.

\section*{Acknowledgements}

We thank the organisers of the Les Houches Workshop Series "Physics at TeV Colliders" 
for providing a most stimulating atmosphere for discussions and research.  
SK and UL acknowledge support by the ``Investissements d'avenir, Labex ENIGMASS'', the ANR project DMASTROLHC, ANR-12-BS05-0006, and the Research Executive Agency (REA) of the European Union under the Grant Agreement PITN-GA2012-316704 (HiggsTools). 
MEK and WP  have been supported by  the DFG, project nr.\ PO-1337/3-1. MEK further acknowledges support from
the BMBF grant 00160287. SuK is supported by the New Frontiers program of the Austrian Academy of Sciences. JT would like to thank Indiana University and Emilie Passemar for kind
hospitality while part of this project was completed.

\AddToContent{S.~Kraml, M.E.~Krauss, S.~Kulkarni, U.~Laa, W.~Porod and J.~Tattersall}
\renewcommand{\thesection}{\arabic{section}}

\superpart{ The Higgs Boson }

\graphicspath{{ggh/}}

\chapter{Higgs boson production via gluon fusion within HEFT}

{\it T.~Schmidt and M.~Spira}


\begin{abstract}
Higgs boson production via gluon fusion is discussed up to the NNLO
level including dimension 6 operators within the SILH and the non-linear
realizations in Higgs effective theories (HEFT). Particular emphasis is
set on the consistent treatment of the new Wilson coefficients at higher
orders, i.e.~including the proper scale dependence and merging with the
SM part.
\end{abstract}

\section{Introduction}
The Higgs boson production cross section via gluon-fusion is known up to
N$^3$LO QCD \cite{Djouadi:1991tka, Dawson:1990zj, Graudenz:1992pv,
Spira:1995rr, Harlander:2005rq, Anastasiou:2009kn, Catani:2001ic,
Harlander:2001is, Harlander:2002wh, Anastasiou:2002yz, Ravindran:2003um,
Gehrmann:2011aa, Anastasiou:2013srw, Anastasiou:2013mca,
Kilgore:2013gba, Li:2014bfa, Anastasiou:2014lda, Anastasiou:2015ema,
Anastasiou:2015yha, Anastasiou:2016cez} and NLO electroweak
\cite{Djouadi:1994ge, Chetyrkin:1996wr, Chetyrkin:1996ke,
Aglietti:2004nj, Degrassi:2004mx, Aglietti:2006yd, Actis:2008ug,
Actis:2008ts, Anastasiou:2008tj} within the SM supplemented by soft and
collinear gluon resummation up to the N$^3$LL level \cite{Kramer:1996iq,
Catani:2003zt, Moch:2005ky, Ravindran:2005vv, Ravindran:2006cg,
Idilbi:2005ni, Ahrens:2008nc, deFlorian:2009hc, deFlorian:2012yg,
deFlorian:2014vta, Bonvini:2014joa, Bonvini:2014tea, Catani:2014uta,
Schmidt:2015cea}. Starting from these results the contributions of
dimension-6 operators beyond the SM are discussed up to NNLO QCD. This
extension of previous work inside the SM is based on the effective
Lagrangian (here in the heavy top limit for the SM part)
\begin{equation}
{\cal L}_{eff} = \frac{\alpha_s}{\pi} \left\{ \frac{c_t}{12} (1 +
\delta) + c_g \right\} G^{a\mu\nu} G^a_{\mu\nu} \frac{H}{v}
\label{eq:leff}
\end{equation}
with the QCD corrections \cite{Kramer:1996iq, Chetyrkin:1997sg,
Chetyrkin:1997un, Spira:1997dg}
\begin{eqnarray}
\delta & = & \delta_1 \frac{\alpha_s}{\pi} + \delta_2 \left(
\frac{\alpha_s}{\pi} \right)^2 + \delta_3 \left(
\frac{\alpha_s}{\pi} \right)^3 + {\cal O}(\alpha_s^4) \nonumber \\
\delta_1 & = & \frac{11}{4} \nonumber \\
\delta_2 & = & \frac{2777}{288} + \frac{19}{16} L_t + N_F
\left(\frac{L_t}{3}-\frac{67}{96} \right) \nonumber \\
\delta_3 & = & \frac{897943}{9216} \zeta_3 - \frac{2761331}{41472} +
\frac{209}{64} L_t^2 + \frac{2417}{288} L_t \nonumber \\
& + & N_F \left(\frac{58723}{20736} - \frac{110779}{13824} \zeta_3 +
\frac{23}{32} L_t^2 + \frac{91}{54} L_t \right) + N_F^2
\left(-\frac{L_t^2}{18} + \frac{77}{1728} L_t - \frac{6865}{31104}
\right)
\end{eqnarray}
where $L_t = \log (\mu_R^2/M_t^2)$ with $\mu_R$ denoting the
renormalization scale and $M_t$ the top quark pole mass. The gluon field
strength tensor is represented by $G^{a\mu\nu}$, the strong coupling
constant by $\alpha_s$ with five active flavours, the electroweak vacuum
expectation value by $v$ and the physical Higgs field by $H$. The
contributions of dimension-6 operators are absorbed in the rescaling
factor $c_t$ for the top Yukawa coupling and the point-like coupling
$c_g$, i.e.~deviations of $c_t$ and $c_g$ from their SM values $c_t=1$
and $c_g=0$ originate from dimension-6 operators.  The contribution of
the chromomagnetic dipole operator \cite{Choudhury:2012np,
Degrande:2012gr} is not included. The extended LO cross section is then
given by
\begin{eqnarray}
\sigma_{LO}(pp\to H) & = & \sigma_0 \tau_H \frac{d{\cal
L}^{gg}}{d\tau_H} \nonumber \\
\mbox{non-linear:} \qquad \sigma_0^{NL} & = &
\frac{G_F\alpha_s^2}{288\sqrt{2}\pi} \left| \sum_Q c_Q A_Q(\tau_Q) + 12
c_g \right|^2 \nonumber \\
\mbox{SILH:} \qquad \sigma_0^{SILH} & = &  \sigma_0^{NL} -
\frac{G_F\alpha_s^2}{288\sqrt{2}\pi} \left| \sum_Q (c_Q-1) A_Q(\tau_Q) +
12 c_g \right|^2
\label{eq:cxnlo}
\end{eqnarray}
with $\tau_H=M_H^2/s$, $\tau_Q = 4m_Q^2/M_H^2$ and the LO form factors
\begin{eqnarray}
 A_Q\left(\tau\right) & = & \frac{3}{2}
\tau\left[1+\left(1-\tau\right)f\left(\tau\right)\right] \\ 
f\left(\tau\right) & = & \left\lbrace \begin{array}{ll}
\displaystyle \arcsin^2\frac{1}{\sqrt{\tau}} & \quad \tau\geq1 \\[0.5cm]
\displaystyle
-\frac{1}{4}\left[\ln\frac{1+\sqrt{1-\tau}}{1-\sqrt{1-\tau}}-i\pi\right]^2
& \quad \tau<1\, . \end{array} \right.
\end{eqnarray}
and $G_F = (\sqrt{2} v^2)^{-1}$ denoting the Fermi constant, while the
gluon-gluon parton luminosity is displayed as ${\cal L}^{gg}$. Here,
rescaling factors $c_Q$ have been introduced for all contributing
quarks, i.e.~the top, bottom and charm quark. The cross section in
Eq.~(\ref{eq:cxnlo}) is shown for two different cases, the non-linear
parametrization of New Physics effects, where the squares of the
deviations from the SM are taken into account, and the SILH
approximation \cite{Giudice:2007fh}, where the observable is systematically
expanded to the dimension-6 level.

\section{Calculation}
The Wilson coefficient $c_g$ does not receive QCD corrections within the
effective Lagrangian, but develops a scale dependence according to the
renormalization group equation
\begin{eqnarray}
\mu^2 \frac{\partial c_g(\mu^2)}{\partial \mu^2} & = & - \left\{\beta_1
\left(\frac{\alpha_s(\mu^2)}{\pi}\right)^2
+ 2 \beta_2 \left(\frac{\alpha_s(\mu^2)}{\pi}\right)^3\right\}
  c_g(\mu^2) \nonumber \\
\beta_0 & = & \frac{33-2 N_F}{12} \nonumber \\
\beta_1 & = & \frac{153-19 N_F}{24} \nonumber \\
\beta_2 & = & \frac{1}{128} \left( 2857 - \frac{5033}{9}N_F +
\frac{325}{27} N_F^2 \right)
\end{eqnarray}
This renormalization group equation can be derived either from the
scale-invariant trace anomaly term $\beta(\alpha_s)/(2\alpha_s) G^{a\mu\nu}
G^a_{\mu\nu}$ \cite{Callan:1970yg, Symanzik:1970rt, Coleman:1970je,
Crewther:1972kn, Chanowitz:1972vd, Chanowitz:1972da} or from the scale
dependence of the factor $(1+\delta)$ of the effective Lagrangian of
Eq.~(\ref{eq:leff}), since both coefficients, $c_t (1+\delta)$ and
$c_g$, have to develop the same scale dependence. The solution of the
RGE for $c_g$ up to the NNLL level can be cast into the form
\begin{eqnarray}
c_g(\mu^2) = c_g(\mu_0^2)~\frac{\beta_0 + \beta_1
\frac{\alpha_s(\mu^2)}{\pi} + \beta_2 \left( \frac{\alpha_s(\mu^2)}{\pi}
\right)^2}{\beta_0 + \beta_1 \frac{\alpha_s(\mu_0^2)}{\pi} + \beta_2
\left( \frac{\alpha_s(\mu_0^2)}{\pi} \right)^2}
\end{eqnarray}
In order to compute the modified cross section up to NNLO the mismatch
of the individual terms of the effective Lagrangian of
Eq.~(\ref{eq:leff}) with respect to the $\delta$ term has been taken
into account properly supplemented by the NNLL scale dependence of the
Wilson coefficient $c_g$, while the finite NLO quark mass terms have
been added at fixed NLO to the SM part and the interference terms
between the quark loops and the novel coupling $c_g(\mu_R^2)$. This
yields a consistent determination of the gluon-fusion cross section up
to NNLO including the dimension-6 operators that lead to a rescaling of
the top, bottom and charm Yukawa couplings and the point-like $Hgg$
coupling parametrized by $c_g(\mu_R^2)$.

The results are implemented in the present version 4.34 of {\tt Higlu}
\cite{Spira:1995mt, Spira:1996if} which is linked to {\tt Hdecay}
\cite{Djouadi:1997yw, Djouadi:2006bz} (version
6.51) and allows to choose the usual SM Higgs input values in the
separate input files {\tt higlu.in} and {\tt hdecay.in}.  In addition in
{\tt higlu.in} the rescaling factors $c_{t,b,c}$ and the point-like
Wilson coefficient $c_g(\mu_0^2)$ can be chosen with the corresponding
input scale $\mu_0$. In this way {\tt Higlu} provides a consistent
calculation of the gluon-fusion cross section up to NNLO QCD including
dimension-6 operators. More detailed information about the input files
{\tt higlu.in} and {\tt hdecay.in} can be found as comment lines at the
beginning of the main Fortran files {\tt higlu.f} and {\tt hdecay.f}.
The pole masses for the bottom and charm quarks are computed from the
$\overline{\rm MS}$ input values $\overline{m}_b(\overline{m}_b)$ and
$\overline{m}_c(3~{\rm GeV})$ with N$^3$LO accuracy internally according
to the recent recommendation of the LHC Higgs Cross Section Working
Group \cite{Denner:2015hwgsmp}.

\section{Results}
\begin{figure}[hbt]
\begin{centering}
    \includegraphics[width=0.80\textwidth]{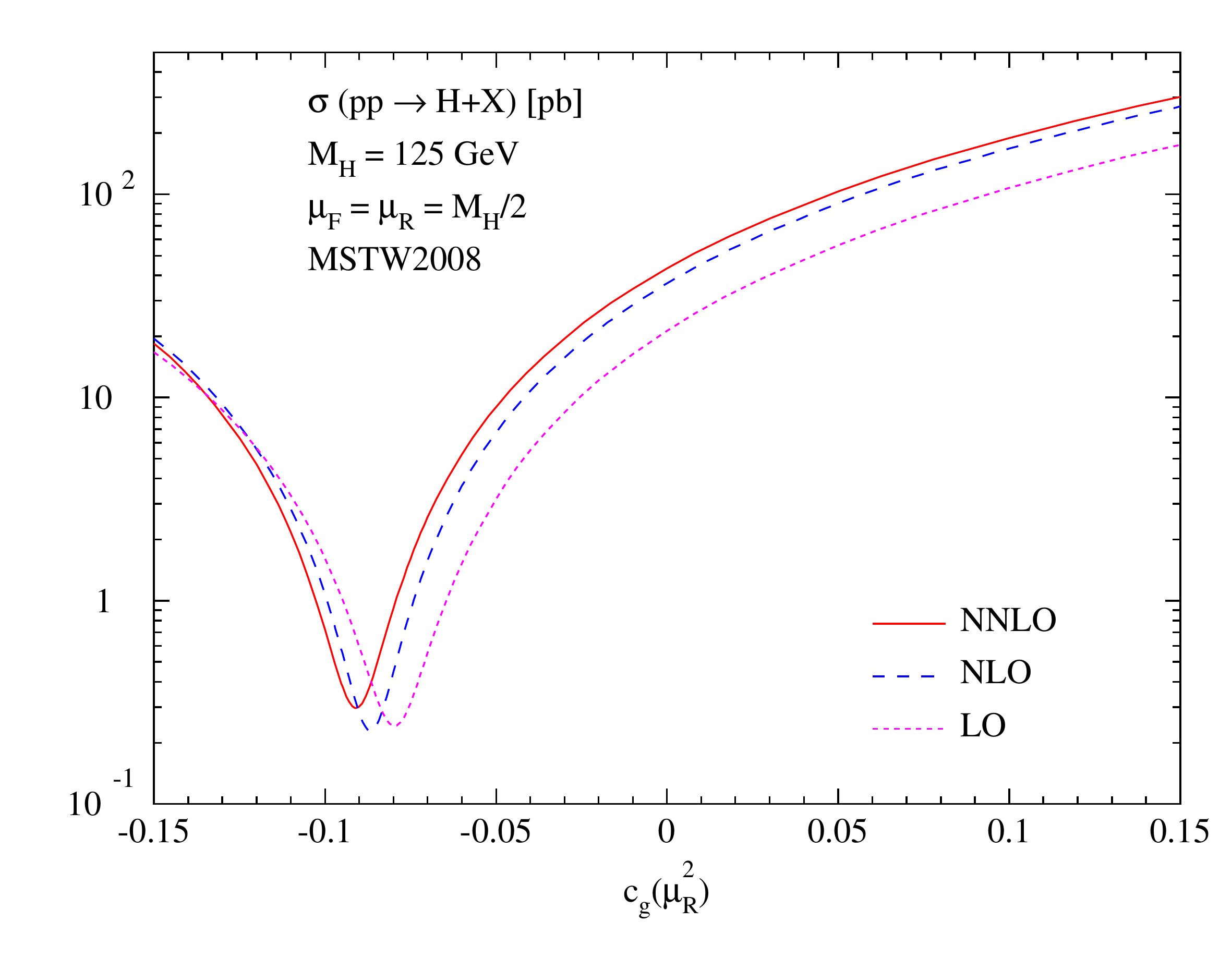}
\caption{Total Higgs boson production cross section via gluon fusion for
a non-linear parametrization of New Physics effects as a function of the
novel point-like Higgs coupling to gluons $c_g(\mu_R^2)$. MSTW2008
parton densities have been adopted with $\alpha_s(M_Z) = 
0.13939~(\mbox{LO}), 0.12018~(\mbox{NLO}), 0.11707~(\mbox{NNLO})$.}
\label{fg:cxn}
\end{centering}
\end{figure}

\begin{figure}[hbt]
\begin{centering}
    \includegraphics[width=0.80\textwidth]{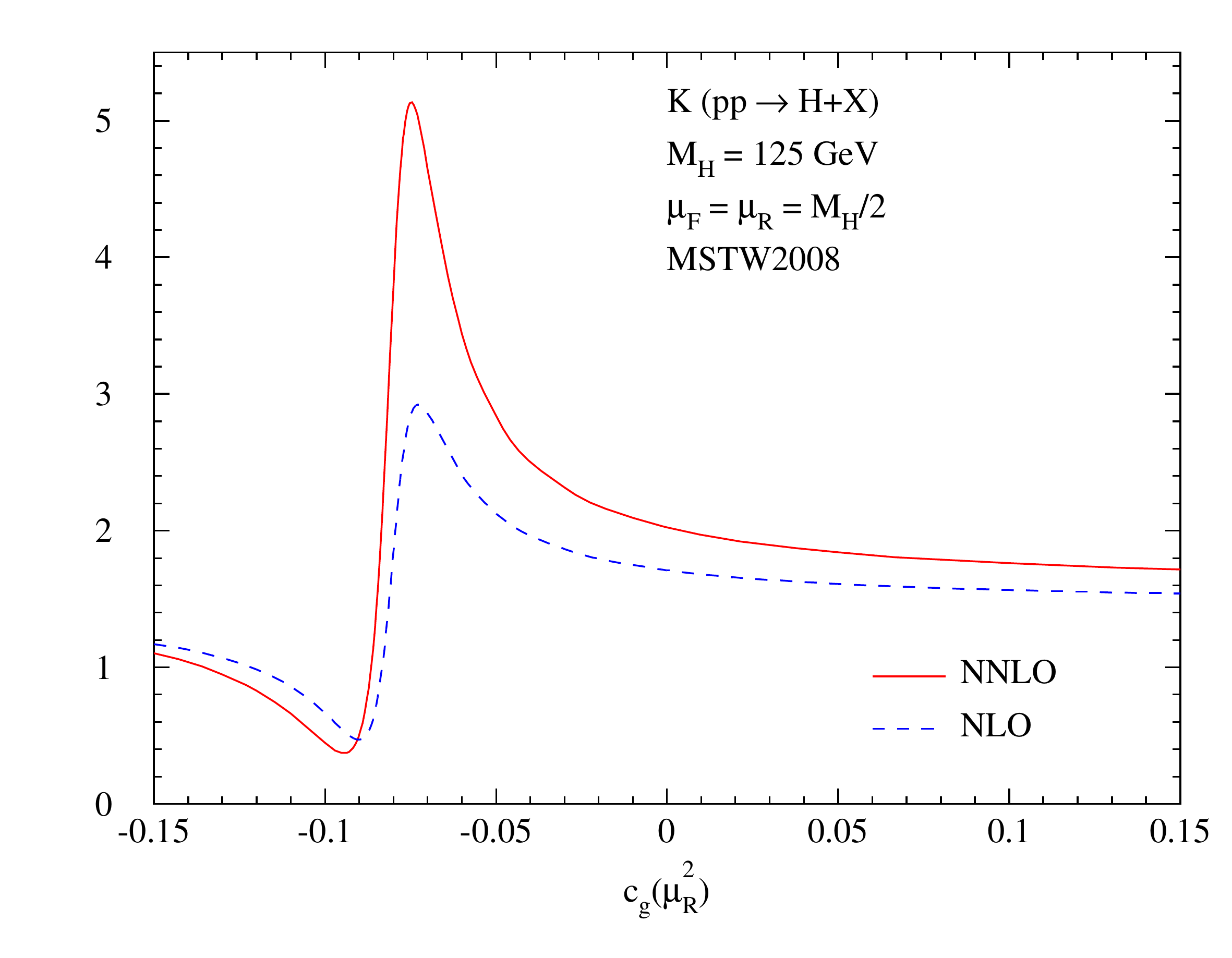}
\caption{K factor of the Higgs boson production cross section via gluon
fusion for a non-linear parametrization of New Physics effects as a
function of the novel point-like Higgs coupling to gluons
$c_g(\mu_R^2)$. MSTW2008 parton densities have been adopted with
$\alpha_s(M_Z) = 0.13939~(\mbox{LO}), 0.12018~(\mbox{NLO}), 
0.11707~(\mbox{NNLO})$.}
\label{fg:kfac} 
\end{centering}
\end{figure}

The results for the total cross sections at LO, NLO and NNLO are shown
in Fig.~\ref{fg:cxn} as a function of the novel point-like Higgs
coupling $c_g(\mu_R^2)$ where we identified the scale with the
renormalization scale. The renormalization and factorization scales
$\mu_R$ and $\mu_F$ have been identified with $M_H/2$ which for
simplicity is also chosen as the input scale $\mu_0$ of the chosen value
of $c_g(\mu_0)$. For the parton densities the MSTW2008 sets have been
adopted according to the order of the calculation. The strong coupling
constants have been chosen accordingly, i.e.~$\alpha_s(M_Z) =
0.13939~(\mbox{LO}), 0.12018~(\mbox{NLO}), 0.11707~(\mbox{NNLO})$. The
quark masses have been taken as $m_t = 172.5$ GeV,
$\overline{m}_b(\overline{m}_b) = 4.18$ GeV and $\overline{m}_c(3~{\rm
GeV}) = 0.986$ GeV. They are converted to the corresponding pole masses
used in our calculation with N$^3$LO accuracy. Fig.~\ref{fg:cxn} shows
that there is a strong dependence of the cross section on $c_g$ and
large destructive or constructive interference effects depending on the
sign and size of $c_g$. The total cross section is minimal where $c_g$
nearly cancels the quark-loop contributions of the SM Higgs case. This
minimum shifts from order to order due to the different QCD corrections
taken into account in the effective Lagrangian of Eq.~(\ref{eq:leff})
and from the residual mass terms of the explicit calculation. This is
reflected by a large variation of the K factor, defined as the (N)NLO
cross section divided by the LO one and shown in Fig.~\ref{fg:kfac},
close to the range of minimal cross sections.  In addition the K factor
varies with $c_g$ due to the mismatch of QCD corrections in the
effective Lagrangian of Eq.~(\ref{eq:leff}) and the finite quark mass
effects taken into account at NLO. The SM value can be read off for
$c_g(\mu_R^2)=0$.

\section*{Acknowledgments}
The work of M.S.~is supported in part by the Research Executive Agency
(REA) of the European Union under the Grant No.~PITN-GA-2012-316704
(Higgstools).



\AddToContent{T.~Schmidt and M.~Spira}
\renewcommand{\thesection}{\arabic{section}}

\graphicspath{{heft-ho/}}

\newcommand{\lsim}{\raisebox{-0.13cm}{~\shortstack{$<$ \\[-0.07cm] $\sim$}}~}
\newcommand{\gsim}{\raisebox{-0.13cm}{~\shortstack{$>$ \\[-0.07cm] $\sim$}}~}


\chapter{HEFT at higher orders for LHC processes}

{\it M.~M\"uhlleitner, V.~Sanz and M.~Spira}


\begin{abstract}
We discuss Higgs Effective Theories (HEFT) for LHC processes with
respect to higher-order QCD and electroweak corrections. Particular
attention is payed to the impact of residual uncertainties on the
accuracies that can be achieved in measurements of the Wilson
coefficients of the HEFT operators at the LHC.
\end{abstract}

\section{Introduction}
The properties of the scalar particle discovered at the LHC
\cite{Aad:2012tfa, Chatrchyan:2012xdj} are consistent \cite{CMS:yva,
ATLAS:2013sla, Aad:2015mxa, Khachatryan:2014kca, Aad:2015gba,
Khachatryan:2014jba} with the Standard Model (SM) Higgs boson
\cite{Higgs:1964ia, Higgs:1964pj, Englert:1964et, Guralnik:1964eu,
Higgs:1966ev}. In order to introduce deviations from the SM Higgs boson
properties either explicit models beyond the SM can be analyzed, or in a
systematic and model-independent way higher-dimension operators can be
introduced that parametrize New Physics (NP) at high-energy scales that
is integrated out \cite{Burges:1983zg, Buchmuller:1985jz,
Hagiwara:1993ck, Grzadkowski:2010es}. This can be performed either in a
weakly interacting case in terms of the SILH Lagrangian
\cite{Giudice:2007fh} or in the case of strongly-interacting NP in terms
of a non-linear effective composite-Higgs Lagrangian
\cite{Terazawa:1976xx, Terazawa:1979pj, Kaplan:1983fs,
Dimopoulos:1981xc, Banks:1984gj, Kaplan:1983sm, Georgi:1984ef,
Georgi:1984af, Dugan:1984hq}. Both lead, after canonical normalization
of all states, couplings and masses, to a phenomenological Lagrangian.
In unitary gauge the part relevant for QCD-induced Higgs production at
the LHC is given by (in the heavy top-quark limit)
\begin{equation}
{\cal L}_{eff} = \frac{\alpha_s}{\pi} \left\{ \left[\frac{c_t}{12} (1 +
\delta) + c_g \right] \frac{H}{v} + \left[\frac{c_{tt} (1+\delta) -
c_t^2 (1+\eta)}{12} + c_{gg} \right] \frac{H^2}{2v^2} \right\}
G^{a\mu\nu} G^a_{\mu\nu} - c_3 \frac{\lambda v}{6} H^3
\label{eq:holeff}
\end{equation}
with the QCD corrections \cite{Chetyrkin:1997sg, Chetyrkin:1997un,
Kramer:1996iq, Spira:1997dg, Grigo:2014jma}
\begin{eqnarray}
\delta & = & \delta_1 \frac{\alpha_s}{\pi} + \delta_2 \left(
\frac{\alpha_s}{\pi} \right)^2 + \delta_3 \left(
\frac{\alpha_s}{\pi} \right)^3 + {\cal O}(\alpha_s^4) \nonumber \\
\delta_1 & = & \frac{11}{4} \nonumber \\
\delta_2 & = & \frac{2777}{288} + \frac{19}{16} L_t + N_F
\left(\frac{L_t}{3}-\frac{67}{96} \right) \nonumber \\
\delta_3 & = & \frac{897943}{9216} \zeta_3 - \frac{2761331}{41472} +
\frac{209}{64} L_t^2 + \frac{2417}{288} L_t \nonumber \\
& + & N_F \left(\frac{58723}{20736} - \frac{110779}{13824} \zeta_3 +
\frac{23}{32} L_t^2 + \frac{91}{54} L_t \right) + N_F^2
\left(-\frac{L_t^2}{18} + \frac{77}{1728} L_t - \frac{6865}{31104}
\right) \nonumber \\
\eta & = & \eta_1 \frac{\alpha_s}{\pi} + \eta_2 \left(
\frac{\alpha_s}{\pi} \right)^2 + {\cal O}(\alpha_s^3) \nonumber \\
\eta_1 & = & \delta_1 \nonumber \\
\eta_2 & = & \delta_2 + \frac{35+16 N_F}{24}
\end{eqnarray}
where $L_t = \log (\mu_R^2/M_t^2)$ with $\mu_R$ denoting the
renormalization scale and $M_t$ the top quark pole mass. The gluon field
strength tensor is represented by $G^{a\mu\nu}$, the strong coupling
constant by $\alpha_s$ with five active flavors, the electroweak vacuum
expectation value by $v$, the trilinear Higgs self-coupling by $\lambda
= 3 M_H^2/v^2$ and the physical Higgs field by $H$. The contributions of
dimension-6 operators are absorbed in the rescaling factors $c_t, c_3$
for the top Yukawa coupling and the trilinear Higgs self-interaction and
the novel couplings $c_g, c_{gg}$ and $c_{tt}$ denoting the point-like
$Hgg$, $HHgg$ and $HHt\bar t$ couplings, i.e.~deviations from their SM
values $c_t=c_3=1$ and $c_g=c_{gg}=c_{tt}=0$ originate from dimension-6
operators.  The contribution of the chromomagnetic dipole operator
\cite{Choudhury:2012np, Degrande:2012gr} is not included.

\section{Higgs transverse-momentum spectrum}
\begin{figure}[htbp]
  \begin{centering}
\vspace*{-1cm}

\hspace{3cm} \includegraphics[width=1.00\textwidth]{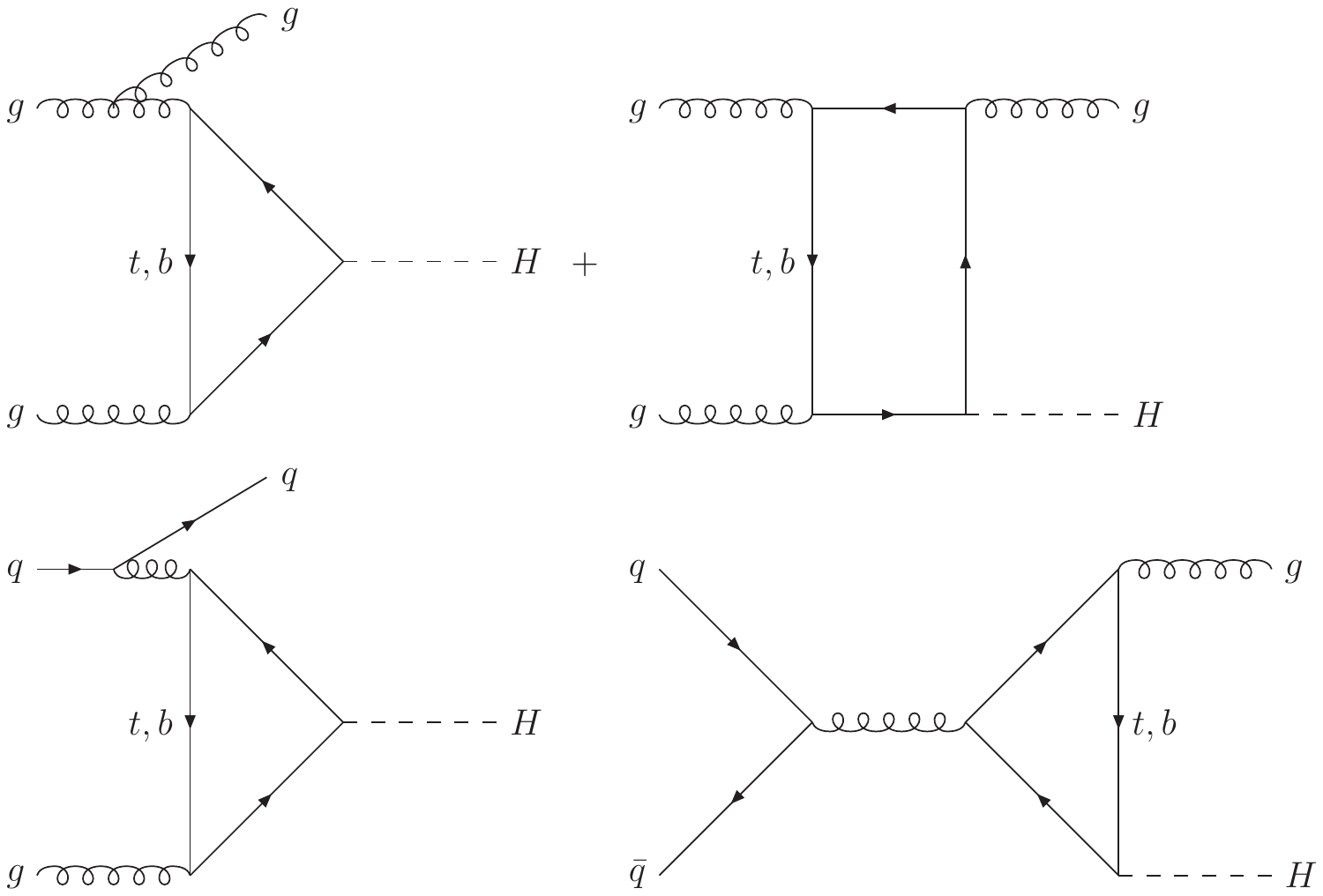} \\[-12cm]
    \caption{Generic diagrams for Higgs production in
association with a jet via gluon fusion at leading order mediated by top
and bottom triangle loops generated by $gg, gq, q\bar q$ initial
states. \label{fg:gg2hg}}
  \end{centering}
\end{figure}

Inclusive Higgs boson production via gluon fusion $gg\to H$ exhibits a
degeneracy between the top Yukawa coupling $c_t$ and the novel
point-like Higgs coupling to gluons, parametrized by $c_g$. Thus, the
consistency of the measured Higgs production rate with the SM prediction
can only constrain a linear combination of these two couplings.  This
degeneracy can be resolved by either measuring the contributions of
$c_t$ and $c_g$ in other production processes, as e.g.~$t\bar tH$
production, or by investigating exclusive distributions. One of the
first observables allowing for a disentanglement of the two
contributions is offered by the Higgs transverse-momentum distribution
at large $p_T$ in gluon fusion that is dominantly mediated by gluon
fusion $gg\to Hg$ (see Fig.~\ref{fg:gg2hg}). First studies in this
direction have been performed in \cite{Harlander:2013oja, Banfi:2013yoa,
Azatov:2013xha, Englert:2013vua, Grojean:2013nya, Schlaffer:2014osa,
Buschmann:2014twa, Buschmann:2014sia, Langenegger:2015lra}.

The NLO corrections to the $p_T$-distribution are only known in the
limit of a heavy top quark \cite{Schmidt:1997wr, deFlorian:1999zd,
Ravindran:2002dc, Glosser:2002gm, Anastasiou:2004xq, Anastasiou:2005qj,
Dawson:2014ora, Hamilton:2015nsa} supplemented by subleading terms in
the inverse top mass at NLO\,\cite{Harlander:2012hf}. As for the
inclusive cross section the QCD corrections are large and positive.
Recently the NNLO QCD corrections to the $p_T$ distribution have been
derived in the heavy top limit yielding a further moderate increase of
$\sim 30\%$ \cite{Gehrmann:2011aa, Boughezal:2013uia, Chen:2014gva,
Boughezal:2015dra, Boughezal:2015aha}, thus corroborating a reliable
perturbative behavior.

Since the pure LO and NLO results diverge for $p_T\to 0$, the small
$p_T$ region requires a soft gluon resummation for a reliable
prediction. This resummation has been performed systematically for the
top quark loops in Refs.\,\cite{Catani:1988vd, Hinchliffe:1988ap,
Kauffman:1991jt, Kauffman:1991cx, Balazs:2000wv, deFlorian:2000pr,
Catani:2000vq, deFlorian:2001zd, Berger:2002ut, Bozzi:2003jy,
Kulesza:2003wi, Watt:2003vf, Kulesza:2003wn, Gawron:2003np,
Lipatov:2005at, Bozzi:2005wk, Bozzi:2007pn, deFlorian:2011xf,
Neill:2015roa}, neglecting finite top mass effects at NLO.  Soft gluon
effects factorize, so that the top mass effects at small $p_T$ are well
approximated by the LO mass dependence for small Higgs
masses\,\cite{Alwall:2011cy,Bagnaschi:2011tu,Mantler:2012bj}. Since the
top-loop contribution dominates the cross section for the SM Higgs
boson, the only limiting factor of the NLO+NNLL result
is thus the heavy-top
approximation of the NLO corrections which affects the whole $p_T$ range
for large Higgs masses and the large $p_T$ region in particular for all
Higgs masses. It has been shown that the subleading NLO terms in the
inverse top mass affect the $p_T$ distribution by less than 10\% for
$p_T \lsim 300$ GeV, if the full LO mass dependence is taken into
account \cite{Harlander:2012hf}. Recently the theoretical predictions
have been extended to the inclusion of dimension-6 operators
\cite{Grazzini:2015gdl}.

\begin{figure}[htbp]
  \begin{centering}
\vspace*{-1cm}

\hspace{8cm} \includegraphics[width=1.20\textwidth]{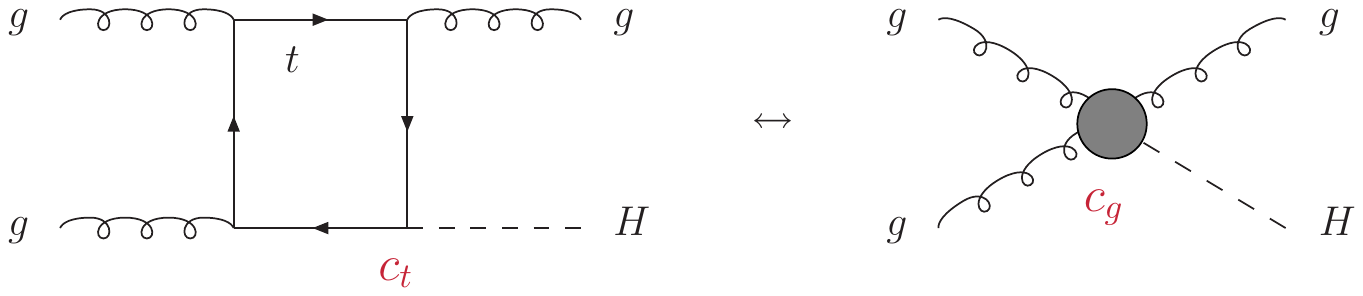} \\[-23cm]
    \caption{Typical diagrams for Higgs production in association with a
jet via gluon fusion at leading order for the different contributions of
top loops and the novel point-like Higgs coupling $c_g$ to gluons.
\label{fg:gg2hgc}}
  \end{centering}
\end{figure}
\begin{figure}[htbp]
  \begin{centering}
    \includegraphics[width=0.49\textwidth]{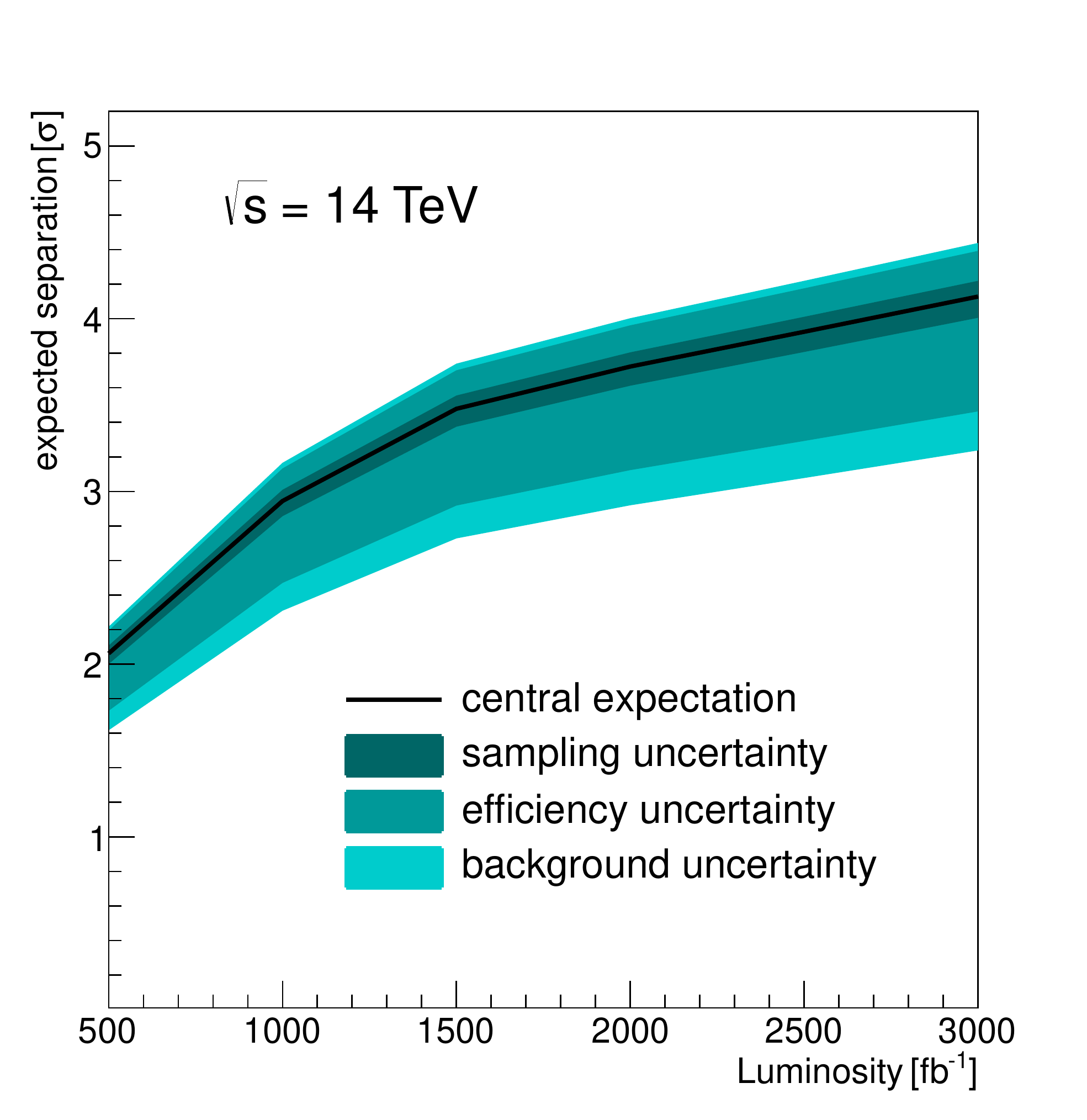}
    \includegraphics[width=0.49\textwidth]{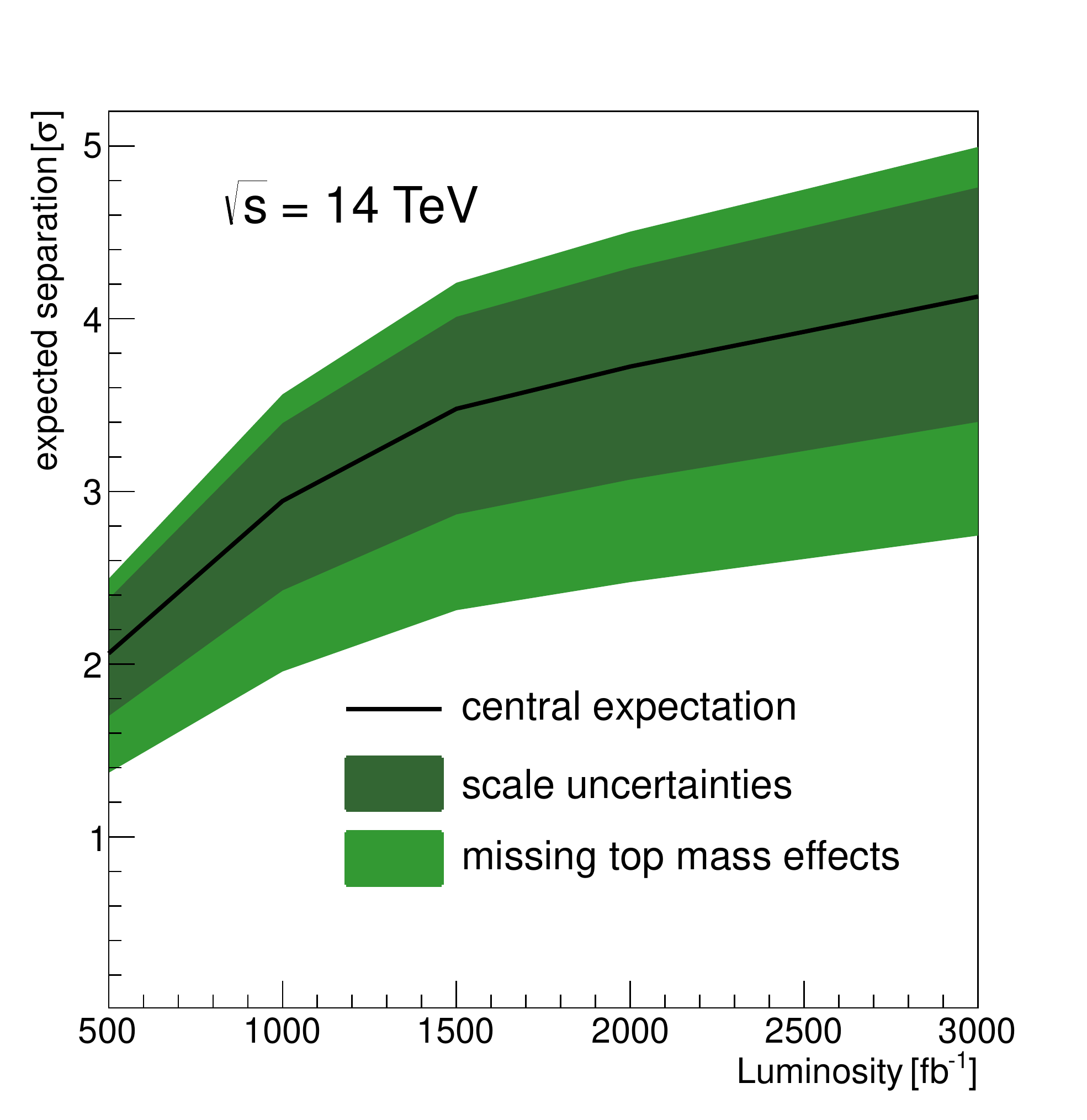}
    \caption{Expected sensitivity vs.~integrated luminosity. The
      experimental uncertainties (left) are combined quadratically,
      the theoretical uncertainties (right) are added linearly
\cite{Langenegger:2015lra}. }
    \label{result-lumi}
  \end{centering}
\end{figure}

As a first step towards the analysis of the future LHC potential to
disentangle the contributions of the top Yukawa coupling and the novel
point-like coupling $c_g$ an investigation about the discrimination
power between a pure top-induced and a pure $c_g$-induced $p_T$
distribution has been performed \cite{Langenegger:2015lra}, as
displayed diagrammatically in Fig.~\ref{fg:gg2hgc}, taking into account
systematic experimental and theoretical uncertainties. The latter are
dominated by scale uncertainties beyond NLO and the missing top mass
effects at NLO. The final result is presented in Fig.~\ref{result-lumi}
for the expected significance of the separation between both scenarios
as a function of the LHC luminosity. First it can be inferred that a
separation of up to $4\sigma$ can be achieved at the HL-LHC, provided,
however, that in particular the significant theoretical uncertainties
will be reduced considerably. It is clearly visible that a full NLO
calculation including top mass effects is required to reach this goal
and the inclusion of the NNLO corrections is necessary to reduce the
residual scale dependence. This signalizes that the uncertainties of the
SM contribution are the limiting factor for the sensitivity of the Higgs
$p_T$ distribution to BSM effects.

\section{Higgs boson pair production}
\begin{figure}[htbp]
  \begin{centering}
\vspace*{-9cm}

\hspace{8cm} \includegraphics[width=1.20\textwidth]{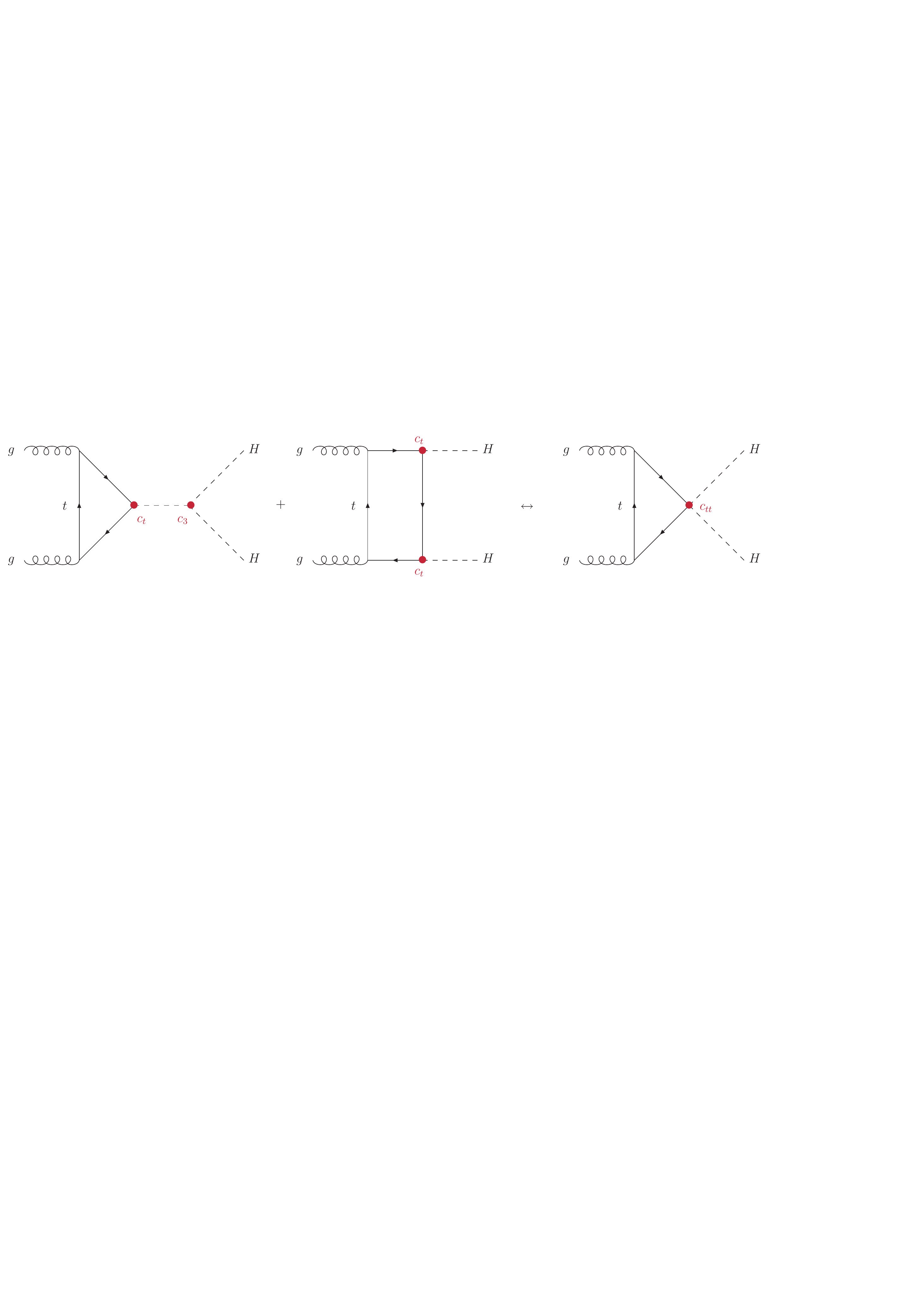} \\[-15.5cm]
    \caption{Typical diagrams for top-loop induced contributions to
Higgs pair production via gluon fusion at leading order for the
different contributions of top and bottom loops for the SM part (first
two diagrams) and the novel point-like Higgs coupling $c_{tt}$ to gluons
in the last diagram.  \label{fg:gg2hh}}
  \end{centering}
\end{figure}
Higgs boson pair production provides the first process sensitive to the
trilinear Higgs self-interaction $\lambda$ at the LHC
\cite{Djouadi:1999rca, Baglio:2012np}. The dominant process is mediated
by gluon fusion $gg\to HH$ (see Fig.~\ref{fg:gg2hh}). The LO cross
section for SM Higgs boson pairs has been calculated a long time ago
\cite{Glover:1987nx, Plehn:1996wb}. The NLO QCD corrections in the limit
of heavy top quarks have been obtained in Ref.~\cite{Dawson:1998py}.
They enhance the cross section considerably, nearly doubling the
production rate.  Subleading NLO top mass effects \cite{Grigo:2013rya,
Frederix:2014hta, Maltoni:2014eza} and NNLO QCD corrections
\cite{deFlorian:2013uza, deFlorian:2013jea, Grigo:2014jma} have been
obtained recently in the heavy top-quark limit.  The NLO heavy top mass
effects have been calculated in terms of an expansion of the total cross
section in inverse powers of the top mass.  They modify the NLO cross
section by about 10\%. At NNLO top mass effects have been estimated to
$\sim 5\%$ \cite{Grigo:2015dia}. The NNLO QCD corrections in the heavy
top-quark limit increase the cross section by about 20\% and signalize a
significant reduction of the theoretical uncertainties as inferred from
the reduced factorization and renormalization scale dependence. Very
recently these calculations have been supplemented by the resummation of
soft and collinear gluon effect up to the NNLL level \cite{Shao:2013bz,
deFlorian:2015moa}.  The total theoretical uncertainty of the production
cross section is estimated in the range of about 10\% for the scale and
PDF+$\alpha_s$ uncertainties \cite{deFlorian:2015moa}. This is increased
to the level of 20\% due to the missing top mass effects at NLO. This
implies that only BSM effects larger than about 20\% on the total cross
section will be visible with the present state of the art. A significant
reduction of this margin is expected by the NLO calculation including
the full top mass effects.

\begin{figure}[htbp]
  \begin{centering}
\vspace*{-0.5cm}

\hspace*{-1.0cm} \includegraphics[width=0.60\textwidth]{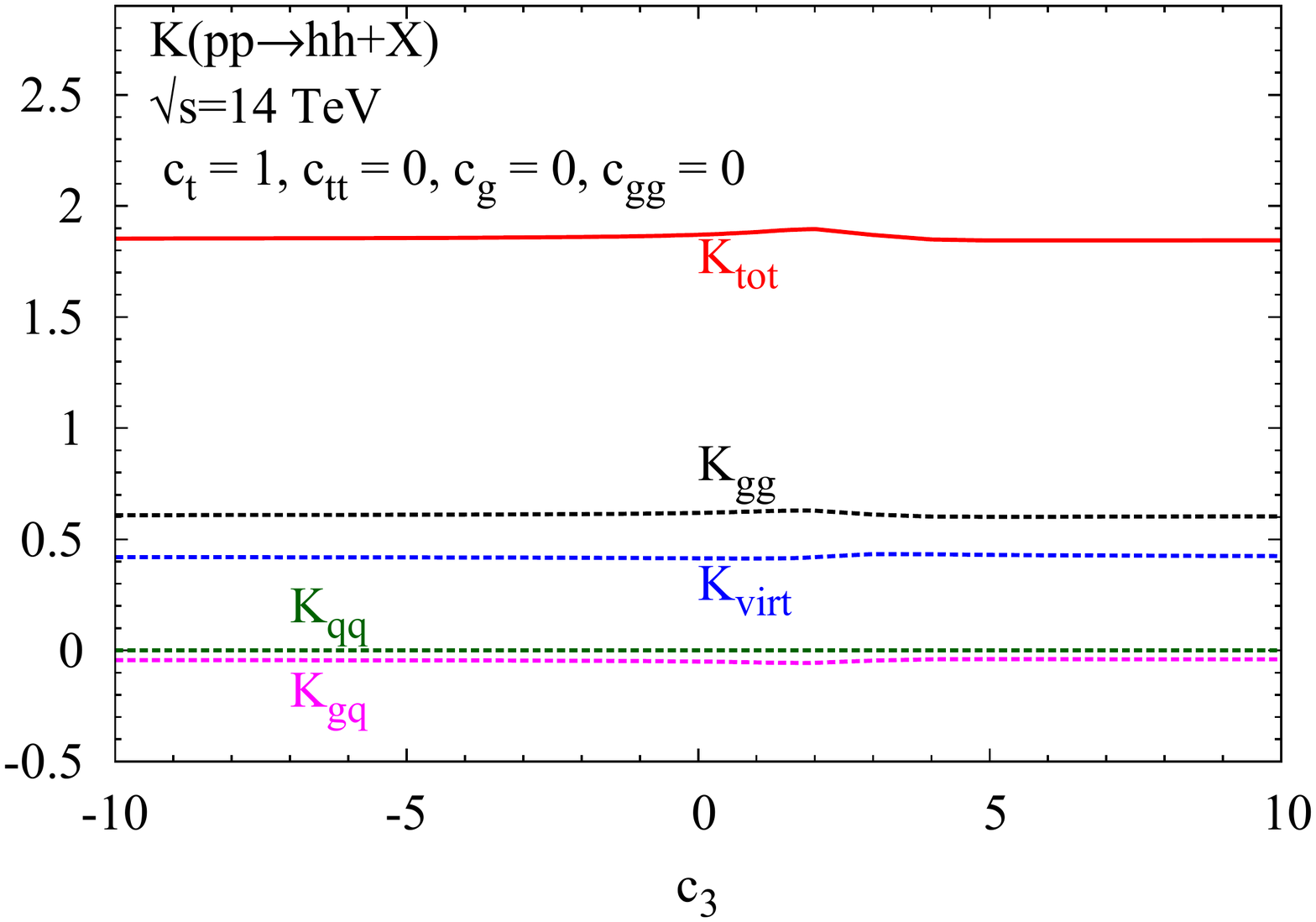}
\hspace*{-2.5cm} \includegraphics[width=0.60\textwidth]{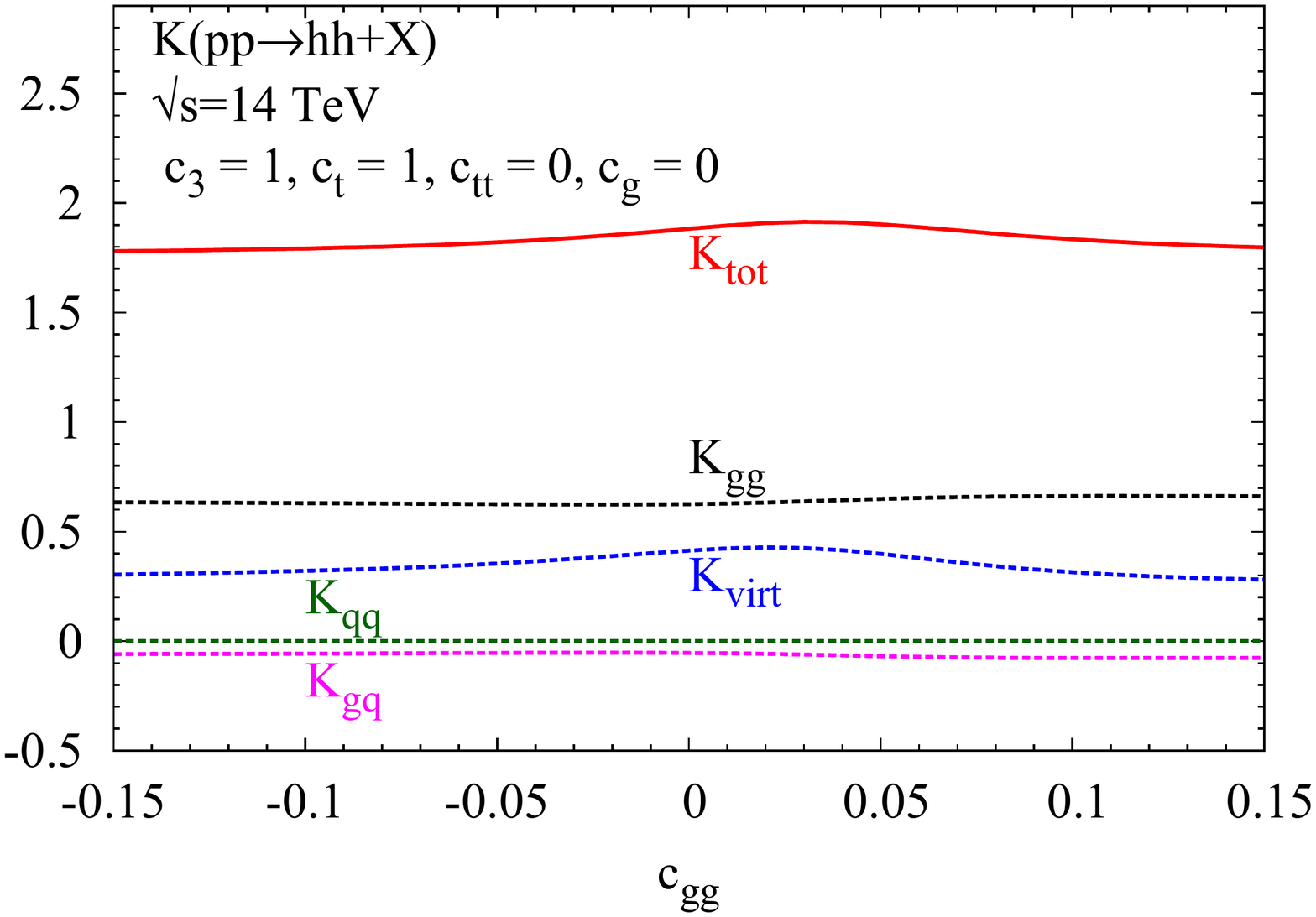}
\\[-1cm]
    \caption{K-factors of Higgs boson pair production via gluon fusion
as function of the dimension-6 Wilson coefficients $c_3$ of the
trilinear Higgs self-coupling and $c_{gg}$ of the novel point-like
$HHgg$ coupling \cite{Grober:2015cwa} for a Higgs mass $M_H=125$ GeV.
\label{fg:hhkfac}}
  \end{centering}
\end{figure}
Starting from the SM Higgs result in the heavy top mass limit the
contributions of dimension-6 operators beyond the SM have been
determined up to NLO QCD \cite{Grober:2015cwa} based on the effective
Lagrangian of Eq.~(\ref{eq:holeff}). There are correlations between all
novel couplings of the Lagrangian in Eq.~(\ref{eq:holeff}) and the
trilinear Higgs coupling $\lambda$, as e.g.~emerging from the last
diagram in Fig.~\ref{fg:gg2hh} due to the novel $HHt\bar t$ coupling
$c_{tt}$. The cross section develops a large dependence on the anomalous
couplings of the dimension-6 operators implemented in the Lagrangian of
Eq.~(\ref{eq:holeff}), i.e.~it can vary by about an order of magnitude
while respecting the present constraints on all dimension-6 Wilson
coefficients. However, the K-factors defined as the ratio between the
NLO and LO cross sections only show a small dependence on the
dimension-6 Wilson coefficients, i.e.~the K-factor agrees with the
corresponding SM K-factor within a couple of per cent as shown for two
examples in Fig.~\ref{fg:hhkfac}. This behavior originates from the
dominance of soft and collinear gluon effects so that this mild
dependence of the K-factor is expected to persist also at higher orders.
On the other hand the results show the necessity to include higher-order
corrections also to the BSM part of the cross sections, since the NLO
QCD corrections are very large for Higgs boson pair production via gluon
fusion. These conclusions are also valid for non-linear composite Higgs
models as demonstrated in the recent work \cite{Grober:2016wmf} at NLO
with the result that the K-factor only develops a small dependence on
the specific Higgs non-linearities for a broad range of the composite
Higgs-model parameters.

\section{Associated Higgs boson production with intermediate vector bosons}

The production of the Higgs in association with a massive vector boson
provides an excellent probe of the structure of the Higgs couplings, and
as a consequence of the sector responsible for electroweak symmetry
breaking. The Higgs in this channel recoils against a vector boson,
which extends the range of energy flowing into the coupling with respect
to leading-order gluon fusion, where $\sqrt{\hat s}\simeq M_H$.  In $VH$
production ($V=W,Z$), the invariant mass of the system gives a measure
of the partonic energy $\sqrt{\hat s}\simeq m_{VH}$, hence the most
sensitive probes of new physics are the bins of high invariant mass or,
similarly, high $p_{T,H}$ or $p_{T,V}$. 

The enhanced sensitivity to new physics from the last bins is a
common-place occurrence when dealing with heavy new physics, and suffers
from sources of theoretical and experimental uncertainties in these
kinematic regions\footnote{Note that in regions of high-momentum
transfer the approach of parametrizing new physics effects in terms of
an effective theory may not be valid. Several approaches are taken to
deal with this issue, including comparisons with UV
models~\cite{Gorbahn:2015gxa,Brehmer:2015rna,Biekotter:2016ecg} or a
restriction of the number of bins dedicated to EFT analyses.}. In particular,
higher-order Standard Model effects, which may be sub-leading in
accounts of the total rates, may be enhanced in the kinematic regions
one focuses on to obtain limits on new physics. 

Specifically, at high-$p_T$ and invariant mass, the effects of SM
higher-order QCD and EW effects may mimic the kind of raise one would
expect from a new physics effect. Therefore, we require an understanding
of these SM effects to the same level as the size of new physics effects
we are looking for, as well as higher-order calculations involving mixed
SM and new physics effects. Moreover, these calculations need to be
incorporated in an MC simulation, in order to perform simulations and
compare showering schemes, another source of systematic uncertainties.  

The effect of QCD effects at NLO is rather straightforward
though~\cite{Maltoni:2013sma,Mimasu:2015nqa}. QCD effects in $VH$ production
factorize, and one can sketch the factorization as
follows~\cite{Mimasu:2015nqa}. Consider an amplitude involving two
partons ($p_{1,2}$) leading to a final state with the Higgs $H$ and a
vector boson $V$   
\begin{eqnarray}
\mathcal{A}_j ( p_1+p_2 \rightarrow V_1(\rightarrow H + V_2) + X)
= \mathcal{J}^{\mu}_{SM}(p_1,p_2,V_1,X) {P}^{V_1}_{\mu\nu}(P_{12X})
V^{\nu}_{\Lambda}(V_2,H) \ ,
\label{eq:curfac}
\end{eqnarray}
where $j$ denotes the total number of initial- and final-state particles
including the vector-boson decays and $P_{12X}$ is the total momentum of
the virtual vector boson $V_1$. Here
$\mathcal{J}^{\mu}_{SM}(p_1,p_2,V_1,X)$ represents the production of a
chiral current in the SM including QCD corrections, so that if $j=5$
then $X=0$, whereas for the real emission amplitude $j=6$ and $X$
corresponds to the emission of an additional gluon. The second current
$V^{\nu}_{\Lambda}(V_2,H)$ corresponds to the splitting of the initial
vector boson $V_1$ into $V_2$ and $H$, with the subsequent decays of
$V_2$ to leptons included.  The current $V^{\nu}_{\Lambda}(V_2,H)$
contains the SM plus possible new physics effects at a typical cut-off
scale $\Lambda$.  Finally, the two currents are connected by a vector
boson propagator ${P}^{V_1}_{\mu\nu}$. Note that this procedure is not
valid when considering Higgs or vector boson decays with relevant
higher-order QCD corrections to new physics effects, e.g. when the decay
$H\to b \bar b$ is affected by new physics. 

\begin{figure}[t]
\begin{center}
 \includegraphics[width=.50\textwidth]{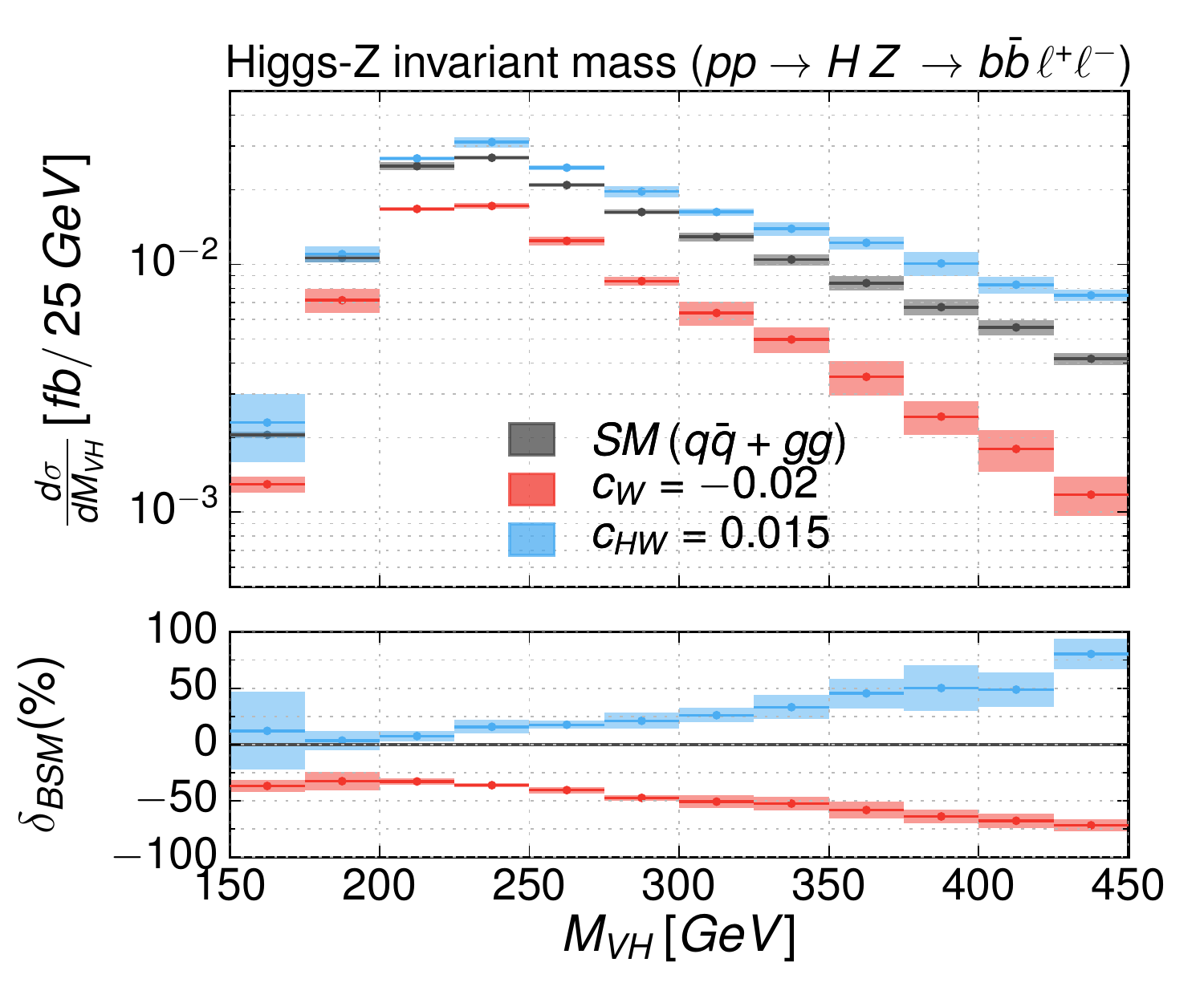}
\hspace*{1cm} \includegraphics[width=.40\textwidth]{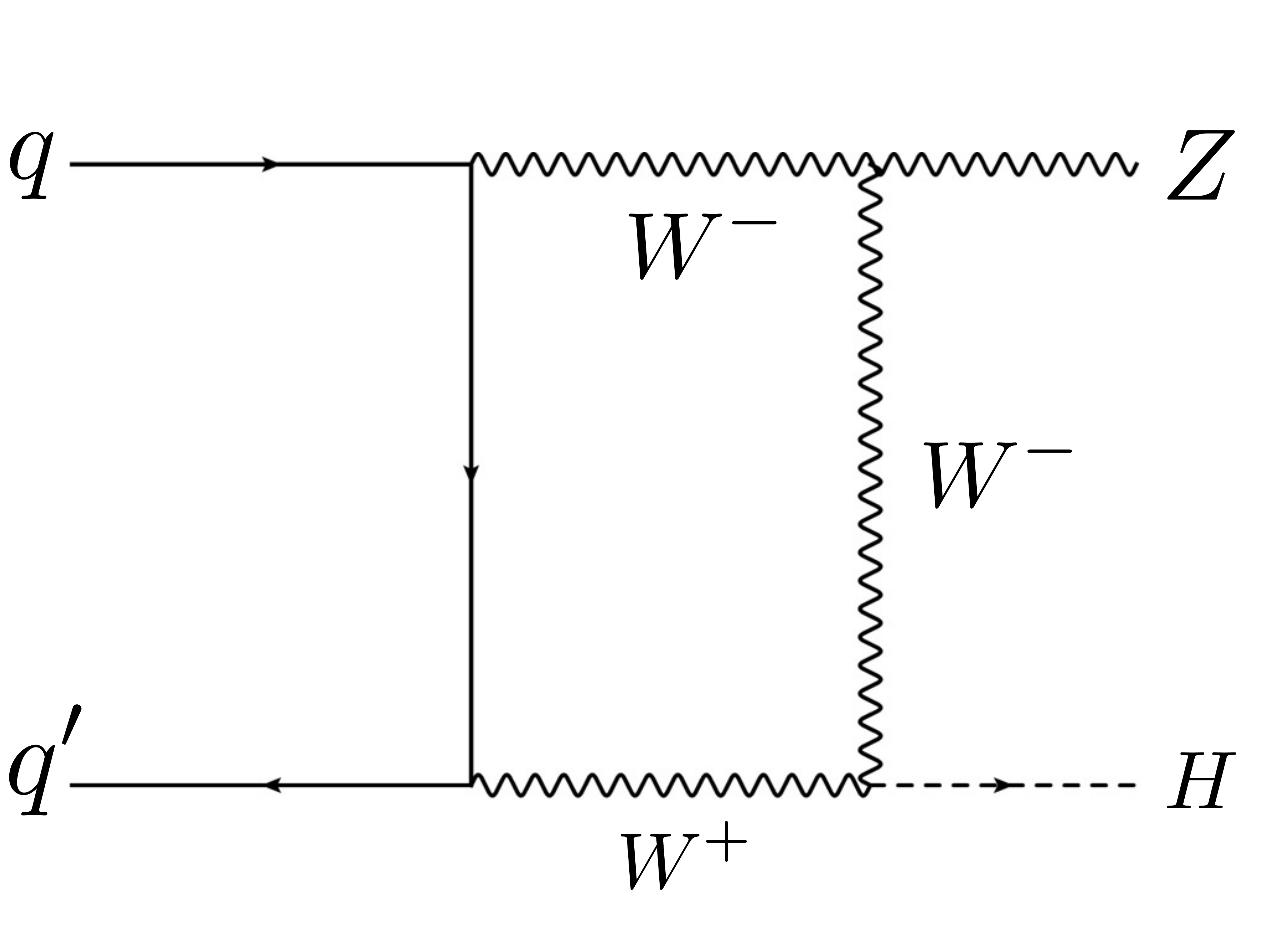}	
    \end{center}
     \caption{Left: Comparison of the SM and EFT predictions with values
of the Wilson coefficients of $\bar c_W=-0.02$ and $\bar c_{HW}=0.015$
in the EFT contribution for the distribution in the Higgs-Z invariant
mass, $m_{VH}$. The relative deviation of the EFT benchmarks from the SM
prediction, $\delta_{BSM}$ in per cent is shown in the lower panel.
Right: A typical diagram developing large Sudakov logarithms in the
high-energy limit.}
 \label{fig:POWHEG}
 \end{figure}

In Fig.~\ref{fig:POWHEG} we show the effect of new physics in the
invariant mass distribution, with the new physics parametrization in
terms of an effective field theory and the coefficients $\bar c_{HW}$
and $\bar c_{W}$ defined in ~\cite{Contino:2013kra}. This figure is an
improved NLO QCD calculation, implemented in  {\sc
Powheg}~\cite{Frixione:2007vw}  and showered through {\sc
Pythia8}~\cite{Sjostrand:2007gs}, and it also contains the
gluon-initiated contribution to $HZ$. Promoting the distribution from LO
to NLO QCD is responsible for a change of the order of ${\cal O}$
(30\%-50\%), see Ref.~\cite{Mimasu:2015nqa} for more details. 

A more subtle question concerns higher-order EW effects, where {\it a
priori} no factorization is possible. Electroweak effects include
diagrams as shown in the right panel of Fig.~\ref{fig:POWHEG}, where
soft and collinear virtual vector bosons are attached to an on-shell
external leg and which lead to the so-called {\it Sudakov logarithms}.
These terms are enhanced by $\log^2 \frac{\hat s}{m_V^2}$, where $m_V$
is the mass of the vector boson and $\sqrt{ \hat s}$  is the parton
energy, and could be parametrically large at large momentum transfer --
the region of interest for exploring new physics. In the case of diboson
production, these corrections have been studied in detail (see e.g.
Ref.~\cite{Biedermann:2016yvs} for a recent study) but more calculations
need to be done for $HV$, as well as the implementation of those in an
MC generator.


\section*{Acknowledgments}
The work of M.S.~is supported in part by the Research Executive Agency
(REA) of the European Union under the Grant No.~PITN-GA-2012-316704
(Higgstools).



\AddToContent{M.~M\"uhlleitner, V.~Sanz and M.~Spira}
\renewcommand{\thesection}{\arabic{section}}

\graphicspath{{VLQ/}}

\newcommand{\beq}{\begin{eqnarray}}
\newcommand{\eeq}{\end{eqnarray}}
\renewcommand {\be} {\begin{equation}}
\renewcommand {\ee} {\end{equation}}
\newcommand\sss{\scriptscriptstyle}
\def\bsp#1\esp{\begin{split}#1\end{split}}

\chapter{Higgs pair production from New Physics at the LHC Run--II}
\label{chap:vlq}
{\it G.~Cacciapaglia, H.~Cai, A.~Carvalho, A.~Deandrea, T.~Flacke, B.~Fuks,
D.~Majumder and H.-S.~Shao}



\begin{abstract}
We consider Higgs pair production at the LHC from New Physics, focusing on two scenarios which are relevant for
various extensions of the Standard Model: those containing vector-like quarks and those with new coloured scalars.
Besides effects that can be described in the effective field theory approach, the Higgses can also be produced via
decays of the heavy states. We present a first detailed study of cross sections and distributions at parton level. 
For the vector-like model, we also introduce a new {\sc MadGraph5\_aMC@NLO} implementation at NLO in QCD.
\end{abstract}

\section{INTRODUCTION}
\label{sec:intro}
The couplings of the Higgs boson ($h$) to other Standard Model (SM) particles and the Higgs boson self-couplings may be modified in presence of physics Beyond the Standard Model (BSM).
In particular, at the LHC, they can give rise to unusual or otherwise rare topologies: in this preliminary study we will focus on pair production of two Higgs bosons which is a rare process within the SM and the only one to be sensitive to Higgs triple couplings (see for example~\cite{Dolan:2012ac}). 
We will focus on two scenarios with
new coloured scalars or new Vector-like Quarks (VLQ). 
These types of states are common in many models of new physics, therefore hunting 
for the associated signals allows one to extract useful information about the possible BSM structure. 
For VLQs we consider in particular the possibility of dominant coupling to the Higgs boson and light quarks, a scenario that is rarely considered in the literature and that may occur in models where near degenerate states are present, or in models with derivative couplings which do
not contribute to the Yukawa couplings of the quarks. We use a {\sc FeynRules} model with implementation that is able to 
simulate processes at NLO in QCD,  introduced and tested in this document. 
For the new coloured scalars, we study in particular the case of colour triplet pair production, focusing on cascade decays that 
give rise to a final state with two Higgs bosons and jets.
In the two scenarios we selected, the dominant couplings to the Higgs implies that production of two Higgses may be the main channel where new physics effects show up.

Both ATLAS and CMS collaborations have conducted searches for di-Higgs production using the LHC proton-proton collision data at a centre-of-mass energy of 8~TeV~\cite{Aad:2015xja,  Aad:2015uka,  Khachatryan:2015yea, Khachatryan:2016cfa, Khachatryan:2015tha, CMS:2015zug, CMS:2014ipa}. The SM prediction for the $hh$ cross section is too small to be observable with the current dataset. However new physics processes can significantly increase the rate of $hh$ signal, therefore they have been the focus of searches by the experimental collaborations. Results have been interpreted in terms of the minimal supersymmetric extensions of the standard model~\cite{Aad:2015xja,Khachatryan:2015tha}, or  two-Higgs-doublet-models~\cite{Aad:2015uka,Khachatryan:2015tha}. More exotic scenarios like radion or Kaluza-Klein graviton~\cite{Agashe:2007zd} and/or radions~\cite{Dominici:2002jv} (that are comon ingredients of Randall--Sundrum models~\cite{Randall:1999ee}) decaying to $hh$ are also probed~\cite{Khachatryan:2015yea,Khachatryan:2016cfa,Aad:2015uka,CMS:2014ipa} over a wide range of masses of the proposed new particles. Both non-resonant and resonant production of $hh$ in BSM would lead to a deviation in the invariant mass of the di-Higgs system, $m_{hh}$, from the SM prediction. In particular, the resonant production leads to a narrow localised excess in the $m_{hh}$ distribution which is easy to detect in a classic ``bump hunt''. The models we consider predict multiple channels which may produce a di-Higgs excess detectable through the $m_{hh}$ spectrum, but not necessarily as a localised bump.

On the theory side, some previous studies considered the possibility of BSM di-Higgs production using a low energy effective field theory (EFT) approach~\cite{Grober:2010yv,Contino:2012xk}. However, if the new particles are light enough to be directly produced at the LHC the Higgs bosons may also come from decays of the heavy coloured states, thus having very different kinematic properties that will impact the search strategies. For comparison, a dedicated EFT treatment was made in~\cite{Dall'Osso:2015aia}.
While the main production mode in the SM is gluon fusion, the main channels in new physics models contain additional particles in the final state.
In the VLQ scenario (Section~\ref{sec:model}), we will assume that the main coupling involves the Higgs and a light SM quark so that  the heavy fermion decays nearly 100\% into a Higgs plus a jet.  We also present first results for cross section based on a QCD-NLO implementation of the VLQ model: for comparison, a recent NLO study in the EFT framework considering top quark partners can be found in~\cite{Grober:2016wmf}.
In the scalar case (Section~\ref{sec:scalar}), we consider scenarios where two such scalars mix via the couplings to the Higgs so that the heavier one dominantly decays into the lighter one plus a Higgs, while the low mass one decays into a pair of jets.
Finally, in Section~\ref{sec:distros} we present preliminary results of distributions obtained via a LO simulation at parton level.

\section{VECTOR-LIKE QUARK MODEL}
\label{sec:model}
\subsection{MODEL DESCRIPTION}
Vector-like quarks play a special role in new physics as they are present in
extensions of the SM such as extra-dimensional~\cite{Antoniadis:1990ew} and Little-Higgs~\cite{ArkaniHamed:2002qx} models, and
composite-Higgs models with partial compositeness~\cite{Kaplan:1991dc}. In general, VLQs interact
with both the $W$ and $Z$ gauge bosons and with the Higgs boson, in vertices
involving the SM quarks~\cite{AguilarSaavedra:2002kr,Cacciapaglia:2010vn}.
At the LHC, the production of the VLQs occurs in pairs, typically through QCD
processes, or singly, through electroweak processes. Pair-production
cross-sections decrease faster with increasing VLQ mass than single production ones, due to parton luminosity and phase space effects. Extensive searches
in the pair production mode have been performed at the LHC Run--I, especially
focusing on decays into third generation quarks~\cite{DeSimone:2012fs}, with
bounds close to 1 TeV for both ATLAS~\cite{Aad:2014efa,Aad:2015mba,Aad:2015kqa}
and CMS~\cite{Chatrchyan:2013uxa,Chatrchyan:2013wfa,Khachatryan:2015axa,%
Khachatryan:2015gza,Khachatryan:2015oba}. Recent searches in the single
production channel have also been performed by ATLAS~\cite{Aad:2015voa,%
Aad:2016qpo}. The preference for third generation couplings is  based on
composite Higgs models (see for instance Refs.~\cite{Contino:2006qr,%
Anastasiou:2009rv,Dissertori:2010ug,Matsedonskyi:2012ym,Berger:2012ec,%
Vignaroli:2012nf,Grojean:2013qca,Matsedonskyi:2015dns}) and the special role
played by the top quark in connection to its large mass and coupling to the
Higgs boson. However, although sizeable couplings to light quarks are still
allowed by indirect constraints~\cite{delAguila:2000rc,Cacciapaglia:2015ixa,%
Ishiwata:2015cga}, only few dedicated searches are available~\cite{Aad:2015tba}.
Furthermore, single production can play a growing important role in the
high-mass range, especially when couplings to light quarks are
involved~\cite{Atre:2011ae}. On the theory side, recent explorations of VLQs can
be found in Refs.~\cite{Cacciapaglia:2011fx,Ellis:2014dza,Bizot:2015zaa}. The
ratios of couplings to the $W$, $Z$ and Higgs bosons typically depend on the
representation under the electroweak symmetry which the VLQ belongs to, however
the simplicity of the coupling structure allows for simple parameterisations. 
In order to describe couplings to all SM families in a handy and model-independent way, a parameterisation was first formulated in a previous Les
Houches meeting~\cite{Brooijmans:2014eja}, and then used to calculate single
production rates at the leading order (LO) accuracy~\cite{Buchkremer:2013bha}.

In this contribution, we study VLQ direct production at the LHC, as well as
effects that are induced when either a pair of Higgs bosons or an associated
pair of a Higgs boson and a VLQ are produced, possibly exchanging virtual VLQ
states. To this aim, we use a {\sc FeynRules}~\cite{Alloul:2013bka}
implementation able to describe the dynamics of usual VLQs carrying the same
electric charge as the Standard Model bottom and top quarks, $B$ and $T$, as
well as of exotically charged $X$ and $Y$ fermionic states whose electric charges
are $Q=5/3$ and $-4/3$ respectively. The (gauge-invariant) kinetic and mass
terms for these fields are given by
\be
  {\cal L}_{\rm kin} =
    i \bar Y  \slashed{D} Y - m_{Y}  \bar Y  Y
 +  i \bar B \slashed{D} B- m_{B} \bar B B
 +  i \bar T \slashed{D} T - m_{T} \bar T T
 +  i \bar X  \slashed{D} X  - m_{X}  \bar X  X \ ,
\ee
where the covariant derivatives are
\be
  D_\mu  = \partial_\mu -i g_s T_a G_\mu^a - i Q  e A_\mu\ .
\ee
In our notation, $g_s$ and $e$ denote the strong and electromagnetic coupling
constants, $G_\mu$ and $A_\mu$ the gluon and photon fields, $T_a$ the
fundamental representation matrices of $SU(3)$ and $Q$ the electric charge
operator. As the new quarks need to belong to complete representations of the
weak isospin group $SU(2)_L$, they also couple, with flavour-violating
couplings involving one VLQ and one SM quark, to the $W$, $Z$ and Higgs bosons.
The most general effective model able to describe the related phenomenology is
parameterised by the Lagrangian
\begin{eqnarray}
  {\cal L}_{\rm eff} &=& \frac{\sqrt{2} g}{2} \bigg[
    \bar Y  \slashed{\bar W} \Big(\kappa_L^{\sss Y} P_L + \kappa_R^{\sss Y} P_R\Big) d
  + \bar B \slashed{\bar W} \Big(\kappa_L^{\sss B} P_L + \kappa_R^{\sss B} P_R\Big) u  \nonumber \\
  &+& \bar T \slashed{W} \Big(\kappa_L^{\sss T} P_L + \kappa_R^{\sss T} P_R\Big) d
  + \bar X  \slashed{W} \Big(\kappa_L^{\sss X} P_L + \kappa_R^{\sss X} P_R\Big) u
   \bigg] \nonumber \\
  & + &\frac{g}{2 c_W} \bigg[
    \bar B \slashed{Z} \Big(\tilde \kappa_L^{\sss B} P_L + \tilde\kappa_R^{\sss B} P_R\Big) d
  + \bar T \slashed{Z} \Big(\tilde \kappa_L^{\sss T} P_L + \tilde\kappa_R^{\sss T} P_R\Big) u
   \bigg]   \nonumber \\
    &-& h \bigg[
    \bar B \Big(\hat\kappa_L^{\sss B} P_L + \hat\kappa_R^{\sss B} P_R\Big) d
  + \bar T \Big(\hat\kappa_L^{\sss T} P_L + \hat\kappa_R^{\sss T} P_R\Big) u
   \bigg] + {\rm h.c.} \ ,
\end{eqnarray} 
where $g$ stands for the weak coupling, $c_W$ for the cosine of the weak mixing
angle and we denote the physical Higgs boson by $h$, and the charge $-1/3$ and $2/3$ partners in the mass eigenbasis by $B$ and $T$. Moreover, all $\kappa$,
$\tilde \kappa$ and $\hat\kappa$ coupling strengths are organised in terms of
tridimensional vectors in flavour space. In our {\sc FeynRules} implementation,
the numerical values of the elements of these vectors can be provided via the
Les Houches blocks {\tt KYLW} ($\kappa_L^{\sss Y}$),
{\tt KYRW} ($\kappa_R^{\sss Y}$), {\tt KBLW} ($\kappa_L^{\sss B}$),
{\tt KBRW} ($\kappa_R^{\sss B}$), {\tt KTLW} ($\kappa_L^{\sss T}$),
{\tt KTRW} ($\kappa_R^{\sss T}$), {\tt KXLW} ($\kappa_L^{\sss X}$),
{\tt KXRW} ($\kappa_R^{\sss X}$), {\tt KBLZ} ($\tilde\kappa_L^{\sss B}$),
{\tt KBRZ} ($\tilde\kappa_R^{\sss B}$), {\tt KTLZ} ($\tilde\kappa_L^{\sss T}$),
{\tt KTRZ} ($\tilde\kappa_R^{\sss T}$), {\tt KBLH} ($\hat\kappa_L^{\sss B}$),
{\tt KBRH} ($\hat\kappa_R^{\sss B}$), {\tt KTLH} ($\hat\kappa_L^{\sss T}$)
and {\tt KTRH} ($\hat\kappa_R^{\sss T}$).

The main difference between the above Lagrangian and the one proposed in
Ref.~\cite{Brooijmans:2014eja} is the absence of a mass dependent factor in the
Higgs couplings. 
This new choice is motivated by the wish of being able to extend the
implementation so that it could be used for calculations at the
next-to-leading order (NLO) accuracy in QCD, in a fully automated framework starting from the sole input of
the model Lagrangian. In the current implementation, and in contrast to the previous
one, the renormalisation of the interactions of the Higgs boson with the new and
SM quarks is independent of the renormalisation of the VLQ mass. Thus, this choice leads to a consistent modelling at the NLO QCD level. 
A mass-dependent coupling
strength would indeed not allow for the cancellation of all ultraviolet
divergences appearing at the one-loop level.

In order to generate a model library that could be used for NLO predictions, we
have made use of the {\sc NloCT}
program~\cite{Degrande:2014vpa} to automatically calculate all the ultraviolet
and $R_2$ counterterms of the model that are necessary for NLO calculations in
QCD within the {\sc MadGraph5\_aMC@NLO} framework~\cite{Alwall:2014hca}. The latter 
provides a general platform for computing (differential) observables within many BSM
 theories~\cite{Degrande:2014sta,Degrande:2015vaa}. More precisely,
virtual contributions are evaluated within {\sc MadLoop} using the
Ossola-Papadopoulos-Pittau technique~\cite{Ossola:2006us,Ossola:2007ax,%
Hirschi:2011pa} and combined with real contributions through {\sc MadFKS},
employing the FKS subtraction method~\cite{Frixione:1995ms,Frederix:2009yq}. Although
this will not be used for this work, the
matching to parton showers can be achieved with the MC@NLO method~\cite{%
Frixione:2002ik}.

\subsection{SINGLE AND PAIR PRODUCTION OF VLQs AT THE NLO ACCURACY IN QCD}
In this section, we make predictions both at the LO and NLO accuracy in QCD for
the single and pair production of VLQs. Our results include the theoretical
uncertainties stemming from scale and parton distribution variations. For the
central values, we set the renormalisation and factorisation scales to the
average transverse mass of the final state particles and use the NLO NNPDF~3.0
set of parton distributions~\cite{Ball:2014uwa} that we have accessed via the
LHAPDF~6 library~\cite{Buckley:2014ana}. Scale uncertainties are derived by
varying both scales independently by factors of two up and down, and the PDF
uncertainties have been extracted following the NNPDF
recommendations~\cite{Demartin:2010er}. Both scale and PDF uncertainties have
then been added in quadrature.

\begin{figure}
  \begin{center}
  \includegraphics[width=0.48\textwidth]{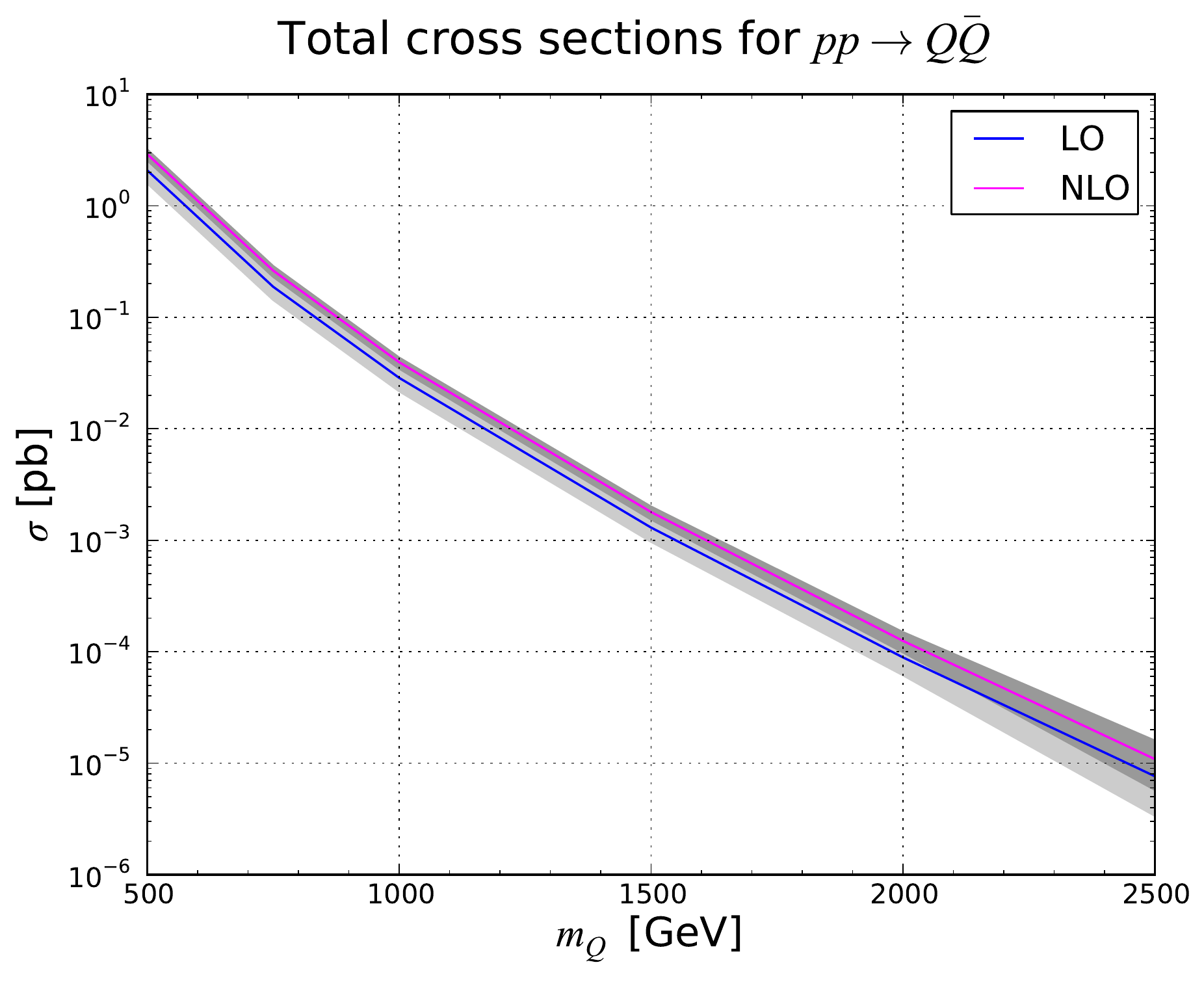}
  \includegraphics[width=0.48\textwidth]{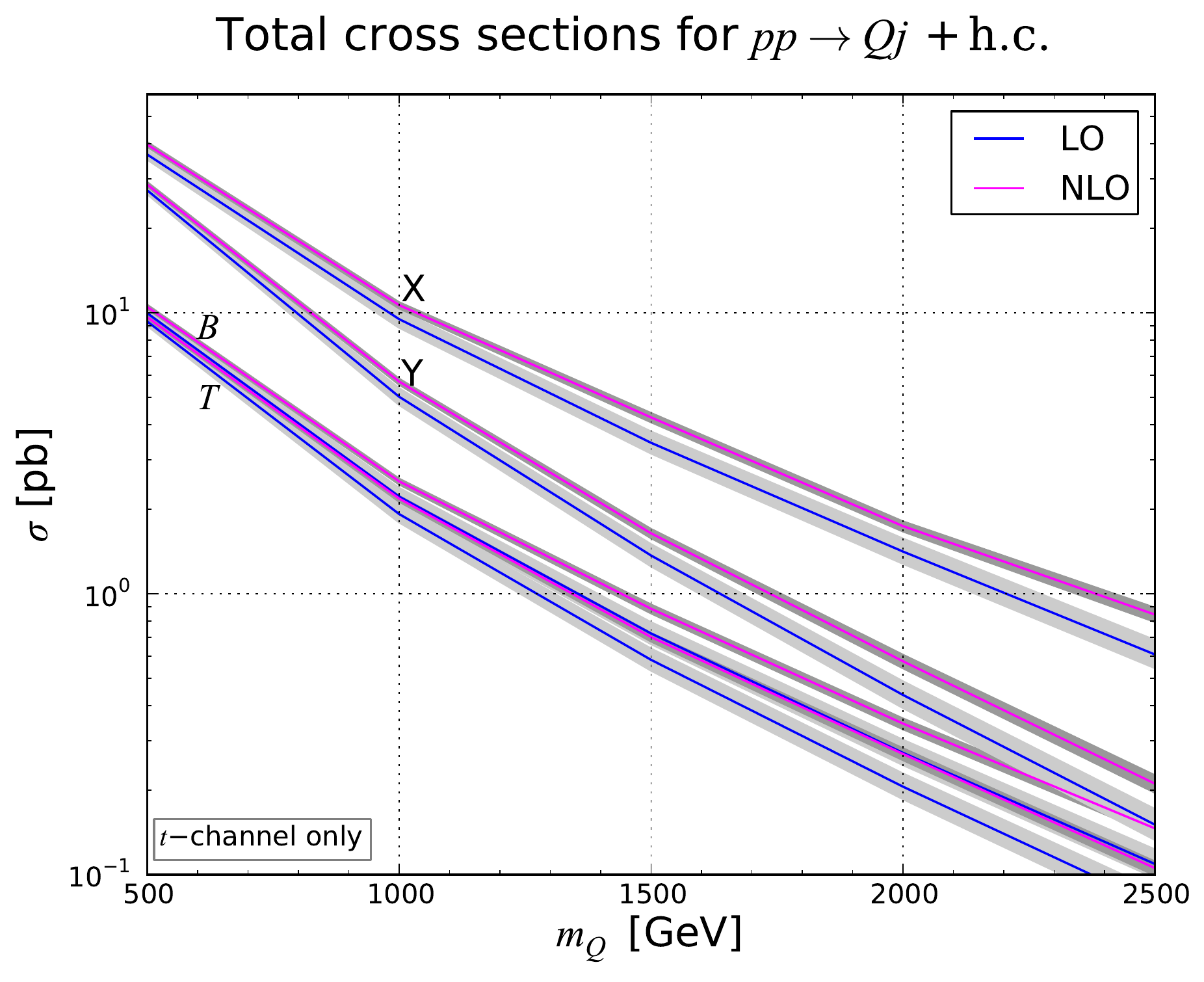}
  \caption{\small LO and NLO QCD inclusive cross sections for different
    production processes involving vector-like quarks in the context of
    proton-proton collisions at a centre-of-mass energy of 13~TeV. The results
    are presented together with the associated theoretical uncertainties
    obtained from a quadratic sum of the scale and PDF uncertainties. We consider
    VLQ pair production (left panel) and single production in association with a
    jet (right panel) for a benchmark scenario defined in the text.
  \label{fig:VLQNLO_QQ_Qj}}
  \end{center}
\end{figure}

In Figure~\ref{fig:VLQNLO_QQ_Qj}, we present results both for the production of
a pair of VLQs (left panel) and the associated production of a single VLQ with
a jet (right panel). In the first case, we are in the context of a pure QCD
process, $p p \to Q \bar Q$, where only QCD vertices are involved for the quark partners $Q\in (Y,B,T,X)$. Therefore, the
cross section is the same for all four VLQ species. 
The results for  the  production of a pair of VLQs can be considered as a validation of 
our implementation.  
The genuine NLO effects can be described within a global $K$-factor,
defined as the ratio of the NLO to the LO predictions, and they yield an
enhancement of the total rate by about 40\% for the entire probed VLQ mass
range. Equivalently, we have roughly $K\sim1.4$ for $M_Q\in [500, 2500]$~GeV.
The corresponding theoretical uncertainties are however sizeably reduced at the
next-to-leading order accuracy, in particular for small VLQ masses where the
precision increases from 20-30\% to about 10\%. For larger values of $M_Q$, the
uncertainties are larger both at the LO (being of about 30-60\%) and at the NLO
(being of about 10-40\%). Their reduction is here tamed by their dominant PDF
component as larger Bjorken-$x$ values are probed, which corresponds to a regime
where the PDFs are not so accurately known. We recall that in both the LO and
NLO cases, the same NLO fit of the NNPDF collaboration has been used. This
choice is a consequence of the poor quality of the NNPDF LO fit that is thus
barely reliable. As the same set of parton density is used in both the LO and
NLO cases, the related uncertainties are comparable in magnitude. 
We have also verified that for $M_Q = M_t$, the predictions obtained using our
model implementation agree with NLO QCD results in the context of top
pair-production.

Turning to single VLQ production, we first setup a benchmark for our study, as
this electroweak process depends on the details of the model. We fix the values
of the $\kappa$ and $\tilde \kappa$ parameters following the guidelines provided
in Ref.~\cite{Buchkremer:2013bha}. The $\hat\kappa$ parameter is
irrelevant in the context of single VLQ production with jets, as the light quark
Yukawa couplings are negligible. Motivated by the fact that the VLQ electroweak
couplings are typically chiral, we start by enforcing that all right-handed
couplings are zero. We then set the remaining couplings so that the branching
ratios of the $B$ and $T$ states into a $Vj$ system are of 30\%, 30\% and 40\%
for $V=W$, $Z$ and $h$. In addition, we normalise the mixing of the VLQ with the
SM model quarks in a way such that the $\zeta_i$ parameters of
Ref.~\cite{Buchkremer:2013bha},
\be
  \zeta_i = \frac{\big|U^{4i}_{L/R}\big|^2}{\sum_{j=1}^3\big|U^{4j}_{L/R}\big|^2}\ ,
\ee
are given by $\zeta_1=\zeta_2=0.3$ and $\zeta_3=0.4$. In this last equation, the
matrix $U$ describes the mixing of the VLQ state with the SM quarks, so that in
our benchmark scenario, the different VLQ states will predominantly couple to
quarks of the third generation and equivalently couple to the lighter quark
species. Finally, the overall coupling strengths to the quarks of the third
generation have been taken equal to unity.
Results for single VLQ production in association with a jet are shown on the
right panel of Figure~\ref{fig:VLQNLO_QQ_Qj}. For this case, only $t$-channel
Born contributions have been retained and no generator-level cut has been
applied on either the jet or the new quark. For low VLQ masses, the $K$-factor is
found to be around unity, so that the NLO cross section is basically equal to
the LO one. The associated theoretical uncertainties are however reduced by a
factor of about two to the 1-2\% level. For larger masses, the $K$-factor
increases and reaches about 1.3 for $M_Q=2$~TeV. The related uncertainties are
reduced to the 5\% level, improving again the accuracy of the results by a
factor of two when comparing with the LO predictions. In the chosen illustrative
benchmark setup, single VLQ production cross sections turn out to be larger than the QCD
pair production ones. However, such a statement is very model-dependent, 
it depends on the VLQ masses and mixing patterns.

\subsection{VLQ-INDUCED DI-HIGGS PRODUCTION}
\begin{figure}
  \begin{center}
  \includegraphics[width=0.2\textwidth, angle =0 ]{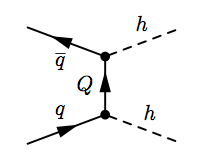}
  \includegraphics[width=0.2\textwidth, angle =0 ]{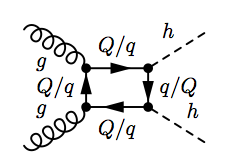}
  \includegraphics[width=0.2\textwidth, angle =0 ]{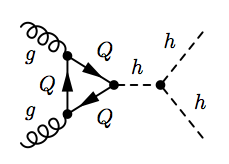}
  \caption{\small Representative Feynman diagrams illustrating the production of
    a pair of Higgs bosons in the presence of VLQs. The first diagram
    corresponds to the tree-level component that has not any counterpart in the
    SM, while the second and third diagrams are similar to the SM contributions
    after having replaced one or more of the internal quark $q$ by a VLQ $Q$.}
  \label{fig:VLQdiag_hh}
  \end{center}
\end{figure}

In the following we focus on a peculiar scenario where the Standard Model is
extended with one VLQ species that couples to a Higgs boson and a light quark.
This situation may occur in models where near-degenerate states are present,
like in the case where two doublets of VLQs are considered~\cite{Atre:2011ae,%
Azatov:2012rj}, or in composite models due to the effect of additional
derivative couplings that do not contribute to the quark
Yukawa interactions~\cite{Backovic:2014ega}. This type of scenarios allows,
therefore, for the production of Higgs-enriched final states at the
LHC~\cite{Atre:2013ap}. In particular, the presence of the VLQ enhances the
production of events featuring a Higgs boson pair, a process that will be widely
searched for at the LHC Run--II. Illustrative Feynman diagrams are depicted in
Figure~\ref{fig:VLQdiag_hh}.

To focus on scenarios in which the only
non-vanishing electroweak VLQ couplings are related uniquely to the Higgs, we
set all the $\kappa$ and $\tilde{\kappa}$ parameters to zero (that are moreover
irrelevant for di-Higgs production), and once again choose $\hat{\kappa}_R = 0$
using the fact that VLQ couplings are typically chiral. The choice of chirality
is  known not to significantly affect the cross sections. In the
following study, the only relevant couplings are the two vectors in flavour space
$\hat{\kappa}^B_L$ and $\hat\kappa^T_L$. We further simplify the modelling by
allowing only one of these six couplings to be non-zero at a time. Such a
coupling is typically connected to the mass of the VLQ via a mixing matrix
entry, $\hat{\kappa} \sim M_Q/v_{\rm SM}\ U$ ($v_{\rm SM}$ being the Standard
Model Higgs vacuum expectation value), whereas the couplings to the gauge bosons
are typically given by the magnitude of the mixing
$\kappa \sim \tilde{\kappa} \sim U$~\cite{Buchkremer:2013bha}. In other words,
our new physics description implies that we will explore the mass dependence of
the cross section by keeping the mixing between the SM quarks and the VLQs
fixed\footnote{Another choice would be to define the mixing matrix as
$U \sim v_{\rm SM}/M_Q$, so that $\hat{\kappa}$ is a constant while the gauge
boson couplings would scale like $1/M_Q$~\cite{Atre:2013ap}. However, typical
corrections to low energy observables are only sensitive to the mixing angle,
which can then be constrained independently of the value of the VLQ mass.},
\begin{equation}
\hat{\kappa}_L^{T/B} = \kappa \ \frac{M_Q}{v_{\rm SM}}\,,
\end{equation}
with $v_{\rm SM} = 246$ GeV, and $\kappa$ being a free parameter of the model.

\begin{figure}
  \begin{center}
  \includegraphics[width=0.48\textwidth]{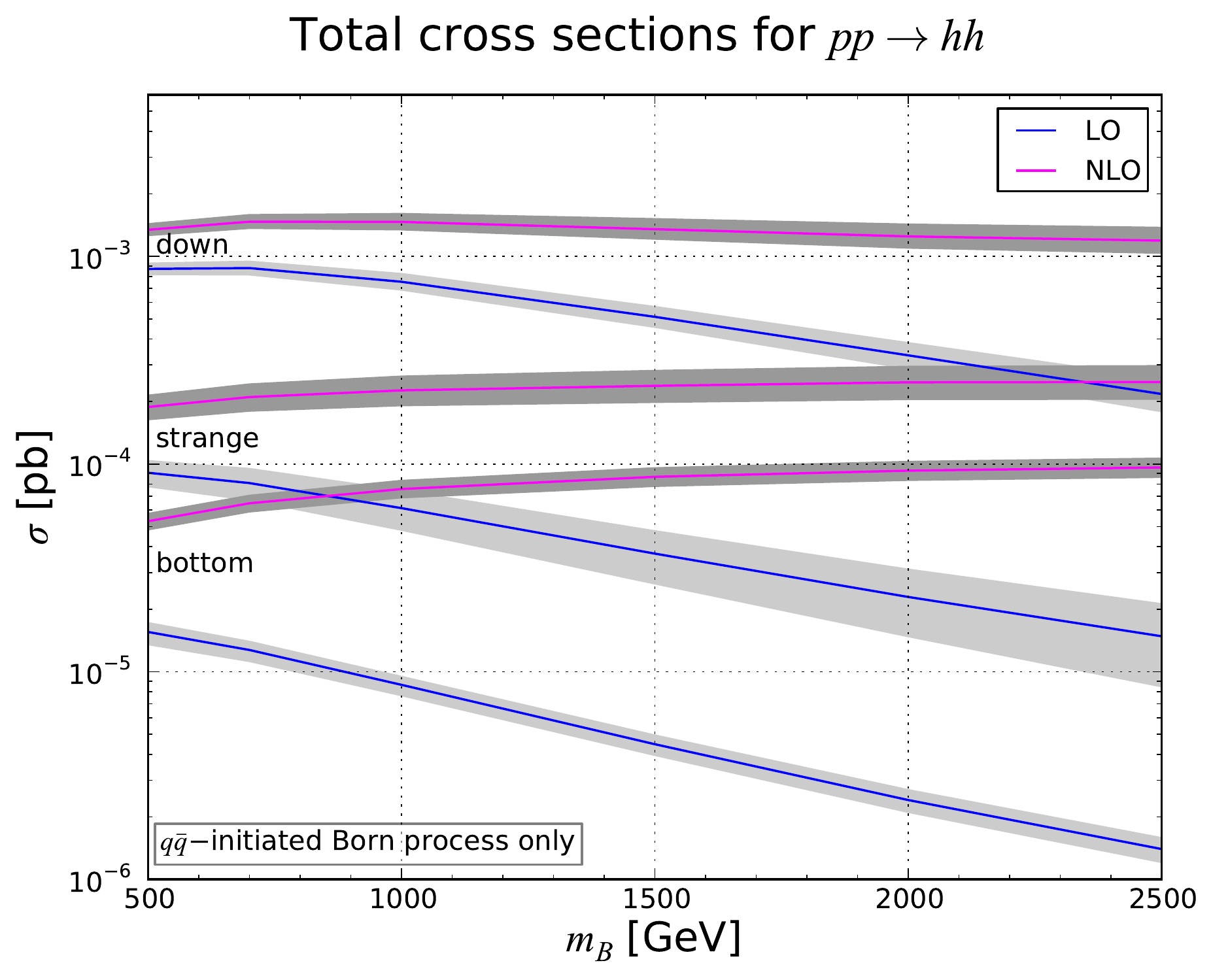}
  \includegraphics[width=0.48\textwidth]{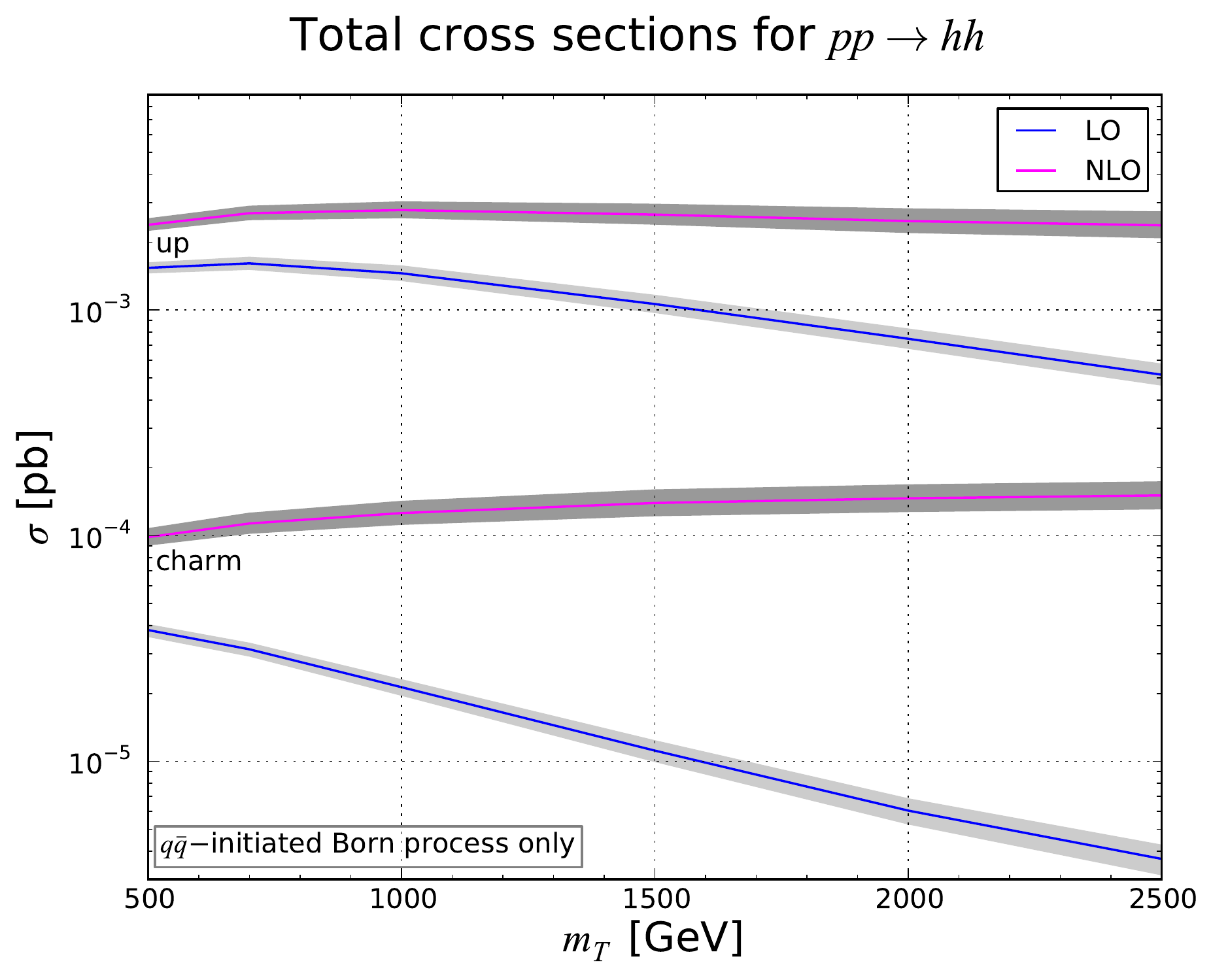}
  \caption{\small LO and NLO QCD inclusive cross sections for Higgs boson
    pair-production in the presence of VLQ and in the context of proton-proton
    collisions at a centre-of-mass energy of 13~TeV. Each curve corresponds to a
    given non-vanishing mixing between the VLQ and the Standard Model quark
    whose flavour is indicated on the figure. The results are presented together
    with the associated theoretical uncertainties obtained from a quadratic sum
    of the scale and PDF uncertainties. We set $\kappa=0.2$. 
  \label{fig:VLQNLO_HH}}
  \end{center}
\end{figure}
The presence of a coupling to light quarks  allows 
for a tree-level $q\bar{q}$-initiated subprocess that exhibits a $t$-channel VLQ
exchange. At the LO, this contribution scales like $\hat{\kappa}^4$ and the
loop-induced SM contribution is usually negligible as the $\hat\kappa$ parameter
can be large. In Figure~\ref{fig:VLQNLO_HH}, we set $\kappa=0.2$ and study the
NLO QCD corrections to this VLQ-induced Higgs pair production subprocess after
removing intermediate resonant diagrams that are accounted for within
the other processes investigated in this contribution. We
observe a huge increase of the cross section when NLO corrections are included,
in particular when the VLQ mass is large. In such parameter space region, the
$K$-factor sometimes even reaches values larger than 50, which is due to several
competing effects. First, the VLQ couplings to the Higgs boson and the relevant
Standard Model quark is proportional to the VLQ mass and is thus much larger
for heavy VLQ. We recall that the cross section depends on the fourth power on
this parameter. Second, a new channel opens at NLO, where the final state is
produced from a gluon and a quark initial state. This component of the NLO cross
section turns to dominate by three orders of magnitude for heavy VLQs due to the
gluon density in the proton. As a result, VLQ-mediated di-Higgs production
total rate is more or less constant with the VLQ mass at the NLO QCD accuracy.

The total cross section turns out to be sizeable in particular for the case
where the VLQ couples to the Higgs and a down-type or up-type quark, which is
due to a PDF enhancement associated with the involvement of valence quarks in
the process. In addition, for all cases, the usage of NLO results for the total
rate is highly recommended, as the LO predictions are incorrect by more than one
order of magnitude in a large fraction of the probed parameter space regions.
Moving on with the study of the associated theoretical uncertainties, we observe
that the latter are similar at the LO and NLO accuracy and range up to about
10\%. This feature is once again related to the $qg$ channel that opens at NLO.

As shown with the diagrams of Figure~\ref{fig:VLQdiag_hh}, we have also
investigated loop-induced gluon fusion diagrams that have no tree-level
counterpart. While
the SM diagrams where only top quarks are running in the loops yield a
negligible contribution compared to the $t$-channel $q\bar q$-initiated
component studied so far, we have calculated the size of the effects related to
the additional
$gg$-initiated loop-diagrams. The results show that these are also
small with respect to the $q\bar q$ subprocess which
contrast with the case where the VLQ are only coupling to the top
quark and where there is no $t$-channel contribution at the tree-level. In this
case, the production of a pair of Higgs bosons indeed proceeds, at the lowest order,
via a loop-induced gluon fusion mechanism which is thus dominant. 
The phase space regions probed within the $gg$- and $q\bar q$-initiated
subprocesses are however different, so
that this channel will be included in the study performed in Section~\ref{ss:DiHiggsIncl}.

\subsection{PRODUCTION OF A PAIR OF HIGGS BOSONS WITH JETS}
\begin{figure}
  \begin{center}
  \includegraphics[width=0.18\textwidth, angle =0 ]{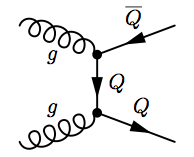}
  \includegraphics[width=0.18\textwidth, angle =0 ]{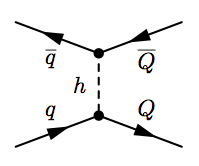}
  \includegraphics[width=0.2\textwidth, angle =0 ]{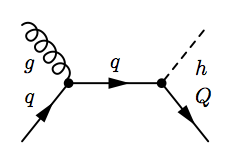}
  \includegraphics[width=0.2\textwidth, angle =0 ]{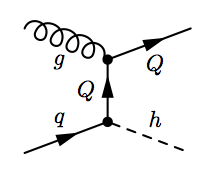}
  \caption{\small Representative Feynman diagram for pair and single VLQ
    production processes that lead to a di-Higgs plus jets final state. These
    extra jets originate from the VLQ decay $Q \to q h$. The adopted benchmark scenario has been
    described in the text. We set $\kappa=0.2$. }
   \label{fig:VLQdiag_hj} 
   \end{center}
\end{figure}

A pair of Higgs bosons could also be produced, in association with a pair of
jets, from the decay of two pair-produced quark partners. In addition to the
pure QCD process studied above, extra contributions could arise due to the
couplings to the Higgs boson, via a $t$-channel Higgs exchange. In particular,
the $qq \to QQ$ process is of interest when the Higgs couplings
involve valence quarks ($u$ or $d$) whose related parton density is larger for
greater $x$-values, which is advantageous for large VLQ masses.
Single VLQ production in association with a Higgs boson can also contribute to
the production of a Higgs pair when the VLQ decays into a Higgs and a jet.
As this channel involves a single heavy state, it is kinematically less
suppressed than the pair production one, and it is thus expected to be dominant
for large mass values. Representative Feynman diagrams are shown in
Figure~\ref{fig:VLQdiag_hj}.

\begin{figure}
  \begin{center}
  \includegraphics[width=0.48\textwidth ]{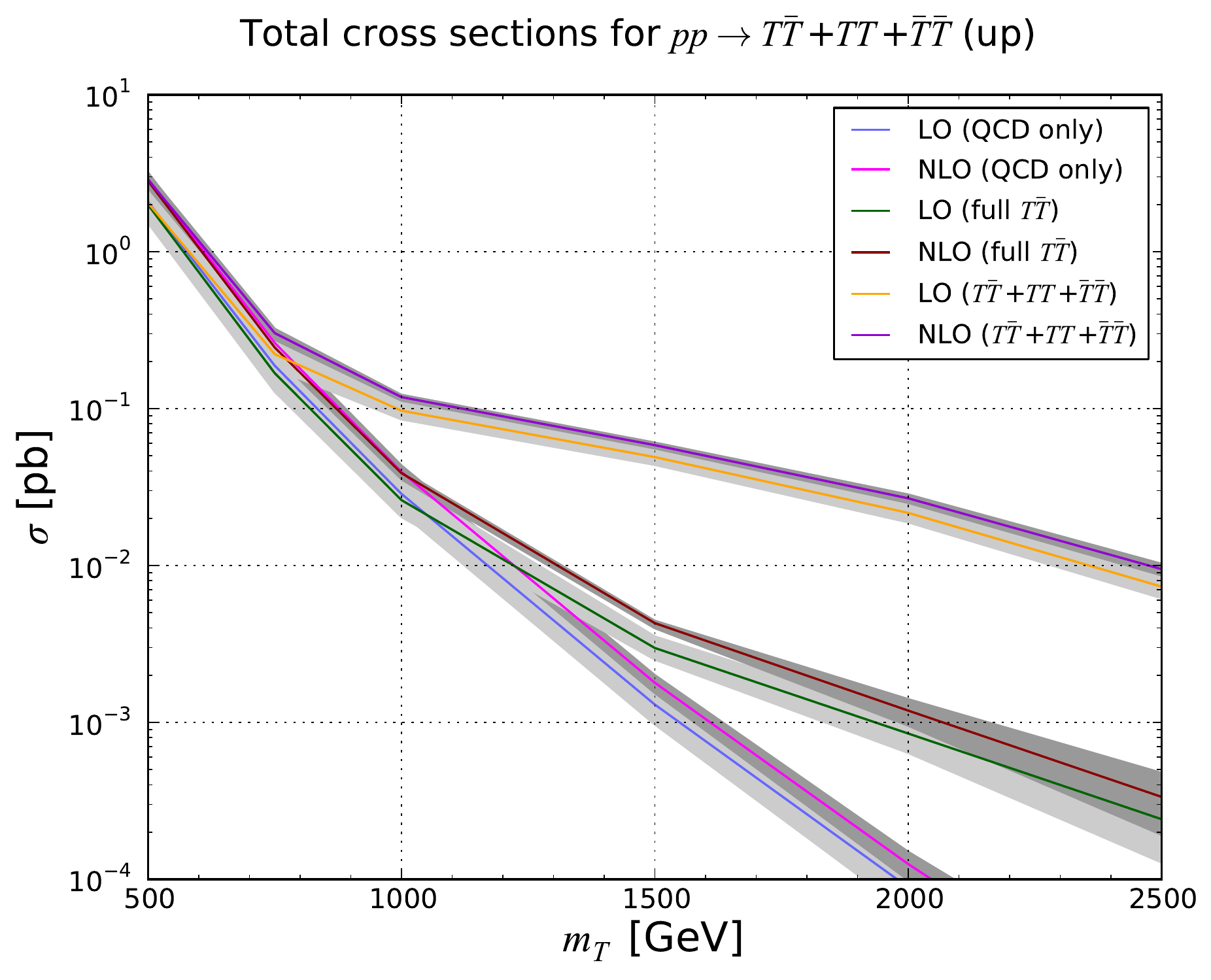}
  \includegraphics[width=0.48\textwidth ]{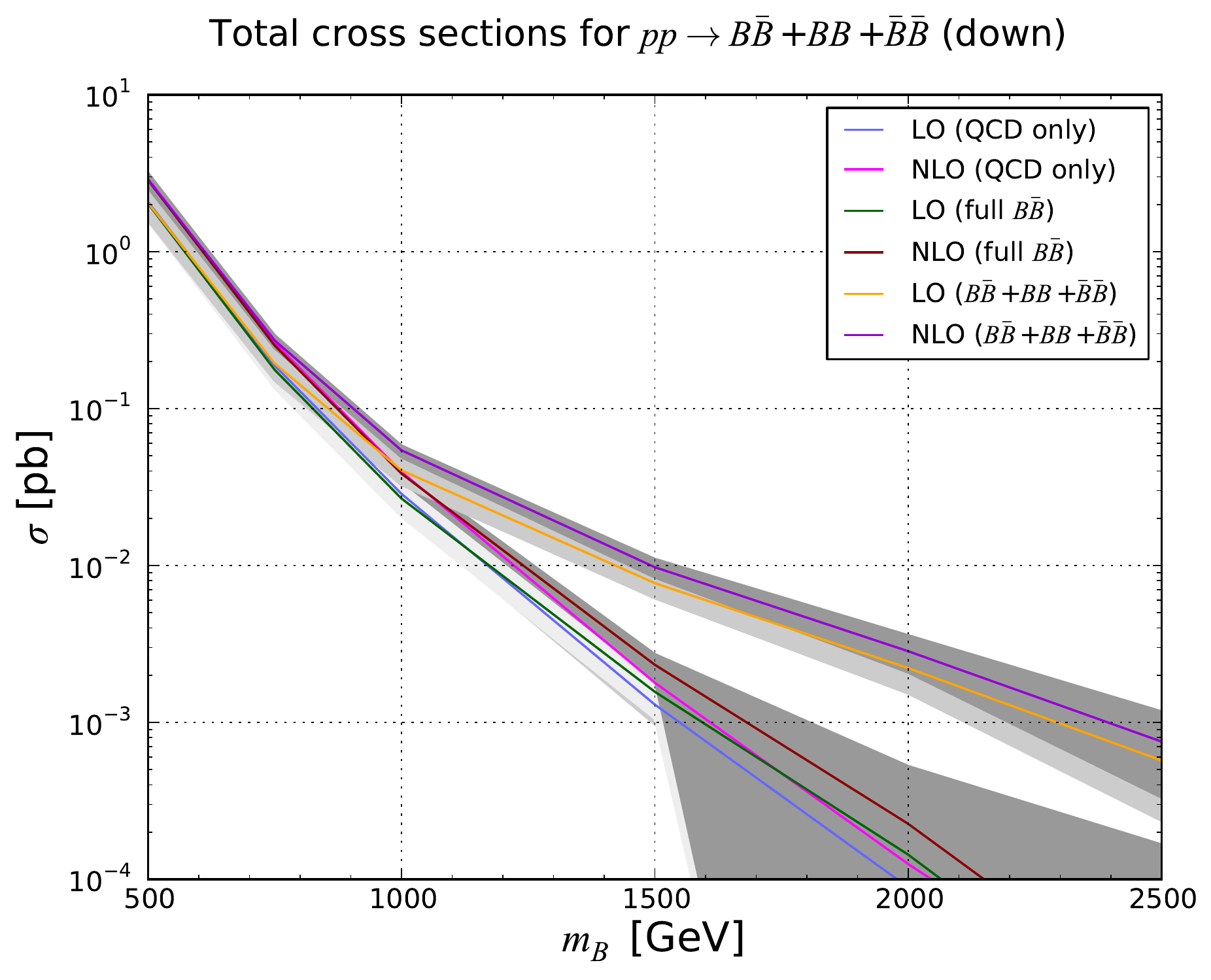}
  \caption{\small LO and NLO QCD inclusive cross sections for VLQ
    pair-production in the context of proton-proton
    collisions at a centre-of-mass energy of 13~TeV. Each curve corresponds to a
    given non-vanishing mixing between the VLQ and the Standard Model quark
    whose flavour is indicated on the figure. The results are presented together
    with the associated theoretical uncertainties obtained from a quadratic sum
    of the scale and PDF uncertainties. The adopted benchmark scenario has been
    described in the text. We set $\kappa=0.2$. 
  \label{fig:VLQLO_hj}}
  \end{center}
\end{figure}

Results  are shown, at the LO and NLO accuracy in QCD, in 
Figure~\ref{fig:VLQLO_hj} for up and down-flavoured partners. In the case of
of strange-, charm- and bottom-flavoured partners, all non-QCD contributions are
negligible as valence quarks are not involved.
For scenarios where the VLQ couples to the valence quarks,
pair-production is dominated by the $t$-channel Higgs exchange diagram, with the
exception of the case of low VLQ masses where the pure QCD contribution dominates.
Sizeable cross sections are also allowed for larger masses,
reaching more than 10~fb for $m_Q$ of about 2~TeV,  for $\kappa = 0.2$. 
Comparing these results with those of the
previous subsection, we observe that the direct di-Higgs channel is always
subdominant even though it is less sensitive to the value of the VLQ mass.
The uncertainties on the predictions are driven by the inaccuracy of the
parton density fits for the large mass case, so that care must be taken with the
interpretation of the predictions. On the other hand, the improvement of the NLO
predictions with respect to the LO ones is very sizeable in the low-mass region
(the 20--30\% of uncertainties are reduced to the 10\% level).

\begin{figure}
  \begin{center}
  \includegraphics[width=0.48\textwidth]{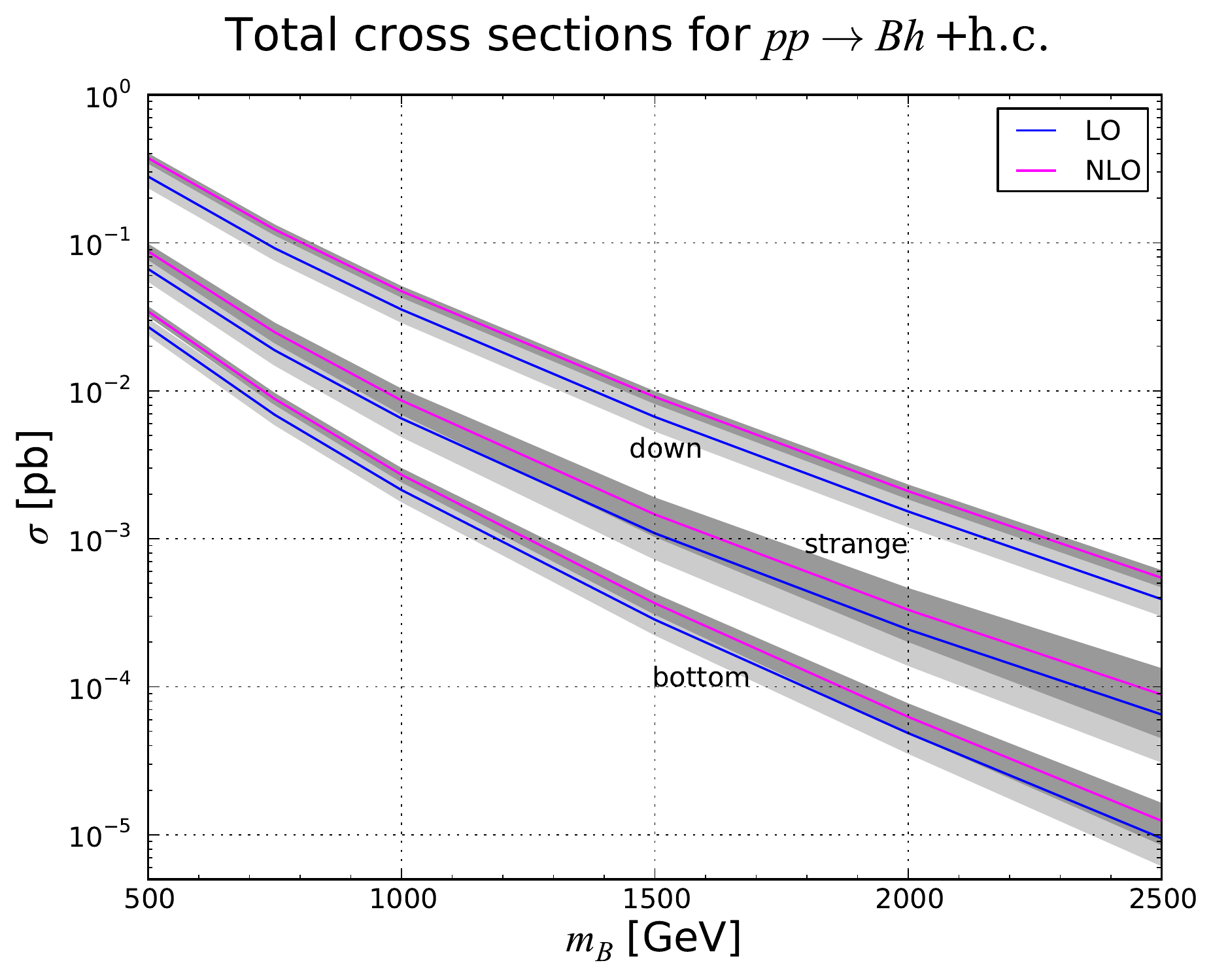}
  \includegraphics[width=0.48\textwidth]{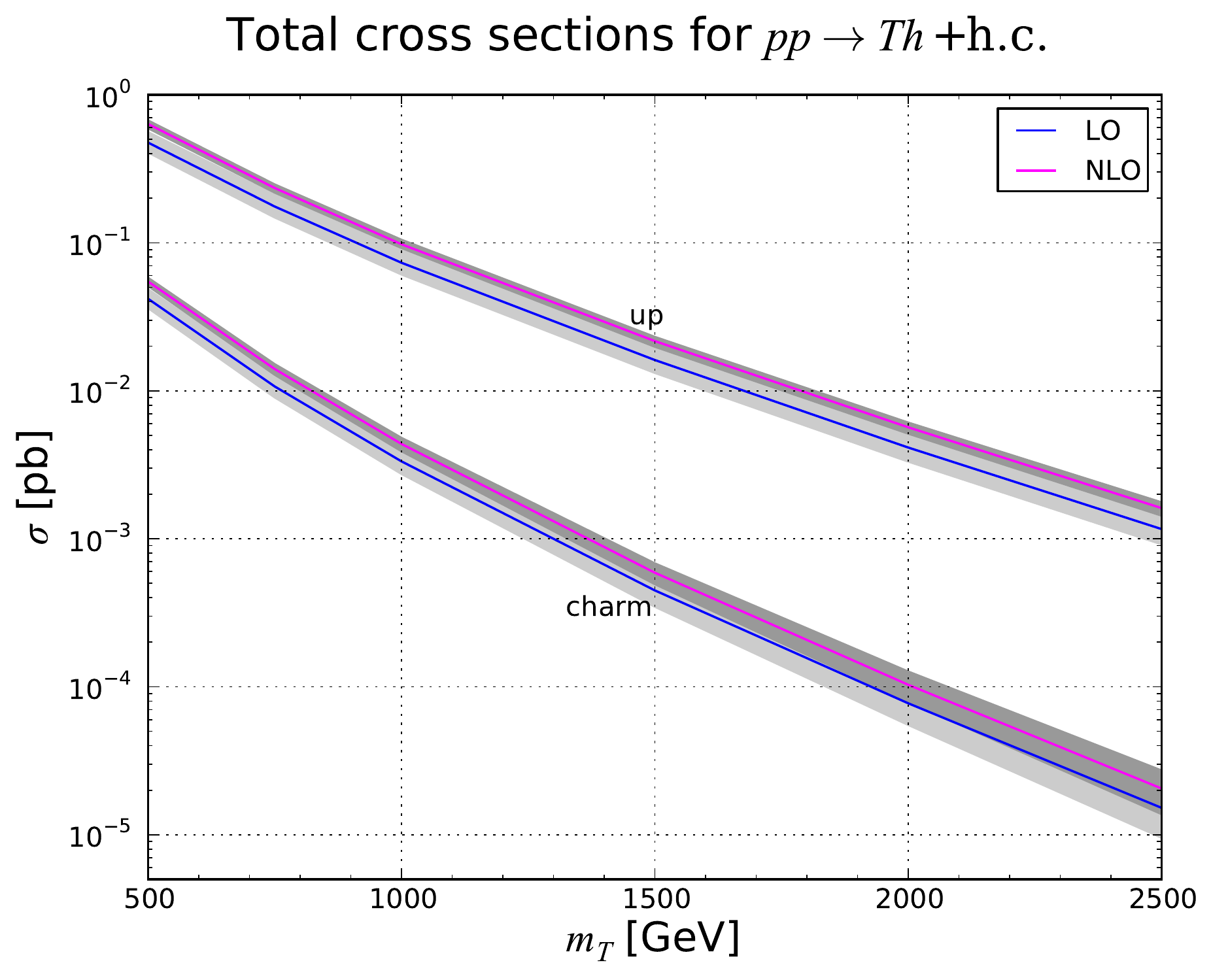}
  \caption{\small LO and NLO QCD inclusive cross sections for the associated production
  of a Higgs boson and a VLQ in the context of proton-proton
    collisions at a centre-of-mass energy of 13~TeV. Each curve corresponds to a
    given non-vanishing mixing between the VLQ and the Standard Model quark
    whose flavour is indicated on the figure. The results are presented together
    with the associated theoretical uncertainties obtained from a quadratic sum
    of the scale and PDF uncertainties. The adopted benchmark scenario has been
    described in the text. We set $\kappa=0.2$.
\label{fig:VLQNLO_QH}}
\end{center}
\end{figure}

Finally, we study in Figure~\ref{fig:VLQNLO_QH} the associated production of a
VLQ and a Higgs boson. In order to be able to coherently split the different
VLQ processes studied in this contribution, we have removed all resonant
diagrams involving the on-shell production of an intermediate new physics
state that are already accounted for in the other processes. Moreover, their
correct treatment requires, \textit{e.g.}, either to use the complex mass
scheme or an appropriate subtraction scheme. We have obtained $K$-factor of
about 1.3 for all investigated cases, with a reduction of the theoretical
uncertainties from 10-20\% to 5-10\% in the case of VLQ lighter than 1~TeV. In
the case where the VLQ are heavier, the uncertainties are controlled by the
quality of the PDF fit, so that the improvement is milder and the total
uncertainties are reduced from 20-35\% to 15-30\%.

\section{COLOURED SCALAR MODEL}
\label{sec:scalar}

\begin{figure*}[htb]\begin{center}
\includegraphics[width=0.9\textwidth, angle =0 ]{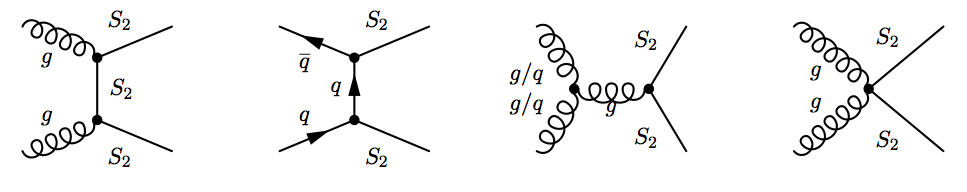}
\caption{\small Pair production of coloured scalars, presumed to cascade decay into two Higgs  plus four jets. The second diagram involves direct couplings to quarks, that may be small.}  \label{fig:scalardiagrams} 
\end{center}\end{figure*}

As a second example for exotic production of Higgs boson pairs, we propose a class of models where the Higgses arise via the decays of coloured scalars, which are produced via QCD interactions. In order for the scalars to finally decay into two jets~\cite{Han:2010rf}, we are limited to 3 possible representations under SU(3)$_c$: triplet~\cite{Han:2009ya}, sextet~\cite{Chen:2008hh,Berger:2010fy} or octet~\cite{Chen:2014haa}. The first two will couple to two quarks (via an anti-symmetric and symmetric tensor respectively), while the octet couples to a quark-antiquark pair. Examples of such states can be found in supersymmetry, where squarks are triplets, and composite models~\cite{Gripaios:2009dq,Cacciapaglia:2015eqa}.
For simplicity, we will focus on the simplest case where the scalars are singlets under the weak isospin, and carry the appropriate hypercharge to allow for couplings to quarks.
The couplings to the Higgs boson arise, in a representation-independent way, via the following potential:
\beq
V_{\rm scalar} = \lambda_1 \Phi^\dag \Phi \tilde{S}_1^\dag \tilde{S}_1  + \lambda_2 \Phi^\dag \Phi  \tilde{S}_2^\dag \tilde{S}_2  +  \Phi^\dag \Phi ( \lambda_3 \tilde{S}_1^\dag \tilde{S}_2 +\lambda_3^\ast \tilde{S}_2^\dag \tilde{S}_1)\,,
\eeq
where we call $\tilde{S}_{1,2}$ the two scalars and $\Phi$ is the Brout-Englert-Higgs iso-doublet.
Once the two states are rotated in the mass basis, due to the presence of gauge-invariant masses not induced by the above potential, there will be a coupling of a single Higgs to the two different scalars, so that the heavier can always decay into the light one plus a Higgs. Therefore, following QCD-pair production and decays, we expect events with two or one Higgs produced, as shown in Figure~\ref{fig:scalardiagrams}.

In the following, we will focus on the specific case of an anti-triplet which couples to two quarks (either up or down-type):
\beq
\mathcal{L}_{\rm NP} &=& (D_\mu \tilde{S}_1)^\dag D_\mu \tilde{S}_1 + (D_\mu \tilde{S}_2)^\dag D_\mu \tilde{S}_2 - m_1^2  \tilde{S}_1^\dag \tilde{S}_1  - m_2^2  \tilde{S}_2^\dag \tilde{S}_2  \nonumber \\  &+& \epsilon^{mnk} \tilde{S}_{1,k} {\bar q_{m, i}}( c_{L, ij}  P_R + c_{R, ij} P_L)q_{n, j}^C + h.c.  \nonumber \\  &+& \epsilon^{mnk} \tilde{S}_{2, k} {\bar q_{m, i}}( \tilde{c}_{L, ij}  P_R + \tilde{c}_{R, ij} P_L)q_{n, j}^C + h.c.\,,
\eeq
where the structure of the couplings to quarks depends uniquely on the charge of the scalars: for $Q_S = -4/3$ ($2/3$) the coupling only involves right-handed up-(down-)type quarks, while for $Q_S = 1/3$ the scalars couple to both left and right handed quarks (one up and one down) due to the conservation of $SU(2)_L \times U(1)_Y$ gauge symmetry.  Couplings with different chiralities can only be generated via electroweak breaking effects. Furthermore, the signature we are interested in does not depend on the couplings to quarks nor on the charge of the scalars.
The mixing induced by the electroweak symmetry breaking for the two coloured scalars  can be diagonalised by a single angle (assuming $\lambda_3$ real)
\beq
\left(
\begin{array}{c}
S_1  \\
 S_2 
\end{array}
\right) = \left(
\begin{array}{cc}
\cos \theta & \sin \theta \\
 - \sin \theta & \cos \theta
\end{array} \right)   \left(
\begin{array}{c}
\tilde{S}_1  \\
 \tilde{S}_2 
\end{array}
\right)\,, \quad \sin 2 \theta = \frac{\lambda_3 v^2}{m_{S_2}^2 - m_{S_1}^2}\,,
\eeq 
where we follow the convention that $m_{S_2} > m_{S_1}$.

Due to the presence of gauge invariant masses for the scalars, the couplings of the Higgs to two coloured scalars will not be aligned with the mass basis in general, i.e. for $m_1 \neq m_2$, thus allowing for the Higgs-producing chain decays.

The parameter space is constrained by other measurements and direct searches at the LHC, which can however be easily escaped:
(a) Loop contributions to gluon fusion production of a single Higgs are only sensitive to the diagonal couplings of the Higgs in the mass basis, furthermore the effect can be small either by choosing small $\lambda$'s or largish masses.
(b) Constraints from di-jet resonance searches~\cite{Han:2009ya} are sensitive to the nature of the couplings to quarks and to their values, which can be chosen small enough to escape all constraints.
(c) Direct search for pairs of di-jet resonances at Run--I~\cite{Khachatryan:2014lpa}: this search applies directly to the mass of the lightest scalar, which is QCD produced and can only decay into a pair of jets (as long as the decay is prompt). The analysis of a stop can be directly reused in our model, and gives a bound $m_{S_1} \geq 350$ GeV.

Following the above mentioned constraints, the parameter space of the model has a viable corner where
\beq
\mbox{BR} (S_2 \to S_1 h) \simeq 100\%\,, \quad \mbox{BR} (S_1 \to j j) \sim 67 \div 100\%\,,
\eeq
which can be achieved for $c_{L, ij} , c_{R, ij}, \tilde{c}_{L, ij} , \tilde{c}_{R, ij} \ll \lambda_i$. Note that $\mbox{BR} (S_1 \to jj) \sim 100\%$ can be obtained assuming small coupling of the scalars to the third generation, while having flavour-independent couplings, so as to avoid flavour bounds, would lead to $\mbox{BR} (S_1 \to jj) \geq 2/3$. The main visible signal will be the pair production of coloured scalars, and cascade decay into two Higgses in association with four jets, as shown in Figure~\ref{fig:scalardiagrams}.
The rates, and kinematic features will depend only on the masses of the two scalars. We leave a more thorough exploration of the parameter space for a forthcoming publication.

\section{DISTRIBUTIONS FROM RESONANT PRODUCTION}
\label{sec:distros}

In Sections~\ref{sec:model} and \ref{sec:scalar}, we presented two models where a pair of Higgs bosons can be produced via decays of heavier coloured spin-1/2 or spin-0 states. In this section we probe the di-Higgs signature of these two models at the LHC Run--II with a centre--of--mass energy of 13~TeV.
Both models predict cross sections in excess of the SM one. Moreover, the event kinematic properties are expected to be very different from the SM processes and from production in the EFT approach. In this section we study the feasibility of the search for new physics via the resonant di-Higgs production channels under these two model assumptions -- vector-like quarks and coloured scalars. Preliminary results are obtained via parton-level simulations done with {\sc MadGraph5\_aMC@NLO}, using model files generated with {\sc FeynRules}~\cite{Alloul:2013bka,Christensen:2008py} (in particular for the VLQ mode we use the implementation described in section~\ref{sec:model}). In both cases we show results at LO in QCD, whereas NLO results are in preparation.

Section~\ref{ss:DiHiggsIncl} discusses the inclusive $hh$ production with VLQ mediation in the loops. The exotic production of $hh$ from direct $QQ$ and $Qh$ production, with subsequent $Q\to hq$ decays are discussed in Sec.~\ref{ss:DiHiggsVLQDecays}. In Sec.~\ref{ss:coloured_scalars} the coloured scalar model is discussed.

\begin{figure*}[!htb]\begin{center}
\begin{tabular}{cccc}
\includegraphics[width=0.32\textwidth, angle =0 ]{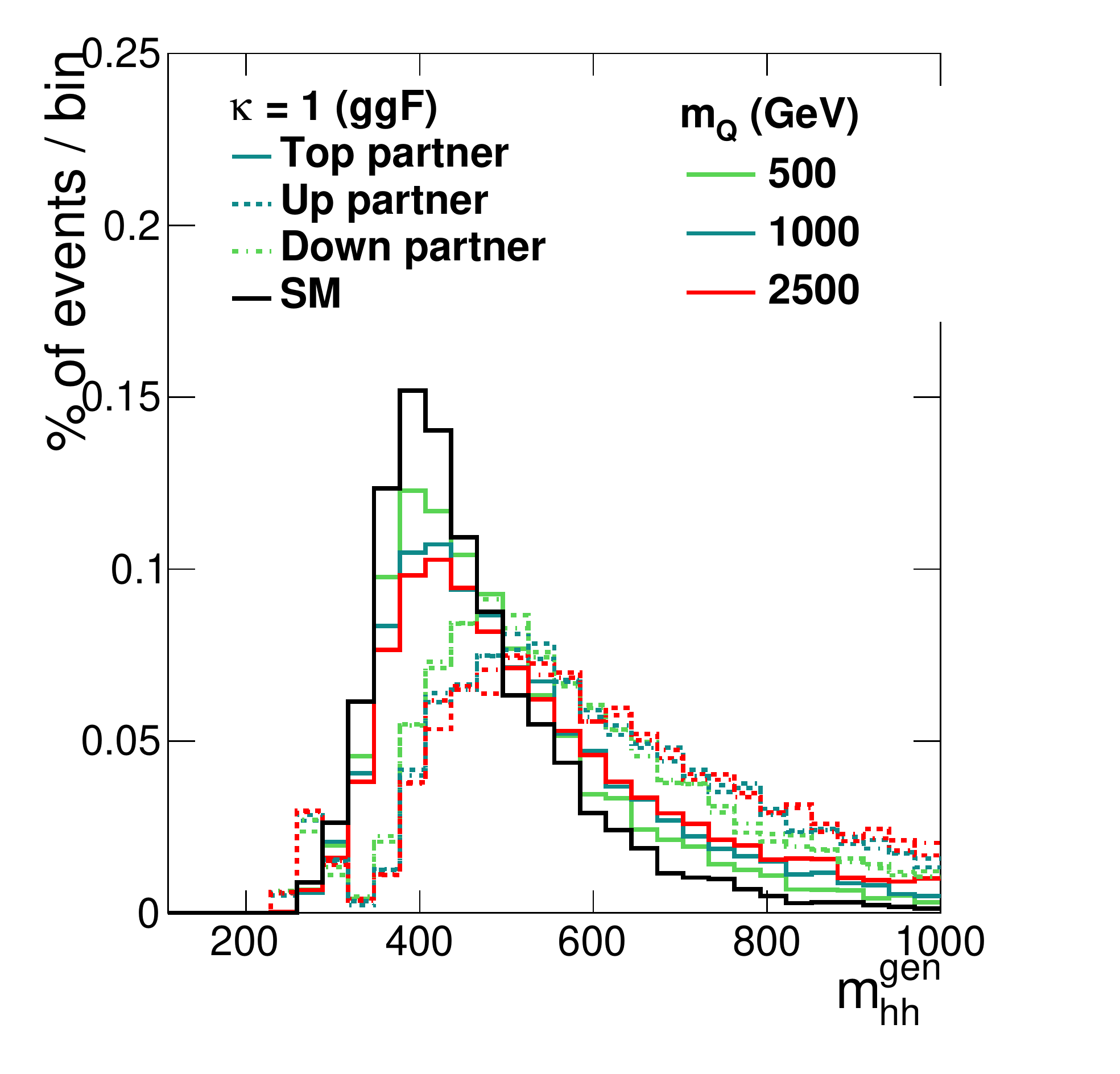} & 
\includegraphics[width=0.32\textwidth, angle =0 ]{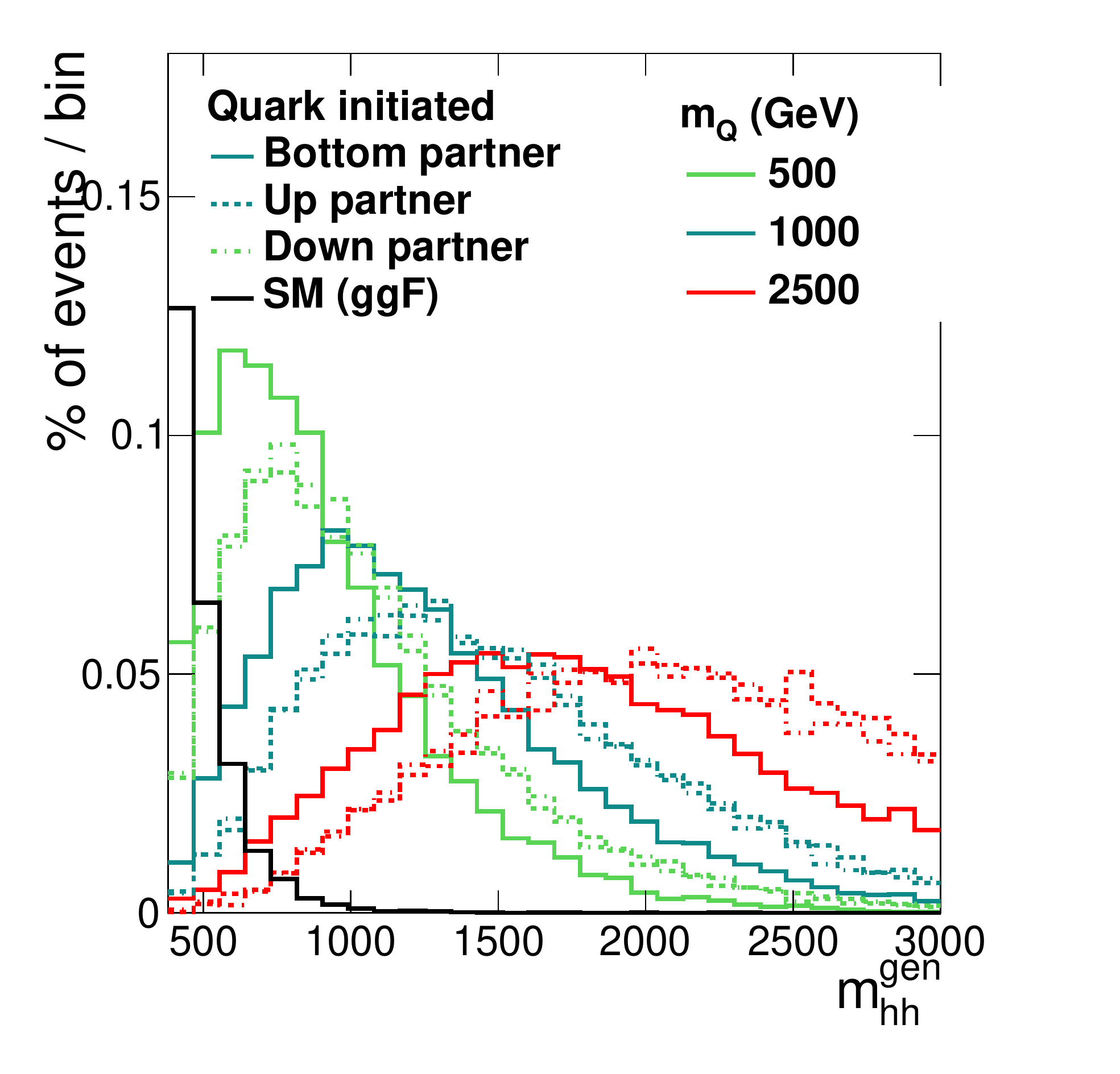} \\
(a) & (b) \\ 
\includegraphics[width=0.32\textwidth, angle =0 ]{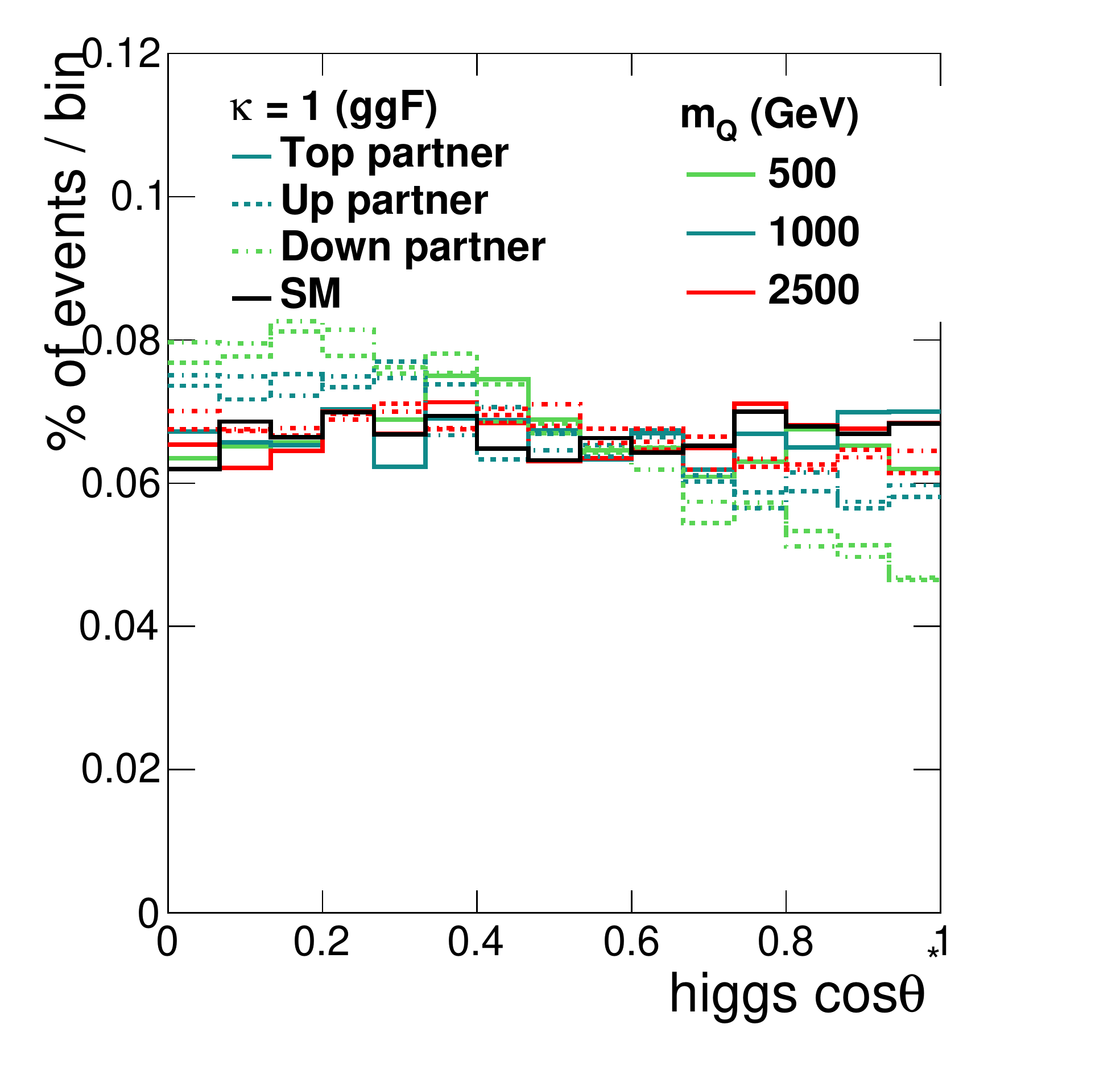} & 
\includegraphics[width=0.32\textwidth, angle =0 ]{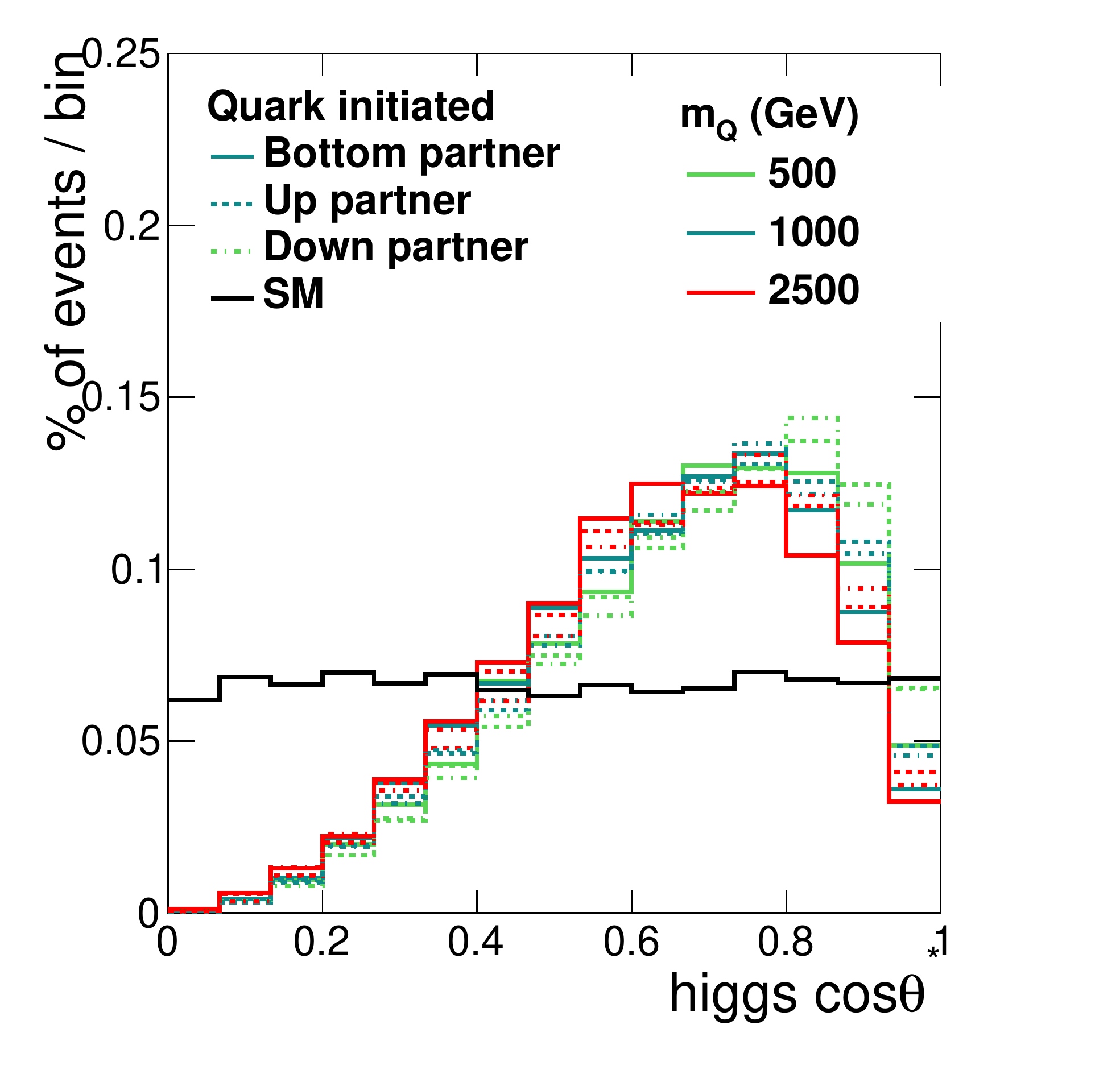} \\
(c) & (d) \\ 
\end{tabular}
\caption{The invariant mass $m_{hh}$ for the ggF process (a) and the quark-induced process (b), together with the $cos\theta^{*}$ for ggF (c) and the quark-induced process (d). All distributions are normalised to unity. The only exception is figure (b), where the SM distribution has been rescaled for visualisation purpose.  We fix $\kappa=0.2$. Note that the value of $\kappa$ does not affect the distributions for the quark-induced process at LO due to the absence of SM diagrams.}
\label{fig:gen_hh_GF_DY_mhh_DY_deta} 
\end{center}
\end{figure*}

\subsection{VLQ-INDUCED DI-HIGGS PRODUCTION\label{ss:DiHiggsIncl}}

This section presents the kinematic distribution of di-Higgs final states originating from the production induced by vector-like quarks (see Fig~\ref{fig:VLQdiag_hh}). Events were generated separately for the gluon-gluon fusion (ggF) and the quark induced processes, and for different VLQ mass values:  500, 800, 1000, 1500, 2000, 2500, and 3000~GeV. Figure~\ref{fig:gen_hh_GF_DY_mhh_DY_deta} (a) shows the invariant mass of the di-Higgs system $m_{hh}$ for the VLQ-mediated production of di-Higgs from gluon-gluon fusion, while Figure~\ref{fig:gen_hh_GF_DY_mhh_DY_deta} (b) shows $m_{hh}$ for the quark-induced di-Higgs production. The difference in signal shape with respect to the SM process, particularly in the high $m_{hh}$ is clearly seen: in particular, the quark initiated channel is peaked to much higher invariant masses.
The angle of the emitted $h$ w.r.t the beam direction $cos\theta^{*}$ in the reference frame of the $hh$ system  is also shown for the ggF, in Fig.~\ref{fig:gen_hh_GF_DY_mhh_DY_deta} (c), and quark-induced process, in Fig.~\ref{fig:gen_hh_GF_DY_mhh_DY_deta} (d). 
The SM process can be seen to have a flat $cos\theta^{*}$ distribution. The ggF process also produces a flat distribution. However the quark-induced process tend to show a more pronounced anisotropy, probable related with the $q\bar{q}$ PDF asymmetry.


\begin{figure*}[!htb]\begin{center}
\begin{tabular}{ccc}
\includegraphics[width=0.32\textwidth, angle =0 ]{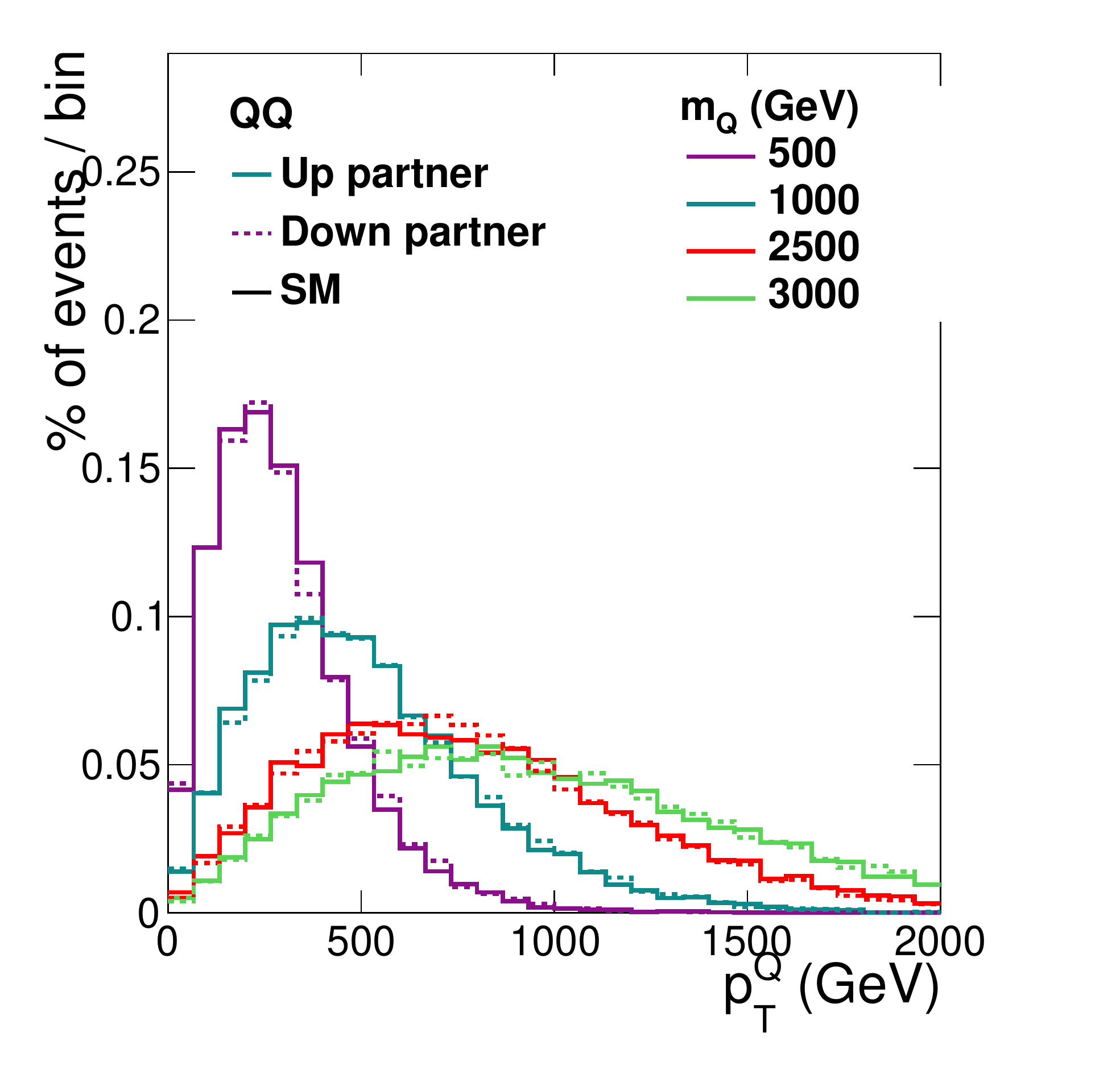} &
\includegraphics[width=0.32\textwidth, angle =0 ]{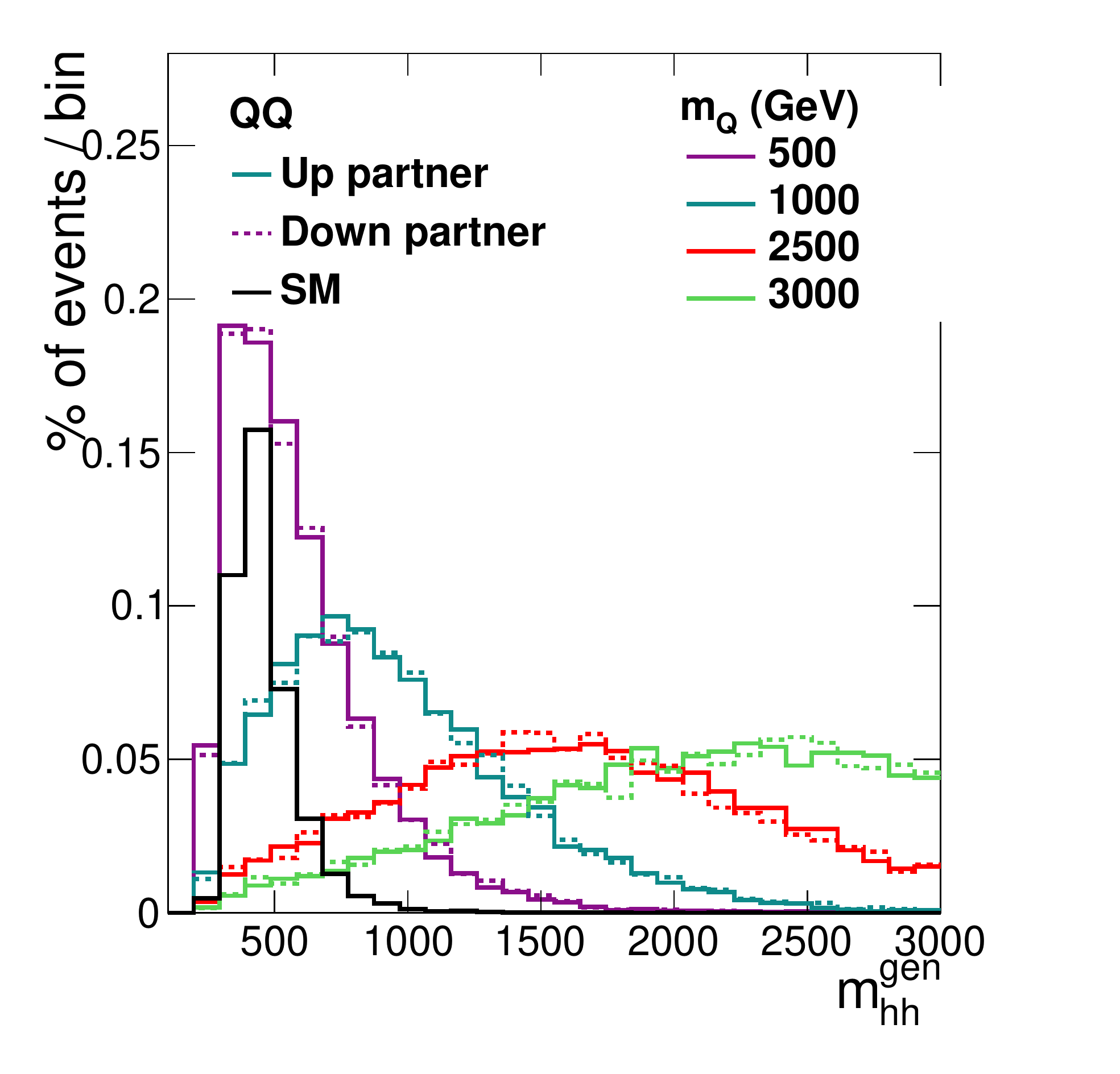} &
\includegraphics[width=0.32\textwidth, angle =0 ]{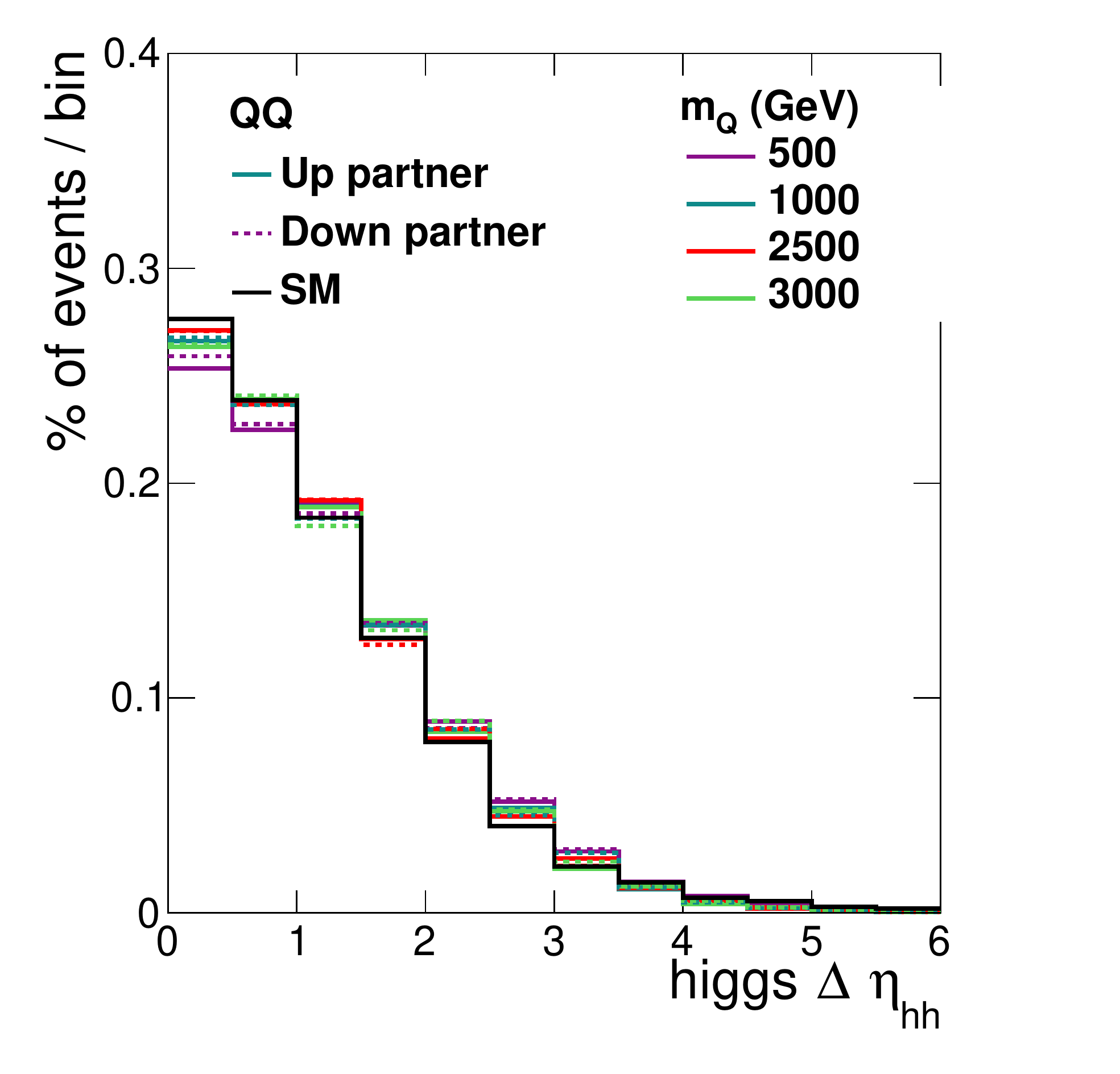} \\
(a) & (b) & (c)\\ 
\includegraphics[width=0.32\textwidth, angle =0 ]{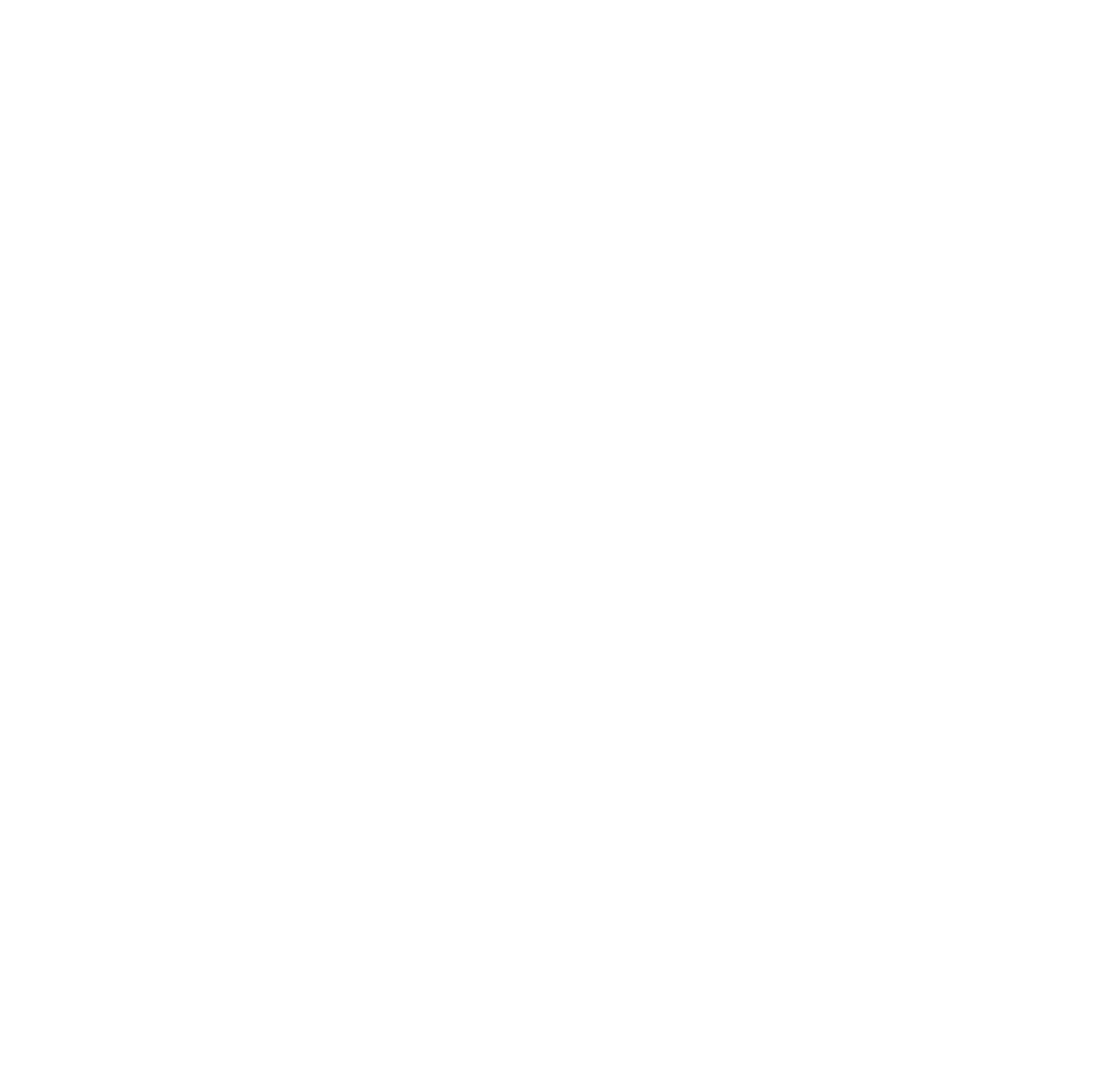} &
\includegraphics[width=0.32\textwidth, angle =0 ]{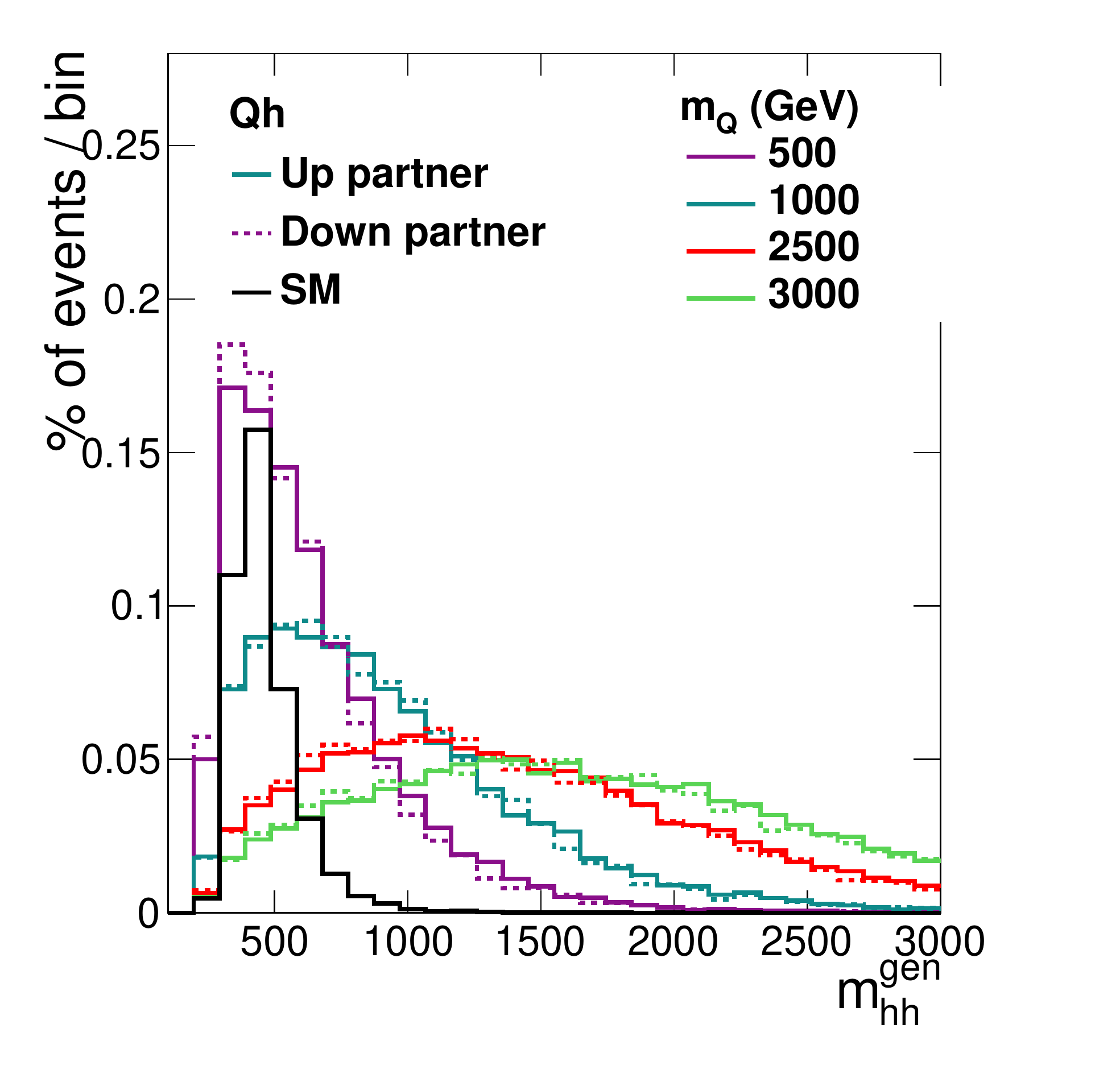} & 
\includegraphics[width=0.32\textwidth, angle =0 ]{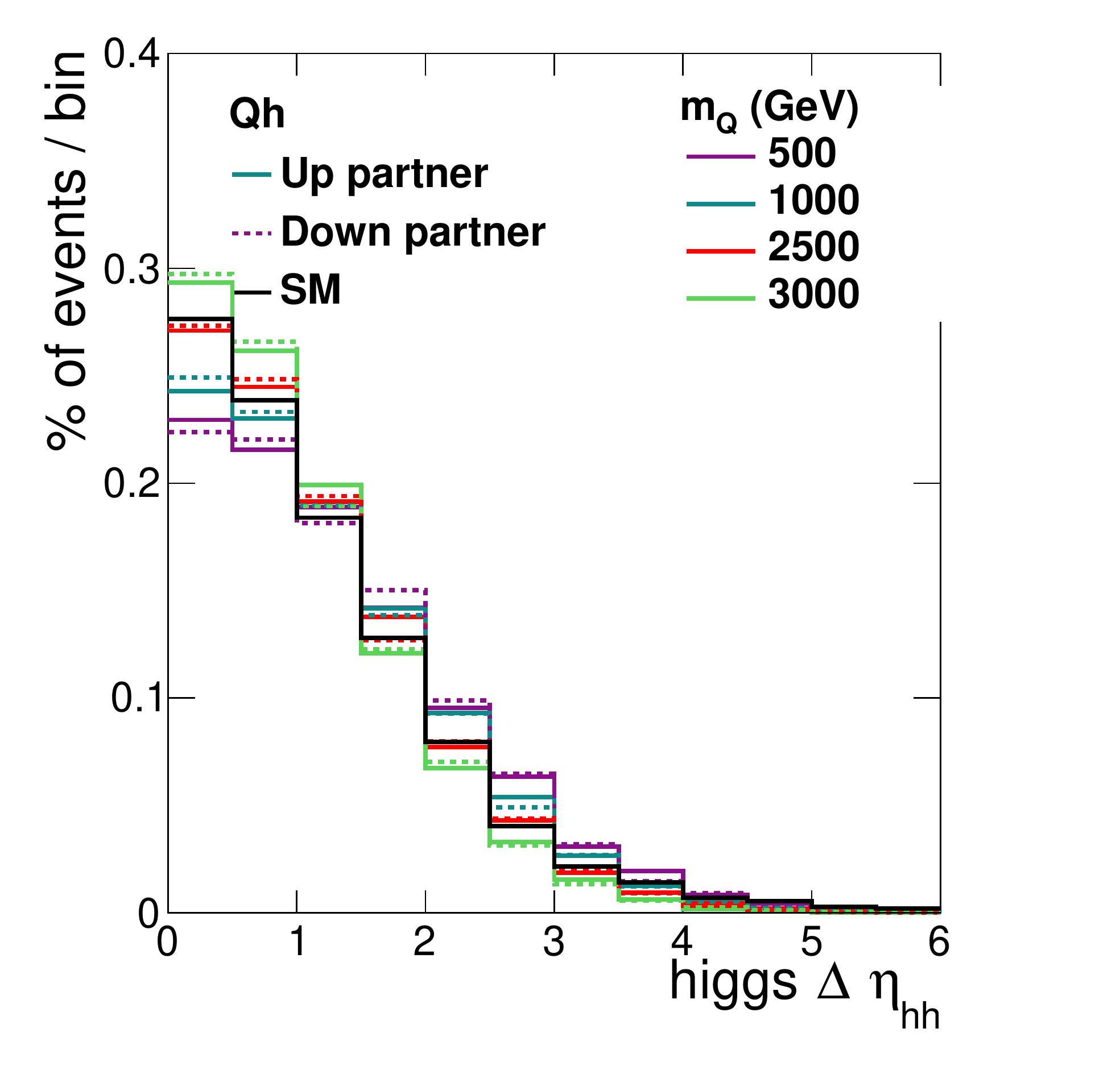} \\
(a') & (b') & (c')\\ 
\end{tabular}
\caption{The $p_{\rm T}$ of the $Q$ quark (a-a'), the invariant mass $m_{hh}$ (b-b'), and the separation $\Delta \eta_{hh}$ (c-c') between the two Higgs bosons in the event for the processes $pp\to QQ\to 2h\,jj$ (top row) and $pp\to Qh\to 2h\,j$ (bottom row). All distributions, for various VLQ masses, are normalised to unity. We fix $\kappa = 0.2$. Note that the value of $\kappa$ does not affect the distributions for the single VLQ process at LO due to the absence of QCD contributions.}
\label{fig:gen_Qh_tppt_njets_mhh_deta} 
\end{center}\end{figure*}

\subsection{PRODUCTION OF A PAIR OF HIGGS BOSONS WITH JETS \label{ss:DiHiggsVLQDecays}}

\begin{figure*}[!htb]\begin{center}
\begin{tabular}{ccc}
\includegraphics[width=0.32\textwidth, angle =0 ]{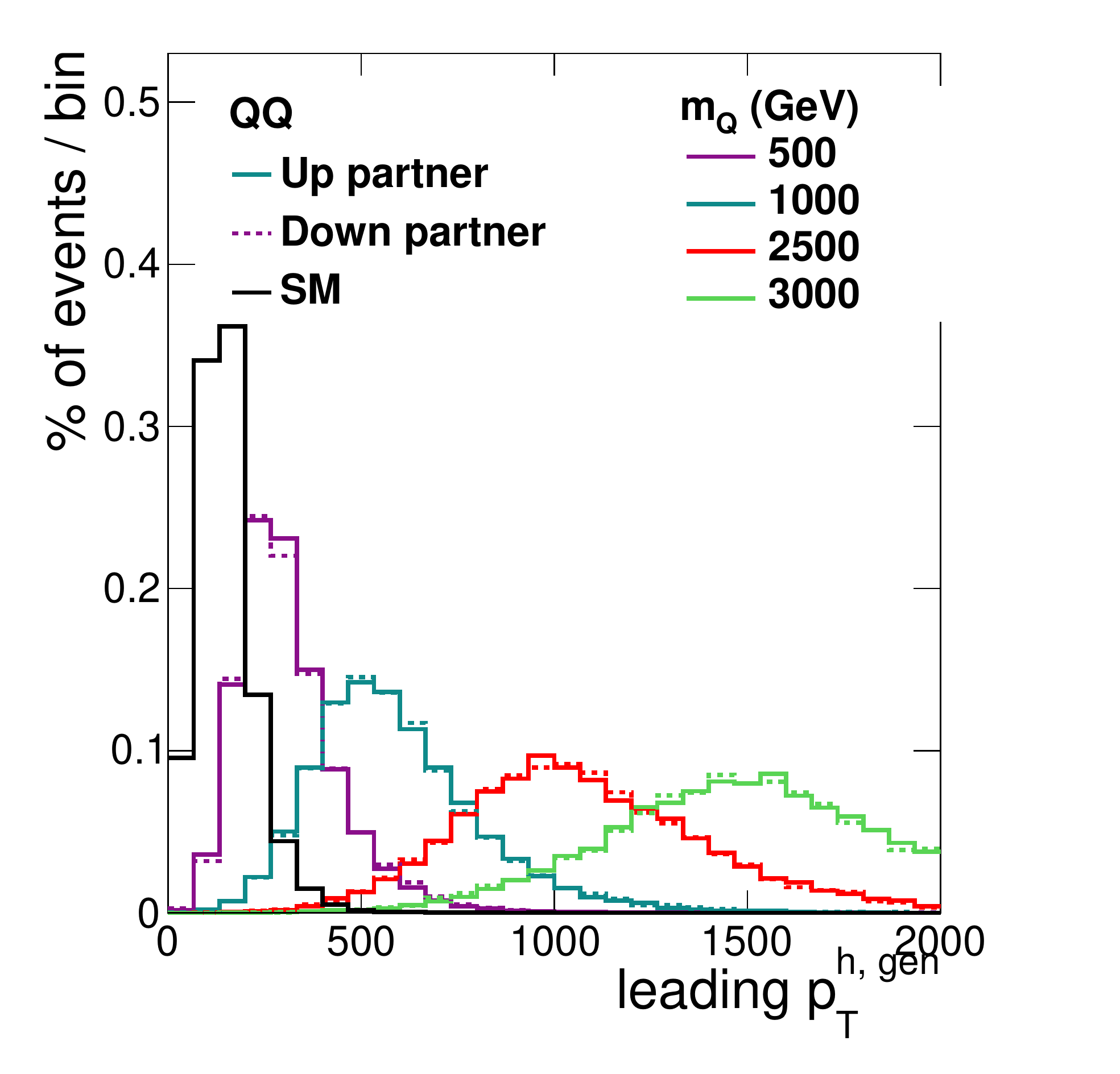} &
\includegraphics[width=0.32\textwidth, angle =0 ]{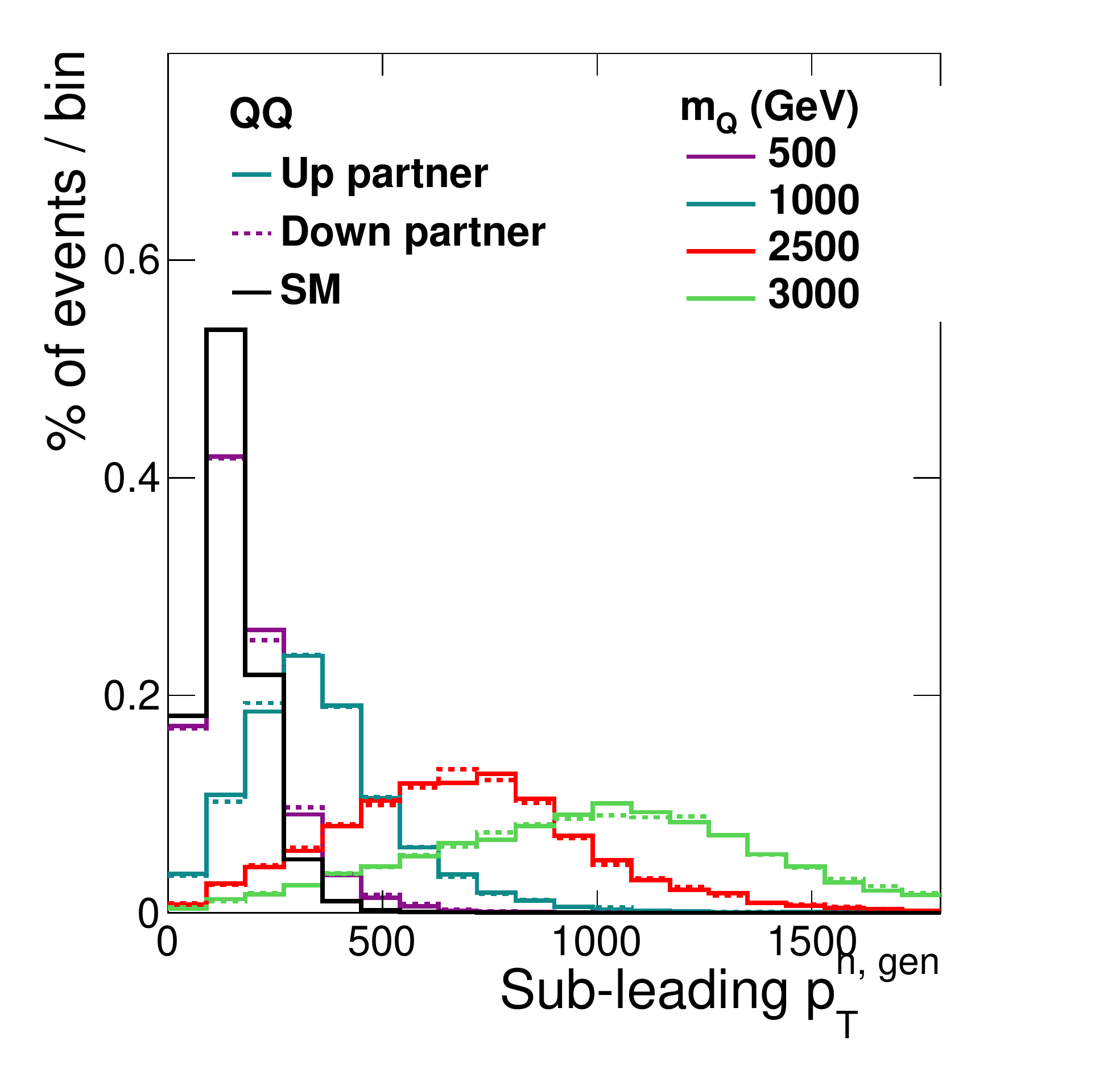} &
\includegraphics[width=0.32\textwidth, angle =0 ]{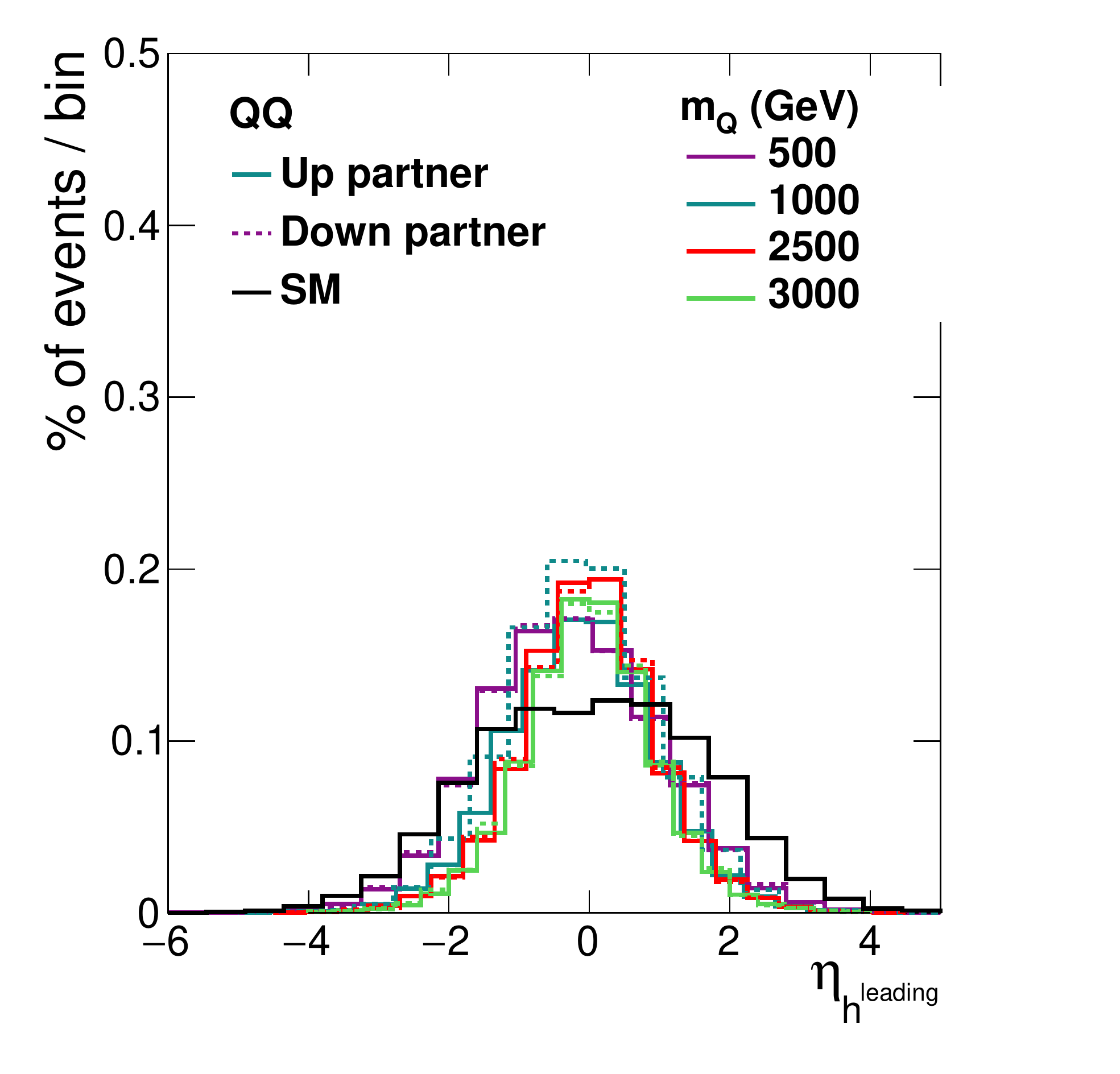} \\
(a) & (b) & (c) \\ 
\includegraphics[width=0.32\textwidth, angle =0 ]{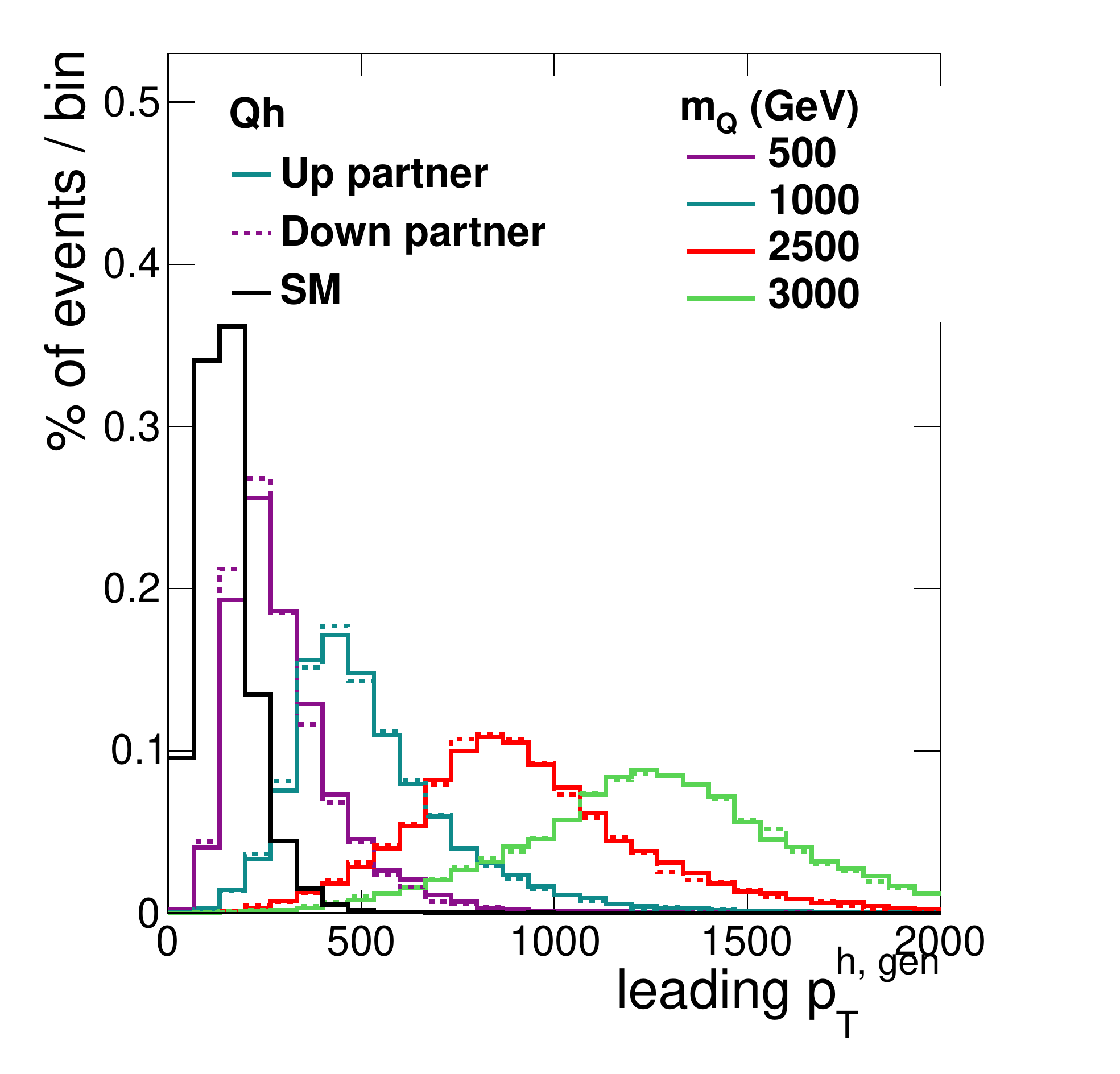} &
\includegraphics[width=0.32\textwidth, angle =0 ]{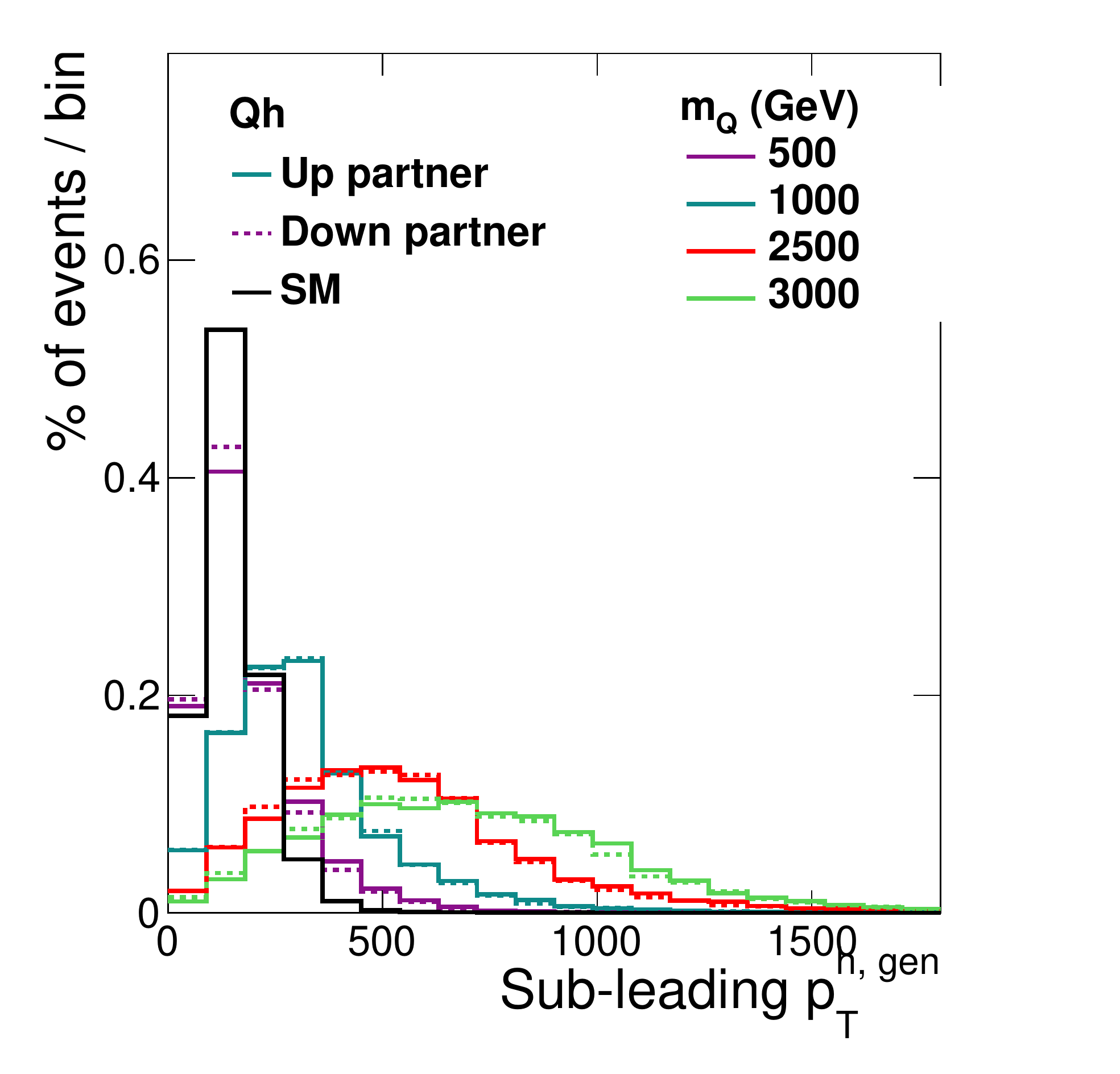} & 
\includegraphics[width=0.32\textwidth, angle =0 ]{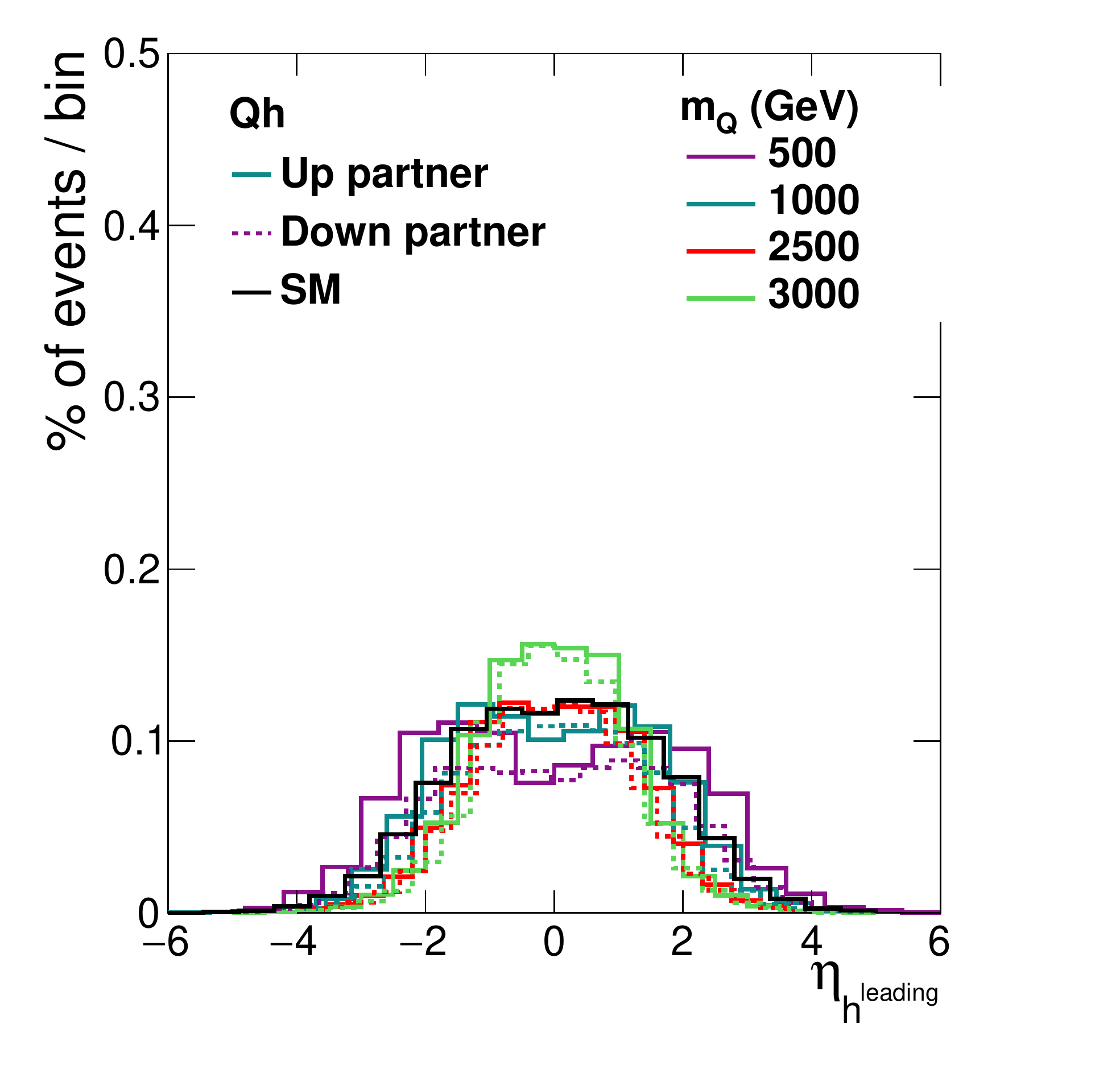} \\
(a') & (b') & (c') \\ 
\end{tabular}
\caption{The $p_{\rm T}$ distribution for the leading (a-a') and sub-leading (b-b') $p_{\rm T}$ Higgs boson, and pseudo-rapidity $\eta_{h}$ for both the Higgs bosons (c-c') in the processes $pp\to QQ\to 2h\,jj$ (top row) and $pp\to Qh\to 2h\,j$ (bottom row). All distributions are normalised to unity.
We fix $\kappa = 0.2$. Note that the value of $\kappa$ does not affect the distributions for the single VLQ process at LO due to the absence of QCD contributions.}
\label{fig:gen_Qh_h_pt_eta} 
\end{center}\end{figure*}

We generate events for the processes of pair VLQ production $pp\to QQ\to 2h\,2j$ and single VLQ production $pp\to Qh\to 2h\,j$, where $Q$ could be vector-like quarks of electric charges either $+2/3$ ($T$) or $-1/3$ ($B$).  The vector-like quark $Q$ is assumed to couple only to the first and second generation SM quarks. The coupling parameter $\kappa=0.2$ is chosen for both production processes. No additional partons were included in the generation of the hard process, and we used the \textsc{NN23LO1}~\cite{Ball:2013hta} parton distribution functions. Events are produced for VLQ masses of 500, 800, 1000, 1500, 2000, 2500, and 3000~GeV. 

The kinematic distributions for the main event are shown in 
Fig.~\ref{fig:gen_Qh_tppt_njets_mhh_deta} for VLQ pair (top row) and single (bottom row) production: panels (a-a') show the $p_{\rm T}$ of the VLQs, panels (b-b') the invariant mass $m_{hh}$ of the $hh$ system, and panel (c-c') the separation in pseudo-rapidity between the two Higgs bosons $\Delta \eta_{hh}$. 
The plots show that, as expected, the VLQs are produced with a modest transverse momentum, peaking at $\sim 500$ GeV for large masses. The two Higgs bosons tend to be back-to-back in both production modes. The main difference can be seen in the invariant mass of the di-Higgs system, panels (b) and (b'), which is more pronounced and peaked to higher values for pair production, where both Higgses come from the decay of a heavy VLQ, with respect to the single production.

\begin{figure*}[!htb]\begin{center}
\begin{tabular}{ccc}
\includegraphics[width=0.32\textwidth, angle =0 ]{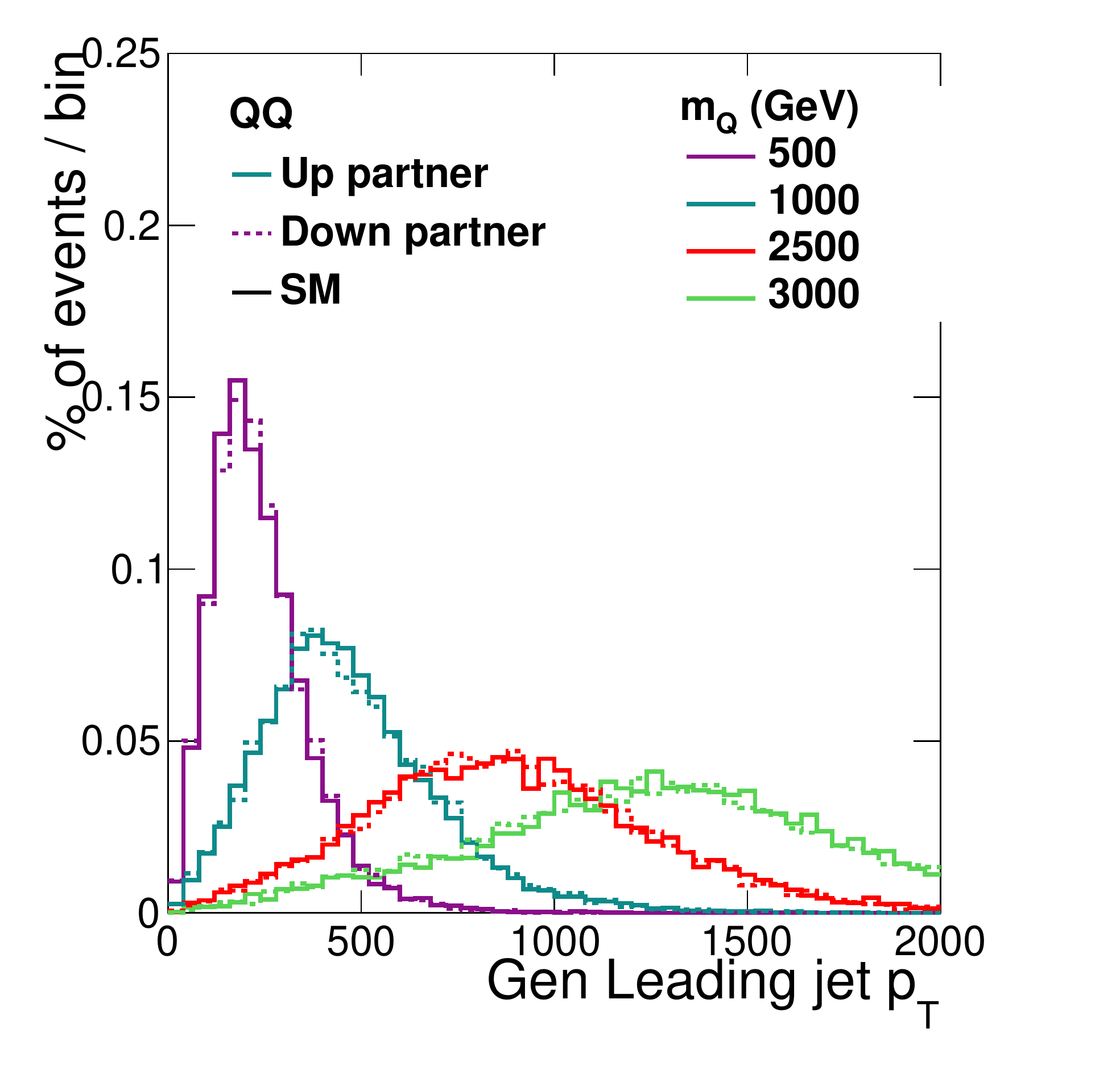} &
\includegraphics[width=0.32\textwidth, angle =0 ]{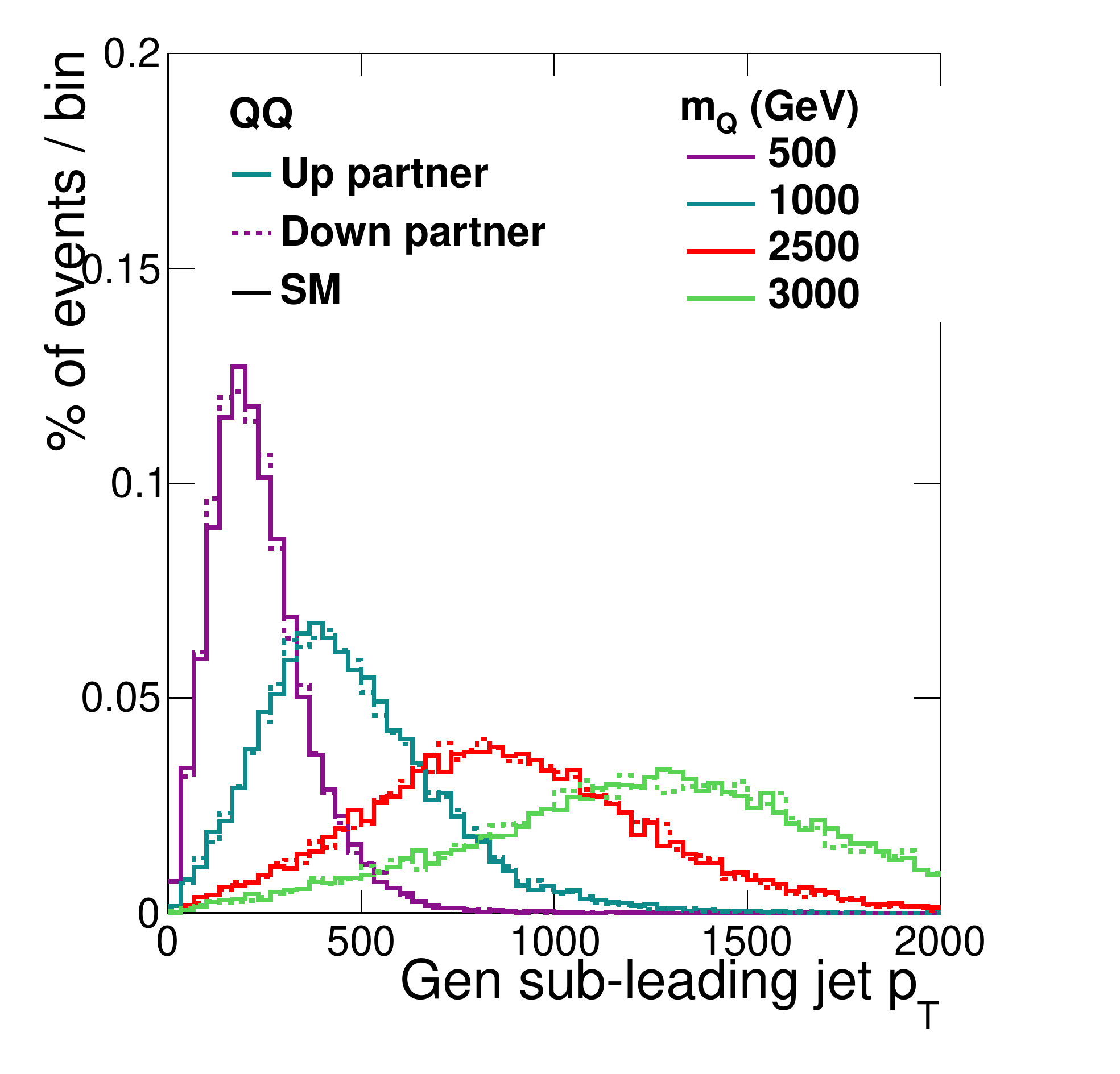} & 
\includegraphics[width=0.32\textwidth, angle =0 ]{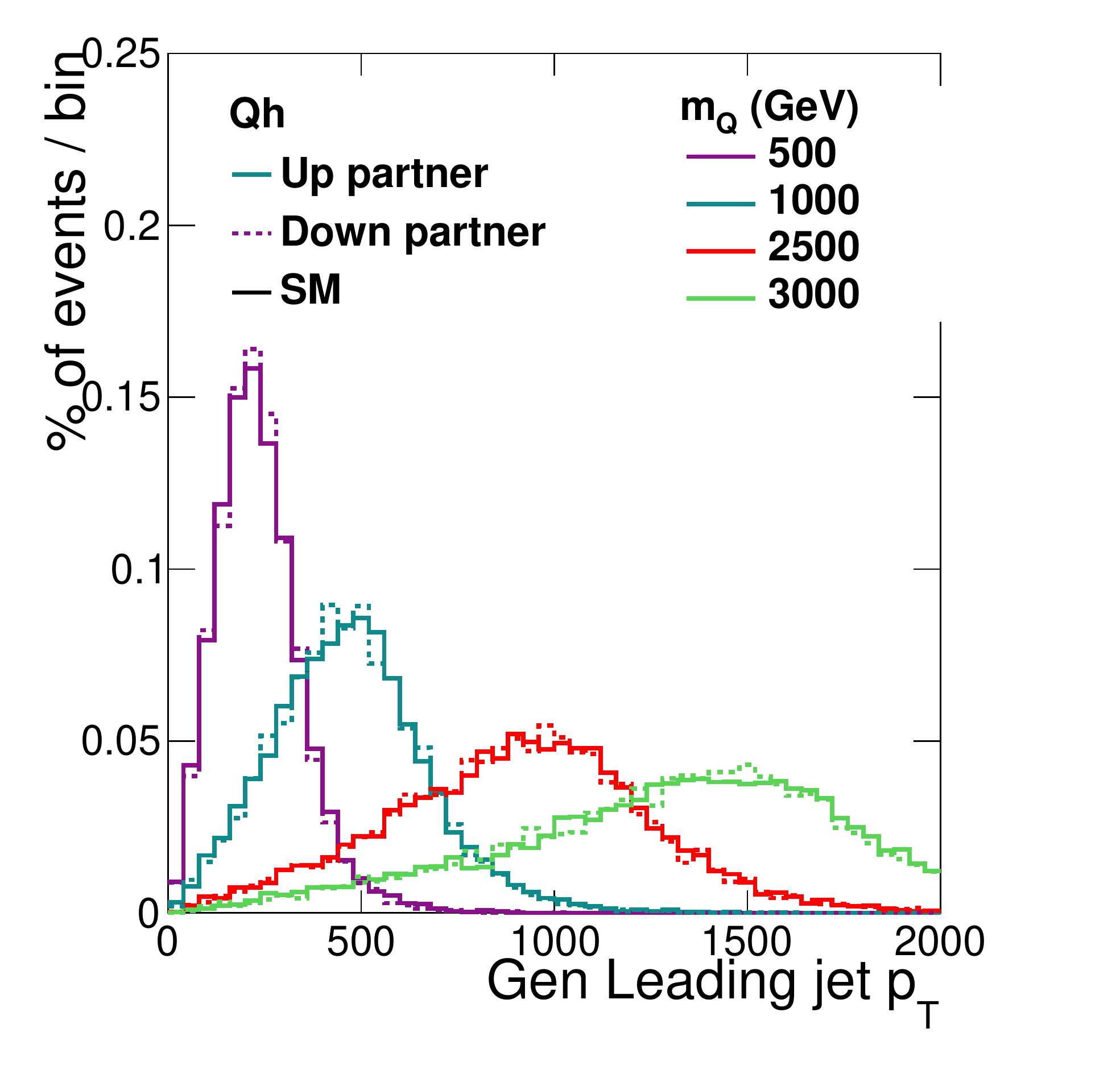}  \\ 
(a) pair & (b) pair & (c) single \\ 
\end{tabular}
\caption{The $p_{\rm T}$ distributions of the leading (a) and sub-leading (b) $p_{\rm T}$ jets in $pp\to QQ\to 2h\,2j$ events. In panel (c), $p_{\rm T}$ distribution of the only jet in  $pp\to Qh\to 2h\,j$ events. All distributions are normalised to unity. 
We fix $\kappa = 0.2$. Note that the value of $\kappa$ does not affect the distributions for the single VLQ process at LO due to the absence of QCD contributions.}
\label{fig:gen_QQ_j_pt_eta} 
\end{center}\end{figure*} 

This behaviour can be understood by looking at the $p_{\rm T}$ of the Higgs bosons, shown in panels (a-a') and (b-b') of Fig.~\ref{fig:gen_Qh_h_pt_eta} for pair and single VLQ production respectively. In the single production channel, the Higgs boson with sub-leading $p_{\rm T}$ tends to be softer as it is not related to the decay of the heavy $Q$. We also observe that the Higgs bosons tend in both cases to be central, as it can be shown in panels (c-c') of the two figures: only for the single production mode, a more spread pseudo-rapidity distribution is observed for low $Q$ masses.

Finally, in Fig.~\ref{fig:gen_QQ_j_pt_eta} we focus on the $p_T$ of the jets produced in the $pp\to QQ\to 2h\,2j$ events, panels (a) and (b), and in $pp\to QQ\to 2h\,j$, panel (c). All jets become harder for increasing VLQ mass, as expected, because they all come from the decays of a heavy fermion. We also observed that they are centrally produced in all cases.

We presented several production modes of $hh$ in BSM scenarios where vector-like quarks are present. A classic di-Higgs search in the collider looks for a bump in the invariant mass spectrum of the two Higgs bosons. We have shown that this can be enhanced by the presence of VLQs in the loops contributing to the inclusive di-Higgs production. An enhancement in the high invariant mass can be seen also in the case of direct pair-production of $QQ$ and associated $Qh$ production, and has the potential to be detected at the LHC. Apart from the invariant mass we also find that angular separation between the Higgs bosons is different in the SM processes from the VLQ-induced process, which is markedly different from the VLQ decay processes. Using this information it is possible to distinguish different production modes, should an enhancement of signal over the SM expectation be observed. 

\begin{figure*}[!htb]
\begin{center}
\begin{tabular}{cc}
\includegraphics[width=0.32\textwidth, angle =0]{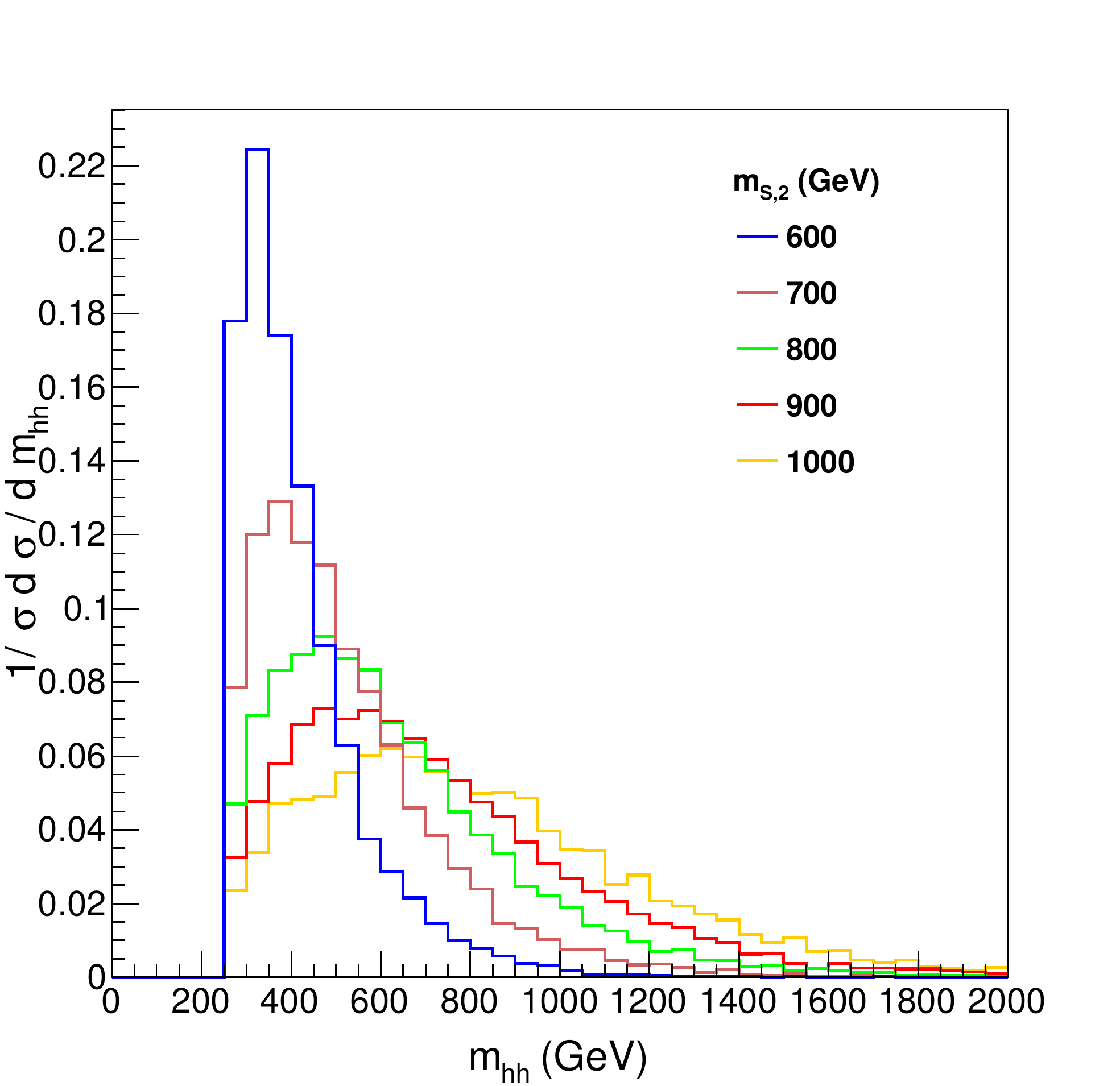} &
\includegraphics[width=0.32\textwidth, angle =0]{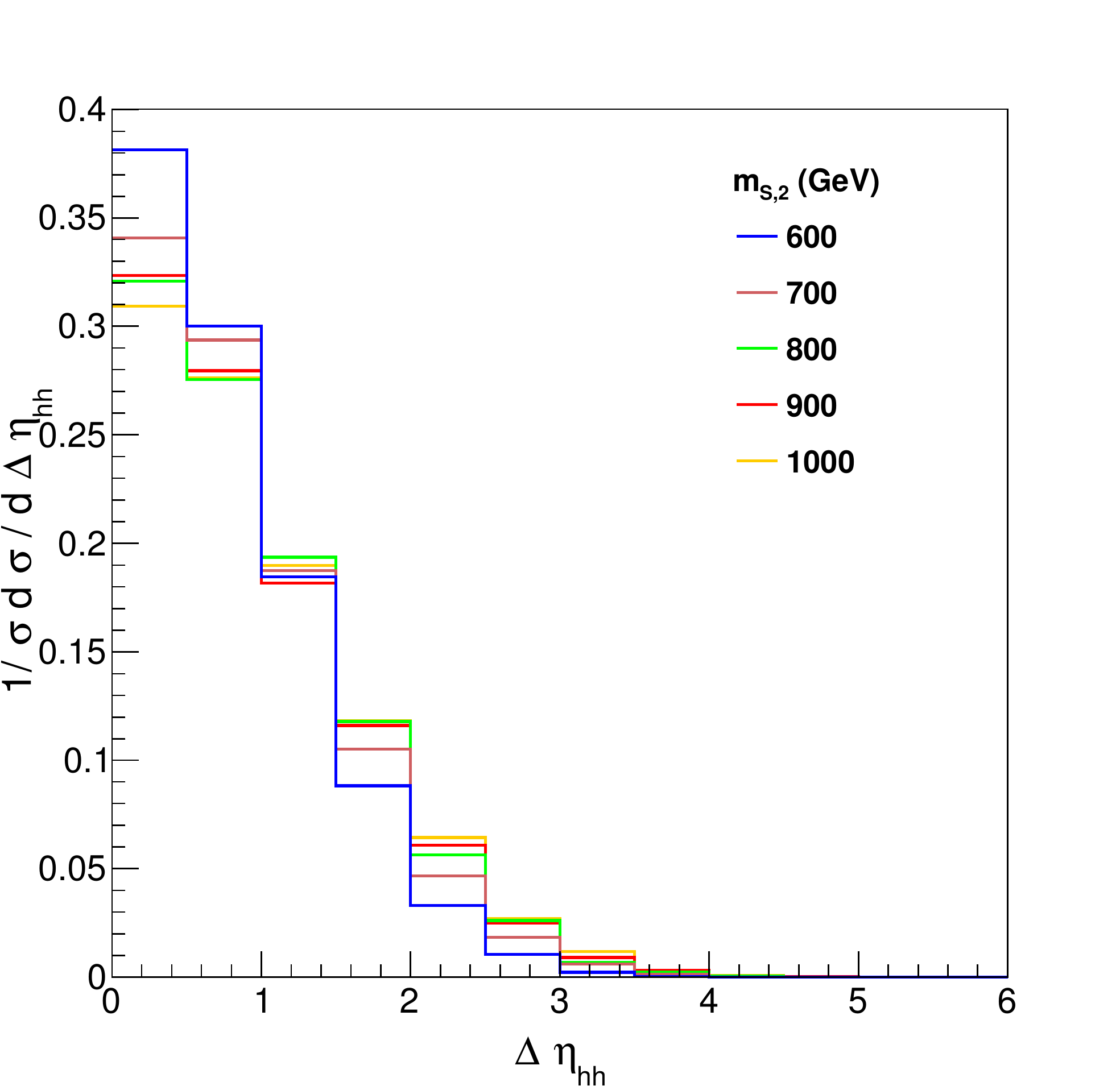}  \\
(a) & (b) \\
\includegraphics[width=0.32\textwidth, angle =0]{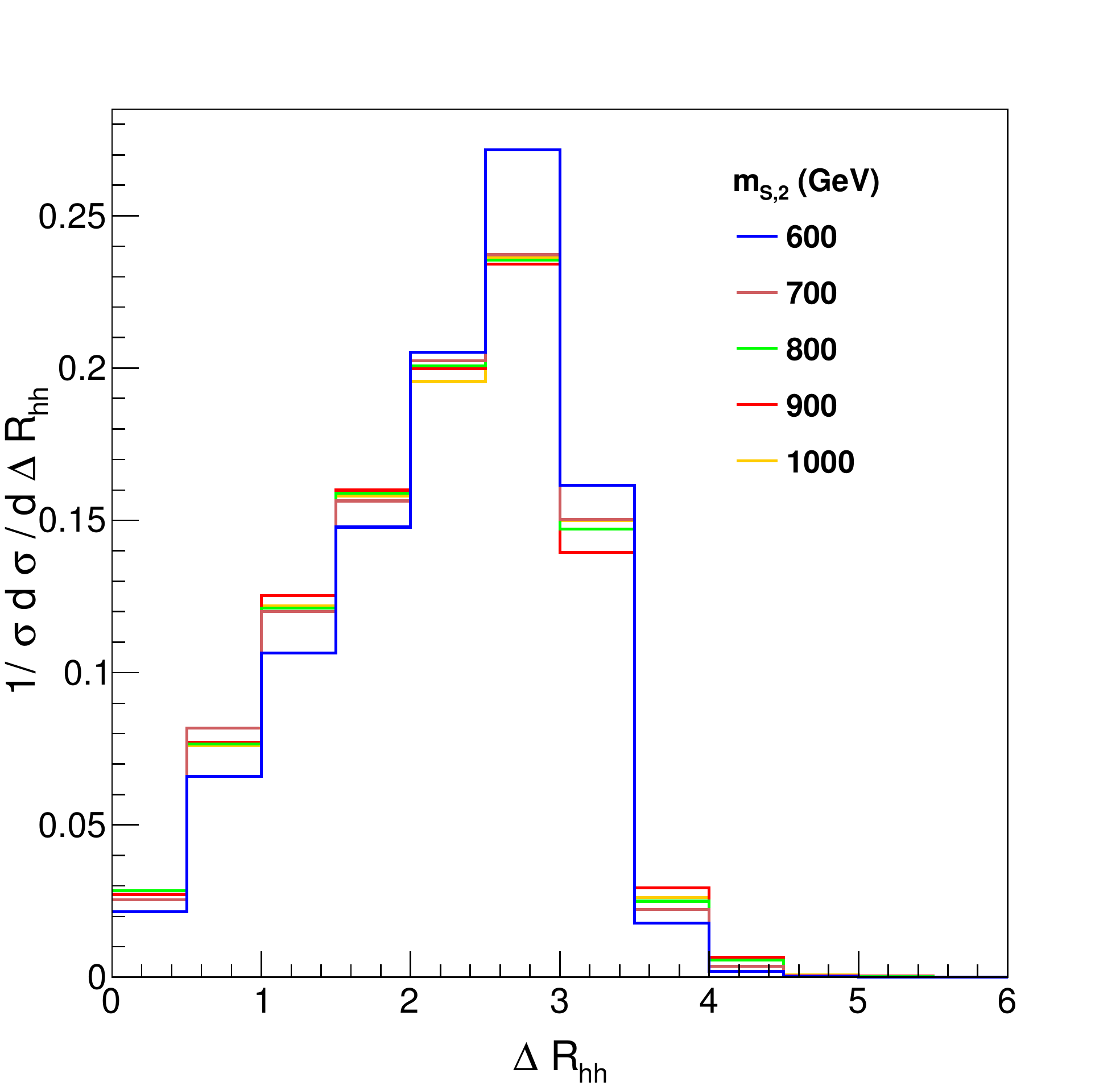} &
\includegraphics[width=0.32\textwidth, angle =0]{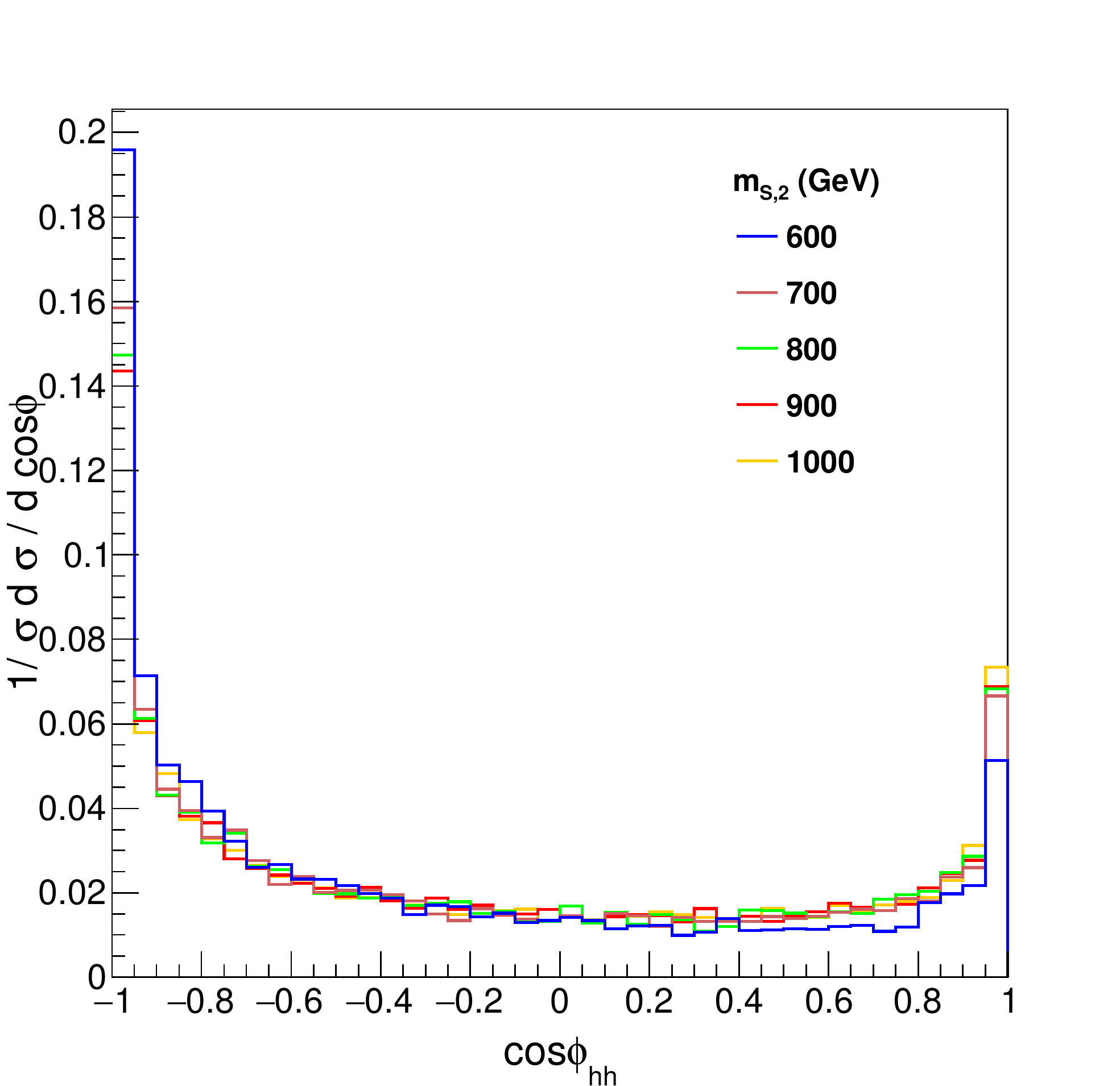} \\
  (c) & (d)  \\
\end{tabular}
\caption{Top row: distributions of  $m_{hh}$  and $\Delta \eta_{hh}$  between the two Higgses. Bottom row: distributions of  $\Delta R_{hh}$,  and $\cos \phi_{hh}$  between the two Higgses. We chose $c_{L,R} =0.1$,  $\lambda_{1,2,3}=1.0$, and $m_{S,1}$ = 400 GeV.}
\label{mdRhh}
\end{center}
\end{figure*}

\subsection{COLOURED SCALARS\label{ss:coloured_scalars}}

In the numerical simulation we chose $ \tilde{c}_{L, ij} = \tilde{c}_{R, ij} = 0$ and $c_{L, ij} = c_{R, ij} = 0.1\ \delta_{ij}$, so that $\mbox{BR} (S_1 \to jj) \sim 75\%$; we further fix $m_{S_1} = 400$ GeV, above the LHC constraint, and scan for various masses of the heavier state.
In  Figure \ref{mdRhh}, we present the distributions of $\Delta m_{hh}$, $\Delta \eta_{hh}$, $\Delta R_{hh}$ and $\cos \phi_{hh}$ of the Higgs pair. For the  invariant mass distribution,  a sharp edge sets a lower invariant mass bound for  $m_{hh} > 200$ GeV, while the peak of  $m_{hh}$ increases with increasing mass of  $m_{S_2}$. Since the Higgs pair is mostly centrally produced, there is a larger percentage of events in the small $\Delta \eta$ region.  Finally, the distributions of $\Delta R_{hh}$ and $\cos \phi_{hh}$ show that the Higgs bosons, being pair produced, tend to move back to back, and that the asymmetry in the backward and forward direction decreases for larger scalar mass.

\begin{figure*}[!htb]
\begin{center}
\begin{tabular}{ccc}
\includegraphics[width=0.32\textwidth, angle =0]{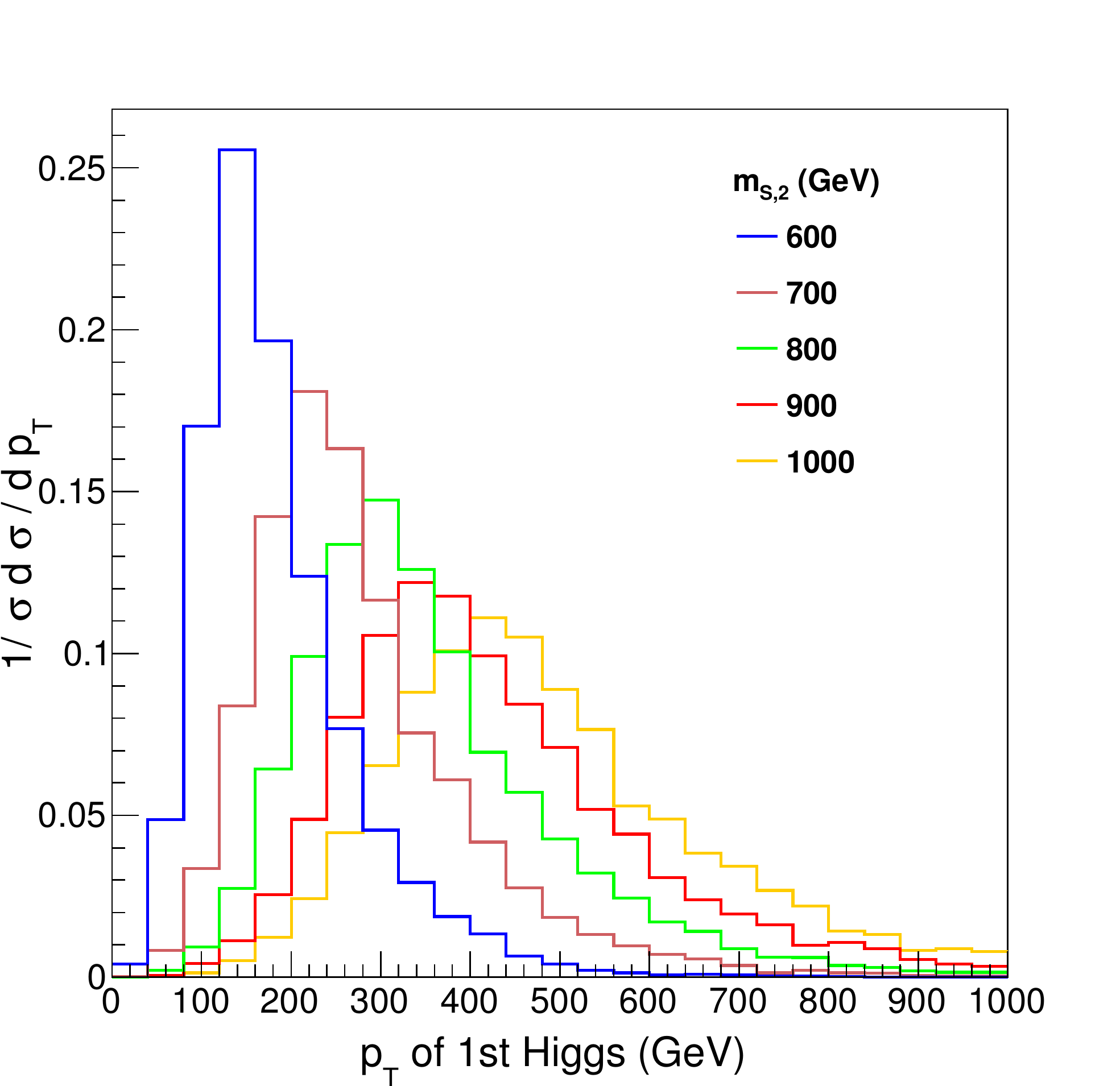} &
\includegraphics[width=0.32\textwidth, angle =0]{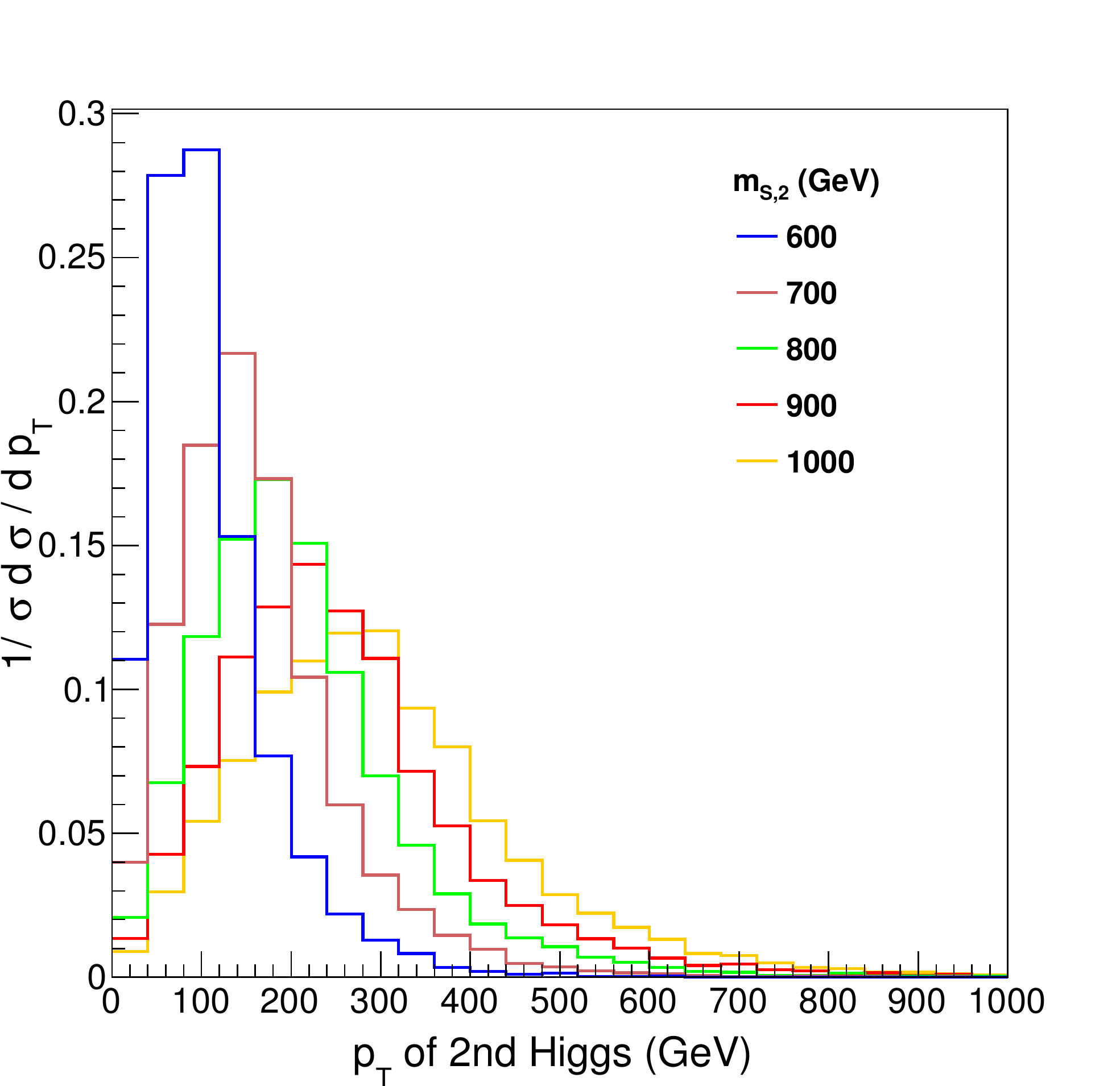} \\
(a) & (b) \\
\includegraphics[width=0.32\textwidth, angle =0]{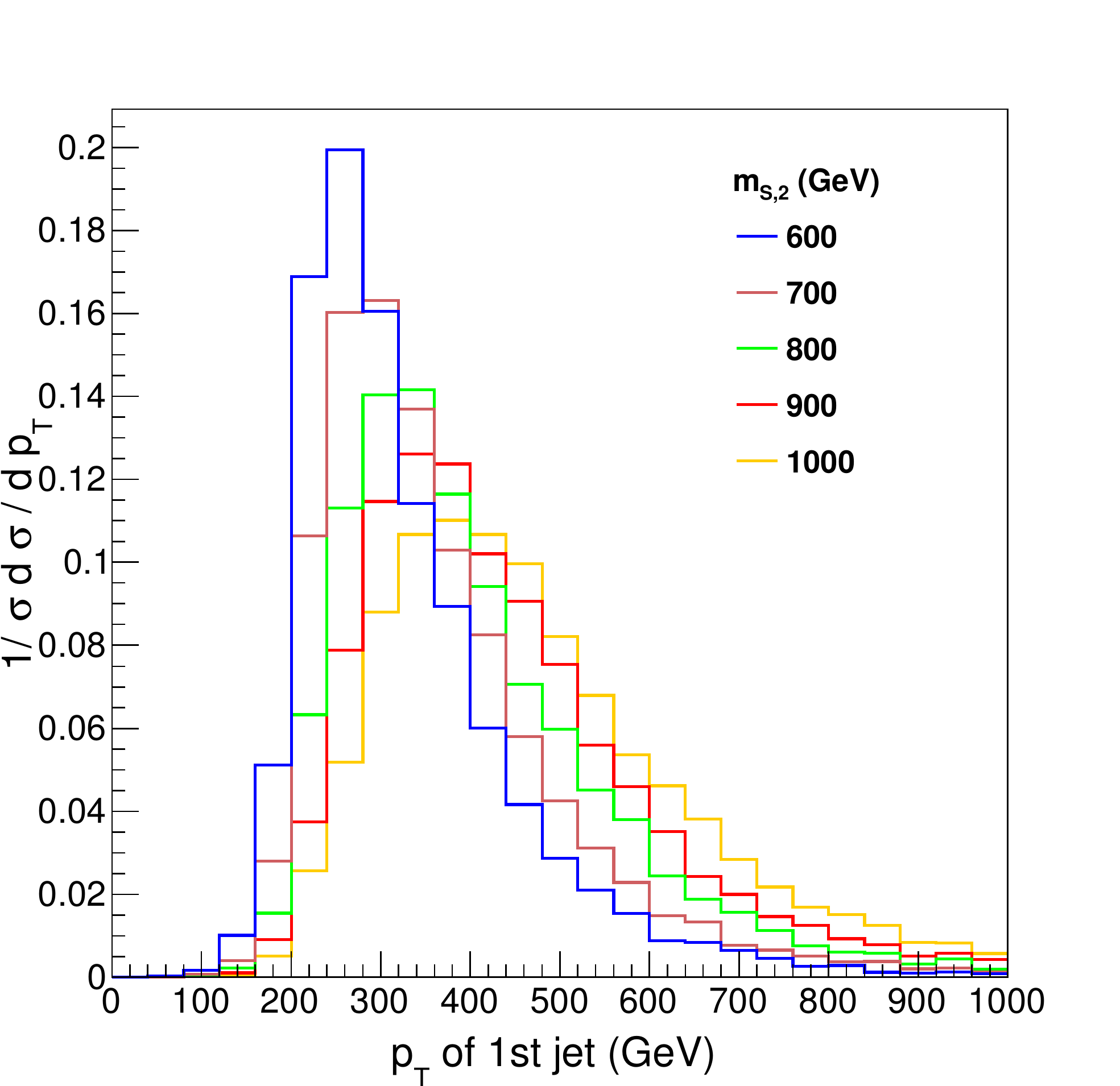}  &
\includegraphics[width=0.32\textwidth, angle =0]{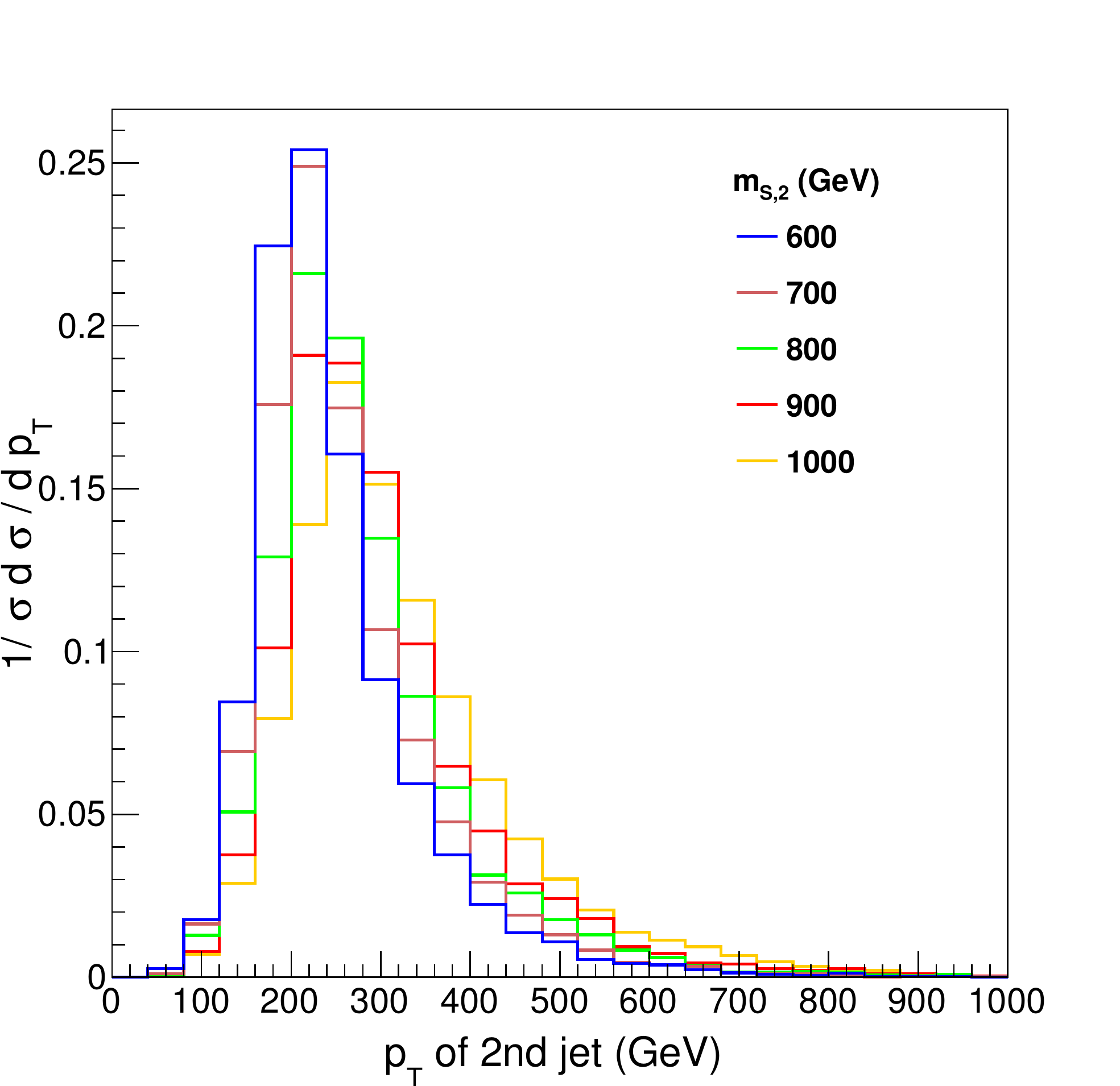} \\
  (c) & (d)  \\
\end{tabular}
\caption{$p_T$ distribution of the leading and sub-leading Higgs (panels (a) and (b)) and jets (panels (c) and (d)), with $c_{L,R} =0.1$,  $\lambda_{1,2, 3} =1.0$, and $m_{S,1}$ = 400 GeV.}
\label{pthjet}
\end{center}
\end{figure*}

Figure~\ref{pthjet} shows  the transverse momentum distribution of the two Higgs bosons and of the leading jets, by ordering them according to the  $p_T$ magnitude of the observed or reconstructed objects. Comparing the $p_T$ distribution  of the Higgses with the one of the jets, we can see that in the cascade decay scenario, where the leading jet comes from decays of the lighter coloured scalar, the jet carries a larger portion of  transverse momenta  than the leading Higgs. Therefore, this  kinematic property can provide a useful criterion to separate the signals from the SM backgrounds after the decay modes of  Higgs pair are specified. Thus, cutting on the $p_T$ of the jets, for instance requiring
$ p_T^{j1} > 200 ~\mbox{GeV}, \quad   p_T^{j2} > 100 ~\mbox{GeV}$, could be a useful discriminating tool.

\section*{CONCLUSIONS}
We performed a preliminary parton-level study of di-Higgs production from physics beyond the standard model in two classes of models, namely those 
containing new vector-like quarks and those with new coloured scalars. When the new particles are close to the present LHC energies, kinematical properties
and search strategies are different from those usually suggested by low energy effective theories. 
In the vector-like quark scenario, we implemented a simple model in {\sc FeynRules} at NLO in QCD interactions, so that it automatically contains one-loop contribution to di-Higgs production and NLO-corrected tree-level processes. 
We first found that NLO corrections to single production of vector-like quarks are sizeable. 
We then specialised to a specific model where the couplings to a Higgs boson and light quarks dominate, so that Higgses can be produced as decay products of the vector-like quarks.
A remarkable result is that, contrary to the common lore, pair production can be enhanced thanks to $t$-channel exchange of the Higgs: we find that such contribution can dominate over single production up to masses of 2~TeV. Studying the kinematic distributions of the di-Higgs events from various production channels, we found that the di-Higgs system tends to peak at high invariant masses. Furthermore, observing other distributions, like angular one or additional hard jets, may allow one to experimentally distinguish the production modes.

For the new coloured scalars, we studied the case of colour triplets: after QCD pair production, the heavier scalar decays into a Higgs plus two jets, giving rise to a final state with two Higgs bosons and jets.  In this cascade decay scenario, the leading jet coming from a coloured scalar has typically a larger transverse momentum
with respect to the leading Higgs boson from the heavier coloured scalar thus allowing for a useful kinematical handle to distinguish the signal from the SM background.

Our preliminary results thus show that di-Higgs production can be enhanced at the LHC Run--II, and that kinematical distributions play a crucial role in identifying the origin of New Physics, if an excess over the expected SM prediction is observed. A more detailed study, including hadronisation and detector effects, as well as NLO QCD effects, will follow to better characterise the signal properties.

\section*{ACKNOWLEDGEMENTS}

GC, HC and AD acknowledges partial support from the Labex-LIO (Lyon Institute of Origins) under grant ANR-10-LABX- 66 and FRAMA (FR3127, F\'ed\'eration de Recherche ``Andr\'e Marie Amp\`ere'').
GC, AD and BF are also grateful to the Th\'eorie LHC France initiative of the
CNRS for partial support. AC is  supported by  MIURFIRB RBFR12H1MW grant. DM is supported by the National Science Foundation under award PHY-1306953. HSS is supported by the ERC grant 291377 ``{\it LHCtheory: Theoretical predictions and analyses of LHC physics: advancing the precision frontier}". TF was supported by the Basic Science Research Program through the National Research Foundation of Korea (NRF) funded by the ministry of Education, Science and Technology (No. 2013R1A1A1062597). TF, GC, HC and AD also thank the Franco-Korean Partenariat Hubert Curien (PHC) STAR 2015, project number 34299VE, and the France-Korea Particle Physics Lab (FKPPL) for support.


\AddToContent{G.~Cacciapaglia, H.~Cai, A.~Carvalho, A.~Deandrea, T.~Flacke, B.~Fuks,
D.~Majumder and H.-S.~Shao\\[.2cm]}
\renewcommand{\thesection}{\arabic{section}}

\graphicspath{{2HDM_BMNS/}}
\chapter{Cornering light scalars in Type II 2HDM}

{\it J.~Bernon, K.~Mimasu, J.~M.~No and D.~Sengupta}

\begin{abstract}
Focusing on a Type II Two-Higgs-Doublet-Model (2HDM) scenario, we explore the possibility of one of the neutral 2HDM states 
being lighter than the 125 GeV Higgs. We investigate the allowed region of parameter space in light of the most recent LHC searches 
for neutral scalars, and discuss prospects to probe these light scalars during LHC Run 2.
\end{abstract}

\section{INTRODUCTION}

The discovery of the Higgs boson at 125 GeV has certainly been the landmark achievement 
of the LHC Run 1~\cite{Aad:2012tfa,Chatrchyan:2012xdj}. While the properties of the Higgs have been found to be consistent with the  
Standard Model (SM) expectations, the possibility that the electroweak (EW) symmetry breaking sector is extended is very much alive. 
The simplest extension is the Two-Higgs-Doublet-Model 
(2HDM), where the SM Higgs doublet is supplemented with a second doublet.  Although a significant amount of work has already been done in this context, interesting features 
and pockets of parameter space relevant for the LHC analysis remain unexplored. Focusing on a Type II 2HDM scenario, we investigate in this work
the possibility of one of the neutral 2HDM states being lighter than the 125 GeV Higgs, in light of the most recent LHC searches for neutral scalars.

We consider a softly-broken $\mathbb{Z}_2$ symmetric, CP-conserving 2HDM, whose scalar potential can be written as 
\begin{equation}
\begin{split}
\label{V2HDM}
\mathcal{V}&=\mu_1^2 |\Phi_1|^2+\mu_2^2 |\Phi_2|^2 -\mu^2 \left[\Phi_1^\dagger \Phi_2+\text{h.c.}\right]+\frac{1}{2}\lambda_1 |\Phi_1|^4+\frac{1}{2}\lambda_2 |\Phi_2|^4\\
&+\lambda_3|\Phi_1|^2|\Phi_2|^2 +\lambda_4\left(\Phi_1^\dagger \Phi_2\right)\left(\Phi_2^\dagger \Phi_1\right)+\frac{1}{2}\lambda_5\left[\left(\Phi_1^\dagger 
\Phi_2\right)^2+\text{h.c.}\right],
\end{split}
\end{equation}
where both Higgs doublets $\Phi_{1,2}$ acquire a vacuum expectation value (vev) $\sqrt 2\langle\Phi_{1,2}\rangle=v_{1,2}$.
The two vevs can be defined to be non-negative since the potential is CP-conserving. Upon diagonalization of the CP-even, CP-odd and charged sectors one finds 
5 physical scalar states: two CP-even states $h$ and $H_0$ with $m_h\leq m_{H_0}$, a CP-odd state $A_0$ and a pair of charged states $H^\pm$. The mixing 
angle of the CP-even sector is denoted as $\alpha$, while the ratio of the two vevs $v_2/v_1$ is noted as $\tan\beta$, with $\beta$ being also 
the mixing angle of the charged and neutral CP-odd sectors. 

\begin{table}[b!]
\begin{center}
\begin{tabular}{|c|c|c|c|}
\hline
State & Coupling to gauge bosons &  Coupling to up-type fermions & Coupling to down-type fermion \cr
\hline
 $h$ & $s_{\beta-\alpha}$ &  $c_\alpha/s_\beta=s_{\beta-\alpha} +c_{\beta-\alpha}/t_\beta$ & $-s_\alpha/c_\beta=s_{\beta-\alpha} -c_{\beta-\alpha}\ t_\beta$   \cr
\hline
 $H_0$ & $c_{\beta-\alpha}$ & $s_\alpha/s_\beta=c_{\beta-\alpha}-s_{\beta-\alpha}/t_\beta$ & \phantom{$-$}$c_\alpha/c_\beta=c_{\beta-\alpha}+ s_{\beta-\alpha}\ t_\beta$ \cr
\hline
 $A_0$ & 0 & $1/t_\beta$  & $t_\beta$ \cr
\hline 
\end{tabular}
\end{center}
\vspace{-4mm}
\caption{Tree-level couplings to gauge bosons, up-type fermions and down-type fermions normalized to their SM values for $h,\ H_0$ and $A_0$ in the Type~II 2HDM. 
We use the shorthand notation $s_{x}\equiv\sin\,x$, $c_{x}\equiv\cos\,x$, $t_{x}\equiv \mathrm{tan}\,x$.}
\label{table:couplings}
\end{table}

We consider the Type II 2HDM scenario, in which one Higgs doublet couples to the up-type quarks, while the other Higgs doublet couples to the down-type quarks and leptons.
The couplings of the neutral scalar states to gauge bosons and fermions can then be expressed in terms of $\alpha$ and $\beta$ as shown in Table~\ref{table:couplings} 
(normalized to the SM values). 

\vspace{3mm}

\noindent We note that for a Type II 2HDM, the following complementary constraints apply:

\vspace{1.5mm}

\begin{enumerate}
\item Weak radiative $B$-meson decays $B \to X_s \gamma$ lead to a severe lower bound on the charged Higgs mass; $m_{H^{\pm}} > 480$~GeV at the 95 \% C.L.~\cite{Misiak:2015xwa}.
 
 \item EW precision observables (EWPO), in particular the $T$-parameter, require either $A_0$ or $H_0$ to be fairly 
 degenerate with $H^{\pm}$~\cite{Gerard:2007kn,Grimus:2007if,Baak:2014ora}.
 
 \item Signal strength measurements of the SM-like Higgs state, identified as $h$ ($H_0$), yield two allowed regions at the 95 \% C.L. in the ($c_{\beta-\alpha},  \, t_{\beta}$) 
 plane: the alignment limit $c_{\beta-\alpha} \simeq~0$ ($s_{\beta-\alpha} \simeq~0$) (see e.g.~\cite{Bernon:2015qea,Bernon:2015wef}) in 
 which the couplings of $h$ ($H_0$) approach their SM values (see Table~\ref{table:couplings}) and the ``wrong-sign" region  
 $s_{\beta+\alpha} \simeq 1$ ($c_{\beta+\alpha} \simeq 1$) in which the coupling of $h$ ($H_0$) to down-type fermions has opposite 
 sign to the gauge bosons and up-type fermions couplings~\cite{Ferreira:2014naa}.
\end{enumerate}

\vspace{1mm}
  
\noindent In addition, the 2HDM scalar potential (\ref{V2HDM}) needs to satisfy the theoretical requirements 
of stability, tree-level unitarity and perturbativity.
 
 \vspace{2mm}
 
In the following we investigate the scenario where a neutral scalar $X$, either the CP-odd scalar $A_0$ or the lightest CP-even scalar $h$, is lighter 
than the $125$ GeV SM-like Higgs boson, identified respectively as $h$ or the heavier CP-even state $H_0$. The presence of such a light scalar $X$, combined with 
the constraints 1. and 2. above, yields a 2HDM spectrum with a large mass splitting $m_Y - m_X \gtrsim 300$ GeV among the light scalar $X=A_0/h$ and the heavy 
neutral scalar $Y = H_0/A_0$. This large splitting provides a key avenue to probe this region of the 2HDM parameter space, as we discuss below.

As already outlined above, we distinguish between two different scenarios, corresponding to the SM-like 125 GeV Higgs observed at the LHC being either 
$h$ or $H_0$. In the former case, we will consider the CP-odd state $A_0$ to be lighter than 
125 GeV. In the latter, $h$ is lighter than 125 GeV, while $A_0$ has to be heavier due to the combination of flavour and EWPO constraints.
In the following, we analyze the two cases separately.

\section{THE $\mathbf{m_h = 125}$ GeV SCENARIO}
\label{mh125}

Here we identify $h$ as the 125 GeV observed state, and explore the 2HDM region of parameter space with the CP-odd state $A_0$ being lighter than $h$.
We concentrate on the alignment limit $c_{\beta-\alpha}=0$ scenario (see e.g.~\cite{Bernon:2015qea}), and note that the combination of flavour bounds and EWPO constraints
leads to a lower bound $m_{H_0} \gtrsim 460$ GeV. This automatically yields a ``hierarchical" 2HDM, 
with a sizeable mass splitting $m_{H_0}-m_{A_0} \gtrsim 300$~GeV and potentially large cross sections in $H_0\to Z A_0$ and $H_0\to A_0 A_0$ 
decay modes, particularly as the stability and unitarity theoretical constraints require $t_{\beta} \sim 1$ in the presence of that sizeable splitting~\cite{Dorsch:2016tab}, 
while perturbativity imposes an upper bound $m_{H_0}\lesssim 620$~GeV.
To simplify the analysis, and satisfy EWPO ($T$-parameter) constraints as well as flavour bounds, we assume degeneracy 
between $H_0$ and $H^\pm$: $m_{H_0}=m_{H^\pm}\equiv m_H$. Below we will thus consider a scenario with $m_H\in[480,620]$~GeV, $c_{\beta-\alpha}=0$ 
and treat the $m_{A_0}<m_h/2$ and $m_h/2<m_{A_0}<m_h$ cases separately.

\subsection{$\mathbf{m_{A_0}< m_h/2}$}

\begin{figure}[!h]\centering
\includegraphics[width=.55\textwidth]{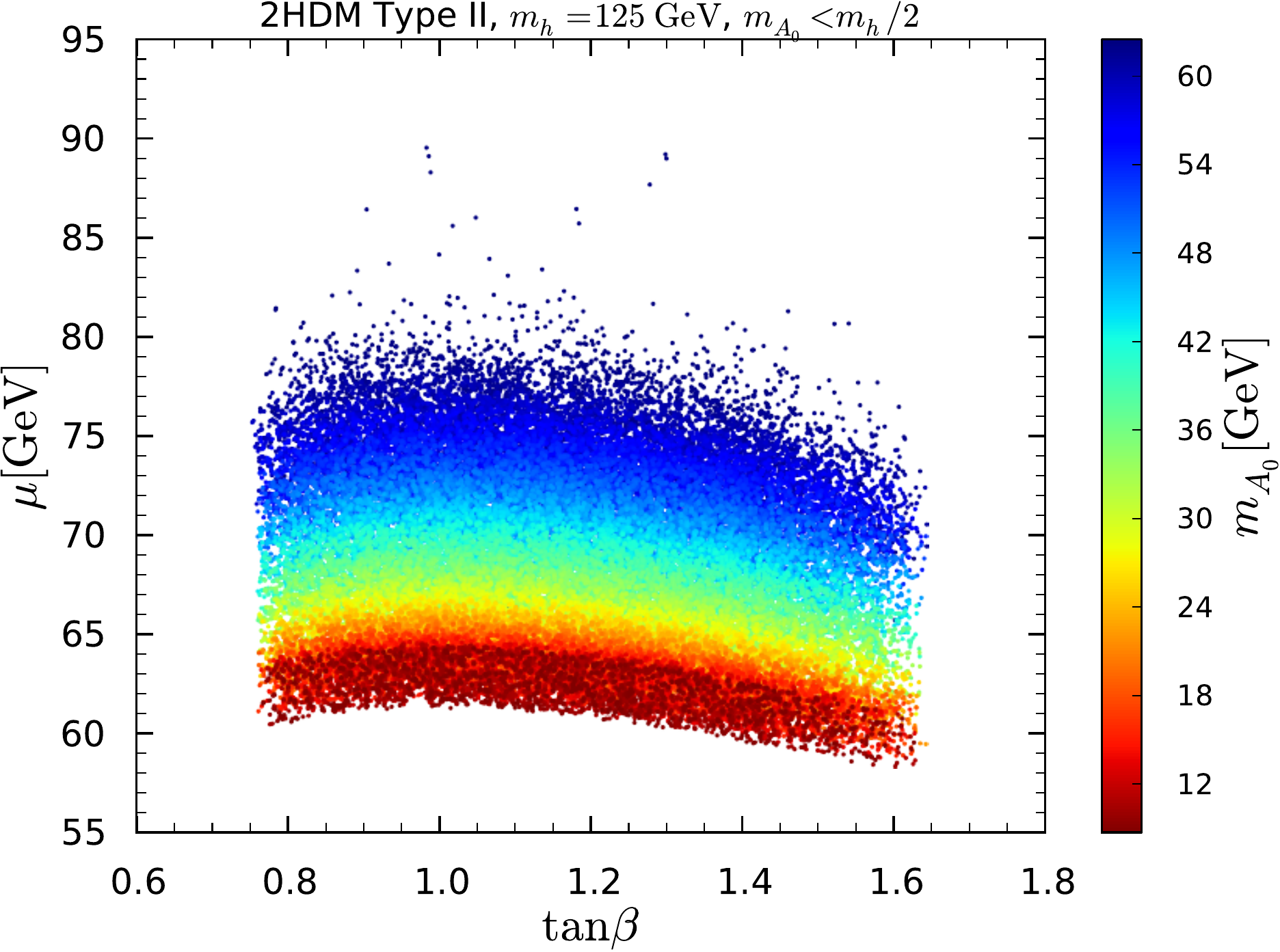}
\vspace{-4mm}
\caption{$\mu\equiv\text{sign}(\mu^2)\sqrt{|\mu^2|}$ versus $\tan\beta$ with $m_{A_0}$ color code. Points are ordered from high to low values of $m_{A_0}$, $\tan\beta$.}
\label{tb_xs13}
\end{figure}

In this case the decay channel $h\to A_0A_0$ is open and may increase dangerously the $h$ width. ATLAS and CMS signal strength measurements indeed 
impose a (model dependent) upper bound on new exotic decay channels of the SM-like state. In the exact alignment limit, the total width of $h$ is 
given by $\Gamma_h^\text{SM}/\left(1-\mathcal{B}(h\to A_0A_0)\right)$, leading to an upper bound on the $h \to A_0 A_0$ branching ratio 
$\mathcal{B}(h\to A_0A_0)\lesssim15\%$. This particularly strong constraint requires subtle adjustments between the 2HDM parameters in order to tune the 
$g_{hA_0A_0}$ coupling down to a few GeV, while it is generically at the TeV level without tuning~\cite{Bernon:2014nxa}. Specifically, in the alignment limit 
the latter coupling reads
\begin{equation}
g_{hA_0A_0}=-\frac{2m_{A_0}^2+m_h^2-4\mu^2/s_{2\beta}}{v},
\end{equation}
and requiring a small $g_{hA_0A_0}$ narrows the possible variation of $\mu^2$ to a small range as a function of $t_{\beta}$,
in particular $4\mu^2\simeq 2m_{A_0}^2+m_h^2$ for $\tan\beta\simeq1$. 
The interrelation between $\mu\equiv\text{sign}(\mu^2)\sqrt{|\mu^2|}$, $\tan\beta$ and $m_{A_0}$ is illustrated in Figure~\ref{tb_xs13}, which 
explicitly shows the strong correlation between $\mu$ and $m_{A_0}$ as a consequence of the required tuning on $g_{hA_0A_0}$.

\begin{figure}[!h]\centering
\includegraphics[width=.492\textwidth]{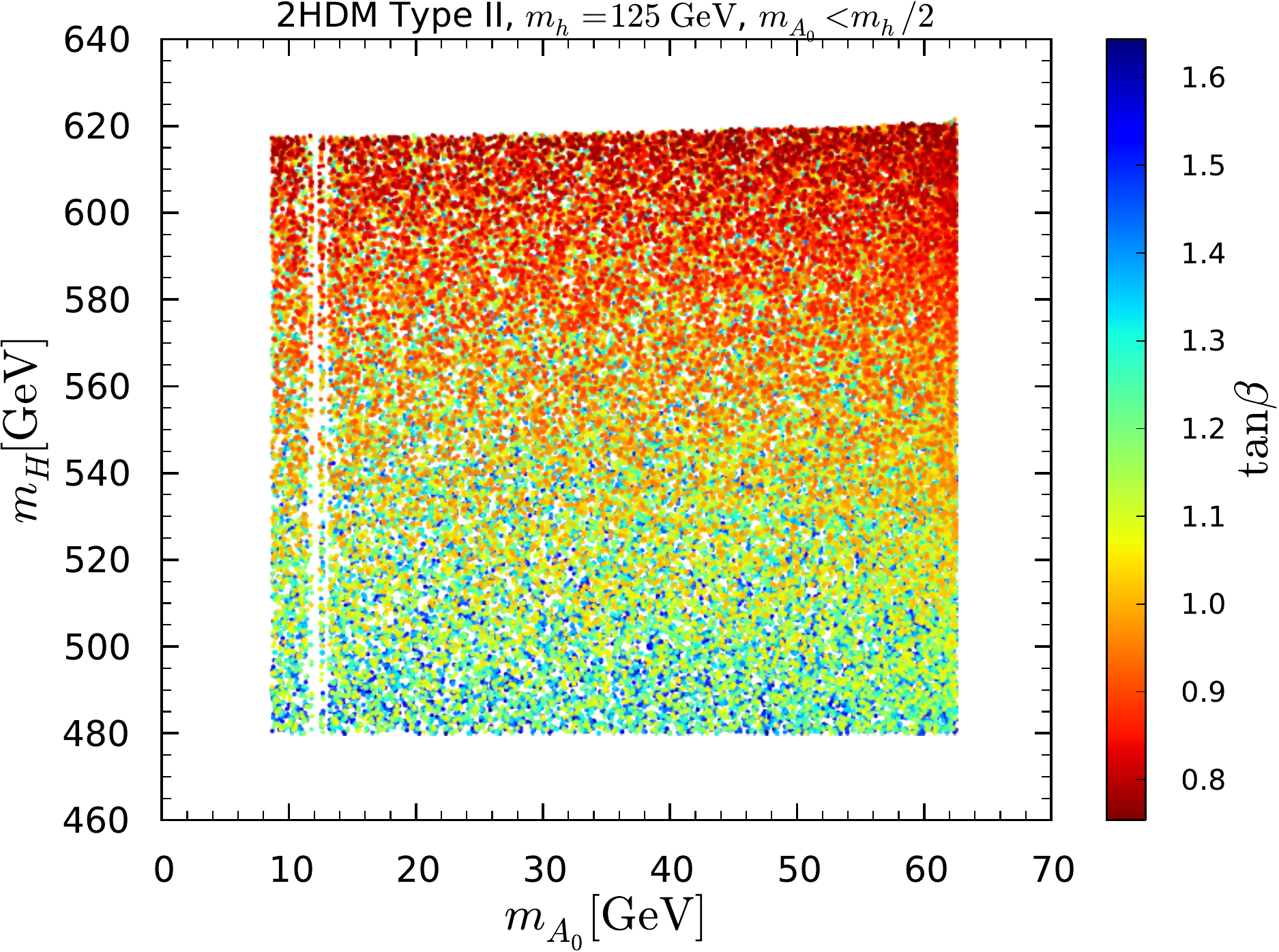}\hspace{2mm}\includegraphics[width=.492\textwidth]{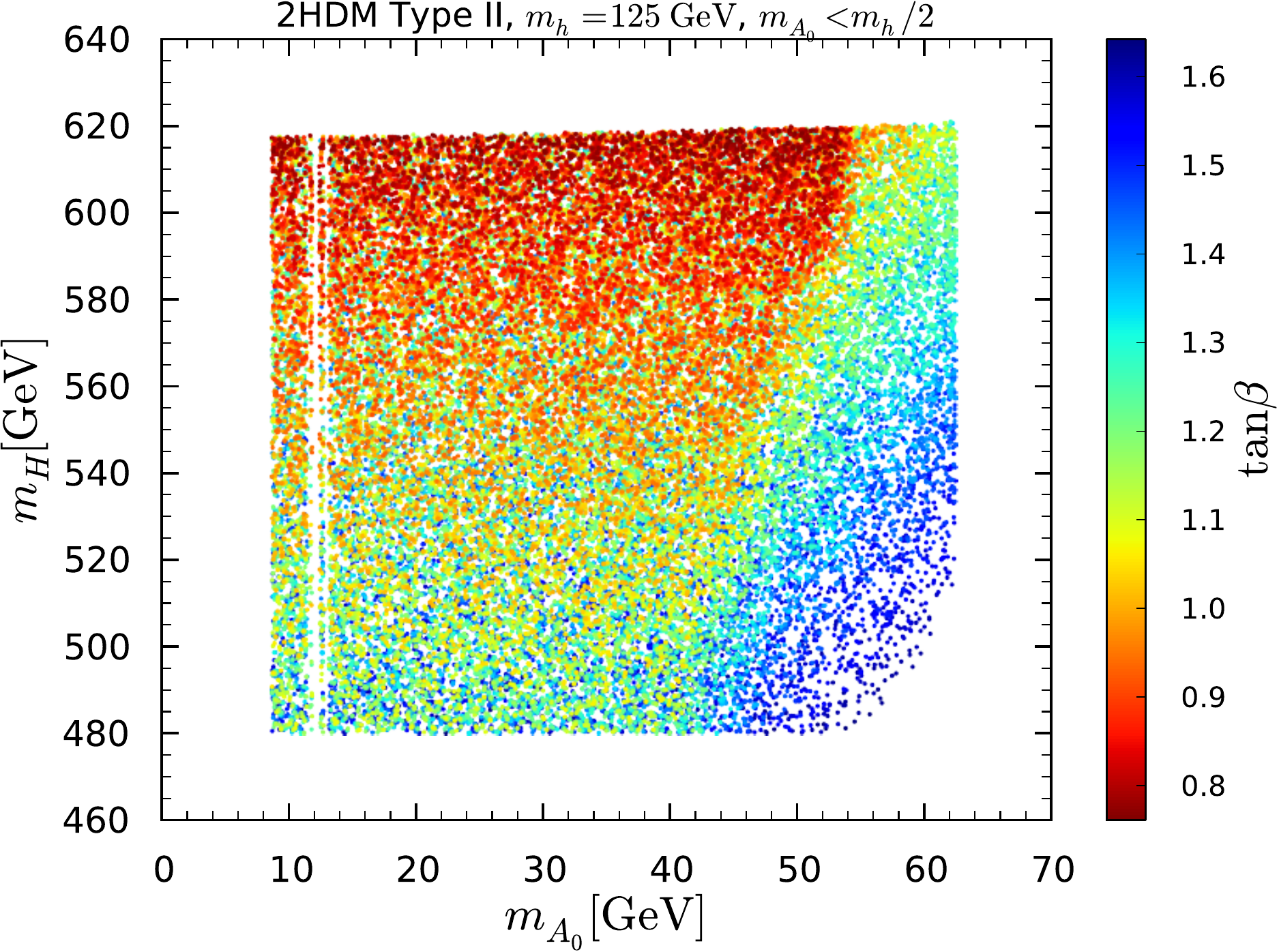}
\vspace{-0.5cm}
  \caption{$m_H$ versus $m_{A_0}$ with $\tan\beta$ color code before [{\it left panel}] and after [{\it right panel}] the implementation of the bound from the CMS 
  search $H_0\to Z\,A_0\to\ell\ell b\bar{b}$~\cite{CMS:2015mba}. Points are ordered from high to low $\tan\beta$ values.}
  \label{exclusion1}
\end{figure}

As discussed above, the large $m_H-m_{A_0}$ splitting results in $H_0\to ZA_0$ and $H_0\to A_0A_0$ being the dominant decay channels for $H_{0}$. 
Concretely, we find $\mathcal{B}(H_0\to ZA_0)\in[0.52,0.77]$ and $\mathcal{B}(H_0\to A_0A_0)\in[0,0.30]$\footnote[1]{We note that the vanishing of the 
branching ratio $\mathcal{B}(H_0\to A_0A_0)$ is due to the fact that in alignment $g_{H_0A_0A_0} \to 0$ for $t_{\beta} \to 1$~\cite{Dorsch:2016tab}.}. 
The $H_0\to ZA_0\to\ell\ell \,b\bar{b}$ channel has recently been the object of a dedicated CMS search~\cite{CMS:2015mba} 
which already constitutes a powerful probe of such hierarchical scenarios, as illustrated in Figure~\ref{exclusion1}: the left (right) panel 
shows the allowed points in the $(m_{A_0},m_H)$ plane without (with) the CMS bound implemented, with a large region above 
$m_{A_0}=40$~GeV considerably affected at low $\tan\beta$, and a small corner being completely excluded at $m_H\simeq 480$~GeV 
and $m_{A_0}\simeq 60$~GeV. This indeed corresponds to the region with the largest cross-section and most stringent CMS exclusion.

We note that the $m_{A_0}\lesssim9$~GeV region is strongly constrained by upsilon radiative decays~\cite{Domingo:2008rr} and by the direct CMS search 
for $A_0\to\mu\mu$~\cite{Chatrchyan:2012am}, and thus leave this region aside in our analysis.

\begin{figure}[!ht]\centering
\includegraphics[width=.55\textwidth]{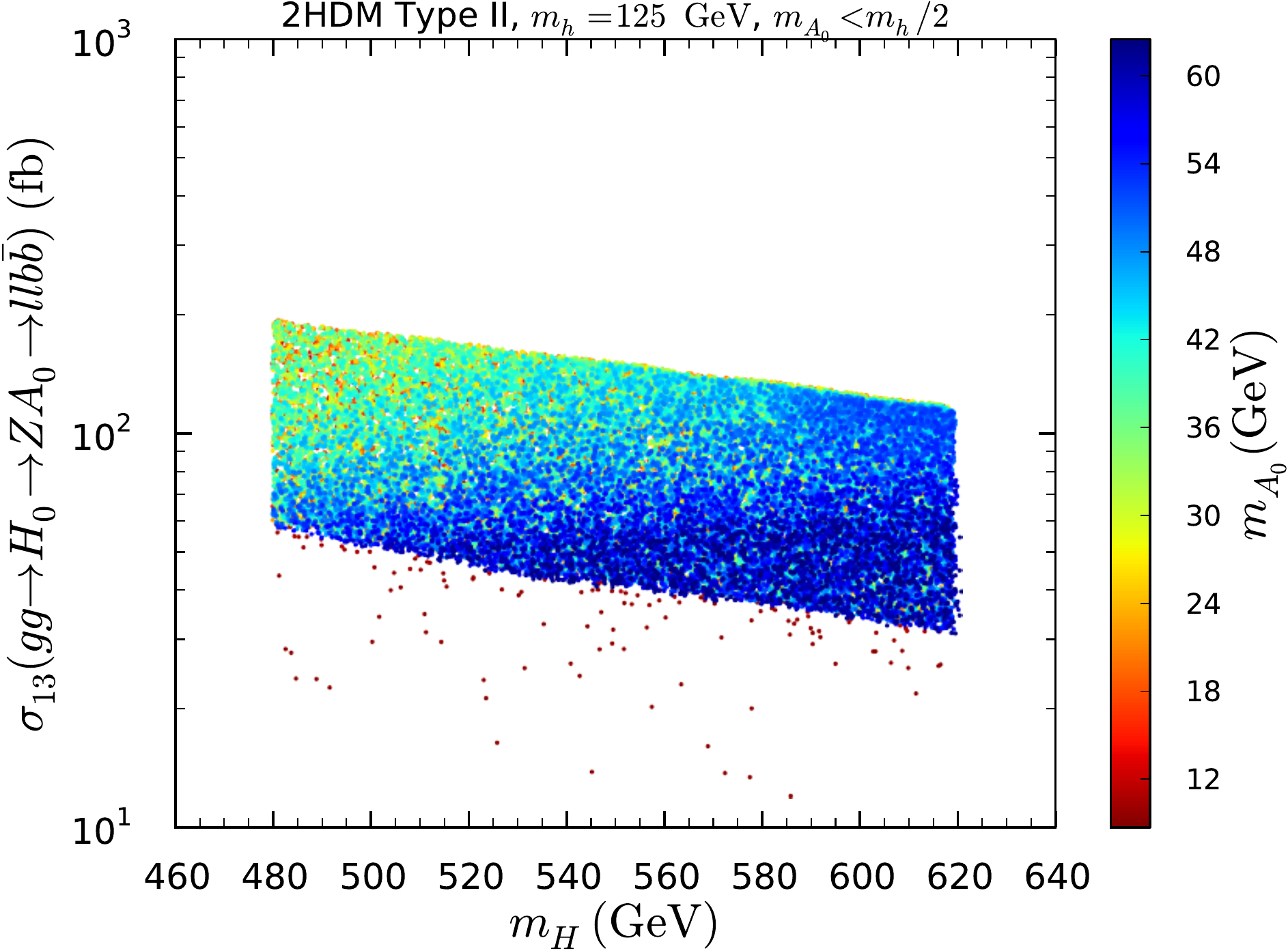}
\vspace{-0.4cm}
  \caption{$gg\to H_0\to ZA_0\to\ell\ell b\bar{b}$ cross-section at the 13 TeV LHC versus $m_H$ with $m_{A_0}$ color code. 
  Note that scattered points with small cross-section have $m_{A_0}<2m_b$. Points are ordered from low to high $m_{A_0}$ values.}
  \label{exclusion}
\end{figure}

The 13~TeV $gg\to H_0\to ZA_0\to\ell\ell b\bar{b}$ cross-sections for the remaining points are shown in Figure~\ref{exclusion}. 
We note that the small allowed range for $t_{\beta}$ around $t_{\beta} \sim 1$, determined by theoretical and flavour constraints, 
results in a lower (as well as an upper) bound on the $gg\to H_0$ cross-section at the LHC\footnote[2]{For such low values of $\tan\beta$, 
the $b\bar{b}H_0$ associated production cross-section is at least a factor 200 smaller than the gluon-fusion cross-section.}, 
which combined with the large values for $\mathcal{B}(H_0\to ZA_0)$ yield 
$gg\to H_0\to ZA_0\to\ell\ell b\bar{b}$ cross-sections as large as 200 fb for $m_H=480$~GeV and $m_{A_0}\lesssim 40$~GeV, as well as minimal 
possible values of 30 fb for $H_0$ and $A_0$ saturating their upper mass bounds. Improved sensitivity from Run~II may probe the entire upper $m_{A_0}$ 
range, and a detailed collider study would be required to precisely estimate this future reach.

\subsection{$\mathbf{m_h/2< m_{A_0} < m_h}$}

\begin{figure}[!ht]\centering
\includegraphics[width=.492\textwidth]{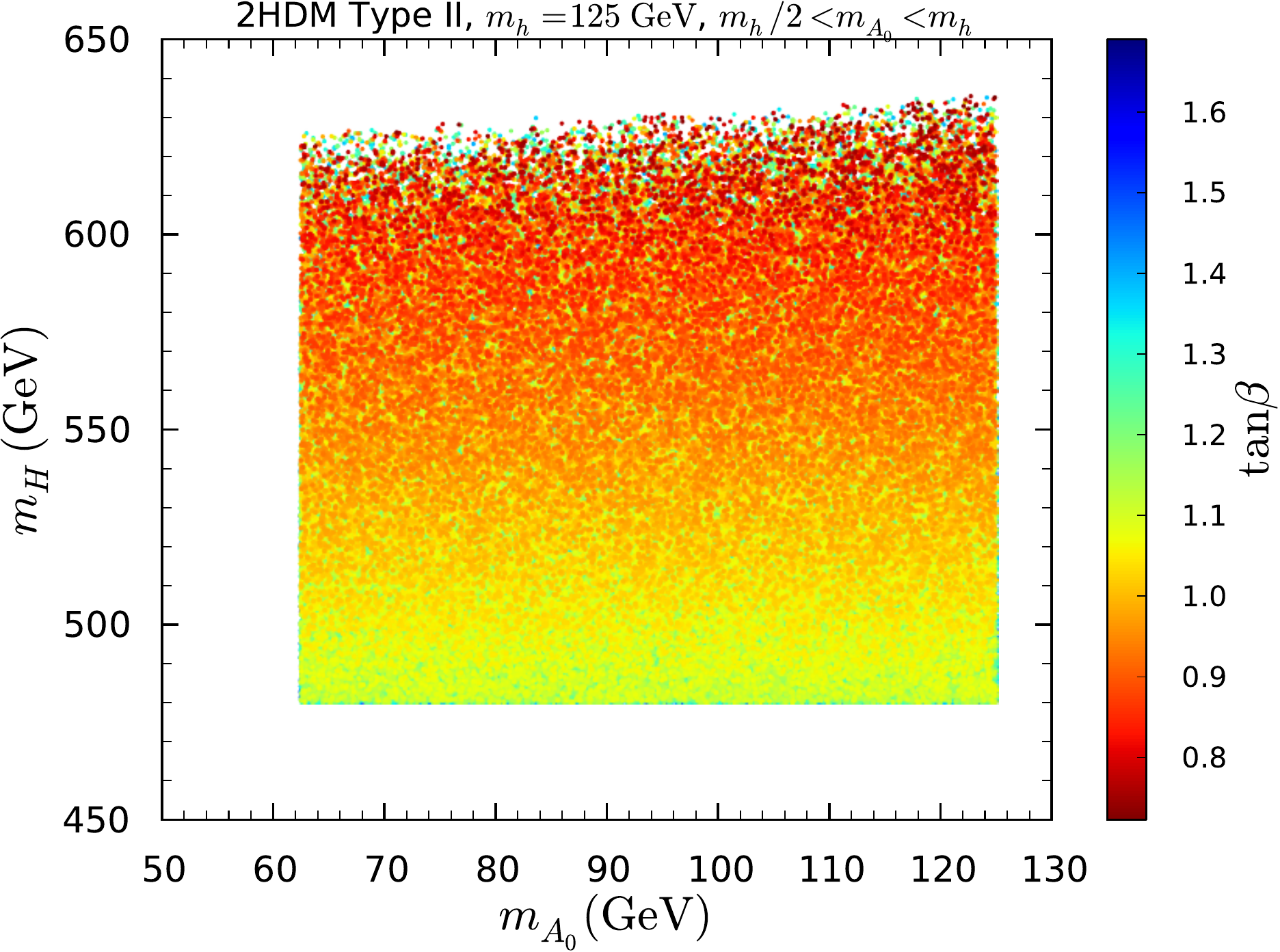}\hspace{2mm}\includegraphics[width=.492\textwidth]{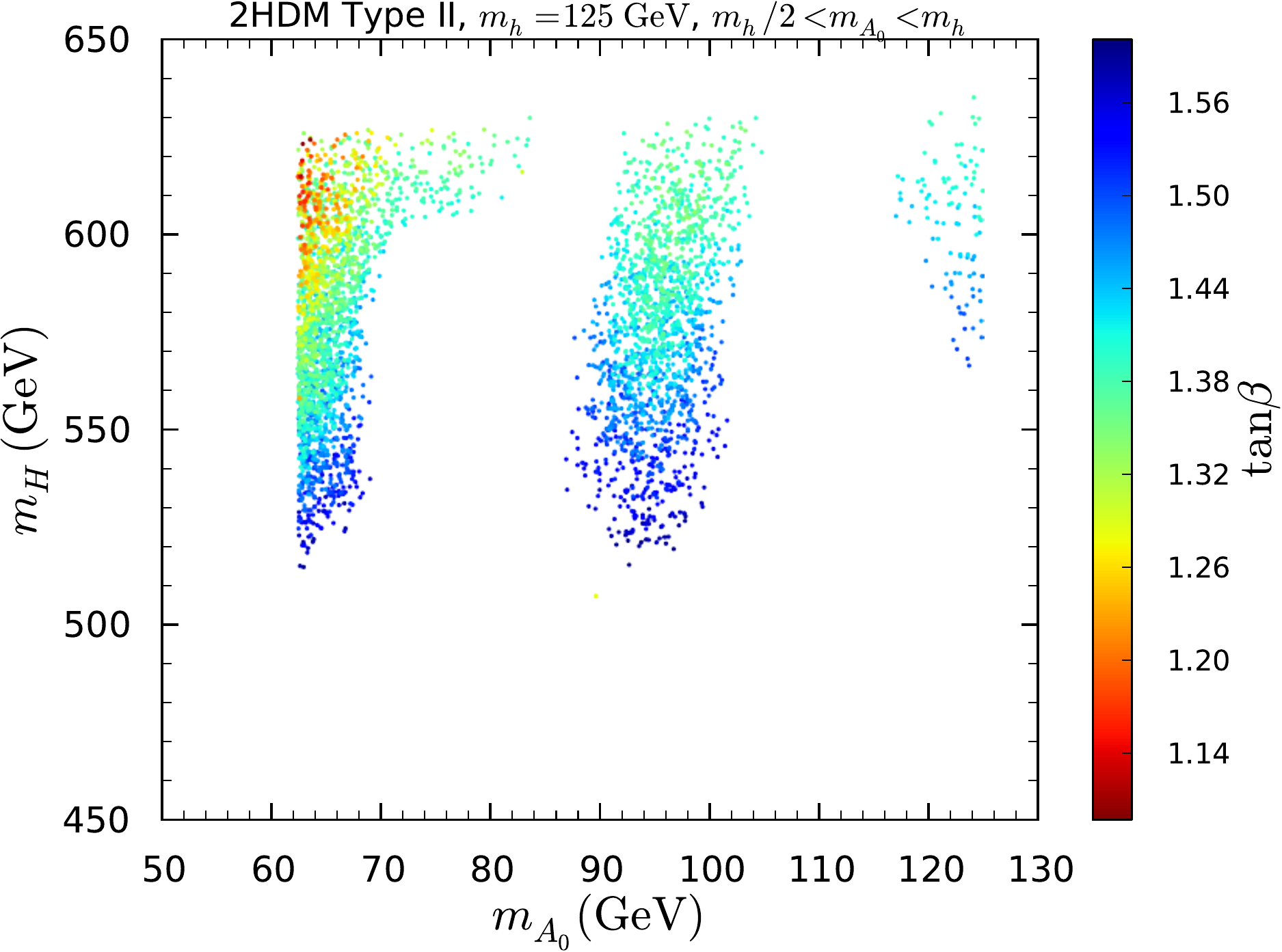}
\vspace{-0.5cm}
  \caption{$m_H$ versus $m_{A_0}$ with $\tan\beta$ color code before [{\it left panel}] and after [{\it right panel}] implementing the CMS search 
  $H_0\to ZA_0\to\ell\ell b\bar{b}$. Points are ordered from high to low $\tan\beta$ values.}
  \label{exclusion_intermediateA}
\end{figure}

Contrary to the $m_{A_0}<m_h/2$ case, $\mu^2$ can now take large negative values and the strong correlation with 
$m_{A_0}$ is lost. The main phenomenological differences with the previous case (aside from the absence of the exotic decay 
channel $h\to A_0A_0$ for the SM-like state) are thus to be found in the behaviour of the triple Higgs couplings 
driven by $\mu^2$, such as $g_{H_0A_0A_0}$. Of crucial interest for this study is the fact that the CMS $H_0\to ZA_0$ 
search has generally more constraining power for heavier $A_0$ at a fixed $H_0$. 
The possible values for the branching ratio $\mathcal{B}(H_0\to ZA_0)$ are still very large: $\mathcal{B}(H_0\to ZA_0)\in[0.35,0.77]$, 
and the $gg\to H_0$ cross-section is identical to the previous case (since the allowed $\tan\beta$ range is identical), 
which results in a stronger exclusion than for the $m_{A_0}<m_h$ scenario as seen explicitly in Figure~\ref{exclusion_intermediateA}. 
The impact of the CMS search~~\cite{CMS:2015mba} is dramatic, with whole regions of the parameter space becoming 
completely excluded, as in particular the full $\tan\beta\lesssim 1.10$ region. There remain three isolated islands of 
allowed points around $m_{A_0}\simeq 65, 95, 123$~GeV.

\begin{figure}[!ht]\centering
\includegraphics[width=.5\textwidth]{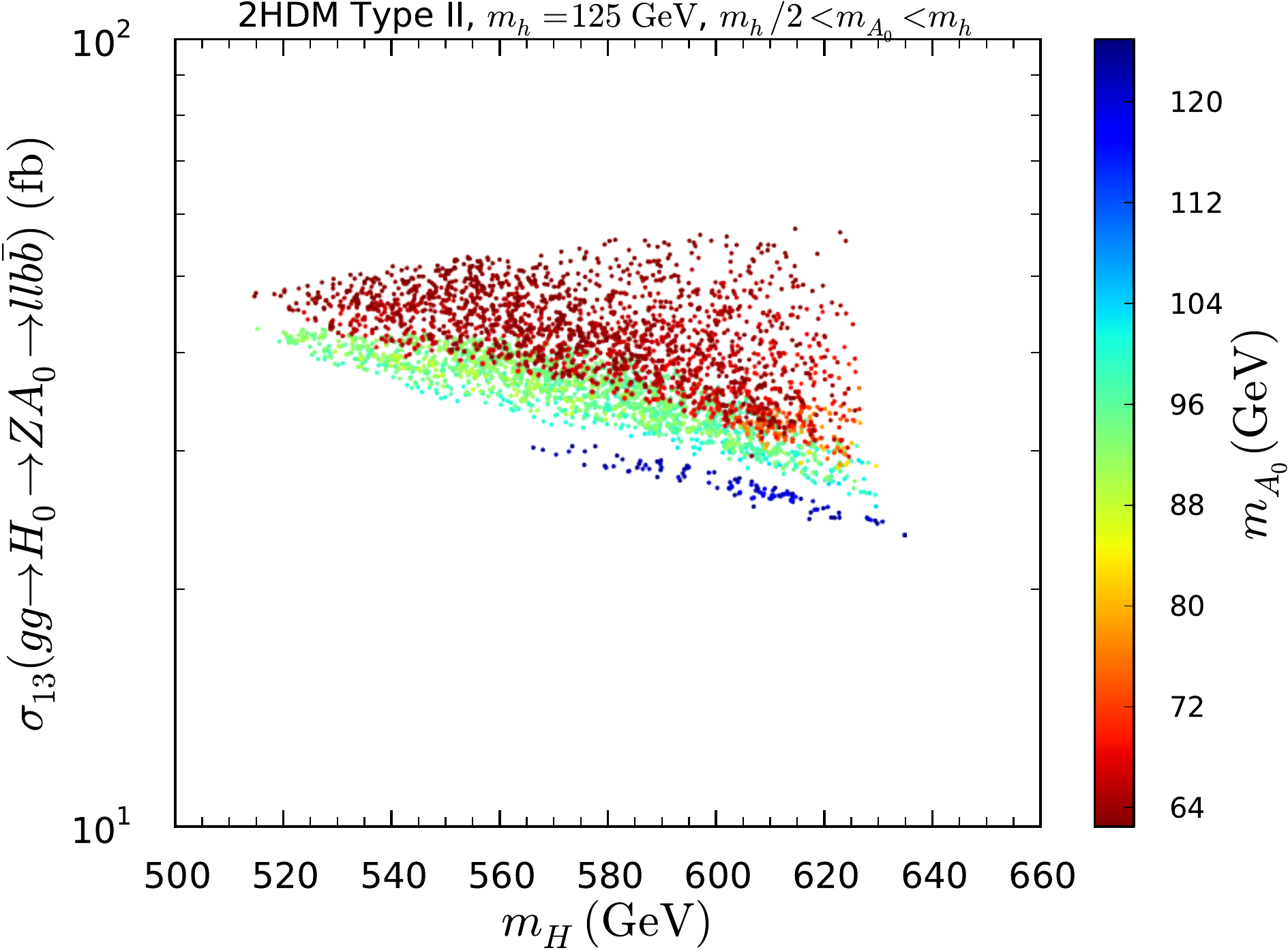}\includegraphics[width=.5\textwidth]{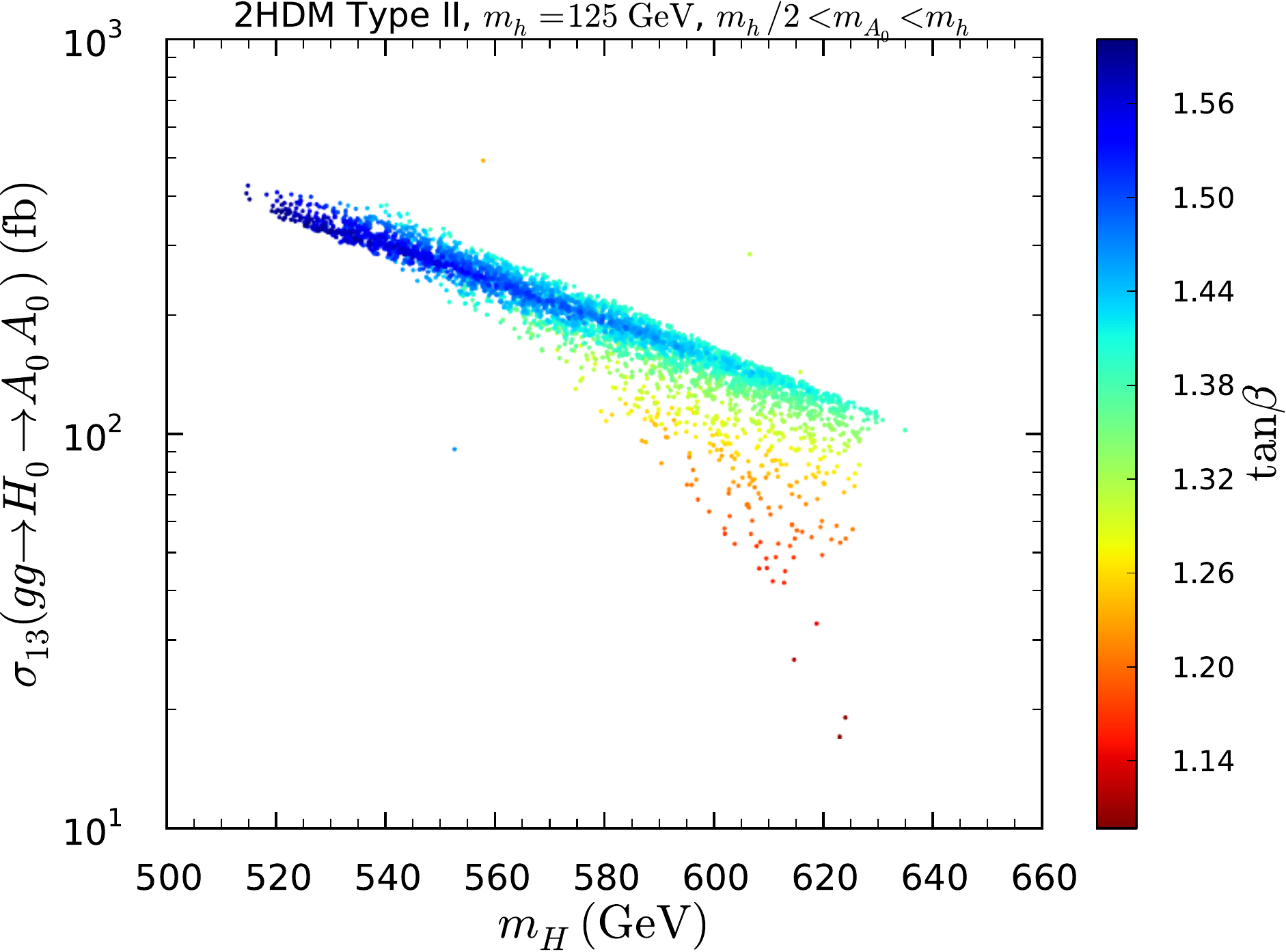}
\vspace{-0.5cm}
  \caption{$gg\to H_0\to ZA_0\to\ell\ell b\bar{b}$~[lower left panel] and $gg\to H_0\to A_0A_0$~[lower right panel] cross-section at the 13 TeV LHC 
versus $m_H$ with $m_{A_0}$ and $\tan\beta$ color code. Points are ordered from low to high $m_{A_0}$ and $\tan\beta$ values.
 [right panel] $\sigma(gg\to H_0)$ at the 
13 TeV LHC versus $m_H$ with $\tan\beta$ color code.}
  \label{exclusion_intermediateA2}
\end{figure}

In the left panel of Figure~\ref{exclusion_intermediateA2} the 13 TeV $gg\to H_0\to ZA_0\to\ell\ell b\bar{b}$ cross-sections for the remaining points are shown. 
Only a narrow cross-section range survives, 20-60 fb, leaving little doubt that future LHC analyses may be able to completely probe this region
(see e.g. \cite{Coleppa:2014hxa,Dorsch:2014qja}). 
The right panel of Figure~\ref{exclusion_intermediateA2} shows the $gg\to H_0\to A_0A_0$ cross-section at 
the 13~TeV LHC for the remaining points, \textit{i.e.} after the CMS $H_0\to ZA_0$ constraint is applied. Interestingly, points with 
high cross-sections, from $\sim 10$~fb to $\sim 400$~fb, remain. Indeed, the $g_{H_0A_0A_0}$ vanishes in the alignment limit for $\tan\beta=1$ while 
$\tan\beta\lesssim 1.10$ is actually excluded by the CMS search as discussed previously. As a result, a strong lower bound on this cross-section is found, 
making the $H_0\to A_0A_0$ channel a complementary powerful probe of this scenario. Note that in the $m_{A_0}<m_h/2$ case analyzed in the previous paragraph, 
$\tan\beta=1$ is still allowed such that the corresponding cross-section can be arbitrarily small.

\section{THE $\mathbf{m_{H_0} = 125}$ GeV SCENARIO}

\subsection{ALIGNMENT LIMIT $s_{\beta-\alpha} \sim 0$}

We begin our analysis of the $m_{H_0} = 125$ GeV case focusing on the alignment limit $s_{\beta-\alpha} \sim 0$ (see e.g.~\cite{Bernon:2015wef}).
As discussed above, the combination of flavour bounds and EWPO constraints yield a lower limit $m_{A_0} \gtrsim 460$ GeV, very weakly dependent on the value 
of $m_h$, resulting in a large mass splitting $m_{A_0} - m_h > 300$ GeV. Comparing the present scenario with that analyzed in Section~\ref{mh125} the CMS 
search for $A_0 \to Z h$ ($Z \to \ell \ell$, $h \to b\bar{b}, \tau\tau$) \cite{CMS:2015mba} also places important constraints in the allowed parameter space, with two 
key differences with respect to the $m_h = 125$ GeV scenario: {\it(i)} For $m_h > m_{H_0}/2$, the theoretical constraints do not restrict the values of $t_{\beta}$,
so that $t_{\beta} \gg 1$ is theoretically allowed. {\it(ii)} There is no competing decay to $A_0 \to Z h$ (an equivalent of $ H_0 \to A_0 A_0$ in the 
$m_{h} = 125$ GeV scenario), such that an $\mathcal{O}(1)$ branching ratio for $A_0 \to Z h$ is generally realized. 

For simplicity, we fix $m_{H^{\pm}} = 480$ GeV for $m_{A_0} < 480$ GeV, and $m_{H^{\pm}} = m_{A_0}$ otherwise throughout our analysis. This choice respects the 
$B \to X_s \gamma$ flavour constraint, as well as EWPO limits, and does not impact our results. As before, the cases $m_h > m_{H_0}/2$ and $m_h < m_{H_0}/2$
are phenomenologically different, and we analyze separately.

\subsubsection{$\mathbf{m_{h} > m_{H_0}/2}$}

\begin{figure}[!ht]\centering
\includegraphics[width=\textwidth]{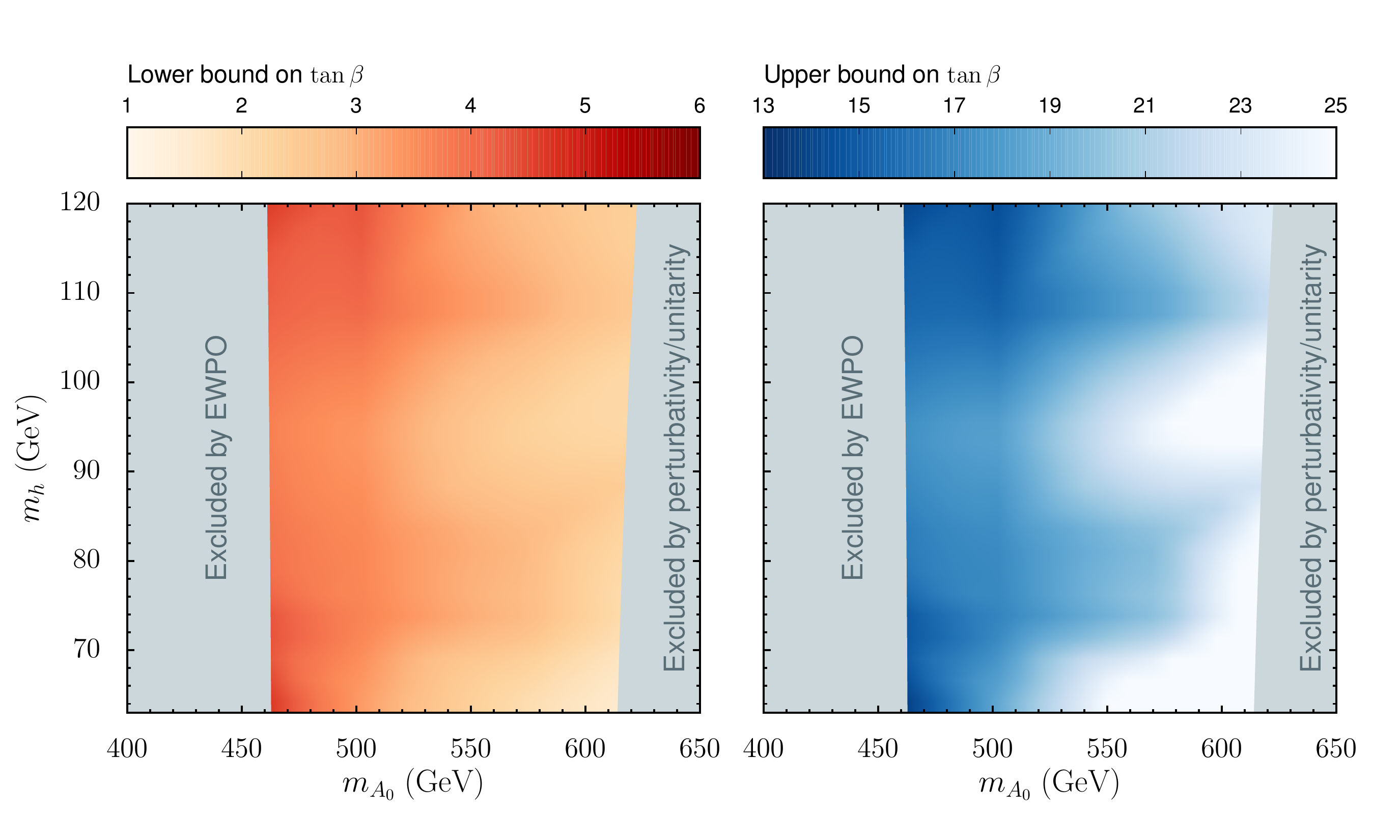}
\vspace{-0.5cm}
\caption{
LEFT: Lower bounds from \cite{CMS:2015mba} on $t_{\beta}$ as a function of ($m_{A_0}$, $m_{h}$) for gluon fusion production of $A_0$.
RIGHT: Upper bounds from \cite{CMS:2015mba} on $t_{\beta}$ as a function of ($m_{A_0}$, $m_{h}$) for $b\bar{b}$-associated production of $A_0$.}
\label{MH125Alignment}
\end{figure}

In this case, the only constraints on the allowed values of $\mu^2$ are theoretical: stability, perturbativity and unitarity. Furthermore,
for positive $m_{A_0} - m_{H_0}$ and $m_{A_0} - m_h$ mass splittings, the value $\mu^2 = m_h^2 s_{\beta} c_{\beta}$ guarantees a stable potential (see e.g.~\cite{Dorsch:2016tab})
independently of the value of $t_{\beta}$. Perturbativity/unitarity constraints also become insensitive to $t_{\beta}$, and are only violated for very large mass splittings, 
occurring for $m_{A_0} \gtrsim 620$ GeV. 
In this setup, we compute the cross section for $A_0 \to Z h$ ($Z \to \ell\ell$, $h \to b\bar{b}$) in gluon fusion and $b\bar{b}$-associated production at 8 TeV as a function of 
$t_{\beta}$ for each mass pair ($m_{A_0}$, $m_{h}$) in $m_h \in [63,\,125]$ GeV, $m_{A_0} \in [460, 620]$ GeV, and 
compare it with the limits from the CMS $A_0 \to Z h$ search \cite{CMS:2015mba} to derive the limits
this search poses on $t_{\beta}$. The results are shown in Figure \ref{MH125Alignment} as lower (LEFT) and upper (RIGHT) bounds on $t_{\beta}$ in the 
($m_{A_0}$, $m_{h}$) plane.

\begin{figure}[!ht]\centering
\includegraphics[width=0.67\textwidth]{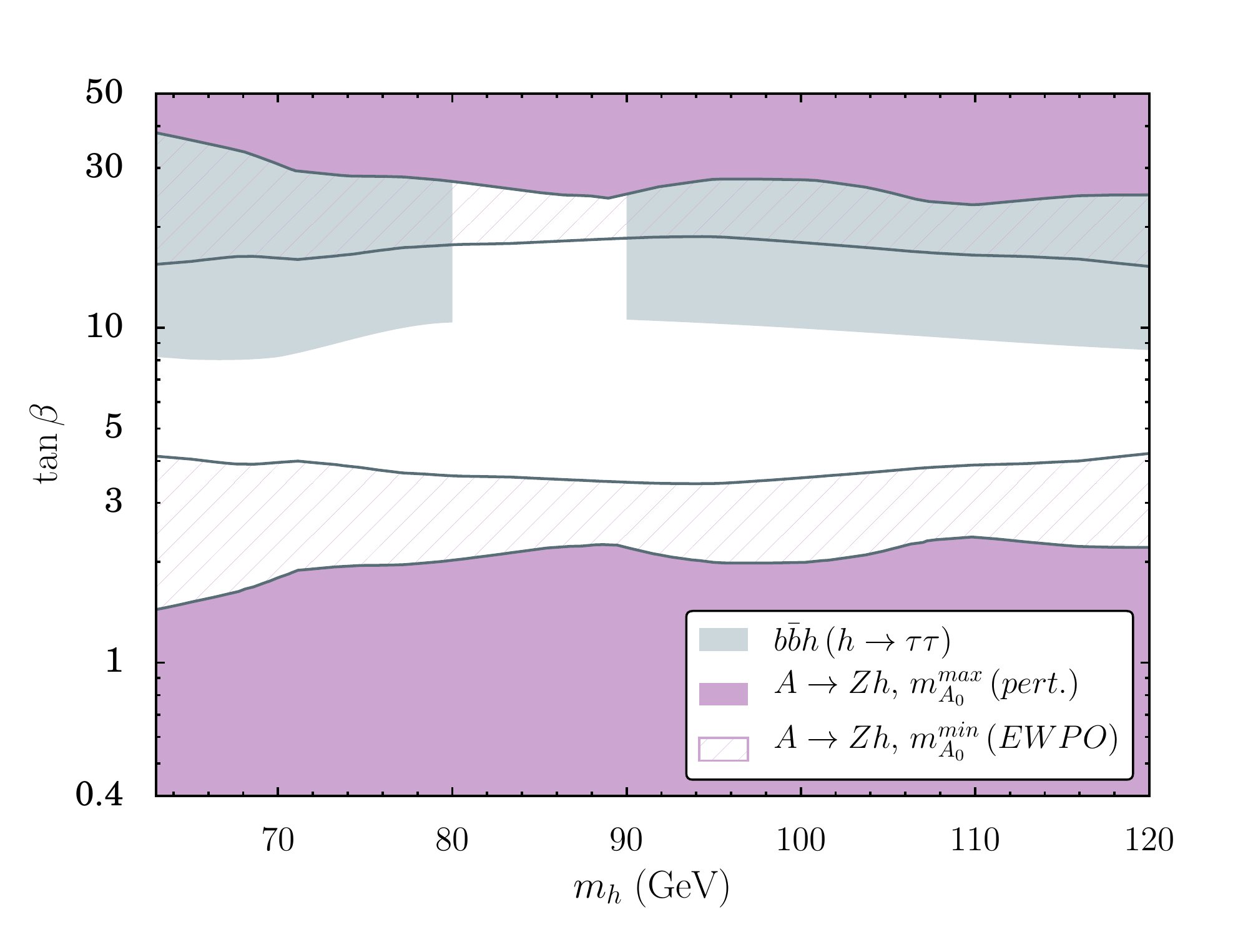}
\vspace{-0.5cm}
\caption{Lower bound on $\tan\beta$ from gluon fusion production of $A_0$ (lines above $\tan\beta=10$) and upper bound from $b\bar{b}$-associated production for 
two extreme $m_{A_0}$ values: $m_{A_0}$(\text{EWPO}) being the minimum allowed by electroweak precision observables and  $m_{A_0}(\text{pert.})$ the maximum allowed 
by stability/perturbative unitarity. The grey regions are excluded by the CMS searches of a light scalar produced in in $bb$-associated production 
decaying to $\tau\tau$~\cite{Khachatryan:2015baw} (covering masses $m_h < 80$ GeV)
and~\cite{Khachatryan:2014wca}  (covering masses $m_h > 90$~GeV).
}
\label{MH125Alignment2}
\end{figure}

As Figure \ref{MH125Alignment} shows, while no mass point ($m_{A_0}$, $m_{h}$) is completely excluded by the CMS $A_0 \to Z h$ search, the allowed range of 
$t_{\beta}$ is quite constrained by this search, to lie approximately within $t_{\beta} \in [4,\,16]$ ($t_{\beta} \in [2,\,30]$) for the lower (upper) 
allowed $m_{A_0}$ region. Moreover, the allowed value of $t_{\beta}$ is also constrained by the ATLAS/CMS searches for 
neutral scalars in $bb$-associated production decaying to $\tau\tau$ for $m_h < 80$ GeV \cite{Khachatryan:2015baw} and $m_h > 90$ GeV 
\cite{Khachatryan:2014wca,Aad:2014vgg}. The combination of various constraints is shown in Figure \ref{MH125Alignment2} in the ($m_h, \,t_{\beta}$) plane, for 
$m_{A_0}$ in the range $m_{A_0} \in [m^{\mathrm{EWPO}}_{A_0},\,m^{\mathrm{pert}}_{A_0}]$, where $m^{\mathrm{EWPO}}_{A_0} \sim 460$ GeV is the lower bound on $m_{A_0}$ from the 
combination of flavour and 
EWPO constraints, and $m^{\mathrm{pert}}_{A_0} \sim 620$ GeV is the upper bound on $m_{A_0}$ from perturbativity (both bounds are weakly dependent on $m_h$).

\subsubsection{$\mathbf{m_{h} < m_{H_0}/2}$}

\begin{figure}[!ht]\centering
\includegraphics[width=0.6\textwidth]{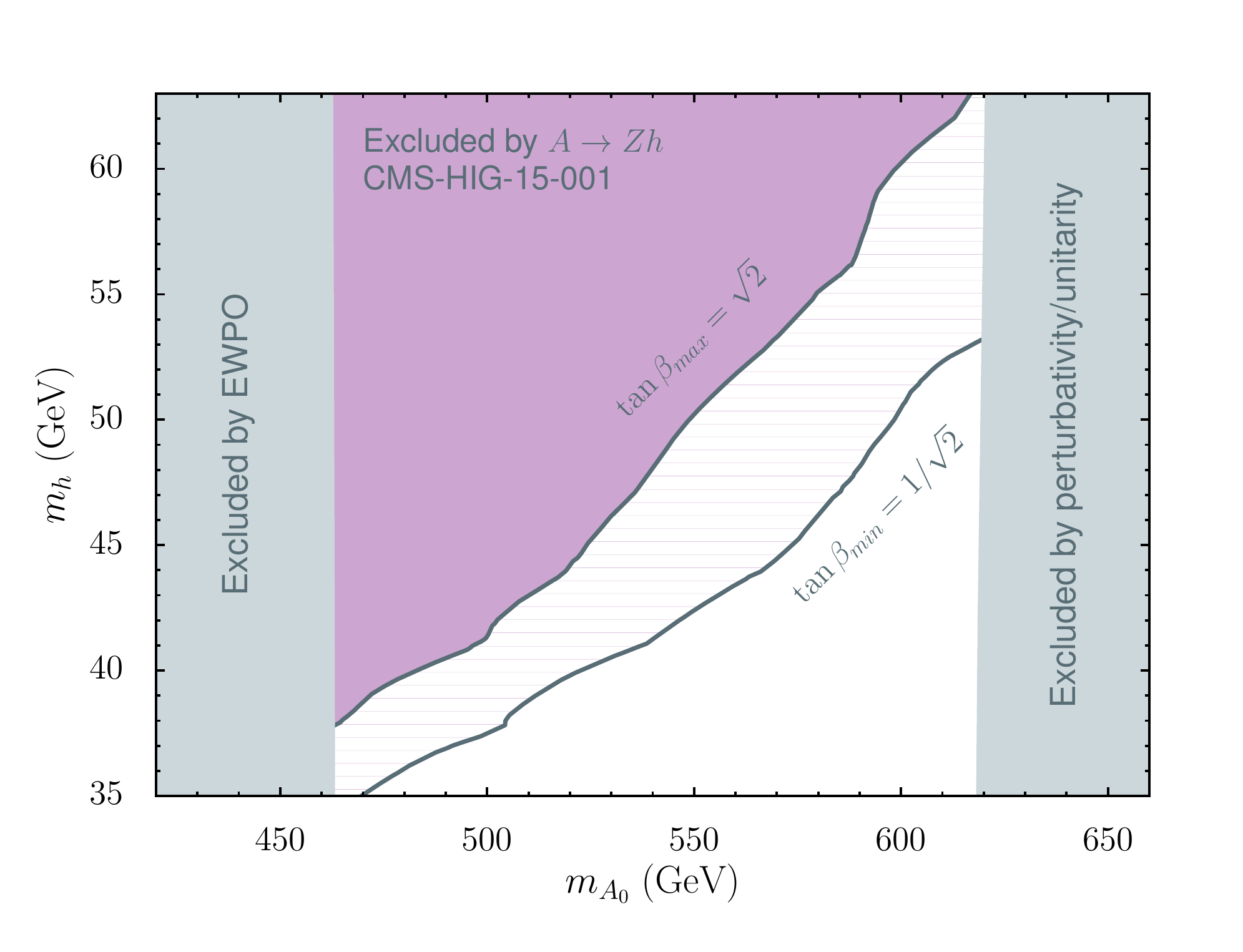}
\vspace{-0.5cm}
\caption{Bounds on a light neutral CP-even Higgs $h$ in the ($m_h,\,t_{\beta}$)-plane from Higgs searches at LEP \cite{Barate:2003sz} and
CMS searches of a light scalar produced in in $bb$-associated production decaying to $\tau\tau$ \cite{Khachatryan:2015baw} (covering masses $m_h < 80$ GeV)
and \cite{Khachatryan:2014wca}  (covering masses $m_h > 90$ GeV). The combination rules out Higgs masses $m_h < 75$ GeV.}
\label{MH125Alignment3}
\end{figure}

For $m_h < m_{H_0}/2$ the decay channel $H_0 \to hh$ becomes kinematically open and could lead to a large modification of the 125 GeV Higgs $H_0$ decay branching ratios, which 
is highly constrained by LHC data. Recalling the discussion in Section \ref{mh125} there is a bound $\mathcal{B}(H_0 \to hh) \lesssim 15$\%, requiring the trilinear 
coupling $g_{H_0hh}$, which in the alignment limit reads
\begin{equation}
g_{H_0hh}=-\frac{2m_{h}^2+m_{H_0}^2-4\mu^2/s_{2\beta}}{v},
\end{equation}
to approximately vanish. This in turn fixes the value of $\mu^2$ as a function of $m_h$ and $t_{\beta}$. The effect of this condition is to restrict the range of $t_{\beta}$ 
allowed by theoretical constraints. In particular we find that stability requires $t_{\beta} \in [1/\sqrt{2},\,\sqrt{2}]$. As only $t_{\beta} \sim 1$ values are allowed for 
$m_h < m_{H_0}/2$, the capability of the CMS $A_0\to Z h$ search~\cite{CMS:2015mba} to probe this region of parameter space greatly increases, completely ruling out a sizable portion 
of the mass plane ($m_{A_0},\, m_h$) as shown in Figure \ref{MH125Alignment3}.

\subsection{WRONG-SIGN LIMIT $c_{\beta+\alpha} = 1$}

\begin{figure}[!ht]\centering
\includegraphics[width=0.6\textwidth]{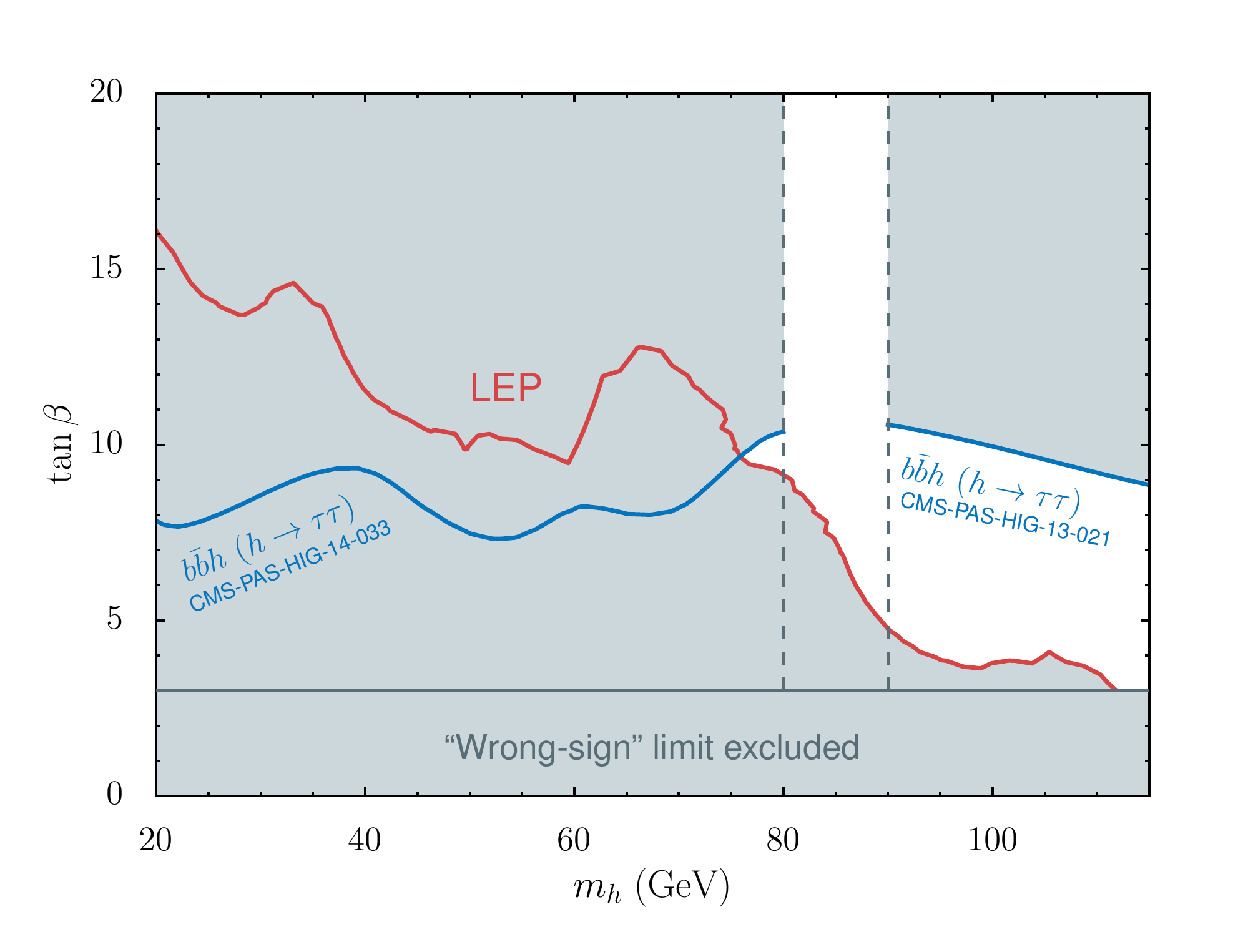}
\vspace{-0.5cm}
\caption{Bounds on a light neutral CP-even Higgs $h$ in the wrong-sign limit in the ($m_h,\,t_{\beta}$)-plane from Higgs searches at LEP \cite{Barate:2003sz} and
CMS searches of a light scalar produced in in $bb$-associated production decaying to $\tau\tau$ \cite{Khachatryan:2015baw} (covering masses $m_h < 80$ GeV)
and \cite{Khachatryan:2014wca}  (covering masses $m_h > 90$ GeV). The combination rules out Higgs masses $m_h < 75$ GeV.}
\label{MH125WS}
\end{figure}

For $m_h < 114.5$ GeV, the wrong-sign limit is strongly constrained by LEP searches for $h$ produced in association 
with a $Z$ boson, which bound the departure from alignment $s_{\beta-\alpha} = 0$ as a function of 
$m_h$ \cite{Barate:2003sz}. For $c_{\beta+\alpha} = 1$ this translates into a 
lower bound on $t_{\beta}$ for each mass $m_h$ in the range $m_h \in [20,\,114.5]$ GeV, shown in Figure \ref{MH125WS}. 
We stress that the wrong-sign limit only yields a 95\% C.L. allowed fit to the 125 GeV Higgs signal strength for 
$t_{\beta} \gtrsim 3$ (see e.g.~\cite{Dumont:2014kna,Bernon:2014vta,Dorsch:2016tab}).

At the same time, searches for neutral scalars produced in association with $b\bar{b}$ and decaying to $\tau\tau$ by ATLAS/CMS 
for $m_h > 90$ GeV \cite{Khachatryan:2014wca,Aad:2014vgg} and by CMS for $m_h \in [20,\,80]$ GeV \cite{Khachatryan:2015baw} yield a corresponding upper bound 
on $t_{\beta}$. As shown in Figure \ref{MH125WS}, the combination of LEP and CMS bounds rules out light Higgs masses up to $m_h = 75$ GeV. 
We note that the value of $\mu^2$ (and the constraint from $H_0\to h h$ decays) is not relevant here, as it does not affect the bounds 
discussed and these rule out the region $m_h < m_{H_0}/2$.

\section*{CONCLUSIONS}
Taking into account the most updated constraints on light neutral Higgs bosons, we have investigated the allowed parameter space in 
Type~II 2HDM with a neutral state below 125 GeV.  Either $h$, the light CP-even state, or $H_0$, the heavy CP-even state, were identified as the 125 GeV state. 

In the $m_h=125$~GeV case, notably, the CMS search for $H_0\to Z A_0\to\ell\ell b\bar{b}$ is very sensitive to the $m_{A_0}<m_h$ region and rules 
out large portions of the previously allowed parameter space, particularly for $m_h>m_{A_0}>m_h/2$ where values of $\tan\beta < 1.10$ are completely excluded. 
As a consequence, the $H_0\to A_0A_0$ channel turns out to be a powerful complementary probe of this region.

In the $m_{H_0}=125$~GeV case, we find that the CMS search for $A_0\to Z h\to\ell\ell b\bar{b}$ rules out a very large portion of the 
$m_{h}<m_{H_0}/2$ region, and highly constrains the allowed values of $\tan\beta$ for $m_{H_0} > m_{h}> m_{H_0}/2$. In addition, the ATLAS and CMS searches 
for $h\to \tau \tau$ in $bb$-associated production provide a strong constraint for high $\tan\beta$. 

In both scenarios, we find that the prospects for the LHC Run 2 at 13 TeV are excellent, and most of the 2HDM parameter space 
characterized by a light neutral scalar could be explored.

\section*{ACKNOWLEDGEMENTS}
This work was supported in part by the Research Executive Agency (REA) of the European Union under the Grant Agreement 
PITN-GA2012-316704 (HiggsTools), the ANR project DMASTROLHC, the ``Investissements d'avenir, Labex ENIGMASS'', 
the UK Science and Technology Facilities Council (STFC) under grant number ST/L000504/1, and the 
People Programme (Marie Curie Actions) of the European Union Seventh Framework Programme (FP7/2007-2013) under REA Grant Agreement PIEF-GA-2013-625809.

\AddToContent{J.~Bernon, K.~Mimasu, J.~M.~No and D.~Sengupta}
\renewcommand{\thesection}{\arabic{section}}

\graphicspath{{fiveplet/}}
\newcommand{\fref}[1]{Fig.~\ref{fig:#1}} 
\newcommand{\eref}[1]{Eq.~\eqref{eq:#1}} 
\newcommand{\aref}[1]{Appendix~\ref{app:#1}}
\newcommand{\sref}[1]{Sec.~\ref{sec:#1}}
\newcommand{\tref}[1]{Table~\ref{tab:#1}}  

\renewcommand{\be}{\begin{equation}}
\renewcommand{\ee}{\end{equation}}

\chapter{LHC diphoton and diboson probes of custodial fiveplet scalars}

{\it A.~Delgado, M.~Garcia-Pepin, M.~Quir\'{o}s, J.~Santiago and R.~Vega-Morales}


\begin{abstract}
We discuss diphoton and diboson probes of custodial fiveplet scalars.~We show for the first time that when the $W$ boson loop dominates the effective couplings to photons, a custodial fiveplet scalar below $\sim 110$~GeV is ruled out by 8 TeV LHC diphoton searches independently of the Higgs triplet VEV.~We also make rough estimates for 13~TeV. 
\end{abstract}

\section{INTRODUCTION}
Both the Georgi-Macachek (GM) model~\cite{Georgi:1985nv} and its supersymmetric generalization, the Supersymmetric Custodial Triplet Model (SCTM)~\cite{Cort:2013foa,Garcia-Pepin:2014yfa}, feature a set of light triplet-like scalars which is well within LHC reach and could thus provide a promising test avenue.~Moreover, for the particular case of the SCTM, it was argued in~\cite{Cort:2013foa} that a light mass for the triplet-like states is tied to a sizeable value of $v_\mathrm{triplet}$.~Therefore, searching for these scalars is also testing the nature of EWSB and probing the interesting properties of custodial Higgs triplet models.~It is then of great importance to perform a collider study searching for these triplet-like states.~The smoking gun of all custodial Higgs triplet models~\cite{Georgi:1985nv,Cort:2013foa,Garcia-Pepin:2014yfa,Chanowitz:1985ug,Gunion:1989ci,Gunion:1990dt,Aoki:2007ah,Chiang:2012cn,Chiang:2013rua,Hartling:2014zca,Logan:2015xpa,Delgado:2015aha,Delgado:2015bwa,Delgado:2015hmy,Garcia-Pepin:2016hvs} is the presence of a $SU(2)_V$ custodial fiveplet ($H_5$) scalar which features a $CP$-even neutral ($H_5^0$), singly ($H_5^\pm$), and doubly ($H_5^{\pm\pm}$) charged component all with degenerate masses.~Here we examine diphoton and diboson probes at the LHC of these fiveplet scalars. 

As the fiveplet does not couple to quarks (it is fermiophobic), production via gluon fusion is not available.~Furthermore, if the VEV of the fermiophobic Higgs is small (as compared to the SM-like Higgs doublet VEV), vector boson fusion (VBF) and associated Higgs vector boson production (VH) quickly become highly suppressed~\cite{Delgado:2016arn}.~Since these are the dominant production mechanisms in the SM, they have been assumed as the production mechanisms in almost all Higgs-like boson searches regardless of if they are fermiophobic or not.~On the other hand, since LHC measurements of the 125~GeV Higgs boson couplings seem to indicate a SM-like Higgs boson, this implies a small VEV for any additional exotic Higgs boson.~As these measurements increase in precision without observing a deviation from the SM prediction, previous collider searches for fermiophobic Higgs bosons, which assumed SM-like production mechanisms, become increasingly obsolete.

However, Drell-Yan (DY) Higgs pair production of the fiveplets is sizable even in the limit of small exotic Higgs vev.~Furthermore, since there is no $b\bar{b}$ decay to compete with, custodial fiveplets can have large branching ratios to vector boson pairs, in particular to photons.~This can be combined with DY pair production to place stringent constraints on the fiveplet Higgs bosons using multiphoton final states.~Actually, the $W$ boson mediated $H_5^{\pm}H_5^0$ production channel (see~\fref{HHprod}), followed by $H_5^{\pm} \to W^{\pm}H^{0}_5$ and $H^{0}_5 \to \gamma\gamma$ decays, leads to a $4\gamma + X$ final state, which has been proposed as a probe~\cite{Akeroyd:2003xi,Akeroyd:2005pr} of fermiophobic Higgs bosons at high energy colliders.~However, the $H^{\pm} \to W^{\pm}H^{0}_F$ decay requires a mass splitting between the charged and neutral Higgs.~In custodial Higgs triplet models, the neutral and charged Higgs scalars are predicted to be degenerate, thus the CDF $4\gamma + X$ search~\cite{Aaltonen:2016fnw} cannot be applied to this case.~We show for the first time that, when the $W$ boson loop dominates the effective couplings to photons, a custodial fiveplet scalar below $\sim 110$~GeV is ruled out by 8 TeV LHC diphoton searches independently of the Higgs triplet VEV.~Larger masses possibly up to $\sim 150$~GeV can also be ruled out if charged scalar loops produce large constructive contributions to the effective photon couplings.~We also find that diboson searches, and in particular $ZZ$ searches, may be useful for higher masses allowing us to potentially obtain limits again for custodial fiveplet masses up to $\sim 250$~GeV independently of the Higgs triplet VEV.

\section{THE MODEL}
The differences between the GM model and the SCTM are not relevant for our current study.~The crucial feature that both share, in addition to being easily made to satisfy constraints from electroweak precision data, is that after EWSB the Higgs triplets decompose into representations of the custodial $SU(2)_V$ global symmetry.~In particular, all custodial Higgs triplet models contain the aforementioned fermiophobic scalar ($H_5$) that transforms as a fiveplet under the custodial symmetry $SU(2)_V$ and can be very light.~We will thus only introduce the non-supersymmetric GM model since it is the most minimal realization of this situation and the results are directly extended to the supersymmetric SCTM.~We only focus on the fiveplet and the relevant couplings for our study.~Further details of these models can be found in~\cite{Georgi:1985nv,Cort:2013foa,Garcia-Pepin:2014yfa,Chanowitz:1985ug,Gunion:1989ci,Gunion:1990dt,Aoki:2007ah,Chiang:2012cn,Chiang:2013rua,Hartling:2014zca,Logan:2015xpa,Delgado:2015aha,Delgado:2015bwa,Delgado:2015hmy,Garcia-Pepin:2016hvs}.

In the GM model, two $SU(2)_L$ triplets scalars are added to the SM in such a way that the Higgs potential preserves a global $SU(2)_L\otimes SU(2)_R$ symmetry which is broken to the vector custodial subgroup $SU(2)_V$ after EWSB, predicting $\rho = 1$ at the tree-level~\cite{Georgi:1985nv}.~More specifically, on top of the SM Higgs doublet $H=(H^+,H^0)^T$, one real $SU(2)_L$ triplet scalar with hypercharge $Y=0$, $\phi=(\phi^+,\phi^0,\phi^-)^T$, and one complex triplet scalar with $Y = 1$,  $\chi=(\chi^{++},\chi^+,\chi^0)^T$, are added.~In terms of representations of $SU(2)_L\otimes SU(2)_R$ they transform as $({\bf 2,2})$ and $({\bf 3,3})$, respectively.

If EWSB proceeds such that $v_H\equiv \langle H^0\rangle\equiv v_{doublet}$, $v_\phi\equiv\langle \phi^0\rangle=v_\chi\equiv\langle \chi^0\rangle\equiv v_{triplet}$,~i.e.~the triplet VEVs are aligned, then $SU(2)_L\otimes SU(2)_R$ will be broken to the custodial subgroup $SU(2)_V$, which ensures that the $\rho$ parameter is equal to one at tree-level as in the SM.~We can also schematically define the SM doublet-exotic Higgs `VEV mixing angles' ($c_\theta \equiv \cos\theta,~s_\theta \equiv \sin\theta$),
\be
c_\theta \equiv \frac{v_{doublet}}{v}\, ,~
s_\theta\equiv \frac{v_{triplet}}{v}~~~(v = 246~\rm{GeV})\, .
\label{eq:sth}
\ee

If one neglects the tiny breaking generated by the hypercharge, all mass eigenstates are classified into $SU(2)_V$ multiplets degenerate in mass and in particular a singlet, triplet, and fiveplet.~The fiveplet, which we focus on, is a pure triplet-like state with no mixing with the doublet sector.~This ensures that the fiveplet is fermiophobic, i.e. it does not couple to fermions at tree level.~Due to hypercharge and Yukawa interactions, custodial breaking effects are introduced at one loop and can spoil the fermiophobic and degenerate mass conditions along with introducing dangerous deviations from $\rho=1$.~However, these effects are naturally small, allowing for these conditions to be maintained to a good approximation.

\subsection{Pair production of $H_5^0$}

The main focus of this study will be the $pp \to W^{\pm} \to H_5^0 H^{\pm}_5$ production channel shown in~\fref{HHprod}.~The relevant vertex for the Drell-Yan pair production~(\fref{HHprod}) reads
\be\label{eq:gwhh}
V_{WH_5H_5} \equiv i g \frac{\sqrt{3}}{2} (p_1 - p_2)^\mu\ ,
\ee
where we can see that the coupling does not depend on the triplet VEV and therefore DY is not suppressed even in the case when the triplet VEV is small.~We show in~\fref{Hprod} the cross section (solid blue) as a function of $H_5^0$ for the $H_5^0 H_5^{\pm}$ channel.~We see that it can be $\sim \mathcal{O}(100)$~fb all the way up to $\sim 200$~GeV at 8 TeV (dashed blue curve) while at 13 TeV (solid blue curve) it will be increased by roughly a factor of $\sim 2$.~If the fiveplet is instead produced in pair with a custodial triplet which is $100$~GeV heavier (dotted blue) the cross section is considerably reduced.~Note that there are also NLO contributions which may generate $\sim \mathcal{O}(1)$ K-factors for Higgs pair production~\cite{Eichten:1984eu,Dawson:1998py,Degrande:2015xnm}, but we do not explore this issue here as it does not qualitatively affect the discussion. 
%
\begin{figure}[tbh]
\begin{center}
\includegraphics[scale=.6]{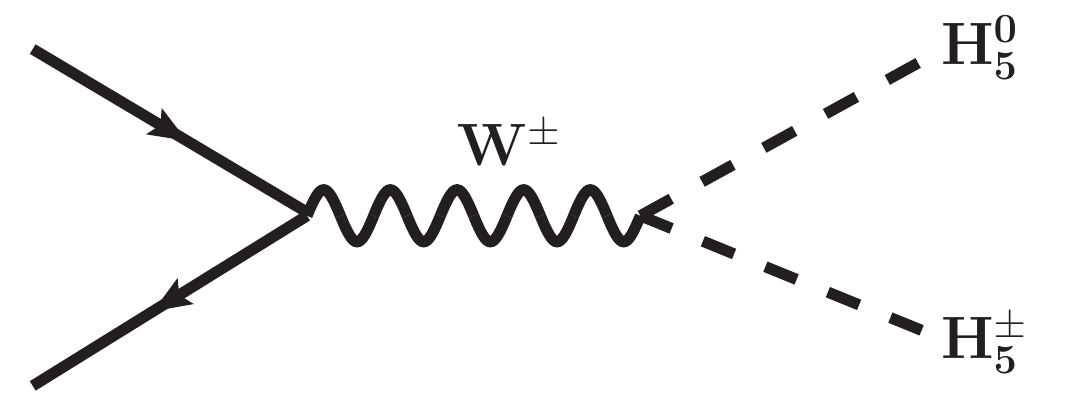}
\caption{The dominant contribution to custodial fiveplet scalar pair production.}
\label{fig:HHprod}
\end{center}
\end{figure}
\begin{figure}[tbh]
\begin{center}
\includegraphics[scale=.5]{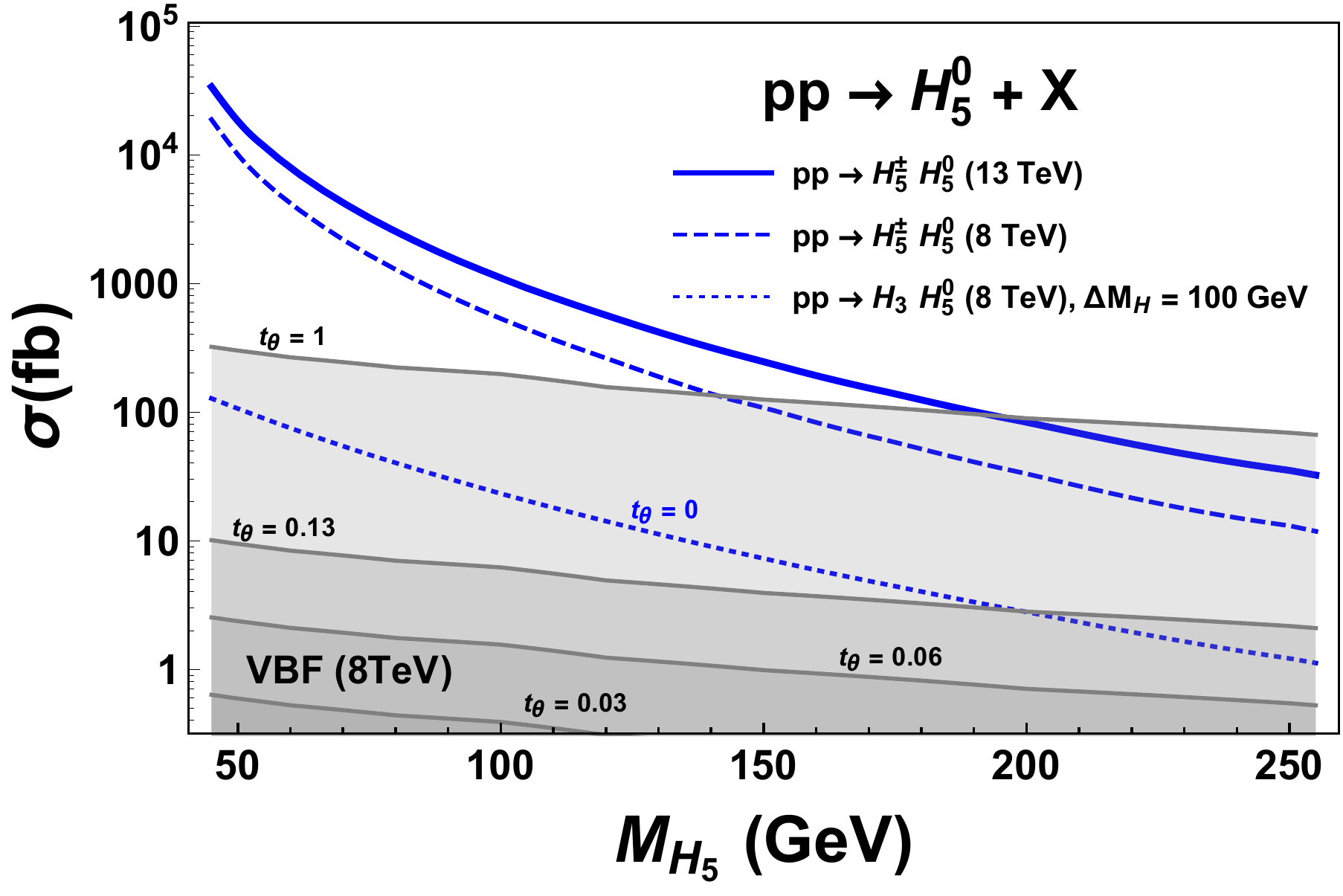}
\caption{Drell-Yan Higgs pair production cross sections for a custodial fiveplet scalar at the LHC with $\sqrt{s} = 8$ TeV (dashed blue curve) and $\sqrt{s} = 13$ TeV (thick blue solid curve).~We also show the case where the fiveplet is produced along with a custodial triplet (blue dotted) which is 100~GeV more massive (see text for more information).} 
\label{fig:Hprod}
\end{center}
\end{figure}
 
To demonstrate the utility of the DY Higgs pair production mechanism we show for comparison results for VBF single $H_5^0$ production.~We see clearly that once the measurements of the Higgs boson at 125~GeV constrain $s_\theta \ll 1$, the VBF production channel quickly becomes highly suppressed relative to the DY Higgs pair production.~Similar behavior can be seen for the $VH$ production channels which are typically smaller than the VBF cross sections except at very low masses~\cite{Dittmaier:2011ti,Dittmaier:2012vm,Heinemeyer:2013tqa}.

To summarize, we see that $\sim\mathcal{O}(100)$~fb cross sections are obtained for the $pp \to H_5^0 H_5^{\pm}$ Higgs pair production channel in the mass range $45 - 250$~GeV.~Crucially, this production mechanism is independent of the Higgs triplet VEV unlike VBF and VH production.~As we will see, diphoton and diboson searches at the 8 TeV LHC are sensitive to  $\sim\mathcal{O}(100)$~fb cross section times branching ratios.~Thus if the branching ratios to dibosons are large, searches at the LHC for pairs of photons or $Z$ and $W$ bosons should be able to probe the fiveplet in this mass range.

\subsection{Decay of $H_5^0$}
\begin{figure}[tbh]
\begin{center}
\includegraphics[scale=.6]{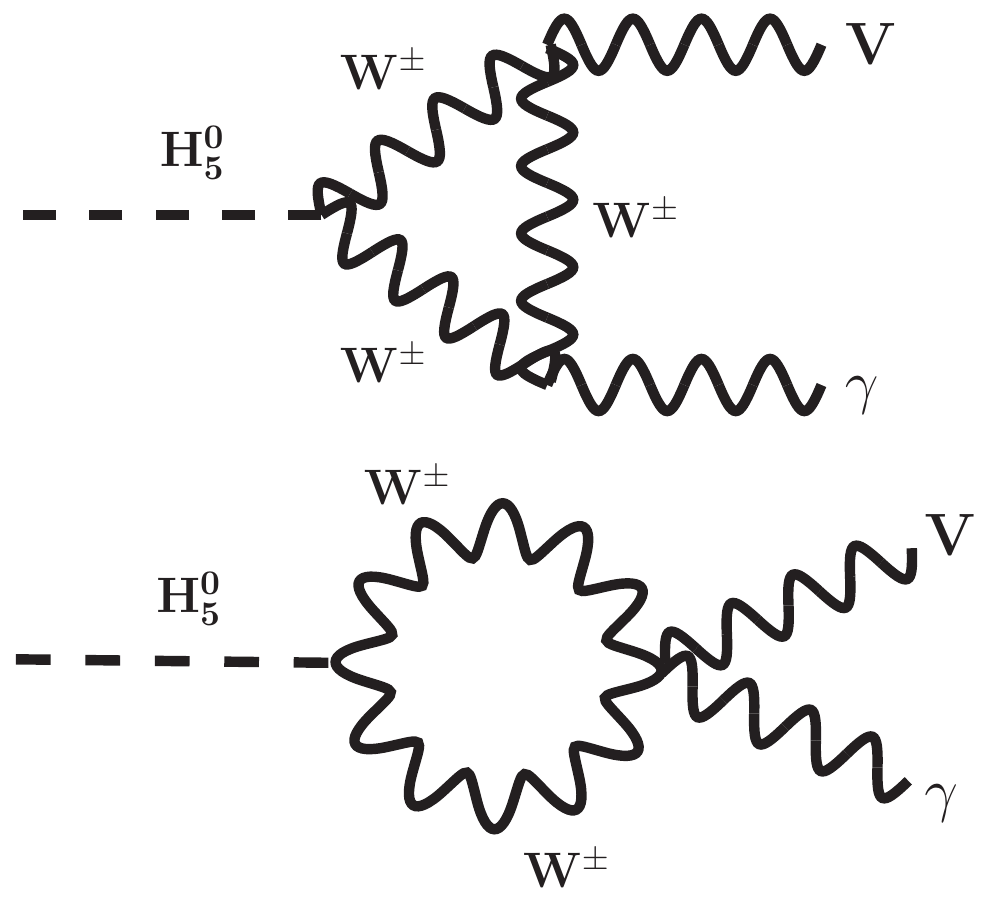}
\caption{One loop contributions from $W$ boson loops to the $H_5^0 \to V\gamma$ decays ($V = Z, \gamma$).}
\label{fig:HtoVA}
\end{center}
\end{figure}

In addition to the $WH_5H_5$ vertex of Eq.~(\ref{eq:gwhh}) , $H_5^0$ will have tree level couplings to $WW$ and $ZZ$ pairs which are generated during EWSB and which will be proportional to the triplet Higgs VEV.~These can be parametrized as,
\be
\label{eq:LZW}
\mathcal{L}
\supset
s_\theta
\frac{H_5^0}{v} 
\Big( g_{Z} m_Z^2 Z^\mu Z_\mu + 2 g_{W} m_W^2 W^{\mu+} W^-_{\mu} 
\Big) ,
\ee
where $g_Z = 4/\sqrt{3}$ and $g_W = -2/\sqrt{3}$.~The ratio $\lambda_{WZ} = g_{Z}/g_{W}$ is fixed by custodial symmetry to be $\lambda_{WZ} = -1/2$~\cite{Low:2010jp}.~At one loop the couplings in Eq.~(\ref{eq:LZW}) will also generate effective couplings to $\gamma\gamma$ and $Z\gamma$ pairs via the $W$ boson loops shown in Fig.~\ref{fig:HtoVA}.~We can parametrize these couplings with the effective operators,
\be
\label{eq:LZA}
\mathcal{L}
\supset
\frac{H_5^0}{v} 
\Big( 
\frac{c_{\gamma\gamma} }{4} 
F^{\mu\nu} F_{\mu\nu} +
\frac{c_{Z\gamma}}{2} 
Z^{\mu\nu} F_{\mu\nu} 
\Big) \, ,
\ee
where $V_{\mu\nu}=\partial_\mu V_\nu - \partial_\nu V_\mu$.~We again define similar ratios,
\be
\lambda_{V\gamma} = c_{V\gamma}/g_{Z}\, ,
\label{eq:lamVA}
\ee
where $V = Z, \gamma$ and we have implicitly absorbed a factor of $s_\theta$ into $c_{V\gamma}$.~There are also contributions to the effective couplings in Eq.~(\ref{eq:LZA}) from the additional charged Higgs bosons which are necessarily present in the GM model and the SCTM.~These contributions can be large or small depending on the model and parameter choice.~They can in principle lead to large enhancements~\cite{Akeroyd:2012ms} when there is constructive interference with the $W$ boson loop, or suppressions if there are cancellations between the different contributions~\cite{Arhrib:2011vc,Kanemura:2012rs}, leading to small $c_{V\gamma}$ effective couplings.
\begin{figure}[tbh]
\begin{center}
\includegraphics[scale=.6]{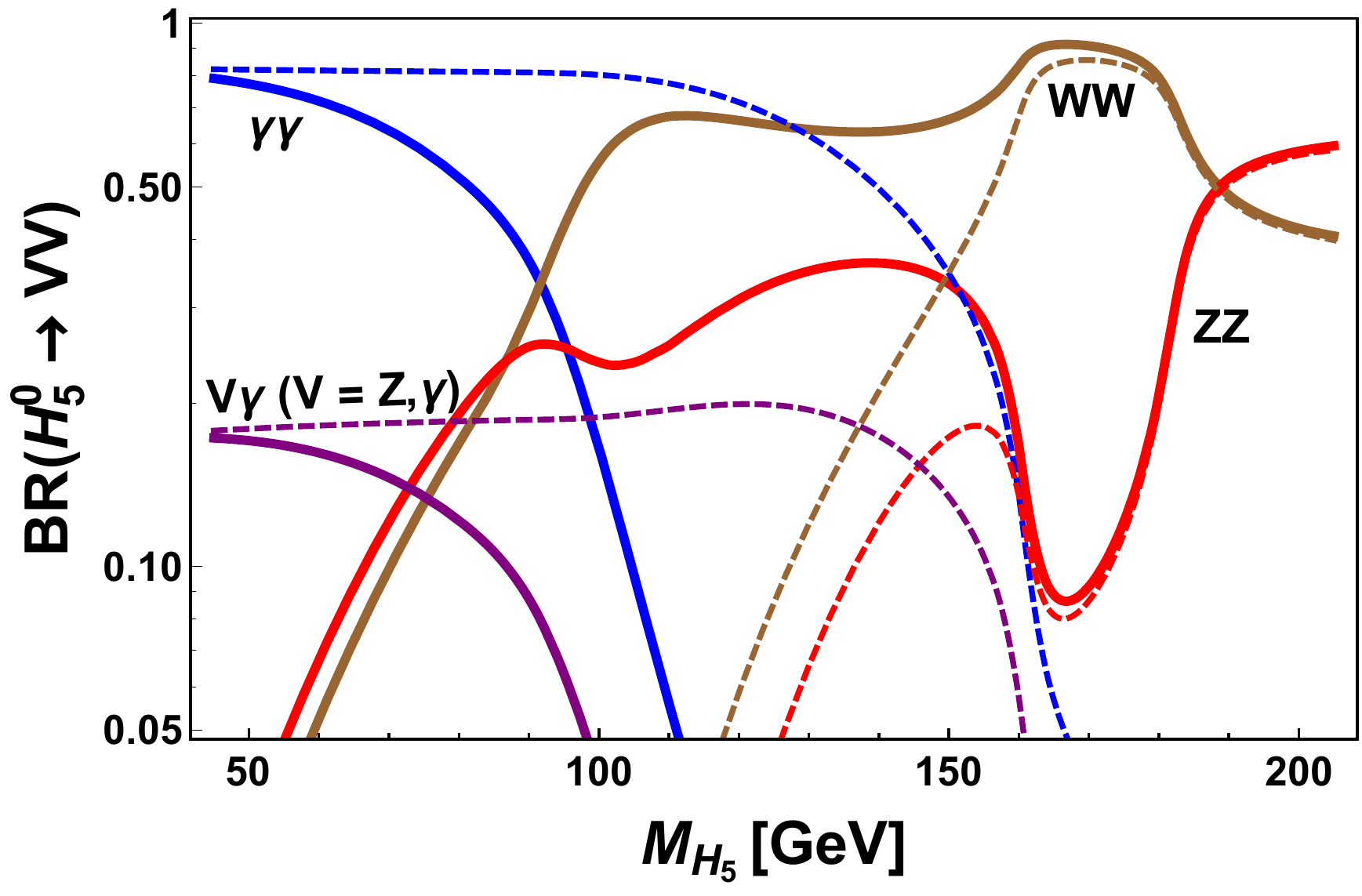}
\caption{Branching ratios for $H_5^0$ as a function of its mass.}
\label{fig:HBR}
\end{center}
\end{figure}

We show the branching ratios of $H_5^0$ in~\fref{HBR}.~To obtain the three and four body decays we have integrated the analytic expressions for the $H_5^0 \to V\gamma \to 2\ell\gamma$ and $H_5^0 \to VV \to 4\ell$ fully differential decay widths computed and validated in~\cite{Chen:2012jy,Chen:2013ejz,Chen:2014ona}.~For the explicit $W$ loop functions which contribute to the effective couplings we use the parametrization and implementation found in~\cite{Chen:2015rha}.~The branching ratios will only depend on the ratios $\lambda_{WZ}$ and $\lambda_{V\gamma}$, and in some cases only on $\lambda_{WZ}$ if the $W$ loop (see Fig.~\ref{fig:HtoVA}) dominates the $H_5^0V\gamma$ effective couplings (solid curves).~In this case any $s_\theta$ dependence in $\lambda_{V\gamma}$ cancels explicitly.~At low masses, below $\sim 100$~GeV, the branching ratio into pairs of photons starts to become significant and quickly dominant below the $W$ mass, or at higher masses if the couplings to photons are enhanced (dashed lines).~We note that these branching ratios include the $\gamma^\ast\gamma$ contribution which, as shown in~\fref{HBR}, can be sizeable at low masses.~At larger masses the three and four body decays involving $W$ and $Z$ bosons become relevant and eventually completely dominant above the $WW$ and $ZZ$ thresholds.

\section{Closing the `Fiveplet Window'}
In~\fref{prodxbr} we show the $pp \to W^{\pm} \to H_5^{\pm} H_5^0$ production cross section times branching ratio for a custodial fiveplet decay into photon (blue), $WW$ (brown), and $ZZ$ (red) pairs at 8 TeV (top) and 13 TeV (bottom).~We also show the limits (dashed lines) coming from ATLAS diphoton searches at 8 TeV~\cite{Aad:2014ioa} (blue) as well as CMS $7 + 8$~TeV searches~\cite{Khachatryan:2015cwa} for decays to $WW$ (brown) and $ZZ$ (red).~To estimate the limits at 13 TeV we have simply rescaled the 8 TeV limits by a factor of 2 which is roughly the increase in Higgs pair production cross section.~Our leading order results for the $pp \to W^{\pm} \to H_5^{\pm} H_5^0$ production cross sections are calculated using the Madgraph/GM model implementation from~\cite{Alwall:2014hca,Hartling:2014xma}.~The branching ratios are obtained from the partial widths into $\gamma\gamma, V^\ast\gamma~(V = Z, \gamma), WW$, and $ZZ$ which are computed for the mass range $45 - 250$~GeV.
%
\begin{figure}
\begin{center}
\includegraphics[scale=.7]{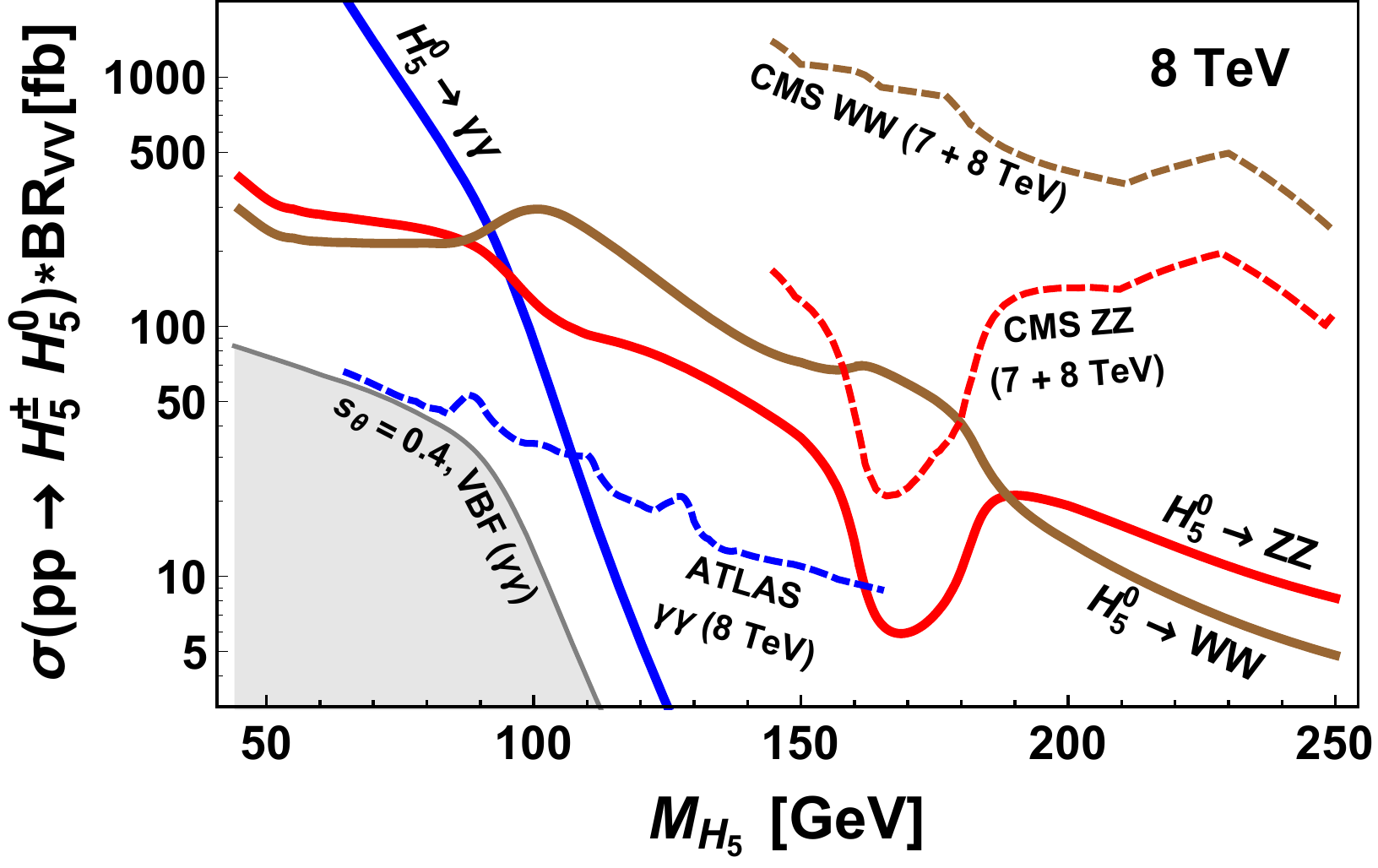}
\includegraphics[scale=.7]{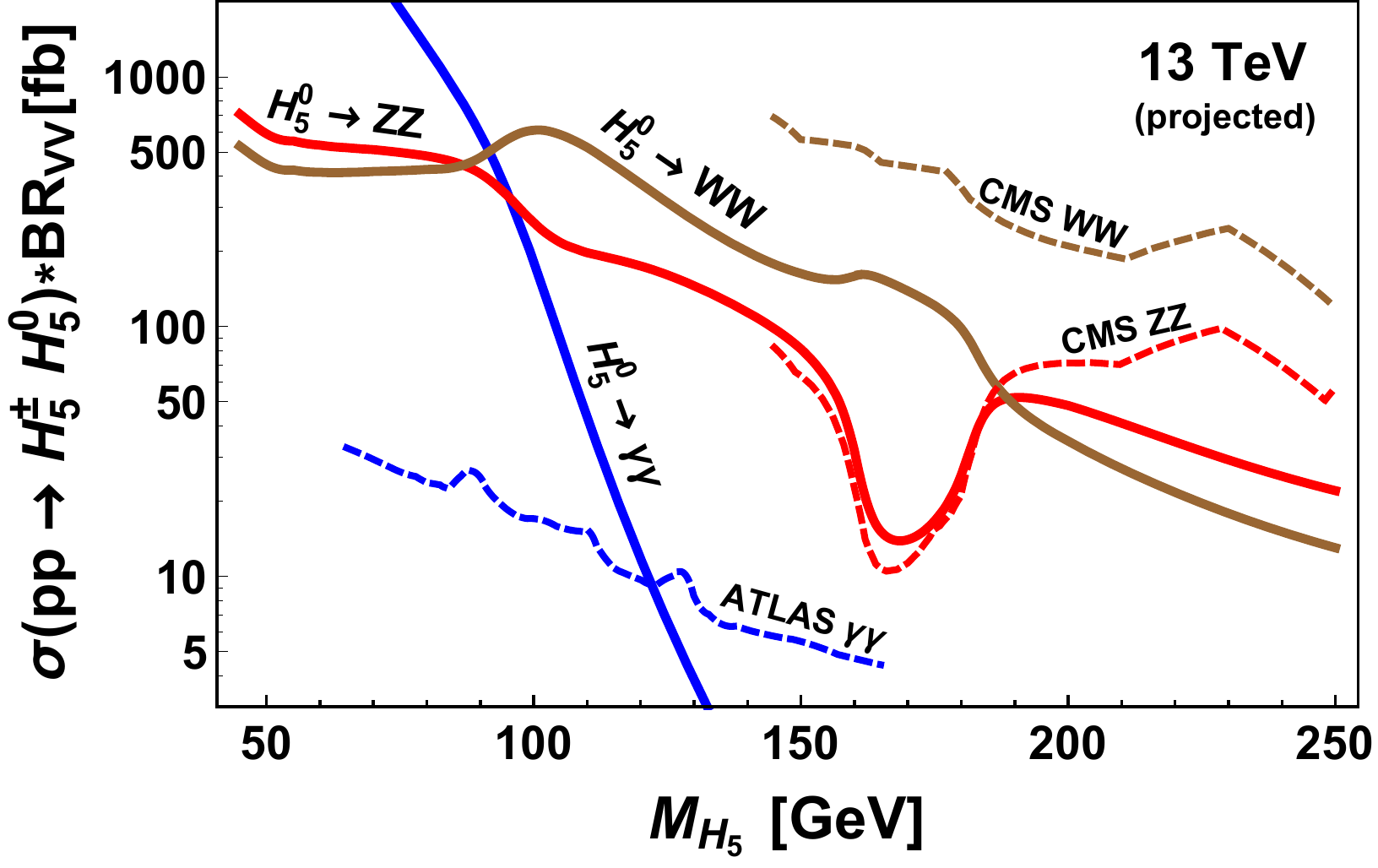}
\caption{{\bf Top:~}Drell-Yan $H^{0}_5 H^{\pm}_5$ production cross sections times branching ratio at 8 TeV (solid curves) into $\gamma\gamma$ (blue), $ZZ$ (red), and $WW$ (brown) for the fermiophobic fiveplet found in custodial Higgs triplet models.~The $95\%$ exclusion limits (dashed curves) from diphoton~8 TeV ATLAS~\cite{Aad:2014ioa} and $7 + 8$~TeV CMS $WW$ and $ZZ$ searches~\cite{Khachatryan:2015cwa} are also shown for each channel.~In the gray shaded region we show for comparison the $s_\theta = 0.4$ contour for single $H_5^0$ VBF production (see text).~{\bf Bottom:~}The same for the 13 TeV LHC.~For the 13 TeV limits we have simply rescaled 8 TeV limits by a factor of 2.}
\label{fig:prodxbr}
\end{center}
\end{figure}
%

We focus on the regime where the effective couplings of the fiveplet to $\gamma\gamma$ and $Z\gamma$ are dominated by the $W$ loop contribution shown in~\fref{HtoVA}.~The effects of the charged scalar sector could in principle be large~\cite{Akeroyd:2012ms} leading to enhanced or suppressed effective couplings to photons.~As discussed above, this can affect the upper limit of masses which can be ruled out and could in principle allow for masses up to the $WW$ threshold to be ruled out by diphoton searches.~Since these effects are more model dependent we do not consider them here.

We see in the top of~\fref{prodxbr} that by exploiting the $H^{0}_5 H^{\pm}_5$ Higgs pair production mechanism, custodial fiveplet scalars with masses $\sim 107$~GeV can be ruled out by 8 TeV diphoton searches, independently of the Higgs triplet VEV.~These are the first such limits on custodial fiveplet scalars and in particular, since the charged and neutral components are degenerate, limits from Tevatron $4\gamma + X$ searches~\cite{Aaltonen:2016fnw} do not apply.~This is because, for the cases of the custodial fiveplet where the masses are degenerate, the $H_5^{\pm} \to H_5^0 W^{\pm}$ decay is not available.~In this case the one loop $H_5^{\pm} \to W^{\pm} \gamma$ decay can become dominant leading instead to a $3\gamma + W$ signal.~Examining this decay as well should improve the sensitivity relative to LHC diphoton searches, but we do not explore that here.

To emphasize the utility of the DY pair production mechanism,~we also show in the top of~\fref{prodxbr} the cross section times branching ratio assuming the VBF production (gray shaded region) mechanism at 8 TeV.~We have fixed $s_\theta = 0.4$ for the doublet-triplet VEV mixing angle as defined in~\cite{Hartling:2014xma} and schematically in Eq.~(\ref{eq:sth}).~The value $s_\theta = 0.4$ is close to the upper limit of values still allowed by electroweak precision and 125~GeV Higgs data~\cite{Hartling:2014aga,Chiang:2015amq,Fabbrichesi:2016alj,Chiang:2016ydx}, but we can see in~\fref{prodxbr} that this already renders diphoton searches for custodial fiveplet scalars based on VBF (and similarly for VH) production irrelevant.

We also emphasize that ruling out a custodial fiveplet below $\sim 110$~GeV independently of the VEV allows us to unambiguously close the fiveplet `window' at masses below $\sim 100$~GeV~\cite{Logan:2015xpa} which is still allowed by electroweak precisions data~\cite{Englert:2013zpa} and essentially unconstrained by other LEP, Tevatron, and LHC direct searches.~Thus we are able to rule out an interesting region of parameter space of custodial Higgs triplet models which would otherwise be difficult to constrain directly.~We estimate in the bottom of~\fref{prodxbr} that 13 TeV diphoton searches will be sensitive to scalar masses up to $\sim 125$~GeV in the regime of dominant $W$ boson loop, though NLO Higgs pair production effects~\cite{Degrande:2015xnm} may allow this to be extended further.~The diphoton search discussed here may of course be useful for other scalars which are found in custodial Higgs triplet models, but we do not explore this here. 

Finally, we also see in~\fref{prodxbr} that $WW$ and $ZZ$ searches may be useful for probing custodial fiveplet scalars independently of the Higgs triplet VEV as well.~Though 8 TeV searches are not quite sensitive, larger Higgs pair production cross sections at 13 TeV (see~\fref{HHprod}) should allow for fiveplet masses well above diphoton limits to be probed and possibly as high as $\sim 250$~GeV.~In particular, as we can see in the bottom of~\fref{prodxbr}, the $ZZ$ channel should become sensitive with early 13 TeV data for masses around the $ZZ$ threshold.~This also serves as a useful compliment to $W^+W^+$ searches for the doubly charged component of the custodial fiveplet~\cite{Englert:2013wga}.

\section*{CONCLUSIONS}
We have examined the particular case of a custodial fiveplet scalar found in all incarnations of custodial Higgs triplet models~\cite{Georgi:1985nv,Hartling:2014zca,Cort:2013foa} in which the neutral and charged component are predicted to be degenerate.~We have shown for the first time that a custodial fiveplet scalar below $\sim 110$~GeV is ruled out by 8 TeV diphoton searches and possibly up to higher masses if charged scalar loops produce large constructive contributions to the effective photon couplings.~These limits are also largely independent of the Higgs triplet VEV and so robustly close the `fiveplet window' at masses below $\sim 110$~GeV~\cite{Logan:2015xpa}, still allowed by electroweak precision and 125~GeV Higgs boson data.~We also find that diboson searches, and in particular $ZZ$ searches, may be useful for larger fiveplet masses, allowing us to potentially obtain limits again independently of the Higgs triplet VEV.

\section*{ACKNOWLEDGEMENTS}
We would like to thank  the Ecole de Physique des Houches for creating a stimulating atmosphere where this project was started.

\AddToContent{A.~Delgado, M.~Garcia-Pepin, M.~Quir\'{o}s, J.~Santiago and R.~Vega-Morales}
\renewcommand{\thesection}{\arabic{section}}

\graphicspath{{HtoWW/}}

\renewcommand{\fref}[1]{Fig.~\ref{fig:#1}} 
\renewcommand{\eref}[1]{Eq.\eqref{eq:#1}} 
\renewcommand{\aref}[1]{Appendix~\ref{app:#1}}
\renewcommand{\sref}[1]{Sec.~\ref{sec:#1}}
\newcommand{\ssref}[1]{Sec.~\ref{subsec:#1}}
\renewcommand{\tref}[1]{Table~\ref{tab:#1}}

\renewcommand{\beq}{\begin{equation}}
\renewcommand{\eeq}{\end{equation}}
\renewcommand{\bea}{\begin{eqnarray}}
\renewcommand{\eea}{\end{eqnarray}}
\newcommand{\nn}{\nonumber}
\newcommand{\gev}{{\mathrm GeV}}
\newcommand{\hc}{\mathrm{h.c.}}
\newcommand{\eps}{\epsilon}

\chapter{Probing effective Higgs couplings at the LHC in $h \to 2\ell2\nu$ decays with multi-dimensional matrix element and likelihood methods}
{\it Y.~Chen, A.~Falkowski and R.~Vega-Morales}


\begin{abstract}
We examine the possibility of probing anomalous Higgs couplings at the LHC in $h \to WW \to 2\ell2\nu$ decays using multi-dimensional matrix element methods.~We describe broadly a likelihood framework which can be used to probe effective Higgs couplings to $WW$ pairs at the LHC or future colliders.~As part of this we compute the $h\to WW \to 2\ell2\nu$ fully differential decay width and discuss the observables available in the $2\ell2\nu$ final state.
\end{abstract}

\section{INTRODUCTION}
The discovery of a 125 GeV boson at the Large Hadron Collider (LHC)~\cite{Aad:2012tfa,Chatrchyan:2012xdj} completes the search for all propagating degrees of freedom predicted by the SM.~The quest to uncover the precise properties of this `Higgs' boson is now underway.~Some One of the most important properties of the Higgs boson to pin down precisely consists of the nature of its couplings to the electroweak gauge bosons.~These couplings contain important information about electroweak symmetry breaking and may offer clues towards beyond the standard model physics.~Thus probing them with the highest precision possible is crucial and a vigorous program for studying them at the LHC is already underway.

At present most of the progress in probing effective couplings, in particular their CP properties, has come via the fully reconstructable Higgs decay to four leptons for which many studies~\cite{Nelson:1986ki,Soni:1993jc,Chang:1993jy,Barger:1993wt,Arens:1994wd,Choi:2002jk,Buszello:2002uu,Godbole:2007cn,Kovalchuk:2008zz,Cao:2009ah,Gao:2010qx,DeRujula:2010ys,Gainer:2011xz,Campbell:2012cz,Campbell:2012ct,Belyaev:2012qa,Coleppa:2012eh,Bolognesi:2012mm,Boughezal:2012tz,Stolarski:2012ps,Avery:2012um,Chen:2012jy,Modak:2013sb,Gainer:2013rxa,Grinstein:2013vsa,Sun:2013yra,Anderson:2013fba,Chen:2013waa,Buchalla:2013mpa,Chen:2013ejz,Gainer:2014hha,Chen:2014gka,Chen:2014pia,Chen:2015iha,Bhattacherjee:2015xra,Gonzalez-Alonso:2015bha} have been conducted.~In four lepton decays, one can probe directly the effective couplings of the Higgs boson to photons and $Z$ bosons.~These studies have shown that Higgs to four lepton decays are useful probes of the tensor structure and CP nature of these effective couplings.

Less emphasized has been Higgs decays mediated by $WW$ pairs as a probe of its effective couplings to $W$ bosons.~This channel is more difficult than the four lepton final state since, due to the presence of neutrinos, it is not fully reconstructable.~There has been some progress so far in the opposite flavor $h \to WW \to e\nu\mu\nu$ channel using kinematic discriminants~\cite{Bolognesi:2012mm,Artoisenet:2013puc,Khachatryan:2014kca} or boosted decision trees (BDT)~\cite{ATLAS:2014aga,Aad:2015rwa}.~These studies indicate that $h \to WW \to 2\ell2\nu$ decays can in principle also be used as a way of probing anomalous Higgs couplings at the LHC.

In this note we examine the possibility of adapting analytic based matrix element methods (MEM) which have been developed for $h\to4\ell$ and $h\to2\ell\gamma$ decays~\cite{Gainer:2011xz,Stolarski:2012ps,Chen:2012jy,Chen:2014gka,Chen:2014pia,Chen:2015iha,Chen:2015rha,Khachatryan:2014kca} to the process $h \to WW \to 2\ell2\nu$.~We initiate this by broadly outlining the construction of a multi-dimensional likelihood analysis framework which can be used to perform s multi-parameter extraction and establish the CP properties of the effective Higgs couplings to $W$ bosons.~In addition we discuss the parametrization of the effective Higgs couplings as well as the kinematic observables available in the $2\ell2\nu$ final state.

\begin{figure}
\begin{center}
\includegraphics[scale=.6]{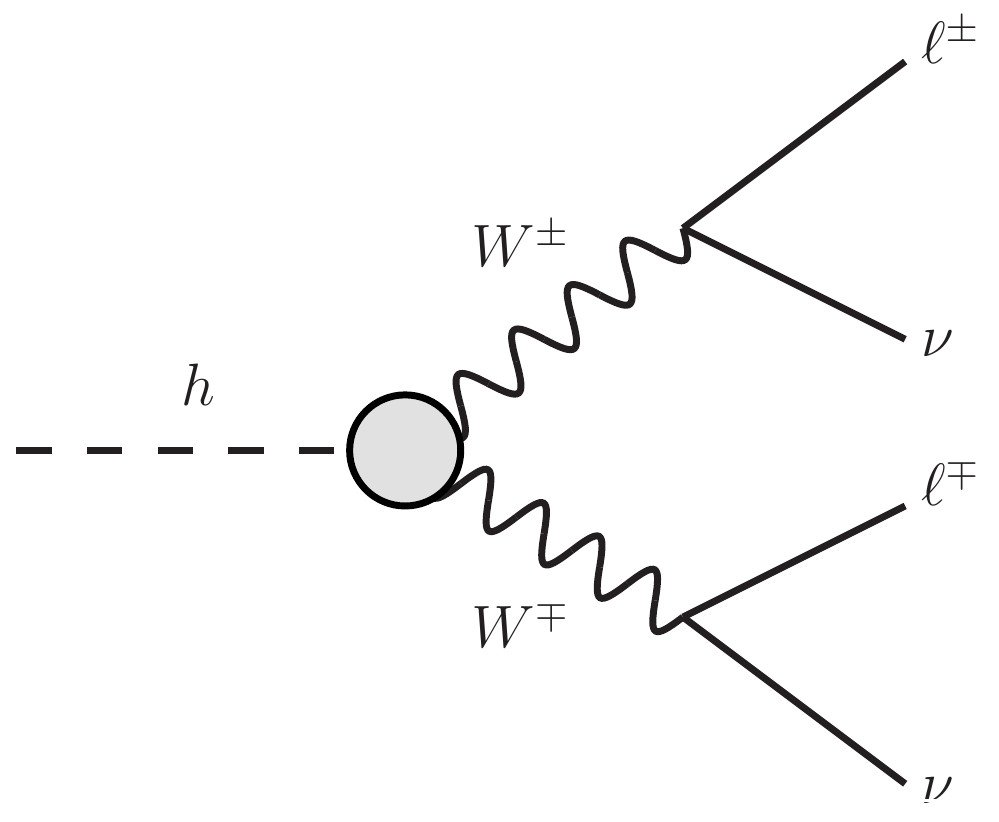}
\end{center}
\caption{Schematic representation of the effective Higgs couplings to $WW$ pairs which generate $h\to2\ell2\nu$ decays.}
\label{fig:htoww}
\end{figure}

\section{EFFECTIVE COUPLINGS AND DIFFERENTIAL DECAY WIDTH}

Here we discuss the parametrization of the effective Higgs couplings to $W$ boson pairs.~We also describe the analytic computation of the parton level $h\to WW\to 2\ell2\nu$ fully differential cross section. 

\subsection{TENSOR STRUCTURE AND EFFECTIVE VERTEX}

The couplings of the Higgs boson to a pair of $W$ bosons can be parameterized with the following tensor structure,
\bea
\label{eq:vert}
 \Gamma_{hWW}^{\mu \nu}  &=&  {1 \over v} 
\Big(  
 A_1^{WW} m_W^2 g_{\mu\nu}  +   A_2^{WW} (k_1^\nu k_2^\mu - k_1\cdot k_2 g^{\mu\nu}) \nn\\
&+&  A_3^{WW} \epsilon^{\mu\nu\alpha\beta} k_{1\alpha} k_{2\beta}
+ (A_4^{WW} k_1^2  + \bar A_4^{WW} k_2^2 )g_{\mu \nu}  \Big) . 
\eea  
The variables $k_1$ and $k_2$ represent the four momentum of the intermediate $W$ bosons and $v$ is the Higgs vacuum expectation value (vev) which we have chosen as our overall normalization.~The $A_{WW}^{i}$ are dimensionless and at this point, they are arbitrary complex form factors with a possible momentum dependence (or more precisely a $\hat{s}, k_1^2, k_2^2$ dependence), making the effective vertex of~Eq.~\eqref{eq:vert} completely general.

\subsection{DIFFERENTIAL DECAY WIDTH AND PARTIAL SUB-RATES}

With the tensor structure in~Eq.~\eqref{eq:vert}, we then analytically compute the fully differential decay width for $h\to 2\ell 2\nu$ as a sum of terms which are quadratic in the $A_{n}^{WW}$ form factors.~This can be written as,
\be
\label{eq:diffwidth}
  \frac{ d\Gamma_{h\rightarrow 2\ell2\nu} }
    { d\vec{\mathcal{P}} }
= \Pi_{2\ell2\nu}
|\mathcal{M}(h\to2\ell2\nu)|^2 ,
\ee
where $\Pi_{2\ell2\nu}$ is the $2\ell2\nu$ four body phase space and $d\vec{\mathcal{P}}$ is the four body fully differential volume element which will be discussed more below.

We have defined the matrix element squared as,
\bea\label{eq:mufmuf}
|\mathcal{M}(h\to2\ell2\nu)|^2
= 
\sum_{nm} 
A_n^{WW} A_{m}^{WW\ast} 
\times
|\mathcal{M}(h\to2\ell2\nu)|_{nm}^2 ,
\eea
where $n,m = 1, 2, 3, 4$.~Each term in this sum makes up a differential `sub-matrix element' squared, but note they need not be positive in the case of interference terms.~We will define these sub-matrix elements as,
\bea
\label{eq:subMM}
|\mathcal{M}|_{nm}^2 
\equiv 
A_n^{WW} A_{m}^{WW} 
\times
|\mathcal{M}(h\to2\ell2\nu)|_{nm}^2 .
\eea

At the level of the fully differential decay width, the decay $h \to W W \to 2\ell2\nu$ is similar to the $h \to ZZ \to 2e2\mu$ fully differential one due to the identical tensor structures of the effective Higgs couplings.~Of course the $Z$ propagator and couplings are now replaced by those for the $W$ boson.

With analytic expressions for the fully differential decay width in~Eq~\eqref{eq:vert} in hand, we could in principle define the same set of center-of-mass (CM) observables as in $h\to4\ell$~\cite{Chen:2012jy} and conduct similar MEM analyses of effective couplings as done in~\cite{Gainer:2011xz,Stolarski:2012ps,Chen:2012jy,Chen:2014gka,Chen:2014pia,Chen:2015iha,Chen:2015rha,Khachatryan:2014kca}.~However, as will be discussed further below, the fully differential decay width in~Eq.~\eqref{eq:diffwidth} must be integrated over the invisible neutrino momenta.~As will also be discussed below, this will require combining the fully differential decay width with an appropriate production spectrum followed by a four dimensional integration over invisible degrees of freedom.

\subsection{EFFECTIVE FIELD THEORY CONSTRAINTS}

We can demand that the effective couplings in~Eq.~\eqref{eq:vert} arise from an $SU(2)_L\times U(1)_Y$ invariant theory at higher energies.~Then we can write the effective lagrangian consisting of operators up to dimension five (in the `Higgs' basis~\cite{HXSWGbasis,Gupta:2014rxa}) as,
\begin{eqnarray}
\label{eq:hvvlag}
\mathcal{L}  &=& \frac{h}{v} 
\Big(
2 \delta c_w   m_W^2 W_\mu W^{\mu}    
+ c_{ww}  \frac{g_L^2}{2} W_{\mu \nu}  W^{\mu\nu}  \nonumber\\
&+& \tilde c_{ww}  \frac{g_L^2}{2} W_{\mu \nu}  
\widetilde{W}^{\mu\nu} + c_{wB} g_L^2 
\left( W_\mu \partial_\nu W^{\mu \nu} + h.c. \right)
\Big) , 
\end{eqnarray}
where the $c_n$ are now real and momentum independent.~This implies the relation between the effective vertex couplings of the tensor structures in Eq.~\eqref{eq:vert} and the lagrangian effective couplings in
Eq.~\eqref{eq:hvvlag},
\bea
\label{eq:AWW}
A_1^{WW} = 2 (1 + \delta c_w),~~
A_2^{WW} = g_L^2 c_{ww},~~  
A_3^{WW} =  g_L^2 \tilde c_{ww},~~  
A_4^{WW} =   \bar A_4^{WW}  = g_L^2 c_{wB}~.
\eea
The couplings in~Eq.~\eqref{eq:hvvlag} can also be expressed in terms of the so called `Warsaw' basis~\cite{Grzadkowski:2010es}.~Once the fully differential decay width is obtained as the sum in~Eq.~\eqref{eq:mufmuf} transforming from basis to another is trivial.~This is especially useful for parameter extraction of effective Higgs couplings.

\section{OBSERVABLES}
\begin{figure}
\begin{center}
\includegraphics[width=0.42\textwidth]{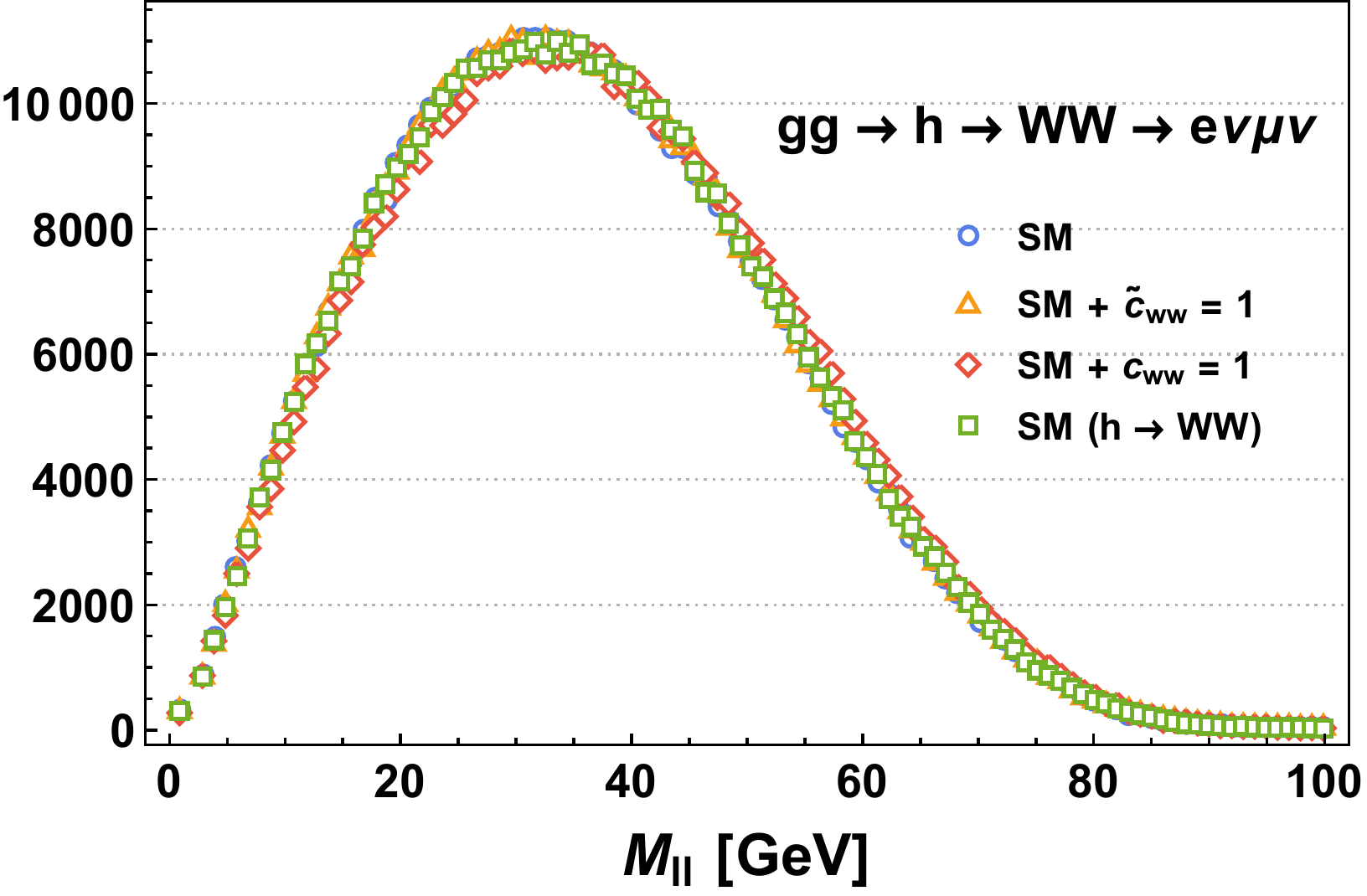}
\includegraphics[width=0.42\textwidth]{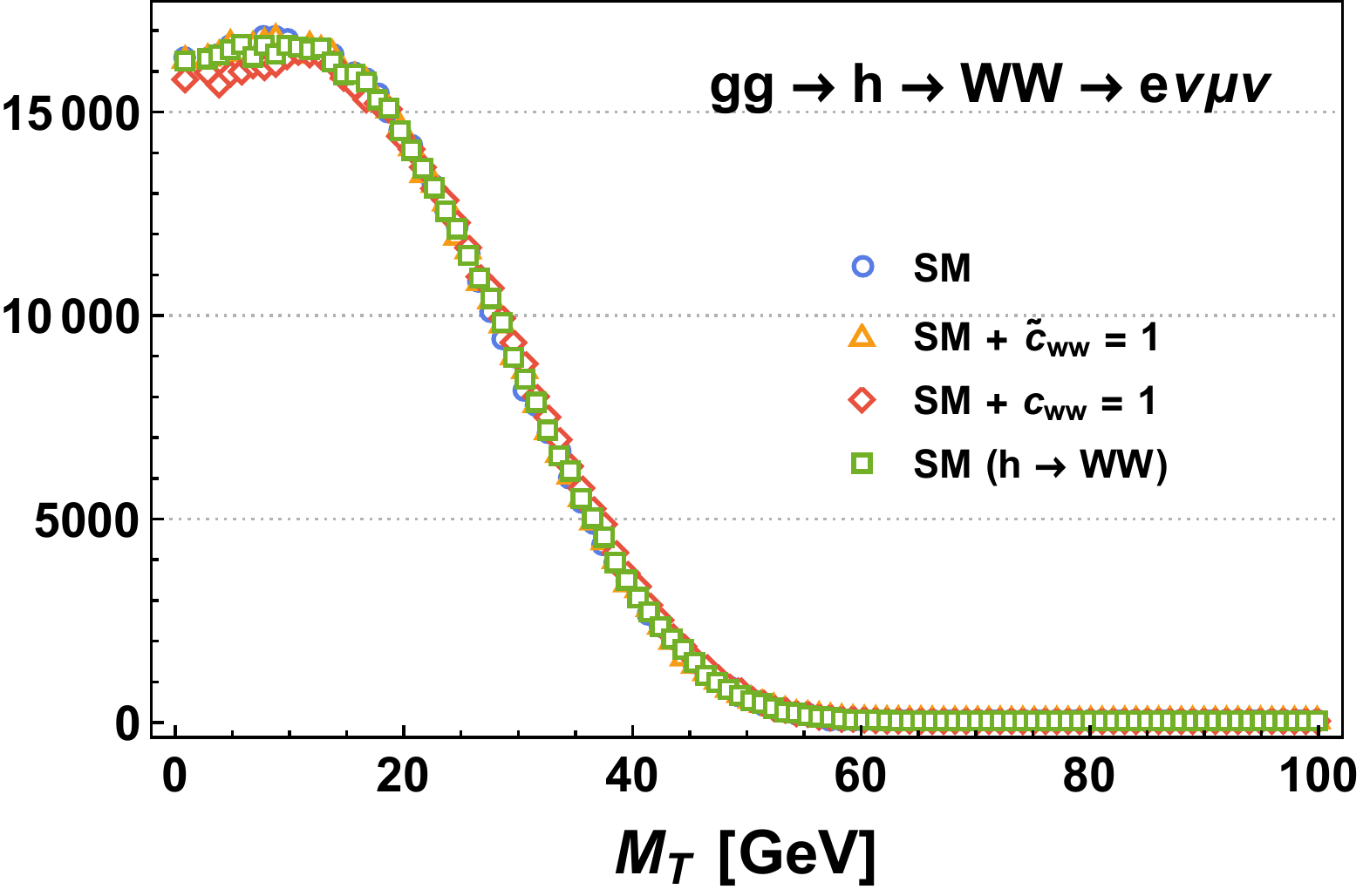}
\includegraphics[width=0.42\textwidth]{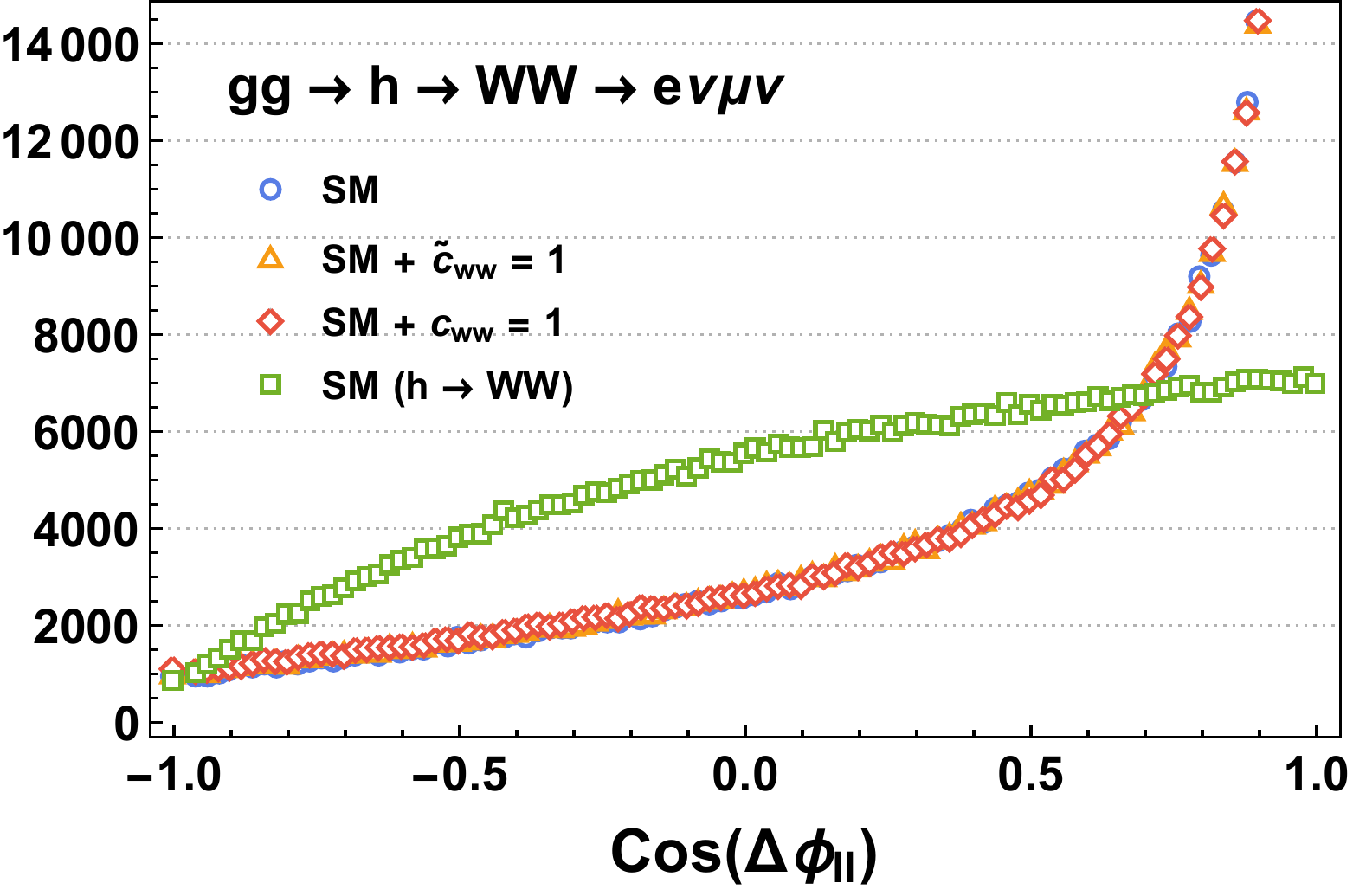}
\includegraphics[width=0.42\textwidth]{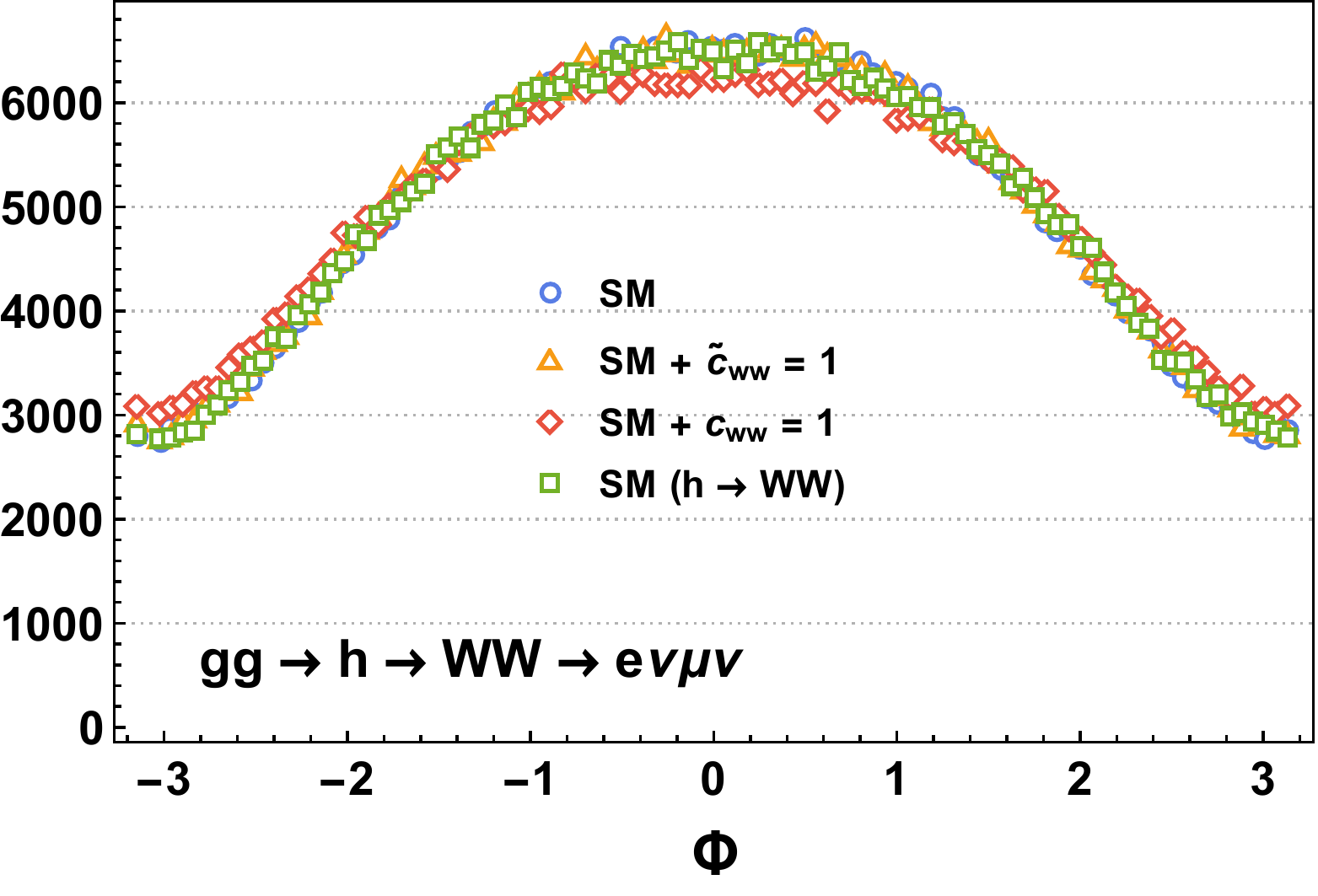}
\caption{Kinematic observables available in $h \to WW \to e\nu\mu\nu$ decays.}
\label{search}
\end{center}
\end{figure}

As discussed, we must integrate over the invisible phase space of the two neutrinos.~Before the integration, as in $h\to4\ell$, there are twelve observables corresponding to the momenta of the four massless leptons.~After integration over the invisible neutrino phase space we are left with eight observables.~Note that we are left with eight observables after integration and not the naively expected six since the total missing $\vec{p}_T$ is `observable'.~A subset of these observables have been identified~\cite{Khachatryan:2014kca,ATLAS:2014aga,Aad:2015rwa} as useful for ascertaining the CP properties of the Higgs boson to $W$ pairs.~These are defined as,
\begin{itemize}
\item $m_{\ell\ell}$ -- invariant mass of the lepton pair system. 
\item $m_{T}$ -- missing transverse mass. 
\item $\Delta\phi_{\ell\ell}$ -- opening angle between charged leptons.
\item $\Phi$ -- azimuthal angle between charged leptons.
\end{itemize}
They are shown in~Fig.~\ref{search} for various cases of effective couplings in~Eq.~\eqref{eq:hvvlag} for the process $gg \to h \to WW \to e\nu\mu\nu$.~To compute this process we have utilized a Higgs effective implementation~\cite{Falkowski:2015fla} in Madgraph~\cite{Alwall:2014hca}.~We have also shown for comparison the distributions for only the $h \to WW \to e\nu\mu\nu$ decay as opposed to those including in addition gluon fusion production (and in particular parton density) effects.~We see that gluon density effects have the largest effect on the dilepton opening angle $\Delta\phi_{\ell\ell}$.

We also see that for the parameter values shown in Fig.~\ref{search} the distributions are similar.~Of course these projections mask the full information contained when all variables including their correlations are used in a MEM analysis.~To date only analyses using BDTs~\cite{ATLAS:2014aga,Aad:2015rwa} have utilized all observables availabe in $gg \to h \to WW \to e\nu\mu\nu$.~Here we initiate an attempt to utilize all observables in the construction of a MEM analysis framework based on the methods explored in~\cite{Gainer:2011xz,Stolarski:2012ps,Chen:2012jy,Chen:2014gka,Chen:2014pia,Chen:2015iha,Chen:2015rha,Khachatryan:2014kca} for $h\to4\ell$ decays.

\section{CONSTRUCTING LIKELIHOOD ANALYSIS}
As discussed, the invisible neutrinos are not reconstructable and so we are not able to boost to the CM frame (or Higgs frame) as we could in $h\to 4\ell$ decays.~We therefore must integrate over the invisible phase space of the two neutrinos.~In order to do this, one must include the production since during this phase space integration, both decay and production variables are integrated over.~The problem then becomes finding a suitable frame in which to perform the integration over the neutrino momenta.~Also note that, unlike $h\to4\ell$, this integration is required at the \emph{parton} level.

We define the full set of twelve observables as $\vec{\mathcal{P}} \equiv (\vec{\mathcal{P}}_V, \vec{\mathcal{P}}_I)$, where $\vec{\mathcal{P}}_V$ indicates the `visible' observables, while $\vec{\mathcal{P}}_I$ represents the `invisible' observables associated with the neutrinos.~We are interested in the fully differential cross section for the \emph{visible} observables.~This can be obtain schematically as,
\bea
\label{eq:invint}
\frac{d\sigma}{d\vec{\mathcal{P}}_V} =
\int \frac{d\sigma}{d\vec{\mathcal{P}}_V d\vec{\mathcal{P}}_I} d\vec{\mathcal{P}}_I
= \int \frac{d\sigma}{d\vec{\mathcal{P}}} |\vec{J}| d\vec{\mathcal{P}}_I ,
\eea
where $ |\vec{J}|$ is a multi-dimensional (in this case four) Jacobian which parametrizes the transformation from the lepton four momenta to visible ($V$) and invisible ($I$) observables defined in a certain frame.~The fully differential cross section can be written as,
\bea
\frac{d\sigma}{d\vec{\mathcal{P}}} = \frac{d\sigma_{gg\to h}}{d\vec{\mathcal{P}}_{prod}} \frac{d\sigma_{h\to 2\ell2\nu}}{d\vec{\mathcal{P}}_{decay}}
\eea
where we have factored production and decay observables.~As discussed, the $h\to2\ell2\nu$ fully differential decay is computed analytically before performing the integration over the invisible degrees of freedom.

A crucial step in the integration will be to obtain the Jacobian which parametrizes the change of variables in~Eq.~\eqref{eq:invint}.~This four dimensional Jacobian is obtained numerically using the methods described in~\cite{Chen:2013ejz,Chen:2014pia,Chen:2014hqs}.~After obtaining the fully differential cross section in terms of visible momenta as in~Eq.~\eqref{eq:invint}, we can go on to perform likelihood analyses as done for $h\to4\ell$ decays in~\cite{Gainer:2011xz,Stolarski:2012ps,Chen:2012jy,Chen:2014gka,Chen:2014pia,Chen:2015iha,Chen:2015rha,Khachatryan:2014kca}.~An in depth exploration of these possibilies is ongoing.

At the `detector' level, a second integration in which the parton level differential cross section is convoluted with a suitable transfer function parametrizing the relevant detector effects will be needed.~This is a much more challenging problem which has been solved for $h\to4\ell$ decays~\cite{Chen:2014pia,Chen:2014hqs,Khachatryan:2014kca}.~Here we initiate a similar endeavor in $h \to WW \to 2\ell2\nu$.~Since as discussed, this channel requires an integration even at the parton level, in this current study we focus on this initial step.~We leave the construction a detector level likelihood for $h \to WW \to 2\ell2\nu$ to ongoing work.

\section*{CONCLUSIONS}
We have outlined a framework for extracting effective couplings in $h \to WW \to 2\ell2\nu$ decays.~In particular we have discussed the possibility of using multi-dimensional matrix element methods to ascertain the CP properties of the Higgs couplings to $W$ bosons.~This framework is based on analytic expressions for the $h \to WW \to 2\ell2\nu$ fully differential decay width which we have presented here.~We have discussed the need for integration over the invisible degrees of freedom in order to obtain the proper $gg\to h \to WW \to 2\ell2\nu$ likelihood and some of the technical difficulties involved.~Further development of this framework is ongoing.

\section*{ACKNOWLEDGEMENTS}
We would like to thank  the Ecole de Physique des Houches for creating a stimulating atmosphere where this project was started.

\AddToContent{Y.~Chen, A.~Falkowski and R.~Vega-Morales}
\renewcommand{\thesection}{\arabic{section}}

\superpart{ Tools and Methods}

\graphicspath{{falcon/}}
\def\bsp#1\esp{\begin{split}#1\end{split}}

\chapter{Falcon: towards an ultra fast non-parametric detector simulator}

{\it S.~Gleyzer, R.~D.~Orlando, H.~B.~Prosper, S.~Sekmen and O.~A.~Zapata}


\begin{abstract}
We describe preliminary work towards a self-tuning non-parametric detector simulator that
maps events at the generator level directly to events at the 
reconstruction level. The idea is not new. One such tool,  {\tt Tur\-bo\-Sim},
was 
developed
at D0 and CDF more than a decade ago. What is
new is the scope of what is proposed and the opportunity to capitalize 
on new 
algorithms for creating the mapping. The ultimate goal is to increase 
substantially the rate at which events can be simulated
relative to that offered by state-of-the-art programs such as {\tt Delphes}, while 
eliminating the need to implement the mapping by hand.
\end{abstract}

\section{INTRODUCTION}
Until compelling evidence of new physics is found that focuses the scope of theoretical models, we shall continue
to face the daunting task of comparing thousands of experimental results with the predictions of thousands of theoretical models, a challenge that is being addressed by a number of groups in a variety of ways (see, for example, Refs.~\cite{Kim:2015wza,Papucci:2014rja,Dumont:2014tja,Conte:2012fm,Waugh:2006ip}). 

Broadly speaking, there are two approaches to compare experimental results and 
theoretical predictions. One can either unfold detector effects from experimental results
and compare the unfolded results directly with the predictions or fold the theoretical predictions with detector effects and 
compare the folded predictions with experimental results. There are pros and cons for both
approaches. On the whole, however, folding results is preferred if only because it is technically easier to fold than to unfold when experimental results are
multidimensional. However, the price to be paid is computation time and the inconvenience
of needing codes  that are generally not publicly available. Moreover, 
even if the codes were readily available, their use typically requires knowledge and expertise not available to those
outside the experimental collaborations. 

The basic task in the folding approach is to approximate
the multidimensional function
\begin{equation}
\label{eq:pdata}\bsp
p(\textrm{r-particles} | \theta) = \int R(\textrm{r-particles} | \textrm{particles}) H(\textrm{particles} | \textrm{partons}) \,
\\\times P(\textrm{partons} | \theta) \, d\textrm{particles} \, d\textrm{partons}, 
\esp\end{equation}
the probability density to observe a collection of reconstructed particles (r-particles) given  a point $\theta$ in the
parameter space of the physics model under investigation.
The probability density $P(\textrm{partons} | \theta)$ represents the theoretical prediction at the parton level
for a given $\theta$, $H(\textrm{particles} | \textrm{partons})$ represents the mapping from the parton to the particle level, that is, the hadronization, and $R(\textrm{r-particles} | \textrm{particles})$ represents the detector response 
to, and reconstruction of, the particles that enter the detector.

Sometimes it is computationally feasible to approximate Eq.~(\ref{eq:pdata}) semi-analytically, in
the so-called matrix element methods\footnote{``So-called" because \emph{all} our methods are
matrix element methods!}. However, routine use of this method requires highly
parallel computing systems~\cite{Grasseau:2015qhg}. Furthermore, current implementations
approximate the detector response
function with empirical functions that may not fully capture non-Gaussian effects.  In practice, 
if an accurate rendering
of detector effects is needed, the only
feasible method is  simulating the detector effects in detail using a Monte Carlo method. Unfortunately, the Monte Carlo approach can become
prohibitive in terms of computation time if the detector response and event reconstruction must be simulated for tens to hundreds of thousands of events at thousands to hundreds of thousands of points in the parameter
space of a multi-parameter model (see for example, Refs.~\cite{Aad:2015baa,Sekmen:2011cz}). 
Moreover, as noted above, the required codes typically remain out of reach of physicists who are not members of the 
experimental collaborations.

These difficulties have spurred the development of fast, publicly available, detector simulators 
in which, as in the matrix element method, the detector response  function
$R$ is approximated  parametrically.  But, in contrast to the
matrix element method, the detector response function is
used to create
simulated \emph{events} at the reconstruction level. The 
{\tt Delphes} package~\cite{deFavereau:2013fsa} is generally regarded as the state-of-the-art
in this approach. {\tt Delphes}, as
well as the fast simulators internal to the experimental collaborations, starts with 
simulated events at the particle level and replace the detailed time-consuming Monte Carlo simulation of the detector response by random sampling from 
$R$, which is a considerably faster procedure.

The principal difficulty with this approach is the need to hand-code the \emph{form} of the detector 
response function.  Should the detector change because of upgrades or changing experimental conditions, or if non-Gaussian effects become important, the response function will have to be re-coded to reflect these changes. Moreover, the form of the response function could differ from one experiment to
another. 

However, it is possible to create a program like {\tt Delphes} that does not require the hand-coding of the
detector response function. Such programs, {\tt Falcon} and before that {\tt TurboSim}, 
 capitalize on the fact
that the millions, and indeed billions, of events that are fully simulated by an experimental collaboration
collectively encode the detector response function. The task is to extract a non-parametric
representation of it.

\section{FALCON}
\subsection{Introduction}
 The basic idea of the non-parametric approach is to represent the detector
response function as a huge, highly optimized, lookup table that maps objects at the parton or particle level to
objects at the reconstruction level. To the best of our knowledge, the first successful example of this
general
approach, which was used to speed up the simulation of particle showers in the D0 calorimeter, was pioneered by the late Rajendran Raja~\cite{Raja:1988fq,Graf:1990ys}. Similar approaches have
been implemented in other experiments~\cite{Sun:1996iw,Arena:1996je,Raicevic:2013vka}.
The first application of this approach, this time to the subject of this paper, namely the fast simulation of detector responses to particles, was pioneered by Bruce Knuteson~\cite{Knuteson:2004nj} who developed a program called
{\tt TurboSim}. The program we propose to build, {\tt Falcon}, can be viewed as
an updated version of {\tt TurboSim}.

\subsection{The design of {\tt Falcon}}
The {\tt Falcon} package comprises two components. The first, the {\tt builder},  abstracts the detector response function from existing fully simulated events and creates a database containing a non-parametric representation of the function. The second, the \emph{simulator}, uses this database to simulate events at  the reconstruction level from events at the parton level;  that is, the simulator
approximates the product $R(\textrm{r-particles} | \textrm{particles})$ $\times$ $H(\textrm{particles} | \textrm{partons})$.
A key assumption, which underlies the matrix element method, all current fast simulators, as well
as {\tt Falcon}, is that
the function which maps events from the parton level to the reconstruction level factorizes into a product of functions each of which map individual objects from one level to the other. 

The first design question to be settled for {\tt Falcon}, which was discussed at the Les Houches meeting, was whether it makes physical sense to map from partons directly to objects at the reconstruction level. The point is that the
quantum nature of the particle interactions places a limit on the validity of the strictly classical notion
of a well-defined parton-to-particle history. Nevertheless, as discussed at the meeting, it is possible to effect a mapping from partons to reconstructed particles provided that the partons are first clustered using any infra-red safe algorithm~\cite{Buckley:2015gua}. Once clustered, the parton jets can then 
be matched to jets at the reconstruction level. An important design feature of {\tt Falcon} is that these jets can be of any flavor: electron, muon, tau, W and Z bosons, Higgs boson, top, bottom, charm, or light quark. The ability to map a parton jet of any flavor to its reconstructed 
counterpart will become increasingly important as more and more analyses
at the LHC make use of boosted objects.  

The second design question to be addressed is at what level should the parton jets be formed? Here the answer is clear: the jets should be formed at the pre-hadronization stage, but after the partons have been
showered. However, a key design feature of the {\tt Falcon} builder is that it should be agnostic with respect to the stage to which
the event has been simulated. That is up to the user. What is key is  that a jet algorithm must be run on the 
event in 
order to create a physically well-defined final state, which prompts a third design question. Should the execution of the jet algorithm be the responsibility of the {\tt Falcon} simulator or of the program that generates the parton level events? We are inclined to argue that it should be the responsibility of the event generator to provide parton-level events with well-defined final states. After all, these final states
together with the associated reconstructed events are the
inputs to the {\tt Falcon} builder.

The fourth  design question is how are parton jets to be matched to their reconstruction level counterparts? For the first version of  {\tt Falcon}, we propose a simple proximity criterion: a parton level object and a reconstruction level object are
matched if
\mbox{$\Delta R = \sqrt{\Delta \eta^2 + \Delta \phi^2} < R_\textrm{cut}$}, where $\Delta\eta$ and $\Delta\phi$ are the differences,
respectively,
between the pseudo-rapidities\footnote{$\eta = -\ln\tan\theta/2$} and
azimulthal angles of the parton and reconstruction level
objects (e.g., jets) and $R_\textrm{cut}$ is a cut-off that may be flavor dependent. 

\subsection{A proof of principle}
{\tt Falcon} does not yet exist as a useable program.
However, we have exercised a prototype of the lookup table to reconfirm that the idea works and thus
reprised the encouraging results obtained with
{\tt TurboSim} a decade ago. 

Any fast simulator for the LHC is expected to do a good job
simulating electrons and
muons since these particles are measured with high precision at CMS and ATLAS. Therefore,
in our preliminary study, we focus on jets; in particular, on bottom and tau jets. We consider a heavy neutral scalar Higgs boson of mass of 2.9\,TeV created in proton-proton
collisions at 13\,TeV, which subsequently decays to bottom quarks 50\% of the time and to
taus 12\% of the time. 
 The goal of the exercise is to reproduce the transverse momentum ($p_T$) spectra
of the three highest $p_T$ jets using {\tt Falcon}.

Three sets of events are generated (without pileup) at 13\,TeV: 10,000 $p + p \rightarrow t\bar{t}$ events and two sets of 10,000  $p + p \rightarrow H \rightarrow f \bar{f}$ using {\tt Pythia 8.2.09}~\cite{Sjostrand:2014zea} and
its default settings. We use
 {\tt Delphes 3.3.0}~\cite{deFavereau:2013fsa} to mimic a full-scale 
Monte Carlo simulation of the
response of the CMS detector. The 
$t\bar{t}$ sample and one heavy Higgs sample are used to create a map between the {\tt Delphes} objects
{\tt GenJet}s and {\tt Jet}s\footnote{A {\tt GenJet} is a jet constructed at the particle level,
while a {\tt Jet} is a jet at the reconstruction level.}. 
Different samples are used in an attempt to populate a large range of jet transverse momenta.
A {\tt GenJet} is matched to a {\tt Jet} if $\Delta R < R_\textrm{cut} = 0.35$. This results in a table
with approximately 100000 {\tt GenJet} objects most of which are matched
to {\tt Jet}s.  In a realistic application, such a table would be populated with millions of jets.

The core of Falcon is one or more lookup tables. The lookup table in our exercise comprises two components. The first is a $k$-d tree~\cite{kdtree}, binned in 
the {\tt GenJet} quantities $(p_T, \eta, \phi)$, which associates a unique index to every {\tt GenJet}. The
second component is a map, which given a {\tt GenJet} index maps a {\tt GenJet} to the associated {\tt Jet}. We assess how well the lookup table works by running
a mockup of the {\tt Falcon} simulator on the 
second sample of heavy Higgs boson events. The detector response is simulated as follows. For
each {\tt GenJet}, its closest match is found in the $k$-d tree together with its index.  Given the index
of the {\tt GenJet}, we 
retrieve the associated reconstruction level jet from the map.

\begin{figure}
\begin{center}
\includegraphics[width=0.32\textwidth]{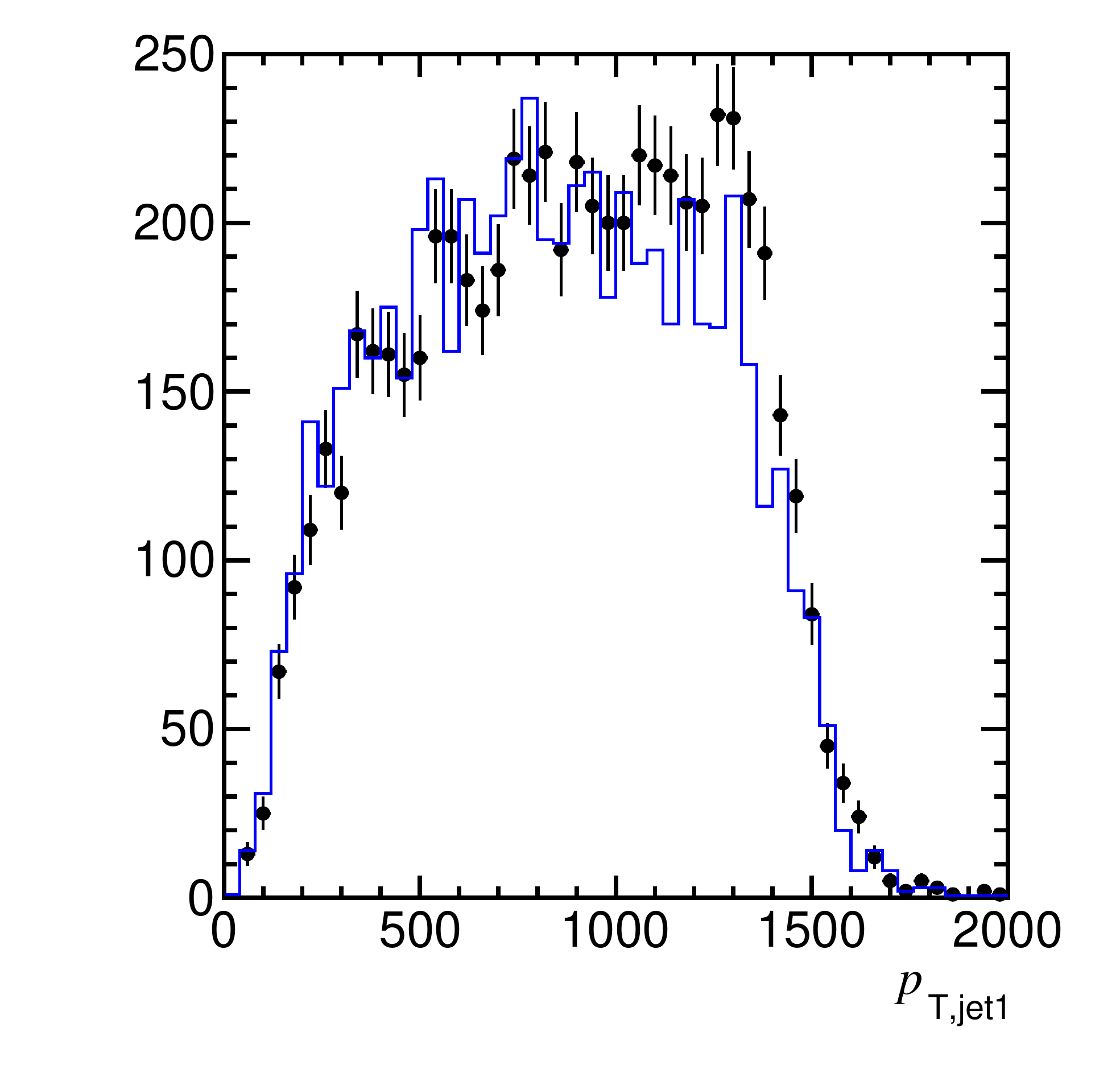}
\includegraphics[width=0.32\textwidth]{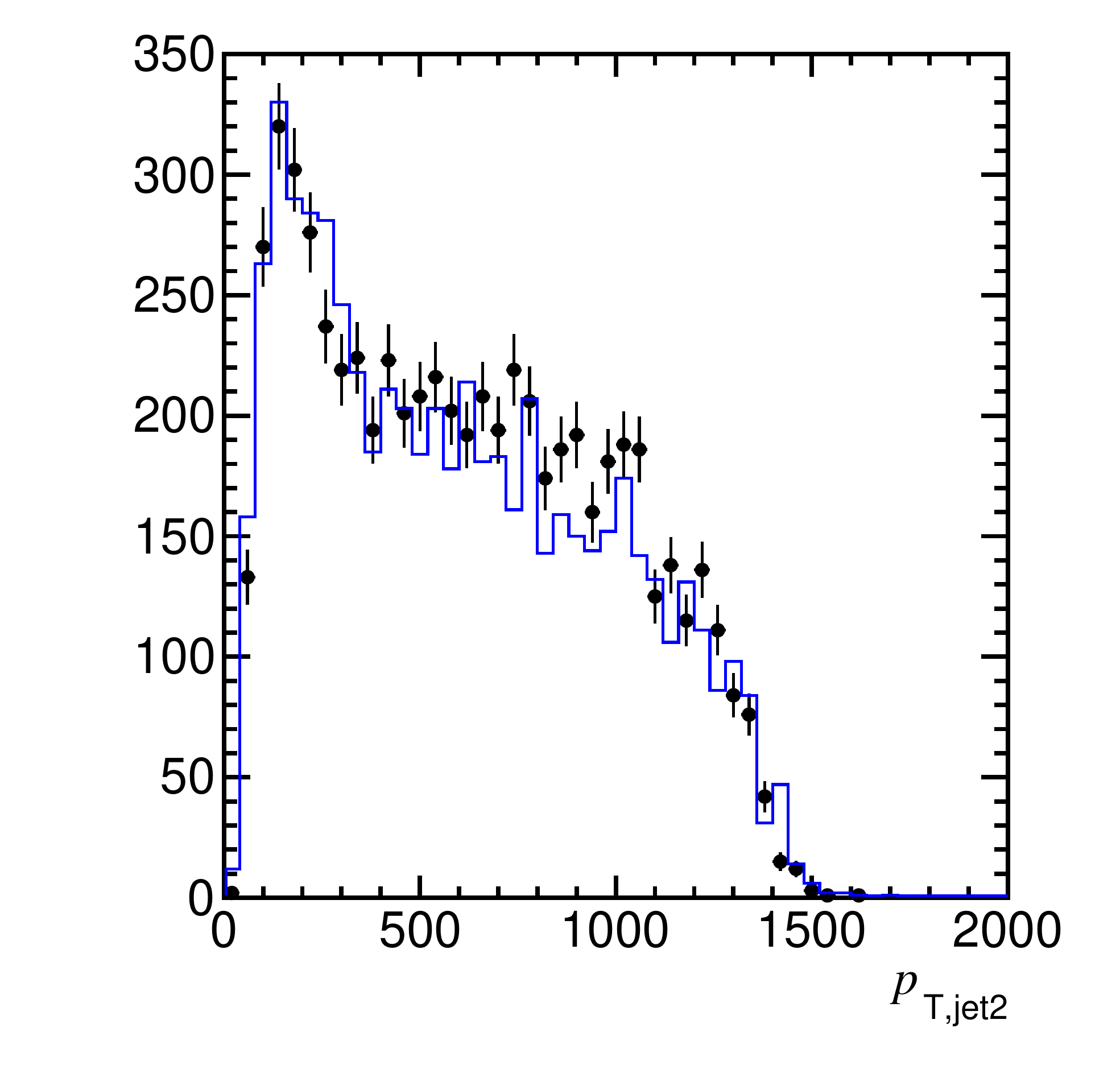}
\includegraphics[width=0.32\textwidth]{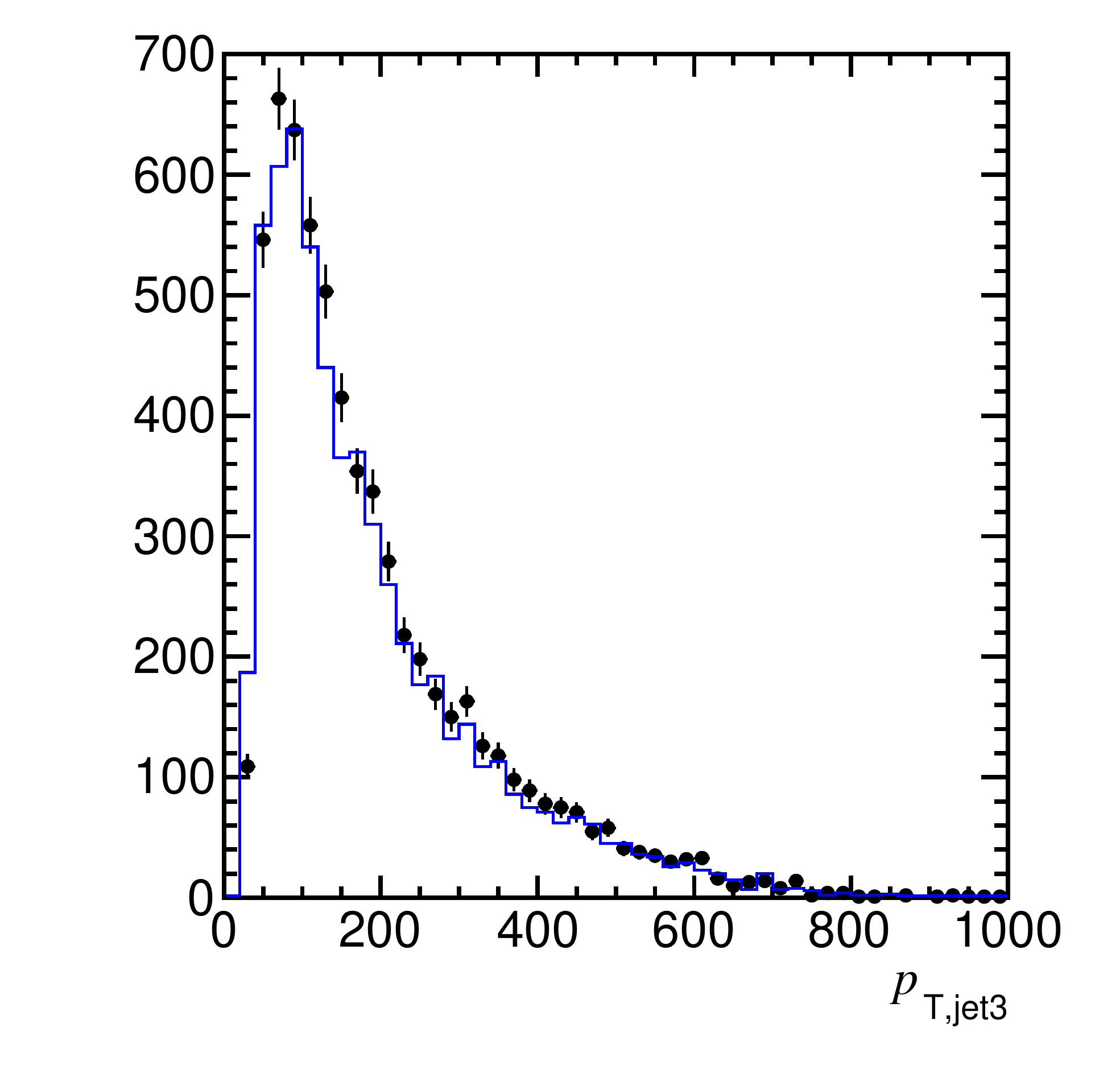}
 \caption{Transverse momentum ($p_\textrm{T}$) distributions of the three highest $p_\textrm{T}$ jets in $p + p \rightarrow H \rightarrow f\bar{f}$,
 where $H$ is the heavy neutral scalar Higgs boson (with mass 2.9\,TeV) and $f$, more than 60\% of
 the time, is
 either a tau or a bottom quark. As expected
 from kinematic considerations, the transverse momentum cuts off at approximately half the
 mass of the parent particle. The distribution depicted with the points is obtained with the 
 full-blown simulation (mimicked using {\tt Delphes}), while the histogram is obtained using
 {\tt Falcon}.}
\label{fig:pT}
\end{center}
\end{figure}

Figure~\ref{fig:pT} shows the result of this exercise. This primitive version of {\tt Falcon} is seen to
do a reasonable job of reproducing the reconstruction level transverse momentum distributions of
the three leading jets. As expected, the transverse momentum of the jets from the heavy
Higgs boson cuts off at roughly half the boson mass.

\section*{SUMMARY}
It was clear, more than decade ago, from the experience with the
{\tt TurboSim} program that a non-parametric, self-tuning, simulator is not only possible but also
very fast and at least as accurate as parametric, hand-coded, simulators.
We have argued for the creation of a modern version of this program, which we have dubbed
{\tt Falcon}. A preliminary study of a crude prototype has yielded encouraging results. However, much
remains to be done before {\tt Falcon} is available to the community. 

\section*{ACKNOWLEDGEMENTS}
We thank the Les Houches organizers  for hosting a very
inspiring workshop, which stimulated fruitful discussions and led to the work
described here.

\AddToContent{S.~Gleyzer, R.~D.~Orlando, H.~B.~Prosper, S.~Sekmen and O.~A.~Zapata}
\renewcommand{\thesection}{\arabic{section}}

\chapter{Towards an analysis description accord for the LHC}
\label{sec:LHADAdoc1}

{\it D.~Barducci$^1$, A.~Buckley$^2$, G.~Chalons$^3$, E.~Conte$^4$, N.~Desai$^5$, N.~de~Filippis$^6$, B.~Fuks$^{7,8}$, P.~Gras$^9$, S.~Kraml$^3$, S.~Kulkarni$^{10}$, U.~Laa$^{1,3}$, M.~Papucci$^{11,12}$, C.~Pollard$^2$, H.~B.~Prosper$^{13}$, K.~Sakurai$^{14}$, D.~Schmeier$^{15}$, S.~Sekmen$^{16}$, D.~Sengupta$^2$, J.~Sonneveld$^{17}$, J.~Tattersall$^{18}$, G.~Unel$^{19}$, W.~Waltenberger$^{10}$, A.~Weiler$^{20}$}

\begin{abstract}
We discuss the concept of an ``analysis description accord" for LHC \mbox{analyses}, a format capable of describing the contents of an analysis in a standard and unambiguous way.
We present the motivation for such an accord, the requirements upon it, and  an initial discussion of the merits of several implementation approaches. With this, we hope to initiate a community-wide discussion that will yield, in due course, an actual accord.
\end{abstract}

Searches for new physics continue fervently at the LHC, using a large variety of final states.
Hundreds of searches are composed and performed by the LHC collaborations, while there is a constant flow of ideas back and forth between the phenomenology and experimental communities about how to probe new models and non-trivial signatures, and improve the sensitivity of existing searches.
The ultimate goal of this 
effort is to discover new physics if such exists within the
range of the LHC, and to 
test the widest possible range of hypothetical new physics models. 

The experimental searches are realized through 
analyses that attempt to extract information from a given set of data
in the form of collider events, both real and simulated.  A typical analysis defines quantities that aid in classifying the event as signal or background:
for example the properties of analysis objects such as jets, electrons, muons, etc., or global event variables such as object multiplicities, transverse momenta, transverse masses, etc.

An analysis can be very complex and feature many intricate definitions of object and event variables,
some of which cannot be expressed in closed algebraic form and
must be defined algorithmically. This complexity renders the task of visualizing, understanding, developing and interpreting analyses increasingly challenging.  One obvious way to cope with the complexity is to devise ways to enforce absolute clarity in the description of analyses.

A discussion was started in the Les Houches PhysTe\kern-0.2exV workshop in 2011, and continued thereafter within a wider group of LHC physicists, in order to determine what information is crucial for describing an analysis.  The outcome of this discussion was reported in the ``Recommendations for Presentation of LHC Results''~\cite{Brooijmans:2012yi, Kraml:2012sg}, and has been embraced by many LHC physicists.

The current practice in our community is to write an analysis in non-public computer codes, which often
rely on event objects specific to the experimental collaboration in question, and then make public a
description of the analysis via journal publications or other documents.  Many analysts take great care in providing pertinent details of their analyses in publications.  Some take this effort further by publishing code snippets that include object or event variable definitions\footnote{For example, code implementations for the \emph{razor} variable used in Ref.~\cite{Chatrchyan:2012uea} can be found in Ref.~\cite{RazorTwiki}.} or
Rivet~\cite{Waugh:2006ip,Buckley:2010ar} implementations of full analyses.  These efforts, 
which merit our great appreciation, have
significantly increased the scientific value of many important experimental
results.  However, more can be done to build on this practice, ensuring better
long term preservation of both experimental data and results and the analyses that yielded
them, through the development of a community standard for expressing data analysis
information.

There is significant precedent for the effectiveness of such community standards.  
Several accords have been established to standardize the communication of physics modeling information, notably
the Les~Houches Event Accord (LHE)~\cite{Boos:2001cv, Alwall:2006yp} and the SUSY Les Houches
Accord (SLHA)~\cite{Skands:2003cj, Allanach:2008qq}. These, respectively, standardize the
description of hard-process particles in simulated collision events,
and the details of all the parameters that define a
BSM model point.  Both accords are widely used in high-energy physics and have greatly helped to simplify and make more efficient the communication between physicists.

In this report, we underscore the need for a standardized format--- an
``analysis description accord'' --- capable of describing the contents of an
analysis in an unambiguous way, which can be fully exploited by the whole
particle physics community.
The accord must be capable of describing all object and event selections, as well as quantities such as efficiencies, analytic and algorithmic
observables, and advanced multivariate selections. 

In the Les Houches PhysTeV workshop in 2015, we have initiated a dedicated discussion on how such an accord can be realized.  In the following sections we shall detail the use cases and design requirements of such
an accord, and the general pros and cons of several approaches.

\section*{MOTIVATIONS AND USE-CASES}

We envisage several important motivations and use-cases for a standard analysis description accord:

\begin{description}
\item[Analysis preservation:] A great deal of thought and work goes into
  designing an analysis, which makes the analysis information valuable, and its
  preservation important.  Usually, the full details of an analysis exist only
  in each experiment's analysis software; there are many such frameworks even
  within a single LHC experiment, and their validity and code compatibility are
  tied to specific phases of LHC data-taking and analysis. It is therefore far too
  easy for crucial information to be lost, especially when personnel turnover
  means that the original analysts are no longer available as ``consultants''.
  A well-designed standard mechanism for encoding analysis details will hence
  have the benefit of universality and stability, making it possible to
  resurrect LHC-era data analyses even decades after the end of the LHC experimental programme.
  Therefore, a serious effort to preserve
  analyses through a universal accord will help ensure a strong LHC legacy.

\item[Analysis design:] A standard means for expressing analysis
  logic will help in the early stages of analysis development, by
  abstracting the analysis from the technicalities of various analysis systems
  and providing a convenient language in which to express and discuss analysis
  ideas. It hence may provide a powerful route by which to explore analysis ideas, and develop and execute analyses, 
  both within experimental collaborations and between experimentalists and theorists when discussing future plans.

\item[Analysis review and communication:] An accord can also serve as a convenient and 
universally understood means by
  which reviewers -- both internal to the experiments and outside -- can rapidly
  understand an analysis procedure without ambiguity. Data analyses
  such as searches for new physics are most commonly communicated in
  publications.
  While authors often do
  attempt to tabulate all pertinent information, constraints on publication
  length (and simply the challenge of expressing detailed algorithms in natural
  language) lead to ambiguities, loss of important details, and general
  difficulty in understanding.\footnote{We note that while the experimental collaborations provide Rivet~routines for many Standard Model measurements, this is not the case for BSM analyses.} 
  A universal language for analysis description,
  familiar to the whole community, would make the process of understanding and reviewing an
  analysis easier.


\item[Interpretation studies and analysis reimplementation:] We are all eager to
  see how results from the LHC will change our understanding of Nature,
  or direct the design of the \emph{next} generation of scientific instruments.
  All those with an interest in LHC results may wish to be fully engaged in
  their interpretation, especially in the case of a discovery. One of the goals
  of the analysis accord is to help make this possible.  While it is relatively
  easy for the analysts who have designed an analysis to interpret its results
  in terms of different theoretical models, it is generally a nontrivial task
  for someone outside, or even within, the experimental collaboration to obtain
  sufficient information to reproduce the analysis.  A standard
  accord will make it easier for scientists of all backgrounds to understand an
  analysis in the detail needed for replication. 
  However, to ensure reliable reimplementation, it is essential that an exact
  analysis description be accompanied by all relevant analysis validation
  material, including precisely defined benchmark points, exact configuration of
  Monte Carlo tools\footnote{Although MC setup information will surely become
    obsolete as MC tools and their configuration interfaces evolve, it will
    still be exceptionally useful in the short- to medium-term following
    publication.}, detailed cut flows and kinematical
  distributions. 
  This will most likely be best achieved by providing this material
    on the HepData system~\cite{Buckley:2010jn}. 

\item[Easier comparison of analyses:]  A large diversity of analyses obtain results from the LHC data.  
Experience shows that combining multiple experimental results enhances LHC sensitivity to the
  target signals.  The obvious way to obtain maximal sensitivity is to design
  analyses with disjoint search regions, but many LHC analyses are designed by
  independent groups, and overlaps between search regions are hard to avoid.  A standard 
  analysis description accord by definition offers a practical and reliable way to
  compare definitions of search regions in different analyses, determine which
  are disjoint and, thus can be easily combined within and across
  collaborations.  Facilitating the comparison of analyses would also serve to spot the final states not yet explored and design new analyses based on them.

\end{description}

\section*{PROPERTIES OF AN ANALYSIS DESCRIPTION ACCORD}


The motivations and use-cases discussed above place several essential requirements on any accord. In addition, there are several desirable features which would further improve the utility of the accord, which, however, may be nontrivial to simultaneously fulfill. Therefore, here we list all the desirable features, leaving it to the community discussion (and to individual prototypes) to decide which of these desiderata should inform an eventual accord. We start by listing the features we believe essential to the success of such an accord:

\vspace{-0.3cm}
\subsection*{\textsc{Basic requirements:}}
\vspace*{1em}

\begin{description}
\item[Public availability:] By definition, a format used for communication between all
  sectors of the LHC community must not depend on tools or information not
  publicly available to all parties. This obviously excludes use of
  experiment-specific tools either for the whole accord or its subcomponents. 
 
  The codes and publications that host our 
  analyses are generally stored in different, scattered, locations.  Our
  ultimate goal would be to preserve all LHC analyses in a robust, long-term
  supported public database that can be accessed easily by the whole community.
  The HepData or Inspire systems are obvious candidates because of
  their existing
  long-term community roles and integration with HEP publication infrastructure,
  but a new repository could also be established for this purpose.  

  The ATOM~\cite{Papucci:2014rja}, CheckMate~\cite{Drees:2013wra,Kim:2015wza},
  MadAnalysis\,5~\cite{Conte:2012fm,Conte:2014zja,Dumont:2014tja} and
  Rivet\footnote{At the time of writing, Rivet's analysis collection is mainly
    composed of comparisons to unfolded data, where no detector effects are
    applied.  However, Rivet now provides sufficient machinery for including
    detector effects so that BSM analyses, in which unfolding is not
    usually an option, can be implemented.}~\cite{Waugh:2006ip,Buckley:2010ar} collaborations have
  already made considerable moves in the direction of comprehensive analysis
  collections, but in many cases the analyses in these databases do not always
  contain all the details of experimental object definitions -- this is only
  realistically achievable in analyses that are directly provided by the experimental
  collaborations and that
  able to encode sufficient additional information beyond the bare analysis cut
  logic.
  Achieving universality in analysis description at all sophistication levels through an accord will be a
  robust step towards developing a persistent analysis database.

\item[Completeness:] The accord must be able to host all information needed 
  for an accurate understanding and reproduction of an analysis. For the development of
  analyses, means are needed to express event and object selection cuts in an
  efficient way for all signal regions -- including complex signal regions, which
  are effectively observable bins, and definitions of custom selection variables in a
  form applicable to experimental physics objects at all levels of definition.   
  The accord may ultimately be expanded to include analyses regions used for background estimation.
  However, it should enable to provide sufficient information about background process estimates in order to
  allow data re-interpretation by BSM specialists without requiring difficult
  explicit simulation of many SM processes.  
  For the purposes of analysis preservation and re-interpretation, the accord must, in particular, be able to
  parameterize the efficiencies and resolutions of physics objects such as jets
  (and their $\{b,c,\tau\}$ tags), charged leptons and photons (with custom jet
  isolation definitions), and missing transverse momentum.





\item[Longevity:] The accord must be encoded in a robust format that
  has a good chance of still be supported decades later, and whose evolution will be
  backward-compatible. In the interest of human readability, and since neither storage nor bandwidth 
  are likely to be
  an issue for the required information transfer, a plain text format
  (potentially compressed using standard tools) seems most appropriate. If
  implementation \textit{via} an established programming language is desired,
  it must enjoy wide acceptance (not just in HEP, although this is
  obviously key), and be adaptable to unforeseen
  developments in analysis techniques.

\item[Correctness and validatability:] 
  It is well known that even the most attentive programmers or
  data encoders will make errors.  The same will apply to
  analyses encoded following this accord: whatever the format, mistakes
  will be made. It is therefore crucial for the prototyping and
  preservation use-cases, as well as the more obvious re-interpretation one,
  that the accord data be programmatically parseable and runnable, producing
  output that can be compared with expected outcomes. The major problem
  here is custom definitions, especially algorithmic ones: for anything beyond
  the simplest logical constructs, the accord execution code will need to be a
  fully-fledged interpreted language that can provide identical functionality
  to different client frameworks.  This should be achievable 
  but is likely to impose more
  technical hurdles.
\end{description}

\noindent
Next, we list the features which are desirable, however may be difficult to achieve (simultaneously) in the design of the accord:

\vspace{-0.3cm}
\subsection*{\textsc{Desirable features:}}
\vspace*{1em}

\begin{description}

\item[Human readability and writeability:] Since an important role is envisioned
  for this accord in analysis design, discussion and review, it is crucial 
  that it be reasonably easy to read and write by humans.  
  This implies a clean syntax with minimal repetition and decoration (i.e.\,minimal use of brackets and if possible, no semicolons or other end-of-line constructs) as well as clear structuring of the different classes of information. To encourage the ready adoption
of the format, it should avoid jarring mental disassociation from
  existing HEP tools, by, e.g., building on the syntax of established tools,
  languages, or data formats. The implicit requirement here is that the accord
  format be simpler and more human-readable than typical analysis
  code for the equivalent procedure.

\item[Self-contained:] In order to be robust against changes, it is best
  that the accord hosts all the above information in a single file or
  bundled collection of files that can be easily archived in a single location
  or transmitted in a single message. 
  Reliance on the persistence of external references introduces a potential failure mode, 
  which could undermine the integrity of the preservation plan. There must, however, be reasonable
  limits to this desired feature -- complex and performance-critical algorithms such
  as jet clustering cannot reasonably be reproduced ``inline'' in each analysis
  accord\footnote{Here we refer to the clustering sequence evaluation, jet area
    calculation, filtering methods, etc. Experiment- and analysis-specific
    details such as the use of muons and neutrinos in jet construction must be
    encoded in the analysis description.}, and so some agreement on a standard
  base set of observables will be required.
  
\item[Language independence:] A related constraint is that of programming
  language. It seems likely too that programming languages will evolve, and
  current paradigms will be replaced. The robustness of the accord format would
  therefore be enhanced if expressed in a programming language-independent form, 
  an analysis language optimized for the
  expression of physics concepts. But the limits of this requirement must be
  noted: a format sufficiently complete to express algorithmic observable
  definitions \emph{is} a \textit{de facto} programming language. 
 It is unlikely that the HEP community, on its own, has the resources to
 engineer such a language, especially one as
  capable as those from mainstream software engineering. However, a hybrid solution may
  be possible, such as embedding snippets of an established programming language
  in an otherwise physics-specific language.

\item[Framework independence:] Different analysis groups, be they theorists or
  experimentalists, use different software frameworks for implementing
  analyses. Over the coming decades it is inevitable that these frameworks will
  change. It is therefore highly desirable that the accord be represented in a
  framework-independent form, as far as possible.
    
\item[Support for combination of analyses:]  The importance of combining analyses for increasing LHC sensitivity, and how an accord can facilitate this task has been discussed earlier.  In this regard, the accord should also permit expression of how the non-disjoint search regions (including between analyses) are correlated, so
 that a more complete statistical treatment can extract maximal 
 information from the LHC data.

\end{description}

\section*{DISCUSSION}

The above requirements and features each have their own motivations and benefits, however they are not wholly compatible with each other.  The main conflict may arise while trying to simultaneously satisfy the desires for 
an easily readable (and writeable) format for physicists, with the need
for algorithmic completeness and avoidance of ambiguity.  Expressing and encapsulating some analysis techniques, notably multivariate functions, such as trained neural networks and boosted decision trees, may be potentially problematic.  However, they are fundamentally no different in kind to functions that calculate,
for example, the razor variables. The challenge is to establish good practice:
to encourage the encapsulation of these functions in exactly the
same way as those that calculate other complicated observables or that implement,
for example, jet-finding.

\noindent We expect that many conversations will follow on the merits and shortcomings of
various approaches, so here we briefly discuss the pros and cons of a few
potential solutions to the challenge set out above:

\begin{description}
\item[Analysis description language:] One approach, demonstrated in another contribution to
  these proceedings, is to follow the path taken for previous data-oriented
  HEP accord formats, and define a new language tailored for analysis description.
  The base syntax could be a standard format such as YAML 
  or some other simple dedicated syntax invented for the purpose.  
  In order to enable running analyses on events, this approach requires parsers that would interpret the accord. Therefore the syntax should be chosen as one that would ease the writing of parser logic.  As not many physicsts are able to write parsers for a new language, public parsers should be made available.
  The benefits of this approach are easiness in obtaining control to reach a standard, ability to target the design
  to the physics application of interest, and natural inclination to human readibility.
  The downside, however, is the difficulty of expressing algorithmic detail (e.g. complex
  observable or object definitions) in a format which is not itself a general
  programming language.  
  A route forward can be found by embedding a subset
  of a standard language into blocks of a data format to allow integration
  with analysis software frameworks.
  

\item[Pseudocode (or real standalone code):] The other end of the spectrum is to
  express the analysis detail in pseudocode, or even in actual executable
  code. This clearly solves the problem of algorithmic completeness, but
  pseudocode is not runnable and hence not amenable to automated validation. It
  is also not necessarily any clearer or less ambiguous than textual 
  descriptions in publications. Meanwhile real code, unless carefully watched,
  could easily fail the requirements of universality and readability. Many
  components of the accord, e.g. reference data, correlations, and background
  histograms are not obviously best expressed in a pure code form. However,
  pseudocode definitions of complex non-standard observables would already be a
  useful complement to the analysis auxiliary information supplied to HepData.

\item[Analysis framework code + metadata:] Despite the desire to be
  framework-independent, the presence of unambiguous and analysis
  logic that can be validated and encoded in a standard public analysis framework would provide a useful
  reference for other implementations to follow and so should not be entirely
  discounted. The lack of universality is a significant downside, however, and
  it is impossible to guarantee that the chosen framework will remain actively
  maintained and available in perpetuity. The comments above regarding
  readability of standalone real code also clearly apply to this approach. But
  the definiteness of the framework interface and physics objects, and the resulting
  connection to validation and reinterpretation uses, are beneficial,
  and typically the implementation is in mainstream programming languages with
  which many (but not all) HEP users already have familiarity. Unlike a pure
  programming-language solution, the frameworks store reference data and other
  auxiliary material outside the analysis code, so more appropriate formats can
  be chosen for each.

\end{description}

\section*{CONCLUSIONS}

We have described the potential benefits of an analysis description accord, which
provide considerable motivation to mount a concerted community-wide effort to develop such an accord.
We have listed the basic requirements and desired features which would guide the design of this accord.  
Simultaneous implementation of these properties is nontrivial, as not all listed properties are wholly compatible with each other, but we leave the decision on the best solution to the community.  With this report, we hope to start a discussion in the LHC community on how such an accord could be best realized.

\section*{ACKNOWLEDGEMENTS}
We thank the organisers of the Les Houches PhysTe\kern-0.2exV workshop for the very
inspiring atmosphere that ignited many fruitful discussions.  To continue work
on the analysis description accord proposal, a follow-up workshop was organised at LPSC Grenoble in February 2016.  We thank the LPSC Grenoble and its theory group for
their hospitality. This work has been partially supported by THEORIE-LHC-France
initiative of CNRS-IN2P3, Investissements d'avenir -- Labex ENIGMASS, the
ANR project DMASTROLHC ANR-12-BS05-0006 and the BATS@LHC ANR-12-JS05-002-01.
AB is supported by a Royal Society
University Research Fellowship and University of Glasgow Leadership Fellow
award. SuK is supported by the New Frontiers program of the Austrian Academy of
Sciences.  The work of JS is supported by the collaborative research centre
SFB676 ``Particles, Strings, and the Early Universe'' by the German Science
Foundation (DFG) and by the German Federal Ministry of Education and Research
(BMBF).

\AddToContent{D.~Barducci, A.~Buckley, G.~Chalons, E.~Conte, N.~Desai, N.~de~Filippis, B.~Fuks, P.~Gras,
  S.~Kraml, S.~Kulkarni, U.~Laa, M.~Papucci, C.~Pollard, H.~B.~Prosper, K.~Sakurai, D.~Schmeier, S.~Sekmen,
  D.~Sengupta, J.~Sonneveld, J.~Tattersall, G.~Unel, W.~Waltenberger and A.~Weiler}
\renewcommand{\thesection}{\arabic{section}}

\chapter{A proposal for a Les Houches Analysis Description Accord}

{\it D.~Barducci$^1$, G.~Chalons$^2$, N.~Desai$^3$, N.~de~Filippis$^4$, P.~Gras$^5$, S.~Kraml$^2$, S.~Kulkarni$^6$, U.~Laa$^{1,2}$, M.~Papucci$^{7,8}$, H.~B.~Prosper$^9$, K.~Sakurai$^{10}$, D.~Schmeier$^{11}$, S.~Sekmen$^{12}$, D.~Sengupta$^2$, J.~Sonneveld$^{13}$, J.~Tattersall$^{14}$, G.~Unel$^{15}$, W.~Waltenberger$^6$, A.~Weiler$^{16}$}

\begin{abstract}
We present the first draft of a proposal for ``a Les Houches Analysis Description Accord" for LHC \mbox{analyses}, a formalism that is capable of describing the contents of an analysis in a standard and unambiguous way independent of any computing framework.  This proposal serves as a starting point for discussions among LHC physicists towards an actual analysis description accord for use by the LHC community.
\end{abstract}

\def\JT#1{{ \textcolor{purple}{{[JT:~#1]}}}}
\def\DS#1{{ \textcolor{blue}{{#1}}}}
\def\ND#1{{ \textcolor{magenta}{{[ND: #1]}}}}

\newcommand{\XML}{\texttt{XML}}

The concept of an ``analysis description accord" -- a software framework-independent universal formalism, which fully describes the components of an analysis -- has been presented in section~\ref{sec:LHADAdoc1}.  Such an accord would provide valuable benefits to the LHC community ranging from analysis preservation that goes beyond the lifetimes of experiments or analysis software, to facilitating the abstraction, visualization, validation, combination, reproduction, interpretation and overall communication of the contents of LHC analyses.  Fostering open discussion of the contents of an analysis, within or outside the team that has designed it, is important as it helps avoid ambiguities that can lead to misunderstandings, and it can help render the description of the analysis complete by identifying undocumented elements of it.  Completeness is clearly a necessary condition for analysis preservation.

The benefits described above motivated our attempt at the 2015 Les Houches PhysTeV workshop to define the ingredients and generic structure for a possible realization for such an accord~\footnote{Previous attempts at defining uniform analysis language frameworks (based on the LHCO format) were made in Refs~\cite{Walker:2012vf,Martin:2015hra}.}. This attempt yielded a first draft of a proposal for a Les Houches Analysis Description Accord (\texttt{LHADA}), including a preliminary implementation based on a new, simple language specifically designed to describe the components of an analysis in a human-readable, unambiguous, framework-independent way. It is expected that this first draft will encourage vigorous discussion and debate within the LHC community that will lead to an actual accord.

We propose an accord that consists of text files fully describing an analysis, accompanied by self-contained functions encapsulating variables that are nontrivial to express.  The text files describing the analysis use a dedicated language with a strict set of syntax rules and a limited number of operators.  In our opinion, limiting the flexibility by using a dedicated language provides a clear advantage for an unambiguous analysis description with respect to existing programming languages such as \texttt{C++} and \texttt{Python}.  Unless one is highly disciplined, and attentive to readability,  the expressiveness of these powerful languages all too easily obscures the structure and algorithmic content of analyses. Therefore, these languages could easily obscure the analysis description.  Moreover, using a flexible language would make the translation between different frameworks, even if they are based on the same language, difficult if not entirely impossible.

Inspired by the previous Les Houches Accords, we suggest that the analysis description files at the center of the accord consist of easy-to-read blocks with a \texttt{key value} structure.  Such a format is capable of  incorporating all the relevant steps of an analysis in a straightforward and clear manner.  Furthermore, the suggested syntax makes possible the automatic translation of the text files of the accord into existing and future analysis frameworks and thus ensures long-term preservation.  For complicated functions (for example, multivariate discriminants) that cannot be reduced to a simple, physical quantity and therefore simple notation, linking to external source code as well as to its documentation is permitted.  However, this functionality is to be used only when no other avenue is open.  

In the following, we define and describe the concept and syntax of our proposal for a Les Houches Analysis Description Accord, and present use cases with several examples.  Again, we stress, that the proposal is a starting point for detailed discussions among the LHC physicists towards an actual accord.

\section{INITIAL SYNTAX PROPOSAL FOR A LES HOUCHES ANALYSIS DESCRIPTION ACCORD}

The proposal for a Les Houches Analysis Description Accord (LHADA) presented here is a first draft, a proof of principle, which shows the possibility to define a rather small set of syntactic rules that are flexible enough to describe complex analyses in a human readable form.  The structure outlined here is open to modifications and extensions based on experience and further proposals by the LHC physics communities.

\subsection{General Structure}
The {\sc LHADA} proposal consists of a plain text file containing easy-to read blocks with a \texttt{key value} structure. Any nesting is implemented using spaces only and we do not recommend more than one level of nesting for reasons of readability and simplicity.

\begin{verbatim}
blocktype blockname
  # general comment
  key value
  key2 value2
  key3 value3 # comment about value3
\end{verbatim}

This structure can be converted into any format of interest.  For example, the conversion to a markup language format would be as follows:
\begin{verbatim}
<blocktype name="blockname" comment="general comment">
<key>value</key>
<key2>value 2</key2>
<key3 comment="comment about value 3">value 3</key3></blocktype>
\end{verbatim}

This structure requires a clear separation of the individual modules of an analysis prescription which makes it easier to extract respective sources of information. Contrary to the similar looking, well known \texttt{SLHA} layout, this format allows for the same key to appear multiple times and a well-defined order of the individual keys, which are both important aspects as will be apparent in the later description. 

\noindent
For the description of an analysis, we propose five types of blocks:
\begin{description}
\item[\texttt{info}:] Provides information about the analysis, e.g.\ publication information, benchmark scenarios and event generators used.
\item[\texttt{function}:] Defines all non-trivial operations that are calculated during the analysis. These usually include advanced kinematic variables, e.g.\ transverse mass or variables
created using machine learning methods, and filtering algorithms, e.g.\ lepton isolation definitions. 
\item[\texttt{object}:] Defines all reconstructed objects that are used in event selection. These are the fundamental building blocks of the analysis and have to be defined in such a way that different input data streams (real, or simulated after no/fast/full detector simulation) can be defined in terms of these objects. Examples are leptons with or without specific identification requirements, jets of any flavor resulting from well-defined clustering algorithms or the missing transverse momentum vector.
\item[\texttt{cut}:] Defines criteria that are applied to a given event in order to define analysis regions,
e.g.\ signal or control. 
\item[\texttt{table}:] Lists the analysis results numerically (e.g.\ observed numbers of events---there could be multiple bins, associated background estimates and individual or total systematic uncertainties, estimated signal counts and uncertainties for signal models specified in the \texttt{info} block, and
statistical summaries such as a p-value) for each signal or control region. Numerical information for cutflows, histograms etc. can also be provided here.
\end{description}

\subsection{The \texttt{info} Block}

This block type specifies general (meta) information about the analysis and its validation material. For now, there are no required or optional keys, nor specific requirements imposed, for this block type. 
We defer to a later date the specification of the minimal required meta information~\footnote{information on MC signals could be a part of this block.  However we prefer to defer this decision to a later point.}.  In
the examples below, we show the kind of meta information we anticipate could become part of the
accord. 
\paragraph{Examples}
\begin{verbatim}
info analysis
  # Minimal required details about analysis: 
  # a searchable ID (report number, internal note number,..) 
  # and a link to documentation
  id CERN-PH-EP-2014-143
  doc http://cds.cern.ch/record/1714148

info analysis
  # Details about experiment 
  experiment ATLAS
  id SUSY-2013-15
  publication JHEP11(2014)118
  sqrtS 8.0
  lumi 20.0
  arXiv 1407.0583
  hepdata http://hepdata.cedar.ac.uk/view/ins1304456

info units
  # Details about units for dimensional quantities
  energy GeV
  length mm
  xsec fb
  phirange -pitopi

\end{verbatim}


\subsection{The \texttt{function} Block}
Most analyses require calculations, such as those for advanced kinematical variables, that go beyond simple arithmetic operations. To improve readability, it is recommended to decouple the description of these calculations from the actual cutflow of an analysis. A similar argument holds for functions that change the properties of an object or sets of objects.  For this reason, we propose separating the definitions of all functions into individual blocks. A general function block has the following layout:
\begin{verbatim}
function function_name
  arg    name1
  arg    name2
  code   link-to-code-repository
  doc    link-to-documentation
\end{verbatim}
The \texttt{arg} keywords declare the arguments of the function, which later in the proposed {\sc LHADA}{} file are accessed by their respective names.  
The \texttt{code} keyword refers to a public database in which an implementation of the function, in at least one commonly used programming language (e.g.\ \texttt{C++}, \texttt{Python}), can be found. 
The value of the  \texttt{doc} keyword is a reference to documentation (a publication or a note) that describes the function.  

Given the anticipated time-scales of high-energy physics experiments, it is likely that new programming methodologies will find their place in high-energy physics. Should this happen, the code database 
will have to be updated. However, crucially, the proposed {\sc LHADA} file itself remains valid even when  the software paradigm shifts. This is an important design goal of the {\sc LHADA} proposal and follows the ideals of ``analysis preservation'' and ``analysis reimplementation'' mentioned in the introduction. 

The principal purpose of the code database is to preserve the code needed to render a {\sc LHADA} analysis reproducible, given its associated validation materials. A secondary, though far-reaching, purpose is to be a growing archive of reusable analysis functions. It will then become possible to simplify this part
of the {\sc LHADA} by referencing functions from lists of reusable functions (like the bibtex files used for citations in papers) that can simply be included in the analysis description without having to manually write their descriptions every time.  Until that level of automation is attained, however, every analysis 
description should be self-contained with all functions explicitly defined. 

Note that standard properties of an object such as the transverse momentum {\tt (pt)}, pseudorapidity {\tt (eta)}, azimuthal angle {\tt (phi)}, mass {\tt (m)} and cartesian momentum components {\tt (px, py, pz, e)} are always assumed to be available without further definition.  

Below is a list of examples for arbitrary use cases:

\paragraph{Examples}
\begin{verbatim}
function mT2
  # stransverse mass
  arg  vis1   # First visible 4-momentum vector
  arg  vis2   # Second visible 4-momentum vector
  arg  invis  # Invisible transverse 4-momentum vector
  arg  mass   # Assumed mass of the invisible particle
  doc  http://inspirehep.net/record/617472?ln=en  # original publ. 
  code http://goo.gl/xLyfN0  # code example from oxbridge package

%function antikt
%  # Standard jet clustering algorithm, returns a list of jets
%  arg  input     # list of objects to be clustered
%  arg  dR        # cone size
%  arg  ptmin     # minimum momentum
%  arg  etamax    # maximum rapidity
%  doc http://inspirehep.net/record/779080  # original publ.
%  code http://goo.gl/y9PQjF  # FastJet implementation    

function isol
  # Sums up activity in the vicinity of a given candidate
  arg  cand    # object whose isolation is to be computed
  arg  src     # "calo", "tracks", "eflow"
  arg  dR      # dR cone to be probed
  arg  relIso  # divide by candidate's pt?
  code ...
  doc  ...

function overlaps
  # returns true if the candidate is too close to any neighbour
  arg  cand   # Tested object
  arg  neighs # Set of objects to which overlap test is applied
  arg  dR  # deltaR distance for overlap test
  doc  ...
  code ...

function detector_muon
  # converts a list of true muons into a list of detector muons
  #  if the input is a MC object
  arg  cands         # cand muons
  arg  workingpoint  # "combined", "standalone"
  doc  ...
  code ... # The code contains efficiency maps for muons

function detector_electron
  # turns a list of true electrons into a list of detected electrons
  #  if the input is a MC object
  arg  cands         # cand electrons
  arg  workingpoint  # "loose", "medium"
  doc  ...
  code ... # The code contains efficiency maps for electrons.
\end{verbatim}

\subsection{The \texttt{object} Block}

These blocks define the reconstructed sets of objects on which the event selection is based. Some of these are assumed to be provided from an external source, and some are processed versions of others.  In cases where this accord is used by experiments to preserve full details of an experimental analysis, the object blocks would serve to host all object definition information, and when needed, advanced mathematical functions that define the object, with the help of the function blocks.  In cases where the accord is used for reimplementation of the experimental analyses by the phenomenology community, the object blocks could directly host object efficiencies.

The first item of each {\tt object} block must be \texttt{take <X>}, where \texttt{<X>} is either the identifier \texttt{external} or the name of another {\tt object} type already defined within the proposed {\sc LHADA} file. In the \texttt{external} case, it is assumed that the source for that particular set of objects is provided from an outside program.  These can be reconstructed detector objects (from experimental data, full simulation or fast simulation), in case where experimental analyses are implemented, or even particles from an event generator, in case an analysis proposal is sketched.  
The reference linked in the \texttt{doc} parameter should therefore carefully define what these objects are supposed to contain.  If not \texttt{external}, \texttt{<X>} can alternatively be a previously defined \texttt{object} type upon which a tighter selection is to be imposed.  Multiple \texttt{take} keys can be used for defining combined object sets. 

After the \texttt{take} key(s), members of the set \texttt{<X>} are removed if they fail any boolean statements which follow a \texttt{select} key or if they pass any condition which follows a \texttt{reject} key.  

Much of what happens in an analysis can be modeled as a pipeline that transforms one set
of objects to another.  For example, a jet algorithm transforms a set of particles to a set of jets, or smearing functions  convert a set of objects into a set that incorporates detector effects.  In a $H \rightarrow ZZ \rightarrow 4\ell$ analysis, reconstructed lepton objects are
transformed into a pair of reconstructed Z boson objects. In an analysis using boosted objects, jets are transformed
into their constituent particles and the latter are then transformed into boosted objects.
Such transformations can be specified  
 using the \texttt{apply} key followed by the function to be applied. Clearly, these functions must be specified within the previously defined \texttt{function} blocks. 

The exact set of accessible object properties needs to be defined at a later stage of the {\sc LHADA} proposal.  For the time being, a sufficiently generic assumption could be to consider each object to be of \texttt{ROOT TLorentzVector} type, extended by the particle's PDG ID. For example, one can
envisage adopting a generic Les Houches particle type, \texttt{LHParticle}, which inherits the functionality of
\texttt{TLorentzVector} and adds extra behaviour and functionality.



%
%
%
%

\paragraph{Examples}
\begin{verbatim}  
objects mu
  # Muons
  take external
  apply detector_muons(workingpoint=combined)
  select pt > 10
  select |eta| < 1.5
  select isol(src=tracks, dR=0.4, reliso=true)<0.1
  doc ...

objects e_l
  # loose electrons
  take external
  apply detector_electrons(workingpoint=loose)
  select pt > 5
  select |eta| < 2.5
  select isol(src=tracks, dR=0.4, relIso=true)<0.1
  reject overlaps(neighs=mu, dR=0.4)
  doc 10.1140/epjc/s10052-014-2941-0 # doi to ATLAS ID def.

objects e_m 
  # medium electrons
  take e_l
  apply detector_electrons(workingpoint=medium)
  select pt > 20
  doc 10.1140/epjc/s10052-014-2941-0 # doi to ATLAS ID def.

objects lep 
  # leptons contain hard medium electrons and loose muons
  take e_m 
  take mu
  select pt > 50
  
objects jets
  # clustered jets from the calorimeter cells
  take external
  apply antikt(dR=0.4, ptmin=20, etamax=2.5)
  code ...
  doc ...

objects met
  # missing energy
  take external
  code ...
  doc ...
\end{verbatim}

\subsection{The \texttt{cut} Block}

The event selection is specified by a one or more of \texttt{cut} blocks, which contain sets of constraints that would classify the event into a certain analysis region.  
To be counted, an event must pass all boolean statements which follow a \texttt{select} key and analogously fail all which follow a \texttt{reject} key. To make branchings in the cutflow procedure comfortable to implement, a whole cut block can itself be considered as a boolean constraint, which returns true if all its contained constraints are true. 

Event reweighting can be possible by the \texttt{weight} key, which multiplies each event by the number which follows after the key.  If unspecified, the weight is assumed to be 1.

A binning of one or more parameters to define different analysis regions can conveniently be achieved by a special \texttt{bin} keyword. This keyword can either be followed by \texttt{nbins:min:max} or, to account for variable bin sizes, explicitly by \texttt{bin1;bin2;bin3;...}, where each \texttt{binX} refers to the left boundary of that bin. 

The selections specified in \texttt{select} and \texttt{reject} can consist of complicated combinations of logical statements.  Logical operators \texttt{and}, \texttt{or} and \texttt{not} can be used.  


\paragraph{Examples}
\begin{verbatim}
cut preselect
  # Pre-selection cuts
  weight triggerefficiency(leptonpt = lep[1].pt)
  reject lep.size > 1
  select lep[1].pt > 75
  select jets.size > 2

cut leadjets_1
  select jets[1].pt >= 60
  select jets[2].pt >= 40
 
cut leadjets_2
  select jets[1].pt >= 40
  select jets[2].pt >= 20
  
cut SRA
  select preselect
  select leadjets_1
  select mT2(vis1=jets[1], vis2=jets[2], invis=met, mass=0) > 100

cut SRBtoF
  select preselect
  select leadjets_2
  bin met.pt = 100,125,150,200
  
cut noZ
  # define a region outside the Z mass range
  select mll < 70 or mll > 100

cut razor
  # Define the ladder-like razor region
  select (MR>100 and R2>0.8) or (MR>300 and R2>0.5) or (MR>500 and R2>500)
\end{verbatim}

The example block {\tt cut SRBtoF} shows how it is possible to partition a region in four bins, in this case with $\mathrm{MET} >= 100,~125,~150$ and $250$ respectively.  This will automatically split the region {\tt SRBtoF} into four that can be refered to as {\tt SRBtoF[0], ... SRBtoF[3]} which can then be used further on for additional cuts if required. 

The blocks {\tt cut noZ} and {\tt cut razor} show two cases for logical operators, where in {\tt noZ}, a selection is defined outside the Z mass range (which is common in dilepton SUSY searches), and in {\tt razor}, a ladder-like selection region using razor kinematic variables is defined.

\subsection{The \texttt{table} Block}

Although the main purpose of {\sc LHADA} proposal is to describe the analysis prescription, it would act as a more complete accord if the results for an analysis are also provided in the {\sc LHADA} proposal.  To enable this, we propose the \texttt{table} block, which is simply a tabular collection of results.

For simplicity, only two keys are associated with the \texttt{table} block, which are \texttt{columns} and \texttt{entry}.  The \texttt{columns} key is used for assigning names to the information columns, whereas each \texttt{entry} key corresponds an entry row in the table, which comprises values corresponding to each column.  The \texttt{table} block can be used for hosting any information that can be stated in a tabular form as long as it is accompanied by an unambiguous description.  The \texttt{type} keyword also specifies the type or source of the result.  This aims to facilitate the automatic processing of the results information. The five results types we propose are the following:
\begin{itemize}
\item {\bf \texttt{events:}} Denotes a simple table consisting of the name of the signal region, expected (and/or observed) signal events, expected background events, and error on the background events (see the first example below).  This is the minimum amount of information that would describe an analysis result.  This simple format can also be used for providing histograms where each entry corresponds to a bin. 
\item {\bf \texttt{limits:}} Denotes upper limits for a given signal region.  
\item {\bf \texttt{cutflow:}} Denotes cutflow tables.  In this case, the name of the table corresponds to the signal region name and each row gives the number of events after a given cut.
\item {\bf \texttt{corr:}} Denotes correlation matrices.  The first row and columns correspond to the names of one of the signal regions described earlier (see the second example below).
\item {\bf \texttt{bkg:}} Specifies individual contributions to the total background (see the third example below). Even though it may be unfeasible for the LHC experiments to compute a correlation matrix for all possible combinations of signal regions, an estimate for such a matrix can be calculated if a breakdown of the different sources of the Standard Model background is detailed along with the corresponding error. The names again match with the \texttt{cut} blocks and each individual background is given together with its error. We hope that a common background naming convention can eventually be agreed upon, such as done for the LHC Higgs analyses, in order to allow easy combination of different analyses.
\end{itemize}

\noindent Finally, if the user wishes to provide an external URL for a HEPDATA table, it can be done with the keyword \texttt{hepdata} followed by the HEPDATA URL.



\paragraph{Examples}
\begin{verbatim}
table results_events
  # Table for basic observed-signal and background events
  type events
  columns  name       obs   bkg   dbkg
  entry    SRA        3452  3452  59
  entry    SRBtoF[0]  1712  1720  161    
  entry    SRBtoF[1]  313   295   50
  entry    SRBtoF[2]  201   235   34
  entry    SRBtoF[3]  2018  2018  45 
  
table result_corr
  # Correlation matrix for signal regions
  type corr
  columns  name      SRA  SRBtoF[0] SRBtoF[1] SRBtoF[2] SRBtoF[3] 
  entry    SRA       1    0.2       0.1       0.15      0.14
  entry    SRBtoF[0] 0.2  1         0.5       0.4       0.3    
  entry    SRBtoF[1] 0.1  0.5       1         0.3       0.2      
  entry    SRBtoF[2] 0.15 0.4       0.3       1         0.7
  entry    SRBtoF[3] 0.14 0.2       0.2       0.7       1
  
table result_bkg
  # Breakdown of background in different signal regions
  type bkg
  columns  name       Z_jets Z_jets_err  W_jets  W_jets_err ...
  entry    SRA        1726   254         1151    178        ...
  entry    SRBtoF[0]  856    89          571     76         ...
  entry    SRBtoF[1]  157    27          105     18         ...
  entry    SRBtoF[2]  101    19          67      12         ...
  entry    SRBtoF[3]  1009   156         674     56         ...
  
table signal_results
  hepdata http://hepdata.cedar.ac.uk/view/ins1304456/d2
\end{verbatim}

\noindent To summarize, the current {\sc LHADA} proposal consists of a simple, plain text documentation of an analysis using a set of eighteen reserved keywords: \\
{\tt
\begin{tabular}{llllll}
apply & arg & bin  & 	code & columns & cut \\
doc & entry & function & hepdata & info & object \\
reject & select & table & take & type & weight  \\
\end{tabular}
}

\section{CONCLUSIONS}
We have presented the first draft of a proposal for the implementation of a Les Houches Analysis Description Accord.    
The proposed accord is a software-framework-independent, human readable formalism that consists of simple text files describing an analysis, accompanied by self-contained functions encapsulating variables that are nontrivial to express.  This formalism would fulfill the requirements expected of an analysis description accord, including analysis preservation and facilitating the abstraction, visualization, validation, combination, reproduction, interpretation and communication of the contents of LHC analyses.  The proposal presented here, inspired by previous Les Houches accords, would allow easy parsing and usage by any analysis framework.  We intend to engage in detailed discussions of the draft proposal with LHC physicists from all backgrounds with the goal of collectively arriving at an analysis description accord that is broadly acceptable to the whole LHC community.  We recognize, and accept, the fact that not every aspect of the final accord will be acceptable to everyone. However, the benefits of such an accord are such that we firmly believe it is a community effort that is both worthwhile and timely. 
The design of this accord would benefit enormously from efforts to implement real-world analysis test cases and by interfacing the accord with existing analysis frameworks. Work in this direction is already underway.

\section*{ACKNOWLEDGEMENTS}
We thank the organizers of the Les Houches PhysTeV workshop for the very inspiring atmosphere that ignited many fruitful discussions.  To continue work on the analysis description accord proposal, a follow-up workshop was organized at LPSC Grenoble in February 2016.  We thank the LPSC Grenoble and its theory group for their hospitality. This work has been partially supported by THEORIE-LHC-France initiative of CNRS-IN2P3, Investissements d'avenir - Labex ENIGMASS, and the ANR project DMASTROLHC, ANR-12-BS05-0006.  SuK is supported by the New Frontiers program of the Austrian Academy of Sciences.  The work of JS is supported by the collaborative research center SFB676  ?Particles, Strings, and the Early Universe? by the German Science Foundation (DFG) and by the German Federal Ministry of Education and Research (BMBF).

\AddToContent{D.~Barducci, G.~Chalons, N.~Desai, N.~de~Filippis, P.~Gras,
  S.~Kraml, S.~Kulkarni, U.~Laa, M.~Papucci, H.~B.~Prosper, K.~Sakurai, D.~Schmeier, S.~Sekmen,
  D.~Sengupta, J.~Sonneveld, J.~Tattersall, G.~Unel, W.~Waltenberger and A.~Weiler}
\renewcommand{\thesection}{\arabic{section}}

\graphicspath{{rosetta/}}
\renewcommand{\beq}{\begin{eqnarray}}
\renewcommand{\eeq}{\end{eqnarray}}
\newcommand{\rosetta}{\textsc{Ro\-set\-ta}}
\newcommand{\lilith}{\textsc{Li\-li\-th}}
\newcommand{\ehdecay}{e\textsc{HDECAY}}
\newcommand{\code}[1]{\texttt{#1}}
\newcommand{\flo}[1]{{\color{blue} #1}}
\newcommand{\tn}{\tabularnewline}

\newcommand{\sigmaHH}{\sigma_{\rm HH}}
\newcommand{\sigmaHHMC}{\sigma_{\rm HH, MC}}
\newcommand{\RHH}{R_{\rm HH}}
\newcommand{\mH}{m_{\rm H}}
\newcommand{\mHH}{m_{\rm HH}}
\newcommand{\HH}{\rm HH}

\chapter{Basis-independent constraints on Standard Model Effective Field Theories with \rosetta}

{\it J.~Bernon, A.~Carvalho, A.~Falkowski, B.~Fuks, F.~Goertz, K.~Mawatari, K.~Mimasu and T.~You}

\begin{abstract}
The \rosetta\ package is an operator basis translation framework that aims for basis-%
independent Standard Model Effective Field Theory analysis. In this note, we present some recent developments of the program to provide information on the compatibility of a particular set of Wilson coefficients in a given operator basis with existing data from low-energy precision, LHC Higgs signal-strength and non-resonant diHiggs production at the LHC. Based on and extending an array of previous work, we implement analytical formulae for relevant observables at Leading Order in the Wilson coefficients which allows \rosetta\ to extract likelihood information for arbitrary points in parameter space. The compatibility with Higgs signal-strength measurements is obtained via an interface to the \lilith\ program and also takes advantage of an existing \ehdecay\ interface within \rosetta.
\end{abstract}

\section{Introduction\label{sec:intro}}
The observation in 2012 of a new scalar particle at the Large 
Hadron Collider (LHC), with characteristics consistent with those of the 
Higgs boson of the Standard Model (SM) of particle physics, has led to a fruitful, ongoing research programme on the 
nature of electroweak (EW) symmetry breaking. A major element of this 
programme is the precise measurement of the properties of the Higgs boson, with 
the aim of pushing the SM hypothesis to its limits. The first run of the LHC 
brought a host of rate or signal-strength measurements concerning the various 
production and decay modes of the Higgs boson as well as the first differential 
distributions probing in particular the high energy behaviour of the Higgs beyond the scale of 
the breaking of the EW symmetry set by the Higgs vacuum expectation value $v$. The 
continuous stream of data from LHC brings complementary information to the 
existing precision measurements of the EW sector performed at, \emph{e.g}, 
the Large Electron-Positron (LEP) collider. In order to extract the maximum amount of information from the available 
data, a global approach is necessary, combining data spanning a wide range of 
energies and involving statistical and systematic uncertainties of diverse 
origins. Furthermore, a general, model independent way to describe deviations 
from the SM couplings is necessary.

The interpretation of the experimental measurements in the context of an Effective 
Field Theory (EFT) is an example of such an approach. Along with the field 
content and symmetries of an
established Quantum Field Theory -- in this case the SM -- an EFT is 
characterised by a well 
defined operator expansion in the ratio of the energy scale at which 
processes of interest occur, and a generic `cut-off' scale
$\Lambda$ where one expects new physics to appear.
The SM is supplemented with new operators respecting its local and global 
symmetries, each with a Wilson coefficient controlling the size of their 
contribution to physical processes.\footnote{Note that respecting the SM
{\it global} symmetries is a \emph{choice}, which we follow henceforth.} The leading canonical dimension at which 
these can appear is six, meaning a suppression by $\Lambda^2$, 
where 59 independent structures 
arise~\cite{Buchmuller:1985jz,Grzadkowski:2010es}\footnote{Relaxing flavor 
universality, the number of
independent dimension-six operators grows to 2499~\cite{Alonso:2013hga}.}. EFT 
interpretations of LHC data are then designed to capture small effects from 
heavy particles living at energies beyond the current experimental reach. The fact that 
direct searches for exotic states have, so far, been 
unsuccessful further reinforces EFTs as essential elements of the LHC 
programme. An EFT not only parametrises
physics beyond the SM but is also useful in studying the origin of 
observed and potential deviations. Correlations among observables can moreover help to 
connect the Wilson coefficients predicted by specific high energy or 
ultraviolet (UV) models. 

The issue of operator basis choice in an EFT arises from the possibility of
relating certain sets of operators via field redefinitions, equations of motion
and Fierz identities. Taking from now on the case of the SM EFT,
contributions from EFT to observables can come
from fixed combinations of Wilson coefficients up to these basis
`rotations'. A number of popular bases exist in the literature, such as the
the so-called \emph{Warsaw} basis~\cite{Grzadkowski:2010es}, \emph{SILH}
(strongly interacting light Higgs) basis~\cite{Giudice:2007fh,Contino:2013kra}
and \emph{BSM primaries} basis~\cite{Gupta:2014rxa,Masso:2014xra, Pomarol:2014dya},
from which the recent works on the~\emph{Higgs} basis~\cite{Falkowski:2015fla}
are inspired. All of these represent equivalent descriptions of deviations from
SM interactions due to dimension-six operators, each with their own conveniences
and inconveniences. The subject of these proceedings is a recently introduced
tool, \rosetta~\cite{Falkowski:2015wza}, which provides a translation framework
between different EFT bases. The program serves to emphasise the notion of basis
independence, unifying basis descriptions concerning a number of research
purposes. While the basic functionality of \rosetta\ is to simply take a set of
input values for the Wilson coefficients in a particular basis and translate
them to another basis, this opens the possibility for users of any particular
basis choice to take advantage of the array of EFT-related third-party software,
such as e{\sc HDecay}~\cite{Contino:2014aaa} and
{\sc MadGraph5\_aMC@NLO}~\cite{Alwall:2011uj}, assuming a certain basis for
input, as well as to make use of theoretical results obtained in a particular
basis choice. These could be, for example, the results of a global fit of EFT
parameters from current measurements~\cite{Pomarol:2013zra, Falkowski:2015krw,
Efrati:2015eaa,Han:2004az,Ellis:2014jta,Ellis:2014dva,Buckley:2015lku} and
projections at future colliders~\cite{Ellis:2015sca, Ge:2016zro}, or the recent
calculation of the complete one-loop anomalous dimension matrix for the
renormalisation group evolution/mixing of the Wilson coefficients and their
effects~\cite{Grojean:2013kd,Elias-Miro:2013eta,Alonso:2013hga,Jenkins:2013zja,%
Jenkins:2013wua, Wells:2015cre, Elias-Miro:2013gya, Elias-Miro:2013mua}, as well
as other EFT-related studies.

In the following we describe some new interfaces developed for \rosetta\ for
determining compatibility of EFT parameters with existing data in a
basis-independent way. We focus here on data concerning electroweak precision
observables (EWPO) on- and off-$Z$ peak obtained at LEP complemented by a few
hadron collider processes as well as selected measurements at low energy scattering
experiments. In addition, we also consider the Higgs signal strength measurements performed thus far at the LHC, and the double-Higgs production rate measurement~\cite{Aad:2015xja} using cluster analysis~\cite{Dall'Osso:2015aia}. In all cases, we make use of previous
works~\cite{Falkowski:2015fla,Falkowski:2014tna,Falkowski:2015jaa,Falkowski:2015krw,Efrati:2015eaa}, determining the Leading Order (LO) parametric dependence of EWPO and Higgs production cross sections and branching fractions assuming either the \emph{Higgs} basis or an anomalous coupling description akin to the BSM Characterisation Lagrangian (BSMC) of \rosetta. The operator content of the BSMC is identical to that of the \emph{Higgs} basis apart from the non-imposition of the relations for `dependent' parameters in terms of `independent' ones that connects it to the other dimension-six descriptions. This allows for the calculation of a likelihood given a specific input of EFT parameters. The EWPO interface will compute a first
likelihood given an internally parametrised $\chi^2$-measure and covariance matrix in the Gaussian limit, while the Higgs signal strengths likelihood will be obtained via an interface to \lilith~\cite{lilith}. The two features are to be included in a new version of the code released concurrently with this document.

The remaining sections of this document describe the implementations of the three new \rosetta\ interfaces. In all cases, the EFT is mapped to the physical observables in terms of the \emph{Higgs} basis parametrisation at the tree-level in the EFT power-counting. We refer the reader
to Ref.~\cite{Falkowski:2015fla} for a comprehensive overview of this
description. In Sec.~\ref{sec:EWPO} we summarise the method by which the
numerical likelihood from the EWPO has been obtained, briefly describing its implementation in \rosetta. In Sec.~\ref{sec:Higgs}, we describe the connection to the Higgs signal strengths and the \rosetta\ interface to \lilith. In Sec.~\ref{sec:diHiggs}, the implementation of the anomalous coupling description and the LHC double-Higgs analysis is described. Sec.~\ref{sec:rosetta} presents a summary of the new features of \rosetta, how to invoke them from the command-line as well various options. Finally, in Sec.~\ref{sec:conclusion} we conclude.

\section{Electroweak Precision Observables\label{sec:EWPO}}
SM EFT descriptions assuming a linearly realised EW symmetry breaking contain a large number of parameters. One may easily get lost in the parameter space when applying the formalism, for example, to Higgs physics at the LHC. Fortunately, it turns out that many directions have already been constrained at leading order by various precision measurements prior to or at the LHC. Several works have derived these constraints for the general dimension-six EFT and obtained likelihood functions in the multidimensional Wilson coefficient space~\cite{Falkowski:2015krw,Efrati:2015eaa,Han:2004az,Ellis:2014jta} assuming various degrees of generality with respect to
the flavour structure of the theory. For the \rosetta\ interface, we make use of results from Refs.~\cite{Falkowski:2015krw,Efrati:2015eaa}, in which a characterisation of a likelihood function using a large number of precision measurements is presented assuming an arbitrary flavour structure. This way, the bounds are more robust, and hold for a larger class of new physics scenarios. In addition, one can identify certain weakly constrained  directions in the flavour space of dimension-six operators, that may be interesting targets for future explorations. 

In the analyses, tree-level contributions of dimension-six operators to the precision observables are taken into account.
While the SM predictions for these observables are state-of-art and typically include one- and/or two-loop contributions, all loop suppressed contributions proportional to the Wilson coefficients of higher-dimensional operators are ignored.  
Furthermore, possible contributions from dimension-eight operators and higher are neglected as the analysis is performed at $\mathcal{O}(v^2/\Lambda^2)$ in the EFT counting. 
This means that corrections to observables are {\em linear} in the dimension-six Wilson coefficients and stem from interferences with the tree-level SM contributions.
On the other hand,  corrections proportional to squares of the Wilson coefficients are 
of $\mathcal{O}(v^4/\Lambda^4)$, much like interference between the SM and the
neglected dimension-eight operators and are thus not taken into account. 
In any case, the experimental constraints force the Wilson coefficients to be small such that their squares are negligible. 
It may, however, matter if experimental constraints leave flat or weakly constrained directions in the parameter space, as along such directions  squares may be numerically significant. 
This issue emerges in the context of LEP-2 constraints on anomalous triple gauge couplings (ATGCs)~\cite{Falkowski:2014tna}, although the flat direction is lifted once constraint from LHC Higgs searches are taken into account~\cite{Falkowski:2015fla}.  
One consequence of neglecting the squares is that the likelihood is Gaussian, which is convenient for practical purposes.  

The experimental results used in this analysis originate from the LEP-1, LEP-2, Tevatron and LHC colliders, as well as from low-energy precision experiments.
Based on that input, constraints on diagonal fermion-fermion-gauge boson vertex corrections $\delta g$,
on ATGCs, and on the Wilson coefficients of the four-lepton operators $c_{4f}$ are derived. 
The analysis assumes that new physics corrections to the SM are well approximated by tree-level effects of dimension-six operators.
In particular, all dimension-six operators can be present simultaneously with arbitrary coefficients (within the EFT validity range).  
The results are quoted as $1\sigma$ confidence intervals for the EFT parameters together with the full correlation matrix, which allows us to reproduce the Gaussian likelihood function in the space of these parameters.

\subsection{Observables \label{sec:EWPO:obs}}
The set of observables used in this fit can be summarised in the following categories, referring to the tables
provided in Appendix~\ref{app:EWPO} (and references therein) for the central values, uncertainties and SM predictions: 
\begin{itemize}
    \item On-shell $Z$- or $W$- boson production and subsequent decays to the SM fermions: The measurements have been
largely performed at LEP-1 for the $Z$-boson, and at LEP-2 for the $W$-boson, and are also complemented by some Tevatron and LHC results for what concerns the couplings to light and top quarks (Tables~\ref{tab:EWPT_zpole} and~\ref{tab:EWPT_wpole}). 
    \item Measurements above the $Z$-pole: The measurements have been performed by the LEP-2 experiments
and constrain ATGCs and certain four-fermion operators.
    \item Low energy measurements: These provide additional sensitivity to four-lepton operators by means of low-energy muon-neutrino--electron scattering at the CHARM, CHARM-II and BNL-E734 experiments (Tab.~\ref{tab:EWPT_lowe}), parity-violating electron scattering at the SLAC-E158 experiment and the decays of the $\tau$ and muon.
\end{itemize}
For the pole observables, the we have used SM predictions quoted in Tables~\ref{tab:EWPT_zpole} and~\ref{tab:EWPT_wpole} which represent the state-of-art theoretical calculations.
Whenever available, the central value quoted in the Table~2 of Ref.~\cite{Baak:2014ora} is used.
The theoretical errors on the SM predictions are subleading compared to the experimental errors on the measured quantities, and are therefore neglected.
Above the $Z$-pole, pseudo-observables for differential cross section measurements of $W^+W^-$ pair production are used along with total cross sections and forward-backward asymmetries measured from
fermion pair production, all performed at LEP-2 at several centre-of-mass energies~\cite{Schael:2013ita}. The low energy neutrino scattering results are presented in terms of measurements of the vector and axial couplings of the $Z$-boson to electrons, for which the SM predictions are also given in Tab.~\ref{tab:EWPT_lowe}. The measurement of the parity-violating asymmetry in M\"o ller scattering was performed at the SLAC-E158 experiment~\cite{Anthony:2005pm} and is quoted in terms of the value of the weak mixing angle at $Q^2=0.026 \,\text{GeV}^2$
\begin{equation}
s^2_\theta(Q^2=0.026 \,\text{GeV}^2)=0.2397\pm 0.0013, 
\end{equation}
for which the SM prediction~\cite{Czarnecki:1995fw} is $0.2381\pm 0.0006$. The observables probing $\tau$ decays correspond to the ratio of the effective Fermi constant of the muonic to the electronic decay of the $\tau$, parametrised by the four-lepton operator
\begin{align}
\mathcal{L}= -\frac{4G_{\tau f}}{\sqrt{2}}(\bar{\nu}_\tau \bar \sigma_\rho \tau)(\bar f \bar \sigma_\rho \nu_f),
\,+\,{\rm h.c.}\,\,\text{with}\,\,f=\mu, e,
\end{align}
and with the Particle Data Group (PDG) values
\begin{align}
A_e &\equiv  \frac{G_{\tau e}^2}{G_F^2}=1.0029 \pm 0.0046,\\ 
A_\mu &\equiv  \frac{G_{\tau \mu}^2}{G_F^2}= 0.981 \pm 0.018. 
\end{align}
Finally, for muon decays, the partial width $\Gamma(\mu \to \nu_\mu\, e \,\nu_e)$ defines the SM input parameter $v$ and is therefore unaffected by any EFT parameter. However, certain differential distributions in polarized decays can offer additional information, usually presented in terms of the \emph{Michel parameters}~\cite{Bouchiat:1957zz}. From the EFT perspective the most interesting are the so-called $\eta$ and $\beta'/A$ parameters, 
because they are the only ones that may receive contributions at $\mathcal{O}(1/\Lambda^2)$ \cite{Gonzalez-Alonso:2014bga, martinthesis}.
These parameters have been measured in an experiment at PSI \cite{Danneberg:2005xv}: 
\begin{align}
\eta = -0.0021 \pm 0.0071, 
\qquad 
\beta'/A =  -0.0013 \pm 0.0036. 
\end{align}
\subsection{Implementation \label{sec:EWPO:imp}}
The multidimensional likelihood in the \emph{Higgs} basis parameter space is built in using analytical expressions for the contribution of the EFT parameters to each observable
considered in the fit. This is a synthesis of the work contained in several papers~\cite{Falkowski:2015fla,Falkowski:2014tna,Efrati:2015eaa, Falkowski:2015krw}. The global $\chi^2$ function is constructed via the generic expression
\begin{align}
O_{i,\rm th}  = O_{i,\rm SM}  +  \vec{c} \cdot \vec \delta O_{i,\rm LO \, EFT}, 
\end{align}
where  $O_{i,\rm SM}$ is the SM prediction, while the leading order EFT corrections are noted by
$\vec{c} \cdot \vec \delta O_{i,\rm LO \, EFT}$. The latter are linear in the
vector of dimension-six parameters $\vec c$ and are computed analytically.
Then, given a set of measurements, $O_{i,\rm exp} \pm \delta O_{i}$, the likelihood  function is  constructed as
\begin{align}
\chi^2 = \sum_{ij}\left [ O_{i,\rm exp}  -  O_{i,\rm th} \right ] \sigma^{-2}_{ij}  \left [ O_{j,\rm exp}  -  O_{j,\rm th} \right ],
\end{align}
where  $\sigma^{-2}_{ij} =  [\delta O_{i} \rho_{ij, {\rm exp}} \delta O_{j} ]^{-1}$ is calculated from the experimental errors and their correlations, $\rho_{ij, \rm exp}$ whenever known. Otherwise, no correlations (for $i\neq j$)
are included.

The central values of the fit to the relevant \emph{Higgs} basis parameters
$\vec c_0$ obtained by minimising the likelihood function are summarised in
Appendix~\ref{app:EWPO}, along with their $1\sigma$ uncertainties $\delta\vec c$
and the associated correlation matrix $\rho$. These form the basis of our
implementation of the \rosetta\ interface. From these quantities, the dependence of the global $\chi^2$ function on the full parameter space can be reconstructed as a multivariate Gaussian,
\begin{align}
\chi^2= \sum_{ij}[\delta c - c_0]_{i} \sigma^{-2}_{ij} [\delta c - c_0]_{j},
\end{align}
where $\sigma^{-2}_{ij} = [\delta c_{i} \rho_{ij} \delta c_{j} ]^{-1}$. This function is implemented in the \code{EWPO} interface of \rosetta, receiving an instance of a basis class, performing the required translation to the \emph{Higgs} basis, and outputting the value of the likelihood for that set of EFT parameters. This can also be interpreted as a $p$-value given 36 fitted parameters via the survival function of a $\chi^2$ of that many degrees of freedom. The fit is performed assuming both a general flavour structure and minimal flavour violation (MFV). 



\section{Single Higgs measurements\label{sec:Higgs}}
\subsection{Signal-strengths}
Given a particular EFT basis choice, it is relatively straightforward to map a given set of Wilson coefficients into predictions for Higgs production cross sections and branching fractions, such that the compatibility with existing measurements can be assessed. The simplest set of such measurements are the so-called Higgs signal-strengths, presented in the $\kappa$-formalism as ratios of predicted rates in a particular production channel $X$  and decay mode $Y$ to the SM expectation,
\begin{align}
    \mu_{X,Y} = \frac{\sigma_X\times \mathrm{BR}(h\to Y)\times\mathcal{A}\times\varepsilon}{\left[\sigma_X\times \mathrm{BR}(h\to Y)\times\mathcal{A}\times\varepsilon\right]_{SM}}.
\end{align}
In this expression, $\sigma_X$ denotes the production cross section and $\rm{BR}$ the decay branching fraction,
while an experimental acceptance times efficiency factor $\mathcal{A}\times\varepsilon$
is also taken into account. Assuming that $\mathcal{A}\times\varepsilon$ for the new physics signal modelling is not significantly different from the SM case, these experimental factors can be simplified to a
`reduced efficiency' $\epsilon_{X,Y}$ such that the combination of production cross section and branching ratio can be directly constrained from the data
\begin{align}
    \mu_{X,Y} = \epsilon_{X,Y}\frac{\sigma_X\times \mathrm{BR}(h\to Y)}{\left[\sigma_X\times \mathrm{BR}(h\to Y)\right]_{SM}}
    = \epsilon_{X,Y}\times\frac{\sigma_X}{\sigma^{SM}_X}\times
                     \frac{\Gamma(h\to Y)}{\Gamma^{SM}(h\to Y)}\times
                     \frac{\Gamma^{SM}_{\mathrm{tot.}}}{\Gamma_{\mathrm{tot.}}}.\label{eq:ratios}
\end{align} 
This setup is then able to constrain models that predict a simple rescaling of the SM Higgs rates using a combination of measurements of the Higgs production and decay products to date. 

It goes without saying that the EFT framework contains far more information than can be extracted by only measuring these rates, since the presence of new Lorentz structures and, in particular, higher derivative interactions induce momentum-dependent vertices not present in the dimension-four SM Lagrangian. Consequently, the signal-strengths do not represent the optimal way to search for deviations due to higher-dimensional operators in the Higgs sector or any other sector. However, it remains true that constraints from these measurements must be satisfied to begin with when considering EFT effects in other observables. Furthermore, there is already an established industry in making use of this data to provide limits on beyond the SM scenarios with several, user-friendly, software packages available. For this reason, incorporating this information into \rosetta\ via an interface to \lilith, was deemed a useful addition to the growing functionality of the package, which can now provide the compatibility with Higgs signal-strength data to the EFT framework in a basis-independent manner.

\subsection{Implementation}
The theoretical ingredient for the rate measurement interface is a map from an
EFT description to predictions for Higgs signal-strengths computed relative to
the SM expectation, \emph{i.e.}, the three ratios of equation~\eqref{eq:ratios}.
More precisely, this necessitates the calculation of production cross sections, decay
branching fractions and total widths. These have been computed semi-analytically and reported in Ref.~\cite{Falkowski:2015fla} as linear functions of the parameters of the BSMC assuming MFV. \rosetta\ incorporates these results such that input from any basis implementation is translated to the BSMC (via the \emph{Higgs} basis) and fed into the \lilith\ machinery to build the likelihood and extract a $\chi^2$ measure of a given parameter point. \lilith\ keeps its own repository of experimental signal-strength results that is periodically updated and from which the program derives the likelihood. Extending the work documented in Ref.~\cite{Falkowski:2015fla}, the parametric dependence is retained up to quadratic order in the EFT parameters, \emph{i.e.}, up to order $1/\Lambda^4$. The question of whether to include such terms in a given analysis will not be addressed here (see Ref.~\cite{Biekotter:2016ecg} for a recent discussion on this) and will ultimately be left up to the user.  

The implementation of these formul\ae, summarised in Appendix~\ref{app:Higgs}, is contained in the \code{SignalStrengths} package found in \code{Rosetta/interfaces}. The key functions implemented are \code{production()} and \code{decay()}, which are contained in the {\sc Python} modules of the same name. \rosetta's \code{Lilith} interface makes use of these functions to create the \code{xml} input string feeding the likelihood calculation function of \lilith.

\section{DiHiggs channel\label{sec:diHiggs}}
One of the key search modes for unveiling the nature of EW symmetry breaking at colliders is the non-resonant production 
of a pair of Higgs bosons,
that  provides the first window on the trilinear Higgs self-coupling. 
Of the several diHiggs production modes the most promising for detection at the LHC is in the gluon-fusion channel~\cite{Plehn:1996wb}. 
Searches for diHiggs production using LHC proton-proton collision data at a centre-of-mass energy of 8~TeV~\cite{Aad:2015xja, Aad:2014yja, Aad:2015uka, Khachatryan:2016cfa, Khachatryan:2015yea, Khachatryan:2016cfa, Khachatryan:2015tha, CMS:2015zug, CMS:2014ipa} have been performed by both ATLAS and CMS collaborations. The few results for non-resonant diHiggs production have assumed the signal to have a SM-like topology,
although the SM cross section for gluon-fusion-induced diHiggs production is too low to be accessed with LHC Run--I data. 

As noted in several works, the signal topology is very sensitive to the details
of the model both in the SM, considering both the leading order and higher-order
calculation cases~\cite{deFlorian:2015moa, Shao:2013bz}, as well as when considering new physics.
This second case has been investigated in the anomalous couplings/EFT approach, as shown for instance
in Refs.~\cite{Glover:1987nx,Azatov:2015oxa,Dall'Osso:2015aia}, and also in explicit models as
illustrated in the works of Refs.~\cite{Dawson:2015oha,Grober:2016wmf,Agostini:2016vze} or in
Chapter~\ref{chap:vlq} of the current proceedings, where diHiggs production is considered in
the context of the presence of vector-like quarks partners.


The number of anomalous couplings needed to define a point of the parameter space (relevant for the production of a Higgs pair) is four or five, depending on whether the Higgs is assumed to be part of a weak doublet. It is non-trivial to identify different regions of parameter space with specific signal topologies, with relevant differences compared to experimental resolution.  
In particular in Ref.~\cite{Dall'Osso:2015aia}, a statistical classification was designed, defining a list of clusters of parameter points, grouped together purely based on their kinematics. 
The approach relies on the fact that at LO, gluon-fusion
diHiggs production can be captured with only two kinematical variables (the
invariant mass of the diHiggs system and the angle between the Higgs bosons and the beam pipe), as any $2\to2$ process.
Moreover, all of the features that come on top of the LO hard scattering matrix element (including the Higgs decay
process that is assumed to be Standard-Model-like), from parton showering and hadronization effects to detector effects, are described by stochastic processes that do not carry new physics information.

Twelve benchmark points (representing the different clusters) are defined to be experimentally studied, that can also 
be found in the Yellow Report 4 of the LHC Higgs Cross Section Working Group~\cite{YR4}, and represent targets of the next generation of experimental results. 
Each parameter space point analysed is associated with one cluster, characterised (through its benchmark) by a specific set of anomalous couplings, as shown in Table~\ref{tab:bench}. These anomalous couplings are defined by the Lagrangian

\begin{table}[h]
\centering
\small{
\begin{tabular}{|c|c|c|c|c|c|}
\hline
Benchmark & $\kappa_{\lambda}$ & $\kappa_{t}$ & $c_{2}$	& $c_{g}$ & $c_{2g}$ \\
\hline
\hline
1 &	7.5	 & 1.0	 &	-1.0	& 0.0	& 0.0 \\
2 &	1.0	 & 1.0	 &	0.5		& -0.8	& 0.6 \\
3 &	1.0	 & 1.0	 &	-1.5	& 0.0	& -0.8 \\
4 &	-3.5 & 1.5  &	-3.0	& 0.0	& 0.0 \\
5 &	1.0	 & 1.0	 &	0.0		& 0.8	& -1 \\
6 &	2.4	 & 1.0	 &	0.0		& 0.2	& -0.2 \\
7 &	5.0	 & 1.0	 &	0.0		& 0.2	& -0.2 \\
8 &	15.0 & 1.0	 &	0.0		& -1	& 1 \\
9 &	1.0	 & 1.0	 &	1.0		& -0.6	& 0.6 \\
10 &	10.0 & 1.5   &	-1.0	& 0.0	& 0.0 \\
11 &	2.4	 & 1.0	 &	0.0		& 1		& -1 \\
12 &	15.0 & 1.0	 &	1.0		& 0.0	& 0.0 \\ 
\hline
\hline
SM &	1.0 & 1.0	 &	0.0		& 0.0	& 0.0 \\
\hline
\end{tabular}
}
\caption{\small Parameter values of the final benchmarks selected by the clustering procedure with the choice of $N_{clus} = 12$ (Table 1. of Ref.~\cite{Dall'Osso:2015aia}). The third cluster is the one that contains the SM sample. \label{tab:bench}}
\end{table}

\begin{eqnarray}
{\cal L}_h &=& 
\frac{1}{2} \partial_{\mu}\, h \partial^{\mu} h - \frac{1}{2} m_h^2 h^2 -
  {\kappa_{\lambda}}\,  \lambda_{SM} v\, h^3 
- \frac{ m_t}{v}(v+   {\kappa_t} \,   h  +  \frac{c_{2}}{v}   \, h\,  h ) \,( \bar{t_L}t_R + h.c.) \nonumber  \\ 
&&+ \frac{1}{4} \frac{\alpha_s}{3 \pi v} (   c_g \, h -  \frac{c_{2g}}{2 v} \, h\, h ) \,  G^{\mu \nu}G_{\mu\nu}\,,
\label{eq:lagHH}
\end{eqnarray}
which has a one-to-one mapping to the BSM Characterisation Lagrangian of \rosetta.




Each of the twelve benchmark points has a very characteristic topology at the generation level, see the Figure~6 of Ref.~\cite{Dall'Osso:2015aia}. 
For instance, there are benchmarks for which the $p_T$ of the Higgs boson peaks at 50 GeV, typically related
to the domination of the Higgs trilinear coupling in the amplitude causing on-threshold production.
In other clusters the Higgs boson $p_T$ can peak around 150 GeV or more, when the interferences between the different diagrams leads to a cancellation on the threshold. Clusters with intermediate behaviour in the Higgs boson $p_T$  and/or double-peaked structures are also found, depending
on the region of the parameter space being probed.
As and example of the usage of such a kinematical mapping, an analyzer that compares the experimental data to the benchmark of cluster 3 (that happens to be the cluster that contains the SM point) concludes that at 8 TeV the experiment excludes a signal of 0.69 pb with this shape at the 95\% C.L. (as the ATLAS limit~\cite{Aad:2015xja} in the SM case). 
If one has in hand the map of the parameter space points, such that one can identify those that are well described by the SM-like signal topology, as well as the corresponding cross sections, on can determine the level of compatibility of the points with the current data. 

\begin{figure*}\begin{center}
\includegraphics[width=0.32\textwidth, angle =0 ]{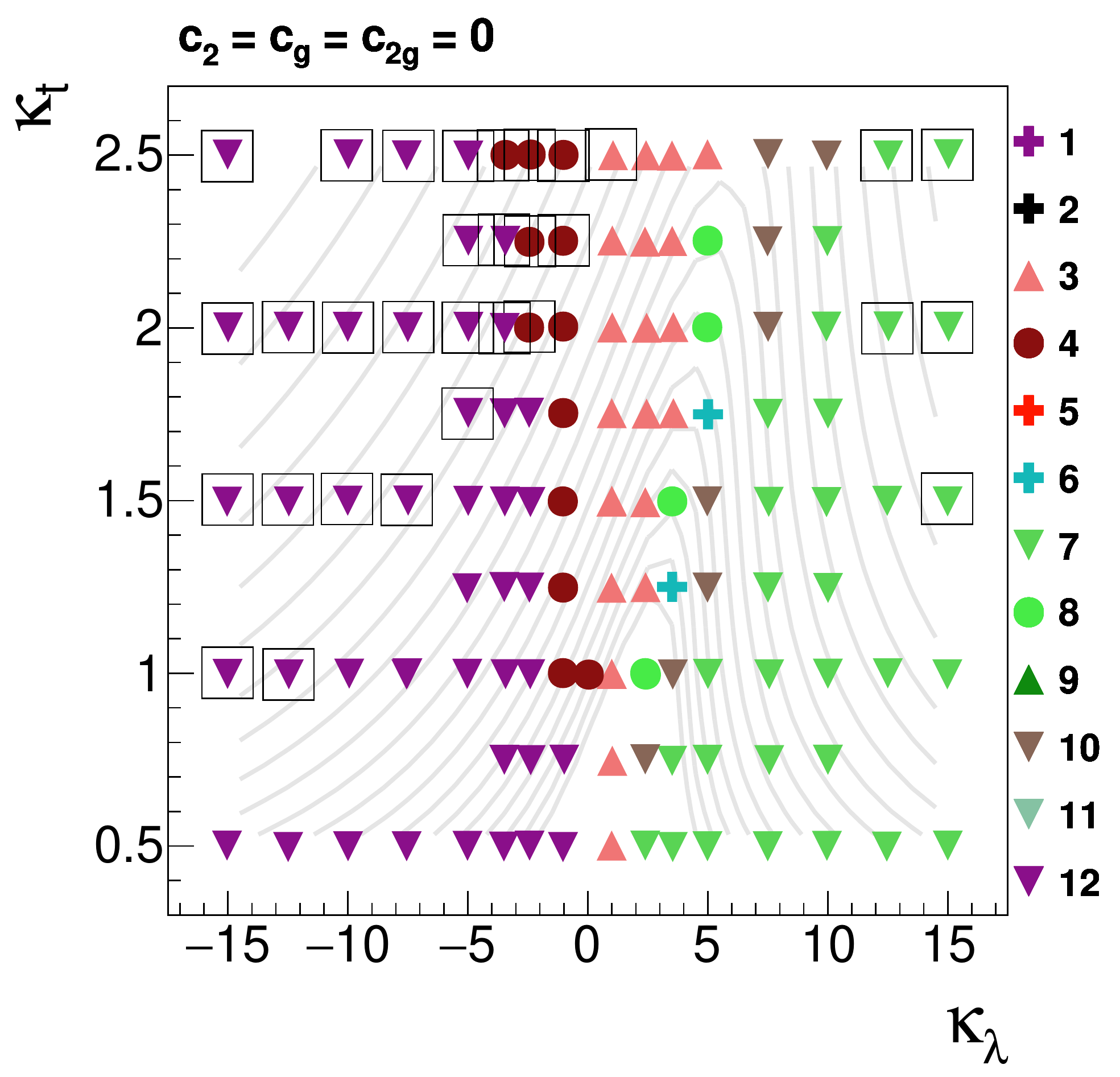}
\includegraphics[width=0.32\textwidth, angle =0 ]{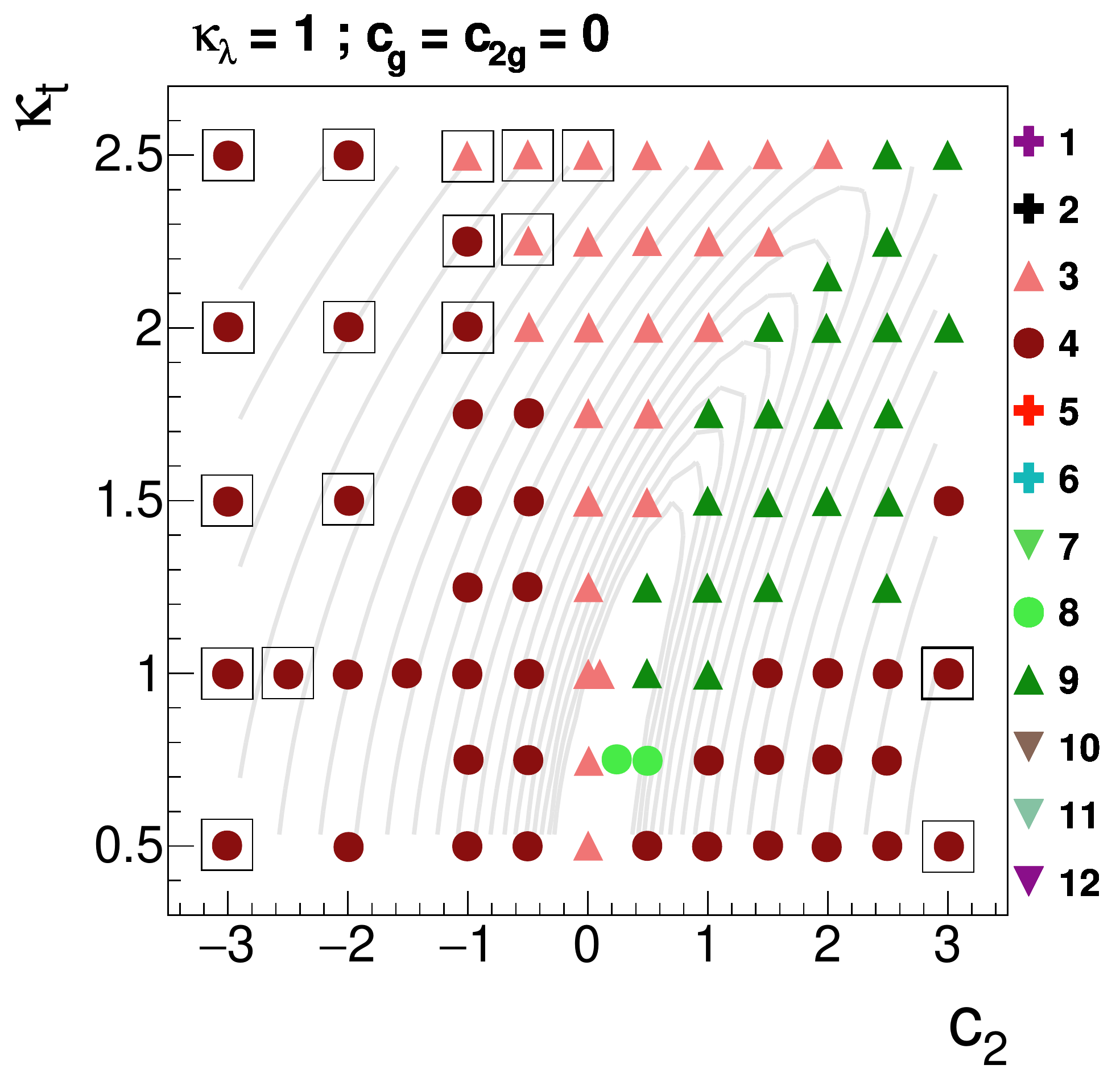}
\includegraphics[width=0.32\textwidth, angle =0 ]{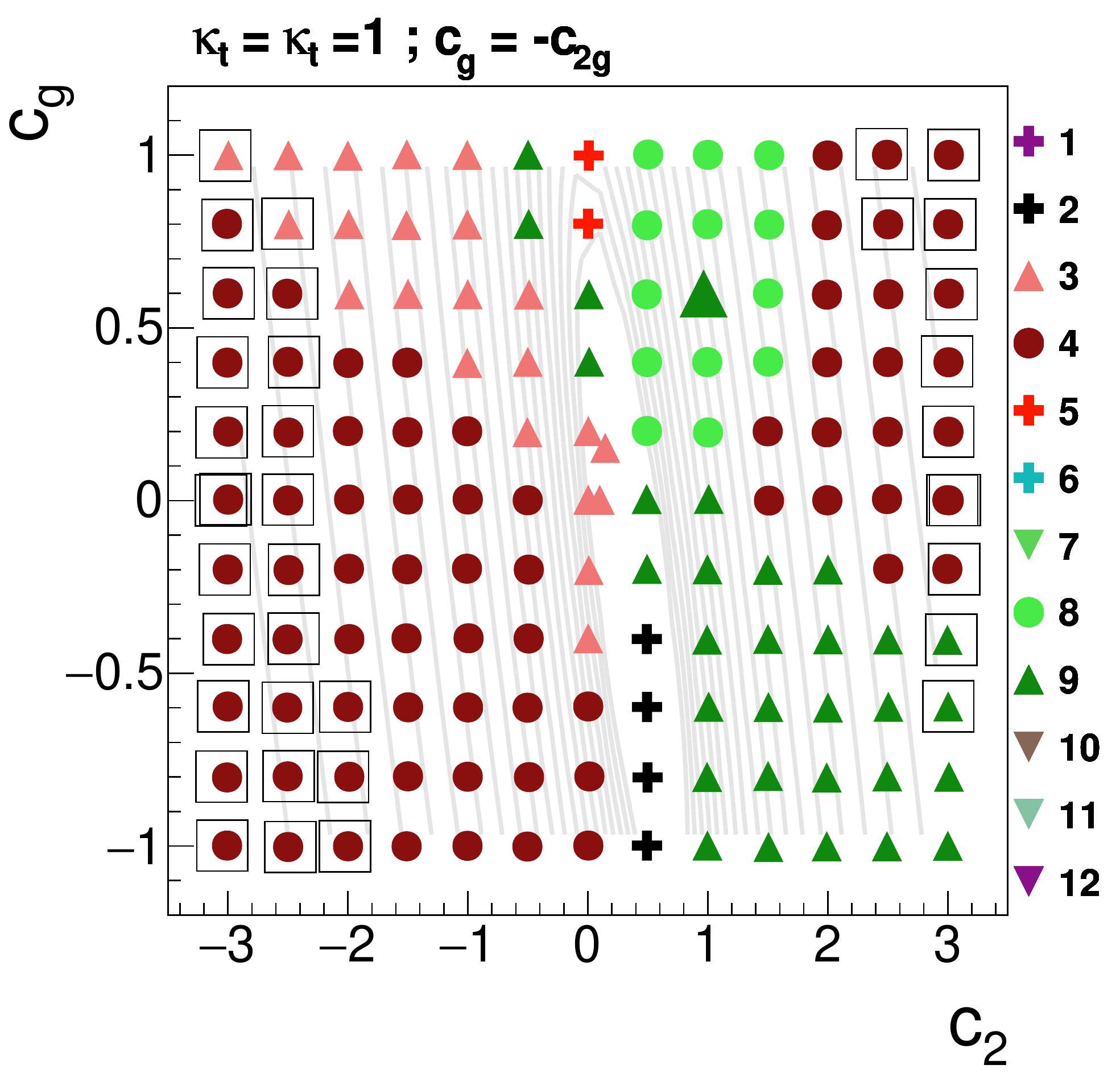}
\caption{\small
Distribution of nodes in various planes that contain the SM point. 
The different markers represent the twelve clusters (see Table~\ref{tab:bench}). The contours correspond to isolines of cross section. 
Cluster 3 represents the SM-like topology. 
The framed points are excluded if we consider the ATLAS limit for the SM benchmark  and assume the 
the Higgs boson decays SM-like. We extend the experimental limit to all the clusters for illustrational purposes, with the caveat that this extension may be not valid, in particular for
threshold like clusters (cluster 7 and 12, see text). 
\label{fig:tempSM} }
\end{center}\end{figure*}

In Figure~\ref{fig:tempSM} we show the result of the above exercise. 
We present three slices of the parameter space, where the left panel shows the  $(\kappa_{\lambda}, \kappa_t)$
plane, when $\kappa_t$ is varied respecting all
constraints from single Higgs measurements. The middle panel shows the $(c_2, \kappa_t)$  plane, when all the other parameters are fixed to their SM  value. Finally, the right panel corresponds to the $(c_2, c_g)$ plane, when the linear-EFT condition $c_g = - c_{2g}$ is imposed and all the other parameters are fixed to their SM value. 
The contours are isolines of cross sections. The cross section grows when $\kappa_t$ increases with respect to its SM value. Downward-pointing triangles symbolise clusters where the benchmark has a Higgs $p_T$ peaking at around 50 GeV or at a smaller value (threshold-like clusters). Circles describe clusters whose benchmark has a Higgs $p_T$ peaking around 100 GeV.  
Upward-pointing triangles describe clusters where the benchmark of the node has a Higgs $p_T$ peaking around 150 GeV or more. Finally, crosses describe clusters that show a double-peaking structure in the $m_{hh}$ distribution. 

Black frames around the parameter space points signify that the point would be excluded by the limit of Ref.~\cite{Aad:2015xja}, assuming SM-like Higgs boson branching ratios and an SM-like topology. The signal cross-section is calculated using the analytical parametrisation as presented in the LHCHXSWG document~\cite{CarvalhoAntunesDeOliveira:2130724}, employing the SM diHiggs cross section at NNLO+NNLL order in QCD~\cite{deFlorian:2015moa,YR4,deFlorian:2013jea}. Cluster 3 is the one that represents the SM-like topology. As an illustration, we extend the experimental limit to all the clusters. 
This approximation is rather non-trivial, as we extrapolate the experimental limit derived for one kinematical topology (SM-like) to clusters that describe different topologies. In this sense this figure is merely an illustration of the method.

The representation of results in this fashion is complicated and
not very elucidating, especially if one wants to extract EFT constraints in more than four dimensions 
(considering also the Higgs boson branching fractions). In particular, one can make use of the \ehdecay\ interface provided by \rosetta\ to get accurate predictions for the production cross-section times branching fraction for each of the diHiggs channels analysed, in order to capture the full analytical dependence of the rates predicted by the EFT. This procedure is implemented in the \code{dihiggs} interface of \rosetta. The non-trivial aspect of determining the compatibility of a given point with the ATLAS analysis is to capture the effect of the modified kinematics on the actual cross-section limit one should use. We therefore begin by simply approximating this quantity by determining the nearest point in the anomalous coupling parameter space for which the clustering procedure was performed in the scan of Ref.~\cite{Dall'Osso:2015aia}. The cluster to which that particular point was associated informs us on the limit to which to compare the computed production cross-section times branching fractions, as the full analysis chain of the ATLAS analysis has been performed for these points. By this method, the \code{dihiggs} interface informs the user in a convenient way whether the non-resonant diHiggs analysis likely excludes a given point in the EFT parameter space. We envisage, in the future a more detailed method of actually determining/approximating the actual likelihood of a given point, rather than a simple exclusion flag.

\section{\rosetta\ command line interface\label{sec:rosetta}}
With the addition of several new interfaces, which could be useful both on their own and in combination with others, we have created a new command-line
interpreter to better suit the various possible uses of \rosetta. Each
interface can be called via a particular command of the type
\begin{verbatim}
  bin/rosetta [OPTIONS1] INTERFACE [OPTIONS2] ARG
\end{verbatim}
In addition, the original core translation method of \rosetta\ can now also be
called in this way. Two possible sets of options can be defined. The first set
defines the global options associated with a particular \rosetta\ run. This
mostly controls the level of output to be printed to the screen.
\begin{center}
\begin{tabular}{lp{10.1cm}}
    \code{[OPTIONS1]}&\tn
    \code{-h} or \code{--help}      &This displays a help message and exits the
                                     program.\tn
    \code{-s} or \code{--silent}    &Suppress all output and take default                                            answers to all questions.\tn

    \code{-d} or \code{--debug}     &Activate debug setting for verbose program 
                                     output.\tn
    \code{--force}                  &Take default answers to all questions.\tn  
\end{tabular}
\end{center}
\noindent
After a given run, \rosetta\ now creates two log files that are named
\code{rosetta.log} and \code{ro\-set\-ta.sup\-pres\-sed.log}. The first of these
files contains the output of the programme as printed to the screen. In contrast,
the second of these files includes the output such as it would have been printed
to the screen in the case where the programme would have been run with the
\code{--debug} option.

With the \code{INTERFACE} keyword above, the user specifies the particular
interface he/she wishes to make use of. While each choice (namely \code{translate},
\code{ehdecay}, \code{ewpo}, \code{dihiggs}, \code{si\-gnal\-strengths} and \code{defaultcard}) includes a \code{-h} (or
\code{--help}) option to inform users of the features that have been implemented,
dedicated options are available and will be detailed below. Unless stated otherwise,
the interfaces receive a single argument \code{ARG} that refer to a path to an SLHA
parameter card compatible with an EFT basis implemented in \rosetta\ (the \code{defaultcard} interface is a utility to generate a template for such a card).\\

\noindent\textbf{\code{translate}}\\
The core translation functionality of \rosetta\ can now be called from the command-line
interface. It allows for the same options as the original \code{translate} executable
that have been documented in Ref.~\cite{Falkowski:2015wza}.
\begin{center}
\begin{tabular}{lp{10.1cm}}
    \code{[OPTIONS2]}&\tn
    \code{-o} or \code{--output}    &This allows for the specification of the
                                     name of the output file, that is by default
                                     \code{PARAMCARD\_new.dat}.\tn
    \code{-w} or \code{--overwrite} &This allows the program to overwrite any
                                     pre-existing output file.\tn
    \code{--target}                 &This allows for providing the name of the
                                     basis into which the translation occurs,                                        the default being \code{bsmc}.
                                     \tn
%
    \code{--ehdecay}                &This allows to use the interface with the
                                     e{\sc HDecay}
                                     program~\cite{Contino:2013kra} for the
                                     calculation of the Higgs boson width and
                                     branching fractions.\tn
    \code{--flavor}                 &This allows to specify the treatment of the
                                     flavour structure relevant for the 
                                     two-fermion operators, the default being
                                     \code{general} and the other acceptable
                                     choices being \code{universal}
                                     and \code{diagonal}.\tn
    \code{--dependent}              &This allows the program to also write out 
                                     any dependent parameters calculated by the 
                                     translation function to the output file.
\end{tabular}
\end{center}

\noindent\textbf{\code{ehdecay}}\\
The e{\sc HDecay} interface may now be used on its own, to simply compute and report the Higgs width and branching fraction in the form of an SLHA block from a given SLHA parameter card in an implemented basis. This interface is also compatible with the \code{--flavor} option.

\begin{center}
\begin{tabular}{lp{10.1cm}}
    \code{[OPTIONS2]}&\tn
    \code{-o} or \code{--output}    & Specify the name for a copy of the input 
                                      SLHA card including the decay block.\tn
    \code{-w} or \code{--overwrite} &This allows the program to overwrite any
                                     pre-existing output file.\tn

    \code{--flavor}                 &This allows to specify the treatment of the
                                     flavour structure relevant for the 
                                     two-fermion operators, the default being
                                     \code{general} and the other acceptable
                                     choices being \code{universal}
                                     and \code{diagonal}.\tn
\end{tabular}
\end{center}

\noindent\textbf{\code{ewpo}}\\
The computation of the multidimensional likelihood described in
Sec.~\ref{sec:EWPO} can be called with the above command, using \code{ewpo}
as the desired \code{INTERFACE}. As a result, the programme prints to the screen
the value of the likelihood together with the associated $p$-value of the parameter
point specified in the input SLHA card. It is also compatible with the 
\code{--flavor} option above described, as the implementation of the calculation of
the likelihood function has been achieved assuming both a general and an MFV flavour
structure. However, since the considered observables can only constrain the diagonal
elements of the weak boson--fermion--fermion vertices, the flavour-violating entries
of the Wilson coefficients will not affect the likelihood.\\

\noindent\textbf{\code{signalstrengths}}\\
The evaluation of the Higgs signal strengths as described in Sec.~\ref{sec:Higgs} can also be invoked on its own. The \code{--flavor} option is once again accepted but observables are restricted in sensitivity to the diagonal elements of the flavour matrices, as with the \code{ewpo} interface above. The signal strengths are computed including effects linear in the EFT coefficients by default. This can be modified by \code{--squares} option which includes the full quadratic dependence.\\

\noindent\textbf{\code{dihiggs}}\\
The \rosetta\ programme can also be used to determine the cross section
times branching fraction related to non-resonant diHiggs production. This
proceeds first by translating a given input point into the BSM Characterisation
Lagrangian (via the \emph{Higgs} basis). It next combines this information with
the Higgs branching fractions as computed by the \ehdecay\ interface.
Finally, the values of the rates corresponding to the production of
$b\bar{b}\gamma\gamma$, $b \bar b b\bar b$, $b\bar{b}\tau\tau$ and $\gamma\gamma W W^{\ast}$
final states are reported or can be selected individually using the available options. The \code{--flavor} option is once again accepted, as well as the possibility of including the squares of the EFT contributions via the
\code{--squares} option.
\begin{center}
\begin{tabular}{lp{10.1cm}}
    \code{[OPTIONS2]}&\tn
    \code{--squares}                &This tells \rosetta\ to include the
                                     quadratic dependence of the non-resonant 
                                     diHiggs production cross section on the 
                                     \emph{BSMC} Lagrangian parameters.\tn
    \code{--channel}                &This selects a particular channel for 
                                     which to report the cross section times 
                                     branching fraction. The allowed values
                                     are `4b',`bbaa', `bbtautau' and  `aaWW'.
\end{tabular}
\end{center}

\section{Conclusion\label{sec:conclusion}}
The work presented in these proceedings represents the first steps towards confronting SM EFT at LO with data in a basis-independent way via \rosetta. We have covered the implementations of a global fit to Electroweak Precision observables (including select low-energy scattering experiments and complemented by ATGC data from LEP, Tevatron and LHC), global Higgs signal-strength constraints via an interface to \lilith, and constraints on non-resonant di-Higgs production at the LHC. These were made possible making use of numerous works deriving the parametric dependence of the observables in question on the SM EFT parameters at LO. These are wrapped into interfaces of \rosetta\ in order to extract the results of each analysis starting from any basis implementation present in the program. In doing so, we have also presented a new command-line interface for \rosetta, through which all of these features can be accessed. The culmination of the work reported here will be a new version of \rosetta, which we anticipate will be released at the same time as these proceedings. Looking ahead, it is envisaged that the work documented here will evolve into a more general framework for data compatibility for EFT using \rosetta\ into which new measurements such as the essential differential observables can be continually added.

\appendix
\clearpage
\section{ Measured values and SM predictions for observables\label{app:EWPO}}
Tab.~\ref{tab:EWPT_zpole}, ~\ref{tab:EWPT_wpole} and ~\ref{tab:EWPT_lowe}
summarise the measured values and the SM predictions for the precision
observables included in the fit introduced Sec.~\ref{sec:EWPO:obs} and that were
not mentioned in the main text.
\begin{table}[h!]
 \begin{center}
 \begin{tabular}{|c|c|c|c|c|}
\hline
{Observable} & {Experimental value}   &   {Ref.}   &  {{SM prediction}}    &  {Definition}
  \\  \hline   \hline
$\Gamma_{Z}$ [GeV]  & $2.4952 \pm 0.0023$ & \cite{ALEPH:2005ab} & $ 2.4950$    & $\sum_f \Gamma (Z \to f \bar f)$
 \\  \hline
$\sigma_{\rm had}$ [nb]  & $41.541\pm 0.037$ &\cite{ALEPH:2005ab} &  $41.484$ &  ${12 \pi \over m_Z^2} {\Gamma (Z \to e^+ e^-) \Gamma (Z \to q \bar q) \over \Gamma_Z^2}$
  \\  \hline
 $R_{e}$  & $20.804\pm 0.050$ & \cite{ALEPH:2005ab} &  $20.743$  &    $ {\sum_{q} \Gamma(Z \to q \bar q) \over  \Gamma(Z \to e^+ e^-)} $
   \\  \hline
 $R_{\mu}$  & $20.785 \pm 0.033$ & \cite{ALEPH:2005ab} &  $20.743$  &    $ {\sum_{q} \Gamma(Z \to q \bar q) \over  \Gamma(Z \to \mu^+ \mu^-)} $
   \\  \hline
 $R_{\tau}$  & $20.764\pm 0.045$ & \cite{ALEPH:2005ab} &  $20.743$  &    $ {\sum_{q} \Gamma(Z \to q \bar q) \over  \Gamma(Z \to \tau^+ \tau^-)} $
 \\  \hline
 $A_{\rm FB}^{0,e}$ & $0.0145\pm 0.0025$ &\cite{ALEPH:2005ab} &  $0.0163$   &  ${3 \over 4} A_e^2$
  \\  \hline
 $A_{\rm FB}^{0,\mu}$ & $0.0169\pm 0.0013$ &\cite{ALEPH:2005ab} &  $0.0163$   &  ${3 \over 4} A_e A_\mu $
  \\  \hline
 $A_{\rm FB}^{0,\tau}$ & $0.0188\pm 0.0017$ &\cite{ALEPH:2005ab} &  $0.0163$   &  ${3 \over 4} A_e A_\tau$
  \\ \hline \hline
$R_b$ & $0.21629\pm0.00066$ & \cite{ALEPH:2005ab} & $0.21578$  &     ${ \Gamma(Z \to b \bar b) \over \sum_q \Gamma(Z \to q \bar q)}$
   \\  \hline
$R_c$ & $0.1721\pm0.0030$  & \cite{ALEPH:2005ab}  & $0.17226$   & ${ \Gamma(Z \to c \bar c) \over \sum_q \Gamma(Z \to q \bar q)} $
\\  \hline
$A_{b}^{\rm FB}$ & $0.0992\pm 0.0016$ & \cite{ALEPH:2005ab}  & $0.1032$  & ${3 \over 4} A_e A_b$
 \\  \hline
 $A_{c}^{\rm FB}$ & $0.0707\pm 0.0035$  & \cite{ALEPH:2005ab} &  $0.0738$  & ${3 \over 4} A_e A_c$
  \\ \hline \hline
 $A_e$ & $0.1516 \pm 0.0021$ &\cite{ALEPH:2005ab} &  $0.1472$ &  ${\Gamma(Z \to e_L^+ e_L^-) - \Gamma(Z \to e_R^+ e_R^-) \over \Gamma(Z \to e^+  e^-) }$
 \\ \hline
  $A_\mu$ & $0.142 \pm 0.015$ &\cite{ALEPH:2005ab} &  $0.1472$ &  ${\Gamma(Z \to \mu_L^+ \mu_L^-) - \Gamma(Z \to \mu_R^+ \mu_R^-) \over \Gamma(Z \to \mu^+  \mu^-) }$
 \\ \hline
 $A_\tau$ & $0.136 \pm 0.015$ &\cite{ALEPH:2005ab} &  $0.1472$ &  ${\Gamma(Z \to \tau_L^+ \tau_L^-) - \Gamma(Z \to \tau_R^+ \tau_R^-) \over \Gamma(Z \to \tau^+  \tau^-) }$
 \\ \hline \hline
  $A_e$ & $0.1498 \pm 0.0049$ & \cite{ALEPH:2005ab} &  $0.1472$ &  ${\Gamma(Z \to e_L^+ e_L^-) - \Gamma(Z \to e_R^+ e_R^-) \over \Gamma(Z \to \tau^+  \tau^-) }$
 \\ \hline \hline
 $A_\tau$ & $0.1439 \pm  0.0043$ & \cite{ALEPH:2005ab} &  $0.1472$ &  ${\Gamma(Z \to \tau_L^+ \tau_L^-) - \Gamma(Z \to \tau_R^+ \tau_R^-) \over \Gamma(Z \to \tau^+  \tau^-) }$
 \\ \hline \hline
 $A_b$ & $0.923\pm 0.020$ & \cite{ALEPH:2005ab} & $0.935$  &
 ${ \Gamma(Z \to b_L \bar b_L)  -   \Gamma(Z \to b_R \bar b_R)  \over  \Gamma(Z \to b \bar b)   }$
 \\  \hline
$A_c$ & $0.670 \pm 0.027$ & \cite{ALEPH:2005ab} & $0.668$
 &  ${ \Gamma(Z \to c_L \bar c_L)  -   \Gamma(Z \to c_R \bar c_R)  \over  \Gamma(Z \to c \bar c)   }$
 \\ \hline \hline
$A_s$ & $0.895 \pm 0.091$ & \cite{Abe:2000uc} & $0.935$
&  ${ \Gamma(Z \to s_L \bar s_L)  -   \Gamma(Z \to s_R \bar s_R)  \over  \Gamma(Z \to s\bar s)   }$
\\ \hline \hline
$R_{uc}$ & $0.166 \pm 0.009$  & \cite{Beringer:1900zz}  & $0.1724$   & ${ \Gamma(Z \to u \bar u) +  \Gamma(Z \to c \bar c) \over 2 \sum_q \Gamma(Z \to q \bar q)} $
\\ \hline 
\end{tabular}
\end{center}
\caption{
 $Z$-boson pole observables. The experimental errors within a set of observables separated by
 two double lines are correlated, which is taken into account in the fit. $A_e$
 and $A_\tau$ are listed twice:  the first number originates from the combination of leptonic
 polarisation and left-right asymmetry measurements at the SLC collider, while the second number
 is issued from tau polarisation measurements at LEP-1. The table also includes
 the model-independent measurement of the on-shell $Z$-boson couplings to light quarks
 performed by D0~\cite{Abazov:2011ws} and the measurements related to inclusive Drell-Yan
 $Z$-boson production in CMS~\cite{Chatrchyan:2014mua}. For what concerns the theoretical predictions,
 we have used the best fit SM values obtained from {\sc GFitter}~\cite{Baak:2014ora}.
}
\label{tab:EWPT_zpole}
 \end{table}

\begin{table}
 \begin{center}
 \begin{tabular}{|c|c|c|c|c|}
 \hline
{{Observable}} & {{Experimental value}}   &   {{Ref.}}   &  {{SM prediction}}    &  {Definition}
 \\  \hline  \hline
$m_{W}$ [GeV]  & $80.385 \pm 0.015$ &\cite{Group:2012gb}    &  $80.364$   &  ${g_L v \over 2} \left ( 1 + \delta m \right )$
\\ \hline  \hline
$\Gamma_{W}$ [GeV]  & $ 2.085 \pm 0.042$  & \cite{Beringer:1900zz} &  $2.091$  &  $ \sum_f  \Gamma(W \to f f')$
\\ \hline \hline
${\rm Br} (W \to e \nu)$ & $ 0.1071 \pm 0.0016$ &\cite{Schael:2013ita} &  $0.1083$  &  $ { \Gamma(W \to e \nu) \over  \sum_f  \Gamma(W \to f f')}$
\\ \hline
${\rm Br} (W \to \mu \nu)$ & $ 0.1063 \pm 0.0015$ &\cite{Schael:2013ita} &  $0.1083$  &  $ { \Gamma(W \to \mu \nu) \over  \sum_f  \Gamma(W \to f f')}$
\\ \hline
${\rm Br} (W \to \tau \nu)$ & $ 0.1138 \pm 0.0021$ &\cite{Schael:2013ita} &  $0.1083$  &  $ {\Gamma(W \to \tau \nu) \over  \sum_f  \Gamma(W \to f f')}$
 \\ \hline \hline
 $R_{Wc}$ & $ 0.49 \pm 0.04$ & \cite{Beringer:1900zz}  &  $0.50$  &  $ { \Gamma(W \to c s) \over  \Gamma(W \to u d) + \Gamma(W \to c s) }$
 \\ \hline \hline
 $R_{\sigma}$ & $0.998 \pm 0.041$  & \cite{Khachatryan:2014iya} & 1.000 & $g^{Wq_3}_L/g^{Wq_3}_{L,\rm SM} $\tabularnewline
\hline 
\end{tabular}
\end{center}
\caption{$W$-boson pole observables. The table also includes the measurement of
  $V_{tb}$ from $t$-channel single-top production, and the one of the inclusive
  Drell-Yan $W$-boson production cross section at the LHC~\cite{Chatrchyan:2014mua}.
  Measurements of the three leptonic branching ratios of the $W$-boson are correlated.
  For what concerns the theoretical predictions of the leptonic branching ratios, we use
  the value quoted in Ref.~\cite{Schael:2013ita}.
}
\label{tab:EWPT_wpole}
 \end{table}

 \begin{table}
\begin{center} 
 \begin{tabular}{|l|c|c|c|}
    \hline
     Experiment & Ref. & $g_V$ & $g_A$ \\
    \hline
    \hline
    SM prediction & \cite{Erler:2013xha} & $-0.0396$  &  $0.5064$\\
     \hline
    \hline
   CHARM-II & \cite{Vilain:1994qy} & $-0.035\pm0.017$  &  $-0.503\pm0.0017$\\
    \hline
   CHARM & \cite{Dorenbosch:1988is} & $-0.06\pm 0.07$  & $-0.54\pm0.07$\\
    \hline
   BNL-E734 & \cite{Ahrens:1990fp} & $-0.107\pm 0.045$  & $-0.514\pm0.036$\\
    \hline
  \end{tabular}
\end{center}
  \caption{Summary of the SM predictions and the experimental measurements of
    the vector ($g_V$) and axial ($g_A$) coupling strengths of the $Z$-boson
    to electrons as reported by the CHARM, CHARM-II and BNL-E734 experiments\label{tab:EWPT_lowe}}
 \end{table}

 \clearpage

\subsection{Central values, uncertainties and correlation matrix of precision observable fit}
We present below the central values resulting from the fit of the precision
observables considered, to the relevant \emph{Higgs} basis parameters $\vec c_0$,
together with the associated $1\sigma$ uncertainties $\delta \vec c$ and
correlation matrix $\rho$. We refer to Sec.~\ref{sec:EWPO:imp} for more information.
 {\footnotesize
 \beq
 \nonumber
 \label{eq:EWPT_generalfit}
 \begin{pmatrix} 
 \delta g^{We}_L \\ 
 \delta g^{W\mu}_L \\ 
 \delta g^{W\tau}_L \\ 
 \delta g^{Ze}_L \\ 
 \delta g^{Z\mu}_L \\ 
 \delta g^{Z\tau}_L \\ 
 \delta g^{Ze}_R  \\ 
 \delta g^{Z\mu}_R \\ 
 \delta g^{Z\tau}_R \\ 
 \delta g^{Zu}_L \\ 
 \delta g^{Zc}_L \\ 
 \delta g^{Zt}_L \\ 
 \delta g^{Zu}_R \\ 
 \delta g^{Zc}_R \\ 
 \delta g^{Zd}_L \\ 
 \delta g^{Zs}_L \\ 
 \delta g^{Zb}_L \\ 
 \delta g^{Zd}_R \\ 
 \delta g^{Zs}_R\\ 
 \delta g^{Zb}_R \\ 
 \hline 
 \delta g_{1,z} \\ 
 \delta \kappa_\gamma \\ 
 \lambda_z \\ 
 \, [c_{\ell \ell}]_{1111} \\ 
  \, [c_{\ell e}]_{1111} \\     
 \,  [c_{e e}]_{1111}   \\    
 \, [c_{\ell \ell}]_{1221} \\ 
 \, [c_{\ell \ell}]_{1122} \\ 
  \, [c_{\ell e}]_{1122} \\   
  \, [c_{\ell e}]_{2211} \\    
 \,  [c_{e e}]_{1122}  \\     
 \, [c_{\ell \ell}]_{1331} \\ 
 \, [c_{\ell \ell}]_{1133} \\ 
  \, [c_{\ell e}]_{1133}+  [c_{\ell e}]_{3311} \\   
 \,  [c_{e e}]_{1133}     \\ 
 \, [c_{\ell \ell}]_{2332} \\ 
 \end{pmatrix} 
 = 
 \begin{pmatrix}
  -0.37 \pm 0.43 \\   -1.43  \pm 0.59 \\  1.46 \pm 0.70 \\
  -0.029 \pm 0.028 \\   0.01  \pm 0.11 \\  0.016  \pm 0.058 \\  
  -0.035 \pm 0.027 \\   0.00  \pm 0.13 \\  0.037  \pm 0.062 \\ 
  -0.6 \pm 3.0 \\ -0.16 \pm 0.36  \\ -0.3 \pm 3. 8 \\ 
  1.3 \pm 5.0 \\ -0.37 \pm 0.51 \\ 
  -1.0 \pm 3.7 \\  1.2 \pm 1.7 \\ 0.33 \pm 0.16 \\ 
  3  \pm 15 \\  2.9 \pm 4.8 \\ 2.3 \pm 0.8 \\  
  \hline 
  -62 \pm 37 \\   -23 \pm 23 \\ 65 \pm 40 \\
 1.00 \pm 0.39 \\  -0.23  \pm  0.22 \\ 0.23 \pm 0.39 \\ 
 -3.7 \pm 1.4 \\  2.0  \pm  2.3 \\ 1.0 \pm 2.3  \\-0.9 \pm 2.2 \\  1.5 \pm 2.6 \\ 
 1.8 \pm 1.3 \\  140  \pm  170 \\ -0.55  \pm 0.64  \\  -150 \pm 180 \\  1.9 \pm 2.1 
 \end{pmatrix} 
  \times 10^{-2}. 
 \eeq }

\noindent We split the symmetric correlation matrix $\rho$ as
 \beq
 \nonumber
 \rho = 
 \begin{pmatrix} 
   \rho_{\delta g}  &  \rho_x
  \\
 \rho_x^T & \rho_{\rm 4f+tgc}
 \end{pmatrix} 
 , 
 \eeq 
where $\rho_{\delta g}$ consists of the correlations among the weak boson-fermion-fermion
vertex corrections, $\rho_{\rm 4f+tgc}$ of the correlations among the four-lepton effective
couplings and the ATGCs, and $\rho_x$ of the cross-correllation between the two above sets
of measurements.

 \clearpage
 {\tiny
 \beq
 \nonumber
 \hspace{-1.5cm}
 \rho_{\delta g} 
 = \left(
 \begin{array}{cccccccccccccccccccc}
  1. & \text{-}0.08 & \text{-}0.48 & \text{-}0.07 & \text{-}0.02 & 0. & 0.05 & \text{-}0.04 & \text{-}0.03 & \text{-}0.01 & 0. & 0. & \text{-}0.02 & 0. & \text{-}0.01 & \text{-}0.03 & 0.02 & \text{-}0.03 & \text{-}0.03 & 0. \\
  . & 1. & \text{-}0.68 & \text{-}0.12 & \text{-}0.04 & 0.01 & 0.09 & \text{-}0.07 & \text{-}0.05 & \text{-}0.02 & 0. & 0. & \text{-}0.03 & 0.01 & \text{-}0.02 & \text{-}0.05 & 0.03 & \text{-}0.05 & \text{-}0.04 & 0. \\
  . & . & 1. & \text{-}0.15 & \text{-}0.05 & 0.01 & 0.11 & \text{-}0.08 & \text{-}0.06 & 0.01 & \text{-}0.01 & 0. & 0.01 & 0. & 0.01 & 0.03 & 0.04 & 0.02 & 0.02 & 0.01 \\
  . & . & . & 1. & \text{-}0.11 & \text{-}0.07 & 0.17 & \text{-}0.06 & 0.03 & 0.01 & 0.08 & \text{-}0.02 & 0.03 & 0.09 & 0.02 & 0.04 & \text{-}0.38 & 0.04 & 0.04 & \text{-}0.37 \\
  . & . & . & . & 1. & 0.07 & \text{-}0.05 & 0.9 & \text{-}0.04 & 0. & \text{-}0.02 & 0. & 0. & \text{-}0.01 & 0. & 0.01 & 0.08 & 0. & 0. & 0.05 \\
  . & . & . & . & . & 1. & 0.02 & \text{-}0.03 & 0.41 & \text{-}0.01 & \text{-}0.02 & 0. & \text{-}0.01 & 0. & \text{-}0.01 & 0. & 0.08 & \text{-}0.01 & \text{-}0.01 & 0.01 \\
  . & . & . & . & . & . & 1. & \text{-}0.08 & \text{-}0.04 & \text{-}0.01 & 0.07 & \text{-}0.02 & \text{-}0.01 & 0.12 & \text{-}0.01 & \text{-}0.01 & \text{-}0.36 & \text{-}0.02 & \text{-}0.01 & \text{-}0.41 \\
  . & . & . & . & . & . & . & 1. & 0.04 & 0.01 & 0. & 0. & 0.01 & \text{-}0.02 & 0.01 & 0.01 & 0.02 & 0.02 & 0.02 & 0.05 \\
  . & . & . & . & . & . & . & . & 1. & 0.01 & 0.02 & 0. & 0.01 & \text{-}0.01 & 0.01 & 0.01 & \text{-}0.04 & 0.02 & 0.02 & 0.01 \\
  . & . & . & . & . & . & . & . & . & 1. & \text{-}0.07 & 0. & 0.74 & 0.05 & 0.92 & \text{-}0.14 & \text{-}0.01 & 0.75 & \text{-}0.09 & 0. \\
  . & . & . & . & . & . & . & . & . & . & 1. & 0. & 0.02 & 0.29 & \text{-}0.06 & 0.19 & \text{-}0.11 & 0.03 & 0.02 & \text{-}0.15 \\
  . & . & . & . & . & . & . & . & . & . & . & 1. & 0. & \text{-}0.01 & 0. & 0. & 0.04 & 0. & 0. & 0.04 \\
  . & . & . & . & . & . & . & . & . & . & . & . & 1. & 0.03 & 0.71 & \text{-}0.08 & \text{-}0.01 & 0.94 & \text{-}0.18 & \text{-}0.01 \\
  . & . & . & . & . & . & . & . & . & . & . & . & . & 1. & 0.05 & 0.01 & \text{-}0.19 & 0.06 & 0.06 & \text{-}0.15 \\
  . & . & . & . & . & . & . & . & . & . & . & . & . & . & 1. & \text{-}0.44 & \text{-}0.01 & 0.73 & \text{-}0.08 & 0. \\
  . & . & . & . & . & . & . & . & . & . & . & . & . & . & . & 1. & \text{-}0.02 & 0. & 0.23 & \text{-}0.02 \\
  . & . & . & . & . & . & . & . & . & . & . & . & . & . & . & . & 1. & \text{-}0.02 & \text{-}0.02 & 0.89 \\
  . & . & . & . & . & . & . & . & . & . & . & . & . & . & . & . & . & 1. & \text{-}0.32 & \text{-}0.01 \\
  . & . & . & . & . & . & . & . & . & . & . & . & . & . & . & . & . & . & 1. & \text{-}0.01 \\
  . & . & . & . & . & . & . & . & . & . & . & . & . & . & . & . & . & . & . & 1. \\
 \end{array}
 \right),  
 \eeq 
 }

 {\tiny
 \beq
 \nonumber
 \hspace{-1.5cm}
 \rho_{\rm 4f+tgc}  = 
 \left(
 \begin{array}{cccccccccccccccc}
  1. & 0.18 & -0.98 & 0. & 0. & 0. & -0.16 & -0.01 & -0.01 & 0.01 & 0.1 & -0.04 & 0. & 0. & 0. & 0.11 \\
  . & 1. & -0.38 & 0. & -0.01 & 0. & 0.46 & 0.02 & 0.02 & -0.02 & -0.28 & 0.12 & 0. & 0. & 0. & -0.31 \\
  . & . & 1. & 0. & 0. & 0. & 0.06 & 0. & 0. & 0. & -0.04 & 0.01 & 0. & 0. & 0. & -0.04 \\
  . & . & . & 1. & -0.53 & -0.08 & 0.01 & 0. & 0. & 0. & 0. & 0.01 & 0. & 0. & 0. & 0.01 \\
  . & . & . & . & 1. & -0.52 & -0.01 & 0. & 0. & 0. & 0. & -0.02 & 0. & 0. & 0. & -0.02 \\
  . & . & . & . & . & 1. & 0. & 0. & 0. & 0. & 0. & 0. & 0. & 0. & 0. & 0. \\
  . & . & . & . & . & . & 1. & -0.22 & -0.19 & 0.19 & -0.38 & -0.55 & 0. & 0. & 0. & -0.14 \\
  . & . & . & . & . & . & . & 1. & 0.04 & -0.04 & -0.81 & 0.21 & 0. & 0. & 0. & -0.03 \\
  . & . & . & . & . & . & . & . & 1. & -0.98 & 0.06 & 0.18 & 0. & 0. & 0. & -0.03 \\
  . & . & . & . & . & . & . & . & . & 1. & -0.07 & -0.19 & 0. & 0. & 0. & 0.03 \\
  . & . & . & . & . & . & . & . & . & . & 1. & 0.12 & 0. & 0. & 0. & 0.11 \\
  . & . & . & . & . & . & . & . & . & . & . & 1. & -0.01 & 0. & 0. & 0.07 \\
  . & . & . & . & . & . & . & . & . & . & . & . & 1. & 0.01 & -1. & 0. \\
  . & . & . & . & . & . & . & . & . & . & . & . & . & 1. & -0.01 & 0. \\
  . & . & . & . & . & . & . & . & . & . & . & . & . & . & 1. & 0. \\
  . & . & . & . & . & . & . & . & . & . & . & . & . & . & . & 1. \\
 \end{array}
 \right),
 \eeq  
 }
 \vspace{0.5cm}
 {\tiny
 \beq
 \nonumber
 \hspace{-1.5cm}
 \rho_x  = 
 \left(
 \begin{array}{cccccccccccccccc}
  -0.29 & 0.83 & 0.1 & 0. & -0.01 & 0. & 0.55 & 0.02 & 0.02 & -0.02 & -0.34 & 0.14 & 0. & 0. & 0. & -0.37 \\
  0.02 & -0.07 & -0.01 & 0. & -0.01 & 0. & 0.79 & -0.28 & -0.24 & 0.24 & -0.2 & -0.77 & 0.01 & 0. & 0. & 0.11 \\
  0.14 & -0.4 & -0.05 & 0.01 & -0.01 & 0. & -0.86 & 0.19 & 0.16 & -0.16 & 0.33 & 0.74 & -0.01 & 0. & 0. & 0.29 \\
  0.02 & -0.05 & -0.01 & -0.09 & 0.04 & 0.08 & -0.14 & 0.06 & 0.03 & -0.03 & 0.03 & -0.21 & 0. & -0.01 & 0. & -0.17 \\
  0.01 & -0.02 & 0. & 0.01 & 0. & -0.01 & -0.05 & -0.04 & -0.03 & 0.03 & 0.07 & -0.07 & 0. & 0. & 0. & -0.05 \\
  0. & 0. & 0. & 0.01 & 0. & 0. & 0.01 & -0.01 & -0.01 & 0.01 & 0. & 0.02 & 0.01 & -0.01 & -0.01 & 0.01 \\
  -0.02 & 0.05 & 0.01 & -0.08 & -0.05 & 0.08 & 0.1 & -0.02 & -0.04 & 0.04 & -0.05 & 0.14 & 0. & 0.01 & 0. & 0.12 \\
  0.01 & -0.03 & 0. & 0.01 & 0. & -0.01 & -0.08 & -0.03 & -0.02 & 0.02 & 0.07 & -0.12 & 0. & 0. & 0. & -0.09 \\
  0.01 & -0.02 & 0. & 0. & 0. & 0. & -0.06 & 0.02 & 0.01 & -0.01 & 0.02 & -0.08 & 0. & 0.02 & 0. & -0.07 \\
  0. & -0.01 & 0. & 0. & 0. & 0. & -0.02 & 0.01 & 0. & 0. & 0.01 & 0. & 0. & 0. & 0. & -0.01 \\
  0. & 0. & 0. & -0.01 & 0. & 0.01 & -0.01 & 0. & 0. & 0. & 0. & -0.02 & 0. & 0. & 0. & -0.01 \\
  0. & 0. & 0. & 0. & 0. & 0. & 0. & 0. & 0. & 0. & 0. & 0. & 0. & 0. & 0. & 0. \\
  0.01 & -0.01 & 0. & 0. & 0. & 0. & -0.04 & 0.01 & 0.01 & -0.01 & 0.01 & 0. & 0. & 0. & 0. & -0.01 \\
  0. & 0.01 & 0. & -0.01 & 0. & 0.01 & 0.01 & 0. & 0. & 0. & -0.01 & 0. & 0. & 0. & 0. & 0. \\
  0. & -0.01 & 0. & 0. & 0. & 0. & -0.03 & 0.01 & 0.01 & -0.01 & 0.01 & 0. & 0. & 0. & 0. & -0.01 \\
  0.01 & -0.02 & 0. & 0. & 0. & 0. & -0.06 & 0.01 & 0.01 & -0.01 & 0.02 & 0.02 & 0. & 0. & 0. & -0.01 \\
  0. & 0.01 & 0. & 0.05 & 0. & -0.05 & 0.03 & -0.02 & 0. & 0. & 0. & 0.05 & 0. & 0. & 0. & 0.04 \\
  0.01 & -0.02 & 0. & 0. & 0. & 0. & -0.06 & 0.02 & 0.01 & -0.01 & 0.02 & 0.01 & 0. & 0. & 0. & -0.01 \\
  0.01 & -0.02 & 0. & 0. & 0. & 0. & -0.05 & 0.01 & 0.01 & -0.01 & 0.02 & 0.01 & 0. & 0. & 0. & -0.01 \\
  0. & 0. & 0. & 0.06 & 0. & -0.05 & 0.01 & -0.01 & 0.01 & -0.01 & 0.01 & 0.01 & 0. & 0. & 0. & 0.01 \\
 \end{array}
 \right). 
 \eeq 
 }

\clearpage
\section{Numerical formulae for Higgs production and decay modes\label{app:Higgs}}
We summarise here the numerical formul\ae\ used for the calculation of the Higgs
signal strengths as implemented in \rosetta. These have been calculated in
Ref.~\cite{Falkowski:2015fla} and have necessitated in some cases a numerical
integration over the parton density functions (PDFs). We have employed the leading
order set of PDFs provided by the NNPDF collaboration~\cite{Ball:2012cx}. The relevant quantities
are the ratio of the Higgs production cross sections and partial widths in each
channel to the SM predictions. For simplicity, we do not report the coefficients
related to the quadratic dependence of the observables on the Wilson coefficients,
but they will be provided in the forthcoming \rosetta~1.1 manual.

\subsection{Higgs production}
\noindent
\textbf{Gluon fusion (ggh)} $g g \to h $:
\beq
{\sigma_{ggh} \over \sigma_{ggh}^{\rm SM}} \simeq \left | 1 + {\hat c_{gg} \over  c_{gg}^{\rm SM} } \right |^2 , 
\eeq 
where the coefficients and the relevant loop functions are defined by
\begin{align}
\label{eq:cgghat}
\hat c_{gg} &\simeq   
c_{gg} +  {1 \over 12 \pi^2}  \left [  \delta y_u  A_f \left (m_h^2 \over 4 m_t^2 \right ) + \delta y_d A_f \left (m_h^2 \over 4 m_b^2 \right ) \right ] ,
\\
c_{gg}^{\rm SM} &\simeq   
{1 \over 12 \pi^2}  \left [ A_f \left (m_h^2 \over 4 m_t^2 \right ) + A_f \left (m_h^2 \over 4 m_b^2 \right ) \right ] ,
 \\
A_f(\tau) &\equiv \frac{3}{2\tau^2} \left [  (\tau-1)f(\tau)  + \tau \right ], 
\\
f(\tau) &\equiv  \left\{ \begin{array}{lll} 
{\rm arcsin}^2\sqrt{\tau} && \tau \le 1 \\ -\frac{1}{4}\left[\log\frac{1+\sqrt{1-\tau^{-1}}}{1-\sqrt{1-\tau^{-1}}}-i\pi\right]^2 && \tau > 1 \end{array}\right. .
\end{align}
\noindent
As discussed in Ref.~\cite{Gori:2013mia}, it is appropriate to calculate
$c_{gg}^{\rm SM}$ at the leading order accuracy in QCD as the large $K$-factors are
approximately common for $c_{gg}$ and $\delta y_u$ and cancel in the ratio. Numerically,
this gives
\beq
\hat c_{gg}  \simeq   c_{gg} + \left ( 8.7 \delta y_u  - (0.3 - 0.3i) \delta y_d \right ) \times 10^{-3}, 
\qquad c_{gg}^{\rm SM} \simeq  (8.4+0.3 i) \times 10^{-3}, 
\eeq  
\beq
\label{eq:ggh}
{\sigma_{ggh} \over \sigma_{ggh}^{\rm SM}} \simeq 1  +  237 c_{gg}  + 2.06 \delta y_u - 0.06  \delta y_d. 
\eeq\\
\clearpage

\noindent
\textbf{Vector boson fusion (VBF)} $q q \to h qq$: 
\label{eq:vbf}
\begin{align}\nonumber
{\sigma_{VBF} \over \sigma_{VBF}^{\rm SM}} &\simeq 
1 + 1.49 \delta c_w + 0.51 \delta c_z   - \begin{pmatrix} 1.08 \\ 1.11  \\ 1.23  \end{pmatrix} c_{w\Box}  - 0.10  c_{ww}  - \begin{pmatrix} 0.35 \\ 0.35 \\  0.40 \end{pmatrix}   c_{z \Box} 
 \\ &  
-0.04 c_{zz}  -0.10 c_{\gamma \Box} -  0.02  c_{z\gamma}.
 \end{align}  
The three numbers appearing in the multiplying factors for $c_{w\Box}$ and
$c_{z\Box}$ refer to the three LHC collision energies of $\sqrt{s} =$7, 8, and
13~TeV. The dependence of the other parameters on the collision energy being weaker,
it can be safely ignored. Each LHC Higgs analysis in the VBF channel uses a somewhat
different selection strategy to isolate the VBF signal. The resulting cross section
therefore slightly depends on it. In the equation above, we have computed the
cross section numerically with {\sc MadGraph5\_aMC@NLO}~\cite{Alwall:2014hca} and
generator selections on the parton-level jets so that their transverse momentum
satisfies $p_T > 20$~GeV, their pseudorapidity $|\eta| < 5$ and the invariant-mass
of a jet pair is greater than 250~GeV. Replacing the last bound by 500~GeV affects
the above numbers at the level of $5\%$.\\

\noindent
\textbf{Vector boson associated production (Vh)} $q \bar q \to Vh$ with $V=W,Z$:
\begin{align}
\label{eq:vhrates}
{\sigma_{W h} \over \sigma_{Wh}^{\rm SM}} & \simeq  
1 + 2 \delta c_w  + \begin{pmatrix}  6.39 \\ 6.51  \\ 6.96  \end{pmatrix} c_{w\Box}   +  \begin{pmatrix} 1.49\\ 1.49 \\ 1.50 \end{pmatrix} c_{ww}  
\\
{\sigma_{Z h} \over \sigma_{Z h}^{\rm SM}} & \simeq 
1 + 2 \delta c_z 
+ \begin{pmatrix}  5.30 \\ 5.40 \\ 5.72 \end{pmatrix} c_{z\Box}   +  \begin{pmatrix} 1.79 \\ 1.80 \\ 1.82 \end{pmatrix} c_{zz}  +  \begin{pmatrix} 0.80 \\ 0.82 \\ 0.87 \end{pmatrix} c_{\gamma \Box} +  \begin{pmatrix} 0.22 \\ 0.22 \\ 0.22 \end{pmatrix} c_{z \gamma}.
\end{align}\\
The triplet of numbers are once again related to the different LHC centre-of-mass energies.\\

\noindent
\textbf{Top pair associated production (tth)} $g g \to h t \bar t$:
\beq
{\sigma_{tth} \over \sigma_{tth}^{\rm SM}} \simeq  1 + 2 \delta y_u.  
\eeq
\subsection{Higgs decays}
\noindent
\textbf{Two-fermions} $h \to f \bar f$.\\
\noindent Higgs boson decays into two fermions occur at the tree level in the SM via the Yukawa couplings.
In the presence of dimension-six operators, these are affected via corrections to the Yukawa couplings,
\beq
{\Gamma_{cc} \over \Gamma_{cc}^{\rm SM}}   \simeq 1 + 2 \delta y_u, 
\qquad 
{ \Gamma_{bb}\over \Gamma_{bb}^{\rm SM} }  \simeq 1 + 2 \delta y_d,  
\qquad 
{\Gamma_{\tau \tau} \over \Gamma_{\tau \tau}^{\rm SM}}   \simeq   1 + 2 \delta y_e,  
\eeq 
where the abbreviation $\Gamma(h \to Y) \equiv \Gamma_Y$ is used.\\ 

\noindent
\textbf{Vector bosons} $h \to V V$\\ 
\noindent In the SM, Higgs decays into on-shell gauge bosons (namely into gluon pairs $gg$,
photon pairs $\gamma \gamma$, and $Z \gamma$ associated pairs) occur only at the one-loop level.
In the presence of dimension-six operators, these decays are corrected already at the tree-level
due to the existence of two-derivative contact interactions of the Higgs boson with two vector bosons.
The relative decay widths are given by 
\beq
\label{eq:gammavv}
{ \Gamma_{VV} \over  \Gamma_{VV}^{\rm SM}}  \simeq  \left |1 +  {\hat c_{vv} \over c_{vv}^{\rm SM}}  \right |^2, 
\qquad    
vv  \in  \{ gg, \gamma \gamma, z \gamma \} ,  
\eeq 
where 
\begin{align}
\label{eq:cvvsm}
\hat c_{\gamma \gamma} &\approx   c_{\gamma \gamma} -  0.11   \delta c_w  + 0.02 \delta y_u, \qquad  c_{\gamma \gamma}^{\rm SM} \simeq -8.3  \times 10^{-2},  
\\
\hat c_{z \gamma} &\approx  c_{z \gamma}  -0.06 \delta c_w + 0.003 \delta y_t , \quad c_{z \gamma}^{\rm SM} \simeq -5.9  \times 10^{-2}, 
\end{align}  
while $\hat c_{gg}$ and $c_{gg}^{\rm SM}$ are defined in Eq.~\eqref{eq:cgghat}.\\

\noindent
\textbf{Four fermions} $h \to 4 f$\\
\noindent The decay process $h\to 2 \ell 2 \nu$ (where $\ell$  here stands for any
charged lepton) proceeds via intermediate $W$ bosons. The related partial width is given by
\begin{align}
{\Gamma_{2 \ell 2\nu} \over  \Gamma_{2\ell 2\nu}^{\rm SM}} 
&\simeq 
1 + 2 \delta c_w  + 0.46 c_{w \Box}   - 0.15 c_{ww}  
\\
&\to  
1 + 2 \delta c_z   + 0.67 c_{z\Box} + 0.05 c_{zz}  - 0.17 c_{z\gamma} -  0.05 c_{\gamma\gamma}.  
\end{align}   
\noindent
In the SM, the decay process $h \to 4 \ell$ proceeds at the tree-level via intermediate $Z$-bosons.
In presence of dimension-six operators, intermediate photon contributions may also arise at the tree level.
If that is the case, the decay width diverges since the photon is massless.
The relative width $\bar \Gamma(h \to 4 \ell)$ is therefore regulated by imposing a selection
on the dilepton invariant mass of $m_{\ell \ell} > 12$~GeV where the two leptons carry the same flavour.
The width is given by
\begin{align}
\hspace{-1cm}
{\bar \Gamma_{4 \ell} \over  \bar \Gamma_{4 \ell}^{\rm SM}} 
&\simeq 
1 + 2 \delta c_z   +  
\begin{pmatrix} 0.41  c_{z\Box}-  0.15 c_{zz} + 0.07  c_{z \gamma} - 0.02 c_{\gamma \Box}  + 0.00 c_{\gamma \gamma} 
\\ 0.39  c_{z\Box} - 0.14  c_{zz} + 0.05  c_{z \gamma}  -0.02 c_{\gamma \Box} + 0.03 c_{\gamma \gamma}    \end{pmatrix}    
\\  &\to 
1+ 2 \delta c_z +  \begin{pmatrix} 0.35 \\  0.32 \end{pmatrix}  c_{z \Box} -  \begin{pmatrix} 0.19 \\ 0.19 \end{pmatrix}  c_{zz} +  \begin{pmatrix}  0.09 \\ 0.08  \end{pmatrix}  c_{z\gamma} +  \begin{pmatrix} 0.01 \\ 0.02 \end{pmatrix} c_{\gamma\gamma} .
\end{align}
The numbers in the columns correspond to the $2 e 2 \mu$ and $4 e/\mu$ final states, respectively. 
The dependence on the $m_{\ell \ell}$ selection is found to be weak. Very similar numbers are obtained
if we replace the 12~GeV threshold by a 4~GeV threshold.

Given these partial widths, the associated branching fractions can be computed as
${\rm BR}_Y= \Gamma_Y/\Gamma(h \to {\rm all})$, while the total Higgs decay width is given by 
\begin{align}
\frac{\Gamma(h \to \mathrm{all})}{ \Gamma(h \to \mathrm{all})_\mathrm{SM} }
 \simeq  
 \sum_{Y} \frac{\Gamma_{Y}}{\Gamma_{Y}^\mathrm{SM} }. \mathrm{BR}_{Y}^\mathrm{SM}
\end{align}


\AddToContent{J.~Bernon, A.~Carvalho, A.~Falkowski, B.~Fuks, F.~Goertz,
  K.~Mawatari, K.~Mimasu and T.~You}
\renewcommand{\thesection}{\arabic{section}}

\clearpage

\bibliography{LH_NewPhysics_Biblio}

\providecommand{\href}[2]{#2}\begingroup\raggedright\begin{thebibliography}{100}

\bibitem{Chivukula:2015zma}
R.~S. Chivukula, P.~Ittisamai, K.~Mohan, and E.~H. Simmons, {\em Phys. Rev.}
  {\bf D92} (2015), no.~7 075020,
  [\href{http://xxx.lanl.gov/abs/1507.06676}{{\tt 1507.06676}}].

\bibitem{ATLAS:2015nsi}
{\bf ATLAS} Collaboration, {\em Phys. Lett.} {\bf B754} (2016) 302--322,
  [\href{http://xxx.lanl.gov/abs/1512.01530}{{\tt 1512.01530}}].

\bibitem{Khachatryan:2015dcf}
{\bf CMS} Collaboration, \href{http://xxx.lanl.gov/abs/1512.01224}{{\tt
  1512.01224}}.

\bibitem{Atre:2013mja}
A.~Atre, R.~S. Chivukula, P.~Ittisamai, and E.~H. Simmons, {\em Phys.Rev.} {\bf
  D88} (2013) 055021, [\href{http://xxx.lanl.gov/abs/1306.4715}{{\tt
  1306.4715}}].

\bibitem{Chivukula:2014npa}
R.~Sekhar~Chivukula, P.~Ittisamai, and E.~H. Simmons, {\em Phys.Rev.} {\bf D91}
  (2015), no.~5 055021, [\href{http://xxx.lanl.gov/abs/1406.2003}{{\tt
  1406.2003}}].

\bibitem{Chivukula:2014pma}
R.~Sekhar~Chivukula, E.~H. Simmons, and N.~Vignaroli, {\em Phys.Rev.} {\bf D91}
  (2015), no.~5 055019, [\href{http://xxx.lanl.gov/abs/1412.3094}{{\tt
  1412.3094}}].

\bibitem{Harris:2011bh}
R.~M. Harris and K.~Kousouris, {\em Int.J.Mod.Phys.} {\bf A26} (2011)
  5005--5055, [\href{http://xxx.lanl.gov/abs/1110.5302}{{\tt 1110.5302}}].

\bibitem{Chivukula:1987py}
R.~S. Chivukula and H.~Georgi, {\em Phys. Lett.} {\bf B188} (1987) 99.

\bibitem{Arnold:2009ay}
J.~M. Arnold, M.~Pospelov, M.~Trott, and M.~B. Wise, {\em JHEP} {\bf 1001}
  (2010) 073, [\href{http://xxx.lanl.gov/abs/0911.2225}{{\tt 0911.2225}}].

\bibitem{Ma:1998pi}
E.~Ma, M.~Raidal, and U.~Sarkar, {\em Eur.Phys.J.} {\bf C8} (1999) 301--309,
  [\href{http://xxx.lanl.gov/abs/hep-ph/9808484}{{\tt hep-ph/9808484}}].

\bibitem{Han:2009ya}
T.~Han, I.~Lewis, and T.~McElmurry, {\em JHEP} {\bf 01} (2010) 123,
  [\href{http://xxx.lanl.gov/abs/0909.2666}{{\tt 0909.2666}}].

\bibitem{Harris:1999ya}
R.~M. Harris, C.~T. Hill, and S.~J. Parke,
  \href{http://xxx.lanl.gov/abs/hep-ph/9911288}{{\tt hep-ph/9911288}}.

\bibitem{Maltoni:2002mq}
F.~Maltoni, K.~Paul, T.~Stelzer, and S.~Willenbrock, {\em Phys. Rev.} {\bf D67}
  (2003) 014026, [\href{http://xxx.lanl.gov/abs/hep-ph/0209271}{{\tt
  hep-ph/0209271}}].

\bibitem{Kilian:2012pz}
W.~Kilian, T.~Ohl, J.~Reuter, and C.~Speckner, {\em JHEP} {\bf 10} (2012) 022,
  [\href{http://xxx.lanl.gov/abs/1206.3700}{{\tt 1206.3700}}].

\bibitem{Curtin:2012rm}
D.~Curtin, R.~Essig, and B.~Shuve, {\em Phys. Rev.} {\bf D88} (2013) 034019,
  [\href{http://xxx.lanl.gov/abs/1210.5523}{{\tt 1210.5523}}].

\bibitem{Gallicchio:2010sw}
J.~Gallicchio and M.~D. Schwartz, {\em Phys. Rev. Lett.} {\bf 105} (2010)
  022001, [\href{http://xxx.lanl.gov/abs/1001.5027}{{\tt 1001.5027}}].

\bibitem{Aad:2014nra}
{\bf ATLAS} Collaboration, {\em Phys. Rev.} {\bf D90} (2014), no.~5 052008,
  [\href{http://xxx.lanl.gov/abs/1407.0608}{{\tt 1407.0608}}].

\bibitem{Aad:2015zva}
{\bf ATLAS} Collaboration, {\em Eur. Phys. J.} {\bf C75} (2015), no.~7 299,
  [\href{http://xxx.lanl.gov/abs/1502.01518}{{\tt 1502.01518}}]. [Erratum: Eur.
  Phys. J.C75,no.9,408(2015)].

\bibitem{Khachatryan:2014rra}
{\bf CMS} Collaboration, {\em Eur. Phys. J.} {\bf C75} (2015), no.~5 235,
  [\href{http://xxx.lanl.gov/abs/1408.3583}{{\tt 1408.3583}}].

\bibitem{Khachatryan:2015wza}
{\bf CMS} Collaboration, {\em JHEP} {\bf 06} (2015) 116,
  [\href{http://xxx.lanl.gov/abs/1503.08037}{{\tt 1503.08037}}].

\bibitem{Frigerio:2012uc}
M.~Frigerio, A.~Pomarol, F.~Riva, and A.~Urbano, {\em JHEP} {\bf 07} (2012)
  015, [\href{http://xxx.lanl.gov/abs/1204.2808}{{\tt 1204.2808}}].

\bibitem{Marzocca:2014msa}
D.~Marzocca and A.~Urbano, {\em JHEP} {\bf 07} (2014) 107,
  [\href{http://xxx.lanl.gov/abs/1404.7419}{{\tt 1404.7419}}].

\bibitem{Fonseca:2015gva}
N.~Fonseca, R.~Z. Funchal, A.~Lessa, and L.~Lopez-Honorez, {\em JHEP} {\bf 06}
  (2015) 154, [\href{http://xxx.lanl.gov/abs/1501.05957}{{\tt 1501.05957}}].

\bibitem{Brivio:2015kia}
I.~Brivio, M.~B. Gavela, L.~Merlo, K.~Mimasu, J.~M. No, R.~del Rey, and
  V.~Sanz, \href{http://xxx.lanl.gov/abs/1511.01099}{{\tt 1511.01099}}.

\bibitem{Aad:2015pla}
{\bf ATLAS} Collaboration, {\em JHEP} {\bf 11} (2015) 206,
  [\href{http://xxx.lanl.gov/abs/1509.00672}{{\tt 1509.00672}}].

\bibitem{Khachatryan:2015vta}
{\bf CMS} Collaboration, {\em Phys. Lett.} {\bf B753} (2016) 363--388,
  [\href{http://xxx.lanl.gov/abs/1507.00359}{{\tt 1507.00359}}].

\bibitem{Khachatryan:2014jba}
{\bf CMS} Collaboration, {\em Eur. Phys. J.} {\bf C75} (2015), no.~5 212,
  [\href{http://xxx.lanl.gov/abs/1412.8662}{{\tt 1412.8662}}].

\bibitem{lacroix}
S.~Lacroix, {\em https://perso.ens-lyon.fr/sylvain.lacroix/Rapport.pdf}.

\bibitem{Craig:2014lda}
N.~Craig, H.~K. Lou, M.~McCullough, and A.~Thalapillil,
  \href{http://xxx.lanl.gov/abs/1412.0258}{{\tt 1412.0258}}.

\bibitem{Carpenter:2013xra}
L.~Carpenter, A.~DiFranzo, M.~Mulhearn, C.~Shimmin, S.~Tulin, and D.~Whiteson,
  {\em Phys. Rev.} {\bf D89} (2014), no.~7 075017,
  [\href{http://xxx.lanl.gov/abs/1312.2592}{{\tt 1312.2592}}].

\bibitem{Panico:2015jxa}
G.~Panico and A.~Wulzer, {\em Lect. Notes Phys.} {\bf 913} (2016) pp.1--316,
  [\href{http://xxx.lanl.gov/abs/1506.01961}{{\tt 1506.01961}}].

\bibitem{Alwall:2014bza}
J.~Alwall, C.~Duhr, B.~Fuks, O.~Mattelaer, D.~G. Ozturk, and C.-H. Shen, {\em
  Comput. Phys. Commun.} {\bf 197} (2015) 312--323,
  [\href{http://xxx.lanl.gov/abs/1402.1178}{{\tt 1402.1178}}].

\bibitem{Alloul:2013bka}
A.~Alloul, N.~D. Christensen, C.~Degrande, C.~Duhr, and B.~Fuks, {\em Comput.
  Phys. Commun.} {\bf 185} (2014) 2250--2300,
  [\href{http://xxx.lanl.gov/abs/1310.1921}{{\tt 1310.1921}}].

\bibitem{ATLAS-CONF-2015-081}
{\bf ATLAS} Collaboration, {\em ATLAS-CONF-2015-081} (2015).

\bibitem{CMS:2015dxe}
{\bf CMS} Collaboration, {\em CMS-PAS-EXO-15-004} (2015).

\bibitem{Alitti:1993pn}
J.~Alitti {\em et.~al.},, {\bf UA2} Collaboration, {\em Nucl. Phys.} {\bf B400}
  (1993) 3--24.

\bibitem{Aaltonen:2008dn}
T.~Aaltonen {\em et.~al.},, {\bf CDF} Collaboration, {\em Phys. Rev.} {\bf D79}
  (2009) 112002, [\href{http://xxx.lanl.gov/abs/0812.4036}{{\tt 0812.4036}}].

\bibitem{Khachatryan:2015sja}
{\bf CMS} Collaboration, {\em Phys. Rev.} {\bf D91} (2015), no.~5 052009,
  [\href{http://xxx.lanl.gov/abs/1501.04198}{{\tt 1501.04198}}].

\bibitem{Aad:2014aqa}
{\bf ATLAS} Collaboration, {\em Phys. Rev.} {\bf D91} (2015), no.~5 052007,
  [\href{http://xxx.lanl.gov/abs/1407.1376}{{\tt 1407.1376}}].

\bibitem{Hisano:2010ct}
J.~Hisano, K.~Ishiwata, and N.~Nagata, {\em Phys. Rev.} {\bf D82} (2010)
  115007, [\href{http://xxx.lanl.gov/abs/1007.2601}{{\tt 1007.2601}}].

\bibitem{Chu:2012qy}
X.~Chu, T.~Hambye, T.~Scarna, and M.~H.~G. Tytgat, {\em Phys. Rev.} {\bf D86}
  (2012) 083521, [\href{http://xxx.lanl.gov/abs/1206.2279}{{\tt 1206.2279}}].

\bibitem{Shifman:1978zn}
M.~A. Shifman, A.~I. Vainshtein, and V.~I. Zakharov, {\em Phys. Lett.} {\bf
  B78} (1978) 443.

\bibitem{Belanger:2014vza}
G.~Bélanger, F.~Boudjema, A.~Pukhov, and A.~Semenov, {\em Comput. Phys.
  Commun.} {\bf 192} (2015) 322--329,
  [\href{http://xxx.lanl.gov/abs/1407.6129}{{\tt 1407.6129}}].

\bibitem{Akerib:2015rjg}
D.~S. Akerib {\em et.~al.},, {\bf LUX} Collaboration,
  \href{http://xxx.lanl.gov/abs/1512.03506}{{\tt 1512.03506}}.

\bibitem{Belanger:2008sj}
G.~Belanger, F.~Boudjema, A.~Pukhov, and A.~Semenov, {\em Comput. Phys.
  Commun.} {\bf 180} (2009) 747--767,
  [\href{http://xxx.lanl.gov/abs/0803.2360}{{\tt 0803.2360}}].

\bibitem{Ade:2015xua}
P.~A.~R. Ade {\em et.~al.},, {\bf Planck} Collaboration,
  \href{http://xxx.lanl.gov/abs/1502.01589}{{\tt 1502.01589}}.

\bibitem{Feng:2005gj}
J.~L. Feng, S.~Su, and F.~Takayama, {\em Phys. Rev. Lett.} {\bf 96} (2006)
  151802, [\href{http://xxx.lanl.gov/abs/hep-ph/0503117}{{\tt
  hep-ph/0503117}}].

\bibitem{Busoni:2014gta}
G.~Busoni, A.~De~Simone, T.~Jacques, E.~Morgante, and A.~Riotto, {\em JCAP}
  {\bf 1503} (2015), no.~03 022, [\href{http://xxx.lanl.gov/abs/1410.7409}{{\tt
  1410.7409}}].

\bibitem{Conte:2012fm}
E.~Conte, B.~Fuks, and G.~Serret, {\em Comput. Phys. Commun.} {\bf 184} (2013)
  222--256, [\href{http://xxx.lanl.gov/abs/1206.1599}{{\tt 1206.1599}}].

\bibitem{Conte:2014zja}
E.~Conte, B.~Dumont, B.~Fuks, and C.~Wymant, {\em Eur. Phys. J.} {\bf C74}
  (2014), no.~10 3103, [\href{http://xxx.lanl.gov/abs/1405.3982}{{\tt
  1405.3982}}].

\bibitem{atlasmonojet}
D.~Sengupta and G.~Chalons, {\em
  http://doi.org/10.7484/INSPIREHEP.DATA.RB33.M3CD}.

\bibitem{Dumont:2014tja}
B.~Dumont, B.~Fuks, S.~Kraml, S.~Bein, G.~Chalons, E.~Conte, S.~Kulkarni,
  D.~Sengupta, and C.~Wymant, {\em Eur. Phys. J.} {\bf C75} (2015), no.~2 56,
  [\href{http://xxx.lanl.gov/abs/1407.3278}{{\tt 1407.3278}}].

\bibitem{Alwall:2014hca}
J.~Alwall, R.~Frederix, S.~Frixione, V.~Hirschi, F.~Maltoni, O.~Mattelaer,
  H.~S. Shao, T.~Stelzer, P.~Torrielli, and M.~Zaro, {\em JHEP} {\bf 07} (2014)
  079, [\href{http://xxx.lanl.gov/abs/1405.0301}{{\tt 1405.0301}}].

\bibitem{Degrande:2011ua}
C.~Degrande, C.~Duhr, B.~Fuks, D.~Grellscheid, O.~Mattelaer, and T.~Reiter,
  {\em Comput. Phys. Commun.} {\bf 183} (2012) 1201--1214,
  [\href{http://xxx.lanl.gov/abs/1108.2040}{{\tt 1108.2040}}].

\bibitem{Sjostrand:2006za}
T.~Sjostrand, S.~Mrenna, and P.~Z. Skands, {\em JHEP} {\bf 05} (2006) 026,
  [\href{http://xxx.lanl.gov/abs/hep-ph/0603175}{{\tt hep-ph/0603175}}].

\bibitem{deFavereau:2013fsa}
J.~de~Favereau, C.~Delaere, P.~Demin, A.~Giammanco, V.~Lemaître, A.~Mertens,
  and M.~Selvaggi,, {\bf DELPHES 3} Collaboration, {\em JHEP} {\bf 02} (2014)
  057, [\href{http://xxx.lanl.gov/abs/1307.6346}{{\tt 1307.6346}}].

\bibitem{Cacciari:2008gp}
M.~Cacciari, G.~P. Salam, and G.~Soyez, {\em JHEP} {\bf 04} (2008) 063,
  [\href{http://xxx.lanl.gov/abs/0802.1189}{{\tt 0802.1189}}].

\bibitem{Cacciari:2011ma}
M.~Cacciari, G.~P. Salam, and G.~Soyez, {\em Eur. Phys. J.} {\bf C72} (2012)
  1896, [\href{http://xxx.lanl.gov/abs/1111.6097}{{\tt 1111.6097}}].

\bibitem{Read:2000ru}
A.~L. Read, in {\em {Workshop on confidence limits, CERN, Geneva, Switzerland,
  17-18 Jan 2000: Proceedings}}, 2000.

\bibitem{Read:2002hq}
A.~L. Read, {\em J. Phys.} {\bf G28} (2002) 2693--2704.

\bibitem{Sjostrand:2014zea}
T.~Sj{\"o}strand, S.~Ask, J.~R. Christiansen, R.~Corke, N.~Desai, P.~Ilten,
  S.~Mrenna, S.~Prestel, C.~O. Rasmussen, and P.~Z. Skands, {\em Comput. Phys.
  Commun.} {\bf 191} (2015) 159--177,
  [\href{http://xxx.lanl.gov/abs/1410.3012}{{\tt 1410.3012}}].

\bibitem{Lonnblad:2011xx}
L.~Lonnblad and S.~Prestel, {\em JHEP} {\bf 03} (2012) 019,
  [\href{http://xxx.lanl.gov/abs/1109.4829}{{\tt 1109.4829}}].

\bibitem{Aad:2009wy}
{\bf ATLAS} Collaboration, {\em SLAC-R-980, CERN-OPEN-2008-020} (2009)
  [\href{http://xxx.lanl.gov/abs/0901.0512}{{\tt 0901.0512}}].

\bibitem{Moneta:2010pm}
L.~Moneta, K.~Belasco, K.~S. Cranmer, S.~Kreiss, A.~Lazzaro, D.~Piparo,
  G.~Schott, W.~Verkerke, and M.~Wolf, {\em PoS} {\bf ACAT2010} (2010) 057,
  [\href{http://xxx.lanl.gov/abs/1009.1003}{{\tt 1009.1003}}].

\bibitem{Abercrombie:2015wmb}
D.~Abercrombie {\em et.~al.}, \href{http://xxx.lanl.gov/abs/1507.00966}{{\tt
  1507.00966}}.

\bibitem{Kilic:2008ub}
C.~Kilic, S.~Schumann, and M.~Son, {\em JHEP} {\bf 04} (2009) 128,
  [\href{http://xxx.lanl.gov/abs/0810.5542}{{\tt 0810.5542}}].

\bibitem{Schumann:2011ji}
S.~Schumann, A.~Renaud, and D.~Zerwas, {\em JHEP} {\bf 09} (2011) 074,
  [\href{http://xxx.lanl.gov/abs/1108.2957}{{\tt 1108.2957}}].

\bibitem{Aad:2011yh}
{\bf ATLAS} Collaboration, {\em Eur. Phys. J.} {\bf C71} (2011) 1828,
  [\href{http://xxx.lanl.gov/abs/1110.2693}{{\tt 1110.2693}}].

\bibitem{ATLAS:2012ds}
{\bf ATLAS} Collaboration, {\em Eur. Phys. J.} {\bf C73} (2013), no.~1 2263,
  [\href{http://xxx.lanl.gov/abs/1210.4826}{{\tt 1210.4826}}].

\bibitem{Khachatryan:2014lpa}
{\bf CMS} Collaboration, {\em Phys. Lett.} {\bf B747} (2015) 98--119,
  [\href{http://xxx.lanl.gov/abs/1412.7706}{{\tt 1412.7706}}].

\bibitem{Aaltonen:2013hya}
T.~Aaltonen {\em et.~al.},, {\bf CDF} Collaboration, {\em Phys. Rev. Lett.}
  {\bf 111} (2013), no.~3 031802,
  [\href{http://xxx.lanl.gov/abs/1303.2699}{{\tt 1303.2699}}].

\bibitem{Kribs:2007ac}
G.~D. Kribs, E.~Poppitz, and N.~Weiner, {\em Phys. Rev.} {\bf D78} (2008)
  055010, [\href{http://xxx.lanl.gov/abs/0712.2039}{{\tt 0712.2039}}].

\bibitem{Plehn:2008ae}
T.~Plehn and T.~M.~P. Tait, {\em J. Phys.} {\bf G36} (2009) 075001,
  [\href{http://xxx.lanl.gov/abs/0810.3919}{{\tt 0810.3919}}].

\bibitem{Choi:2008ub}
S.~Y. Choi, M.~Drees, J.~Kalinowski, J.~M. Kim, E.~Popenda, and P.~M. Zerwas,
  {\em Phys. Lett.} {\bf B672} (2009) 246--252,
  [\href{http://xxx.lanl.gov/abs/0812.3586}{{\tt 0812.3586}}].

\bibitem{ATL-PHYS-PUB-2014-021}
{\bf ATLAS} Collaboration, Tech. Rep. ATL-PHYS-PUB-2014-021, CERN, Geneva, Nov,
  2014.

\bibitem{GoncalvesNetto:2012yt}
D.~Gon\c{c}alves-Netto, D.~L\'opez-Val, K.~Mawatari, T.~Plehn, and I.~Wigmore,
  {\em Phys. Rev.} {\bf D87} (2013) 014002,
  [\href{http://xxx.lanl.gov/abs/1211.0286}{{\tt 1211.0286}}].

\bibitem{GoncalvesNetto:2012nt}
D.~Goncalves-Netto, D.~Lopez-Val, K.~Mawatari, T.~Plehn, and I.~Wigmore, {\em
  Phys. Rev.} {\bf D85} (2012) 114024,
  [\href{http://xxx.lanl.gov/abs/1203.6358}{{\tt 1203.6358}}].

\bibitem{Komatsu:2010fb}
{\bf WMAP} Collaboration, {\em Astrophys. J. Suppl.} {\bf 192} (2011) 18,
  [\href{http://xxx.lanl.gov/abs/1001.4538}{{\tt 1001.4538}}].

\bibitem{Ade:2013zuv}
P.~A.~R. Ade {\em et.~al.},, {\bf Planck} Collaboration, {\em Astron.
  Astrophys.} {\bf 571} (2014) A16,
  [\href{http://xxx.lanl.gov/abs/1303.5076}{{\tt 1303.5076}}].

\bibitem{Buckley:2014fba}
M.~R. Buckley, D.~Feld, and D.~Goncalves, {\em Phys. Rev.} {\bf D91} (2015)
  015017, [\href{http://xxx.lanl.gov/abs/1410.6497}{{\tt 1410.6497}}].

\bibitem{Haisch:2015ioa}
U.~Haisch and E.~Re, {\em JHEP} {\bf 06} (2015) 078,
  [\href{http://xxx.lanl.gov/abs/1503.00691}{{\tt 1503.00691}}].

\bibitem{Sjostrand:2007gs}
T.~Sjostrand, S.~Mrenna, and P.~Z. Skands, {\em Comput. Phys. Commun.} {\bf
  178} (2008) 852--867, [\href{http://xxx.lanl.gov/abs/0710.3820}{{\tt
  0710.3820}}].

\bibitem{Aad:2014qaa}
{\bf ATLAS} Collaboration, {\em JHEP} {\bf 06} (2014) 124,
  [\href{http://xxx.lanl.gov/abs/1403.4853}{{\tt 1403.4853}}].

\bibitem{Alioli:2010xd}
S.~Alioli, P.~Nason, C.~Oleari, and E.~Re, {\em JHEP} {\bf 06} (2010) 043,
  [\href{http://xxx.lanl.gov/abs/1002.2581}{{\tt 1002.2581}}].

\bibitem{Lester:1999tx}
C.~G. Lester and D.~J. Summers, {\em Phys. Lett.} {\bf B463} (1999) 99--103,
  [\href{http://xxx.lanl.gov/abs/hep-ph/9906349}{{\tt hep-ph/9906349}}].

\bibitem{Barr:2003rg}
A.~Barr, C.~Lester, and P.~Stephens, {\em J. Phys.} {\bf G29} (2003)
  2343--2363, [\href{http://xxx.lanl.gov/abs/hep-ph/0304226}{{\tt
  hep-ph/0304226}}].

\bibitem{Barr:2005dz}
A.~Barr, {\em JHEP} {\bf 0602} (2006) 042,
  [\href{http://xxx.lanl.gov/abs/hep-ph/0511115}{{\tt hep-ph/0511115}}].

\bibitem{Linnemann:2003vw}
J.~T. Linnemann, {\em eConf} {\bf C030908} (2003) MOBT001,
  [\href{http://xxx.lanl.gov/abs/physics/0312059}{{\tt physics/0312059}}].

\bibitem{Glashow:1961tr}
S.~L. Glashow, {\em Nucl. Phys.} {\bf 22} (1961) 579--588.

\bibitem{Weinberg:1967tq}
S.~Weinberg, {\em Phys. Rev. Lett.} {\bf 19} (1967) 1264--1266.

\bibitem{salam}
A.~Salam, in {\em Elementary particle theory} (N.~Svartholm, ed.),
  pp.~367--377, Almquist \& Wiksell.

\bibitem{Aad:2012tfa}
{\bf ATLAS} Collaboration, {\em Phys. Lett.} {\bf B716} (2012) 1--29,
  [\href{http://xxx.lanl.gov/abs/1207.7214}{{\tt 1207.7214}}].

\bibitem{Chatrchyan:2012xdj}
{\bf CMS} Collaboration, {\em Phys. Lett.} {\bf B716} (2012) 30--61,
  [\href{http://xxx.lanl.gov/abs/1207.7235}{{\tt 1207.7235}}].

\bibitem{Englert:1964et}
F.~Englert and R.~Brout, {\em Phys. Rev. Lett.} {\bf 13} (1964) 321--323.

\bibitem{Higgs:1964ia}
P.~W. Higgs, {\em Phys. Lett.} {\bf 12} (1964) 132--133.

\bibitem{Higgs:1964pj}
P.~W. Higgs, {\em Phys. Rev. Lett.} {\bf 13} (1964) 508--509.

\bibitem{Guralnik:1964eu}
G.~S. Guralnik, C.~R. Hagen, and T.~W.~B. Kibble, {\em Phys. Rev. Lett.} {\bf
  13} (1964) 585--587.

\bibitem{Higgs:1966ev}
P.~W. Higgs, {\em Phys. Rev.} {\bf 145} (1966) 1156--1163.

\bibitem{PhysRev.155.1554}
T.~W.~B. Kibble, {\em Phys. Rev.} {\bf 155} (Mar, 1967) 1554--1561.

\bibitem{Berlin:2014cfa}
A.~Berlin, T.~Lin, and L.-T. Wang, {\em JHEP} {\bf 06} (2014) 078,
  [\href{http://xxx.lanl.gov/abs/1402.7074}{{\tt 1402.7074}}].

\bibitem{Berlin:2015wwa}
A.~Berlin, S.~Gori, T.~Lin, and L.-T. Wang, {\em Phys. Rev.} {\bf D92} (2015)
  015005, [\href{http://xxx.lanl.gov/abs/1502.06000}{{\tt 1502.06000}}].

\bibitem{No:2015xqa}
J.~M. No, {\em Phys. Rev.} {\bf D93} (2016) 031701,
  [\href{http://xxx.lanl.gov/abs/1509.01110}{{\tt 1509.01110}}]. [Phys.
  Rev.D93,031701(2016)].

\bibitem{Nomura:2008ru}
Y.~Nomura and J.~Thaler, {\em Phys. Rev.} {\bf D79} (2009) 075008,
  [\href{http://xxx.lanl.gov/abs/0810.5397}{{\tt 0810.5397}}].

\bibitem{Branco:2011iw}
G.~C. Branco, P.~M. Ferreira, L.~Lavoura, M.~N. Rebelo, M.~Sher, and J.~P.
  Silva, {\em Phys. Rept.} {\bf 516} (2012) 1--102,
  [\href{http://xxx.lanl.gov/abs/1106.0034}{{\tt 1106.0034}}].

\bibitem{Ipek:2014gua}
S.~Ipek, D.~McKeen, and A.~E. Nelson, {\em Phys. Rev.} {\bf D90} (2014), no.~5
  055021, [\href{http://xxx.lanl.gov/abs/1404.3716}{{\tt 1404.3716}}].

\bibitem{Nason:2004rx}
P.~Nason, {\em JHEP} {\bf 0411} (2004) 040,
  [\href{http://xxx.lanl.gov/abs/hep-ph/0409146}{{\tt hep-ph/0409146}}].

\bibitem{Frixione:2007vw}
S.~Frixione, P.~Nason, and C.~Oleari, {\em JHEP} {\bf 11} (2007) 070,
  [\href{http://xxx.lanl.gov/abs/0709.2092}{{\tt 0709.2092}}].

\bibitem{Alioli:2008tz}
S.~Alioli, P.~Nason, C.~Oleari, and E.~Re, {\em JHEP} {\bf 04} (2009) 002,
  [\href{http://xxx.lanl.gov/abs/0812.0578}{{\tt 0812.0578}}].

\bibitem{Nason:2009ai}
P.~Nason and C.~Oleari, {\em JHEP} {\bf 02} (2010) 037,
  [\href{http://xxx.lanl.gov/abs/0911.5299}{{\tt 0911.5299}}].

\bibitem{Luisoni2013}
G.~Luisoni, P.~Nason, C.~Oleari, and F.~Tramontano, {\em JHEP} {\bf 10} (2013)
  083, [\href{http://xxx.lanl.gov/abs/1306.2542}{{\tt 1306.2542}}].

\bibitem{Heinemeyer:2013tqa}
{LHC Higgs Cross Section Working Group}, S.~Heinemeyer, C.~Mariotti,
  G.~Passarino, and R.~Tanaka~(Eds.), {\em CERN-2013-004} (CERN, Geneva, 2013)
  [\href{http://xxx.lanl.gov/abs/1307.1347}{{\tt 1307.1347}}].

\bibitem{Aad:2014zya}
{\bf ATLAS} Collaboration, {\em Eur. Phys. J.} {\bf C74} (2014) 3034,
  [\href{http://xxx.lanl.gov/abs/1404.4562}{{\tt 1404.4562}}].

\bibitem{Aad:2014fxa}
{\bf ATLAS} Collaboration, {\em Eur. Phys. J.} {\bf C74} (2014) 2941,
  [\href{http://xxx.lanl.gov/abs/1404.2240}{{\tt 1404.2240}}].

\bibitem{Chatrchyan:2012xi}
{\bf CMS} Collaboration, {\em JINST} {\bf 7} (2012) P10002,
  [\href{http://xxx.lanl.gov/abs/1206.4071}{{\tt 1206.4071}}].

\bibitem{Khachatryan:2015hwa}
{\bf CMS} Collaboration, {\em JINST} {\bf 10} (2015), no.~06 P06005,
  [\href{http://xxx.lanl.gov/abs/1502.02701}{{\tt 1502.02701}}].

\bibitem{Aad:2014efa}
{\bf ATLAS} Collaboration, {\em JHEP} {\bf 11} (2014) 104,
  [\href{http://xxx.lanl.gov/abs/1409.5500}{{\tt 1409.5500}}].

\bibitem{Aad:2015gdg}
{\bf ATLAS} Collaboration, {\em JHEP} {\bf 10} (2015) 150,
  [\href{http://xxx.lanl.gov/abs/1504.04605}{{\tt 1504.04605}}].

\bibitem{Aad:2015kqa}
{\bf ATLAS} Collaboration, {\em JHEP} {\bf 08} (2015) 105,
  [\href{http://xxx.lanl.gov/abs/1505.04306}{{\tt 1505.04306}}].

\bibitem{Aad:2015mba}
{\bf ATLAS} Collaboration, {\em Phys. Rev.} {\bf D91} (2015) 112011,
  [\href{http://xxx.lanl.gov/abs/1503.05425}{{\tt 1503.05425}}].

\bibitem{Aad:2015tba}
{\bf ATLAS} Collaboration, {\em Phys. Rev.} {\bf D92} (2015) 112007,
  [\href{http://xxx.lanl.gov/abs/1509.04261}{{\tt 1509.04261}}].

\bibitem{Aad:2015voa}
{\bf ATLAS} Collaboration, \href{http://xxx.lanl.gov/abs/1510.02664}{{\tt
  1510.02664}}.

\bibitem{Aad:2016shx}
{\bf ATLAS} Collaboration, \href{http://xxx.lanl.gov/abs/1602.06034}{{\tt
  1602.06034}}.

\bibitem{Aad:2016qpo}
{\bf ATLAS} Collaboration, \href{http://xxx.lanl.gov/abs/1602.05606}{{\tt
  1602.05606}}.

\bibitem{Chatrchyan:2013uxa}
{\bf CMS} Collaboration, {\em Phys. Lett.} {\bf B729} (2014) 149--171,
  [\href{http://xxx.lanl.gov/abs/1311.7667}{{\tt 1311.7667}}].

\bibitem{Chatrchyan:2013wfa}
{\bf CMS} Collaboration, {\em Phys. Rev. Lett.} {\bf 112} (2014) 171801,
  [\href{http://xxx.lanl.gov/abs/1312.2391}{{\tt 1312.2391}}].

\bibitem{Khachatryan:2015axa}
{\bf CMS} Collaboration, {\em JHEP} {\bf 06} (2015) 080,
  [\href{http://xxx.lanl.gov/abs/1503.01952}{{\tt 1503.01952}}].

\bibitem{Khachatryan:2015gza}
{\bf CMS} Collaboration, \href{http://xxx.lanl.gov/abs/1507.07129}{{\tt
  1507.07129}}.

\bibitem{Khachatryan:2015oba}
{\bf CMS} Collaboration, {\em Phys. Rev.} {\bf D93} (2016) 012003,
  [\href{http://xxx.lanl.gov/abs/1509.04177}{{\tt 1509.04177}}].

\bibitem{Serra:2015xfa}
J.~Serra, {\em JHEP} {\bf 09} (2015) 176,
  [\href{http://xxx.lanl.gov/abs/1506.05110}{{\tt 1506.05110}}].

\bibitem{Anandakrishnan:2015yfa}
A.~Anandakrishnan, J.~H. Collins, M.~Farina, E.~Kuflik, and M.~Perelstein,
  \href{http://xxx.lanl.gov/abs/1506.05130}{{\tt 1506.05130}}.

\bibitem{Cacciapaglia:2010vn}
G.~Cacciapaglia, A.~Deandrea, D.~Harada, and Y.~Okada, {\em JHEP} {\bf 11}
  (2010) 159, [\href{http://xxx.lanl.gov/abs/1007.2933}{{\tt 1007.2933}}].

\bibitem{Cai:2012ji}
H.~Cai, {\em JHEP} {\bf 02} (2013) 104,
  [\href{http://xxx.lanl.gov/abs/1210.5200}{{\tt 1210.5200}}].

\bibitem{Contino:2011np}
R.~Contino, D.~Marzocca, D.~Pappadopulo, and R.~Rattazzi, {\em JHEP} {\bf 10}
  (2011) 081, [\href{http://xxx.lanl.gov/abs/1109.1570}{{\tt 1109.1570}}].

\bibitem{ArkaniHamed:2002qx}
N.~Arkani-Hamed, A.~G. Cohen, E.~Katz, A.~E. Nelson, T.~Gregoire, and J.~G.
  Wacker, {\em JHEP} {\bf 08} (2002) 021,
  [\href{http://xxx.lanl.gov/abs/hep-ph/0206020}{{\tt hep-ph/0206020}}].

\bibitem{Burdman:2002ns}
G.~Burdman, M.~Perelstein, and A.~Pierce, {\em Phys. Rev. Lett.} {\bf 90}
  (2003) 241802, [\href{http://xxx.lanl.gov/abs/hep-ph/0212228}{{\tt
  hep-ph/0212228}}]. [Erratum: Phys. Rev. Lett.92,049903(2004)].

\bibitem{Dugan:1984hq}
M.~J. Dugan, H.~Georgi, and D.~B. Kaplan, {\em Nucl. Phys.} {\bf B254} (1985)
  299.

\bibitem{Katz:2005au}
E.~Katz, A.~E. Nelson, and D.~G.~E. Walker, {\em JHEP} {\bf 08} (2005) 074,
  [\href{http://xxx.lanl.gov/abs/hep-ph/0504252}{{\tt hep-ph/0504252}}].

\bibitem{Gripaios:2009pe}
B.~Gripaios, A.~Pomarol, F.~Riva, and J.~Serra, {\em JHEP} {\bf 04} (2009) 070,
  [\href{http://xxx.lanl.gov/abs/0902.1483}{{\tt 0902.1483}}].

\bibitem{Kearney:2013oia}
J.~Kearney, A.~Pierce, and J.~Thaler, {\em JHEP} {\bf 08} (2013) 130,
  [\href{http://xxx.lanl.gov/abs/1304.4233}{{\tt 1304.4233}}].

\bibitem{Kearney:2013cca}
J.~Kearney, A.~Pierce, and J.~Thaler, {\em JHEP} {\bf 10} (2013) 230,
  [\href{http://xxx.lanl.gov/abs/1306.4314}{{\tt 1306.4314}}].

\bibitem{Barnard:2013zea}
J.~Barnard, T.~Gherghetta, and T.~S. Ray, {\em JHEP} {\bf 02} (2014) 002,
  [\href{http://xxx.lanl.gov/abs/1311.6562}{{\tt 1311.6562}}].

\bibitem{Cacciapaglia:2015eqa}
G.~Cacciapaglia, H.~Cai, A.~Deandrea, T.~Flacke, S.~J. Lee, and A.~Parolini,
  {\em JHEP} {\bf 11} (2015) 201,
  [\href{http://xxx.lanl.gov/abs/1507.02283}{{\tt 1507.02283}}].

\bibitem{Ferretti:2014qta}
G.~Ferretti, {\em JHEP} {\bf 06} (2014) 142,
  [\href{http://xxx.lanl.gov/abs/1404.7137}{{\tt 1404.7137}}].

\bibitem{Pati:1974yy}
J.~C. Pati and A.~Salam, {\em Phys. Rev.} {\bf D10} (1974) 275--289. [Erratum:
  Phys. Rev.D11,703(1975)].

\bibitem{Wudka:1985ef}
J.~Wudka, {\em Phys. Lett.} {\bf B167} (1986) 337.

\bibitem{Gripaios:2009dq}
B.~Gripaios, {\em JHEP} {\bf 02} (2010) 045,
  [\href{http://xxx.lanl.gov/abs/0910.1789}{{\tt 0910.1789}}].

\bibitem{Giacchino:2015hvk}
F.~Giacchino, A.~Ibarra, L.~L. Honorez, M.~H.~G. Tytgat, and S.~Wild, {\em
  JCAP} {\bf 1602} (2016), no.~02 002,
  [\href{http://xxx.lanl.gov/abs/1511.04452}{{\tt 1511.04452}}].

\bibitem{Aad:2015wqa}
{\bf ATLAS} Collaboration, {\em Eur. Phys. J.} {\bf C75} (2015), no.~7 318,
  [\href{http://xxx.lanl.gov/abs/1503.03290}{{\tt 1503.03290}}]. [Erratum: Eur.
  Phys. J.C75,no.10,463(2015)].

\bibitem{Aad:2014mda}
{\bf ATLAS} Collaboration, {\em JHEP} {\bf 01} (2015) 049,
  [\href{http://xxx.lanl.gov/abs/1410.7238}{{\tt 1410.7238}}].

\bibitem{Aad:2015xja}
{\bf ATLAS} Collaboration, {\em Phys. Rev.} {\bf D92} (2015) 092004,
  [\href{http://xxx.lanl.gov/abs/1509.04670}{{\tt 1509.04670}}].

\bibitem{Chatrchyan:2013gia}
{\bf CMS} Collaboration, {\em Phys. Lett.} {\bf B730} (2014) 193--214,
  [\href{http://xxx.lanl.gov/abs/1311.1799}{{\tt 1311.1799}}].

\bibitem{Brehmer:2015dan}
J.~Brehmer {\em et.~al.}, \href{http://xxx.lanl.gov/abs/1512.04357}{{\tt
  1512.04357}}.

\bibitem{dec15seminar}
{ATLAS and CMS physics results from Run 2, speakers Marumi Kado and Jim Olsen,
  CERN seminar, 15 Dec 2016}.

\bibitem{ATLAS-CONF-2015-062}
{\bf ATLAS} Collaboration, Tech. Rep. ATLAS-CONF-2015-062, CERN, Geneva, Dec,
  2015.

\bibitem{CMS-PAS-SUS-15-003}
{\bf CMS} Collaboration, Tech. Rep. CMS-PAS-SUS-15-003, CERN, Geneva, 2015.

\bibitem{Aad:2015zhl}
{\bf ATLAS, CMS} Collaboration, {\em Phys. Rev. Lett.} {\bf 114} (2015) 191803,
  [\href{http://xxx.lanl.gov/abs/1503.07589}{{\tt 1503.07589}}].

\bibitem{Ellwanger:2006rm}
U.~Ellwanger and C.~Hugonie, {\em Mod. Phys. Lett.} {\bf A22} (2007)
  1581--1590, [\href{http://xxx.lanl.gov/abs/hep-ph/0612133}{{\tt
  hep-ph/0612133}}].

\bibitem{Ma:2011ea}
E.~Ma, {\em Phys. Lett.} {\bf B705} (2011) 320--323,
  [\href{http://xxx.lanl.gov/abs/1108.4029}{{\tt 1108.4029}}].

\bibitem{Hirsch:2011hg}
M.~Hirsch, M.~Malinsky, W.~Porod, L.~Reichert, and F.~Staub, {\em JHEP} {\bf
  02} (2012) 084, [\href{http://xxx.lanl.gov/abs/1110.3037}{{\tt 1110.3037}}].

\bibitem{Mohapatra:1999vv}
R.~N. Mohapatra, in {\em {Particle physics. Proceedings, Summer School,
  Trieste, Italy, June 21-July 9, 1999}}, pp.~336--394, 1999.
\newblock \href{http://xxx.lanl.gov/abs/hep-ph/9911272}{{\tt hep-ph/9911272}}.

\bibitem{Hirsch:2015fvq}
M.~Hirsch, M.~E. Krauss, T.~Opferkuch, W.~Porod, and F.~Staub, {\em JHEP} {\bf
  03} (2016) 009, [\href{http://xxx.lanl.gov/abs/1512.00472}{{\tt
  1512.00472}}].

\bibitem{Hirsch:2012kv}
M.~Hirsch, W.~Porod, L.~Reichert, and F.~Staub, {\em Phys. Rev.} {\bf D86}
  (2012) 093018, [\href{http://xxx.lanl.gov/abs/1206.3516}{{\tt 1206.3516}}].

\bibitem{DeRomeri:2011ie}
V.~De~Romeri, M.~Hirsch, and M.~Malinsky, {\em Phys. Rev.} {\bf D84} (2011)
  053012, [\href{http://xxx.lanl.gov/abs/1107.3412}{{\tt 1107.3412}}].

\bibitem{Dreiner:2012gx}
H.~K. Dreiner, M.~Kramer, and J.~Tattersall, {\em Europhys. Lett.} {\bf 99}
  (2012) 61001, [\href{http://xxx.lanl.gov/abs/1207.1613}{{\tt 1207.1613}}].

\bibitem{Arina:2007tm}
C.~Arina and N.~Fornengo, {\em JHEP} {\bf 11} (2007) 029,
  [\href{http://xxx.lanl.gov/abs/0709.4477}{{\tt 0709.4477}}].

\bibitem{Arina:2015uea}
C.~Arina, M.~E.~C. Catalan, S.~Kraml, S.~Kulkarni, and U.~Laa, {\em JHEP} {\bf
  05} (2015) 142, [\href{http://xxx.lanl.gov/abs/1503.02960}{{\tt
  1503.02960}}].

\bibitem{Belanger:2015cra}
G.~B\'elanger, J.~Da~Silva, U.~Laa, and A.~Pukhov, {\em JHEP} {\bf 09} (2015)
  151, [\href{http://xxx.lanl.gov/abs/1505.06243}{{\tt 1505.06243}}].

\bibitem{Mohapatra:1986bd}
R.~N. Mohapatra and J.~W.~F. Valle, {\em Phys. Rev.} {\bf D34} (1986) 1642.

\bibitem{Krauss:2013jva}
M.~E. Krauss, W.~Porod, and F.~Staub, {\em Phys. Rev.} {\bf D88} (2013), no.~1
  015014, [\href{http://xxx.lanl.gov/abs/1304.0769}{{\tt 1304.0769}}].

\bibitem{Dev:2009aw}
P.~S.~B. Dev and R.~N. Mohapatra, {\em Phys. Rev.} {\bf D81} (2010) 013001,
  [\href{http://xxx.lanl.gov/abs/0910.3924}{{\tt 0910.3924}}].

\bibitem{BhupalDev:2010he}
P.~S. Bhupal~Dev and R.~N. Mohapatra, {\em Phys. Rev.} {\bf D82} (2010) 035014,
  [\href{http://xxx.lanl.gov/abs/1003.6102}{{\tt 1003.6102}}].

\bibitem{Ibanez:1991hv}
L.~E. Ibanez and G.~G. Ross, {\em Phys. Lett.} {\bf B260} (1991) 291--295.

\bibitem{Dreiner:2005rd}
H.~K. Dreiner, C.~Luhn, and M.~Thormeier, {\em Phys. Rev.} {\bf D73} (2006)
  075007, [\href{http://xxx.lanl.gov/abs/hep-ph/0512163}{{\tt
  hep-ph/0512163}}].

\bibitem{Fonseca:2011vn}
R.~M. Fonseca, M.~Malinsky, W.~Porod, and F.~Staub, {\em Nucl. Phys.} {\bf
  B854} (2012) 28--53, [\href{http://xxx.lanl.gov/abs/1107.2670}{{\tt
  1107.2670}}].

\bibitem{Krauss:2012ku}
M.~E. Krauss, B.~O'Leary, W.~Porod, and F.~Staub, {\em Phys. Rev.} {\bf D86}
  (2012) 055017, [\href{http://xxx.lanl.gov/abs/1206.3513}{{\tt 1206.3513}}].

\bibitem{Staub:2016dxq}
F.~Staub {\em et.~al.}, \href{http://xxx.lanl.gov/abs/1602.05581}{{\tt
  1602.05581}}.

\bibitem{atlasdilepton}
{\bf ATLAS} Collaboration,
  \href{http://xxx.lanl.gov/abs/ATLAS-CONF-2015-070}{{\tt
  ATLAS-CONF-2015-070}}.

\bibitem{GonzalezGarcia:1988rw}
M.~C. Gonzalez-Garcia and J.~W.~F. Valle, {\em Phys. Lett.} {\bf B216} (1989)
  360.

\bibitem{Basso:2012ew}
L.~Basso, A.~Belyaev, D.~Chowdhury, M.~Hirsch, S.~Khalil, S.~Moretti,
  B.~O'Leary, W.~Porod, and F.~Staub, {\em Comput. Phys. Commun.} {\bf 184}
  (2013) 698--719, [\href{http://xxx.lanl.gov/abs/1206.4563}{{\tt 1206.4563}}].

\bibitem{Abada:2012mc}
A.~Abada, D.~Das, A.~M. Teixeira, A.~Vicente, and C.~Weiland, {\em JHEP} {\bf
  02} (2013) 048, [\href{http://xxx.lanl.gov/abs/1211.3052}{{\tt 1211.3052}}].

\bibitem{Abada:2014kba}
A.~Abada, M.~E. Krauss, W.~Porod, F.~Staub, A.~Vicente, and C.~Weiland, {\em
  JHEP} {\bf 11} (2014) 048, [\href{http://xxx.lanl.gov/abs/1408.0138}{{\tt
  1408.0138}}].

\bibitem{Staub:2008uz}
F.~Staub, \href{http://xxx.lanl.gov/abs/0806.0538}{{\tt 0806.0538}}.

\bibitem{Staub:2010jh}
F.~Staub, {\em Comput.Phys.Commun.} {\bf 182} (2011) 808--833,
  [\href{http://xxx.lanl.gov/abs/1002.0840}{{\tt 1002.0840}}].

\bibitem{Staub:2012pb}
F.~Staub, {\em Computer Physics Communications} {\bf 184} (2013) pp.
  1792--1809, [\href{http://xxx.lanl.gov/abs/1207.0906}{{\tt 1207.0906}}].

\bibitem{Staub:2013tta}
F.~Staub, {\em Comput. Phys. Commun.} {\bf 185} (2014) 1773--1790,
  [\href{http://xxx.lanl.gov/abs/1309.7223}{{\tt 1309.7223}}].

\bibitem{Staub:2015kfa}
F.~Staub, {\em Adv. High Energy Phys.} {\bf 2015} (2015) 840780,
  [\href{http://xxx.lanl.gov/abs/1503.04200}{{\tt 1503.04200}}].

\bibitem{Porod:2003um}
W.~Porod, {\em Comput.Phys.Commun.} {\bf 153} (2003) 275--315,
  [\href{http://xxx.lanl.gov/abs/hep-ph/0301101}{{\tt hep-ph/0301101}}].

\bibitem{Porod:2011nf}
W.~Porod and F.~Staub, {\em Comput.Phys.Commun.} {\bf 183} (2012) 2458--2469,
  [\href{http://xxx.lanl.gov/abs/1104.1573}{{\tt 1104.1573}}].

\bibitem{Goodsell:2015ira}
M.~Goodsell, K.~Nickel, and F.~Staub, {\em Eur. Phys. J.} {\bf C75} (2015),
  no.~6 290, [\href{http://xxx.lanl.gov/abs/1503.03098}{{\tt 1503.03098}}].

\bibitem{Allanach:2008qq}
B.~C. Allanach {\em et.~al.}, {\em Comput. Phys. Commun.} {\bf 180} (2009)
  8--25, [\href{http://xxx.lanl.gov/abs/0801.0045}{{\tt 0801.0045}}].

\bibitem{Alwall:2011uj}
J.~Alwall, M.~Herquet, F.~Maltoni, O.~Mattelaer, and T.~Stelzer, {\em JHEP}
  {\bf 06} (2011) 128, [\href{http://xxx.lanl.gov/abs/1106.0522}{{\tt
  1106.0522}}].

\bibitem{Beenakker:1996ch}
W.~Beenakker, R.~Hopker, M.~Spira, and P.~M. Zerwas, {\em Nucl. Phys.} {\bf
  B492} (1997) 51--103, [\href{http://xxx.lanl.gov/abs/hep-ph/9610490}{{\tt
  hep-ph/9610490}}].

\bibitem{Beenakker:1997ut}
W.~Beenakker, M.~Kramer, T.~Plehn, M.~Spira, and P.~M. Zerwas, {\em Nucl.
  Phys.} {\bf B515} (1998) 3--14,
  [\href{http://xxx.lanl.gov/abs/hep-ph/9710451}{{\tt hep-ph/9710451}}].

\bibitem{Kulesza:2008jb}
A.~Kulesza and L.~Motyka, {\em Phys. Rev. Lett.} {\bf 102} (2009) 111802,
  [\href{http://xxx.lanl.gov/abs/0807.2405}{{\tt 0807.2405}}].

\bibitem{Kulesza:2009kq}
A.~Kulesza and L.~Motyka, {\em Phys. Rev.} {\bf D80} (2009) 095004,
  [\href{http://xxx.lanl.gov/abs/0905.4749}{{\tt 0905.4749}}].

\bibitem{Beenakker:2010nq}
W.~Beenakker, S.~Brensing, M.~Kramer, A.~Kulesza, E.~Laenen, and I.~Niessen,
  {\em JHEP} {\bf 08} (2010) 098,
  [\href{http://xxx.lanl.gov/abs/1006.4771}{{\tt 1006.4771}}].

\bibitem{Beenakker:2011fu}
W.~Beenakker, S.~Brensing, M.~n. Kramer, A.~Kulesza, E.~Laenen, L.~Motyka, and
  I.~Niessen, {\em Int. J. Mod. Phys.} {\bf A26} (2011) 2637--2664,
  [\href{http://xxx.lanl.gov/abs/1105.1110}{{\tt 1105.1110}}].

\bibitem{Pumplin:2002vw}
J.~Pumplin, D.~R. Stump, J.~Huston, H.~L. Lai, P.~M. Nadolsky, and W.~K. Tung,
  {\em JHEP} {\bf 07} (2002) 012,
  [\href{http://xxx.lanl.gov/abs/hep-ph/0201195}{{\tt hep-ph/0201195}}].

\bibitem{Drees:2013wra}
M.~Drees, H.~Dreiner, D.~Schmeier, J.~Tattersall, and J.~S. Kim, {\em Comput.
  Phys. Commun.} {\bf 187} (2014) 227--265,
  [\href{http://xxx.lanl.gov/abs/1312.2591}{{\tt 1312.2591}}].

\bibitem{Kim:2015wza}
J.~S. Kim, D.~Schmeier, J.~Tattersall, and K.~Rolbiecki, {\em Comput. Phys.
  Commun.} {\bf 196} (2015) 535--562,
  [\href{http://xxx.lanl.gov/abs/1503.01123}{{\tt 1503.01123}}].

\bibitem{Cacciari:2005hq}
M.~Cacciari and G.~P. Salam, {\em Phys. Lett.} {\bf B641} (2006) 57--61,
  [\href{http://xxx.lanl.gov/abs/hep-ph/0512210}{{\tt hep-ph/0512210}}].

\bibitem{Cao:2015ara}
J.~Cao, L.~Shang, J.~M. Yang, and Y.~Zhang, {\em JHEP} {\bf 06} (2015) 152,
  [\href{http://xxx.lanl.gov/abs/1504.07869}{{\tt 1504.07869}}].

\bibitem{TheATLAScollaboration:2013tha}
{\bf ATLAS} Collaboration,
  \href{http://xxx.lanl.gov/abs/ATLAS-CONF-2013-061}{{\tt
  ATLAS-CONF-2013-061}}.

\bibitem{Aad:2014wea}
{\bf ATLAS} Collaboration, {\em JHEP} {\bf 09} (2014) 176,
  [\href{http://xxx.lanl.gov/abs/1405.7875}{{\tt 1405.7875}}].

\bibitem{Aad:2014kra}
{\bf ATLAS} Collaboration, {\em JHEP} {\bf 11} (2014) 118,
  [\href{http://xxx.lanl.gov/abs/1407.0583}{{\tt 1407.0583}}].

\bibitem{Aad:2013wta}
{\bf ATLAS} Collaboration, {\em JHEP} {\bf 10} (2013) 130,
  [\href{http://xxx.lanl.gov/abs/1308.1841}{{\tt 1308.1841}}]. [Erratum:
  JHEP01,109(2014)].

\bibitem{Aad:2014pda}
{\bf ATLAS} Collaboration, {\em JHEP} {\bf 06} (2014) 035,
  [\href{http://xxx.lanl.gov/abs/1404.2500}{{\tt 1404.2500}}].

\bibitem{ColliderReach}
G.~P. Salam and A.~Weiler,
  \href{http://xxx.lanl.gov/abs/http://collider-reach.web.cern.ch/collider-reach/}{{\tt
  http://collider-reach.web.cern.ch/collider-reach/}}.

\bibitem{Salam:2008qg}
G.~P. Salam and J.~Rojo, {\em Comput. Phys. Commun.} {\bf 180} (2009) 120--156,
  [\href{http://xxx.lanl.gov/abs/0804.3755}{{\tt 0804.3755}}].

\bibitem{Djouadi:1991tka}
A.~Djouadi, M.~Spira, and P.~M. Zerwas, {\em Phys. Lett.} {\bf B264} (1991)
  440--446.

\bibitem{Dawson:1990zj}
S.~Dawson, {\em Nucl. Phys.} {\bf B359} (1991) 283--300.

\bibitem{Graudenz:1992pv}
D.~Graudenz, M.~Spira, and P.~M. Zerwas, {\em Phys. Rev. Lett.} {\bf 70} (1993)
  1372--1375.

\bibitem{Spira:1995rr}
M.~Spira, A.~Djouadi, D.~Graudenz, and P.~M. Zerwas, {\em Nucl. Phys.} {\bf
  B453} (1995) 17--82, [\href{http://xxx.lanl.gov/abs/hep-ph/9504378}{{\tt
  hep-ph/9504378}}].

\bibitem{Harlander:2005rq}
R.~Harlander and P.~Kant, {\em JHEP} {\bf 12} (2005) 015,
  [\href{http://xxx.lanl.gov/abs/hep-ph/0509189}{{\tt hep-ph/0509189}}].

\bibitem{Anastasiou:2009kn}
C.~Anastasiou, S.~Bucherer, and Z.~Kunszt, {\em JHEP} {\bf 10} (2009) 068,
  [\href{http://xxx.lanl.gov/abs/0907.2362}{{\tt 0907.2362}}].

\bibitem{Catani:2001ic}
S.~Catani, D.~de~Florian, and M.~Grazzini, {\em JHEP} {\bf 05} (2001) 025,
  [\href{http://xxx.lanl.gov/abs/hep-ph/0102227}{{\tt hep-ph/0102227}}].

\bibitem{Harlander:2001is}
R.~V. Harlander and W.~B. Kilgore, {\em Phys. Rev.} {\bf D64} (2001) 013015,
  [\href{http://xxx.lanl.gov/abs/hep-ph/0102241}{{\tt hep-ph/0102241}}].

\bibitem{Harlander:2002wh}
R.~V. Harlander and W.~B. Kilgore, {\em Phys. Rev. Lett.} {\bf 88} (2002)
  201801, [\href{http://xxx.lanl.gov/abs/hep-ph/0201206}{{\tt
  hep-ph/0201206}}].

\bibitem{Anastasiou:2002yz}
C.~Anastasiou and K.~Melnikov, {\em Nucl. Phys.} {\bf B646} (2002) 220--256,
  [\href{http://xxx.lanl.gov/abs/hep-ph/0207004}{{\tt hep-ph/0207004}}].

\bibitem{Ravindran:2003um}
V.~Ravindran, J.~Smith, and W.~L. van Neerven, {\em Nucl. Phys.} {\bf B665}
  (2003) 325--366, [\href{http://xxx.lanl.gov/abs/hep-ph/0302135}{{\tt
  hep-ph/0302135}}].

\bibitem{Gehrmann:2011aa}
T.~Gehrmann, M.~Jaquier, E.~W.~N. Glover, and A.~Koukoutsakis, {\em JHEP} {\bf
  02} (2012) 056, [\href{http://xxx.lanl.gov/abs/1112.3554}{{\tt 1112.3554}}].

\bibitem{Anastasiou:2013srw}
C.~Anastasiou, C.~Duhr, F.~Dulat, and B.~Mistlberger, {\em JHEP} {\bf 07}
  (2013) 003, [\href{http://xxx.lanl.gov/abs/1302.4379}{{\tt 1302.4379}}].

\bibitem{Anastasiou:2013mca}
C.~Anastasiou, C.~Duhr, F.~Dulat, F.~Herzog, and B.~Mistlberger, {\em JHEP}
  {\bf 12} (2013) 088, [\href{http://xxx.lanl.gov/abs/1311.1425}{{\tt
  1311.1425}}].

\bibitem{Kilgore:2013gba}
W.~B. Kilgore, {\em Phys. Rev.} {\bf D89} (2014), no.~7 073008,
  [\href{http://xxx.lanl.gov/abs/1312.1296}{{\tt 1312.1296}}].

\bibitem{Li:2014bfa}
Y.~Li, A.~von Manteuffel, R.~M. Schabinger, and H.~X. Zhu, {\em Phys. Rev.}
  {\bf D90} (2014), no.~5 053006,
  [\href{http://xxx.lanl.gov/abs/1404.5839}{{\tt 1404.5839}}].

\bibitem{Anastasiou:2014lda}
C.~Anastasiou, C.~Duhr, F.~Dulat, E.~Furlan, T.~Gehrmann, F.~Herzog, and
  B.~Mistlberger, {\em JHEP} {\bf 03} (2015) 091,
  [\href{http://xxx.lanl.gov/abs/1411.3584}{{\tt 1411.3584}}].

\bibitem{Anastasiou:2015ema}
C.~Anastasiou, C.~Duhr, F.~Dulat, F.~Herzog, and B.~Mistlberger, {\em Phys.
  Rev. Lett.} {\bf 114} (2015) 212001,
  [\href{http://xxx.lanl.gov/abs/1503.06056}{{\tt 1503.06056}}].

\bibitem{Anastasiou:2015yha}
C.~Anastasiou, C.~Duhr, F.~Dulat, E.~Furlan, F.~Herzog, and B.~Mistlberger,
  {\em JHEP} {\bf 08} (2015) 051,
  [\href{http://xxx.lanl.gov/abs/1505.04110}{{\tt 1505.04110}}].

\bibitem{Anastasiou:2016cez}
C.~Anastasiou, C.~Duhr, F.~Dulat, E.~Furlan, T.~Gehrmann, F.~Herzog,
  A.~Lazopoulos, and B.~Mistlberger,
  \href{http://xxx.lanl.gov/abs/1602.00695}{{\tt 1602.00695}}.

\bibitem{Djouadi:1994ge}
A.~Djouadi and P.~Gambino, {\em Phys. Rev. Lett.} {\bf 73} (1994) 2528--2531,
  [\href{http://xxx.lanl.gov/abs/hep-ph/9406432}{{\tt hep-ph/9406432}}].

\bibitem{Chetyrkin:1996wr}
K.~G. Chetyrkin, B.~A. Kniehl, and M.~Steinhauser, {\em Phys. Rev. Lett.} {\bf
  78} (1997) 594--597, [\href{http://xxx.lanl.gov/abs/hep-ph/9610456}{{\tt
  hep-ph/9610456}}].

\bibitem{Chetyrkin:1996ke}
K.~G. Chetyrkin, B.~A. Kniehl, and M.~Steinhauser, {\em Nucl. Phys.} {\bf B490}
  (1997) 19--39, [\href{http://xxx.lanl.gov/abs/hep-ph/9701277}{{\tt
  hep-ph/9701277}}].

\bibitem{Aglietti:2004nj}
U.~Aglietti, R.~Bonciani, G.~Degrassi, and A.~Vicini, {\em Phys. Lett.} {\bf
  B595} (2004) 432--441, [\href{http://xxx.lanl.gov/abs/hep-ph/0404071}{{\tt
  hep-ph/0404071}}].

\bibitem{Degrassi:2004mx}
G.~Degrassi and F.~Maltoni, {\em Phys. Lett.} {\bf B600} (2004) 255--260,
  [\href{http://xxx.lanl.gov/abs/hep-ph/0407249}{{\tt hep-ph/0407249}}].

\bibitem{Aglietti:2006yd}
U.~Aglietti, R.~Bonciani, G.~Degrassi, and A.~Vicini, in {\em {TeV4LHC
  Workshop: 2nd Meeting Brookhaven, Upton, New York, February 3-5, 2005}},
  2006.
\newblock \href{http://xxx.lanl.gov/abs/hep-ph/0610033}{{\tt hep-ph/0610033}}.

\bibitem{Actis:2008ug}
S.~Actis, G.~Passarino, C.~Sturm, and S.~Uccirati, {\em Phys. Lett.} {\bf B670}
  (2008) 12--17, [\href{http://xxx.lanl.gov/abs/0809.1301}{{\tt 0809.1301}}].

\bibitem{Actis:2008ts}
S.~Actis, G.~Passarino, C.~Sturm, and S.~Uccirati, {\em Nucl. Phys.} {\bf B811}
  (2009) 182--273, [\href{http://xxx.lanl.gov/abs/0809.3667}{{\tt 0809.3667}}].

\bibitem{Anastasiou:2008tj}
C.~Anastasiou, R.~Boughezal, and F.~Petriello, {\em JHEP} {\bf 04} (2009) 003,
  [\href{http://xxx.lanl.gov/abs/0811.3458}{{\tt 0811.3458}}].

\bibitem{Kramer:1996iq}
M.~Kramer, E.~Laenen, and M.~Spira, {\em Nucl. Phys.} {\bf B511} (1998)
  523--549, [\href{http://xxx.lanl.gov/abs/hep-ph/9611272}{{\tt
  hep-ph/9611272}}].

\bibitem{Catani:2003zt}
S.~Catani, D.~de~Florian, M.~Grazzini, and P.~Nason, {\em JHEP} {\bf 07} (2003)
  028, [\href{http://xxx.lanl.gov/abs/hep-ph/0306211}{{\tt hep-ph/0306211}}].

\bibitem{Moch:2005ky}
S.~Moch and A.~Vogt, {\em Phys. Lett.} {\bf B631} (2005) 48--57,
  [\href{http://xxx.lanl.gov/abs/hep-ph/0508265}{{\tt hep-ph/0508265}}].

\bibitem{Ravindran:2005vv}
V.~Ravindran, {\em Nucl. Phys.} {\bf B746} (2006) 58--76,
  [\href{http://xxx.lanl.gov/abs/hep-ph/0512249}{{\tt hep-ph/0512249}}].

\bibitem{Ravindran:2006cg}
V.~Ravindran, {\em Nucl. Phys.} {\bf B752} (2006) 173--196,
  [\href{http://xxx.lanl.gov/abs/hep-ph/0603041}{{\tt hep-ph/0603041}}].

\bibitem{Idilbi:2005ni}
A.~Idilbi, X.-d. Ji, J.-P. Ma, and F.~Yuan, {\em Phys. Rev.} {\bf D73} (2006)
  077501, [\href{http://xxx.lanl.gov/abs/hep-ph/0509294}{{\tt
  hep-ph/0509294}}].

\bibitem{Ahrens:2008nc}
V.~Ahrens, T.~Becher, M.~Neubert, and L.~L. Yang, {\em Eur. Phys. J.} {\bf C62}
  (2009) 333--353, [\href{http://xxx.lanl.gov/abs/0809.4283}{{\tt 0809.4283}}].

\bibitem{deFlorian:2009hc}
D.~de~Florian and M.~Grazzini, {\em Phys. Lett.} {\bf B674} (2009) 291--294,
  [\href{http://xxx.lanl.gov/abs/0901.2427}{{\tt 0901.2427}}].

\bibitem{deFlorian:2012yg}
D.~de~Florian and M.~Grazzini, {\em Phys. Lett.} {\bf B718} (2012) 117--120,
  [\href{http://xxx.lanl.gov/abs/1206.4133}{{\tt 1206.4133}}].

\bibitem{deFlorian:2014vta}
D.~de~Florian, J.~Mazzitelli, S.~Moch, and A.~Vogt, {\em JHEP} {\bf 10} (2014)
  176, [\href{http://xxx.lanl.gov/abs/1408.6277}{{\tt 1408.6277}}].

\bibitem{Bonvini:2014joa}
M.~Bonvini and S.~Marzani, {\em JHEP} {\bf 09} (2014) 007,
  [\href{http://xxx.lanl.gov/abs/1405.3654}{{\tt 1405.3654}}].

\bibitem{Bonvini:2014tea}
M.~Bonvini and L.~Rottoli, {\em Phys. Rev.} {\bf D91} (2015), no.~5 051301,
  [\href{http://xxx.lanl.gov/abs/1412.3791}{{\tt 1412.3791}}].

\bibitem{Catani:2014uta}
S.~Catani, L.~Cieri, D.~de~Florian, G.~Ferrera, and M.~Grazzini, {\em Nucl.
  Phys.} {\bf B888} (2014) 75--91,
  [\href{http://xxx.lanl.gov/abs/1405.4827}{{\tt 1405.4827}}].

\bibitem{Schmidt:2015cea}
T.~Schmidt and M.~Spira, {\em Phys. Rev.} {\bf D93} (2016), no.~1 014022,
  [\href{http://xxx.lanl.gov/abs/1509.00195}{{\tt 1509.00195}}].

\bibitem{Chetyrkin:1997sg}
K.~G. Chetyrkin, B.~A. Kniehl, and M.~Steinhauser, {\em Phys. Rev. Lett.} {\bf
  79} (1997) 2184--2187, [\href{http://xxx.lanl.gov/abs/hep-ph/9706430}{{\tt
  hep-ph/9706430}}].

\bibitem{Chetyrkin:1997un}
K.~G. Chetyrkin, B.~A. Kniehl, and M.~Steinhauser, {\em Nucl. Phys.} {\bf B510}
  (1998) 61--87, [\href{http://xxx.lanl.gov/abs/hep-ph/9708255}{{\tt
  hep-ph/9708255}}].

\bibitem{Spira:1997dg}
M.~Spira, {\em Fortsch. Phys.} {\bf 46} (1998) 203--284,
  [\href{http://xxx.lanl.gov/abs/hep-ph/9705337}{{\tt hep-ph/9705337}}].

\bibitem{Choudhury:2012np}
D.~Choudhury and P.~Saha, {\em JHEP} {\bf 08} (2012) 144,
  [\href{http://xxx.lanl.gov/abs/1201.4130}{{\tt 1201.4130}}].

\bibitem{Degrande:2012gr}
C.~Degrande, J.~M. Gerard, C.~Grojean, F.~Maltoni, and G.~Servant, {\em JHEP}
  {\bf 07} (2012) 036, [\href{http://xxx.lanl.gov/abs/1205.1065}{{\tt
  1205.1065}}]. [Erratum: JHEP03,032(2013)].

\bibitem{Giudice:2007fh}
G.~F. Giudice, C.~Grojean, A.~Pomarol, and R.~Rattazzi, {\em JHEP} {\bf 06}
  (2007) 045, [\href{http://xxx.lanl.gov/abs/hep-ph/0703164}{{\tt
  hep-ph/0703164}}].

\bibitem{Callan:1970yg}
C.~G. Callan, Jr., {\em Phys. Rev.} {\bf D2} (1970) 1541--1547.

\bibitem{Symanzik:1970rt}
K.~Symanzik, {\em Commun. Math. Phys.} {\bf 18} (1970) 227--246.

\bibitem{Coleman:1970je}
S.~R. Coleman and R.~Jackiw, {\em Annals Phys.} {\bf 67} (1971) 552--598.

\bibitem{Crewther:1972kn}
R.~J. Crewther, {\em Phys. Rev. Lett.} {\bf 28} (1972) 1421.

\bibitem{Chanowitz:1972vd}
M.~S. Chanowitz and J.~R. Ellis, {\em Phys. Lett.} {\bf B40} (1972) 397.

\bibitem{Chanowitz:1972da}
M.~S. Chanowitz and J.~R. Ellis, {\em Phys. Rev.} {\bf D7} (1973) 2490--2506.

\bibitem{Spira:1995mt}
M.~Spira, \href{http://xxx.lanl.gov/abs/hep-ph/9510347}{{\tt hep-ph/9510347}}.

\bibitem{Spira:1996if}
M.~Spira, {\em Nucl. Instrum. Meth.} {\bf A389} (1997) 357--360,
  [\href{http://xxx.lanl.gov/abs/hep-ph/9610350}{{\tt hep-ph/9610350}}].

\bibitem{Djouadi:1997yw}
A.~Djouadi, J.~Kalinowski, and M.~Spira, {\em Comput. Phys. Commun.} {\bf 108}
  (1998) 56--74, [\href{http://xxx.lanl.gov/abs/hep-ph/9704448}{{\tt
  hep-ph/9704448}}].

\bibitem{Djouadi:2006bz}
A.~Djouadi, M.~M. Muhlleitner, and M.~Spira, {\em Acta Phys. Polon.} {\bf B38}
  (2007) 635--644, [\href{http://xxx.lanl.gov/abs/hep-ph/0609292}{{\tt
  hep-ph/0609292}}].

\bibitem{Denner:2015hwgsmp}
A.~Denner {\em et.~al.},
  \href{http://xxx.lanl.gov/abs/LHCHXSWG-INT-2015-006}{{\tt
  LHCHXSWG-INT-2015-006}}.

\bibitem{CMS:yva}
{\bf CMS} Collaboration, \href{http://xxx.lanl.gov/abs/CMS-PAS-HIG-13-005}{{\tt
  CMS-PAS-HIG-13-005}}.

\bibitem{ATLAS:2013sla}
{\bf ATLAS} Collaboration,
  \href{http://xxx.lanl.gov/abs/ATLAS-CONF-2013-034}{{\tt
  ATLAS-CONF-2013-034}}.

\bibitem{Aad:2015mxa}
{\bf ATLAS} Collaboration, {\em Eur. Phys. J.} {\bf C75} (2015), no.~10 476,
  [\href{http://xxx.lanl.gov/abs/1506.05669}{{\tt 1506.05669}}].

\bibitem{Khachatryan:2014kca}
{\bf CMS} Collaboration, {\em Phys. Rev.} {\bf D92} (2015), no.~1 012004,
  [\href{http://xxx.lanl.gov/abs/1411.3441}{{\tt 1411.3441}}].

\bibitem{Aad:2015gba}
{\bf ATLAS} Collaboration, {\em Eur. Phys. J.} {\bf C76} (2016), no.~1 6,
  [\href{http://xxx.lanl.gov/abs/1507.04548}{{\tt 1507.04548}}].

\bibitem{Burges:1983zg}
C.~J.~C. Burges and H.~J. Schnitzer, {\em Nucl. Phys.} {\bf B228} (1983) 464.

\bibitem{Buchmuller:1985jz}
W.~Buchmuller and D.~Wyler, {\em Nucl. Phys.} {\bf B268} (1986) 621--653.

\bibitem{Hagiwara:1993ck}
K.~Hagiwara, S.~Ishihara, R.~Szalapski, and D.~Zeppenfeld, {\em Phys. Rev.}
  {\bf D48} (1993) 2182--2203.

\bibitem{Grzadkowski:2010es}
B.~Grzadkowski, M.~Iskrzynski, M.~Misiak, and J.~Rosiek, {\em JHEP} {\bf 10}
  (2010) 085, [\href{http://xxx.lanl.gov/abs/1008.4884}{{\tt 1008.4884}}].

\bibitem{Terazawa:1976xx}
H.~Terazawa, K.~Akama, and Y.~Chikashige, {\em Phys. Rev.} {\bf D15} (1977)
  480.

\bibitem{Terazawa:1979pj}
H.~Terazawa, {\em Phys. Rev.} {\bf D22} (1980) 184.

\bibitem{Kaplan:1983fs}
D.~B. Kaplan and H.~Georgi, {\em Phys. Lett.} {\bf B136} (1984) 183.

\bibitem{Dimopoulos:1981xc}
S.~Dimopoulos and J.~Preskill, {\em Nucl. Phys.} {\bf B199} (1982) 206.

\bibitem{Banks:1984gj}
T.~Banks, {\em Nucl. Phys.} {\bf B243} (1984) 125.

\bibitem{Kaplan:1983sm}
D.~B. Kaplan, H.~Georgi, and S.~Dimopoulos, {\em Phys. Lett.} {\bf B136} (1984)
  187.

\bibitem{Georgi:1984ef}
H.~Georgi, D.~B. Kaplan, and P.~Galison, {\em Phys. Lett.} {\bf B143} (1984)
  152.

\bibitem{Georgi:1984af}
H.~Georgi and D.~B. Kaplan, {\em Phys. Lett.} {\bf B145} (1984) 216.

\bibitem{Grigo:2014jma}
J.~Grigo, K.~Melnikov, and M.~Steinhauser, {\em Nucl. Phys.} {\bf B888} (2014)
  17--29, [\href{http://xxx.lanl.gov/abs/1408.2422}{{\tt 1408.2422}}].

\bibitem{Harlander:2013oja}
R.~V. Harlander and T.~Neumann, {\em Phys. Rev.} {\bf D88} (2013) 074015,
  [\href{http://xxx.lanl.gov/abs/1308.2225}{{\tt 1308.2225}}].

\bibitem{Banfi:2013yoa}
A.~Banfi, A.~Martin, and V.~Sanz, {\em JHEP} {\bf 08} (2014) 053,
  [\href{http://xxx.lanl.gov/abs/1308.4771}{{\tt 1308.4771}}].

\bibitem{Azatov:2013xha}
A.~Azatov and A.~Paul, {\em JHEP} {\bf 01} (2014) 014,
  [\href{http://xxx.lanl.gov/abs/1309.5273}{{\tt 1309.5273}}].

\bibitem{Englert:2013vua}
C.~Englert, M.~McCullough, and M.~Spannowsky, {\em Phys. Rev.} {\bf D89}
  (2014), no.~1 013013, [\href{http://xxx.lanl.gov/abs/1310.4828}{{\tt
  1310.4828}}].

\bibitem{Grojean:2013nya}
C.~Grojean, E.~Salvioni, M.~Schlaffer, and A.~Weiler, {\em JHEP} {\bf 05}
  (2014) 022, [\href{http://xxx.lanl.gov/abs/1312.3317}{{\tt 1312.3317}}].

\bibitem{Schlaffer:2014osa}
M.~Schlaffer, M.~Spannowsky, M.~Takeuchi, A.~Weiler, and C.~Wymant, {\em Eur.
  Phys. J.} {\bf C74} (2014), no.~10 3120,
  [\href{http://xxx.lanl.gov/abs/1405.4295}{{\tt 1405.4295}}].

\bibitem{Buschmann:2014twa}
M.~Buschmann, C.~Englert, D.~Goncalves, T.~Plehn, and M.~Spannowsky, {\em Phys.
  Rev.} {\bf D90} (2014), no.~1 013010,
  [\href{http://xxx.lanl.gov/abs/1405.7651}{{\tt 1405.7651}}].

\bibitem{Buschmann:2014sia}
M.~Buschmann, D.~Goncalves, S.~Kuttimalai, M.~Schonherr, F.~Krauss, and
  T.~Plehn, {\em JHEP} {\bf 02} (2015) 038,
  [\href{http://xxx.lanl.gov/abs/1410.5806}{{\tt 1410.5806}}].

\bibitem{Langenegger:2015lra}
U.~Langenegger, M.~Spira, and I.~Strebel,
  \href{http://xxx.lanl.gov/abs/1507.01373}{{\tt 1507.01373}}.

\bibitem{Schmidt:1997wr}
C.~R. Schmidt, {\em Phys. Lett.} {\bf B413} (1997) 391--395,
  [\href{http://xxx.lanl.gov/abs/hep-ph/9707448}{{\tt hep-ph/9707448}}].

\bibitem{deFlorian:1999zd}
D.~de~Florian, M.~Grazzini, and Z.~Kunszt, {\em Phys. Rev. Lett.} {\bf 82}
  (1999) 5209--5212, [\href{http://xxx.lanl.gov/abs/hep-ph/9902483}{{\tt
  hep-ph/9902483}}].

\bibitem{Ravindran:2002dc}
V.~Ravindran, J.~Smith, and W.~L. Van~Neerven, {\em Nucl. Phys.} {\bf B634}
  (2002) 247--290, [\href{http://xxx.lanl.gov/abs/hep-ph/0201114}{{\tt
  hep-ph/0201114}}].

\bibitem{Glosser:2002gm}
C.~J. Glosser and C.~R. Schmidt, {\em JHEP} {\bf 12} (2002) 016,
  [\href{http://xxx.lanl.gov/abs/hep-ph/0209248}{{\tt hep-ph/0209248}}].

\bibitem{Anastasiou:2004xq}
C.~Anastasiou, K.~Melnikov, and F.~Petriello, {\em Phys. Rev. Lett.} {\bf 93}
  (2004) 262002, [\href{http://xxx.lanl.gov/abs/hep-ph/0409088}{{\tt
  hep-ph/0409088}}].

\bibitem{Anastasiou:2005qj}
C.~Anastasiou, K.~Melnikov, and F.~Petriello, {\em Nucl. Phys.} {\bf B724}
  (2005) 197--246, [\href{http://xxx.lanl.gov/abs/hep-ph/0501130}{{\tt
  hep-ph/0501130}}].

\bibitem{Dawson:2014ora}
S.~Dawson, I.~M. Lewis, and M.~Zeng, {\em Phys. Rev.} {\bf D90} (2014), no.~9
  093007, [\href{http://xxx.lanl.gov/abs/1409.6299}{{\tt 1409.6299}}].

\bibitem{Hamilton:2015nsa}
K.~Hamilton, P.~Nason, and G.~Zanderighi, {\em JHEP} {\bf 05} (2015) 140,
  [\href{http://xxx.lanl.gov/abs/1501.04637}{{\tt 1501.04637}}].

\bibitem{Harlander:2012hf}
R.~V. Harlander, T.~Neumann, K.~J. Ozeren, and M.~Wiesemann, {\em JHEP} {\bf
  08} (2012) 139, [\href{http://xxx.lanl.gov/abs/1206.0157}{{\tt 1206.0157}}].

\bibitem{Boughezal:2013uia}
R.~Boughezal, F.~Caola, K.~Melnikov, F.~Petriello, and M.~Schulze, {\em JHEP}
  {\bf 06} (2013) 072, [\href{http://xxx.lanl.gov/abs/1302.6216}{{\tt
  1302.6216}}].

\bibitem{Chen:2014gva}
X.~Chen, T.~Gehrmann, E.~W.~N. Glover, and M.~Jaquier, {\em Phys. Lett.} {\bf
  B740} (2015) 147--150, [\href{http://xxx.lanl.gov/abs/1408.5325}{{\tt
  1408.5325}}].

\bibitem{Boughezal:2015dra}
R.~Boughezal, F.~Caola, K.~Melnikov, F.~Petriello, and M.~Schulze, {\em Phys.
  Rev. Lett.} {\bf 115} (2015), no.~8 082003,
  [\href{http://xxx.lanl.gov/abs/1504.07922}{{\tt 1504.07922}}].

\bibitem{Boughezal:2015aha}
R.~Boughezal, C.~Focke, W.~Giele, X.~Liu, and F.~Petriello, {\em Phys. Lett.}
  {\bf B748} (2015) 5--8, [\href{http://xxx.lanl.gov/abs/1505.03893}{{\tt
  1505.03893}}].

\bibitem{Catani:1988vd}
S.~Catani, E.~D'Emilio, and L.~Trentadue, {\em Phys. Lett.} {\bf B211} (1988)
  335--342.

\bibitem{Hinchliffe:1988ap}
I.~Hinchliffe and S.~F. Novaes, {\em Phys. Rev.} {\bf D38} (1988) 3475--3480.

\bibitem{Kauffman:1991jt}
R.~P. Kauffman, {\em Phys. Rev.} {\bf D44} (1991) 1415--1425.

\bibitem{Kauffman:1991cx}
R.~P. Kauffman, {\em Phys. Rev.} {\bf D45} (1992) 1512--1517.

\bibitem{Balazs:2000wv}
C.~Balazs and C.~P. Yuan, {\em Phys. Lett.} {\bf B478} (2000) 192--198,
  [\href{http://xxx.lanl.gov/abs/hep-ph/0001103}{{\tt hep-ph/0001103}}].

\bibitem{deFlorian:2000pr}
D.~de~Florian and M.~Grazzini, {\em Phys. Rev. Lett.} {\bf 85} (2000)
  4678--4681, [\href{http://xxx.lanl.gov/abs/hep-ph/0008152}{{\tt
  hep-ph/0008152}}].

\bibitem{Catani:2000vq}
S.~Catani, D.~de~Florian, and M.~Grazzini, {\em Nucl. Phys.} {\bf B596} (2001)
  299--312, [\href{http://xxx.lanl.gov/abs/hep-ph/0008184}{{\tt
  hep-ph/0008184}}].

\bibitem{deFlorian:2001zd}
D.~de~Florian and M.~Grazzini, {\em Nucl. Phys.} {\bf B616} (2001) 247--285,
  [\href{http://xxx.lanl.gov/abs/hep-ph/0108273}{{\tt hep-ph/0108273}}].

\bibitem{Berger:2002ut}
E.~L. Berger and J.-w. Qiu, {\em Phys. Rev.} {\bf D67} (2003) 034026,
  [\href{http://xxx.lanl.gov/abs/hep-ph/0210135}{{\tt hep-ph/0210135}}].

\bibitem{Bozzi:2003jy}
G.~Bozzi, S.~Catani, D.~de~Florian, and M.~Grazzini, {\em Phys. Lett.} {\bf
  B564} (2003) 65--72, [\href{http://xxx.lanl.gov/abs/hep-ph/0302104}{{\tt
  hep-ph/0302104}}].

\bibitem{Kulesza:2003wi}
A.~Kulesza and W.~J. Stirling, {\em JHEP} {\bf 12} (2003) 056,
  [\href{http://xxx.lanl.gov/abs/hep-ph/0307208}{{\tt hep-ph/0307208}}].

\bibitem{Watt:2003vf}
G.~Watt, A.~D. Martin, and M.~G. Ryskin, {\em Phys. Rev.} {\bf D70} (2004)
  014012, [\href{http://xxx.lanl.gov/abs/hep-ph/0309096}{{\tt
  hep-ph/0309096}}]. [Erratum: Phys. Rev.D70,079902(2004)].

\bibitem{Kulesza:2003wn}
A.~Kulesza, G.~F. Sterman, and W.~Vogelsang, {\em Phys. Rev.} {\bf D69} (2004)
  014012, [\href{http://xxx.lanl.gov/abs/hep-ph/0309264}{{\tt
  hep-ph/0309264}}].

\bibitem{Gawron:2003np}
A.~Gawron and J.~Kwiecinski, {\em Phys. Rev.} {\bf D70} (2004) 014003,
  [\href{http://xxx.lanl.gov/abs/hep-ph/0309303}{{\tt hep-ph/0309303}}].

\bibitem{Lipatov:2005at}
A.~V. Lipatov and N.~P. Zotov, {\em Eur. Phys. J.} {\bf C44} (2005) 559--566,
  [\href{http://xxx.lanl.gov/abs/hep-ph/0501172}{{\tt hep-ph/0501172}}].

\bibitem{Bozzi:2005wk}
G.~Bozzi, S.~Catani, D.~de~Florian, and M.~Grazzini, {\em Nucl. Phys.} {\bf
  B737} (2006) 73--120, [\href{http://xxx.lanl.gov/abs/hep-ph/0508068}{{\tt
  hep-ph/0508068}}].

\bibitem{Bozzi:2007pn}
G.~Bozzi, S.~Catani, D.~de~Florian, and M.~Grazzini, {\em Nucl. Phys.} {\bf
  B791} (2008) 1--19, [\href{http://xxx.lanl.gov/abs/0705.3887}{{\tt
  0705.3887}}].

\bibitem{deFlorian:2011xf}
D.~de~Florian, G.~Ferrera, M.~Grazzini, and D.~Tommasini, {\em JHEP} {\bf 11}
  (2011) 064, [\href{http://xxx.lanl.gov/abs/1109.2109}{{\tt 1109.2109}}].

\bibitem{Neill:2015roa}
D.~Neill, I.~Z. Rothstein, and V.~Vaidya, {\em JHEP} {\bf 12} (2015) 097,
  [\href{http://xxx.lanl.gov/abs/1503.00005}{{\tt 1503.00005}}].

\bibitem{Alwall:2011cy}
J.~Alwall, Q.~Li, and F.~Maltoni, {\em Phys. Rev.} {\bf D85} (2012) 014031,
  [\href{http://xxx.lanl.gov/abs/1110.1728}{{\tt 1110.1728}}].

\bibitem{Bagnaschi:2011tu}
E.~Bagnaschi, G.~Degrassi, P.~Slavich, and A.~Vicini, {\em JHEP} {\bf 02}
  (2012) 088, [\href{http://xxx.lanl.gov/abs/1111.2854}{{\tt 1111.2854}}].

\bibitem{Mantler:2012bj}
H.~Mantler and M.~Wiesemann, {\em Eur. Phys. J.} {\bf C73} (2013), no.~6 2467,
  [\href{http://xxx.lanl.gov/abs/1210.8263}{{\tt 1210.8263}}].

\bibitem{Grazzini:2015gdl}
M.~Grazzini, A.~Ilnicka, M.~Spira, and M.~Wiesemann, in {\em {Proceedings, 2015
  European Physical Society Conference on High Energy Physics (EPS-HEP 2015)}},
  2015.
\newblock \href{http://xxx.lanl.gov/abs/1511.08059}{{\tt 1511.08059}}.

\bibitem{Djouadi:1999rca}
A.~Djouadi, W.~Kilian, M.~Muhlleitner, and P.~M. Zerwas, {\em Eur. Phys. J.}
  {\bf C10} (1999) 45--49, [\href{http://xxx.lanl.gov/abs/hep-ph/9904287}{{\tt
  hep-ph/9904287}}].

\bibitem{Baglio:2012np}
J.~Baglio, A.~Djouadi, R.~Grober, M.~M. Muhlleitner, J.~Quevillon, and
  M.~Spira, {\em JHEP} {\bf 04} (2013) 151,
  [\href{http://xxx.lanl.gov/abs/1212.5581}{{\tt 1212.5581}}].

\bibitem{Glover:1987nx}
E.~N. Glover and J.~van~der Bij, {\em Nucl.Phys.} {\bf B309} (1988) 282.

\bibitem{Plehn:1996wb}
T.~Plehn, M.~Spira, and P.~M. Zerwas, {\em Nucl. Phys.} {\bf B479} (1996)
  46--64, [\href{http://xxx.lanl.gov/abs/hep-ph/9603205}{{\tt
  hep-ph/9603205}}]. [Erratum: Nucl. Phys.B531,655(1998)].

\bibitem{Dawson:1998py}
S.~Dawson, S.~Dittmaier, and M.~Spira, {\em Phys. Rev.} {\bf D58} (1998)
  115012, [\href{http://xxx.lanl.gov/abs/hep-ph/9805244}{{\tt
  hep-ph/9805244}}].

\bibitem{Grigo:2013rya}
J.~Grigo, J.~Hoff, K.~Melnikov, and M.~Steinhauser, {\em Nucl. Phys.} {\bf
  B875} (2013) 1--17, [\href{http://xxx.lanl.gov/abs/1305.7340}{{\tt
  1305.7340}}].

\bibitem{Frederix:2014hta}
R.~Frederix, S.~Frixione, V.~Hirschi, F.~Maltoni, O.~Mattelaer, P.~Torrielli,
  E.~Vryonidou, and M.~Zaro, {\em Phys. Lett.} {\bf B732} (2014) 142--149,
  [\href{http://xxx.lanl.gov/abs/1401.7340}{{\tt 1401.7340}}].

\bibitem{Maltoni:2014eza}
F.~Maltoni, E.~Vryonidou, and M.~Zaro,
  \href{http://xxx.lanl.gov/abs/1408.6542}{{\tt 1408.6542}}.

\bibitem{deFlorian:2013uza}
D.~de~Florian and J.~Mazzitelli, {\em Phys. Lett.} {\bf B724} (2013) 306--309,
  [\href{http://xxx.lanl.gov/abs/1305.5206}{{\tt 1305.5206}}].

\bibitem{deFlorian:2013jea}
D.~de~Florian and J.~Mazzitelli, {\em Phys. Rev. Lett.} {\bf 111} (2013)
  201801, [\href{http://xxx.lanl.gov/abs/1309.6594}{{\tt 1309.6594}}].

\bibitem{Grigo:2015dia}
J.~Grigo, J.~Hoff, and M.~Steinhauser, {\em Nucl. Phys.} {\bf B900} (2015)
  412--430, [\href{http://xxx.lanl.gov/abs/1508.00909}{{\tt 1508.00909}}].

\bibitem{Shao:2013bz}
D.~Y. Shao, C.~S. Li, H.~T. Li, and J.~Wang, {\em JHEP} {\bf 07} (2013) 169,
  [\href{http://xxx.lanl.gov/abs/1301.1245}{{\tt 1301.1245}}].

\bibitem{deFlorian:2015moa}
D.~de~Florian and J.~Mazzitelli, {\em JHEP} {\bf 09} (2015) 053,
  [\href{http://xxx.lanl.gov/abs/1505.07122}{{\tt 1505.07122}}].

\bibitem{Grober:2015cwa}
R.~Grober, M.~Muhlleitner, M.~Spira, and J.~Streicher, {\em JHEP} {\bf 09}
  (2015) 092, [\href{http://xxx.lanl.gov/abs/1504.06577}{{\tt 1504.06577}}].

\bibitem{Grober:2016wmf}
R.~Grober, M.~Muhlleitner, and M.~Spira,
  \href{http://xxx.lanl.gov/abs/1602.05851}{{\tt 1602.05851}}.

\bibitem{Gorbahn:2015gxa}
M.~Gorbahn, J.~M. No, and V.~Sanz, {\em JHEP} {\bf 10} (2015) 036,
  [\href{http://xxx.lanl.gov/abs/1502.07352}{{\tt 1502.07352}}].

\bibitem{Brehmer:2015rna}
J.~Brehmer, A.~Freitas, D.~Lopez-Val, and T.~Plehn,
  \href{http://xxx.lanl.gov/abs/1510.03443}{{\tt 1510.03443}}.

\bibitem{Biekotter:2016ecg}
A.~Biekotter, J.~Brehmer, and T.~Plehn,
  \href{http://xxx.lanl.gov/abs/1602.05202}{{\tt 1602.05202}}.

\bibitem{Maltoni:2013sma}
F.~Maltoni, K.~Mawatari, and M.~Zaro, {\em Eur. Phys. J.} {\bf C74} (2014),
  no.~1 2710, [\href{http://xxx.lanl.gov/abs/1311.1829}{{\tt 1311.1829}}].

\bibitem{Mimasu:2015nqa}
K.~Mimasu, V.~Sanz, and C.~Williams,
  \href{http://xxx.lanl.gov/abs/1512.02572}{{\tt 1512.02572}}.

\bibitem{Contino:2013kra}
R.~Contino, M.~Ghezzi, C.~Grojean, M.~Muhlleitner, and M.~Spira, {\em JHEP}
  {\bf 07} (2013) 035, [\href{http://xxx.lanl.gov/abs/1303.3876}{{\tt
  1303.3876}}].

\bibitem{Biedermann:2016yvs}
B.~Biedermann, A.~Denner, S.~Dittmaier, L.~Hofer, and B.~Jager,
  \href{http://xxx.lanl.gov/abs/1601.07787}{{\tt 1601.07787}}.

\bibitem{Dolan:2012ac}
M.~J. Dolan, C.~Englert, and M.~Spannowsky, {\em Phys. Rev.} {\bf D87} (2013)
  055002, [\href{http://xxx.lanl.gov/abs/1210.8166}{{\tt 1210.8166}}].

\bibitem{Aad:2015uka}
{\bf ATLAS} Collaboration, {\em Eur. Phys. J.} {\bf C75} (2015) 412,
  [\href{http://xxx.lanl.gov/abs/1506.00285}{{\tt 1506.00285}}].

\bibitem{Khachatryan:2015yea}
{\bf CMS} Collaboration, {\em Phys. Lett.} {\bf B749} (2015) 560--582,
  [\href{http://xxx.lanl.gov/abs/1503.04114}{{\tt 1503.04114}}].

\bibitem{Khachatryan:2016cfa}
{\bf CMS} Collaboration, \href{http://xxx.lanl.gov/abs/1602.08762}{{\tt
  1602.08762}}. Submitted to EPJC.

\bibitem{Khachatryan:2015tha}
{\bf CMS} Collaboration, {\em Phys. Lett.} {\bf B755} (2016) 217--244,
  [\href{http://xxx.lanl.gov/abs/1510.01181}{{\tt 1510.01181}}].

\bibitem{CMS:2015zug}
{\bf CMS} Collaboration, Tech. Rep. CMS-PAS-EXO-15-008, CERN, Geneva, 2015.

\bibitem{CMS:2014ipa}
V.~Khachatryan {\em et.~al.},, {\bf CMS} Collaboration,
  \href{http://xxx.lanl.gov/abs/1603.06896}{{\tt 1603.06896}}.

\bibitem{Agashe:2007zd}
K.~Agashe, H.~Davoudiasl, G.~Perez, and A.~Soni, {\em Phys. Rev. D} {\bf 76}
  (2007) 036006, [\href{http://xxx.lanl.gov/abs/hep-ph/0701186}{{\tt
  hep-ph/0701186}}].

\bibitem{Dominici:2002jv}
D.~Dominici, B.~Grzadkowski, J.~F. Gunion, and M.~Toharia, {\em Nucl. Phys. B}
  {\bf 671} (2003) 243, [\href{http://xxx.lanl.gov/abs/hep-ph/0206192}{{\tt
  hep-ph/0206192}}].

\bibitem{Randall:1999ee}
L.~Randall and R.~Sundrum, {\em Phys. Rev. Lett.} {\bf 83} (1999) 3370,
  [\href{http://xxx.lanl.gov/abs/hep-ph/9905221}{{\tt hep-ph/9905221}}].

\bibitem{Grober:2010yv}
R.~Grober and M.~Muhlleitner, {\em JHEP} {\bf 06} (2011) 020,
  [\href{http://xxx.lanl.gov/abs/1012.1562}{{\tt 1012.1562}}].

\bibitem{Contino:2012xk}
R.~Contino, M.~Ghezzi, M.~Moretti, G.~Panico, F.~Piccinini, and A.~Wulzer, {\em
  JHEP} {\bf 08} (2012) 154, [\href{http://xxx.lanl.gov/abs/1205.5444}{{\tt
  1205.5444}}].

\bibitem{Dall'Osso:2015aia}
M.~Dall'Osso, T.~Dorigo, C.~A. Gottardo, A.~Oliveira, M.~Tosi, and F.~Goertz,
  \href{http://xxx.lanl.gov/abs/1507.02245}{{\tt 1507.02245}}.

\bibitem{Antoniadis:1990ew}
I.~Antoniadis, {\em Phys. Lett.} {\bf B246} (1990) 377--384.

\bibitem{Kaplan:1991dc}
D.~B. Kaplan, {\em Nucl. Phys.} {\bf B365} (1991) 259--278.

\bibitem{AguilarSaavedra:2002kr}
J.~A. Aguilar-Saavedra, {\em Phys. Rev.} {\bf D67} (2003) 035003,
  [\href{http://xxx.lanl.gov/abs/hep-ph/0210112}{{\tt hep-ph/0210112}}].
  [Erratum: Phys. Rev.D69,099901(2004)].

\bibitem{DeSimone:2012fs}
A.~De~Simone, O.~Matsedonskyi, R.~Rattazzi, and A.~Wulzer, {\em JHEP} {\bf 04}
  (2013) 004, [\href{http://xxx.lanl.gov/abs/1211.5663}{{\tt 1211.5663}}].

\bibitem{Contino:2006qr}
R.~Contino, L.~Da~Rold, and A.~Pomarol, {\em Phys. Rev.} {\bf D75} (2007)
  055014, [\href{http://xxx.lanl.gov/abs/hep-ph/0612048}{{\tt
  hep-ph/0612048}}].

\bibitem{Anastasiou:2009rv}
C.~Anastasiou, E.~Furlan, and J.~Santiago, {\em Phys. Rev.} {\bf D79} (2009)
  075003, [\href{http://xxx.lanl.gov/abs/0901.2117}{{\tt 0901.2117}}].

\bibitem{Dissertori:2010ug}
G.~Dissertori, E.~Furlan, F.~Moortgat, and P.~Nef, {\em JHEP} {\bf 09} (2010)
  019, [\href{http://xxx.lanl.gov/abs/1005.4414}{{\tt 1005.4414}}].

\bibitem{Matsedonskyi:2012ym}
O.~Matsedonskyi, G.~Panico, and A.~Wulzer, {\em JHEP} {\bf 01} (2013) 164,
  [\href{http://xxx.lanl.gov/abs/1204.6333}{{\tt 1204.6333}}].

\bibitem{Berger:2012ec}
J.~Berger, J.~Hubisz, and M.~Perelstein, {\em JHEP} {\bf 07} (2012) 016,
  [\href{http://xxx.lanl.gov/abs/1205.0013}{{\tt 1205.0013}}].

\bibitem{Vignaroli:2012nf}
N.~Vignaroli, {\em Phys. Rev.} {\bf D86} (2012) 075017,
  [\href{http://xxx.lanl.gov/abs/1207.0830}{{\tt 1207.0830}}].

\bibitem{Grojean:2013qca}
C.~Grojean, O.~Matsedonskyi, and G.~Panico, {\em JHEP} {\bf 10} (2013) 160,
  [\href{http://xxx.lanl.gov/abs/1306.4655}{{\tt 1306.4655}}].

\bibitem{Matsedonskyi:2015dns}
O.~Matsedonskyi, G.~Panico, and A.~Wulzer,
  \href{http://xxx.lanl.gov/abs/1512.04356}{{\tt 1512.04356}}.

\bibitem{delAguila:2000rc}
F.~del Aguila, M.~Perez-Victoria, and J.~Santiago, {\em JHEP} {\bf 09} (2000)
  011, [\href{http://xxx.lanl.gov/abs/hep-ph/0007316}{{\tt hep-ph/0007316}}].

\bibitem{Cacciapaglia:2015ixa}
G.~Cacciapaglia, A.~Deandrea, N.~Gaur, D.~Harada, Y.~Okada, and L.~Panizzi,
  {\em JHEP} {\bf 09} (2015) 012,
  [\href{http://xxx.lanl.gov/abs/1502.00370}{{\tt 1502.00370}}].

\bibitem{Ishiwata:2015cga}
K.~Ishiwata, Z.~Ligeti, and M.~B. Wise, {\em JHEP} {\bf 10} (2015) 027,
  [\href{http://xxx.lanl.gov/abs/1506.03484}{{\tt 1506.03484}}].

\bibitem{Atre:2011ae}
A.~Atre, G.~Azuelos, M.~Carena, T.~Han, E.~Ozcan, J.~Santiago, and G.~Unel,
  {\em JHEP} {\bf 08} (2011) 080,
  [\href{http://xxx.lanl.gov/abs/1102.1987}{{\tt 1102.1987}}].

\bibitem{Cacciapaglia:2011fx}
G.~Cacciapaglia, A.~Deandrea, L.~Panizzi, N.~Gaur, D.~Harada, and Y.~Okada,
  {\em JHEP} {\bf 03} (2012) 070,
  [\href{http://xxx.lanl.gov/abs/1108.6329}{{\tt 1108.6329}}].

\bibitem{Ellis:2014dza}
S.~A.~R. Ellis, R.~M. Godbole, S.~Gopalakrishna, and J.~D. Wells, {\em JHEP}
  {\bf 09} (2014) 130, [\href{http://xxx.lanl.gov/abs/1404.4398}{{\tt
  1404.4398}}].

\bibitem{Bizot:2015zaa}
N.~Bizot and M.~Frigerio, {\em JHEP} {\bf 01} (2016) 036,
  [\href{http://xxx.lanl.gov/abs/1508.01645}{{\tt 1508.01645}}].

\bibitem{Brooijmans:2014eja}
G.~Brooijmans {\em et.~al.}, \href{http://xxx.lanl.gov/abs/1405.1617}{{\tt
  1405.1617}}.

\bibitem{Buchkremer:2013bha}
M.~Buchkremer, G.~Cacciapaglia, A.~Deandrea, and L.~Panizzi, {\em Nucl. Phys.}
  {\bf B876} (2013) 376--417, [\href{http://xxx.lanl.gov/abs/1305.4172}{{\tt
  1305.4172}}].

\bibitem{Degrande:2014vpa}
C.~Degrande, {\em Comput. Phys. Commun.} {\bf 197} (2015) 239--262,
  [\href{http://xxx.lanl.gov/abs/1406.3030}{{\tt 1406.3030}}].

\bibitem{Degrande:2014sta}
C.~Degrande, B.~Fuks, V.~Hirschi, J.~Proudom, and H.-S. Shao, {\em Phys. Rev.}
  {\bf D91} (2015) 094005, [\href{http://xxx.lanl.gov/abs/1412.5589}{{\tt
  1412.5589}}].

\bibitem{Degrande:2015vaa}
C.~Degrande, B.~Fuks, V.~Hirschi, J.~Proudom, and H.-S. Shao, {\em Phys. Lett.}
  {\bf B755} (2016) 82--87, [\href{http://xxx.lanl.gov/abs/1510.00391}{{\tt
  1510.00391}}].

\bibitem{Ossola:2006us}
G.~Ossola, C.~G. Papadopoulos, and R.~Pittau, {\em Nucl. Phys.} {\bf B763}
  (2007) 147--169, [\href{http://xxx.lanl.gov/abs/hep-ph/0609007}{{\tt
  hep-ph/0609007}}].

\bibitem{Ossola:2007ax}
G.~Ossola, C.~G. Papadopoulos, and R.~Pittau, {\em JHEP} {\bf 03} (2008) 042,
  [\href{http://xxx.lanl.gov/abs/0711.3596}{{\tt 0711.3596}}].

\bibitem{Hirschi:2011pa}
V.~Hirschi, R.~Frederix, S.~Frixione, M.~V. Garzelli, F.~Maltoni, and
  R.~Pittau, {\em JHEP} {\bf 05} (2011) 044,
  [\href{http://xxx.lanl.gov/abs/1103.0621}{{\tt 1103.0621}}].

\bibitem{Frixione:1995ms}
S.~Frixione, Z.~Kunszt, and A.~Signer, {\em Nucl. Phys.} {\bf B467} (1996)
  399--442, [\href{http://xxx.lanl.gov/abs/hep-ph/9512328}{{\tt
  hep-ph/9512328}}].

\bibitem{Frederix:2009yq}
R.~Frederix, S.~Frixione, F.~Maltoni, and T.~Stelzer, {\em JHEP} {\bf 10}
  (2009) 003, [\href{http://xxx.lanl.gov/abs/0908.4272}{{\tt 0908.4272}}].

\bibitem{Frixione:2002ik}
S.~Frixione and B.~R. Webber, {\em JHEP} {\bf 06} (2002) 029,
  [\href{http://xxx.lanl.gov/abs/hep-ph/0204244}{{\tt hep-ph/0204244}}].

\bibitem{Ball:2014uwa}
R.~D. Ball {\em et.~al.},, {\bf NNPDF} Collaboration, {\em JHEP} {\bf 04}
  (2015) 040, [\href{http://xxx.lanl.gov/abs/1410.8849}{{\tt 1410.8849}}].

\bibitem{Buckley:2014ana}
A.~Buckley, J.~Ferrando, S.~Lloyd, K.~Nordström, B.~Page, M.~Rüfenacht,
  M.~Schönherr, and G.~Watt, {\em Eur. Phys. J.} {\bf C75} (2015) 132,
  [\href{http://xxx.lanl.gov/abs/1412.7420}{{\tt 1412.7420}}].

\bibitem{Demartin:2010er}
F.~Demartin, S.~Forte, E.~Mariani, J.~Rojo, and A.~Vicini, {\em Phys. Rev.}
  {\bf D82} (2010) 014002, [\href{http://xxx.lanl.gov/abs/1004.0962}{{\tt
  1004.0962}}].

\bibitem{Azatov:2012rj}
A.~Azatov, O.~Bondu, A.~Falkowski, M.~Felcini, S.~Gascon-Shotkin, D.~K. Ghosh,
  G.~Moreau, and S.~Sekmen, {\em Phys. Rev.} {\bf D85} (2012) 115022,
  [\href{http://xxx.lanl.gov/abs/1204.0455}{{\tt 1204.0455}}].

\bibitem{Backovic:2014ega}
M.~Backovi\'c, T.~Flacke, J.~H. Kim, and S.~J. Lee, {\em JHEP} {\bf 04} (2015)
  082, [\href{http://xxx.lanl.gov/abs/1410.8131}{{\tt 1410.8131}}].

\bibitem{Atre:2013ap}
A.~Atre, M.~Chala, and J.~Santiago, {\em JHEP} {\bf 05} (2013) 099,
  [\href{http://xxx.lanl.gov/abs/1302.0270}{{\tt 1302.0270}}].

\bibitem{Han:2010rf}
T.~Han, I.~Lewis, and Z.~Liu, {\em JHEP} {\bf 12} (2010) 085,
  [\href{http://xxx.lanl.gov/abs/1010.4309}{{\tt 1010.4309}}].

\bibitem{Chen:2008hh}
C.-R. Chen, W.~Klemm, V.~Rentala, and K.~Wang, {\em Phys. Rev.} {\bf D79}
  (2009) 054002, [\href{http://xxx.lanl.gov/abs/0811.2105}{{\tt 0811.2105}}].

\bibitem{Berger:2010fy}
E.~L. Berger, Q.-H. Cao, C.-R. Chen, G.~Shaughnessy, and H.~Zhang, {\em Phys.
  Rev. Lett.} {\bf 105} (2010) 181802,
  [\href{http://xxx.lanl.gov/abs/1005.2622}{{\tt 1005.2622}}].

\bibitem{Chen:2014haa}
C.-Y. Chen, A.~Freitas, T.~Han, and K.~S.~M. Lee, {\em JHEP} {\bf 05} (2015)
  135, [\href{http://xxx.lanl.gov/abs/1410.8113}{{\tt 1410.8113}}].

\bibitem{Christensen:2008py}
N.~D. Christensen and C.~Duhr, {\em Comput. Phys. Commun.} {\bf 180} (2009)
  1614--1641, [\href{http://xxx.lanl.gov/abs/0806.4194}{{\tt 0806.4194}}].

\bibitem{Ball:2013hta}
R.~D. Ball, V.~Bertone, S.~Carrazza, L.~Del~Debbio, S.~Forte, A.~Guffanti,
  N.~P. Hartland, and J.~Rojo,, {\bf NNPDF} Collaboration, {\em Nucl. Phys.}
  {\bf B877} (2013) 290--320, [\href{http://xxx.lanl.gov/abs/1308.0598}{{\tt
  1308.0598}}].

\bibitem{Misiak:2015xwa}
M.~Misiak {\em et.~al.}, {\em Phys. Rev. Lett.} {\bf 114} (2015), no.~22
  221801, [\href{http://xxx.lanl.gov/abs/1503.01789}{{\tt 1503.01789}}].

\bibitem{Gerard:2007kn}
J.~M. Gerard and M.~Herquet, {\em Phys. Rev. Lett.} {\bf 98} (2007) 251802,
  [\href{http://xxx.lanl.gov/abs/hep-ph/0703051}{{\tt hep-ph/0703051}}].

\bibitem{Grimus:2007if}
W.~Grimus, L.~Lavoura, O.~M. Ogreid, and P.~Osland, {\em J. Phys.} {\bf G35}
  (2008) 075001, [\href{http://xxx.lanl.gov/abs/0711.4022}{{\tt 0711.4022}}].

\bibitem{Baak:2014ora}
M.~Baak, J.~Cuth, J.~Haller, A.~Hoecker, R.~Kogler, K.~Moenig, M.~Schott, and
  J.~Stelzer,, {\bf Gfitter Group} Collaboration, {\em Eur. Phys. J.} {\bf C74}
  (2014) 3046, [\href{http://xxx.lanl.gov/abs/1407.3792}{{\tt 1407.3792}}].

\bibitem{Bernon:2015qea}
J.~Bernon, J.~F. Gunion, H.~E. Haber, Y.~Jiang, and S.~Kraml, {\em Phys. Rev.}
  {\bf D92} (2015), no.~7 075004,
  [\href{http://xxx.lanl.gov/abs/1507.00933}{{\tt 1507.00933}}].

\bibitem{Bernon:2015wef}
J.~Bernon, J.~F. Gunion, H.~E. Haber, Y.~Jiang, and S.~Kraml,
  \href{http://xxx.lanl.gov/abs/1511.03682}{{\tt 1511.03682}}.

\bibitem{Ferreira:2014naa}
P.~M. Ferreira, J.~F. Gunion, H.~E. Haber, and R.~Santos, {\em Phys. Rev.} {\bf
  D89} (2014), no.~11 115003, [\href{http://xxx.lanl.gov/abs/1403.4736}{{\tt
  1403.4736}}].

\bibitem{Dorsch:2016tab}
G.~C. Dorsch, S.~J. Huber, K.~Mimasu, and J.~M. No,
  \href{http://xxx.lanl.gov/abs/1601.04545}{{\tt 1601.04545}}.

\bibitem{Bernon:2014nxa}
J.~Bernon, J.~F. Gunion, Y.~Jiang, and S.~Kraml, {\em Phys. Rev.} {\bf D91}
  (2015), no.~7 075019, [\href{http://xxx.lanl.gov/abs/1412.3385}{{\tt
  1412.3385}}].

\bibitem{CMS:2015mba}
{\bf CMS} Collaboration, CMS-PAS-HIG-15-001.

\bibitem{Domingo:2008rr}
F.~Domingo, U.~Ellwanger, E.~Fullana, C.~Hugonie, and M.-A. Sanchis-Lozano,
  {\em JHEP} {\bf 01} (2009) 061,
  [\href{http://xxx.lanl.gov/abs/0810.4736}{{\tt 0810.4736}}].

\bibitem{Chatrchyan:2012am}
{\bf CMS} Collaboration, {\em Phys. Rev. Lett.} {\bf 109} (2012) 121801,
  [\href{http://xxx.lanl.gov/abs/1206.6326}{{\tt 1206.6326}}].

\bibitem{Coleppa:2014hxa}
B.~Coleppa, F.~Kling, and S.~Su, {\em JHEP} {\bf 09} (2014) 161,
  [\href{http://xxx.lanl.gov/abs/1404.1922}{{\tt 1404.1922}}].

\bibitem{Dorsch:2014qja}
G.~C. Dorsch, S.~J. Huber, K.~Mimasu, and J.~M. No, {\em Phys. Rev. Lett.} {\bf
  113} (2014), no.~21 211802, [\href{http://xxx.lanl.gov/abs/1405.5537}{{\tt
  1405.5537}}].

\bibitem{Khachatryan:2015baw}
{\bf CMS} Collaboration, \href{http://xxx.lanl.gov/abs/1511.03610}{{\tt
  1511.03610}}.

\bibitem{Khachatryan:2014wca}
{\bf CMS} Collaboration, {\em JHEP} {\bf 10} (2014) 160,
  [\href{http://xxx.lanl.gov/abs/1408.3316}{{\tt 1408.3316}}].

\bibitem{Aad:2014vgg}
{\bf ATLAS} Collaboration, {\em JHEP} {\bf 11} (2014) 056,
  [\href{http://xxx.lanl.gov/abs/1409.6064}{{\tt 1409.6064}}].

\bibitem{Barate:2003sz}
R.~Barate {\em et.~al.},, {\bf OPAL, DELPHI, LEP Working Group for Higgs boson
  searches, ALEPH, L3} Collaboration, {\em Phys. Lett.} {\bf B565} (2003)
  61--75, [\href{http://xxx.lanl.gov/abs/hep-ex/0306033}{{\tt
  hep-ex/0306033}}].

\bibitem{Dumont:2014kna}
B.~Dumont, J.~F. Gunion, Y.~Jiang, and S.~Kraml,
  \href{http://xxx.lanl.gov/abs/1409.4088}{{\tt 1409.4088}}.

\bibitem{Bernon:2014vta}
J.~Bernon, B.~Dumont, and S.~Kraml, {\em Phys. Rev.} {\bf D90} (2014) 071301,
  [\href{http://xxx.lanl.gov/abs/1409.1588}{{\tt 1409.1588}}].

\bibitem{Georgi:1985nv}
H.~Georgi and M.~Machacek, {\em Nucl.Phys.} {\bf B262} (1985) 463.

\bibitem{Cort:2013foa}
L.~Cort, M.~Garcia, and M.~Quiros, {\em Phys.Rev.} {\bf D88} (2013) 075010,
  [\href{http://xxx.lanl.gov/abs/1308.4025}{{\tt 1308.4025}}].

\bibitem{Garcia-Pepin:2014yfa}
M.~Garcia-Pepin, S.~Gori, M.~Quiros, R.~Vega, R.~Vega-Morales, and T.-T. Yu,
  {\em Phys. Rev.} {\bf D91} (2015), no.~1 015016,
  [\href{http://xxx.lanl.gov/abs/1409.5737}{{\tt 1409.5737}}].

\bibitem{Chanowitz:1985ug}
M.~S. Chanowitz and M.~Golden, {\em Phys.Lett.} {\bf B165} (1985) 105.

\bibitem{Gunion:1989ci}
J.~Gunion, R.~Vega, and J.~Wudka, {\em Phys.Rev.} {\bf D42} (1990) 1673--1691.

\bibitem{Gunion:1990dt}
J.~Gunion, R.~Vega, and J.~Wudka, {\em Phys.Rev.} {\bf D43} (1991) 2322--2336.

\bibitem{Aoki:2007ah}
M.~Aoki and S.~Kanemura, {\em Phys.Rev.} {\bf D77} (2008) 095009,
  [\href{http://xxx.lanl.gov/abs/0712.4053}{{\tt 0712.4053}}].

\bibitem{Chiang:2012cn}
C.-W. Chiang and K.~Yagyu, {\em JHEP} {\bf 01} (2013) 026,
  [\href{http://xxx.lanl.gov/abs/1211.2658}{{\tt 1211.2658}}].

\bibitem{Chiang:2013rua}
C.-W. Chiang, A.-L. Kuo, and K.~Yagyu, {\em JHEP} {\bf 10} (2013) 072,
  [\href{http://xxx.lanl.gov/abs/1307.7526}{{\tt 1307.7526}}].

\bibitem{Hartling:2014zca}
K.~Hartling, K.~Kumar, and H.~E. Logan, {\em Phys.Rev.} {\bf D90} (2014)
  015007, [\href{http://xxx.lanl.gov/abs/1404.2640}{{\tt 1404.2640}}].

\bibitem{Logan:2015xpa}
H.~E. Logan and V.~Rentala, {\em Phys. Rev.} {\bf D92} (2015) 075011,
  [\href{http://xxx.lanl.gov/abs/1502.01275}{{\tt 1502.01275}}].

\bibitem{Delgado:2015aha}
A.~Delgado, M.~Garcia-Pepin, B.~Ostdiek, and M.~Quiros, {\em Phys. Rev.} {\bf
  D92} (2015), no.~1 015011, [\href{http://xxx.lanl.gov/abs/1504.02486}{{\tt
  1504.02486}}].

\bibitem{Delgado:2015bwa}
A.~Delgado, M.~Garcia-Pepin, and M.~Quiros, {\em JHEP} {\bf 08} (2015) 159,
  [\href{http://xxx.lanl.gov/abs/1505.07469}{{\tt 1505.07469}}].

\bibitem{Delgado:2015hmy}
A.~Delgado, M.~Garcia-Pepin, and M.~Quiros, in {\em {18th International
  Conference From the Planck Scale to the Electroweak Scale (Planck 2015)
  Ioannina, Greece, May 25-29, 2015}}, 2015.
\newblock \href{http://xxx.lanl.gov/abs/1511.03254}{{\tt 1511.03254}}.

\bibitem{Garcia-Pepin:2016hvs}
M.~Garcia-Pepin and M.~Quiros, \href{http://xxx.lanl.gov/abs/1602.01351}{{\tt
  1602.01351}}.

\bibitem{Delgado:2016arn}
A.~Delgado, M.~Garcia-Pepin, M.~Quiros, J.~Santiago, and R.~Vega-Morales,
  \href{http://xxx.lanl.gov/abs/1603.00962}{{\tt 1603.00962}}.

\bibitem{Akeroyd:2003xi}
A.~G. Akeroyd, M.~A. Diaz, and F.~J. Pacheco, {\em Phys. Rev.} {\bf D70} (2004)
  075002, [\href{http://xxx.lanl.gov/abs/hep-ph/0312231}{{\tt
  hep-ph/0312231}}].

\bibitem{Akeroyd:2005pr}
A.~G. Akeroyd, A.~Alves, M.~A. Diaz, and O.~J.~P. Eboli, {\em Eur. Phys. J.}
  {\bf C48} (2006) 147--157,
  [\href{http://xxx.lanl.gov/abs/hep-ph/0512077}{{\tt hep-ph/0512077}}].

\bibitem{Aaltonen:2016fnw}
T.~A. Aaltonen {\em et.~al.},, {\bf CDF} Collaboration,
  \href{http://xxx.lanl.gov/abs/1601.00401}{{\tt 1601.00401}}.

\bibitem{Eichten:1984eu}
E.~Eichten, I.~Hinchliffe, K.~D. Lane, and C.~Quigg, {\em Rev. Mod. Phys.} {\bf
  56} (1984) 579--707. [Addendum: Rev. Mod. Phys.58,1065(1986)].

\bibitem{Degrande:2015xnm}
C.~Degrande, K.~Hartling, H.~E. Logan, A.~D. Peterson, and M.~Zaro,
  \href{http://xxx.lanl.gov/abs/1512.01243}{{\tt 1512.01243}}.

\bibitem{Dittmaier:2011ti}
{LHC Higgs Cross Section Working Group}, S.~Dittmaier, C.~Mariotti,
  G.~Passarino, and R.~Tanaka~(Eds.), {\em CERN-2011-002} (CERN, Geneva, 2011)
  [\href{http://xxx.lanl.gov/abs/1101.0593}{{\tt 1101.0593}}].

\bibitem{Dittmaier:2012vm}
{LHC Higgs Cross Section Working Group}, S.~Dittmaier, C.~Mariotti,
  G.~Passarino, and R.~Tanaka~(Eds.), {\em CERN-2012-002} (CERN, Geneva, 2012)
  [\href{http://xxx.lanl.gov/abs/1201.3084}{{\tt 1201.3084}}].

\bibitem{Low:2010jp}
I.~Low and J.~Lykken, {\em JHEP} {\bf 1010} (2010) 053,
  [\href{http://xxx.lanl.gov/abs/1005.0872}{{\tt 1005.0872}}].

\bibitem{Akeroyd:2012ms}
A.~G. Akeroyd and S.~Moretti, {\em Phys. Rev.} {\bf D86} (2012) 035015,
  [\href{http://xxx.lanl.gov/abs/1206.0535}{{\tt 1206.0535}}].

\bibitem{Arhrib:2011vc}
A.~Arhrib, R.~Benbrik, M.~Chabab, G.~Moultaka, and L.~Rahili, {\em JHEP} {\bf
  04} (2012) 136, [\href{http://xxx.lanl.gov/abs/1112.5453}{{\tt 1112.5453}}].

\bibitem{Kanemura:2012rs}
S.~Kanemura and K.~Yagyu, {\em Phys. Rev.} {\bf D85} (2012) 115009,
  [\href{http://xxx.lanl.gov/abs/1201.6287}{{\tt 1201.6287}}].

\bibitem{Chen:2012jy}
Y.~Chen, N.~Tran, and R.~Vega-Morales, {\em JHEP} {\bf 1301} (2013) 182,
  [\href{http://xxx.lanl.gov/abs/1211.1959}{{\tt 1211.1959}}].

\bibitem{Chen:2013ejz}
Y.~Chen and R.~Vega-Morales, {\em JHEP} {\bf 1404} (2014) 057,
  [\href{http://xxx.lanl.gov/abs/1310.2893}{{\tt 1310.2893}}].

\bibitem{Chen:2014ona}
Y.~Chen, A.~Falkowski, I.~Low, and R.~Vega-Morales, {\em Phys.Rev.} {\bf D90}
  (2014) 113006, [\href{http://xxx.lanl.gov/abs/1405.6723}{{\tt 1405.6723}}].

\bibitem{Chen:2015rha}
Y.~Chen, D.~Stolarski, and R.~Vega-Morales, {\em Phys. Rev.} {\bf D92} (2015),
  no.~5 053003, [\href{http://xxx.lanl.gov/abs/1505.01168}{{\tt 1505.01168}}].

\bibitem{Aad:2014ioa}
{\bf ATLAS} Collaboration, {\em Phys. Rev. Lett.} {\bf 113} (2014), no.~17
  171801, [\href{http://xxx.lanl.gov/abs/1407.6583}{{\tt 1407.6583}}].

\bibitem{Khachatryan:2015cwa}
{\bf CMS} Collaboration, {\em JHEP} {\bf 10} (2015) 144,
  [\href{http://xxx.lanl.gov/abs/1504.00936}{{\tt 1504.00936}}].

\bibitem{Hartling:2014xma}
K.~Hartling, K.~Kumar, and H.~E. Logan,
  \href{http://xxx.lanl.gov/abs/1412.7387}{{\tt 1412.7387}}.

\bibitem{Hartling:2014aga}
K.~Hartling, K.~Kumar, and H.~E. Logan, {\em Phys. Rev.} {\bf D91} (2015),
  no.~1 015013, [\href{http://xxx.lanl.gov/abs/1410.5538}{{\tt 1410.5538}}].

\bibitem{Chiang:2015amq}
C.-W. Chiang, A.-L. Kuo, and T.~Yamada, {\em JHEP} {\bf 01} (2016) 120,
  [\href{http://xxx.lanl.gov/abs/1511.00865}{{\tt 1511.00865}}].

\bibitem{Fabbrichesi:2016alj}
M.~Fabbrichesi and A.~Urbano, \href{http://xxx.lanl.gov/abs/1601.02447}{{\tt
  1601.02447}}.

\bibitem{Chiang:2016ydx}
C.-W. Chiang and A.-L. Kuo, \href{http://xxx.lanl.gov/abs/1601.06394}{{\tt
  1601.06394}}.

\bibitem{Englert:2013zpa}
C.~Englert, E.~Re, and M.~Spannowsky, {\em Phys.Rev.} {\bf D87} (2013), no.~9
  095014, [\href{http://xxx.lanl.gov/abs/1302.6505}{{\tt 1302.6505}}].

\bibitem{Englert:2013wga}
C.~Englert, E.~Re, and M.~Spannowsky, {\em Phys.Rev.} {\bf D88} (2013) 035024,
  [\href{http://xxx.lanl.gov/abs/1306.6228}{{\tt 1306.6228}}].

\bibitem{Nelson:1986ki}
C.~A. Nelson, {\em Phys.Rev.} {\bf D37} (1988) 1220.

\bibitem{Soni:1993jc}
A.~Soni and R.~Xu, {\em Phys.Rev.} {\bf D48} (1993) 5259--5263,
  [\href{http://xxx.lanl.gov/abs/hep-ph/9301225}{{\tt hep-ph/9301225}}].

\bibitem{Chang:1993jy}
D.~Chang, W.-Y. Keung, and I.~Phillips, {\em Phys.Rev.} {\bf D48} (1993)
  3225--3234, [\href{http://xxx.lanl.gov/abs/hep-ph/9303226}{{\tt
  hep-ph/9303226}}].

\bibitem{Barger:1993wt}
V.~D. Barger, K.-m. Cheung, A.~Djouadi, B.~A. Kniehl, and P.~Zerwas, {\em
  Phys.Rev.} {\bf D49} (1994) 79--90,
  [\href{http://xxx.lanl.gov/abs/hep-ph/9306270}{{\tt hep-ph/9306270}}].

\bibitem{Arens:1994wd}
T.~Arens and L.~Sehgal, {\em Z.Phys.} {\bf C66} (1995) 89--94,
  [\href{http://xxx.lanl.gov/abs/hep-ph/9409396}{{\tt hep-ph/9409396}}].

\bibitem{Choi:2002jk}
S.~Choi, .~Miller, D.J., M.~Muhlleitner, and P.~Zerwas, {\em Phys.Lett.} {\bf
  B553} (2003) 61--71, [\href{http://xxx.lanl.gov/abs/hep-ph/0210077}{{\tt
  hep-ph/0210077}}].

\bibitem{Buszello:2002uu}
C.~Buszello, I.~Fleck, P.~Marquard, and J.~van~der Bij, {\em Eur.Phys.J.} {\bf
  C32} (2004) 209--219, [\href{http://xxx.lanl.gov/abs/hep-ph/0212396}{{\tt
  hep-ph/0212396}}].

\bibitem{Godbole:2007cn}
R.~M. Godbole, .~Miller, D.J., and M.~M. Muhlleitner, {\em JHEP} {\bf 0712}
  (2007) 031, [\href{http://xxx.lanl.gov/abs/0708.0458}{{\tt 0708.0458}}].

\bibitem{Kovalchuk:2008zz}
V.~Kovalchuk, {\em J.Exp.Theor.Phys.} {\bf 107} (2008) 774--786.

\bibitem{Cao:2009ah}
Q.-H. Cao, C.~Jackson, W.-Y. Keung, I.~Low, and J.~Shu, {\em Phys.Rev.} {\bf
  D81} (2010) 015010, [\href{http://xxx.lanl.gov/abs/0911.3398}{{\tt
  0911.3398}}].

\bibitem{Gao:2010qx}
Y.~Gao, A.~V. Gritsan, Z.~Guo, K.~Melnikov, M.~Schulze, {\em et.~al.}, {\em
  Phys.Rev.} {\bf D81} (2010) 075022,
  [\href{http://xxx.lanl.gov/abs/1001.3396}{{\tt 1001.3396}}].

\bibitem{DeRujula:2010ys}
A.~De~Rujula, J.~Lykken, M.~Pierini, C.~Rogan, and M.~Spiropulu, {\em
  Phys.Rev.} {\bf D82} (2010) 013003,
  [\href{http://xxx.lanl.gov/abs/1001.5300}{{\tt 1001.5300}}].

\bibitem{Gainer:2011xz}
J.~S. Gainer, K.~Kumar, I.~Low, and R.~Vega-Morales, {\em JHEP} {\bf 1111}
  (2011) 027, [\href{http://xxx.lanl.gov/abs/1108.2274}{{\tt 1108.2274}}].

\bibitem{Campbell:2012cz}
J.~M. Campbell, W.~T. Giele, and C.~Williams,
  \href{http://xxx.lanl.gov/abs/1204.4424}{{\tt 1204.4424}}.

\bibitem{Campbell:2012ct}
J.~M. Campbell, W.~T. Giele, and C.~Williams,
  \href{http://xxx.lanl.gov/abs/1205.3434}{{\tt 1205.3434}}.

\bibitem{Belyaev:2012qa}
A.~Belyaev, N.~D. Christensen, and A.~Pukhov,
  \href{http://xxx.lanl.gov/abs/1207.6082}{{\tt 1207.6082}}.

\bibitem{Coleppa:2012eh}
B.~Coleppa, K.~Kumar, and H.~E. Logan,
  \href{http://xxx.lanl.gov/abs/1208.2692}{{\tt 1208.2692}}.

\bibitem{Bolognesi:2012mm}
S.~Bolognesi, Y.~Gao, A.~V. Gritsan, K.~Melnikov, M.~Schulze, {\em et.~al.},
  \href{http://xxx.lanl.gov/abs/1208.4018}{{\tt 1208.4018}}.

\bibitem{Boughezal:2012tz}
R.~Boughezal, T.~J. LeCompte, and F.~Petriello,
  \href{http://xxx.lanl.gov/abs/1208.4311}{{\tt 1208.4311}}.

\bibitem{Stolarski:2012ps}
D.~Stolarski and R.~Vega-Morales, {\em Phys.Rev.} {\bf D86} (2012) 117504,
  [\href{http://xxx.lanl.gov/abs/1208.4840}{{\tt 1208.4840}}].

\bibitem{Avery:2012um}
P.~Avery, D.~Bourilkov, M.~Chen, T.~Cheng, A.~Drozdetskiy, {\em et.~al.},
  \href{http://xxx.lanl.gov/abs/1210.0896}{{\tt 1210.0896}}.

\bibitem{Modak:2013sb}
T.~Modak, D.~Sahoo, R.~Sinha, and H.-Y. Cheng,
  \href{http://xxx.lanl.gov/abs/1301.5404}{{\tt 1301.5404}}.

\bibitem{Gainer:2013rxa}
J.~S. Gainer, J.~Lykken, K.~T. Matchev, S.~Mrenna, and M.~Park, {\em
  Phys.Rev.Lett.} {\bf 111} (2013) 041801,
  [\href{http://xxx.lanl.gov/abs/1304.4936}{{\tt 1304.4936}}].

\bibitem{Grinstein:2013vsa}
B.~Grinstein, C.~W. Murphy, and D.~Pirtskhalava, {\em JHEP} {\bf 1310} (2013)
  077, [\href{http://xxx.lanl.gov/abs/1305.6938}{{\tt 1305.6938}}].

\bibitem{Sun:2013yra}
Y.~Sun, X.-F. Wang, and D.-N. Gao,
  \href{http://xxx.lanl.gov/abs/1309.4171}{{\tt 1309.4171}}.

\bibitem{Anderson:2013fba}
I.~Anderson, S.~Bolognesi, F.~Caola, Y.~Gao, A.~V. Gritsan, {\em et.~al.},
  \href{http://xxx.lanl.gov/abs/1309.4819}{{\tt 1309.4819}}.

\bibitem{Chen:2013waa}
M.~Chen, T.~Cheng, J.~S. Gainer, A.~Korytov, K.~T. Matchev, {\em et.~al.},
  \href{http://xxx.lanl.gov/abs/1310.1397}{{\tt 1310.1397}}.

\bibitem{Buchalla:2013mpa}
G.~Buchalla, O.~Cata, and G.~D'Ambrosio,
  \href{http://xxx.lanl.gov/abs/1310.2574}{{\tt 1310.2574}}.

\bibitem{Gainer:2014hha}
J.~S. Gainer, J.~Lykken, K.~T. Matchev, S.~Mrenna, and M.~Park,
  \href{http://xxx.lanl.gov/abs/1403.4951}{{\tt 1403.4951}}.

\bibitem{Chen:2014gka}
Y.~Chen, R.~Harnik, and R.~Vega-Morales, {\em Phys.Rev.Lett.} {\bf 113} (2014),
  no.~19 191801, [\href{http://xxx.lanl.gov/abs/1404.1336}{{\tt 1404.1336}}].

\bibitem{Chen:2014pia}
Y.~Chen, E.~Di~Marco, J.~Lykken, M.~Spiropulu, R.~Vega-Morales, {\em et.~al.},
  {\em JHEP} {\bf 1501} (2015) 125,
  [\href{http://xxx.lanl.gov/abs/1401.2077}{{\tt 1401.2077}}].

\bibitem{Chen:2015iha}
Y.~Chen, R.~Harnik, and R.~Vega-Morales, {\em JHEP} {\bf 09} (2015) 185,
  [\href{http://xxx.lanl.gov/abs/1503.05855}{{\tt 1503.05855}}].

\bibitem{Bhattacherjee:2015xra}
B.~Bhattacherjee, T.~Modak, S.~K. Patra, and R.~Sinha,
  \href{http://xxx.lanl.gov/abs/1503.08924}{{\tt 1503.08924}}.

\bibitem{Gonzalez-Alonso:2015bha}
M.~Gonzalez-Alonso, A.~Greljo, G.~Isidori, and D.~Marzocca,
  \href{http://xxx.lanl.gov/abs/1504.04018}{{\tt 1504.04018}}.

\bibitem{Artoisenet:2013puc}
P.~Artoisenet {\em et.~al.}, {\em JHEP} {\bf 11} (2013) 043,
  [\href{http://xxx.lanl.gov/abs/1306.6464}{{\tt 1306.6464}}].

\bibitem{ATLAS:2014aga}
{\bf ATLAS} Collaboration, {\em Phys. Rev.} {\bf D92} (2015), no.~1 012006,
  [\href{http://xxx.lanl.gov/abs/1412.2641}{{\tt 1412.2641}}].

\bibitem{Aad:2015rwa}
{\bf ATLAS} Collaboration, {\em Eur. Phys. J.} {\bf C75} (2015), no.~5 231,
  [\href{http://xxx.lanl.gov/abs/1503.03643}{{\tt 1503.03643}}].

\bibitem{HXSWGbasis}
{\bf LHC Higgs Cross Section Working Group 2},
  \href{http://xxx.lanl.gov/abs/LHCHXSWG-INT-2015-001}{{\tt
  LHCHXSWG-INT-2015-001}}.

\bibitem{Gupta:2014rxa}
R.~S. Gupta, A.~Pomarol, and F.~Riva, {\em Phys.Rev.} {\bf D91} (2015), no.~3
  035001, [\href{http://xxx.lanl.gov/abs/1405.0181}{{\tt 1405.0181}}].

\bibitem{Falkowski:2015fla}
A.~Falkowski, \href{http://xxx.lanl.gov/abs/1505.00046}{{\tt 1505.00046}}.

\bibitem{Chen:2014hqs}
Y.~Chen, E.~Di~Marco, J.~Lykken, M.~Spiropulu, R.~Vega-Morales, {\em et.~al.},
  \href{http://xxx.lanl.gov/abs/1410.4817}{{\tt 1410.4817}}.

\bibitem{Papucci:2014rja}
M.~Papucci, K.~Sakurai, A.~Weiler, and L.~Zeune, {\em Eur. Phys. J.} {\bf C74}
  (2014), no.~11 3163, [\href{http://xxx.lanl.gov/abs/1402.0492}{{\tt
  1402.0492}}].

\bibitem{Waugh:2006ip}
B.~M. Waugh, H.~Jung, A.~Buckley, L.~Lonnblad, J.~M. Butterworth, and E.~Nurse,
  in {\em {15th International Conference on Computing in High Energy and
  Nuclear Physics (CHEP 2006) Mumbai, Maharashtra, India, February 13-17,
  2006}}, 2006.
\newblock \href{http://xxx.lanl.gov/abs/hep-ph/0605034}{{\tt hep-ph/0605034}}.

\bibitem{Grasseau:2015qhg}
G.~Grasseau, D.~Chamont, F.~Beaudette, L.~Bianchini, O.~Davignon,
  L.~Mastrolorenzo, C.~Ochando, P.~Paganini, and T.~Strebler, {\em J. Phys.
  Conf. Ser.} {\bf 664} (2015), no.~9 092009.

\bibitem{Aad:2015baa}
{\bf ATLAS} Collaboration, {\em JHEP} {\bf 10} (2015) 134,
  [\href{http://xxx.lanl.gov/abs/1508.06608}{{\tt 1508.06608}}].

\bibitem{Sekmen:2011cz}
S.~Sekmen, S.~Kraml, J.~Lykken, F.~Moortgat, S.~Padhi, L.~Pape, M.~Pierini,
  H.~B. Prosper, and M.~Spiropulu, {\em JHEP} {\bf 02} (2012) 075,
  [\href{http://xxx.lanl.gov/abs/1109.5119}{{\tt 1109.5119}}].

\bibitem{Raja:1988fq}
R.~Raja, in {\em {Workshop on Detector Simulation for the SSC Argonne,
  Illinois, August 24-28, 1987}}, 1988.

\bibitem{Graf:1990ys}
{\bf D0} Collaboration,N.~A. Graf, , in {\em {International Conference on
  Calorimetry in High-energy Physics Batavia, Illinois, October 29-November 1,
  1990}}, 1990.

\bibitem{Sun:1996iw}
Q.~Sun, Tech. Rep. USIP-96-05, Stockholm Univ. Inst. Theor. Phys., Stockholm,
  May, 1996.

\bibitem{Arena:1996je}
V.~Arena, G.~Boca, G.~Bonomi, G.~Gianini, M.~Marchesotti, M.~Merlo, S.~P.
  Ratti, C.~Riccardi, L.~Viola, and P.~Vitulo, in {\em {Calorimetry in
  high-energy physics. Proceedings, 6th International Conference, Frascati,
  Italy, June 8-14, 1996}}, pp.~509--518, 1996.

\bibitem{Raicevic:2013vka}
N.~Raicevic, A.~Glazov, and A.~Zhokin, {\em Nucl. Instrum. Meth.} {\bf A718}
  (2013) 104--106.

\bibitem{Knuteson:2004nj}
B.~Knuteson, {\em Nucl. Instrum. Meth.} {\bf A534} (2004) 7--14,
  [\href{http://xxx.lanl.gov/abs/hep-ex/0402029}{{\tt hep-ex/0402029}}].

\bibitem{Buckley:2015gua}
A.~Buckley and C.~Pollard, {\em Eur. Phys. J.} {\bf C76} (2016), no.~2 71,
  [\href{http://xxx.lanl.gov/abs/1507.00508}{{\tt 1507.00508}}].

\bibitem{kdtree}
``{k-d tree}.''
\newblock http://en.wikipedia.org/wiki/Kd-tree.

\bibitem{Brooijmans:2012yi}
G.~Brooijmans {\em et.~al.}, in {\em {Proceedings, 7th Les Houches Workshop on
  Physics at TeV Colliders}}, pp.~221--463, 2012.
\newblock \href{http://xxx.lanl.gov/abs/1203.1488}{{\tt 1203.1488}}.

\bibitem{Kraml:2012sg}
S.~Kraml {\em et.~al.}, {\em Eur. Phys. J.} {\bf C72} (2012) 1976,
  [\href{http://xxx.lanl.gov/abs/1203.2489}{{\tt 1203.2489}}].

\bibitem{Chatrchyan:2012uea}
S.~Chatrchyan {\em et.~al.},, {\bf CMS} Collaboration, {\em Phys. Rev. Lett.}
  {\bf 111} (2013) 081802, [\href{http://xxx.lanl.gov/abs/1212.6961}{{\tt
  1212.6961}}].

\bibitem{RazorTwiki}
{ Maurizio Pierini, Javier Duarte},, ``Reproducing the razor limits in your
  susy study.''
\newblock [Online; accessed May 2 2016].

\bibitem{Buckley:2010ar}
A.~Buckley, J.~Butterworth, L.~Lonnblad, D.~Grellscheid, H.~Hoeth, J.~Monk,
  H.~Schulz, and F.~Siegert, {\em Comput. Phys. Commun.} {\bf 184} (2013)
  2803--2819, [\href{http://xxx.lanl.gov/abs/1003.0694}{{\tt 1003.0694}}].

\bibitem{Boos:2001cv}
E.~Boos {\em et.~al.}, in {\em {Physics at TeV colliders. Proceedings, Euro
  Summer School, Les Houches, France, May 21-June 1, 2001}}, 2001.
\newblock \href{http://xxx.lanl.gov/abs/hep-ph/0109068}{{\tt hep-ph/0109068}}.

\bibitem{Alwall:2006yp}
J.~Alwall {\em et.~al.}, {\em Comput. Phys. Commun.} {\bf 176} (2007) 300--304,
  [\href{http://xxx.lanl.gov/abs/hep-ph/0609017}{{\tt hep-ph/0609017}}].

\bibitem{Skands:2003cj}
P.~Z. Skands {\em et.~al.}, {\em JHEP} {\bf 07} (2004) 036,
  [\href{http://xxx.lanl.gov/abs/hep-ph/0311123}{{\tt hep-ph/0311123}}].

\bibitem{Buckley:2010jn}
A.~Buckley and M.~Whalley, {\em PoS} {\bf ACAT2010} (2010) 067,
  [\href{http://xxx.lanl.gov/abs/1006.0517}{{\tt 1006.0517}}].

\bibitem{Walker:2012vf}
J.~W. Walker, \href{http://xxx.lanl.gov/abs/1207.3383}{{\tt 1207.3383}}.

\bibitem{Martin:2015hra}
T.~A.~W. Martin, \href{http://xxx.lanl.gov/abs/1503.03073}{{\tt 1503.03073}}.

\bibitem{Alonso:2013hga}
R.~Alonso, E.~E. Jenkins, A.~V. Manohar, and M.~Trott, {\em JHEP} {\bf 1404}
  (2014) 159, [\href{http://xxx.lanl.gov/abs/1312.2014}{{\tt 1312.2014}}].

\bibitem{Masso:2014xra}
E.~Masso, {\em JHEP} {\bf 1410} (2014) 128,
  [\href{http://xxx.lanl.gov/abs/1406.6376}{{\tt 1406.6376}}].

\bibitem{Pomarol:2014dya}
A.~Pomarol, in {\em {2014 European School of High-Energy Physics (ESHEP 2014)
  Garderen, The Netherlands, June 18-July 1, 2014}}, 2014.
\newblock \href{http://xxx.lanl.gov/abs/1412.4410}{{\tt 1412.4410}}.

\bibitem{Falkowski:2015wza}
A.~Falkowski, B.~Fuks, K.~Mawatari, K.~Mimasu, F.~Riva, and V.~sanz, {\em Eur.
  Phys. J.} {\bf C75} (2015), no.~12 583,
  [\href{http://xxx.lanl.gov/abs/1508.05895}{{\tt 1508.05895}}].

\bibitem{Contino:2014aaa}
R.~Contino, M.~Ghezzi, C.~Grojean, M.~Mühlleitner, and M.~Spira, {\em Comput.
  Phys. Commun.} {\bf 185} (2014) 3412--3423,
  [\href{http://xxx.lanl.gov/abs/1403.3381}{{\tt 1403.3381}}].

\bibitem{Pomarol:2013zra}
A.~Pomarol and F.~Riva, {\em JHEP} {\bf 1401} (2014) 151,
  [\href{http://xxx.lanl.gov/abs/1308.2803}{{\tt 1308.2803}}].

\bibitem{Falkowski:2015krw}
A.~Falkowski and K.~Mimouni, {\em JHEP} {\bf 02} (2016) 086,
  [\href{http://xxx.lanl.gov/abs/1511.07434}{{\tt 1511.07434}}].

\bibitem{Efrati:2015eaa}
A.~Efrati, A.~Falkowski, and Y.~Soreq, {\em JHEP} {\bf 07} (2015) 018,
  [\href{http://xxx.lanl.gov/abs/1503.07872}{{\tt 1503.07872}}].

\bibitem{Han:2004az}
Z.~Han and W.~Skiba, {\em Phys.Rev.} {\bf D71} (2005) 075009,
  [\href{http://xxx.lanl.gov/abs/hep-ph/0412166}{{\tt hep-ph/0412166}}].

\bibitem{Ellis:2014jta}
J.~Ellis, V.~Sanz, and T.~You, {\em JHEP} {\bf 03} (2015) 157,
  [\href{http://xxx.lanl.gov/abs/1410.7703}{{\tt 1410.7703}}].

\bibitem{Ellis:2014dva}
J.~Ellis, V.~Sanz, and T.~You, {\em JHEP} {\bf 1407} (2014) 036,
  [\href{http://xxx.lanl.gov/abs/1404.3667}{{\tt 1404.3667}}].

\bibitem{Buckley:2015lku}
A.~Buckley, C.~Englert, J.~Ferrando, D.~J. Miller, L.~Moore, M.~Russell, and
  C.~D. White, \href{http://xxx.lanl.gov/abs/1512.03360}{{\tt 1512.03360}}.

\bibitem{Ellis:2015sca}
J.~Ellis and T.~You, \href{http://xxx.lanl.gov/abs/1510.04561}{{\tt
  1510.04561}}.

\bibitem{Ge:2016zro}
S.-F. Ge, H.-J. He, and R.-Q. Xiao,
  \href{http://xxx.lanl.gov/abs/1603.03385}{{\tt 1603.03385}}.

\bibitem{Grojean:2013kd}
C.~Grojean, E.~E. Jenkins, A.~V. Manohar, and M.~Trott, {\em JHEP} {\bf 04}
  (2013) 016, [\href{http://xxx.lanl.gov/abs/1301.2588}{{\tt 1301.2588}}].

\bibitem{Elias-Miro:2013eta}
J.~Elias-Miró, C.~Grojean, R.~S. Gupta, and D.~Marzocca, {\em JHEP} {\bf 05}
  (2014) 019, [\href{http://xxx.lanl.gov/abs/1312.2928}{{\tt 1312.2928}}].

\bibitem{Jenkins:2013zja}
E.~E. Jenkins, A.~V. Manohar, and M.~Trott, {\em JHEP} {\bf 1310} (2013) 087,
  [\href{http://xxx.lanl.gov/abs/1308.2627}{{\tt 1308.2627}}].

\bibitem{Jenkins:2013wua}
E.~E. Jenkins, A.~V. Manohar, and M.~Trott, {\em JHEP} {\bf 1401} (2014) 035,
  [\href{http://xxx.lanl.gov/abs/1310.4838}{{\tt 1310.4838}}].

\bibitem{Wells:2015cre}
J.~D. Wells and Z.~Zhang, \href{http://xxx.lanl.gov/abs/1512.03056}{{\tt
  1512.03056}}.

\bibitem{Elias-Miro:2013gya}
J.~Elias-Mir\'o, J.~Espinosa, E.~Masso, and A.~Pomarol, {\em JHEP} {\bf 1308}
  (2013) 033, [\href{http://xxx.lanl.gov/abs/1302.5661}{{\tt 1302.5661}}].

\bibitem{Elias-Miro:2013mua}
J.~Elias-Miro, J.~Espinosa, E.~Masso, and A.~Pomarol, {\em JHEP} {\bf 1311}
  (2013) 066, [\href{http://xxx.lanl.gov/abs/1308.1879}{{\tt 1308.1879}}].

\bibitem{Falkowski:2014tna}
A.~Falkowski and F.~Riva, \href{http://xxx.lanl.gov/abs/1411.0669}{{\tt
  1411.0669}}.

\bibitem{Falkowski:2015jaa}
A.~Falkowski, M.~Gonzalez-Alonso, A.~Greljo, and D.~Marzocca, {\em Phys. Rev.
  Lett.} {\bf 116} (2016), no.~1 011801,
  [\href{http://xxx.lanl.gov/abs/1508.00581}{{\tt 1508.00581}}].

\bibitem{lilith}
J.~Bernon and B.~Dumont, {\em Eur. Phys. J.} {\bf C75} (2015), no.~9 440,
  [\href{http://xxx.lanl.gov/abs/1502.04138}{{\tt 1502.04138}}].

\bibitem{Schael:2013ita}
S.~Schael {\em et.~al.},, {\bf ALEPH, DELPHI, L3, OPAL, LEP Electroweak}
  Collaboration, {\em Phys.Rept.} {\bf 532} (2013) 119--244,
  [\href{http://xxx.lanl.gov/abs/1302.3415}{{\tt 1302.3415}}].

\bibitem{Anthony:2005pm}
P.~L. Anthony {\em et.~al.},, {\bf SLAC E158} Collaboration, {\em Phys. Rev.
  Lett.} {\bf 95} (2005) 081601,
  [\href{http://xxx.lanl.gov/abs/hep-ex/0504049}{{\tt hep-ex/0504049}}].

\bibitem{Czarnecki:1995fw}
A.~Czarnecki and W.~J. Marciano, {\em Phys. Rev.} {\bf D53} (1996) 1066--1072,
  [\href{http://xxx.lanl.gov/abs/hep-ph/9507420}{{\tt hep-ph/9507420}}].

\bibitem{Bouchiat:1957zz}
C.~Bouchiat and L.~Michel, {\em Phys. Rev.} {\bf 106} (1957) 170--172.

\bibitem{Gonzalez-Alonso:2014bga}
M.~González-Alonso, {\em Nucl. Part. Phys. Proc.} {\bf 260} (2015) 3--11,
  [\href{http://xxx.lanl.gov/abs/1411.4529}{{\tt 1411.4529}}].

\bibitem{martinthesis}
M.~Gonzalez-Alonso, {\em Phd Thesis} (2010).

\bibitem{Danneberg:2005xv}
N.~Danneberg {\em et.~al.}, {\em Phys. Rev. Lett.} {\bf 94} (2005) 021802.

\bibitem{Aad:2014yja}
{\bf ATLAS} Collaboration, {\em Phys. Rev. Lett.} {\bf 114} (2015), no.~8
  081802, [\href{http://xxx.lanl.gov/abs/1406.5053}{{\tt 1406.5053}}].

\bibitem{Azatov:2015oxa}
A.~Azatov, R.~Contino, G.~Panico, and M.~Son,
  \href{http://xxx.lanl.gov/abs/1502.00539}{{\tt 1502.00539}}.

\bibitem{Dawson:2015oha}
S.~Dawson, A.~Ismail, and I.~Low, {\em Phys. Rev.} {\bf D91} (2015) 115008,
  [\href{http://xxx.lanl.gov/abs/1504.05596}{{\tt 1504.05596}}].

\bibitem{Agostini:2016vze}
A.~Agostini, G.~Degrassi, R.~Gröber, and P.~Slavich,
  \href{http://xxx.lanl.gov/abs/1601.03671}{{\tt 1601.03671}}.

\bibitem{YR4}
S.~Dawnson {\em et.~al.},, ``{LHC Higgs Cross Section Working Group HH
  Cross-group (Higgs Pair Production)}.''.

\bibitem{CarvalhoAntunesDeOliveira:2130724}
A.~Carvalho Antunes De~Oliveira, M.~Dall'Osso, M.~Gouzevitch, F.~Goertz,
  M.~Tosi, P.~De~Castro~Manzano, and T.~Dorigo,
  \href{http://xxx.lanl.gov/abs/LHCHXSWG-INT-2016-001}{{\tt
  LHCHXSWG-INT-2016-001}}.

\bibitem{ALEPH:2005ab}
S.~Schael {\em et.~al.},, {\bf ALEPH, DELPHI, L3 , OPAL, SLD, LEP Electroweak
  Working Group, SLD Electroweak Group, SLD Heavy Flavour Group} Collaboration,
  {\em Phys.Rept.} {\bf 427} (2006) 257--454,
  [\href{http://xxx.lanl.gov/abs/hep-ex/0509008}{{\tt hep-ex/0509008}}].

\bibitem{Abe:2000uc}
K.~Abe {\em et.~al.},, {\bf SLD} Collaboration, {\em Phys. Rev. Lett.} {\bf 85}
  (2000) 5059--5063, [\href{http://xxx.lanl.gov/abs/hep-ex/0006019}{{\tt
  hep-ex/0006019}}].

\bibitem{Beringer:1900zz}
J.~Beringer {\em et.~al.},, {\bf Particle Data Group} Collaboration, {\em
  Phys.Rev.} {\bf D86} (2012) 010001.

\bibitem{Abazov:2011ws}
V.~M. Abazov {\em et.~al.},, {\bf D0} Collaboration, {\em Phys. Rev.} {\bf D84}
  (2011) 012007, [\href{http://xxx.lanl.gov/abs/1104.4590}{{\tt 1104.4590}}].

\bibitem{Chatrchyan:2014mua}
{\bf CMS} Collaboration, {\em Phys. Rev. Lett.} {\bf 112} (2014) 191802,
  [\href{http://xxx.lanl.gov/abs/1402.0923}{{\tt 1402.0923}}].

\bibitem{Group:2012gb}
{Tevatron Electroweak Working Group},, {\bf CDF, D0} Collaboration,
  \href{http://xxx.lanl.gov/abs/1204.0042}{{\tt 1204.0042}}.

\bibitem{Khachatryan:2014iya}
{\bf CMS} Collaboration, {\em JHEP} {\bf 1406} (2014) 090,
  [\href{http://xxx.lanl.gov/abs/1403.7366}{{\tt 1403.7366}}].

\bibitem{Erler:2013xha}
J.~Erler and S.~Su, {\em Prog. Part. Nucl. Phys.} {\bf 71} (2013) 119--149,
  [\href{http://xxx.lanl.gov/abs/1303.5522}{{\tt 1303.5522}}].

\bibitem{Vilain:1994qy}
P.~Vilain {\em et.~al.},, {\bf CHARM-II} Collaboration, {\em Phys. Lett.} {\bf
  B335} (1994) 246--252.

\bibitem{Dorenbosch:1988is}
J.~Dorenbosch {\em et.~al.},, {\bf CHARM} Collaboration, {\em Z. Phys.} {\bf
  C41} (1989) 567. [Erratum: Z. Phys.C51,142(1991)].

\bibitem{Ahrens:1990fp}
L.~A. Ahrens {\em et.~al.}, {\em Phys. Rev.} {\bf D41} (1990) 3297--3316.

\bibitem{Ball:2012cx}
R.~D. Ball {\em et.~al.}, {\em Nucl. Phys.} {\bf B867} (2013) 244--289,
  [\href{http://xxx.lanl.gov/abs/1207.1303}{{\tt 1207.1303}}].

\bibitem{Gori:2013mia}
S.~Gori and I.~Low, {\em JHEP} {\bf 09} (2013) 151,
  [\href{http://xxx.lanl.gov/abs/1307.0496}{{\tt 1307.0496}}].

\end{thebibliography}\endgroup

\end{document}